\documentclass[12pt]{report}
\usepackage{lingmacros}
\usepackage{tree-dvips}
\usepackage{graphicx}
\usepackage{float}
\usepackage{amssymb}
\usepackage{url}
\usepackage{cancel}
\usepackage{morefloats}
\usepackage{harpoon}
\usepackage{caption}
\usepackage{subcaption}
\usepackage{amsmath}
\usepackage{multirow}
\usepackage{enumitem}
\usepackage{cite}
\usepackage[colorlinks=true, linkcolor=black, citecolor=blue]{hyperref}
\hypersetup{linkcolor=black,citecolor=black,filecolor=black,urlcolor=black}
\usepackage{newtxtext,newtxmath} 
\usepackage[protrusion=true,expansion=true]{microtype}
\usepackage{booktabs}
\usepackage[margin=1in]{geometry}

\usepackage{lineno}
\usepackage{amsmath}
\usepackage{bm}
\usepackage{tikz}
\usepackage{titlesec}
\titleformat{\paragraph}
{\normalfont\normalsize\bfseries}{\theparagraph}{1em}{}
\titlespacing*{\paragraph}
{0pt}{3.25ex plus 1ex minus .2ex}{1.5ex plus .2ex}

\newcommand{\Nch}{N_\text{ch}}

\newcommand{\MpT}{\langle [p_T] \rangle}
\newcommand{\SumET}{\Sigma E_{\mathrm{T}}}

\newcommand{\pTbar}{\bar{p}_{\mathrm{T}}}
\newcommand{\pTref}{p^{\mathrm{ref}}_{\mathrm{T}}}
\newcommand{\sqn}{\sqrt{s_\text{NN}}}
\newcommand{\zvtx}{z_{vtx}}
\newcommand{\Deta}{\mbox{$\Delta \eta$}}
\newcommand{\NchR}{N_\text{ch}^\text{rec}}
\newcommand{\pT}{p_\text{T}}

\newcommand{\sumET}{\Sigma E_\text{T}}

\newcommand{\dzero}{d_\text{0}}
\newcommand{\zzsth}{z_\text{0}\sin(\theta)}

\newcommand{\lr}[1]{\left\langle #1\right\rangle}
\newcommand{\llrr}[1]{\langle\!\langle #1 \rangle\!\rangle}
\newcommand{\llangle}{\langle\!\langle}
\newcommand{\rrangle}{\rangle\!\rangle}

\newcommand{\bq} {\textbf{q}}

\newcommand{\GEANT}{\text{Geant4}}

\newcommand{\cov}[1] {\mathrm{cov}(v_{ #1 }^2,[\pT])}
\newcommand{\var}[1] {\mathrm{var}(v_{ #1 }^2)}
\newcommand{\vOpTF}{v^{\prime}_0(\pT)}
\newcommand{\nCov}{\left\langle \delta n(\pT) \delta [\pT] \right\rangle}
\newcommand{\Cov}{\left\langle \delta N(\pT) \delta [\pT] \right\rangle}
\begin{document}
\pagenumbering{gobble}
\title{\bf{Novel Probes of Quark-Gluon Plasma Evolution in Heavy-Ion Collisions With the ATLAS Detector:\\ {
\textit{From Structure of Colliding Nuclei to Collective Expansion of Medium}}}}
\vspace*{3\baselineskip}
\centerline{\bf{Novel Probes of Quark-Gluon Plasma Evolution in Heavy-Ion Collisions With the ATLAS Detector:}}
\centerline{\bf{\textit{From Structure of Colliding Nuclei to Collective Expansion of Medium}}}
\vspace*{1\baselineskip}
\centerline{A Dissertation presented}
\vspace*{1\baselineskip}
\centerline{by} 
\vspace*{1\baselineskip}
\centerline{\bf{Somadutta Bhatta}}
\vspace*{1\baselineskip}
\centerline{to} 
\vspace*{1\baselineskip}
\centerline{The Graduate School}
\vspace*{1\baselineskip}
\centerline{in Partial Fulfillment of the}
\vspace*{1\baselineskip}
\centerline{Requirements}
\vspace*{1\baselineskip}
\centerline{for the Degree of}
\vspace*{1\baselineskip}
\centerline{\bf{Doctor of Philosophy}}
\vspace*{1\baselineskip}
\centerline{in}
\vspace*{1\baselineskip}
\centerline{\bf{Chemistry}}
\vspace*{2\baselineskip}
\centerline{Stony Brook University}
\vspace*{2\baselineskip}
\centerline{\bf{August 2025}}

\newpage
\pagenumbering{gobble}

\newpage
\pagenumbering{roman}
\setcounter{page}{2}

\centerline{\bf{Stony Brook University}}
\vspace*{1\baselineskip}
\centerline{The Graduate School}
\vspace*{2\baselineskip}
\centerline{\bf{Somadutta Bhatta}}
\vspace*{2\baselineskip}
\centerline{We, the dissertation committee for the above candidate for the}
\vspace*{1\baselineskip}
\centerline{Doctor of Philosophy degree, hereby recommend}
\vspace*{1\baselineskip}
\centerline{acceptance of this dissertation}
\vspace*{2\baselineskip}
\centerline{\bf{Jiangyong Jia - Dissertation Advisor}}
\centerline{\bf{Professor, Department of Chemistry}}
\vspace*{1\baselineskip}
\centerline{\bf{Roy Lacey - Chairperson of Defense}}
\centerline{\bf{Professor, Department of Chemistry}}
\vspace*{1\baselineskip}
\centerline{\bf{Christian Aponte-Rivera - Third Member}}
\centerline{\bf{Assistant Professor, Department of Chemistry}} 
\vspace*{1\baselineskip}
\centerline{\bf{Derek Teaney}}
\centerline{\bf{Professor, Department of Physics and Astronomy}}
\vspace*{1\baselineskip}
\centerline{\bf{Dmitry Tsybychev}}
\centerline{\bf{Professor, Department of Physics and Astronomy}}
\vspace*{2\baselineskip}
\centerline{This dissertation is accepted by the Graduate School}
\vspace*{3\baselineskip}
\centerline{\bf{Celia Marshik}}
\centerline{\bf{Dean of the Graduate School}}

\newpage

\centerline{Abstract of the Dissertation}
\vspace*{1\baselineskip}
\centerline{\bf{Novel Probes of Quark-Gluon Plasma Evolution in Heavy-Ion Collisions With the ATLAS Detector:}}
\centerline{\bf{\textit{From Structure of Colliding Nuclei to Collective Expansion of Medium}}}
\vspace*{1\baselineskip}
\centerline{by}
\vspace*{1\baselineskip}
\centerline{\bf{Somadutta Bhatta}}
\vspace*{1\baselineskip}
\centerline{\bf{Doctor of Philosophy}}
\vspace*{1\baselineskip}
\centerline{in}
\vspace*{1\baselineskip}
\centerline{\bf{Chemistry}}
\vspace*{1\baselineskip}
\centerline{Stony Brook University}
\vspace*{1\baselineskip}
\centerline{\bf{2025}}
\vspace*{2\baselineskip}

Understanding the properties of the quark-gluon plasma (QGP) offers insights into the strong interaction and the conditions of the early universe. This deconfined state of quarks and gluons is created by colliding heavy ions at relativistic speeds, generating regions of extreme temperature and density.

Since the QGP cannot be observed directly, its properties must be inferred from the particles emitted as it cools. This dissertation introduces a suite of novel probes based on event-by-event multi-particle correlations allowing us to work backward from final-state particles to constrain the collective evolution, transport properties, and initial conditions of the QGP. The studies utilize Lead-Lead (\(^{208}\mathrm{Pb}+{}^{208}\mathrm{Pb}\)) and Xenon-Xenon (\(^{129}\mathrm{Xe}+{}^{129}\mathrm{Xe}\)) collision data recorded by the ATLAS detector at the Large Hadron Collider.

First, evidence for the collective nature of the QGP's radial expansion is provided using a transverse momentum ($\pT$)-differential observable, $v_0(\pT)$. This observable exhibits genuine long-range correlations and factorizes into a single-particle property. Its strong sensitivity to bulk viscosity providing more direct constraints on this medium property than traditional measures. Second, the sources of event-wise fluctuations in radial flow within the initial state are disentangled by analyzing higher moments of the event-wise mean-$\pT$ distribution. The average of this distribution is shown to constrain the speed of sound in the QGP. Finally, deformations in the colliding nuclei, which control the overlap area geometry, are constrained by comparing correlations between the anisotropic flow, $v_{n}$ and mean-$\pT$ in spherical Pb+Pb and deformed Xe+Xe collisions. This comparison provides the first experimental evidence for triaxial deformation in $^{129}$Xe.

Together, these three complementary analyses significantly improve our understanding of the initial conditions, transport properties, and collective behavior of the QGP. They establish new avenues for precision studies of QGP, the primordial matter that existed microseconds after the Big Bang.

\newpage

\centerline{\bf{Dedication}}
\vspace*{4\baselineskip}
\begin{center}

This work is dedicated to \\
 \vspace*{1\baselineskip}
My parents,\\
    \textit{Bou}, Mrs. Parbati Panda,
    \textit{Nana}, Mr. Dayanidhi Bhatta.\\
    \vspace*{1\baselineskip}
    and,\\
     \vspace*{1\baselineskip}
    My Advisor, \\
    Prof. Jiangyong Jia
\end{center}

\newpage
\tableofcontents

\newpage
\listoffigures
\addtocontents{toc}{\vspace{-3em}}
\addcontentsline{toc}{chapter}{List of Figures}

\newpage
\listoftables
\addtocontents{toc}{\vspace{-0.9em}}
\addcontentsline{toc}{chapter}{List of Tables}

\chapter*{List of Abbreviations}
\addtocontents{toc}{\vspace{-0.9em}}
\addcontentsline{toc}{chapter}{List of Abbreviations}
\vspace*{2\baselineskip}
\begin{table}[H]
\large
\begin{tabular}{ll}
ALICE & A Large Ion Collider Experiment \\
ATLAS & A Toroidal LHC Apparatus \\
BES & Beam Energy Scan \\
CGC & Color Glass Condensate \\
CMS & Compact Muon Solenoid \\
DIS & Deep Inelastic Scattering \\
EM & Electromagnetic \\
EMCal & Electromagnetic Calorimeter \\
EoS & Equation of State \\
FCal & Foward Calorimeter \\
GeV & Giga-electronVolt \\
HIC & Heavy-Ion Collisions\\
ID & Inner Detector \\
LHC & Large Hadron Collider \\
MB & Minimum Bias \\
MeV & Mega-electronVolt \\
MinBias & Minimum Bias \\
MC & Monte Carlo \\
NBD & Negative Binomial Distribution \\
QCD & Quantum Chromodynamics \\
QED & Quantum Electrodynamics \\
QGP & Quark Gluon Plasma \\
RHIC & Relativistic Heavy Ion Collider \\
ZDC & Zero Degree Calorimeter \\
 &
\end{tabular}
\end{table}
\newpage

\chapter*{Acknowledgements}
\addtocontents{toc}{\vspace{-0.9em}}
\addcontentsline{toc}{chapter}{Acknowledgements}
First and foremost, I owe everything I have accomplished to my {\it \bf{parents}}. Thank you for living by the motto ``Work is worship'', for showing me the true meaning of ``simple living, high thinking'', and for always prioritizing my education above all else. Your unwavering love, daily sacrifices, and steadfast belief in me have shaped the person I am today, and I will be forever grateful for the example you have set.

My sincerest thanks go to my advisor, {\it \bf{Prof. Jiangyong Jia}}. He ignited in me a profound passion for research and taught me the importance of leading a balanced life. His incessant support and relentless push not only fueled my interest in science but also provided me with everything necessary for this Ph.D., from sustenance to resources. He has been an exemplary figure, showing me how to navigate everyday life with excitement and an unstoppable interest in research and the joy it brings. I will forever be grateful for his time, his belief in me, and his steadfast support, which will continue to guide me as a researcher.

I am also grateful to {\it \bf{Prof. Roy Lacey}} for all the discussions and sharing his constant love for research. He taught me the true meaning of being deeply interested in one's own work and instilled in me the importance of self-belief and faith in one's own understanding and abilities.

My introduction to the captivating field of heavy ion collision physics was thanks to {\it \bf{Prof. Bedangadas Mohanty}}. He patiently guided me through the initial learning curves during and after my master's degree at NISER. I am deeply grateful for his belief and support from the moment I started doing research until now, and for always being there whenever I reached out for help.

This thesis would not have been possible without the constant belief and tremendous support of {\it \bf{all my teachers and professors, including those not explicitly listed above}}. Their collective wisdom and encouragement have been invaluable.

I extend my heartfelt thanks to {\it \bf{Dr. Prithwish Tribedy}} for our insightful discussions on approaches to science and life as a researcher, both during and after my Ph.D. I will always treasure our Friday afternoon conversations on diverse physics topics and the futuristic aspects of various subfields. Moreover, I am deeply grateful for his moral and mental support during times of need.

I am also thankful to {\it \bf{Prof. Jean-Yves Ollitrault}}, {\it \bf{Prof. Derek Teaney}}, {\it \bf{Prof. Wilke Van der Schee}}, {\it \bf{Dr. Govert Nijs}}, {\it \bf{Dr. Giuliano Giacalone}}, and {\it \bf{Dr. Rupam Samanta}} for their invaluable inputs and enriching discussions, which form an integral part of the material presented in this thesis.

To my colleagues from the ATLAS and STAR collaborations, with whom I had the privilege to work, discuss, and refine my ideas through their valuable inputs: thank you. A special mention goes to {\it \bf{Prof. Tomasz Bold}}, {\it \bf{Dr. Arabinda Behera}}, {\it \bf{Prof. Chunjian Zhang}}, {\it \bf{Dr. Shengli Huang}}, {\it \bf{Dr. Soumya Mohapatra}}, {\it \bf{Dr. Mingliang Zhou}}, {\it \bf{Dr. Niseem Magdy}}, {\it \bf{Aman Dimri}}, {\it \bf{Zhengxi Yan}} and {\it \bf{Souvik Paul}} for their contributions.

Finally, to my dear sister, {\it \bf{Gayatri Bhatta}}, and my beloved wife, {\it \bf{Ankita Swain}}, thank you for showering me with endless love, care, and unwavering support throughout the entire duration of my Ph.D. Your presence made every challenge lighter.

I acknowledge the unwavering perseverance and self-belief that have guided me through every challenge, culminating in the completion of this work.

\newpage
\chapter*{Publications}
\addtocontents{toc}{\vspace{-0.9em}}
\addcontentsline{toc}{chapter}{Publications}

\vspace*{2\baselineskip}
\begin{enumerate}
\item {\bf S. Bhatta} for the ATLAS Collaboration.
``Evidence for the collective nature of radial flow in Pb+Pb collisions with the ATLAS detector'', arXiv:2503.24125, submitted to Phys. Rev. Lett. (2025).
\item {\bf S. Bhatta} for the ATLAS Collaboration.
``Disentangling sources of momentum fluctuations in Xe+Xe and Pb+Pb collisions with the ATLAS detector'', Phys. Rev. Lett. \textbf{133}, 252301 (2024).
\item {\bf S. Bhatta} for the ATLAS Collaboration.
``Correlations between flow and transverse momentum in Xe+Xe and Pb+Pb collisions at the LHC with the ATLAS detector'', Phys. Rev. C {\bf 107}, 054910 (2023).
\item {\bf S. Bhatta}, A. Dimri, J. Jia.
``Disentangling the global multiplicity and spectral shape fluctuations in radial flow'', arXiv:2504.20008 [nucl-th] (2025).
\item {\bf S. Bhatta}, V. Bairathi.
``Experimental method to constrain preferential emission and spectator dynamics in heavy-ion collisions'', arXiv:2407.06977 [nucl-th], submitted to Phys. Rev. C (2024).
\item {\bf S. Bhatta}, C Zhang, J. Jia.
``Energy dependence of heavy-ion initial condition in isobar collisions'', Phys. Lett. B {\bf 858}, 139034 (2024).
\item {\bf S. Bhatta}, C Zhang, J. Jia.
``Higher-order transverse momentum fluctuations in heavy-ion collisions'', Phys. Rev. C {\bf 105}, 024904 (2022).
\item {\bf S. Bhatta}, V. Bairathi.
``An improved method to access initial states in relativistic heavy-ion collisions'', Eur. Phys. Jour. C {\bf 82}, 855 (2022).
\item R. Samanta, {\bf S. Bhatta}, J. Jia, M. Luzum, J-Y Ollitrault.
``Thermalization at the femtoscale seen in high-energy Pb+Pb collisions'', Phys. Rev. C {\bf 109}, L051902 (2024).
\item A. Dimri, {\bf S. Bhatta}, J. Jia.
``Impact of nuclear shape fluctuations in high-energy heavy ion collisions'', Eur. Phys. Jour. A {\bf 59}, 45 (2023).
\item C Zhang, {\bf S. Bhatta}, J. Jia.
``Ratios of collective flow observables in high-energy isobar collisions are insensitive to final state interactions'', Phys. Rev. C {\bf 106}, L031901 (2022).
\item J. Jia, S. Huang, C. Zhang, {\bf S. Bhatta}, ''Sources of longitudinal flow decorrelations in high-energy nuclear collisions'', arXiv:2408.15006 [nucl-th].
\item C. Zhang, A. Behera, {\bf S. Bhatta}, J. Jia, ``Non-flow effects in correlation between harmonic flow and transverse momentum in nuclear collisions'', Phys. Lett. B {\bf 822}, 136702 (2021).
\end{enumerate}

\chapter*{Proceedings and Conference Notes}
\addtocontents{toc}{\vspace{-0.9em}}
\addcontentsline{toc}{chapter}{Proceedings and Conference Notes}

\vspace*{2\baselineskip}
\begin{enumerate}
\item {\bf S. Bhatta} for the ATLAS Collaboration.
``Measurement of collective dynamics in small and large systems with the ATLAS detector'', PoS EPS-HEP2023 {\bf 202} (2024).
\item {\bf S. Bhatta} for the ATLAS Collaboration.
``Measurement of $[\pT]$ Fluctuations in Xe+Xe and Pb+Pb Collisions with ATLAS'', ATLAS-CONF-2023-061 (2023).
\item {\bf S. Bhatta}.
``Deciphering initial states of high energy heavy-ion collisions using spectators'', IJMPCS, (2022).
\item {\bf S. Bhatta} for the ATLAS Collaboration.
``Flow and transverse momentum correlation in Pb+Pb and Xe+Xe collisions with ATLAS: assessing the initial condition of the QGP'', APPB Proc. Suppl. (2022).
\item {\bf S. Bhatta} for the ATLAS Collaboration.
``Measurements of collective behavior in pp, Xe+Xe, and Pb+Pb collisions with the ATLAS detector'', PoS EPS-HEP2021 {\bf 305} (2021).
\item {\bf S. Bhatta} for the ATLAS Collaboration.
``Measurement of flow and transverse momentum correlations in Pb+Pb collisions at $\sqrt{s_{_{\mathrm{NN}}}}$ = 5.02 TeV and Xe+Xe collisions at $\sqrt{s_{_{\mathrm{NN}}}}$ = 5.44 TeV with the ATLAS detector'', ATLAS-CONF-2021-001 (2021).
\end{enumerate}

\newpage
\chapter*{Selected Conference and Seminar Talks}
\addtocontents{toc}{\vspace{-0.9em}}
\addcontentsline{toc}{chapter}{Selected Conference and Seminar Talks}

\vspace*{2\baselineskip}
\begin{enumerate}
\item {\it CERN-TH Seminar 2025}, ``Insights into Nuclear Geometry and Initial Conditions of Heavy-Ion Collisions from ATLAS'' (Invited Talk).
\item {\it HENPIC Seminar 2024}, ``Insights into Nuclear Geometry and Initial Conditions of Heavy-Ion Collisions from ATLAS'' (Invited Talk).
\item {\it SBU HEP Seminar 2024}, ``Leveraging Ultra-Central Collisions to Constrain Shape of the Nucleus and Initial State'' (Invited Talk).
\item {\it Nuclear Physics Seminars at BNL 2024}, ``Disentangling sources of momentum fluctuations in heavy-ion collisions with the ATLAS detector'' (Invited Talk).
\item {\it LHCP 2024}, ``Probing medium properties in ultra-central collisions (ATLAS+ALICE+CMS talk)'' (Invited Talk).
\item {\it INT-Program 2023} Intersection of nuclear structure and high‐energy nuclear collisions, ``Energy dependence of initial condition from isobar'' (Invited Talk).
\item {\it Nuclear Physics Seminars at BNL 2022}, ``Flow and transverse momentum correlations at LHC: a probe of the heavy-ion initial state and nuclear deformation'' (Invited Talk).
\item {\it Quark Matter 2025}, ``Evidence for the collective nature of radial flow in Pb+Pb collisions with the ATLAS detector''.
\item {\it ATHIC 2025}, ``Unveiling initial state fluctuations using $[\pT]$ cumulants with ATLAS''.
\item {\it EPS-HEP 2023}, ``Measurement of collective dynamics in small and large systems with the ATLAS detector''.
\item {\it Initial Stages 2023}, ``Probing initial state using higher order fluctuations: $v_{n}-[\pT]$ and $[\pT]$ correlations in ATLAS''.
\item {\it Quark Matter 2022}, ``Flow and transverse momentum correlation in Pb+Pb and Xe+Xe collisions with ATLAS: assessing the initial condition of the QGP''.
\item {\it EPS-HEP 2021}, ``Measurements of collective behavior in pp, Xe+Xe, and Pb+Pb collisions with the ATLAS detector''.
\item {\it ATLAS TDAQ week at CERN 2020}, ``MinBias and Forward Detector''.
\item {\it Fall Meeting for APS DNP-2022}, ``Probing nuclear deformation at LHC energies using AMPT''.
\item {\it APS April Meeting 2022}, ``Higher order transverse momentum fluctuations in heavy ion collisions''.
\item {\it Fall Meeting for APS DNP-2021}, ``Flow and transverse momentum correlations in Pb+Pb and Xe+Xe collisions with ATLAS''.
\item {\it APS April Meeting-2021}, ``Cumulant analysis of deformed systems using AMPT model''.
\item {\it Quark Matter 2023}, ``Exploring the origin of $[\pT]$ fluctuations in ultra-central heavy ion collisions: Higher order $[\pT]$ correlations in ATLAS'' (Conference Poster).
\item {\it SQM-2022}, ``Flow and transverse momentum correlations in Pb+Pb and Xe+Xe collisions with ATLAS: assessing the initial condition of the QGP'' (Conference Poster).
\item {\it Initial stages 2021}, ``Measurement of flow and transverse momentum correlations in Pb+Pb collisions at $\sqrt{s_{NN}} = $5.02 TeV and Xe+Xe collisions at $\sqrt{s_{NN}} = $5.44 TeV with the ATLAS detectors'' (Conference Poster).
\item {\it APS Fall Meeting 2020}, ``Flow cumulants for multi-particle azimuthal correlations in heavy-ion Collisions'' (Conference Poster).
\end{enumerate}

\newpage
\pagenumbering{arabic}
\chapter{Introduction}
\label{sec:intro}

The quest to decipher the fundamental constituents of the universe has been a driving force behind human curiosity. Antiquated concepts, such as the indivisible {\it anu} (atom/particle) by the Indian Sage Kanada and the atomic theory by Democritus represent early milestones in this pursuit. The 19$^{\mathrm{th}}$ century saw the rise of Dalton's atomic theory, succeeded by the discoveries of the electron (Thomson, 1897), the nucleus and proton (Rutherford, 1909, 1919), and the neutron (Chadwick, 1932).

By the mid-20$^{\mathrm{th}}$ century, cosmic-ray and accelerator studies revealed a ``particle zoo'' of hadrons, leading Gell-Mann and Zweig (1964) to postulate that hadrons are composed of fractionally charged quarks\cite{Gell-Mann:1964ewy, Zweig:1964ruk}. Subsequent deep-inelastic electron-proton scattering experiments at SLAC (1968–73) corroborated the existence of quarks as physical entities by demonstrating scaling behavior consistent with point-like partons \cite{Breidenbach:1969kd}. The progressive development of a cohesive model included the electroweak unification by Glashow, Weinberg, and Salam\cite{Glashow:1961tr, Weinberg:1967tq, Salam:1968rm}, and was further cemented by the experimental observation of weak bosons ($W^\pm$, $Z^0$) \cite{UA1:1983crd, UA2:1983tsx}, charm\cite{SLAC-SP-017:1974ind,E598:1974sol}, bottom \cite{E288:1977xhf}, and top \cite{CDF:1995wbb, D0:1995jca} quarks, culminating in the 2012 detection of the Higgs boson\cite{ATLAS:2012yve, CMS:2012qbp}. These developments culminated into the {\it Standard Model}, a theoretical construct accurately describing all known fundamental particles and their interactions, with the notable exception of gravity.

\section{Standard Model}

The Standard Model represents one of physics' greatest achievements, explaining the fundamental constituents of matter and their interactions through force-carrying particles. This theoretical framework has successfully unified electromagnetic, weak, and strong interactions while providing remarkably accurate predictions about the universe at its fundamental level.

\begin{figure}
    \centering
    \includegraphics[width=0.55\linewidth]{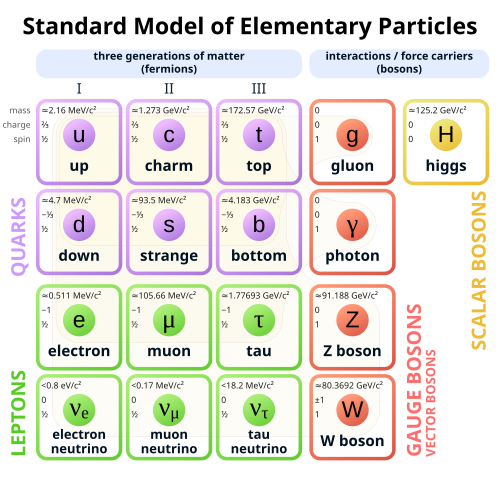}
    \caption{The fundamental particles of the standard model and the force mediating bosons~\cite{wiki:Quantumchromodynamics}.}
    \label{fig:stdmodel}
\end{figure}

Elementary particles in the Standard Model are organized into three generations that differ primarily in mass, as shown in Figure~\ref{fig:stdmodel}. Each generation includes two quarks and two leptons, resulting in six quark flavors: up (u), down (d), charm (c), strange (s), top (t), and bottom (b). The heaviest of them, the top quark (with mass $\sim$ 177 GeV) was discovered much later than the lighter bottom quark (with mass $ \sim$ 4.18 GeV) when sufficiently powerful accelerators became available. Leptons include the electron (e), muon ($\mu$), tau ($\tau$), and their corresponding neutrinos ($\nu_e, \nu_\mu, \nu_\tau$).

The four fundamental forces are: electromagnetic (mediated by photons, $\gamma$), weak (mediated by W and Z bosons, $W^+, W^-, Z^0$), strong (mediated by gluons, g), and gravitational (hypothesized to be mediated by gravitons). While photons and gluons are massless, W and Z bosons are massive. Force mediators have spin 1, except the hypothesized graviton with spin 2. All elementary particles interact via the weak force; charged particles additionally interact electromagnetically, and quarks interact through the strong force.

The Standard Model incorporates electromagnetic and weak forces into the Electroweak theory and describes the strong force using Quantum Chromodynamics (QCD). Each force has a characteristic coupling constant $\alpha$ that varies with the energy scale. At everyday energies, these constants differ from each other: strong interaction $\alpha_s \approx 1$, electromagnetic $\alpha_{em} \approx 10^{-3}$, weak $\alpha_w \approx 10^{-16}$, and gravitational $\alpha_g \approx 10^{-41}$. At higher energies, these coupling constants converge, {\it potentially} allowing a unified description (Grand Unified Theory) at energies around $10^{19}$ GeV~\cite{Georgi:1974sy}.

\section{Quantum Chromodynamics (QCD)}

The strong force binds quarks through gluon exchange. Unlike photons in electromagnetism, gluons carry color charge and can interact with themselves, creating a fundamentally different interaction dynamic. Quarks possess three color charges (red, green, blue), with antiquarks carrying corresponding anticolors~\cite{Griffiths:1987tj}.

Gluons are represented as combinations of eight independent SU(3) hermitian matrices with zero trace, corresponding to eight gluon types:
\begin{align*}
&\frac{r\bar{b} + b\bar{r}}{\sqrt{2}}, \quad \frac{r\bar{g} + g\bar{r}}{\sqrt{2}}, \quad \frac{g\bar{b} + b\bar{g}}{\sqrt{2}}, \quad \frac{r\bar{r} - b\bar{b}}{\sqrt{2}},\\
&-i\frac{r\bar{b} - b\bar{r}}{\sqrt{2}}, \quad -i\frac{r\bar{g} - g\bar{r}}{\sqrt{2}}, \quad -i\frac{b\bar{g} - g\bar{b}}{\sqrt{2}}, \quad \frac{r\bar{r} + b\bar{b} - 2g\bar{g}}{\sqrt{6}}.
\end{align*}

\begin{figure}[htbp]
    \centering
    \includegraphics[width=0.45\linewidth]{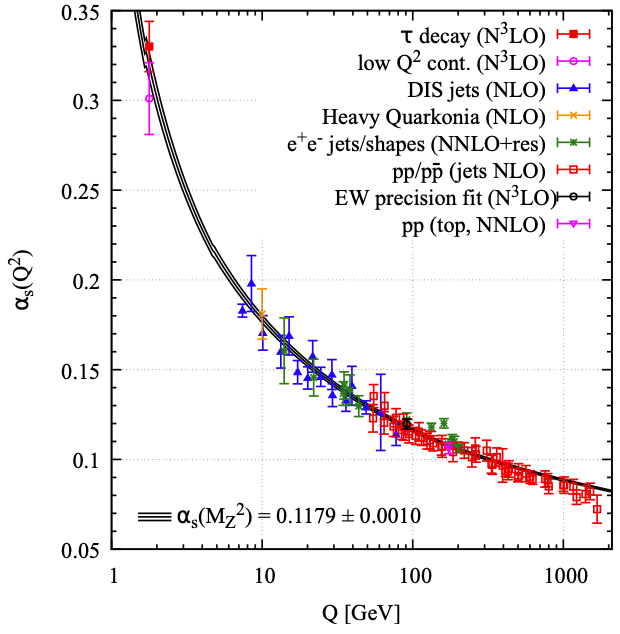}
    \caption{The experimentally measured values of the coupling constant $\alpha_{s}$ vs. energy scale, $Q^{2}$ compared to prediction of the asymptotic freedom in QCD~\cite{ParticleDataGroup:2020ssz}.}
    \label{fig:alphas}
\end{figure}

The strength of the strong interaction is characterized by its running coupling constant, $\alpha_{s}(Q^2)$, given by:
$$ \alpha_s = \frac{12\pi}{(11n - 2f)\ln(|Q^2|/\Lambda^2)}$$
where $Q^2$ denotes the momentum transfer, $n$ the number of colors, $f$ the number of flavors, and $\Lambda \approx 217\,\text{MeV}$ is fixed by experiment~\cite{ParticleDataGroup:2004fcd}. Figure~\ref{fig:alphas} illustrates this functional relationship: for high momentum transfer, $Q^2$ (corresponding to low $\alpha_s$), coupling strength decreases logarithmically.

This leads to QCD exhibiting two distinctive phenomena: {\it confinement} and {\it asymptotic freedom}. As quarks separate, the potential energy between them increases without limit, requiring infinite energy to separate them. This explains why quarks are never observed as isolated particles but always confined within color-neutral hadrons—a property called {\it confinement}. Conversely, when quarks are in close proximity, as within hadrons, they interact weakly, behaving almost like free particles, a phenomenon known as {\it asymptotic freedom}.

\begin{figure}[htbp]
    \centering
    \includegraphics[width=0.45\linewidth]{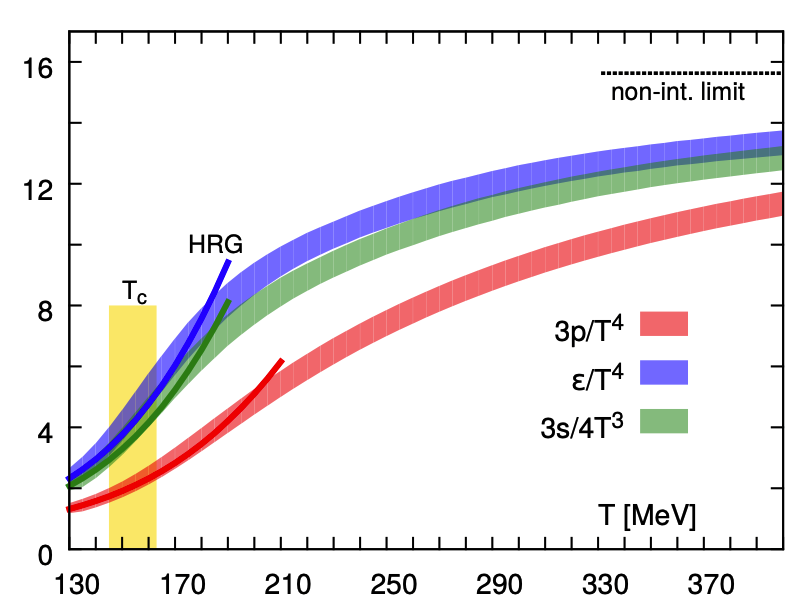}
    \caption{Normalized pressure, $3p/T^4$, energy density, $\epsilon/T^4$, and entropy density, $3s/4T^3$ as a function of the temperature, $T$. The dark lines show the prediction of the Hadron Resonance Gas model~\cite{HotQCD:2014kol}. The horizontal line at the top-right corresponds to ideal gas limit and the vertical band marks the crossover region with $T_c = (154\pm9)$ MeV. The approach of the lattice data to this limit at high temperatures supports the picture of deconfined quarks and gluons becoming the dominant degrees of freedom.}
    \label{fig:phasetr}
\end{figure}

Lattice QCD calculations predict that at sufficiently high energy densities, a phase transition can occur from hadronic matter to a system of deconfined quarks and gluons, revealing higher degrees of freedom, as shown in Figure~\ref{fig:phasetr}. This transition is of a \textit{crossover} nature; for details, see Ref.~\cite{HotQCD:2014kol}.

This deconfined state of matter, known as Quark-Gluon Plasma, bridges particle physics, nuclear physics, and cosmology. Cosmological models suggest that during its first microsecond, the early universe consisted of QGP that cooled and hadronized as it expanded. Lattice QCD results points towards a crossover nature of the transition at vanishing baryon chemical potential, $\mu_B$, provide constraints to big-bang nucleosynthesis and cosmic-background calculations.

\section{Recreating QGP in the Laboratory}

\begin{figure}[htbp]
\centering
\includegraphics[width=0.5\linewidth]{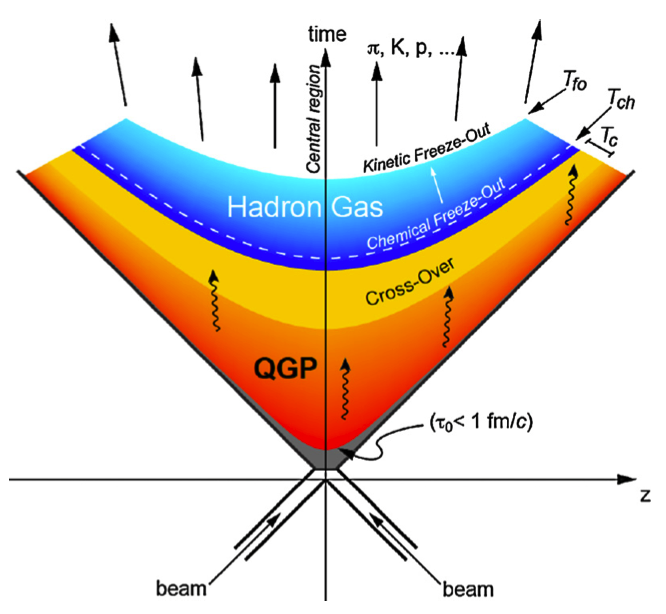}
\caption{Space-time evolution after heavy ion collision. The left scenario represents conditions where no QGP forms (T $< T_c$), while the right shows QGP formation (T $>T_c$)~\cite{Braun-Munzinger:2018hat}.}
\label{fig:xtevolQGP}
\end{figure}

The extreme energy densities necessary for the creation of the QGP are achieved in laboratory settings by colliding heavy nuclei, such as Gold ($^{197}$Au) or Lead ($^{208}$Pb), accelerated to relativistic speeds at major facilities like the Relativistic Heavy Ion Collider (RHIC) at Brookhaven National Laboratory and the Large Hadron Collider (LHC) at CERN. These collisions form a minuscule, transient fireball with temperatures briefly exceeding several times the critical temperature, $T_c$, the threshold for QGP formation. This experimentally produced QGP allows investigation into its evolution and fundamental properties by analyzing the characteristics of the particles produced during its subsequent expansion and cooling.

The space-time evolution following relativistic Heavy-Ion Collisions (HIC) proceeds through a series of distinct stages as illustrated in Figure~\ref{fig:xtevolQGP}. Initially, in sufficiently energetic collisions, a deconfined QGP state is formed. As this extremely hot and dense system expands and its temperature decreases, it undergoes a phase transition, into a state of confined hadronic matter when the temperature drops below $T_c$. The system then cools until it reaches the chemical freeze‑out temperature, $T_{ch}$, at which inelastic collisions between the particles cease and the relative abundances of all particle species become fixed. As the fireball expands further, it ultimately reaches the kinetic freeze‑out temperature, $T_{kin}$, when elastic collisions also end and the particles stream freely outwards. 

However, studying the QGP presents an experimental challenge. Its existence in the laboratory lasts only for an incredibly short time (on the order of $10^{-23}$ of a second), and it cannot be directly observed. Instead, its presence and properties must be inferred indirectly from experimental signatures encoded in the behavior of the particles that emerge from the collision. The strong interactions within the QGP medium imprint information about both the initial conditions and the properties of the medium onto the final distributions of these particles. By carefully analyzing observables in the final state, such as particle spectra, yields, and correlations, it is possible to reconstruct and infer key QGP characteristics, including temperature, energy density, and viscosities.

\section{Kinematics of Heavy-Ion Collisions}\label{subsec:kinematics}
Before discussing the properties of the QGP in detail, it is necessary to introduce the kinematic variables pertinent to its study.

When nuclei are accelerated to ultra-relativistic speeds in HIC, the principles of special relativity govern their kinematics, rendering traditional non-relativistic variables like velocity impractical. The theory of special relativity governs the kinematics at relativistic speeds. Within this theory, the velocity addition under Lorentz transformations is non-linear, complicating analyses involving different reference frames. This necessitates the use of specific kinematic variables, such as \textit{rapidity} and \textit{pseudorapidity}, which are inherently suited to the relativistic regime. 

\subsection{Lorentz Boost}
We start with understanding the Lorentz boost, which describes the relationship between frames moving at a constant relative velocity.

If $x^\mu = (t, x, y, z)$ denotes the spacetime coordinates (a four-vector) in one inertial frame S (using natural units where $c=1$), the coordinates $x'^\mu = (t', x', y', z')$ in a frame S' moving with a constant velocity $\beta$ along the $z$-axis relative to S are related by a {\it Lorentz transformation}. For a boost purely along the $z$-axis, the transformation mixes the time ($t$) and longitudinal spatial ($z$) coordinates, leaving the transverse coordinates ($x, y$) unchanged. The transformation matrix for $(t, z)$ is given by:
\begin{equation}
\begin{pmatrix}t'\\ z'\end{pmatrix} = \gamma \begin{pmatrix} 1 & -\beta\\ -\beta & 1 \end{pmatrix} \begin{pmatrix}t\\ z\end{pmatrix}
\label{eq:lorentz_boost}
\end{equation}
Here, $\beta = v_z/c$ is the relative velocity along the $z$-axis, and $\gamma=\frac{1}{\sqrt{1-\beta^{2}}}$ is the Lorentz factor. $\gamma$ becomes much larger than 1 at RHIC and LHC energies.  For instance, at maximum RHIC energies, the velocity $\beta = v/c$ is approximately 0.999957, corresponding to a Lorentz factor $\gamma = 1/\sqrt{1-\beta^2} \approx 108$. At the LHC, energies reaching 6 TeV per nucleon pair imply $\beta \approx 0.999999943$ and $\gamma \approx 3000$.

\subsubsection{Impact on Colliding Nuclei and Initial Conditions}
One of the consequences of Lorentz boosts in HIC is the \textit{Lorentz contraction} of the colliding nuclei along their direction of motion. In the center-of-mass frame of the collision, each nucleus, approximately spherical in its rest frame with a radius $R_0$ (e.g., about 9 fm for Pb), appears severely flattened into a ``pancake'' shape. The thickness of the nucleus along the beam direction ($z$-axis) is dramatically reduced by the Lorentz factor $\gamma$, becoming approximately $2R_0/\gamma$. Given the large $\gamma$ values in HIC, this longitudinal thickness becomes extremely small, of the order of 0.14 fm at RHIC and about 0.01 fm at LHC energies.

This extreme longitudinal compression helps create the necessary conditions for QGP formation. At the instant the two Lorentz-contracted nuclei overlap, their entire energy is concentrated into a reduced longitudinal volume. This results in an enormous initial energy density within the interaction region. Shortly after a collision at LHC energies, the average energy density can exceed the critical energy density required for QGP formation (typically around $1 \text{ GeV/fm}^3$), far surpassing that of normal nuclear matter ($\sim 0.16 \text{ GeV/fm}^3$ \cite{Busza:2018rrf}).

Beyond enabling QGP formation, Lorentz contraction also dictates the initial geometry of the overlap area. In non-central collisions, the overlap region has a spatially anisotropic shape, typically resembling an almond shape in the transverse plane. This initial spatial anisotropy leads to anisotropic pressure gradients whose subsequent evolution drives the collective expansion of the QGP.

\subsection{Rapidity}
While velocity is intuitive in classical mechanics, its behavior under Lorentz transformations in special relativity is complicated. The relativistic velocity addition is non-linear, motivating the introduction of \textit{rapidity} ($y$), a kinematic variable specifically designed to simplify the description of the relativistic motion of colliding nuclei.

\subsubsection{Definition and Relation to Velocity}
The rapidity $y$ for a particle with energy $E$ and longitudinal momentum component $p_z$ is defined as:
\begin{equation}
y=\frac{1}{2}\ln\left(\frac{E+p_{z}}{E-p_{z}}\right)
\label{eq:rapidity_def_Epz}
\end{equation}
This definition is equivalent to expressing rapidity in terms of the longitudinal component of velocity, $\beta_L = p_z/E$:
\begin{equation}
y = \tanh^{-1}(\beta_L)
\label{eq:rapidity_def_betaL}
\end{equation}
 While $\beta_L$ is confined to the range $[-1, 1]$, rapidity maps this finite interval onto the infinite range $(-\infty, \infty)$. As $\beta_L$ approaches $\pm 1$ (the speed of light), rapidity approaches $\pm \infty$. This property makes rapidity well-suited for describing particles moving at ultra-relativistic speeds.

\subsubsection{The Key Property: Additivity under Boosts}
The advantage of rapidity lies in its simple transformation property under Lorentz boosts. If a particle has rapidity $y$ in frame S, and frame S' moves with a constant velocity with rapidity $Y_{boost}$ relative to S, the particle's rapidity $y'$ in frame S' is simply:
\begin{equation}
y' = y - Y_{boost}
\label{eq:rapidity_boost_additivity}
\end{equation}
This implies that under successive collinear boosts, rapidities simply add linearly.
 HIC involve multiple relevant reference frames, including the laboratory frame, the center-of-mass (CM) frame of the colliding nuclei, and the CM frames of individual constituent partons. The additivity of rapidity enables straightforward transformations between these frames by simply adding or subtracting the relative rapidity boost.

 A direct consequence of additivity is that the shape of the particle distribution as a function of rapidity, $dN/dy$, remains invariant under longitudinal Lorentz boosts. The entire distribution merely undergoes a rigid shift along the rapidity axis, allowing for consistent comparisons of physics information encoded in $dN/dy$ spectra across different frames or models.

 Because rapidities transform additively, the difference in rapidity between any two particles, $\Delta y = y_1 - y_2$, is invariant under longitudinal boosts. This makes $\Delta y$ an intrinsic measure of separation in longitudinal phase space, independent of the observer's motion.

 In the expanding fireball created in HIC, a particle's final rapidity is strongly correlated with its longitudinal position and velocity at the time of decoupling. Measuring particle yields as a function of rapidity, $dN/dy$, provides a direct probe of the system's longitudinal structure and expansion dynamics. The total rapidity range available for particle production increases logarithmically with collision energy, offering a wider window into these dynamics at higher energies.
 
In essence, rapidity serves as the natural language for describing the longitudinal dimension of phase space in relativistic collisions. Its properties, stemming from the hyperbolic geometry of Lorentz transformations, make it the ideal coordinate for analyzing dynamics dominated by boosts along the beam direction.

\subsection{Pseudorapidity}
While rapidity is the theoretically preferred variable for longitudinal kinematics due to its simple behavior under Lorentz boosts, its measurement requires knowledge of both a particle's momentum and its energy (or mass). In experimental settings, determining these quantities for every produced particle can be challenging. This practical constraint motivates the use of a related, more experimentally accessible variable: \textit{pseudorapidity} ($\eta$).

\subsubsection{Definition and Relation to Angle}
$\eta$ is defined solely based on the particle's trajectory angle $\theta$ with respect to the beam axis ($z$-axis):
\begin{equation}
\eta \equiv -\ln\left(\tan\frac{\theta}{2}\right) = \frac{1}{2}\ln\left(\frac{1+\cos\theta}{1-\cos\theta}\right)
\label{eq:pseudorapidity_def}
\end{equation}
This definition maps the physical polar angle range $[0, \pi]$ onto the infinite pseudorapidity range $(-\infty, \infty)$. Particles emitted perpendicular to the beam axis ($\theta = 90^{\circ}$) have $\eta = 0$. Particles emitted along the forward beam direction ($\theta \rightarrow 0^{\circ}$) have $\eta \rightarrow +\infty$, while those emitted along the backward beam direction ($\theta \rightarrow 180^{\circ}$) have $\eta \rightarrow -\infty$.

\subsubsection{Relationship to Rapidity} In the relativistic limit, where a particle's mass $m$ is negligible compared to its momentum, $p$ (i.e., $m \ll p$, implying $E \approx p$) pseudorapidity approaches rapidity. This convergence arises because $y = \tanh^{-1}(p_z/E)$ and $\eta = \tanh^{-1}(p_z/p)$; when $E \approx p$, the expressions become nearly identical. The exact relationship between $y$ and $\eta$ depends on the particle's mass $m$ and transverse momentum $p_T$ (or transverse mass $m_T = \sqrt{p_T^2 + m^2}$). The approximation $\eta \approx y$ is less accurate for massive particles at low $p_T$ and larger $|\eta|$ values.

\subsubsection{Experimental Advantages}
Since $\eta$ depends only on the polar angle $\theta$, it can be determined from the particle's trajectory in tracking detectors or from energy deposition in calorimeters, without requiring individual momentum or energy measurements. For the majority of light, high-momentum particles produced at high energies (e.g., pions), the approximation $\eta \approx y$ holds well, allowing $dN/d\eta$ distributions to serve as reliable proxies for the more fundamental $dN/dy$ distributions.

The differential relationship between distributions in rapidity and pseudorapidity involves a Jacobian factor $p/E$:
\begin{equation}
\frac{dN}{d\eta dp_{T}} = \frac{p}{E}\frac{dN}{dy dp_{T}} = \sqrt{1-\frac{m^{2}}{m_{T}^{2}\cosh^{2}y}}\frac{dN}{dy dp_{T}}
\label{eq:dndeta_vs_dndy}
\end{equation}
The presence of this mass-dependent factor $\sqrt{1-m^{2}/(m_{T}^{2}\cosh^{2}y)}$ (where $m_T = \sqrt{p_T^2 + m^2}$ and $p/E = \sqrt{1-m^2/E^2}$) implies that distributions in $d\eta$ are generally not Lorentz invariant, unlike those constructed using $dy$ and the invariant phase space element $d^3p/E$.

\subsection{Center-of-Mass Energy}
In relativistic mechanics, the total energy available in the system is also different than that in classical mechanics. In classical mechanics, the center-of-mass energy of two colliding particles is defined as their total energy (kinetic plus rest mass) in the reference frame where their net momentum is zero. In special relativity, this concept is extended to the invariant mass of the two-particle system, commonly expressed via the Mandelstam variable $S$:
\begin{equation}
S = (p_{1}+p_{2})^{2} = m_{1}^{2}+m_{2}^{2} + 2\bigl(E_{1}E_{2}-\mathbf{p}_{1}\!\cdot\!\mathbf{p}_{2}\bigr)
\label{eq:mandelstam_S}
\end{equation}
where $p_{i}=(E_{i},\mathbf{p}_{i})$ are the four-momenta and $m_{i}$ are the rest masses of the colliding particles. The quantity $\sqrt{S}$ represents the total energy available for particle production and transformation in the center-of-mass frame (where $\mathbf{p}_{1}+\mathbf{p}_{2}=0$).

For symmetric collisions, the momenta are equal and opposite ($\mathbf{p}_{1}=-\mathbf{p}_{2}$) and the energies are equal ($E_{1}=E_{2}=E$). In this case, the center-of-mass energy simplifies to $\sqrt{S}=2E$. At ultra-relativistic energies, where the rest masses are negligible compared to the particle energies ($m_{i}\ll E_{i}$), the approximation $\sqrt{S} \approx 2\sqrt{E_{1}E_{2}}$ is often used, implying that the center-of-mass energy grows approximately as the geometric mean of the two beam energies. For HIC, this quantity is usually denoted $\sqrt{s_{NN}}$ to emphasize that it is a per nucleon-nucleon pair quantity.

\subsection{Luminosity}
In any collider experiment, the observed rate $R$ of a specific process is directly proportional to its cross-section $\sigma$ and the instantaneous luminosity $L$:
\begin{equation}
R = \sigma L
\label{eq:rate}
\end{equation}
Here, $\sigma$ (with dimensions of area) represents the effective probability or ``size'' for a particular interaction to occur between the incoming particles. The instantaneous luminosity $L$ quantifies the beam intensity, representing the number of particles per unit area per unit time that effectively cross paths. For colliding beams with Gaussian bunch profiles in the transverse plane, $L$ is given by:
\begin{equation}
L = f_{\rm rev}\,n_b\;\frac{N_1\,N_2}{4\pi\sigma_x\,\sigma_y}
\label{eq:luminosity}
\end{equation}
where $f_{\rm rev}$ is the beam revolution frequency, $n_b$ is the number of colliding bunch pairs, $N_{1,2}$ are the numbers of ions per bunch in each beam, and $\sigma_{x,y}$ are the root-mean-square beam widths in the horizontal and vertical directions (such that the effective transverse overlap area is $A_{eff}=4\pi\sigma_x\sigma_y$).

For HIC, the geometric (inelastic) cross-section of the nuclei scales approximately as $\sigma_{\rm geom} \sim \pi R_A^2 \sim \pi\,(r_0 A^{1/3})^2 \sim A^{2/3}$, where $r_0 \approx 1.2 \text{ fm}$ and $A$ is the mass number. High luminosity is paramount in heavy-ion experiments because many of the phenomena under study, such as the formation of the QGP and the production of rare probes (e.g., heavy quarks, jets), have small cross-sections. A larger instantaneous luminosity $L$ (and consequently, a larger integrated luminosity $L_{\rm int}=\int L\,dt$) translates directly to more recorded collisions and therefore better statistical precision for these often subtle signals. Beam parameters are carefully tuned to balance peak luminosity with acceptable background levels and detector conditions, ensuring optimal data quality for measurements of collective observables like flow harmonics and multi-particle correlations.

\subsection{Collision Geometry and Centrality}
\label{sec:geometry}
The \emph{impact parameter} $b$ is a fundamental concept in HIC, defined as the transverse distance between the centers of the two nuclei at their closest approach. Collisions are classified based on $b$: those with $b \approx 0$ are termed \textit{central} collisions, indicating a head-on interaction, while larger values, up to $b \sim 2R$ (where $R$ is the nuclear radius), correspond to \textit{peripheral} (grazing) events.

Since the impact parameter, $b$, cannot be directly measured experimentally, events are instead grouped into \textit{centrality classes}. This classification relies on experimental proxies for the collision's ``event activity'', such as the total charged-particle multiplicity ($N_{\rm ch}$) measured in tracking detectors or the summed transverse energy ($\sum E_T$) deposited in forward calorimeters. Subsequently, a Monte Carlo Glauber model is employed to map each centrality percentile (e.g., 0-5\% central, 50-60\% peripheral) to average geometric quantities: the number of participant nucleons ($N_{\rm part}$), which are the nucleons that undergo at least one inelastic collision, and the number of binary nucleon-nucleon collisions ($N_{\rm coll}$), which are the total number of individual nucleon-nucleon interactions.

\section{Signatures of QGP}
Several key experimental signatures provide strong evidence for QGP formation, such as Jet Quenching, Strangeness Enhancement and Collective Flow.

\subsection{Jet Quenching}
High-momentum partons produced early in HIC typically fragment into collimated sprays of hadrons called jets. When these energetic partons traverse the hot, dense QGP, they lose energy through interactions with the colored medium. This energy loss manifests as a reduced yield of high-momentum hadrons in the final state, known as jet quenching.

\begin{figure}[htbp]
    \centering
    \includegraphics[width=0.6\linewidth]{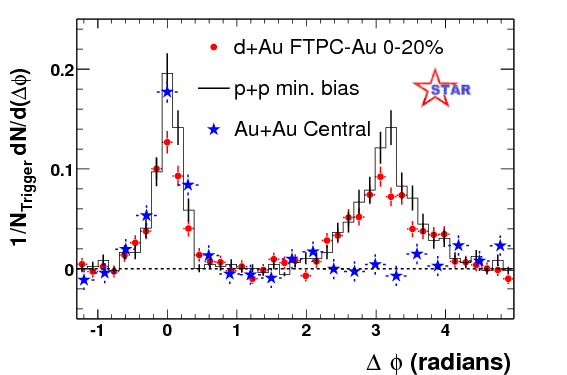}
    \caption{Angular correlations between pairs of high $\pT$ charged particles, referenced to a “trigger” particle that is required to have $\pT > $4 GeV. The $pp$ and $d$+Au data show back-to-back pairs of jets: a peak at $\Delta\phi = 0^{\circ}$ and broadened recoil peak at $180^{\circ}$. In central Au+Au data, the recoil jet is quenched~\cite{STAR:2003pjh}.}
    \label{fig:jetquench}
\end{figure}

The effect becomes evident when examining azimuthal distributions of back-to-back jet pairs, as shown in Figure~\ref{fig:jetquench}. When one parton travels a longer path through the QGP before hadronizing, it loses more energy, resulting in a ``quenched'' jet with lower total energy or even complete suppression in extreme cases. The absence of such quenching in proton-proton collisions strongly indicates QGP formation in HIC.

The nuclear modification factor, $R_{AA}$, for inclusive jet production quantifies this effect~\cite{ATLAS:2014ipv,CMS:2016uxf,ATLAS:2018gwx}:
\begin{equation}
R_{AA}^{jet} = \frac{\frac{1}{N_{evt}} \frac{d^2N_{jet}}{dp_T dy}|_{cent}}{\langle T_{AA} \rangle \frac{d^2\sigma^{jet}}{dp_T dy}|_{pp}}
\end{equation}
where the numerator represents jet yield in nucleus-nucleus collisions for a specific centrality, and the denominator represents the scaled cross-section for jet production in proton-proton collisions, normalized by the nuclear thickness function $\langle T_{AA} \rangle$. Measurements reveal jet suppression ($R_{AA}^{jet} < 1$) in central HIC.

\begin{figure}[htbp]
    \centering
    \includegraphics[width=0.8\linewidth]{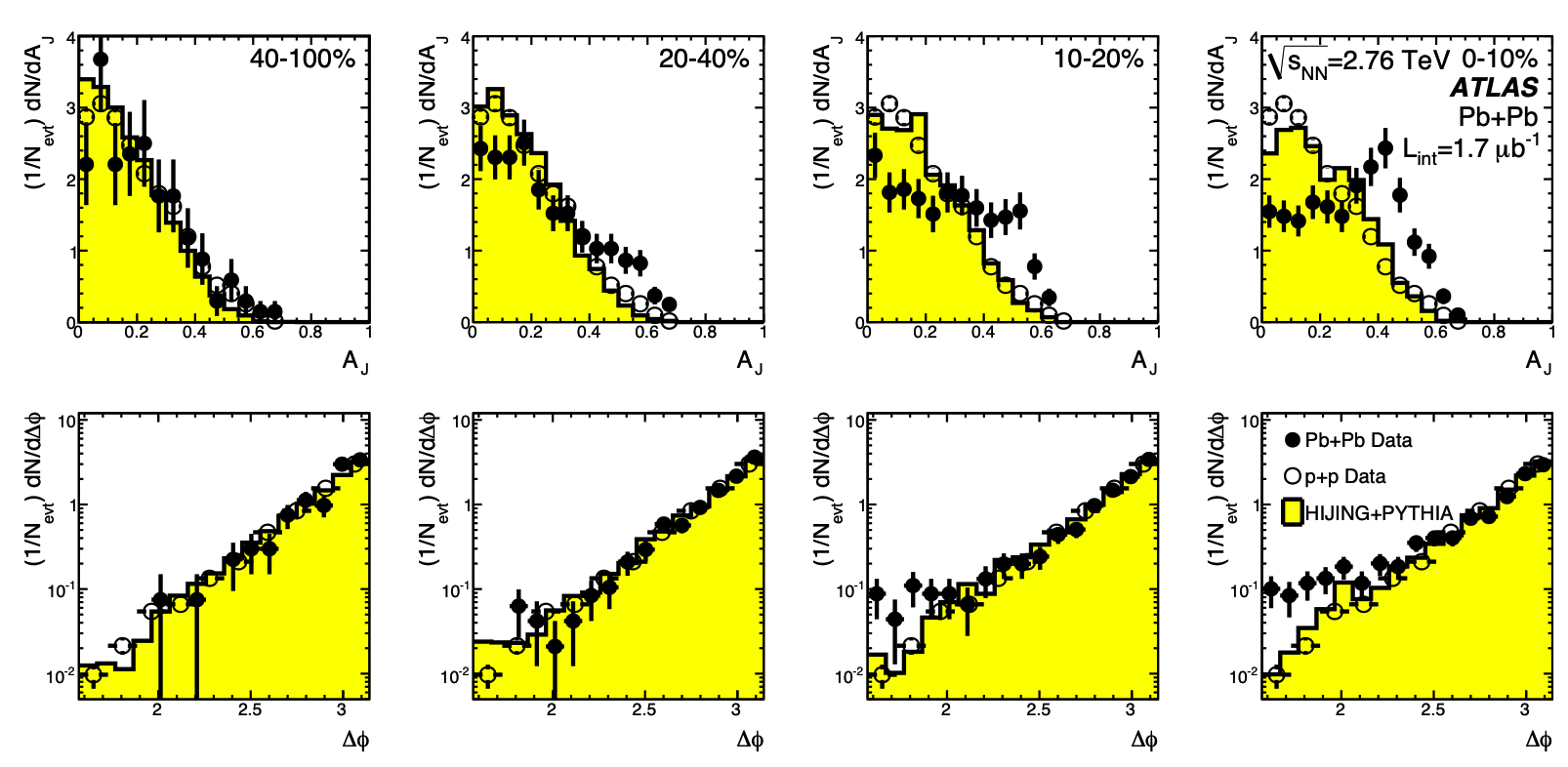}
    \caption{(Top) Dijet asymmetry, $A_J$ distributions for data shown in points compared with unquenched HIJING with superimposed PYTHIA dijets (solid yellow histograms), as a function of collision centrality. $pp$ data shown using open circles are for $\sqn$ = 7 TeV, analyzed with the same jet selection. (Bottom) Distribution of azimuthal angle separation, $\Delta\phi$, between the two jets, for data and HIJING+PYTHIA~\cite{ATLAS:2010isq}.}
    \label{fig:dijetasym}
\end{figure}

Correlated back-to-back jet pairs offer another probe of jet-medium interactions via the energy asymmetry factor, $A_J$~\cite{Milhano:2015mng,CMS:2011iwn,ATLAS:2010isq}.  The energy asymmetry between the leading and subleading jets is defined as
\begin{equation}
A_J \;=\;\frac{E_{T,1} - E_{T,2}}{E_{T,1} + E_{T,2}},
\label{eq:Aj}
\end{equation}
where $E_{T,i}$ is the transverse energy of the $i$th jet.
Figure~\ref{fig:dijetasym} compares the $A_J$ and $\Delta\phi = \bigl|\phi_1 - \phi_2\bigr|$ distributions in four centrality intervals to $pp$ data and HIJING+PYTHIA simulations.  In peripheral Pb+Pb collisions, the $A_J$ spectrum closely follows the $pp$ baseline, whereas central collisions exhibit a pronounced broadening. A Larger asymmetry indicates greater energy loss as jets traverse the QGP. Likewise, the $\Delta\phi$ distribution remains sharply peaked near $\pi$ in peripheral events but develops a substantial low‑$\Delta\phi$ tail in more central collisions.

\subsection{Strangeness Enhancement}
The enhanced production of strange quark-antiquark pairs in HIC compared to proton-proton collisions provides another key QGP signature. This strangeness enhancement results from the increased availability of quarks and gluons in the deconfined QGP environment, facilitating strange quark creation through processes like gluon splitting and light quark annihilation.

Strange quarks are valuable for QGP studies because they are not valence quarks in the initial colliding nuclei, meaning any observed strange quarks signify new particle creation in the hot medium. Additionally, the strange quark mass ($\sim$95 MeV/$c^2$) is comparable to the critical temperature for QGP formation ($T_c \approx$ 150-170 MeV), making strange quark production sensitive to the system's energy density and temperature. These strange quarks decay into lighter quarks on timescales ($\sim$10$^{-10}$ s) much longer than HIC duration ($\sim$10$^{-23}$ s), allowing detection of strange hadrons through their decay products.

\begin{figure}[htbp]
    \centering
    \includegraphics[width=0.5\linewidth]{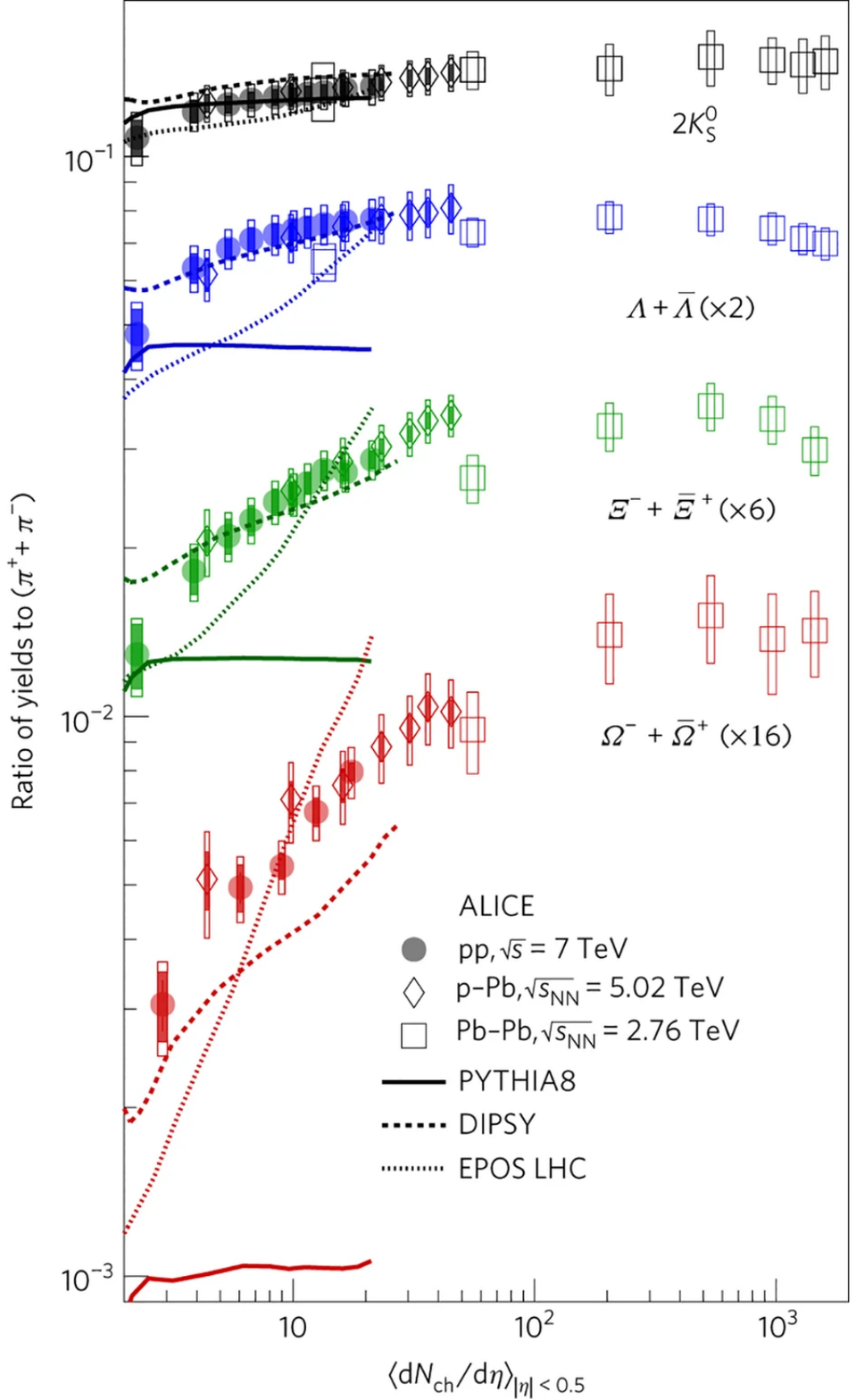}
    \caption{$\pT$-integrated yield ratios to pions ($\pi^+ + \pi^-$) as a function of $\langle dN_{ch}/d\eta\rangle$ measured in $|y|< 0.5$. The values are compared to calculations from MC models and to results obtained in $p+Pb$ and $Pb+Pb$ collisions at the LHC. For Pb-Pb results the ratio $2\Lambda/(\pi^+ + \pi^-)$ is shown. The indicated uncertainties all represent standard deviations~\cite{ALICE:2016fzo}.}
    \label{fig:strangess}
\end{figure}

Enhanced strange particle abundance in HICs compared to scaled proton-proton collisions provides compelling evidence for QGP formation. Measurements show enhancement in strange-to-non-strange hadron production ratios with increasing particle multiplicity, with high-multiplicity proton-proton and proton-lead collisions approaching values similar to lead-lead collisions, as shown in Figure~\ref{fig:strangess}.

\subsection{Collective Flow}
The concept of ``collectivity'' is a cornerstone signature of QGP formation, representing emergent phenomena from the collective behavior of many strongly interacting constituents. Collectivity manifests as coordinated motion patterns among the thousands of particles produced, indicating that the system establishes local thermal equilibrium with a common velocity field.

\begin{figure}[htbp]
\centering
\includegraphics[width=0.8\linewidth]{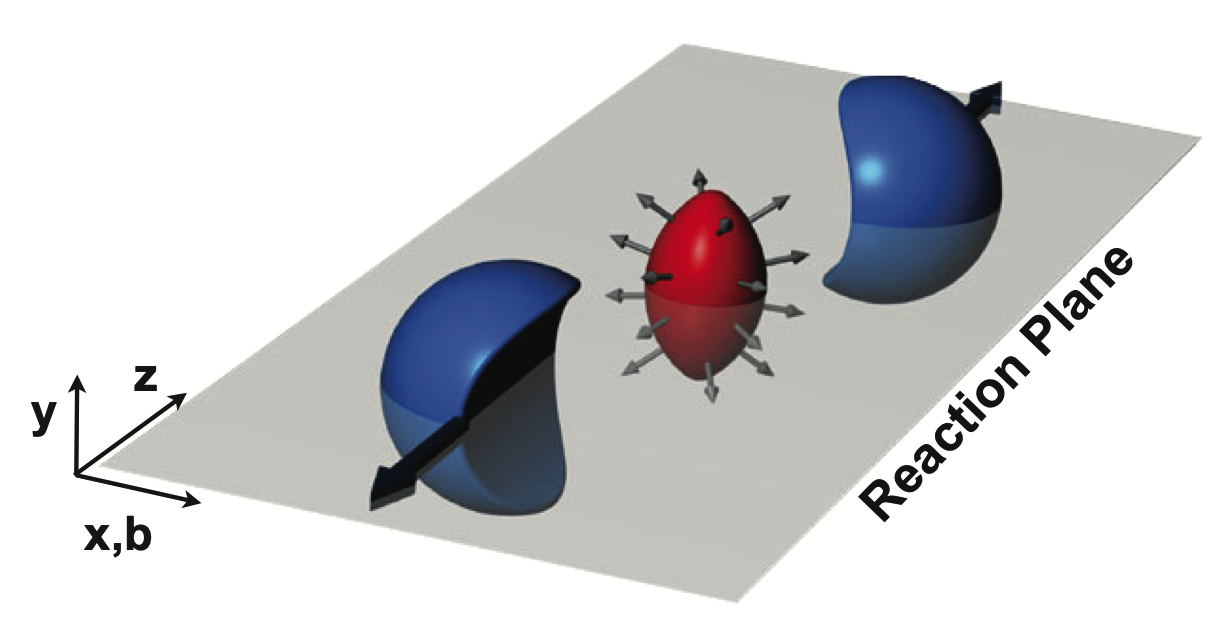}
\caption{Almond-shaped interaction volume after a non-central collision of two nuclei. The initial spatial anisotropy with respect to the reaction plane translates into a momentum anisotropy of the produced particles (anisotropic flow)~\cite{Snellings:2011sz}.}
\label{fig:flowCartoon}
\end{figure}

The origin of collectivity lies in the initial state and the subsequent development of pressure gradients. The pressure differential between the dense central region and the surrounding vacuum generates an outward pressure gradient. In central collisions, this gradient has radial symmetry, creating a collective radial velocity boost for all particles, proportional to their mass. This radial flow affects heavier particles by giving them a larger average $p_T$, characteristically flattening their transverse momentum spectra at low $p_T$. For non-central collisions, the initial interaction region forms an elliptical or irregular shape in the transverse plane, characterized by spatial anisotropies $\varepsilon_n$. The pressure gradients that develop within this asymmetric zone translate the initial spatial anisotropy into momentum anisotropy in the final state particle distributions, termed anisotropic flow, as shown in Figure~\ref{fig:flowCartoon}.

The azimuthal anisotropy in momentum distribution is quantitatively characterized through a Fourier decomposition relative to a specific event plane angle $\Psi_n$:
\begin{equation}
\label{eq:fourier}
\frac{dN}{d\phi} \propto 1 + 2 \sum_{n=1}^{\infty} v_n \cos(n(\phi - \Psi_n))
\end{equation}
where $\phi$ represents the azimuthal angle of an emitted particle, $\Psi_n$ denotes the $n^{th}$ order event plane angle, and $v_n$ is the magnitude of the $n^{th}$ harmonic flow coefficient. Each harmonic corresponds to a specific pattern of anisotropic flow: $v_1$ (directed flow) represents a collective sideward motion, $v_2$ (elliptic flow) characterizes the elliptical asymmetry, and higher harmonics ($v_3$, $v_4$, etc.) capture higher order geometric patterns. $v_2$ directly reflects the initial spatial anisotropy $\varepsilon_2$ and is sensitive to the collision's early stages and the formed matter's properties. Higher-order harmonics, such as $v_3$ and $v_4$ primarily arise from initial-state fluctuations in $\varepsilon_3$ and $\varepsilon_4$, respectively, rather than from the average geometry.

\begin{figure}[htbp]
\centering
\includegraphics[width=0.5\linewidth]{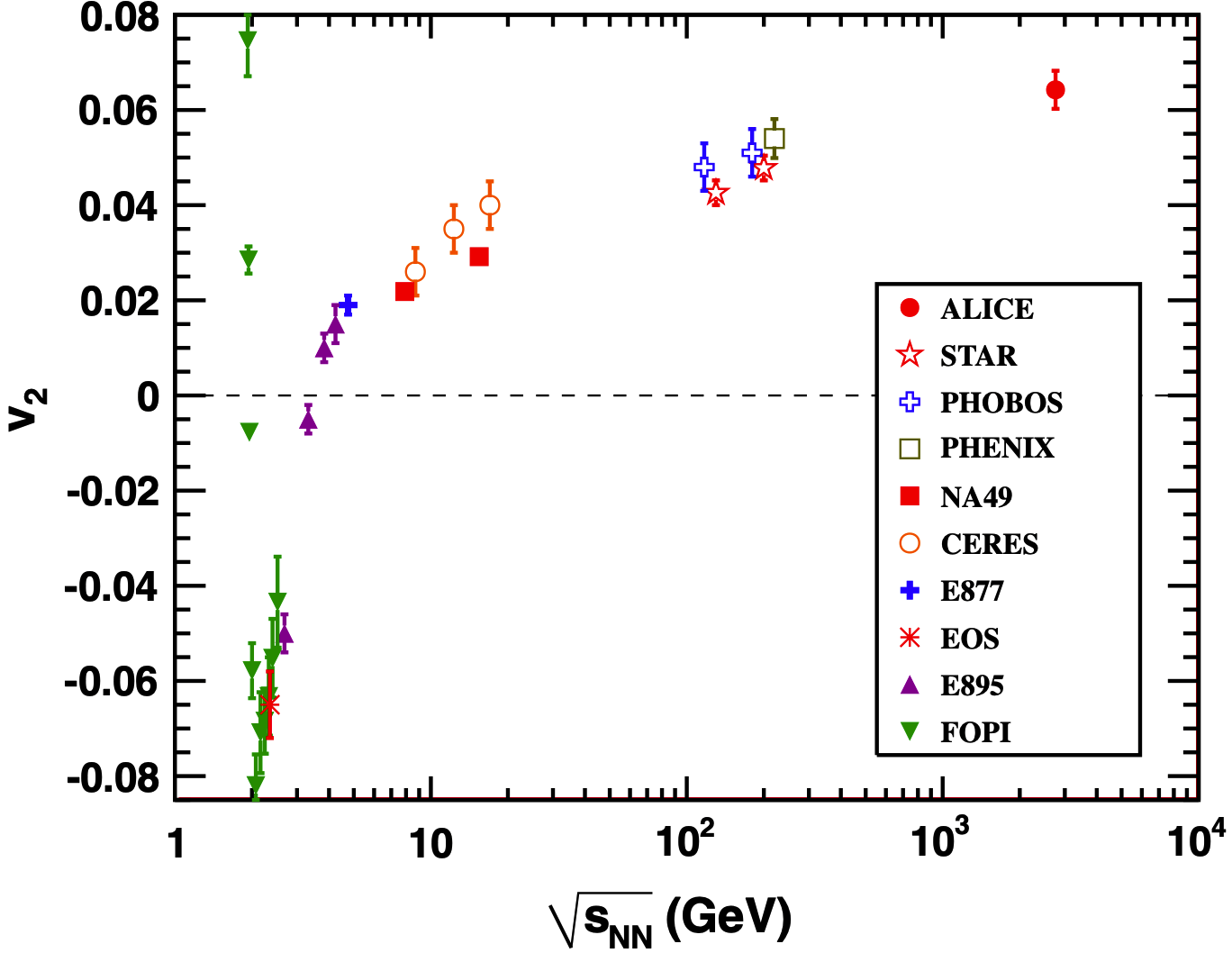}
\caption{Integrated elliptic flow at $\sqn$ = 2.76 TeV in the 20–30\% centrality class compared with results from lower CM energies taken at similar centralities~\cite{ALICE:2010suc}}
\label{fig:v2energydep}
\end{figure}

The observation of strong collective flow provides compelling evidence for QGP formation. Key supporting aspects include: a magnitude of anisotropic flow that cannot be explained by independent particle production or simple hadronic rescattering models; mass ordering of flow coefficients consistent with hydrodynamic predictions; scaling of flow with initial collision eccentricity; and an extraordinarily low shear viscosity to entropy density ratio, $\eta/s$. Furthermore, measurements show that anisotropic flow increases with collision energy~\cite{ALICE:2010suc}, an observation consistent with hydrodynamic predictions and support a collectively expanding medium, as shown in Figure~\ref{fig:v2energydep}.

The study of collective flow phenomena has a rich history, starting with observations of $v_1$ at lower energies (e.g., Bevalac) which provided early evidence of pressure development in compressed nuclear matter. The advent of RHIC and LHC enabled the study of higher energy collisions, revealing substantial $v_2$ and higher harmonics consistent with the hydrodynamic behavior of a nearly perfect fluid, profoundly impacting the theoretical understanding of the QGP.

While anisotropic flow harmonics have provided extensive insights into the QGP's properties, direct experimental evidence for the collective nature of its radial expansion and a precise understanding of its sensitivity to bulk viscosity remained less developed, motivating the novel $v_0(p_T)$ analysis presented in Chapter~\ref{sec:chap4_v0pt}.

\section{Model Description of QGP Evolution}
To interpret the diverse experimental signatures of the QGP discussed previously, and ultimately extract its fundamental properties, sophisticated theoretical models are indispensable. These models connect the initial conditions of a HIC to the experimentally observed final-state particle distributions. Since the QGP is a fleeting and unobservable state, modeling provides the essential framework to interpret experimental signatures and extract fundamental properties like the QGP's equation of state and transport coefficients. The key components of these models, from the initial state to the hydrodynamic evolution are described below.

\subsection{Initial-State Models}\label{sec:ini}
The initial state of a HIC, immediately after the nuclei pass through each other but before expansion, fundamentally determines the subsequent evolution and the resulting collective flow patterns. Due to the inability to directly measure this state, various theoretical models are employed to describe the geometry and energy deposition in the initial state.  

In theoretical models of HICs, the initial geometry of the overlapping region is characterized by its shape and size, which are quantified using two key parameters: the eccentricity $\varepsilon_n$ and the transverse area $S_{\perp}$ (or $R_{\perp}$). The eccentricity $\varepsilon_n$ quantifies the $n$-th order anisotropy in the energy or entropy density distribution and is defined as:
\begin{equation}
\label{eq:eccentricity}
\varepsilon_n = \frac{\sqrt{\langle r^n \cos(n\phi) \rangle^2 + \langle r^n \sin(n\phi) \rangle^2}}{\langle r^n \rangle}
\end{equation}

\begin{figure}
    \centering
    \includegraphics[width=0.5\linewidth]{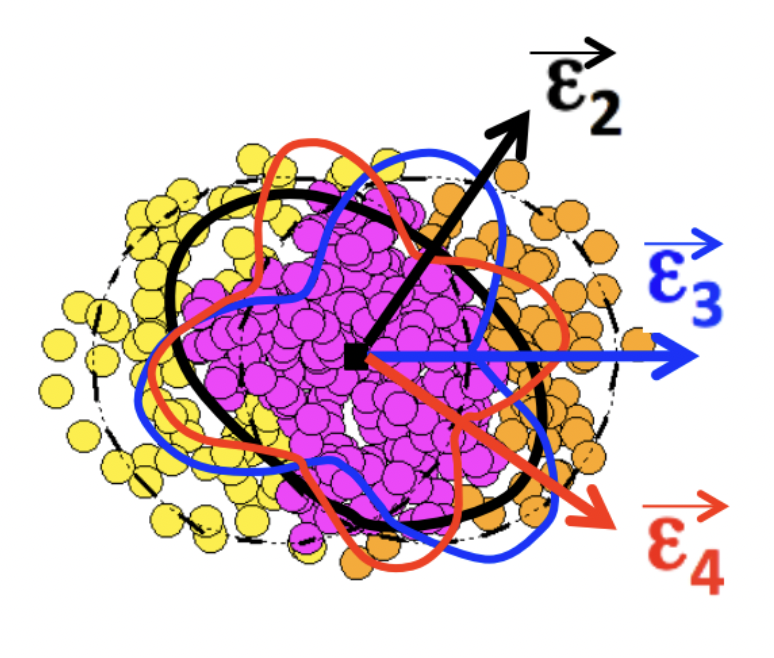}
    \caption{Illustration of $\varepsilon_2$, $\varepsilon_3$, $\varepsilon_4$ in the initial-state overlap region.}
    \label{fig:ecc}
\end{figure}
where $r$ and $\phi$ are the radial coordinate and azimuthal angle in the transverse plane, respectively, and the angular brackets denote an average over the initial density profile. The second-order eccentricity $\varepsilon_2$ typically reflects the almond-shaped overlap in non-central collisions, while higher-order components such as $\varepsilon_3$ and $\varepsilon_4$ primarily arise from initial-state fluctuations due to the positions of individual nucleons, as shown in Figure~\ref{fig:ecc}. 

Complementing the shape, the transverse size of the system is characterized by:

\begin{equation}
\label{eq:size}
S_{\perp} = \pi R^{2}_{\perp} = \pi\sqrt{\lr{x^2}+\lr{y^2}}
\end{equation}
 where $x$ and $y$ represent coordinates in the transverse plane, defined in the rotated center-of-mass frame such that $x$ and $y$ align with the minor and major axes of the participant distribution, respectively. Together, $\varepsilon_n$ and $S_{\perp}$ describe the initial-state geometry that influences the subsequent hydrodynamic evolution.

Several theoretical frameworks exist for modeling the initial state, each with distinct underlying physical assumptions:
\subsubsection{Glauber Model}

\begin{figure}[h!]
    \centering
    \includegraphics[width=0.8\linewidth]{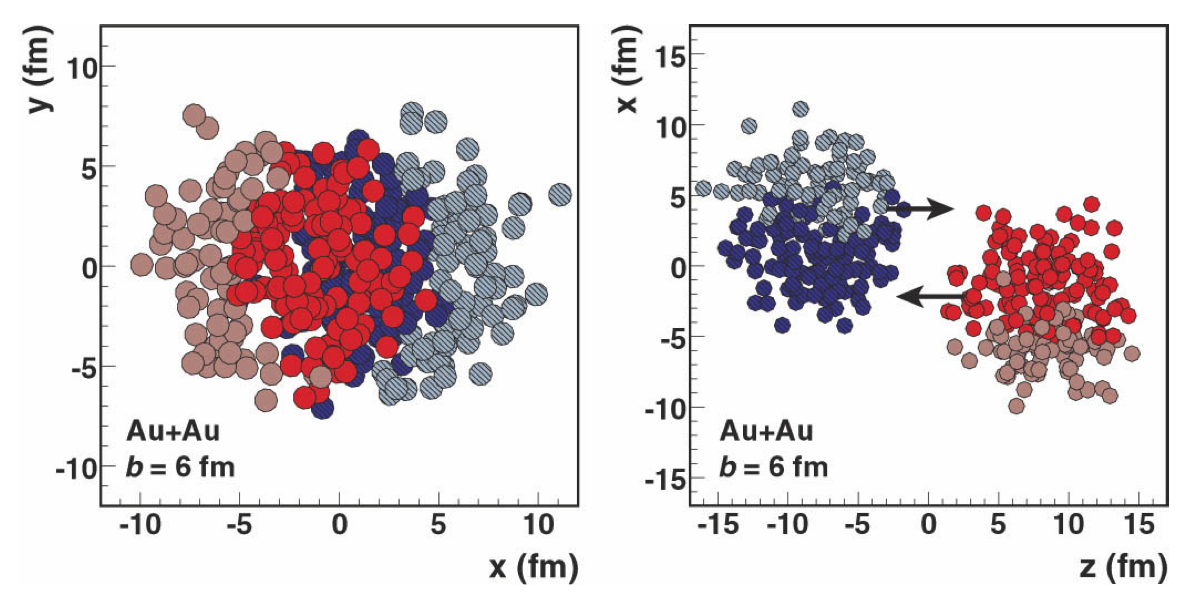}
    \caption{Illustration of a Glauber Monte Carlo Au+Au event at $\sqrt{s_{NN}} = 200$ GeV and $b = 6$ fm, shown in transverse (left) and longitudinal (right) views. Darker circles represent participant nucleons~\cite{Miller:2007ri}.}
    \label{fig:glau}
\end{figure}

The Glauber model~\cite{Miller:2007ri} offers a geometric description of HIC based on independent nucleon-nucleon scatterings. It quantifies the collision geometry using $b$, $N_{\rm part}$, and $N_{\rm coll}$. These quantities are derived from parameterized nuclear density profiles and the measured inelastic nucleon-nucleon cross-section $\sigma_{NN}^{\rm inel}$, assuming straight-line nucleon trajectories and interaction based on transverse proximity, as illustrated in Figure~\ref{fig:glau}. While used for centrality determination, it also provides the initial transverse positions of these interactions, which are foundational for estimating the initial energy or entropy density distribution.

Centrality quantifies the geometric overlap between colliding nuclei and is typically expressed as a percentile of the total inelastic cross-section. More central collisions correspond to smaller $b$, maximal overlap, and generally higher charged-particle multiplicity, while peripheral collisions have larger $b$, minimal overlap, and lower multiplicity. The Glauber model is a standard tool for relating experimental observables, like multiplicity, to these underlying geometric properties and defining centrality classes.

Two main implementations exist: the optical and Monte Carlo Glauber models. The optical model treats nuclei as continuous density distributions, providing analytical results but lacking event-by-event fluctuations. It assumes independent nucleon motion and that nuclear size exceeds the nucleon-nucleon force range.

The Monte Carlo Glauber model provides a more realistic event-by-event simulation, incorporating initial-state fluctuations. Nucleons are randomly positioned within nuclei according to their density distributions. The $b$ is sampled from $d\sigma/db = 2\pi b$. A nucleon-nucleon collision occurs if their transverse separation $d$ satisfies $d \leq \sqrt{\sigma_{\mathrm{NN,inelastic}}/\pi}$, where $\sigma_{\mathrm{NN,inelastic}}$ is the energy-dependent inelastic nucleon-nucleon cross-section.

The accuracy of the Glauber model relies on input parameters, notably the nuclear density distribution, often parameterized by a three-parameter Fermi distribution:
\begin{equation}
\label{eq:three_param_fermi}
\rho(r) = \rho_0 \frac{1 + w(r/R)^2}{1 + \exp\bigl[(r - R)/a\bigr]},
\end{equation}
where $\rho_0$ is the central density, $R$ is the half-density radius, $a$ is the skin depth, and $w$ accounts for non-sphericity. Nucleon positions are sampled from this distribution. The $b$ is sampled from a $b d b$ distribution. The inelastic nucleon-nucleon cross-section, $\sigma_{\mathrm{NN,inelastic}}$, is a crucial energy-dependent input parameter, taken from experimental measurements.

Since geometric quantities like $b$, $N_{\mathrm{part}}$, and $N_{\mathrm{coll}}$ are not directly measurable, the Glauber model is used to relate them to experimentally accessible charged particle multiplicity distributions $P(N_{\mathrm{ch}})$~\cite{Miller:2007ri,Broniowski:2007nz,dEnterria:2020dwq}. This mapping utilizes a particle production model, often a two-component model where particle production scales with both $N_{\mathrm{part}}$ (soft processes) and $N_{\mathrm{coll}}$ (hard processes).

Particle production from sources can be modeled by a negative binomial distribution (NBD). The total event multiplicity $N_{\mathrm{ch}}$ is commonly expressed as:
\begin{equation}
\label{eq:nch_model}
N_{\mathrm{ch}} = n_{pp} \left[ (1-x) \frac{N_{\mathrm{part}}}{2} + x N_{\mathrm{coll}} \right],
\end{equation}
where $n_{pp}$ is the average proton-proton multiplicity, $x$ is the fraction of hard processes, and $(1-x)$ represents the soft component.

\begin{figure}[htbp]
    \centering
    \includegraphics[width=0.45\linewidth]{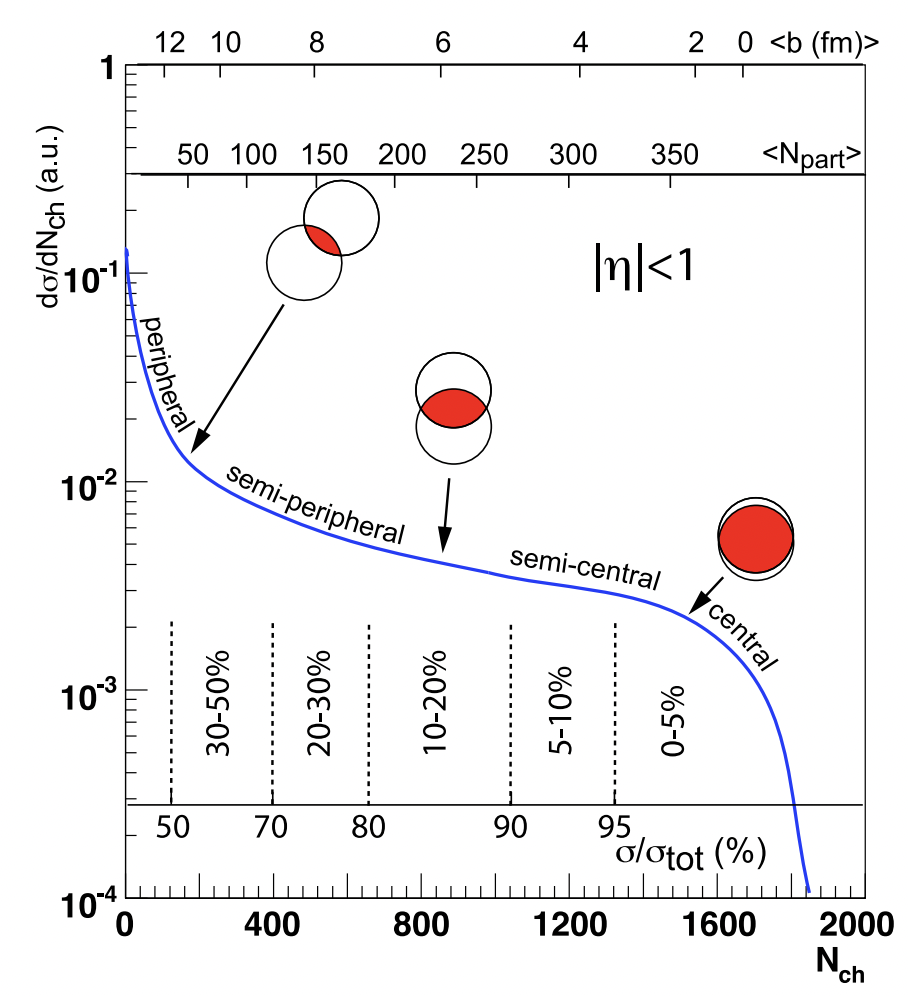}
    \caption{Centrality determination using the Glauber model. The solid line shows a simulated charged-particle multiplicity distribution for Pb+Pb collisions at LHC energies ($|\eta|<1$), mapped to $b$ and $N_{\mathrm{part}}$ using the Glauber model~\cite{Miller:2007ri}.}
    \label{fig:glauberCent}
\end{figure}

Centrality determination involves adjusting particle production model parameters to match the simulated multiplicity distribution to experimental data. Once matched, centrality classes (e.g., 0–5\%) are defined by slicing the cumulative distribution at corresponding percentiles, as shown in Figure~\ref{fig:glauberCent}.

Despite its utility, the standard Glauber model assumes uncorrelated nucleons and neglects short-range correlations and quantum effects. Modern approaches include sub-nucleonic and color fluctuations to improve the agreement of model simulations with data~\cite{Loizides:2016djv,STAR:2022pfn}.

\subsubsection{Color Glass Condensate (CGC) Based Model}

Traditional Glauber models, while successful in describing many global features of HICs based on nucleon-nucleon cross sections and nuclear geometry, treat nucleons as fundamental entities. They do not incorporate the internal partonic structure (quarks and gluons) of nucleons or the dynamical evolution of these partons, especially at the high energies achieved in modern colliders. 

At these energies, hadrons and nuclei are probed at very small Bjorken-$x$, where gluon densities become extremely high. This dense gluonic environment necessitates a more fundamental description based on QCD that accounts for gluon saturation effects. The limitations of linear QCD evolution at high energies and high parton densities motivated the development of the CGC effective field theory. This framework aims to describe high-density gluonic matter, moving beyond the simple superposition of independent nucleon collisions.

\begin{figure}[htbp]
\centering
\includegraphics[width=0.9\linewidth]{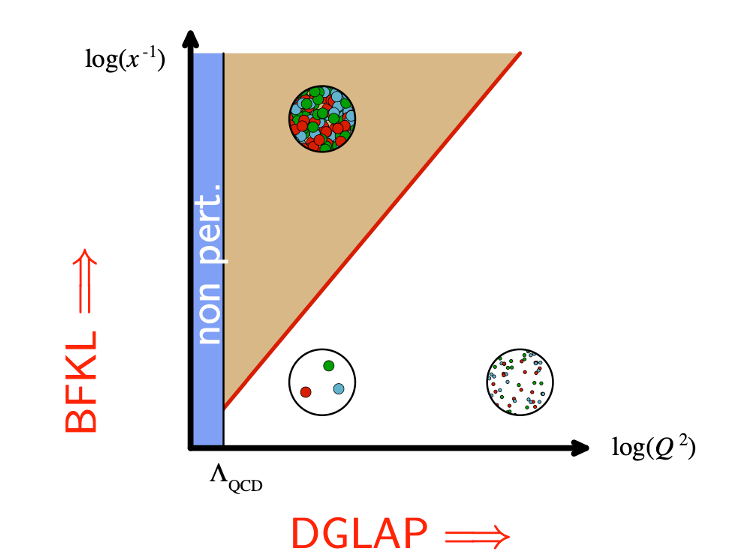}
\caption{Evolution of parton densities in the $(\ln Q^2, \ln(1/x))$ plane. DGLAP evolution predominantly drives the system to larger $Q^2$ (right); BFKL evolution drives it to smaller $x$ (larger $Y$, upward). The saturation boundary, characterized by the saturation scale $Q_s^2(Y)$ (often approximated as $\ln Q_s^2(Y) \approx \lambda Y$), separates the dilute linear regime from the dense nonlinear Color Glass Condensate regime.}
\label{fig:evolution}
\end{figure}

\begin{figure}[htbp]
\centering
\includegraphics[width=0.8\linewidth]{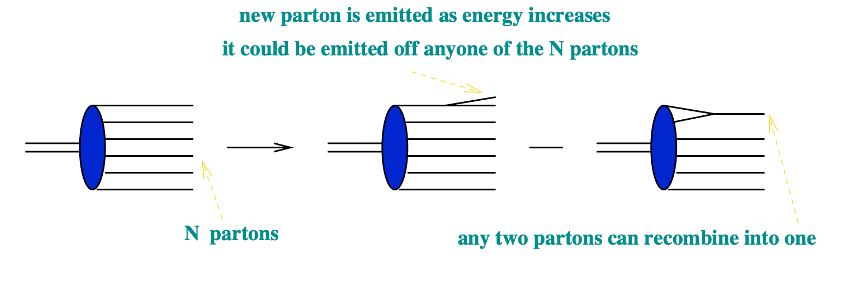}
\caption{Gluon splitting as described by BFKL evolution, showing one gluon splitting into two, increasing gluon number, and gluon recombination, where two gluons merge into one, a key feature of BK/JIMWLK evolution that tames the gluon growth.}
\label{fig:splitting_recombination}
\end{figure}

The dynamics of partonic matter in high-energy QCD can be clearly understood in terms of transverse resolution, parametrized by $\ln Q^2$ versus evolution ``time'', $Y = \ln(1/x)$, where $x$ is the Bjorken scaling variable. This is displayed in Figure~\ref{fig:evolution} which shows that, at fixed $Y$ and increasing $Q^2$ (probing smaller transverse distances $\sim 1/Q$), the Dokshitzer-Gribov-Lipatov-Altarelli-Parisi (DGLAP) equations~\cite{Gribov:1972ri,Dokshitzer:1977sg,Altarelli:1977zs} govern the evolution of parton distribution functions. While the overall parton density tends to increase with $Q^2$, the DGLAP evolution describes a scenario where the gluon-density per unit of transverse area effectively dilutes. This is because each parton occupies an area $\sim 1/Q^2$, and parton splittings predominantly shift radiated partons towards higher $Q^2$ and lower $x$, but with the system becoming sparser at a fixed $x$ as $Q^2$ increases. This evolution is characterized by logarithmic scaling in $Q^2$. Conversely, at fixed $Q^2$ and growing $Y$ (i.e., decreasing $x$), the Balitsky-Fadin-Kuraev-Lipatov (BFKL) evolution~\cite{Lipatov:1976zz,Fadin:1975cb,Balitsky:1978ic} describes a rapid, power-law like increase in the number of gluons due to successive gluon emissions as shown in Figure~\ref{fig:splitting_recombination}. This leads to a growth in gluon density in the transverse plane. 

However, the linear BFKL evolution, if extrapolated to very small $x$ (very high energies), predicts an unabated growth of the gluon density. This leads to theoretical inconsistencies, such as the violation of the Froissart bound, which posits that total hadronic cross-sections cannot grow faster than $\ln^2 s$ (where $\sqrt{s}$ is the center-of-mass energy)~\cite{Froissart:1961ux}, and the black disk limit, $\sigma_{\text{tot}} \le 2\pi R^2$. Such unphysical growth indicates that at very high parton densities, the underlying physics must change.

This breakdown of linear evolution in the high-density, small-$x$ regime necessitates the inclusion of nonlinear effects. The resolution lies in recognizing that when the density of gluons becomes sufficiently high, they can no longer be treated as independent particles. Nonlinear effects, most notably gluon recombination, $gg \rightarrow g$, depicted in Figure~\ref{fig:splitting_recombination}, become important. These recombination processes counteract the gluon splitting. An early framework to describe such high-density gluonic systems is the McLerran-Venugopalan (MV) model~\cite{McLerran:1993ni,McLerran:1993ka}. According to the MV model, the numerous small-$x$ gluons can be described as classical color fields sourced by the large-$x$ partons (valence quarks) of the nucleus, with the initial gluon density scaling with $A^{1/3}$.

The evolution from the dilute regime towards this high-density, saturated state is described by nonlinear evolution equations. The Balitsky-Kovchegov (BK) equation~\cite{Balitsky:1978ic, Balitsky:1995ub,Kovchegov:1999yj} is a mean-field equation that incorporates both gluon splitting (linear BFKL term) and gluon recombination (a quadratic nonlinear term in gluon density or dipole amplitude). More generally, the Jalilian-Marian-Iancu-McLerran-Weigert-Leonidov-Kovner (JIMWLK) equations~\cite{Jalilian-Marian:1997qno,Jalilian-Marian:1997jhx,Iancu:2000hn,Ferreiro:2001qy} provide a more complete description of this evolution, including fluctuations. These nonlinear equations tame the rapid growth of gluon densities predicted by BFKL alone.

This dynamic interplay between gluon splitting and recombination leads to the emergence of a characteristic momentum scale known as the \textit{saturation scale}, $Q_s(x)$. The saturation scale increases with decreasing $x$ (increasing energy) and with nuclear size $A$. It is typically defined as the scale where the gluon occupation number becomes large (of order $1/\alpha_s$), or equivalently, where the probability for a parton from a dilute probe to interact with the dense target becomes of order unity. Below this scale ($k_\perp \lesssim Q_s$), gluon fields are highly occupied and coherent, exhibiting classical field behavior.

This regime of saturated gluon matter, where nonlinear effects are dominant and parton densities are high, is precisely what the CGC describes~\cite{Gelis:2010nm}. The name itself encapsulates key properties of this state: `Color' because it describes the dynamics of colored quarks and gluons; `Glass' due to the separation of time scales, where the large-$x$ partons serving as sources of the gluon field evolve very slowly (akin to the amorphous, slowly evolving structure of a glass) compared to the natural time scale of the small-$x$ gluons they generate, thus appearing `frozen' during the high-energy interaction; and `Condensate' because the gluon fields have very high occupation numbers, reaching $1/\alpha_s$, and behave as coherent classical fields, analogous to a Bose-Einstein condensate. In the CGC framework, the fast color charges (large-$x$ partons) act as random static sources $\rho(\mathbf{x}_\perp)$ for the small-$x$ gluon fields, with a statistical weight functional $W[\rho]$. Event-by-event fluctuations of these sources $\rho$ generate transverse ``lumpy'' color field configurations with a typical transverse size $1/Q_s$.

Unlike simpler nucleon-based Glauber models, modern CGC implementations such as IP-Glasma encode these sub-nucleonic fluctuations directly in the initial conditions for HICs. As a consequence, the effective transverse overlap area $S_\perp$ is reduced, the initial-state eccentricities $\varepsilon_n$ (especially for $n\ge3$) are enhanced even in central collisions, and the energy density distribution becomes more granular and irregular. These realistic initial-state fluctuations are essential for accurately reproducing the observed anisotropic flow harmonics $v_n$ in hydrodynamic simulations of the QGP, which are driven by the spatial anisotropy converting into momentum anisotropy during hydrodynamic expansion.

Crucially, the CGC framework, in the initial stages immediately after a high-energy collision (often referred to as the Glasma stage), naturally produces intrinsic momentum-space anisotropies. This arises from the coherent, highly occupied color fields that exist in the glasma. These fields, generated by the collision of the saturated gluonic wavefunctions of the incoming nuclei, are not isotropically distributed in momentum space. Instead, they exhibit strong, correlated longitudinal components due to the Lorentz contraction of the incoming nuclei and the nature of the classical fields. These anisotropic chromo-fields exist \textit{before} any particle production or thermalization occurs.

The importance of this initial momentum anisotropy is multifaceted. First, the presence of these anisotropic fields means that particles produced directly from the decay or fragmentation of these Glasma fields will inherit a momentum anisotropy, even in the absence of subsequent hydrodynamic evolution. This Glasma-induced initial momentum anisotropy contributes directly to the measured azimuthal anisotropies ($v_n$) of produced particles. This is significant in small collision systems (like $p+A$ or $d+A$) and in peripheral $A+A$ collisions, where the created fireball may have a very short lifetime or be too small for full hydrodynamization. In such scenarios, a substantial fraction, or even the majority, of the final-state anisotropy $v_n$ can be directly attributed to these initial Glasma-induced momentum anisotropies, rather than solely to a collective hydrodynamic flow, thereby challenging the traditional interpretation of $v_n$ as purely a signature of hydrodynamics. Furthermore, these initial momentum correlations can persist and influence the subsequent hydrodynamic evolution, potentially leaving an imprint on the final-state observables even in large systems.

Experimentally, the presence of the saturation boundary depicted in Figure~\ref{fig:evolution} implies that at sufficiently high energies (large $Y$) and for processes probing moderate momentum scales ($Q \sim Q_s$), one should observe clear departures from linear DGLAP or BFKL evolution. Signatures of gluon saturation and CGC dynamics have been actively sought and identified in various experiments. These include observations in deep-inelastic scattering at HERA (e.g., geometric scaling of cross-sections), forward hadron production in $d+\mathrm{Au}$ collisions at RHIC (suppression of yields), and the observation of long-range ``ridge'' correlations in $p+p$ and $p+\mathrm{Pb}$ collisions at the LHC. The onset of geometric scaling, the suppression of back-to-back di-jet yields at forward rapidities, and the emergence of collective-like azimuthal patterns in small systems collectively provide evidence for the relevance of the gluon saturation regime and the CGC picture for describing the initial state of high-energy nuclear collisions.

\subsubsection{TRENTo Model} 
The TRENTo model~\cite{Moreland:2014oya} is a versatile, parametric generator of initial entropy‐density profiles in high‐energy nuclear collisions.  At a chosen thermalization time $\tau_{0}$, participant nucleon positions are sampled (e.g.\ from Woods–Saxon or correlated distributions) and each nucleon is assigned a fluctuated thickness
\begin{equation}
T_{i}(x,y)
= w_{i}\,\int dz\,\rho_{\rm proton}\bigl(x - x_{i},\,y - y_{i},\,z - z_{i}\bigr),
\end{equation}
where the weights $w_{i}$ are drawn independently from a Gamma distribution to control event‐by‐event multiplicity fluctuations.  The projectile thicknesses $T_{A}$ and $T_{B}$ are then combined using the reduced‐thickness ansatz
\begin{equation}
T_{R}(p;T_{A},T_{B})
=\Bigl(\tfrac{T_{A}^{p} + T_{B}^{p}}{2}\Bigr)^{1/p},
\end{equation}
which satisfies scale invariance, $T_{R}(c\,T_{A},c\,T_{B})=c\,T_{R}(T_{A},T_{B})$.  The continuous parameter $p$ interpolates between wounded‐nucleon scaling ($p=1$), geometric‐mean scaling ($p=0$, akin to CGC models), and harmonic‐mean scaling ($p=-1$).

Key ingredients of TRENTo include the inelastic collision probability
\begin{equation}
P_{\rm coll}
=1 - \exp\bigl[-\sigma_{gg}\,T_{pp}(b)\bigr],
\end{equation}
which determines which nucleons are “wounded,” Gamma‐distributed weights $w_{i}\sim\mathrm{Gamma}(k,k)$ that set the variance of multiplicity fluctuations, and a Gaussian transverse profile of width $w$ for each participant, fixing the granularity scale of the initial entropy density.  These components produce an ensemble whose integrated reduced thickness $T_{R}$ reproduces charged‐particle multiplicity distributions in $p$+$p$, $p$+A, and A+A collisions.

TRENTo is widely used to define centrality classes by mapping the event‐by‐event integral of $T_{R}$ onto measured multiplicity percentiles and to compute spatial anisotropies (“eccentricities”)
\begin{equation}
\varepsilon_{n}\,e^{i n \Phi_{n}}
=-\,\frac{\displaystyle\int dx\,dy\;r^{n}e^{i n\phi}\;T_{R}(x,y)}
{\displaystyle\int dx\,dy\;r^{n}\;T_{R}(x,y)},
\end{equation}
which seed the final‐state flow harmonics $v_{n}$.  By tuning the parameters $(p,\,k,\,w,\text{norm})$ in Bayesian analyses, TRENTo simultaneously describes multiplicity distributions across system sizes and constrains both the initial‐state geometry and the medium’s transport coefficients.

\begin{figure}[htbp]
  \centering
  \includegraphics[width=0.4\linewidth]{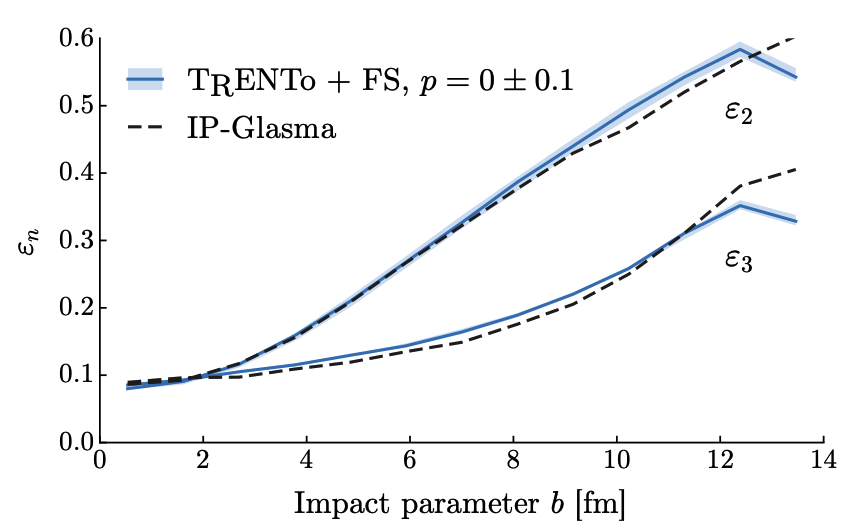}
  \caption{Comparison of $\varepsilon_{2}$ (upper curves) and   $\varepsilon_{3}$ (lower curves) vs. $b$ in Pb+Pb collisions at $\sqrt{s_{NN}}=2.76$ TeV.  Dashed lines: IP‑Glasma evolved with classical Yang–Mills dynamics to $\tau=0.4\,$fm/$c$ \cite{Schenke:2012wb}.  Shaded bands: TRENTo profiles (with $p=0\pm0.1$, $k=1.6$, $w=0.4\,$fm) followed by free‐streaming to the same time \cite{Moreland:2014oya}.}
  \label{fig:trento_ipglasma_ecc}
\end{figure}

Figure~\ref{fig:trento_ipglasma_ecc} compares the ellipticity $\varepsilon_{2}$ and triangularity $\varepsilon_{3}$ as functions of $b$ for Pb+Pb collisions at $\sqrt{s_{NN}}=2.76\,$TeV, using both IP‑Glasma and TRENTo initial conditions.  TRENTo (with $p=0\pm0.1$, $k=1.6$, and $w=0.4\,$fm) reproduces IP‑Glasma results across most of the impact‐parameter range, diverging only in the most peripheral collisions where sub‐nucleonic fluctuations become important. This agreement demonstrates that the geometric‐mean prescription ($p\approx0$) effectively captures the same initial‐geometry scaling as the more detailed CGC‑based IP‑Glasma framework.

\subsection{Relativistic Hydrodynamics}
Once the initial state of a HIC is established, the subsequent evolution of the hot, dense medium is primarily described by relativistic hydrodynamics. The remarkable success of hydrodynamic models in quantitatively reproducing complex collective flow phenomena observed experimentally is a cornerstone of the field, providing strong evidence that the QGP behaves as a nearly perfect fluid.

The applicability of relativistic hydrodynamics to the QGP relies fundamentally on the assumption that the system rapidly reaches and maintains a state of \textit{local thermal equilibrium}. This condition is met when the mean free path of the medium's constituents (quarks and gluons) is smaller than the characteristic size of the system and the length scales over which macroscopic properties like temperature, energy density, and flow velocity vary. Under these circumstances, the microscopic details of particle interactions can be effectively coarse-grained, and the macroscopic evolution of the medium can be described by conservation laws supplemented by constitutive relations.

The basic framework of relativistic hydrodynamics is built upon the local conservation laws for energy, momentum, and conserved charges (such as baryon number, strangeness, and electric charge). These are expressed through the divergence of the energy-momentum tensor $T^{\mu\nu}$ and the currents for conserved charges $J_i^{\mu}$:
\begin{align}
\label{eq:energy_momentum}
\partial_\mu T^{\mu\nu} &= 0 \\
\label{eq:conserved_charge}
\partial_\mu J_i^{\mu} &= 0
\end{align}
where $T^{\mu\nu}$ describes the energy and momentum densities and fluxes within the fluid, and $J_i^{\mu}$ describes the density and flow of conserved charge $i$.

To solve these conservation equations, one needs to specify the properties of the fluid, which are encoded in the Equation of State and the transport coefficients.

The \textit{Equation of State (EOS)} for the QGP establishes the relationship between the pressure ($P$) and the energy density ($\epsilon$), often expressed in the functional form $P = f(\epsilon)$. A key property derived from the EOS is the speed of sound squared, $c_s^2$, which is defined as the derivative of the pressure with respect to the energy density at constant entropy per baryon number ($s/n_B$):
$$c_s^2 = \left( \frac{\partial P}{\partial \epsilon} \right)_{s/n_B}$$
In simplified scenarios, or as a first-order approximation, this is often represented as:
$$c_s^2 \approx \frac{dP}{d\epsilon}$$

The speed of sound, $c_s$, represents the velocity at which pressure disturbances (or sound waves) propagate through the medium. Therefore, $c_s^2$ is a fundamental quantity for constraining the EOS of the QGP, as it directly links the medium's response to compression and expansion with its underlying thermodynamic properties. Historically, constraints on $c_s^2$ were obtained primarily through indirect methods, often relying on Bayesian analyses of various experimental observables compared to hydrodynamic simulations that incorporate different EOS models.

The energy-momentum tensor $T^{\mu\nu}$ quantifies the distribution and flow of energy and momentum within the fluid. In the simplest case of an \textit{ideal fluid}, which possesses no dissipation, $T^{\mu\nu}$ is given by:
\begin{equation}
\label{eq:T_ideal}
T^{\mu\nu}_{\text{ideal}} = (\epsilon + P)u^{\mu}u^{\nu} - Pg^{\mu\nu}
\end{equation}
where $u^{\mu}$ is the local four-velocity field of the fluid element and $g^{\mu\nu}$ is the metric tensor. The pressure $P$ in this equation is the equilibrium pressure determined by the EOS.

\begin{figure}[htbp]
    \centering
    \includegraphics[width=0.4\linewidth]{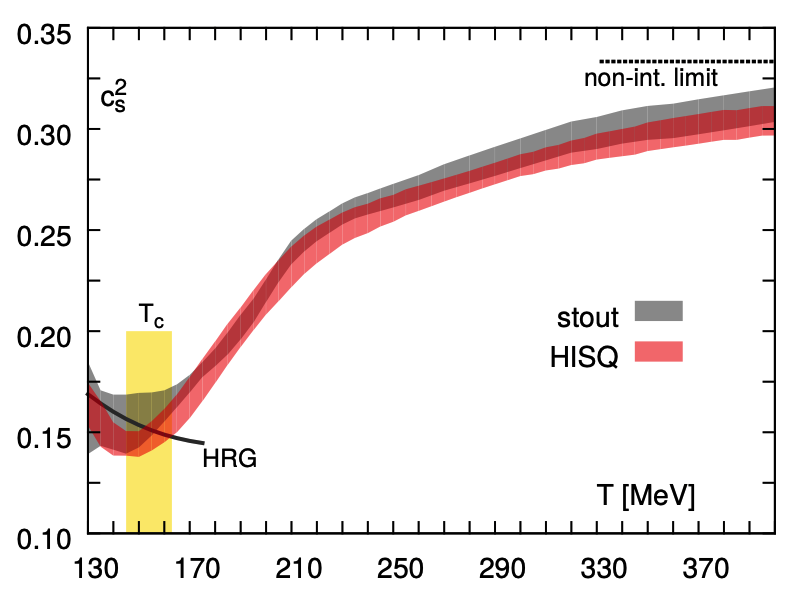}
    \caption{$c^2_{s}$ from lattice QCD model calculations and the Hadron Resonance Gas model versus temperature. The vertical band marks the location of the crossover region $T_c = (154\pm9)$ MeV. The value for ideal massless gas, $c_s^2 = 1/3$, is also shown~\cite{HotQCD:2014kol}.}
\label{fig:csLattice}
\end{figure}

Figure~\ref{fig:csLattice} illustrates the relationship between the speed of sound squared, $c_s^2$, and temperature, $T$, derived from various theoretical calculations \cite{HotQCD:2014kol, Borsanyi:2013bia}. For reference, the value for an ideal massless gas, $c_s^2 = 1/3$, is also depicted. The EoS exhibits its ``softest point'' at the minimum of the speed of sound curve, which occurs around $T \simeq (145-150)~\mathrm{MeV}$. This softest point of the EoS is of particular interest in the phenomenology of heavy-ion collisions, because it signifies the temperature and energy density range where the expansion and cooling of the created matter slows down considerably. Consequently, measurement of observables sensitive to $c^2_s$ in heavy-ion collisions is crucial to constrain the EoS of the system.

Real fluids, however, exhibit dissipative effects such as viscosity. These effects are incorporated into the hydrodynamic framework through viscous corrections to the ideal energy-momentum tensor, resulting in a \textit{viscous fluid}:
\begin{equation}
\label{eq:T_viscous}
T^{\mu\nu} = T^{\mu\nu}_{\text{ideal}} + \pi^{\mu\nu} + \Delta \pi g^{\mu\nu}
\end{equation}
Here, $\pi^{\mu\nu}$ is the \textit{shear stress tensor}, which accounts for momentum transport caused by velocity gradients leading to shearing motions, and $\Delta \pi$ is the \textit{bulk viscous pressure}, which arises from resistance to volume changes (expansion or compression). These viscous terms are directly related to the fluid's transport coefficients: the \textit{shear viscosity} ($\eta$) and the \textit{bulk viscosity} ($\zeta$). Specifically, $\pi^{\mu\nu}$ is proportional to $\eta$ and velocity gradients, while $\Delta \pi$ is proportional to $\zeta$ and the divergence of the four-velocity $\partial_\mu u^\mu$ (representing the expansion rate).


\subsubsection{Shear Viscosity to Entropy Density Ratio ($\eta/s$)}
\begin{figure}[htbp]
\centering
\includegraphics[width=0.5\linewidth]{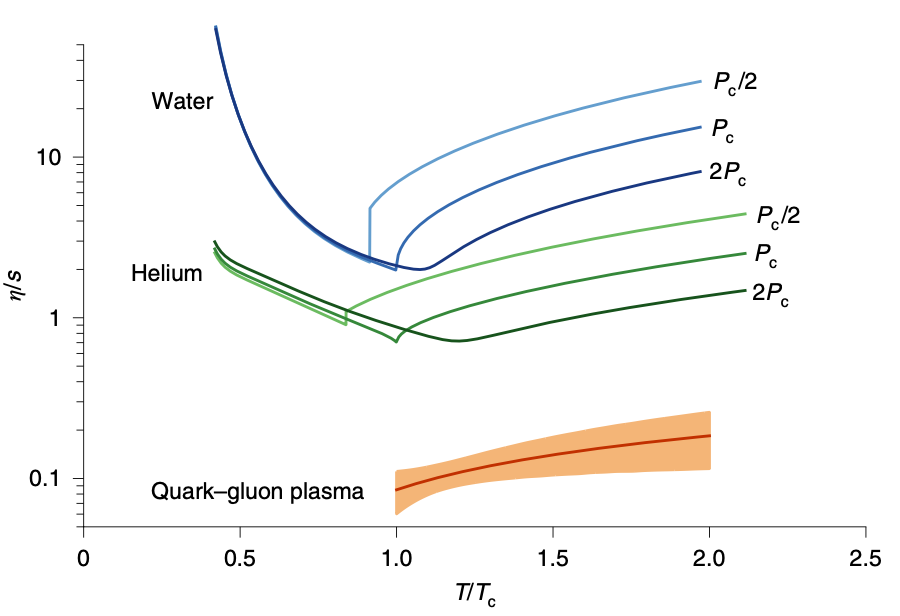}
\caption{Estimated temperature–dependent $\eta/s$ of the QGP compared with common fluids. The orange line and band show the median and 90\% credible region for the QGP $\eta/s(T)$ estimated from Pb--Pb collision data at $\sqrt{s_{\rm NN}}=2.76$ and $5.02\,$TeV. The blue and green lines show $\eta/s(T)$ for water and helium at different pressures relative to their critical pressures, as annotated, calculated from NIST data~\cite{NIST_Fluid}. The temperature dependence is plotted relative to each fluid’s critical temperature, $T/T_{\rm c}$~\cite{Bernhard:2019bmu}.}
\label{fig:shearbulk}
\end{figure}

The ratio $\eta/s$ indicates the fluid's ability to transport momentum (dissipation) relative to its information content (entropy). A smaller value of $\eta/s$ signifies a less dissipative, ``more perfect'' fluid. As shown in Figure~\ref{fig:shearbulk}, the estimated $\eta/s$ of the QGP across a range of temperatures is lower than that of most known substances, including water and helium at various conditions. The exceptionally low value of $\eta/s$ estimated for the QGP from analyses of HIC data and its proximity to the conjectured quantum lower bound of $1/4\pi$~\cite{Kovtun:2004de,Policastro:2001yc}, is a key result supporting the ``nearly perfect fluid'' description. 

\subsubsection{Bulk Viscosity to Entropy Density Ratio ($\zeta/s$)}
The ratio $\zeta/s$ quantifies the fluid's resistance to isotropic expansion or compression. Non-zero bulk viscosity can lead to deviations from equilibrium pressure during the rapid expansion phase of the QGP. Theoretical calculations suggest $\zeta/s$ may be enhanced near the QCD phase transition temperature, and experimental data are also used to constrain its temperature dependence.

\subsubsection{Impact of Viscosity on Observables}
\begin{figure}[htbp]
\centering
\includegraphics[width=0.45\linewidth]{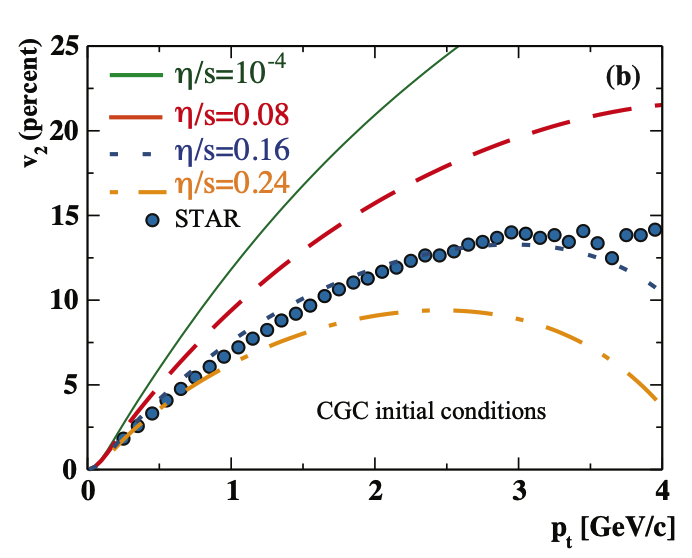}
\caption{The transverse momentum dependence of $v_{2}$ compared to viscous hydrodynamic calculations with different $\eta/s$~\cite{Luzum:2008cw}}
\label{fig:v2depetaS}
\end{figure}

Viscosity acts to dampen the collective flow, meaning that a more viscous fluid will develop less anisotropy in momentum space from a given initial spatial anisotropy. Consequently, observables sensitive to flow, such as the elliptic flow coefficient $v_2$, are directly affected by the fluid's viscosity. Comparing experimental measurements of $v_2$ as a function of transverse momentum and centrality to predictions from viscous hydrodynamic models allows for quantitative constraints on the QGP's transport coefficients, especially $\eta/s$. Figure~\ref{fig:v2depetaS} illustrates how different assumed values of $\eta/s$ in hydrodynamic calculations lead to distinct predictions for $v_2$, highlighting its sensitivity. Constraining the temperature dependence of both $\eta/s$ and $\zeta/s$ is a major goal of the analysis programs at RHIC and the LHC.

The success of hydrodynamics, therefore, lies not only in describing the collective motion but also in providing a framework where the measured flow patterns can be used to extract fundamental transport properties of the QGP, such as its viscosities and the Equation of State, which are key characteristics of this state of matter. However, it is important to note that hydrodynamics is an effective description that applies once the system is sufficiently equilibrated. It does not describe the initial pre-equilibrium stage of the collision or the final stages where the medium becomes too dilute.

\subsection{Particlization and the Hadronic Phase}
As the system expands and cools, its temperature drops towards the critical temperature ($T_c \approx 155 \text{ MeV}$) where the transition from the deconfined QGP phase to a confined state of hadronic matter occurs. Hydrodynamic evolution describes the system up to a certain freeze-out hypersurface, typically characterized by a constant temperature or energy density, where the fluid description is no longer applicable. At this point, the system undergoes \textit{particlization}, meaning the collective fluid degrees of freedom are converted into individual hadrons.

This conversion is commonly described by the \textit{Cooper-Frye formula}, which calculates the yield of particle species $i$ with degeneracy $g_i$ and momentum $p$ on the particlization hypersurface $\Sigma$:
\begin{equation}
E\frac{dN_i}{d^3p} = \frac{g_i}{(2\pi)^3} \int_\Sigma f_i(x,p) p^\mu d\sigma_\mu
\label{eq:cooper_frye}
\end{equation}
where $f_i(x,p)$ is the phase-space distribution function of particle species $i$ on the hypersurface $\Sigma$, and $d\sigma_\mu$ is the Lorentz-invariant element of the freeze-out hypersurface.

Following particlization, the system enters a \textit{hadronic phase}. In this phase, the produced hadrons continue to interact via elastic and inelastic scatterings, and unstable resonances decay. This stage can modify the momentum distributions inherited from the particlization hypersurface. The evolution during the hadronic phase is typically modeled using \textit{hadronic transport approaches} (often referred to as hadronic cascade models, e.g., UrQMD, SMASH), which simulate the interactions of individual particles. Realistic descriptions of HICs at collider energies require coupling the hydrodynamic evolution of the QGP phase to a hadronic transport stage, as the final-state observables are a result of the combined evolution through both stages.

\section{Bayesian Parameter Estimation}
The sophisticated theoretical models described previously involve numerous parameters, such as those related to initial state, transport coefficients, and particlization. To rigorously constrain these parameters against the wide array of experimental observables, modern heavy-ion physics increasingly relies on Bayesian parameter estimation frameworks. Approaches like the Jetscape~\cite{JETSCAPE:2020shq} and Trajectum~\cite{Nijs:2020ors} frameworks use statistical methods to simultaneously constrain multiple model parameters by comparing model predictions to experimental data.

These frameworks typically involve:
\begin{itemize}
    \item Initial-State Models: Using flexible models like TRENTo or IP-Glasma.
    \item Hydrodynamic Evolution: Simulating the space-time evolution using viscous hydrodynamics (e.g., VISH2+1, MUSIC).
    \item Hadronic Transport: Modeling the post-particlization phase (e.g., UrQMD, SMASH).
    \item Emulators: Building fast statistical models (emulators) that mimic the full model chain's output for rapid prediction.
    \item Bayesian Inference: Using techniques like Markov Chain Monte Carlo (MCMC) to sample the parameter space and determine the probability distribution of the parameters given the experimental data.
\end{itemize}
Bayesian analysis provides a rigorous way to quantify uncertainties in the extracted QGP properties and identify correlations between different parameters. This is important for moving beyond qualitative comparisons and obtaining precise, quantitative constraints on the fundamental characteristics of the QGP.

\begin{figure}
    \centering
    \includegraphics[width=1.0\linewidth]{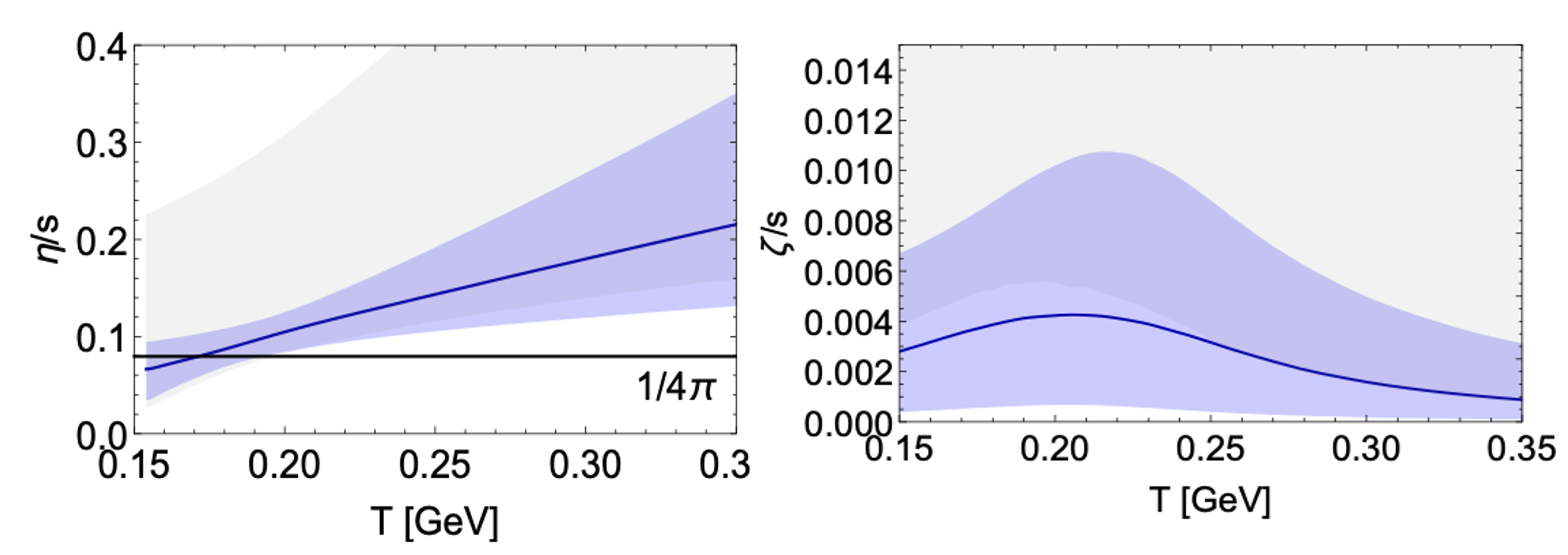}
    \caption{Bayesian estimates of the temperature dependence of $\eta/s$ and $\zeta/s$ from the Trajectum framework.  The shaded bands denote the posterior credible intervals, illustrating the substantial uncertainties in the extracted values~\cite{Nijs:2020ors}.}
    \label{fig:modeluncert}
\end{figure}

Despite the progress offered by these Bayesian frameworks, the extraction of QGP properties, such as the transport coefficients, $\eta/s$ and $\zeta/s$, remains subject to considerable uncertainties, as shown in Figure~\ref{fig:modeluncert}. A primary source of this uncertainty stems from the inherent degeneracy between the initial-state description and the medium's response. Different initial-state models, or different parameter choices within a single flexible model like TRENTo, can lead to similar final-state observables when combined with varying values of transport coefficients. This means that the models can sometimes compensate for different assumptions about the initial geometry with adjustments in the viscosities, leading to a situation where multiple combinations of initial-state parameters and QGP properties can equally well describe the experimental data. This degeneracy makes it challenging for even sophisticated Bayesian analyses, which fit to a wide range of observables, to uniquely determine the values of these fundamental medium properties, resulting in the relatively large uncertainty bands observed in current extractions.

Furthermore, the list of experimental observables measured in the past, while extensive, may not be sufficiently sensitive to disentangle the subtle interplay between initial-state fluctuations, the early time dynamics, and the viscous evolution throughout the QGP lifetime. 

To break these degeneracies and provide tighter constraints on the QGP's transport properties, it is of vital importance to develop and incorporate novel experimental measurements that exhibit strong differential sensitivity to distinct stages of the collision evolution. Specifically, observables that are highly sensitive to the detailed structure of the initial state, as well as those that probe the medium's response are needed. Such measurements can provide orthogonal constraints that limit the allowed parameter space in Bayesian analyses, paving the way for a more precise and robust determination of the QGP's fundamental properties and a deeper understanding of the dynamics governing its behavior.

\clearpage

\clearpage

\newpage
\chapter{Motivation}
\label{sec:motivation}

While the QGP offers profound insights into the strong interaction and the early universe, its properties must be inferred indirectly from final-state particles. As discussed previously, significant challenges remain in precisely constraining these properties due to model degeneracies and experimental uncertainties. 

This goal of this dissertation is to address these limitations by developing and applying a suite of novel experimental probes that utilize event-by-event multi-particle correlations. By working backward from the final-state particles, these probes allow for more stringent constraints on the collective evolution, transport coefficients, and initial-state characteristics of the QGP.

To advance this endeavor, this work presents three complementary analyses that apply multi-particle correlation techniques to data from $^{208}\text{Pb}+^{208}\text{Pb}$ and $^{129}\text{Xe}+^{129}\text{Xe}$ collisions, as recorded by the ATLAS detector at the LHC. Each analysis investigates a different aspect of the QGP by addressing a specific and pressing scientific question:

\begin{enumerate}
    \item \textbf{Establishing Collective Nature of Radial Flow Fluctuations and Constraining Bulk Viscosity:} While anisotropic flow ($v_n$) unambiguously demonstrates the collective, fluid-like behavior, direct experimental evidence for the \emph{collective nature} of its radial expansion remains less established. To address this gap, the {\it first measurement of the $p_T$-differential radial flow fluctuation observable, $v_0(p_T)$} is discussed in Chapter~\ref{sec:chap4_v0pt}. 
    
    This probe quantifies event-by-event deviations in the isotropic transverse boost. The observation of $v_0(p_T)$ correlations persisting across large pseudorapidity gaps, and their factorization into single-particle components, would confirm an early-time, bulk origin, solidifying the collective picture of radial flow. This novel probe provides not only the first comprehensive experimental evidence for the collective nature of the QGP's radial expansion but also offers heightened sensitivity to the bulk viscosity, $\zeta/s$, a transport coefficient that has hitherto been poorly constrained by other flow observables.
    
    \item \textbf{Disentangling Initial-State Fluctuations and Probing the Equation of State:} The QGP's evolution is critically influenced by its initial conditions, including not only the overall collision geometry but also intrinsic quantum variations. Event-by-event fluctuations in the average transverse momentum ($\langle p_T \rangle$) encode information about both these sources. Yet, disentangling their respective contributions experimentally is challenging.
    
     By examining higher moments (e.g., variance and skewness) of the event-wise $\langle p_T \rangle$ distribution in ultra-central collisions where geometric fluctuations are minimal, Chapter~\ref{sec:chap5_ptfluc} provides {\it the first experimental constraints on the contribution of these initial-state variations}. 
     
     Moreover, since the propagation and damping of initial density perturbations depend on the QGP’s equation of state, notably the speed of sound, $c_s$, Chapter~\ref{sec:chap5_ptfluc} further demonstrate the sensitivity of fluctuation measures to $c_s$, providing key insights to constrain this vital medium property.

    \item \textbf{Constraining Nuclear Structure and Initial conditions through Flow-Momentum Correlations:} The precise geometry of the initial nuclear overlap region, dictated primarily by the colliding nuclei's shape and internal sub-nucleonic structure—profoundly influences the QGP’s subsequent evolution and all final-state observables. This motivates using heavy-ion collisions as a unique tool to investigate nuclear structure itself. 

     The event-by-event correlation between anisotropic flow coefficients ($v_n$) and $\langle p_T \rangle$ discussed in Chapter~\ref{sec:chap6_vnpt} serves as a unique lens on the initial collision geometry, providing stringent experimental constraints on initial-condition models and, notably, offering the {\it first experimental evidence for the triaxial deformation of the $^{129}\text{Xe}$ nucleus}.
\end{enumerate}

Collectively, these three complementary analyses, through the application and development of novel multi-particle correlation techniques, are driven by the overarching goal to improve understanding of the QGP's dynamic expansion, its fundamental transport properties, and the complexity of its initial conditions. This work thereby aims to open new avenues for precision studies of this primordial state of matter.

\section{Tool to be used: Multi-Particle Correlations}\label{sec:multcorr}

As outlined in the motivation for this dissertation, characterizing the Quark-Gluon Plasma (QGP) relies on analyzing final-state particle distributions. These distributions in azimuthal angle ($\phi$) and transverse momentum ($p_T$), retain vital information about the initial collision geometry, its fluctuations, the system's subsequent collective evolution, and its hadronization. This section details the multi-particle correlation method, the primary tool employed in this work to extract this information.

\subsubsection{Initial-state fluctuations and final-state observables}
Event-by-event fluctuations in the $\phi$ distribution and $p_T$ spectra in HIC arise from several sources:
\begin{itemize}
    \item Quantum fluctuations in the nucleon positions within colliding nuclei.
    \item Sub-nucleonic fluctuations originating from the internal structure of nucleons.
    \item Initial-state dynamics, including gluon saturation effects.
    \item Fluctuations in the energy deposition process.
    \item Effects from the transport properties of the QGP medium.
\end{itemize}

The mapping between initial-state variables and final-state observables is generally expressed in terms of linear correlations. For example, the final-state anisotropy in the $\phi$ distribution of particles, $v_n$, is proportional to the initial-state eccentricity, $\varepsilon_n$, through a response coefficient $\kappa_n$:
$$v_n \propto \kappa_n \varepsilon_n$$
Here, $\kappa_n$ depends on the medium properties, primarily the $\eta/s$. Beyond this linear response, non-linear coupling between different harmonic orders introduces mode-mixing effects, which are important for higher-order harmonics~\cite{Gardim:2011xv,Teaney:2012ke}. Similarly, fluctuations in the event-wise mean of $p_T$, $\langle p_T \rangle$, and its higher moments, provide complementary information about initial-state energy density fluctuations and the subsequent thermalization process \cite{Broniowski:2009fm,Bozek:2012hy}. These fluctuations of the final-state observables, such as $v_{n}$ and $\langle p_T \rangle$, can be quantified in terms of the moments of their distribution in the final state.

\subsubsection{Mathematical framework of multi-particle correlations}
Multi-particle correlation techniques provide a powerful mathematical framework for systematically quantifying these fluctuations while suppressing unwanted background contributions. In addition, they offer direct access to the moments and cumulants of the event-by-event distribution of an observable $X$ (e.g., $v_n$ or $\langle p_T \rangle$). The relationship between cumulants $\kappa_m$ and central moments $\mu_m = \langle(X-\langle X\rangle)^m\rangle$ is given by:
\begin{align}
\kappa_1 &= \langle X\rangle, \\
\kappa_2 &= \mu_2 = \langle X^2\rangle - \langle X\rangle^2, \\
\kappa_3 &= \mu_3 = \langle (X-\langle X\rangle)^3\rangle, \\
\kappa_4 &= \mu_4 - 3\mu_2^2 = \langle (X-\langle X\rangle)^4\rangle - 3\,\kappa_2^2.
\end{align}

These cumulants characterize the shape of the probability distribution:
\begin{itemize}
    \item $\kappa_1$: Mean (location)
    \item $\kappa_2$: Variance (width)
    \item $\kappa_3/\kappa_2^{3/2}$: Skewness (left-right asymmetry)
    \item $\kappa_4/\kappa_2^2$: Kurtosis (tailedness)
\end{itemize}

In the case of $n$-particle azimuthal correlations, the $2k$-particle cumulant for a given harmonic order $n$ is constructed as:
\begin{equation}
c_n\{2k\} = \bigl\langle\!\bigl\langle e^{\,i n(\phi_1 + \cdots + \phi_k - \phi_{k+1} - \cdots - \phi_{2k})}\bigr\rangle\!\bigr\rangle,
\end{equation}
where the double angular brackets denote averaging over all particles within an event and then over all events in the ensemble. For flow analysis, one typically focuses on configurations where $n_1+n_2+\cdots+n_k=0$ to ensure rotational invariance.

When related to flow harmonics $v_n$ in the absence of non-flow backgrounds, these cumulants yield:
\begin{align}
c_n\{2\} &= v_n^2 \\
c_n\{4\} &= v_n^4 - 2(v_n^2)^2 = -v_n^4 + \mathcal{O}(\sigma_{v_n}^4) \\
c_n\{6\} &= v_n^6 - 9v_n^2v_n^4 + 12(v_n^2)^3 = v_n^6 + \mathcal{O}(\sigma_{v_n}^2)
\end{align}

The corresponding flow harmonics can then be extracted as:
\begin{align}
v_n\{2\} &= \sqrt{c_n\{2\}} \\
v_n\{4\} &= \sqrt[4]{-c_n\{4\}} \\
v_n\{6\} &= \sqrt[6]{c_n\{6\}/4}
\end{align}

Several computational methods have been developed to efficiently calculate multi-particle correlations, such as the generating function method \cite{Borghini:2001vi}, the Q-cumulant method \cite{Bilandzic:2010jr}, and related modifications using the subevent method \cite{Jia:2017hbm}. In this dissertation, we primarily utilize the Q-cumulant method and its modification for use within a subevent framework.

The Q-cumulant method \cite{Bilandzic:2010jr} utilizes flow vectors, $Q_{n,p} = \sum_{j=1}^M w_j^p e^{in\phi_j}$, to efficiently compute multi-particle azimuthal correlations without explicit nested loops. The subevent method divides each event into $m$ rapidity-separated regions and correlates particles across these regions, defined as:
$$c_n\{2k\}^{m-\text{subevent}} = \bigl\langle\!\bigl\langle e^{in(\phi_1^{A_1} + \cdots + \phi_k^{A_k} - \phi_{k+1}^{A_{k+1}} - \cdots - \phi_{2k}^{A_{2k}})}\bigr\rangle\!\bigr\rangle$$
where $A_i$ denotes different subevents. This approach substantially reduces short-range non-flow contributions.

The multi-particle correlation approach has thus emerged as an indispensable toolset in heavy-ion physics, providing model-independent access to collective dynamics. By connecting measurable final-state correlations to initial-state geometry and medium properties, this framework enables precise constraints on the equation of state and transport coefficients of the QGP, ultimately advancing the understanding of strongly interacting matter under extreme conditions.

\subsection{Challenges in using Multi-Particle Correlations}\label{sec:challenge}
One of the primary challenges in multi-particle correlation measurements is background contamination from ``non-flow'' correlations among a small number of particles arising from resonance decays, jets, or multi-jet production. These non-flow contributions are important in smaller collision systems and more peripheral events, where the overall charged-particle multiplicity is low and typically concentrated near mid-rapidity.

\begin{figure}[htbp]
    \centering
    \includegraphics[width=0.5\linewidth]{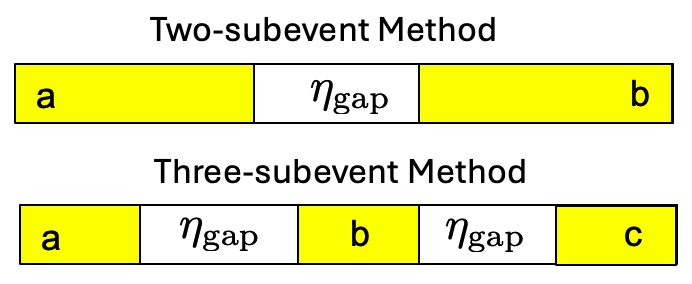}
    \caption{Illustration of typical division of $\eta$-ranges in different subevent methods.}
    \label{fig:subevent}
\end{figure}

The multi-particle method of calculating cumulants provides a first line of defense against such correlations, which typically involve a small number of particles. The combinatorial subtraction of lower-order correlations isolates the genuine $2k$-particle correlations that cannot be factorized into products of lower-order terms. This structure ensures the systematic elimination of non-flow effects order by order.

Another common strategy to suppress non-flow is to introduce a $\eta$-gap between the particles being correlated, as shown in Figure~\ref{fig:subevent}. In the simplest two-particle correlation, placing a single gap around mid-rapidity already reduces much of the non-flow background. However, residual correlations from di-jet fragmentation can persist, as particles from di-jets populate two widely separated opposite $\eta$ ranges. To further mitigate this, we employ a three-subevent method: the event is divided into three consecutive $\eta$ regions, and correlations are formed only between particles in different subevents. By ensuring an $\eta$-gap between each pair of subevents, this approach effectively suppresses di-jet contributions~\cite{Jia:2017hbm,Huo:2017nms}.

To quantify the effectiveness of multi-particle correlation method or subevent method towards suppression of non-flow, the Heavy Ion Jet Interaction Generator (HIJING) model~\cite{Wang:1991hta, Gyulassy:1994ew} is often used. The HIJING is a Monte Carlo event generator widely used in the heavy-ion community. It simulates nucleus-nucleus collisions as a superposition of independent nucleon-nucleon ($pp$) collisions, incorporating effects such as jet production, parton shadowing, and multiple scattering. This feature makes HIJING a valuable tool for establishing baselines in multi-particle correlation studies. For instance, it can provide an estimate of the independent superposition baseline for mean transverse momentum ($\langle p_T \rangle$) fluctuations, allowing for the identification of any excess fluctuations in the data that might signal collective behavior beyond what is expected from uncorrelated sources.

Furthermore, HIJING serves as a tool for characterizing the non-flow baseline, especially in small collision systems and peripheral heavy-ion collisions where non-flow contributions from jets and resonance decays are large. By comparing multi-particle correlation measurements in experimental data with HIJING simulations, one can quantify the degree of non-flow suppression achieved by techniques like the subevent method and ultimately isolate the genuine collective signals arising from the hot, dense matter created in heavy-ion collisions. These comparisons are essential for a robust interpretation of the experimental results and for extracting meaningful information about the properties of the QGP.

\section{Organization of the Dissertation}

This dissertation addresses central challenges in the study of heavy-ion collisions by introducing novel experimental observables and sophisticated analysis methodologies. Building upon the motivation and framework established in the preceding chapters, the subsequent chapters detail the specific measurements, comparisons with theoretical models, and their implications for understanding the QGP. The core analyses and findings are presented in Chapters~\ref{sec:chap4_v0pt}-\ref{sec:chap6_vnpt}.

The structure of this dissertation is as follows:

Chapter~\ref{chap:expt} introduces the experimental facilities employed for collecting the heavy‑ion collision data analyzed here, detailing the LHC configuration and the relevant ATLAS sub‑detectors.  

Chapter~\ref{sec:data} outlines the event cleaning procedures and track selection criteria applied to both Pb+Pb and Xe+Xe collision datasets, and describes the measures taken to mitigate detector and tracking inefficiencies.  

Chapter~\ref{sec:chap4_v0pt} presents a novel observable for probing QGP collectivity via $p_{T}$‑differential radial flow fluctuations.  This chapter motivates the new observable, discusses its theoretical implications and complementarity with existing methods, and then details the ATLAS measurement procedure and results, focusing on their impact on understanding collectivity.

Chapter~\ref{sec:chap5_ptfluc} establishes a method to disentangle geometric from non‑geometric contributions to radial flow fluctuations.  After reviewing limitations in current directed‑flow–fluctuation studies, the chapter demonstrates how mean transverse momentum and its event‑by‑event fluctuations can be used to constrain the speed of sound in QGP. 

Chapter~\ref{sec:chap6_vnpt} reports the first experimental constraints on higher‑order deformations of $^{129}$Xe nuclei using flow–mean‑$p_{T}$ correlations in spherical Pb+Pb and deformed Xe+Xe collisions.  The influence of nuclear geometry and nucleon substructure on initial‑state fluctuations is discussed, followed by an analysis of how these fluctuations propagate to final‑state observables and a discussion of the technique’s potential for nuclear‑deformation measurements.

Chapter~\ref{sec:summary} summarizes the key findings in the context of the scientific questions posed in this chapter, evaluates the advances made, and identifies open directions for future research.  

The Appendices document several technical aspects, including a comprehensive account of the working formulae and systematic uncertainty estimation procedures used in this dissertation.
\clearpage

\newpage
\chapter{Experimental Setup}
\label{chap:expt}

This chapter provides a description of the experimental setup relevant to this work, focusing on the Large Hadron Collider and the ATLAS detector subsystems used in the analyses presented in the subsequent chapters.

The Large Hadron Collider (LHC)~\cite{Evans:2008zzb}, located in CERN’s 27 km underground tunnel near Geneva, Switzerland, is a two-ring superconducting hadron accelerator and collider. Capable of reaching $\sqrt{s}=13$--$14$ TeV for $pp$ collisions with a peak design luminosity of $10^{34}\,\mathrm{cm}^{-2}\mathrm{s}^{-1}$, it also delivers heavy-ion collisions, such as Pb+Pb and Xe+Xe, at center-of-mass energies per nucleon pair $\sqrt{s_{NN}}\approx5$ TeV, with design luminosities up to $10^{27}\,\mathrm{cm}^{-2}\mathrm{s}^{-1}$. The LHC utilizes 1232 superconducting dipole magnets for steering the beams, numerous quadrupole magnets for focusing, and RF cavities for accelerating the particles and maintaining beam stability.

ATLAS is one of the four major experiments situated at LHC interaction points (alongside CMS, ALICE, and LHCb). Its comprehensive design, incorporating an inner tracking system, calorimeters, and a muon spectrometer, provides broad kinematic coverage suitable for both proton and heavy-ion physics programs. Data acquisition involves a multi-level trigger system that selects events in real time. Subsequent offline processing includes event cleaning, pileup suppression, and rigorous track-selection criteria to ensure uniform data quality and acceptance across datasets, detailed in Chapter~\ref{sec:data}.

\begin{figure}[h!]
    \centering
    \includegraphics[width=0.8\linewidth]{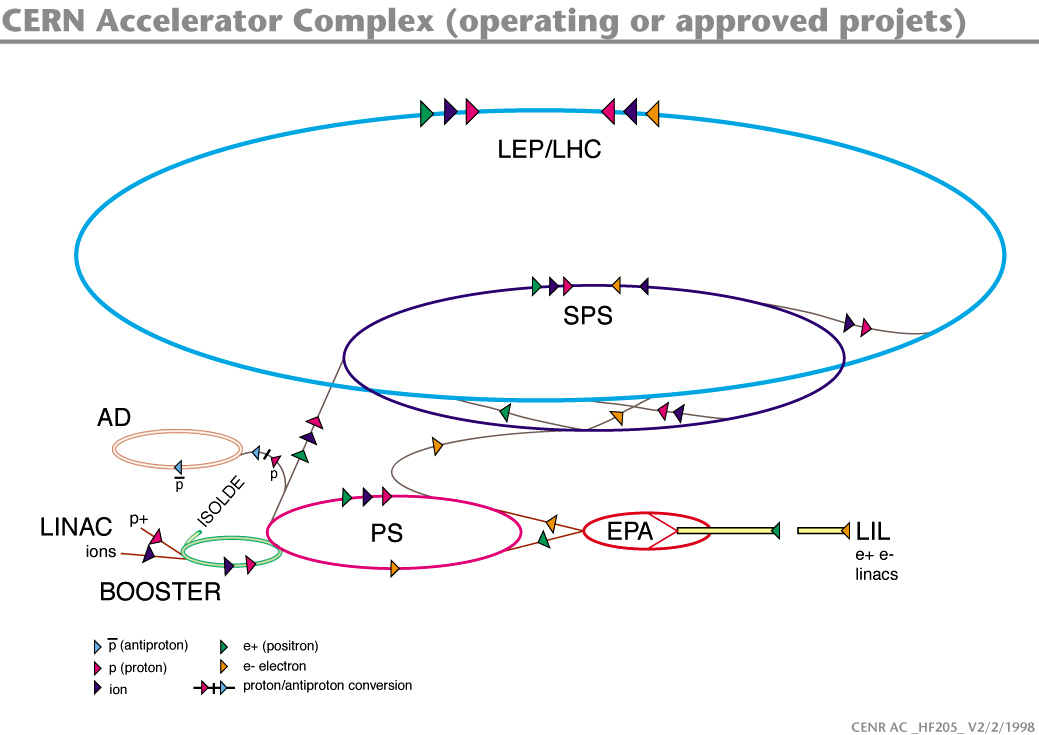}
    \caption{The CERN accelerator sequence~\cite{Caron:1991}.}
    \label{fig:cern}
\end{figure}

Heavy ions are prepared for injection into the LHC through a sequence of pre-accelerators: LINAC3, LEIR, PS, and SPS, as illustrated in Figure.~\ref{fig:cern}. The beam energy is progressively increased at each stage, reaching 4.2 MeV in LINAC3, 72 MeV in LEIR, 6 GeV in PS, 177 GeV in SPS, and finally 5 TeV per nucleon in the LHC.

\section{The ATLAS Detector}\label{sec:detector}

The ATLAS (A Toroidal LHC ApparatuS) detector~\cite{ATLAS:2023mnw,ATLAS:2008xda} is one of the two general-purpose experiments at the LHC, located at Interaction Point 1. Measuring 46 m in length and 25 m in diameter, and weighing approximately 7000 tonnes, ATLAS shares the scientific goals of its counterpart, CMS, at the energy and intensity frontiers. Although primarily optimized for high-luminosity $pp$ operation, its fine granularity and large acceptance are also highly advantageous for studying heavy-ion collisions. The datasets analyzed in this dissertation include Pb+Pb collisions at $\sqrt{s_{NN}} = 5.02\,$TeV and Xe+Xe collisions at $\sqrt{s_{NN}} = 5.44\,$TeV.

\begin{figure}[h!]
  \centering
  \includegraphics[width=0.9\linewidth]{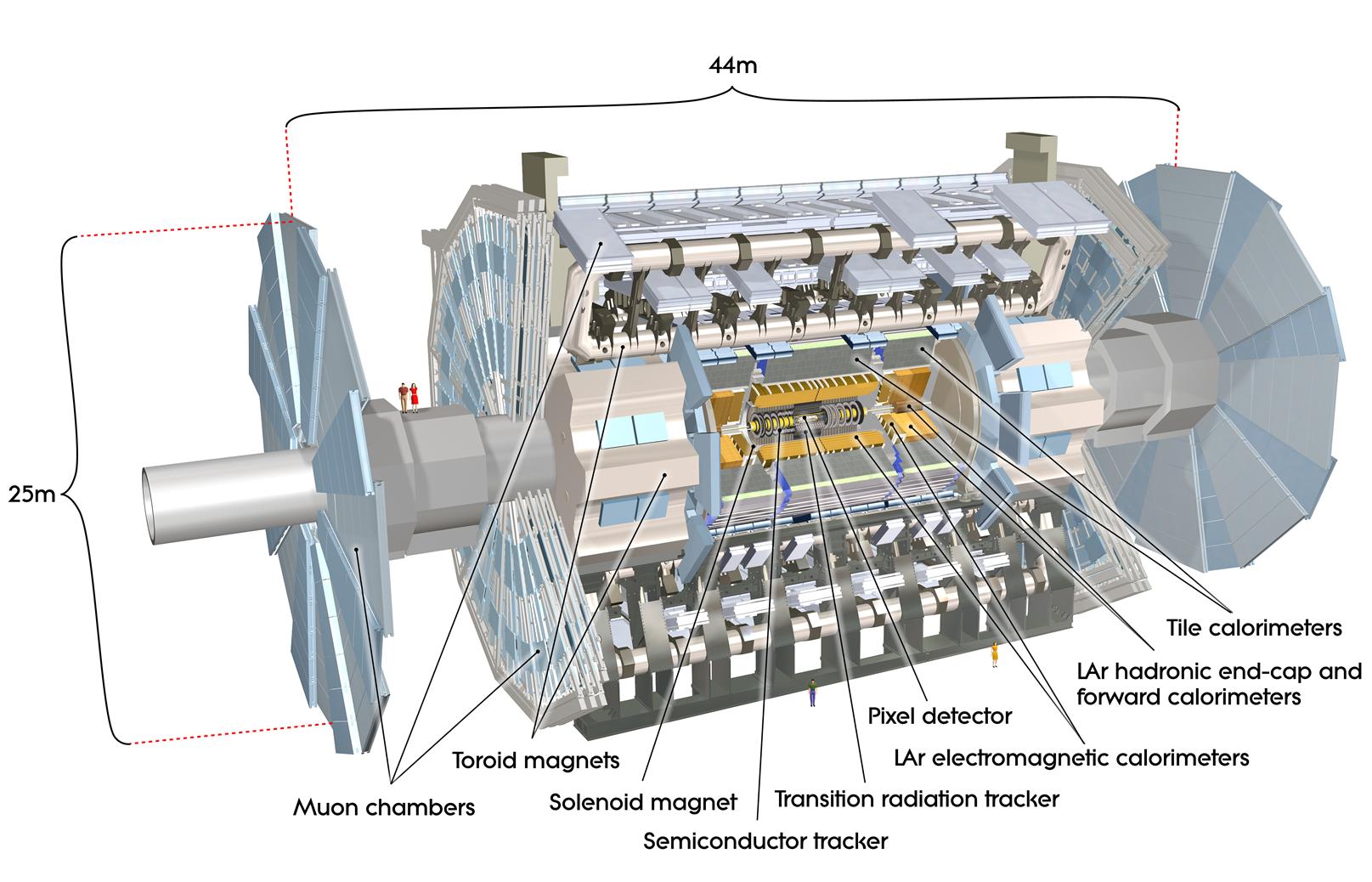}
  \caption{Cut-away view of the ATLAS detector, showing the main subsystems~\cite{ATLAS:2008xda}.}
  \label{fig:atlas_detector_overview}
\end{figure}

The detector exhibits approximate cylindrical symmetry around the beam axis, with full azimuthal coverage. A standard right-handed coordinate system is defined with the interaction point as the origin. The $z$-axis is aligned with the beam direction, with the positive $z$ direction pointing towards the ``A'' side of the detector and the negative $z$ direction towards the ``C'' side. The $x-y$ plane is transverse to the beam, with the positive $x$ direction pointing towards the center of the LHC ring and the positive $y$ direction pointing upwards.

The principal detector subsystems relevant to this work are:
\begin{itemize}
  \item \textbf{Inner Detector (ID)}: Responsible for tracking charged particles in a 2 T solenoidal magnetic field over $|\eta|<2.5$.
  \item \textbf{Forward Calorimeter (FCal)}: Measures forward transverse energy up to $|\eta|<4.9$.
  \item \textbf{Zero-Degree Calorimeter (ZDC)}: Detects spectator neutrons at very forward rapidities, $|\eta|>8.3$.
  \item \textbf{Minimum-Bias Trigger Scintillators (MBTS)}: Provide low-bias triggers, especially for inelastic Pb+Pb events.
  \item \textbf{Trigger and DAQ system}: A two-stage system (hardware L1 and software HLT) that reduces the initial 40 MHz collision rate to a manageable recording rate of a few hundred Hz.
\end{itemize}

\subsection{Inner Detector}
\begin{figure}[h!]
\centering
\includegraphics[width=0.8\linewidth]{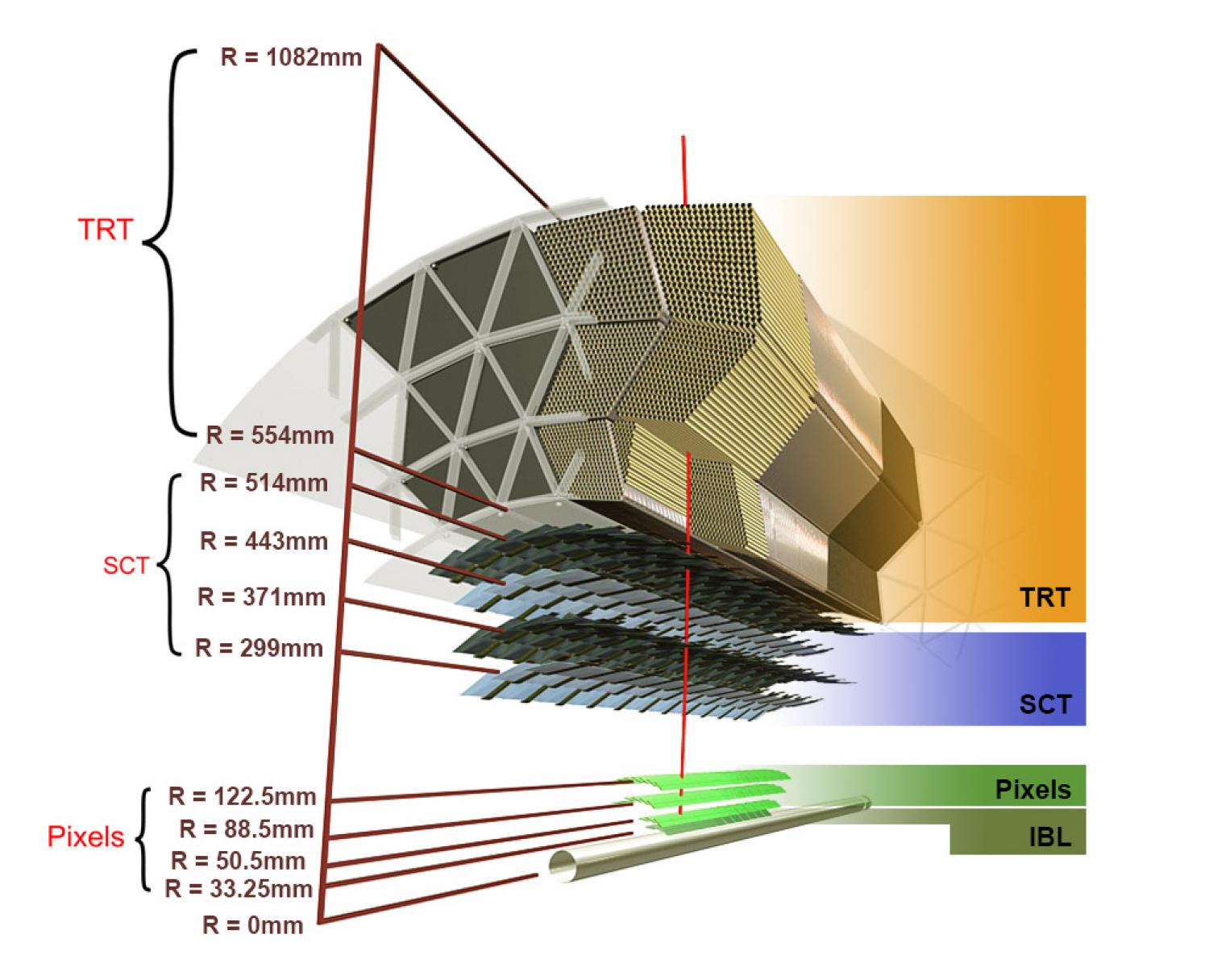}
\includegraphics[width=0.8\linewidth]{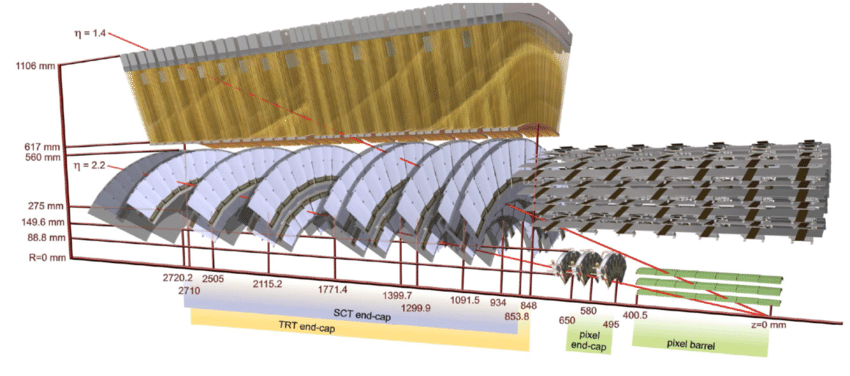}
\caption{Renderings of the ATLAS inner detector barrel (top) and end-cap (bottom) sections, along with SemiConductor Tracker and Transition Radiation Tracker.}
\label{fig:chap2_inner_det}
\end{figure}

The ATLAS Inner Detector (ID) is a composite tracking system situated at the center of the ATLAS detector. Its primary purpose is to reconstruct the trajectories of charged particles based on precision hit information. The reconstructed tracks enable the measurement of each particle's momentum and facilitate the determination of the collision vertex. The ID is designed to offer robust pattern recognition capabilities, excellent momentum resolution, and the ability to identify both primary and secondary vertices. This system operates within a 2 T solenoidal magnetic field that bends charged particle tracks, enabling high-precision momentum measurements typically down to 0.5 GeV. The ID offers full azimuthal coverage and a pseudorapidity range of $|\eta|<2.5$. It comprises three complementary sub-detectors, arranged in layers from innermost to outermost: the Pixel Detector, the SemiConductor Tracker (SCT), and the Transition Radiation Tracker (TRT). The Pixel and SCT detectors cover the region $|\eta| < 2.5$, while the TRT covers $|\eta| < 2$. Each of these sub-detectors is segmented into barrel modules (layers parallel to the beam pipe) and end-cap modules (layers perpendicular to the beam pipe). Figure~\ref{fig:chap2_inner_det} provides graphical representations of the inner detector barrel (top) and end-cap (bottom) sections.

\begin{figure}[h!]
\centering
\includegraphics[width=0.8\linewidth]{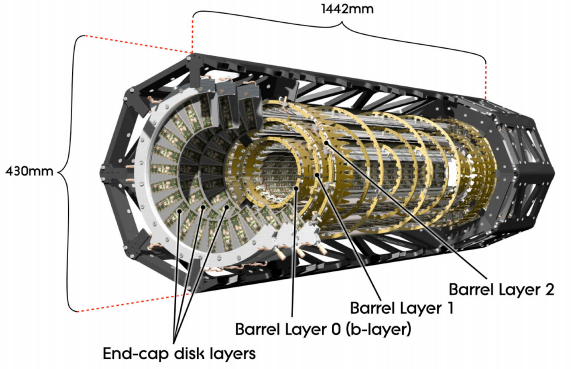}
\caption{A schematic view of the ATLAS Pixel detector with barrel and endcap layers.}
\label{fig:pixel}
\end{figure}

\paragraph{The Pixel Detector:}
The Pixel detector is the innermost ID component, offering the highest granularity and situated closest to the beam pipe. It consists of three cylindrical layers in the barrel region and three perpendicular disk layers forming the end-caps on either side of the interaction point. The detector features very high granularity, with 1744 silicon pixel modules, each containing 47232 pixels. The minimum pixel size in $R$-$\phi \times z$ is $50 \times 400 \, \mu\text{m}^2$, resulting in a total of 80.4 million readout channels. Typically, charged particle tracks traverse three pixel layers. The Pixel detector plays an important role in the precise determination of the impact parameter and the detection of short-lived particles. An additional pixel tracking layer at the innermost radius, the Insertable B-Layer (IBL), was installed for Run 2. This layer adds redundancy and improves precision for displaced vertex finding, important for b-tagging. Its installation necessitated replacing the beam pipe with one of smaller diameter in this region. A schematic diagram of the Pixel detector is shown in Figure~\ref{fig:pixel}.

\paragraph{The SemiConductor Tracker (SCT):}
The SCT surrounds the Pixel detector and utilizes silicon strip technology for precise space-point measurements. The system is designed to provide eight strip measurements (four space-points) per track traversing the intermediate radial range, thereby contributing significantly to the measurement of momentum, impact parameter, and vertex position. The detector comprises 4088 modules of silicon-strip detectors distributed across the barrel regions and two end-caps. The barrel contains four cylindrical SCT layers, each consisting of two perpendicular stereo strip layers. The SCT end-caps are each composed of nine disk layers featuring radially oriented strips and a set of stereo strips angled at 40 mrad. The total SCT readout system has approximately 6.3 million channels. The layout of barrel sensors within an SCT module, along with modules mounted on a barrel structure, is depicted in Figure~\ref{fig:chap2_inner_det}.

\paragraph{The Transition Radiation Tracker (TRT):}
The TRT is a drift-tube system that provides robust tracking information and is depicted in Figure~\ref{fig:chap2_inner_det}. It consists of three parts: a barrel and two end-caps. Its basic detection elements are thin-walled proportional drift tubes, referred to as straws. Straw tubes were selected as detecting elements due to their high degree of mod- ularity and ease of integration into a medium designed to produce transition radiation, without compromising the continuous tracking concept. The barrel comprises longitudinally oriented 144 cm straw tubes arranged in 73 planes, with the wires cut in half around $\eta = 0$. The end-caps consist of 37 cm straws oriented radially in wheels, forming 160 planes. The TRT system has a total of 351,000 readout channels. The detector geometry is such that particles traversing the region $|\eta|< 2.0$ cross 35-40 straws, providing continuous tracking at larger radii within the Inner Detector and enhancing pattern recognition capabilities. 

The TRT was not utilized in the heavy-ion analyses presented in this dissertation due to the challenging high-occupancy environment encountered in central Pb+Pb and Xe+Xe collisions, which limits its tracking and particle identification performance under such conditions.

\subsection{Forward Calorimeter}
\begin{figure}[h!]
\centering
\begin{subfigure}{.8\textwidth}
  \centering
  \includegraphics[width=.99\linewidth]{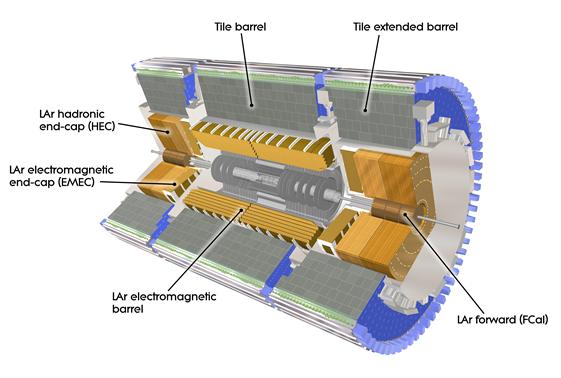}
  \caption{Overview of the ATLAS calorimeter system.}
  \label{fig:chap2_fcal1}
\end{subfigure}
\begin{subfigure}{.45\textwidth}
  \centering
  \includegraphics[width=.99\linewidth]{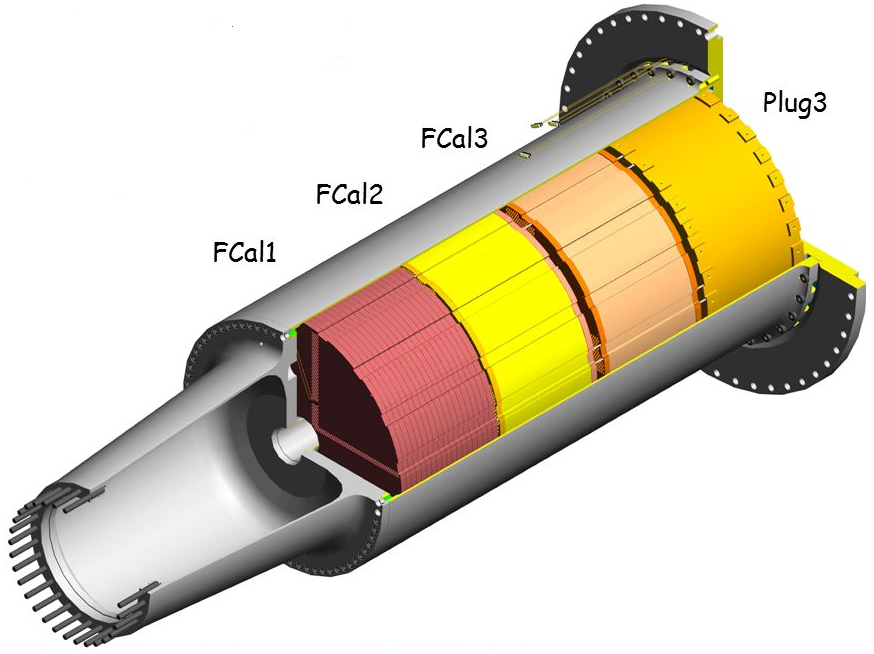}
  \caption{General arrangement and electrode spacing of the ATLAS FCal.}
  \label{fig:chap2_fcal2}
\end{subfigure}
\caption{The ATLAS calorimeter system.}
\label{fig:chap2_fcal}
\end{figure}

A calorimeter is a detector designed to measure the energy deposited by particles traversing it. Calorimeters are typically constructed to fully contain or absorb most of the particles originating from a collision, thereby forcing them to deposit all of their energy within the detector volume. They commonly consist of alternating layers of passive, high-density material (absorbers), such as lead or tungsten, interleaved with layers of an active medium, such as liquid argon or scintillating material. Interactions within the absorber material initiate a ``shower'' of secondary particles, whose energy is then sampled and detected by the active medium. Electromagnetic calorimeters are optimized to measure the energy of electrons and photons, which primarily interact electromagnetically with matter. Hadronic calorimeters are designed to measure the energy of hadrons (particles containing quarks, like protons and neutrons) through their strong interactions with atomic nuclei. Calorimeters can stop most known particles with the exception of muons and neutrinos. Figure~\ref{fig:chap2_fcal1} shows a cut-away view of the ATLAS calorimeter system, which provides nearly full azimuthal coverage across the pseudorapidity range $|\eta|<4.9$. For the analyses presented in this work, only the Forward Calorimeter (FCal) is utilized. A cut-away view of the FCal system is shown in Figure~\ref{fig:chap2_fcal2}.
The FCals are sampling calorimeters that use liquid argon (LAr) as the active medium to measure the transverse energy ($E_\text{T}$) of incident particles. Located at both ends of the ATLAS detector (referred to as FCalA and FCalC), they cover the pseudorapidity range $3.1 < |\eta|< 4.9$. Each of the two Forward Calorimeters is composed of three modules positioned sequentially: FCal1, FCal2, and FCal3, as illustrated in Figure~\ref{fig:chap2_fcal} The FCal modules are cylindrical with a coaxial hole allowing the LHC beams to pass through. The FCal1 module, constructed from copper, is situated closest to the Interaction Point and is optimized for electromagnetic measurements. The FCal2 and FCal3 modules are primarily made of tungsten and are designed for the measurement of hadronic showers. In this work, the energy deposited in the FCal cells is grouped into towers of $0.1 \times 0.1$ in $\eta-\phi$ space. Energy from cells that span multiple towers is divided proportionally between them. These calorimeter towers are employed in the final analyses.

\subsection{Zero-Degree Calorimeter}
\begin{figure}[h!]
\centering
\includegraphics[width=0.8\linewidth]{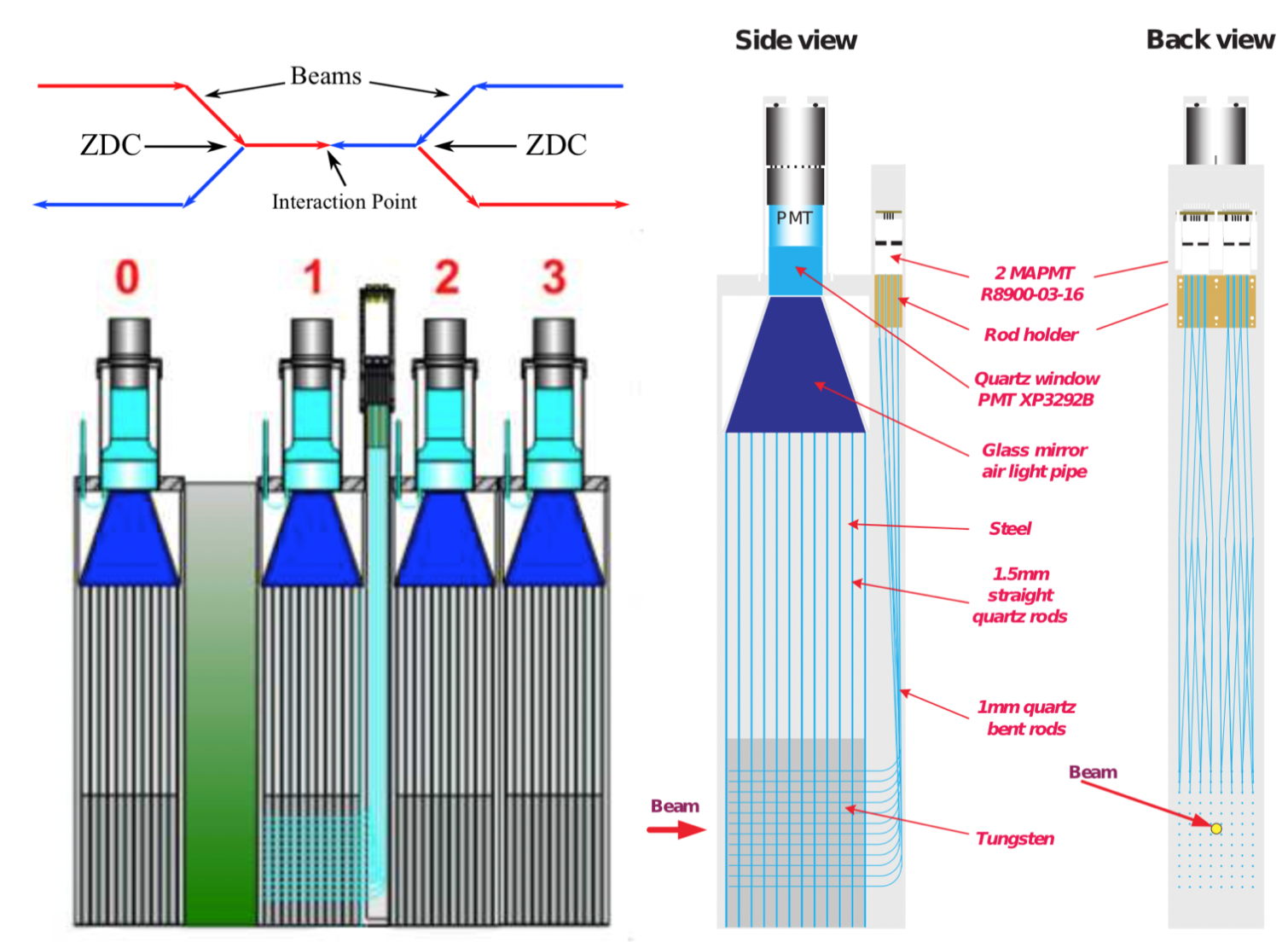}
\caption{Left: The four ZDC modules on the A-side of ATLAS. Right: Details of a single ZDC module design.}
\label{fig:chap2_zdc}
\end{figure}

The ZDC is positioned 140 m from the center of ATLAS on either side of the interaction point, downstream of where the beam pipe splits into two. It covers the very forward pseudorapidity region, $|\eta|> 8.3$. The calorimeter is named the Zero-Degree Calorimeter due to its location along the beam axis. The primary role of the ZDC in heavy-ion collisions is event characterization. Since charged particles are deflected away by the magnetic fields in the beam pipe, only neutral particles from the collision or projectile/target remnants reach the ZDC. Thus, in Pb+Pb collisions, the ZDC provides a measure of the number of spectator neutrons, which is directly related to the collision impact parameter. Each side of the ZDC comprises four modules, as shown in the left panel of Figure~\ref{fig:chap2_zdc}. The detailed design of a single module is depicted in the right panel of the same figure. Each module consists of 11 tungsten plates, 10 mm thick along the beam direction, and steel plates at the front and back, also 10 mm thick. Sandwiched between the plates are 1.5 mm diameter quartz rods that run vertically. These rods collect Cherenkov radiation produced by shower particles and guide the light via light-pipes to photomultiplier tubes (PMTs) located above. Each PMT signal is read out by multiple channels of a Pre Processor Module (PPM). The PPMs are 64-channel, 40 MHz, 10-bit ADCs. The first two ZDC modules on the C side and the second module on the A side are also equipped with quartz rods arranged in an $x-y$ grid along the beam pipe, enabling position measurements of the showers. However, these position measurements were not used in any of the analyses presented here.

\subsection{Minimum-Bias Trigger Scintillators}
\begin{figure}[h!]
\centering
\includegraphics[width=0.9\linewidth]{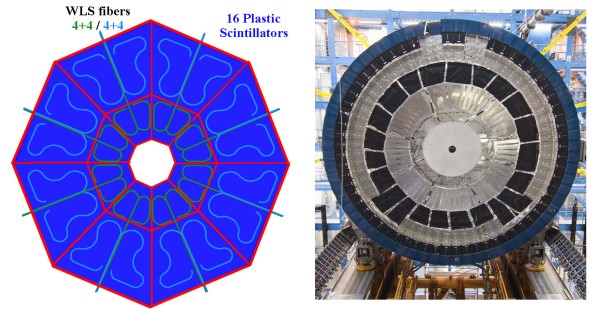}
\caption{Left: Layout of one of the two MBTS disks. Right: A photograph of an actual MBTS disk.}
\label{fig:chap2_mbts}
\end{figure}

The Minimum Bias Trigger Scintillators (MBTS) provide a low-bias and fast readout signal essential for the Level-1 (L1) trigger in minimum bias (primarily non-diffractive inelastic) LHC collisions. The MBTS system consists of 2 cm thick polystyrene scintillator disks mounted on both sides of the interaction point at a distance of approximately 3.6 m along the beam pipe. On each side, the scintillators are arranged in inner and outer rings in $\eta$, with eight counters in azimuthal angle $\phi$ per ring. The pseudorapidity acceptance for the outer counters is $2.08 < |\eta|< 2.78$, while the acceptance for the inner counters is $2.78 < |\eta| < 3.75$. Light collected from each edge of the scintillator is guided to photomultiplier tubes (PMTs) using wavelength-shifting fibers (WLS). The MBTS signals are shaped and amplified such that their pulse amplitude is proportional to the energy deposited in the counter. These shaped pulses are then fed into leading-edge discriminators and sent as 25 ns pulses to the Central Trigger Processor (CTP). Both the total charge collected and the arrival time of the signal are recorded. An MBTS hit is defined as a signal exceeding the discriminator threshold. The layout of one of the two MBTS disks and a photograph of an actual MBTS disk are shown in Figure~\ref{fig:chap2_mbts}.

The excellent momentum resolution and broad pseudorapidity coverage ($|\eta|<2.5$) of the Inner Detector, in particular, the high-granularity Pixel and SCT sub-detectors, were paramount for the precise reconstruction of charged particle tracks. This capability was indispensable for the multi-particle correlation techniques central to this work.

\subsection{Trigger System}
\label{sec:trigger}
A trigger system is of utmost importance to any large-scale high-energy collider experiment. At facilities like the LHC, the vast majority of events produced are considered background, with only a small fraction containing important and interesting physics signatures. Given the high luminosity, typically on the order of $10^{33}\,\mathrm{cm}^{-2}\mathrm{s}^{-1}$ for $pp$ collisions, and even up to $10^{27}\,\mathrm{cm}^{-2}\mathrm{s}^{-1}$ for heavy-ion collisions, it is technologically impossible to record all the data generated by the collision events due to limitations in hardware capabilities and storage capacity. Trigger systems are thus essential to filter out unwanted events, allowing only the events of interest to be stored with high efficiency while simultaneously suppressing background in the recorded sample. A well-designed trigger system must fulfill several key requirements to effectively select interesting physics events while managing data volume. These requirements can be broadly categorized as Physics Requirements and Operational Requirements, and the system itself is composed of two main stages, as described below.

\subsubsection{Physics Requirements for Triggers:}
The trigger system is designed to achieve the following physics-driven goals:
\begin{itemize}
    \item \textbf{High Efficiency:} A high efficiency for selecting desired physics processes is crucial. Efficiency uncertainties, including systematic contributions, directly impact the overall uncertainty of a measurement and must be minimized.
    \item \textbf{Good Background Rejection Power:} Effective background rejection by the HLT is necessary to improve the overall efficiency and optimize the use of detector and computing resources.
    \item \textbf{High Availability:} The fraction of time the trigger and DAQ systems are unable to accept events (due to factors like trigger latency or dead time) should be minimized. This ensures that only a small percentage of collisions are lost due to the detector's inability to cope with the data readout from adjacent bunch crossings.
    \item \textbf{Multichannel Selection:} To be inclusive for the wide variety of physics processes studied, a good trigger system must allow individual selection signatures for different objects and phenomena to be configured independently.
    \item \textbf{Flexible Construction of Complex Signatures:} The trigger must possess the capability to combine signatures from different detector objects within defined time constraints to flexibly reconstruct complex event topologies.
    \item \textbf{Decision Introspection:} The recorded data should retain a history of the trigger decision-making process. This allows analysts to verify the trigger performance and correct for any inefficiencies during offline analysis. The record of the complex decision-making process must include both the decisions made and a description of the reconstructed physics objects (e.g., electrons, muons, jets) used in those decisions.
    \item \textbf{Single Event Reproducibility:} The trigger must select processes deterministically, meaning the decision for a given event is independent of the number and type of events processed before it. This ensures consistent and reproducible event selection.
\end{itemize}

\subsubsection{Operational Requirements for Triggers:}
In addition to physics goals, the ATLAS Trigger system is subject to several operational constraints:
\begin{itemize}
    \item \textbf{Processing Time Requirements:} The trigger must process events quickly enough to avoid missing interesting events between bunch crossings, operating within the constraints of finite CPU resources.
    \item \textbf{Online Monitoring:} The ability to monitor the trigger system's performance while it is actively taking and processing data is essential. This helps ensure the health and stability of a complex system like the ATLAS trigger, which operates under variable detector and data conditions.
    \item \textbf{Input and Output Data Bandwidth:} The system must be capable of handling the flow of data for events that pass the trigger selection. Not all raw data from each processed event can be requested due to hardware and storage limitations. This necessitates a balance between the rate of events that need detailed inspection for interesting physics (which should be as high as possible) and the construction costs of the readout system (which should be as low as possible). Limitations on the final output rate to disk arise from offline data storage and handling costs. Therefore, the HLT process must have the flexibility to utilize the available offline output bandwidth optimally, adapting to changing experimental conditions.
    \item \textbf{Deployment Environments:} The trigger system must function reliably in different deployment environments, from controlled test setups to the complex conditions of the actual experiment.
\end{itemize}

\subsubsection{Key Design Principles}
The High-Level Trigger at ATLAS was designed based on several key principles aimed at optimizing performance and resource usage:
\begin{itemize}
    \item \textbf{Abstracted Data Access:} Data access should be abstracted to allow algorithms to retrieve necessary information without modification, regardless of whether the data is readily available in memory (in a test environment) or stored in readout system buffers (in the deployment environment).
    \item \textbf{Early Rejection:} To improve trigger timing and reduce the data rate, partial selection requirements are applied as early as possible in the event reconstruction process.
    \item \textbf{Step-Wise Validation of Decisions:} For multi-object signatures, further optimization is achieved by grouping reconstruction steps and requiring coherent validation of the hypotheses at each step.
    \item \textbf{Avoidance of Repeated Reconstructions:} Duplicated reconstruction of information from the same detector region is avoided. This helps prevent overuse of CPU resources.
\end{itemize}

The ATLAS trigger system operates in two main steps: a fast, hardware-based Level-1 (L1) trigger and a slower, more precise software-based High-Level Trigger (HLT). This system collectively reduces the event rate from the nominal LHC bunch-crossing rate of 40 MHz to an average recording rate of a few hundred Hz.

\begin{figure}[h!]
\centering
\includegraphics[width=0.99\linewidth]{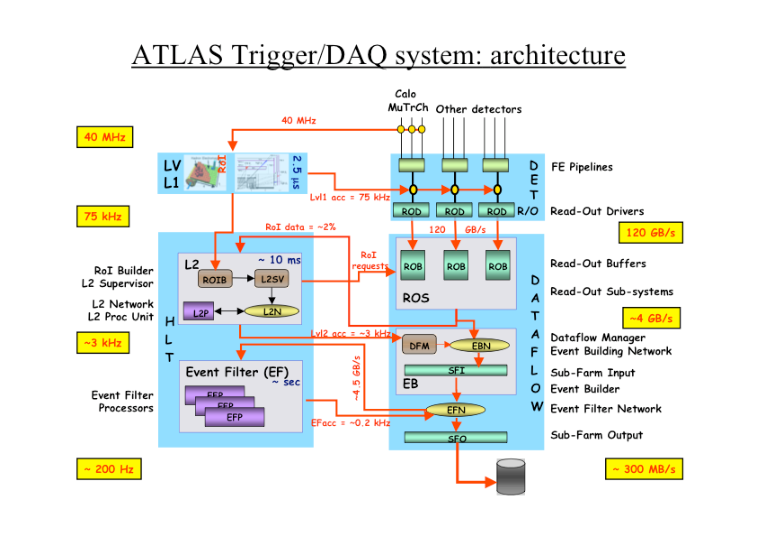}
\caption{A schematic diagram illustrating the architecture of the ATLAS Trigger and DAQ system.}
\label{fig:chap2_trigger}
\end{figure}

\paragraph{Level-1 Trigger (L1):}
The Level-1 trigger is a hardware-based system designed to make rapid decisions. It utilizes signals from fast detector components such as the calorimeters (Level 1 Calorimeter Trigger, L1Calo) and muon detectors (Level 1 Muon Trigger, L1Muon). These components deliver signals sufficiently fast to allow a decision to be made at the LHC bunch crossing frequency of 40 MHz. The L1 decision must be made with a small enough latency ($\approx 2.5\,\mu$s) to prevent readout buffer overflows. L1Calo and L1Muon identify physics signatures within this short time frame. Their decisions, along with signals from the MBTS and other auxiliary detectors, are fed to the Level 1 Central Trigger (L1CT) to contribute to the final event acceptance decision. The Level-1 hardware trigger, utilizing custom-built electronics, reduces the initial event rate from 40 MHz to approximately 100 kHz. A schematic view of the ATLAS Trigger system architecture is shown in Figure~\ref{fig:chap2_trigger}.

\paragraph{High-Level Trigger (HLT):}
Following the Level-1 decision, events accepted by L1 are passed to the High-Level Trigger (HLT). The HLT is a software-based system implemented on a large farm of computing servers, comprising approximately 40,000 physical cores. It performs a more detailed analysis of the event data than L1. Historically, the HLT was conceptually divided into Level-2 (L2) and the Event Filter (EF), although both stages are now implemented in software. The HLT further reduces the event rate from the $\approx 100$ kHz output of L1 to an average recording rate of a few hundred Hz (approximately 1 kHz as mentioned in the original text, this rate can vary depending on running conditions and physics goals). The HLT and DAQ systems manage the interface between the detector readout electronics, the L1 decisions, and the CERN Tier-0 mass storage and computing resources.
\clearpage

\textbf{Note About Figure Reproducibility:} 

Beyond this section, all plots labeled:
\begin{itemize}
    \item ``ATLAS'' are published results and can be freely reproduced.
    \item ``ATLAS preliminary'' are not published results, but exist on ATLAS websites that are open to the public and can be reproduced. 
    \item ``ATLAS Internal'' cannot be reproduced in any way and are only allowed to be shown outside of private ATLAS meetings in the context of Ph.D. thesis and defense.
\end{itemize}
\clearpage

\newpage
\chapter{Dataset, Event, and Track Selection}
\label{sec:data}

 Following the description of the experimental setup in Chapter~\ref{chap:expt}, this chapter provides a detailed description of the datasets, trigger configurations, event cleaning protocols, track selection criteria, and the corrections applied to ensure good data quality used for subsequent measurements in both Pb+Pb and Xe+Xe collisions.

\section{Collision Systems and Datasets}
\label{sec:datasets_systems}
The analyses presented in this thesis utilize data collected by the ATLAS detector~\cite{ATLAS:2008xda} at the Large Hadron Collider (LHC). Two heavy-ion collision systems are studied: Pb+Pb collisions at a center-of-mass energy per nucleon pair of $\sqrt{s_{\mathrm{NN}}} = 5.02$ TeV and Xe+Xe collisions at $\sqrt{s_{\mathrm{NN}}} = 5.44$ TeV.

\subsection{Pb+Pb Collisions at $\sqrt{s_{\mathrm{NN}}} = 5.02$ TeV}
\label{subsec:pbpb_dataset}
The data for Pb+Pb collisions were recorded by the ATLAS detector during the 2015 LHC run. This study employs the \texttt{data15\_hi.*.physics\_MinBias.\\merge.AOD.r7874\_p2580} datasets. Out of 35 total runs, 33 were deemed good based on the Good Run List (GRL)\footnote{http://atlasdqm.web.cern.ch/atlasdqm/grlgen/HeavyIonP/data15\_hi.periodAllYear\_DetStatus-v75-repro20-01\_DQDefects-00-02-02\_PHYS\_HeavyIonP\_All\_Good.xml}. Following the application of event selection criteria (detailed in Section~\ref{sec:event_selection}), a total of 189 million minimum-bias events were analyzed, corresponding to an integrated luminosity of approximately 22 $\mu\text{b}^{-1}$.

\subsection{Xe+Xe Collisions at $\sqrt{s_{\mathrm{NN}}} = 5.44$ TeV}
\label{subsec:xexe_dataset}
The Xe+Xe collision data were collected in October 2017 during ATLAS Run 338037. A single heavy-ion collision fill was recorded in the ``Minimum-Bias'' data stream over a period of approximately 8 hours, yielding an integrated luminosity of 3 $\mu\text{b}^{-1}$. After the application of event-quality and vertex selection criteria (detailed in Section~\ref{sec:event_selection}), 13 million high-quality minimum-bias events were available for analysis. The specific dataset utilized is \texttt{data17\_hi.00338037.physics\_MinBias.merge.AOD.f900\_m1912}

\section{Event Selection Criteria}
\label{sec:event_selection}
To ensure the use of high-quality data, several event selection criteria were applied. These generally included adherence to good lumi-blocks specified in the GRL, event-level cleaning cuts to remove events affected by detector malfunctions, and the requirement of a reconstructed primary vertex.

\subsection{Common Event Selection Criteria}
\begin{itemize}
    \item \textbf{Good Lumi-blocks:} Events were required to belong to good luminosity blocks as defined by the respective GRL for each dataset.
    \item \textbf{Event Cleaning:} Event-level cleaning cuts were implemented to discard events potentially compromised by detector issues.
    \item \textbf{Primary Vertex Requirement:} A reconstructed primary vertex with its $z$-coordinate satisfying $|\zvtx| < 100$ mm was required. This selection minimizes the vertex position dependence of the detector acceptance and reconstruction efficiency.
\end{itemize}

\subsection{Pb+Pb Collision Specifics}
\label{subsec:pbpb_event_selection}
For the Pb+Pb dataset, event selection was consistent with previous ATLAS analyses~\cite{Aad:2019fgl}.
\subsubsection{Minimum-Bias Triggers}
Data from the ``Minimum-Bias'' stream were primarily used. The minimum-bias triggers employed were:
\begin{enumerate}
    \item \verb|HLT_noalg_mb_L1TE50|: This trigger required more than 50 GeV of transverse energy in the calorimeter at Level-1.
    \item \verb|HLT_mb_sptrk_ion_L1ZDC_A_C_VTE50|: This trigger selected events with less than 50 GeV of transverse energy in the calorimeter at Level-1, in conjunction with a signal in both Zero Degree Calorimeters (ZDCs).
\end{enumerate}

\subsubsection{Inclusion of Ultra-Central Collision Events}
\label{subsubsec:pbpb_ucc}
To enhance statistical precision for ultra-central Pb+Pb collisions, events selected by dedicated Ultra-Central Collision (UCC) triggers were incorporated. These triggers selected events based on the total transverse energy in the entire calorimeter system ($\sumET^{tot}$) at Level-1 and the total transverse energy in the Forward Calorimeter (FCal) ($\sumET$) at the High-Level Trigger (HLT). The specific UCC triggers included:
\begin{itemize}
    \item \verb|HLT_hi_th1_ucc_L1TE10000|, \verb|HLT_hi_th2_ucc_L1TE10000|, \verb|HLT_hi_th3_ucc_L1TE10000|
    \item \verb|HLT_hi_th1_ucc_L1TE12000|, \verb|HLT_hi_th2_ucc_L1TE12000|, \verb|HLT_hi_th3_ucc_L1TE12000|
    \item \verb|HLT_hi_th1_ucc_L1TE14000|, \verb|HLT_hi_th2_ucc_L1TE14000|, \verb|HLT_hi_th3_ucc_L1TE14000|
\end{itemize}
Here, \verb|L1TEX| signifies the minimum Level-1 total energy threshold, while \verb|thX| corresponds to distinct online minimum FCal $\sumET$ thresholds at the HLT: \verb|th1| (4.172 TeV), \verb|th2| (4.326 TeV), and \verb|th3| (4.500 TeV). These triggers were fully efficient for the 1.3\%, 0.5\%, and 0.1\% of events with the highest FCal $\sumET$, respectively. The integrated luminosities collected by these triggers were 52 $\mu\text{b}^{-1}$, 140 $\mu\text{b}^{-1}$, and 470 $\mu\text{b}^{-1}$, respectively, improving event statistics in the ultra-central regime by more than 20 times compared to minimum-bias triggers.

\begin{figure}[htbp]
    \centering
    \includegraphics[width=0.9\linewidth]{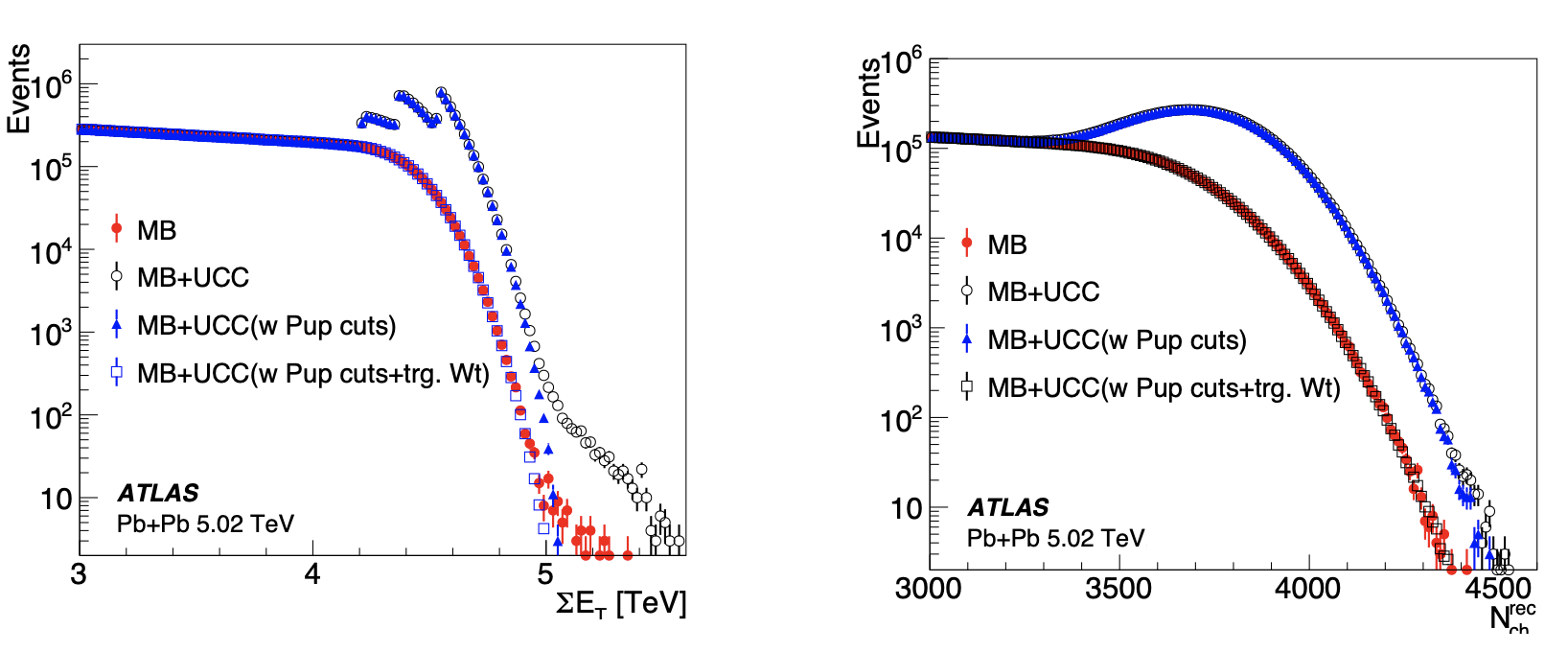}
    \caption{(Left Panel) FCal-$\sumET$ distribution, (Right Panel) $\NchR$ distribution for the MinBias and UCC triggers for the Pb+Pb collision dataset. Both panels show MinBias+UCC events with and without trigger reweighting, compared to the MinBias-only distribution.}
    \label{fig:UccEt}
\end{figure}

Figure~\ref{fig:UccEt} displays the FCal-$\sumET$ and reconstructed charged particle ($\NchR$) distributions for UCC-triggered events. These triggers substantially increase statistics for FCal-$\sumET > 4.2$ TeV. To correct for potential biases introduced by these triggers, an event weight $w_{trig}$ was applied to correlation calculations for events passing the UCC trigger. This weight, a function of FCal-$\sumET$, is defined as:
\begin{equation*}
w_{trig} = \frac{\Sigma(\text{events passing MinBias})}{\Sigma(\text{events passing MinBias or UCC})}
\end{equation*}
The weight $w_{trig}$ is unity for minimum-bias events and decreases in three steps with increasing FCal-$\sumET$, corresponding to the UCC trigger thresholds. As shown in Fig.~\ref{fig:UccEt}, applying $w_{trig}$ makes the combined (Minimum-Bias + UCC) distribution closely match that of minimum-bias events alone. The figure also shows the FCal-$\sumET$ distribution for UCC events before and after pileup cuts (discussed in Section~\ref{sec:pileup_rejection}), demonstrating the efficacy of pileup removal.

\subsection{Xe+Xe Collision Specifics}
\label{subsec:xexe_event_selection}
For the Xe+Xe dataset, event selection criteria were consistent with previous ATLAS analyses~\cite{Aad:2019xmh}.
\begin{itemize}
    \item \textbf{Lumi-block selection:} Only data from lumi-blocks 198-565 of Run 338037 were utilized.
    \item \textbf{Detector-specific cleaning:} Events flagged by monitoring systems for issues related to the Liquid Argon (LAr) calorimeter, Tile calorimeter, SemiConductor Tracker (SCT), or incomplete event records were removed.
\end{itemize}

\subsubsection{Minimum-Bias Triggers}
\begin{figure}[htbp]
\centering
\includegraphics[width=0.6\linewidth]{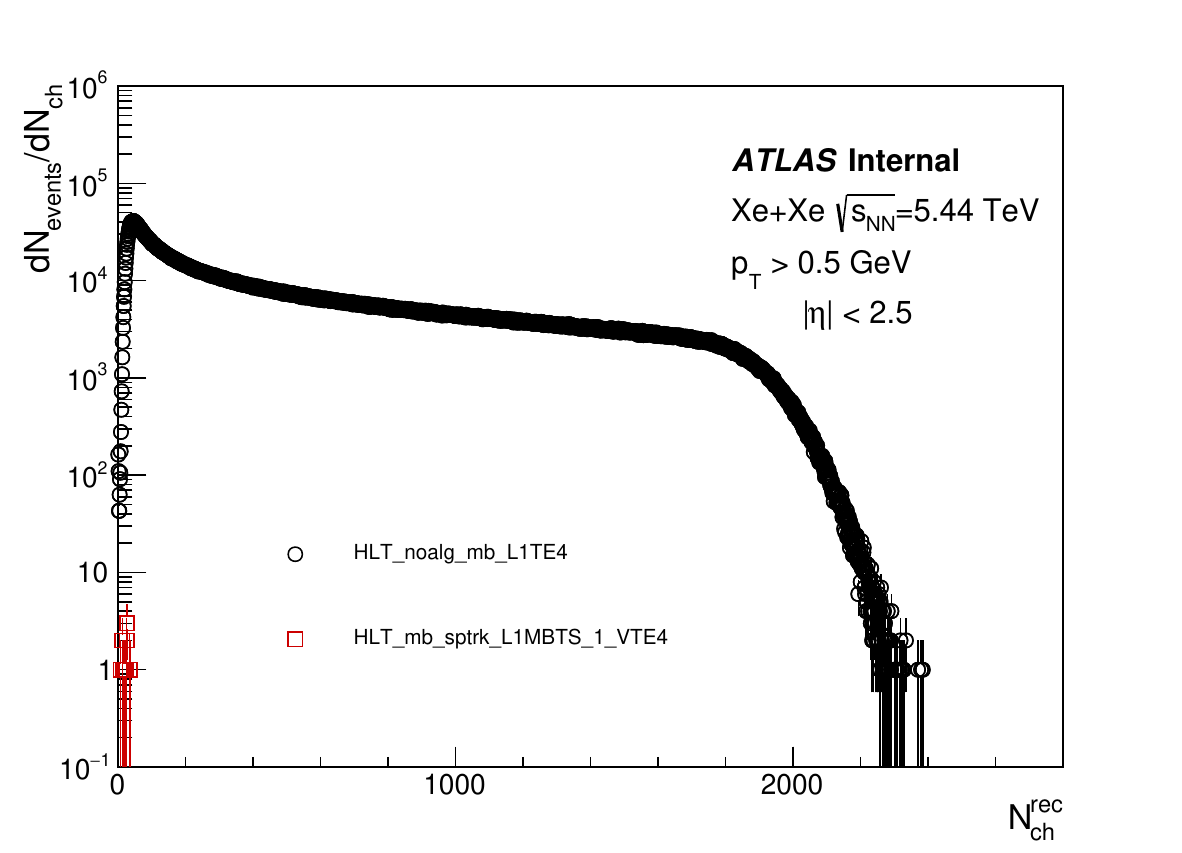}
\caption{Distributions of the number of reconstructed charged particles ($N_\mathrm{ch}^\mathrm{rec}$) for the two minimum-bias triggers used in the Xe+Xe collision dataset.}
\label{fig:evtSel_TrigXe}
\end{figure}

Minimum-bias events were recorded using a combination of two triggers:
\begin{enumerate}
    \item \verb|HLT_noalg_mb_L1TE4|: This trigger required a minimum transverse energy of 4 GeV in the calorimeter at Level-1 (L1).
    \item \verb|HLT_mb_sptrk_L1MBTS_1_VTE4|: This selected events with less than 4 GeV transverse energy in the calorimeter at L1 but required a signal in at least one Minimum-Bias Trigger Scintillator (MBTS) hodoscope. At the HLT stage, it further mandated at least one reconstructed track, enhancing the selection of peripheral events.
\end{enumerate}
Figure~\ref{fig:evtSel_TrigXe} compares the number of reconstructed tracks ($N_\mathrm{ch}^\mathrm{rec}$) for these two triggers, showing the \verb|HLT_noalg_mb_L1TE4| trigger records most events, while \verb|HLT_mb_sptrk_L1MBTS_1_VTE4| primarily contributes to low-multiplicity events.

\section{Pileup Rejection}
\label{sec:pileup_rejection}
Pileup, the occurrence of multiple inelastic nucleon-nucleon interactions within a single or closely spaced bunch crossing, can distort measurements by artificially inflating event activity.

The rigorous pileup rejection procedures detailed herein, utilizing correlations between $\NchR$ and FCal-$\sumET$ as well as ZDC versus FCal-$\sumET$ information for Pb+Pb, are especially critical for the fluctuation-sensitive observables explored in this dissertation. These methods ensure that measurements reported in this dissertation are not biased by spurious signals from pileup events.

\subsection{Pb+Pb Collisions}
\label{sec:pileupPb}
\begin{figure}[htbp]
\centering
\includegraphics[width=0.49\linewidth]{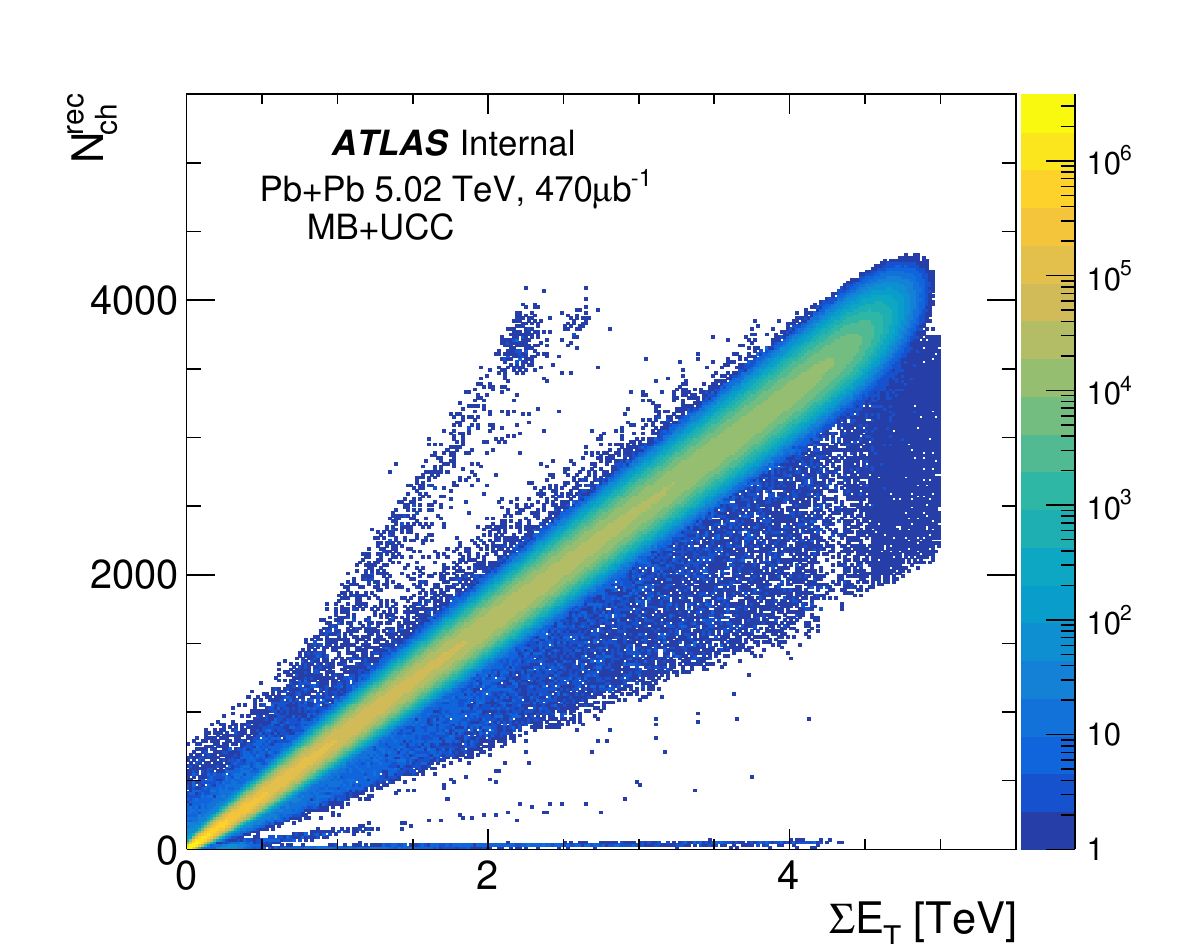}
\includegraphics[width=0.49\linewidth]{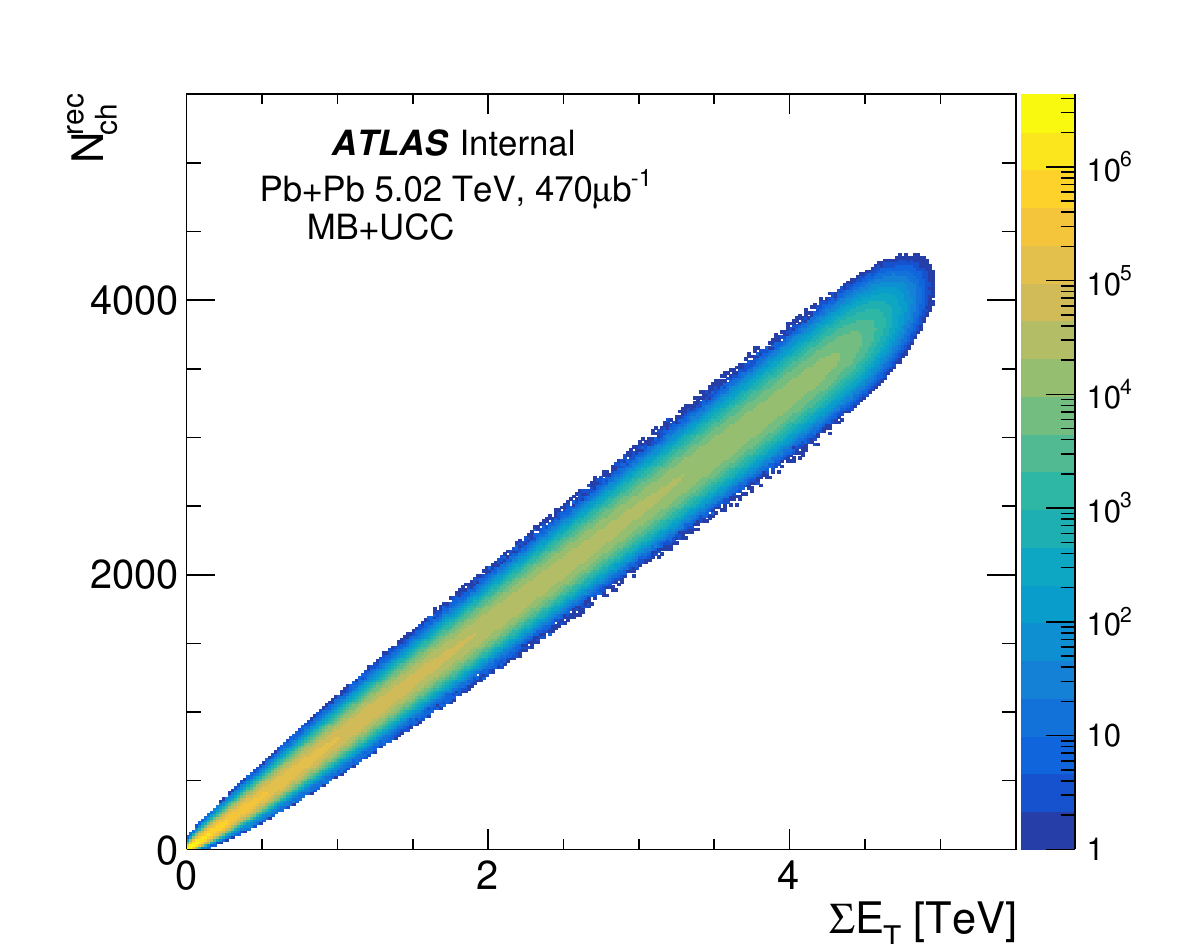}
\caption{$\NchR$-$\sumET$ correlation before pileup removal (left) and after applying a 6$\sigma$ cut around the mean of the primary band to reject pileup events (right) in Pb+Pb collisions.}
\label{fig:nchet_pup}
\end{figure}

During the 2015 Pb+Pb run, the average pileup probability was approximately 0.1\%. The primary pileup mitigation strategy involved rejecting events with more than one reconstructed primary vertex. Two correlation-based methods were implemented:
\begin{enumerate}
    \item \textbf{$\NchR$-$\sumET$ Correlation:} The correlation between FCal-$\sumET$ and the number of reconstructed charged particles ($\NchR$) associated with the primary vertex (for tracks with $0.5 < p_T < 5.0$ GeV and $|\eta| < 2.5$) was utilized. Single primary vertex events form a well-defined diagonal band (Figure~\ref{fig:nchet_pup}, left). Pileup events and spurious signals, deviating from this band, were removed by a 6 standard deviations ($6\sigma$) cut around the mean of the primary band (Figure~\ref{fig:nchet_pup}, right). This removed about 0.026\% of the total event sample.
\end{enumerate}

\begin{figure}[htbp]
\centering
\includegraphics[width=0.6\linewidth]{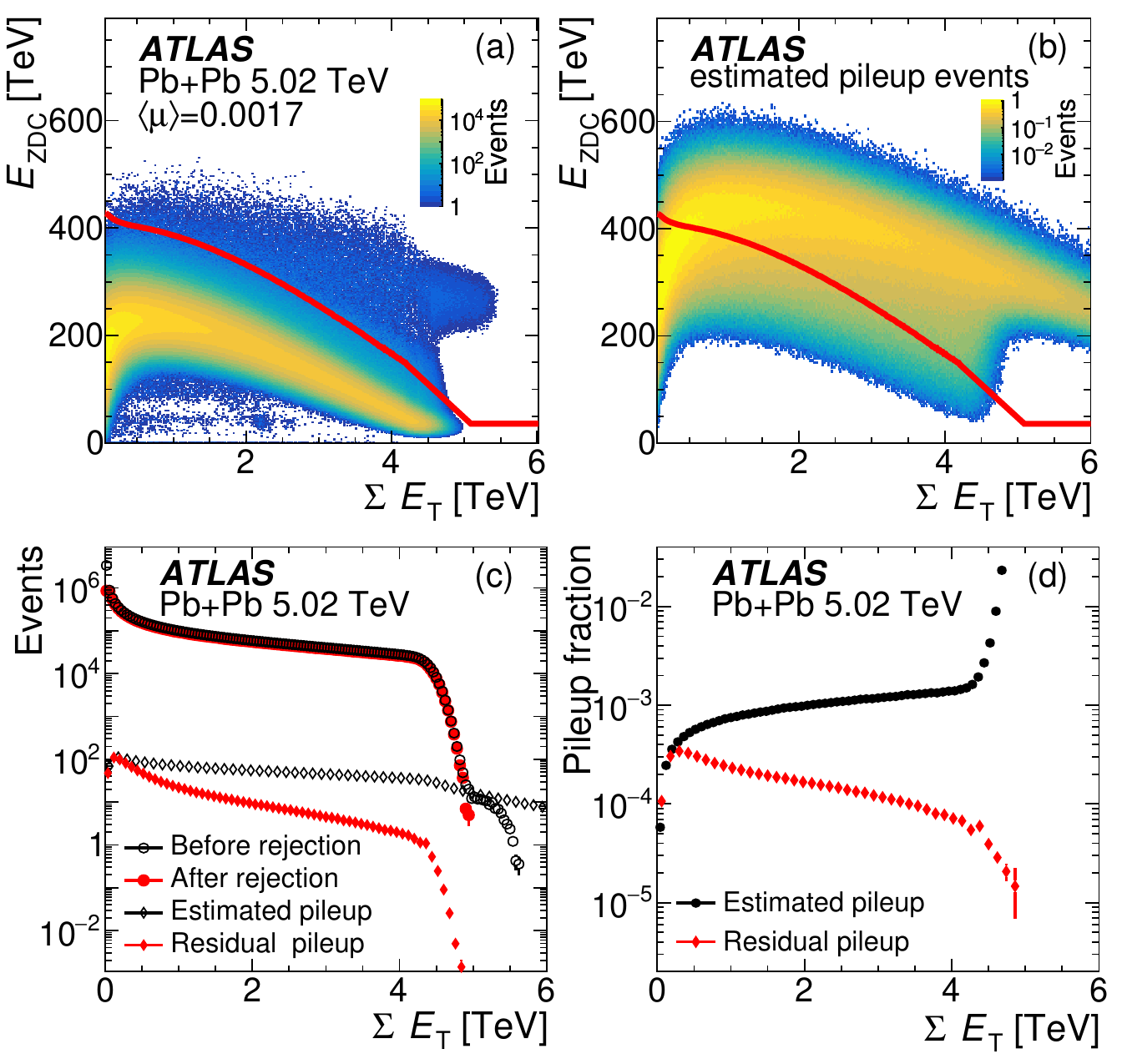}
\caption{Top: Distributions of energy deposited in the ZDCs vs. FCal $\sumET$ for (a) all events, and (b) estimated pileup events; the red line represents the selection for good events. Bottom: As a function of $\sumET$, (c) distributions of all events, pileup events, good events, and residual pileup events, and (d) the fraction of pileup events before and after pileup rejection. Results are for 5.02 TeV Pb+Pb collisions.}
\label{app:1}
\end{figure}

\begin{enumerate}
    \item[2.] \textbf{ZDC Energy vs. $\sumET$ Correlation:} Further pileup rejection leveraged the anti-correlation between ZDC energy and FCal $\sumET$ (Figure~\ref{app:1}(a)). Pileup in UCC can appear as a superposition of a central event (low ZDC, high $\sumET$) and a peripheral event (high ZDC, low $\sumET$), forming a satellite band. Good events were selected within a $6\sigma$ window of the peak ZDC energy at a given $\sumET$. By convolving the good event distribution with the measured pileup probability, the distribution of estimated pileup events was obtained (Figure~\ref{app:1}(b)). Figure~\ref{app:1}(c) shows distributions of all, estimated pileup, good, and residual pileup events versus $\sumET$. This method reduces the pileup fraction to $\lesssim 0.01\%$ in the UCC region (Figure~\ref{app:1}(d)) and provides an independent estimate of residual pileup for systematic uncertainties.
\end{enumerate}

\begin{figure}[htbp]
\centering
\includegraphics[width=0.45\linewidth]{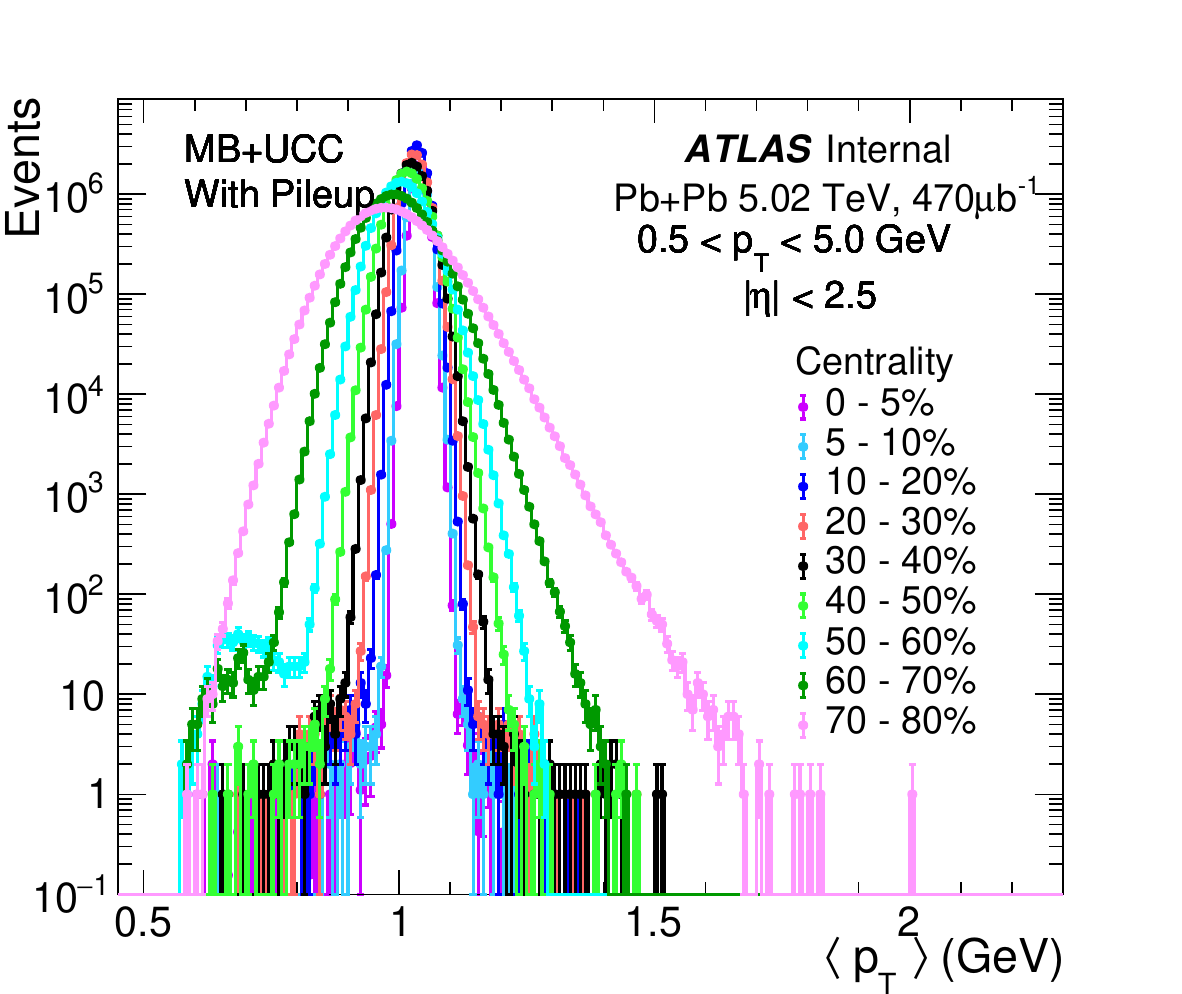}
\includegraphics[width=0.45\linewidth]{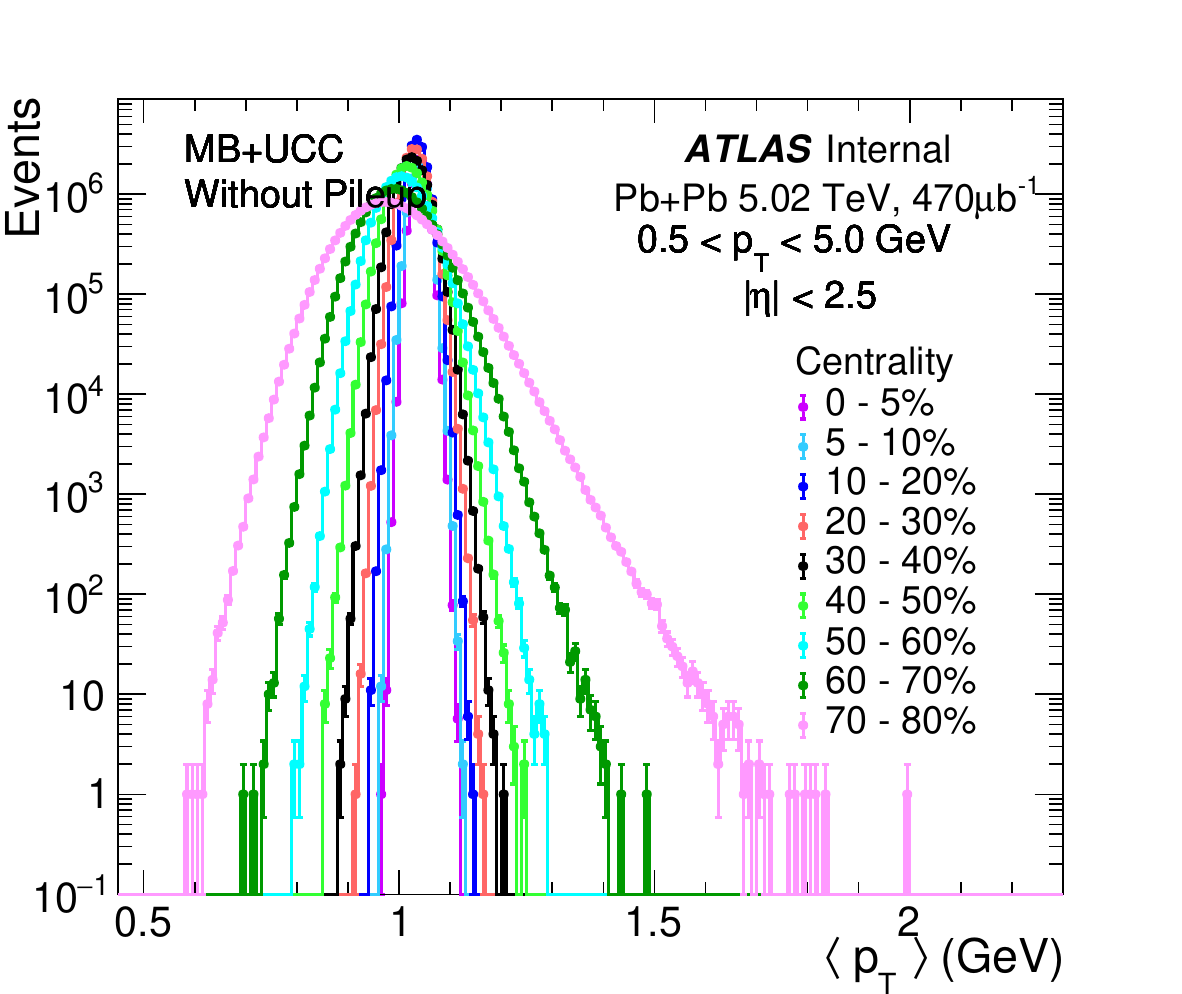}
\caption{Event-by-event $[\pT]$ distributions for various centralities before (left) and after (right) pileup removal using significance cuts in Pb+Pb collisions.}
\label{fig:meanpThistBefAf}
\end{figure}

The combined effectiveness of these procedures is shown in the event-by-event average transverse momentum ($[\pT]$) distributions for tracks ($0.5 < \pT < 5$ GeV) before and after cuts in Figure~\ref{fig:meanpThistBefAf}. The cuts mitigate spurious contributions from pileup, especially in the high-$[\pT]$ tail of peripheral events.

\subsection{Xe+Xe Collisions}
\begin{figure}[htbp]
    \centering
    \includegraphics[width=0.45\linewidth]{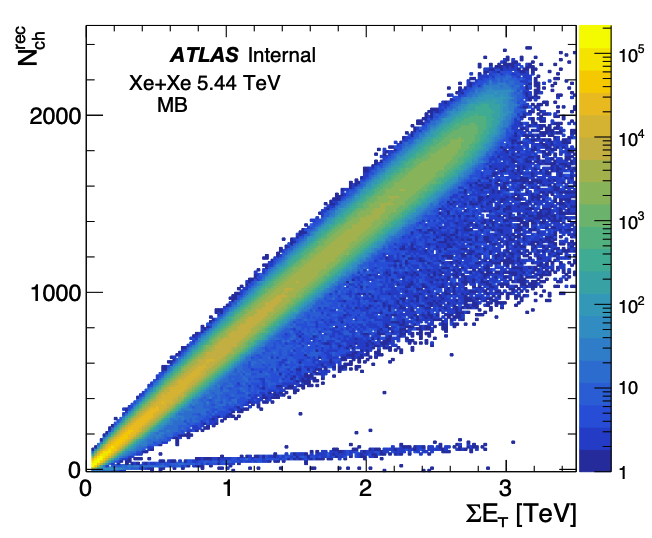}
    \includegraphics[width=0.45\linewidth]{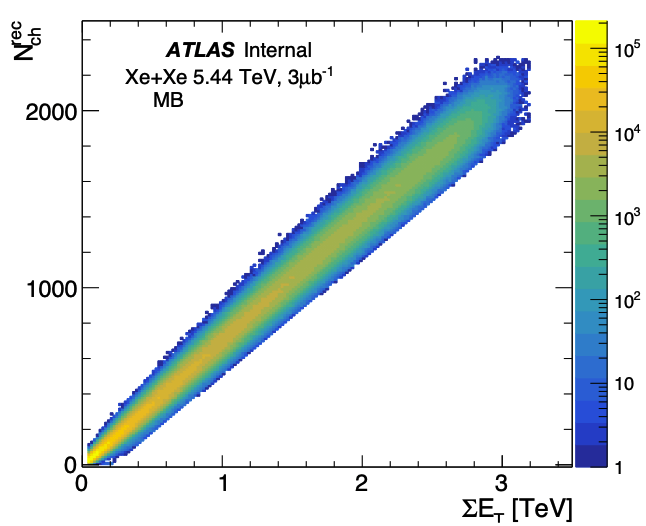}
    \caption{$\NchR$-$\sumET$ correlation plot without any pileup removal (left) and with a 6$\sigma$ cut about the mean of the main band to remove pileup events (right) for Xe+Xe collisions.}
    \label{fig:nchet_NopupXe}
\end{figure}

For the Xe+Xe dataset, the ZDC signal was unavailable for pileup rejection. However, the average number of interactions per bunch crossing ($\mu \approx 0.00019$) was very low, leading to a reduced pileup rate. Residual pileup was suppressed using a significance cut on the correlation between FCal-$\sumET$ and $\NchR$ (tracks with $0.5 < \pT < 5.0$ GeV and $|\eta|<2.5$), analogous to the Pb+Pb procedure.
Figure~\ref{fig:nchet_NopupXe} (left) shows this correlation, where single primary vertex events form a tight diagonal band. Events lying more than $6\sigma$ from the mean of this band were discarded to remove outliers (Figure~\ref{fig:nchet_NopupXe}, right).

\begin{figure}[htbp]
\centering
\includegraphics[width=0.45\linewidth]{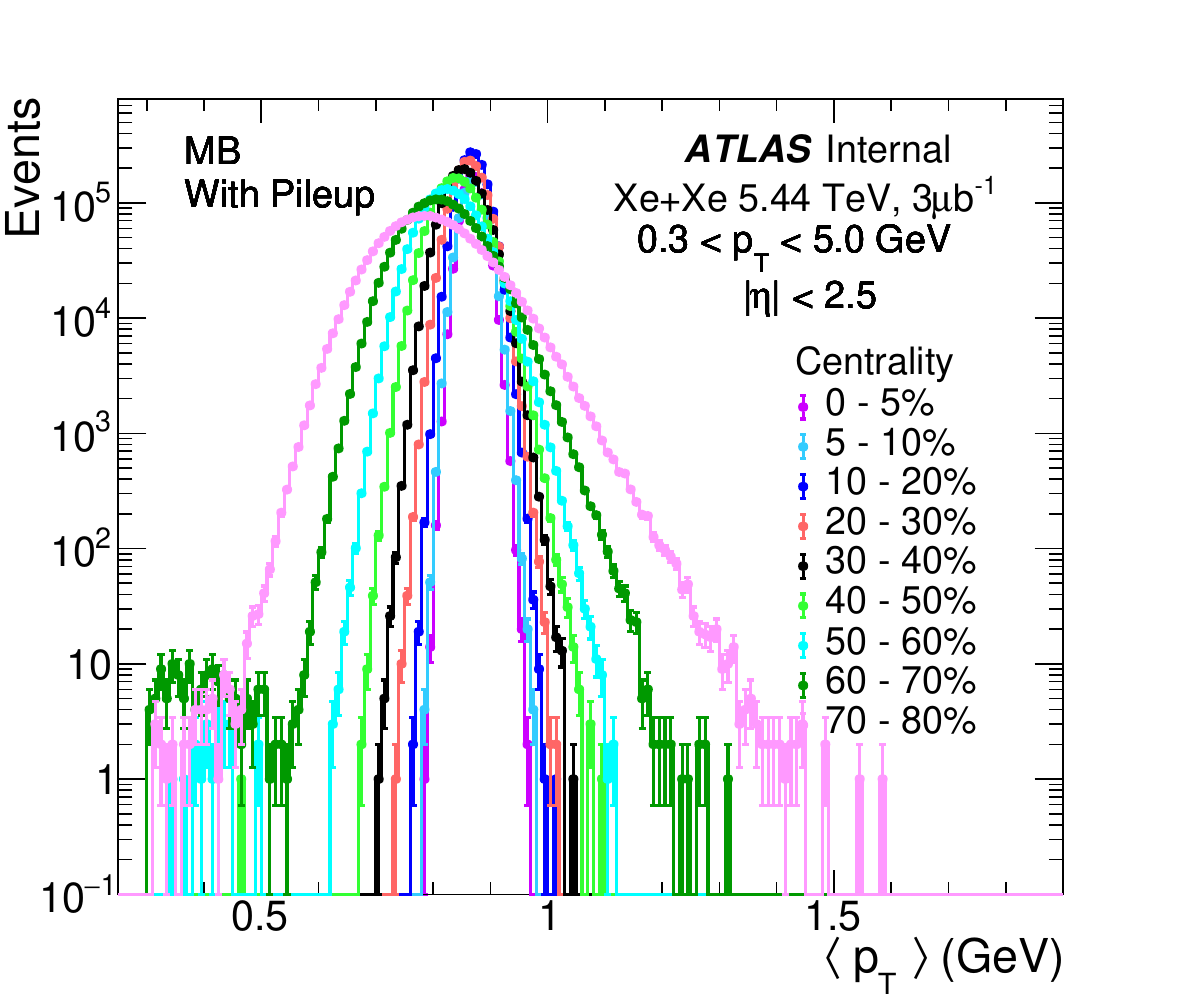}
\includegraphics[width=0.45\linewidth]{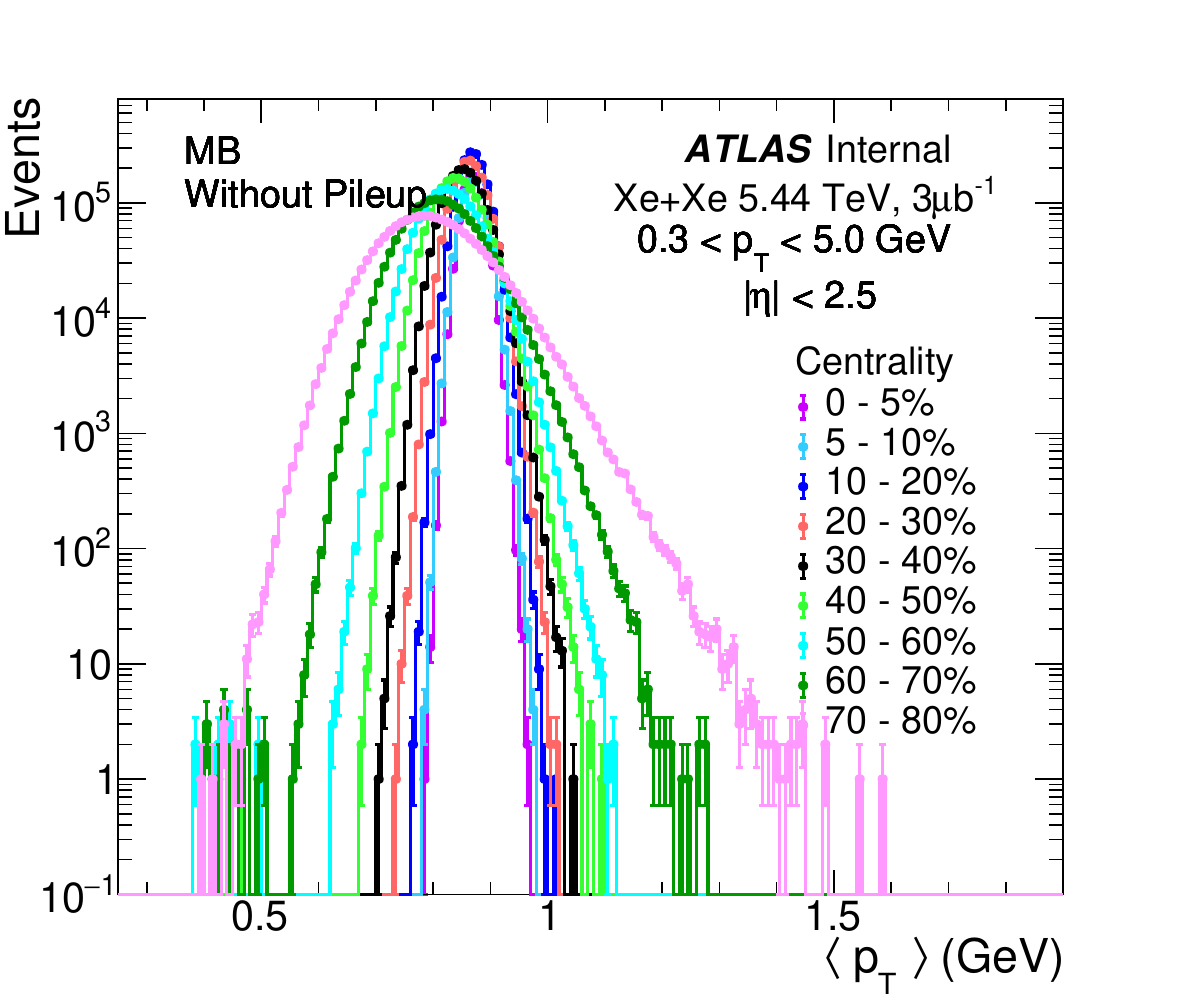}
\caption{Event-by-event $\lr{\pT}$ distributions for different centralities before (left) and after (right) pileup removal using significance cuts for Xe+Xe collisions.}
\label{fig:meanpThistBefAfXe}
\end{figure}

The effectiveness of this pileup removal is demonstrated in the event-by-event average transverse momentum ($\lr{\pT}$) distributions (Figure~\ref{fig:meanpThistBefAfXe}). The cuts reduce pileup contributions, particularly in the high-$\lr{\pT}$ tail of peripheral events.

\clearpage
\section{Track Selection Criteria}
\label{sec:track_selection}
Charged particle tracks were reconstructed using the ATLAS Inner Detector. To ensure high quality, specific selection criteria were applied. Two standard sets of track selection cuts, HITIGHT and HILOOSE, defined in the ATLAS xAOD tool \verb|InDetTrackSelectionTool/InDetTrackSelectionTool.h|, were utilized. The HILOOSE selection served as the default, with HITIGHT used for systematic variations. Tracks were generally required to have transverse momentum $\pT > 0.5$ GeV and pseudorapidity $|\eta| < 2.5$.

\subsection{HITIGHT Selection Criteria}
This selection imposes stringent requirements for high-purity tracks.
\subsubsection{For Pb+Pb Collisions:}
\begin{itemize}
    \item Transverse momentum $\pT > 0.5$ GeV.
    \item Pseudorapidity $|\eta| < 2.5$.
    \item A hit in the Insertable B-Layer (IBL) was required if expected. If an IBL hit was not expected, a hit in the next-to-innermost B-layer was required if expected.
    \item At least 2 Pixel detector hits.
    \item At least 8 SCT detector hits.
    \item If $\pT > 10$ GeV, the $\chi^2$ probability of the track fit was required to be less than 0.01.
    \item Track fit $\chi^2$ per degree of freedom ($\chi^2/\text{n.d.f.}$) < 6.
    \item Distance of closest approach to the beamline in the transverse plane $|\dzero| < 1.0$ mm.
    \item Longitudinal impact parameter with respect to the primary vertex $|\zzsth| < 1.0$ mm.
\end{itemize}

\subsubsection{For Xe+Xe Collisions:}
The criteria are identical to Pb+Pb HITIGHT with one addition:
\begin{itemize}
    \item Transverse momentum $\pT > 0.5$ GeV.
    \item Pseudorapidity $|\eta| < 2.5$.
    \item A hit in the IBL was required if expected. If an IBL hit is not expected or found, a hit in the next-to-innermost B-layer is required if expected.
    \item At least 2 hits in the Pixel detector.
    \item At least 8 hits in the SCT.
    \item At most 1 SCT hole (a detector layer where a hit was expected but not found).
    \item For tracks with $\pT > 10$ GeV, the $\chi^2$ probability of the track fit must be less than 0.01.
    \item The track fit $\chi^2/\text{n.d.f.}$ must be less than 6.
    \item The transverse impact parameter ($|\dzero|$) with respect to the beam-line must be less than 1.0 mm.
    \item The longitudinal impact parameter ($|\zzsth|$) with respect to the primary vertex must be less than 1.0 mm.
\end{itemize}

\subsection{HILOOSE Selection Criteria}
This selection provides a larger track sample with slightly relaxed criteria, applied identically for both Pb+Pb and Xe+Xe collision data:
\begin{itemize}
    \item Transverse momentum $\pT > 0.5$ GeV.
    \item Pseudorapidity $|\eta| < 2.5$.
    \item A hit in the IBL was required if a hit was expected. If an IBL hit was not expected, a hit in the next-to-innermost B-layer was required if expected.
    \item At least 1 Pixel detector hit.
    \item At least 6 SCT detector hits.
    \item If $\pT > 10$ GeV, the $\chi^2$ probability of the track fit was required to be less than 0.01.
    \item Distance of closest approach to the beamline in the transverse plane $|\dzero| < 1.5$ mm.
    \item Longitudinal impact parameter with respect to the primary vertex $|\zzsth| < 1.5$ mm.
\end{itemize}

\begin{figure}[htbp]
\centering
\includegraphics[width=0.6\linewidth]{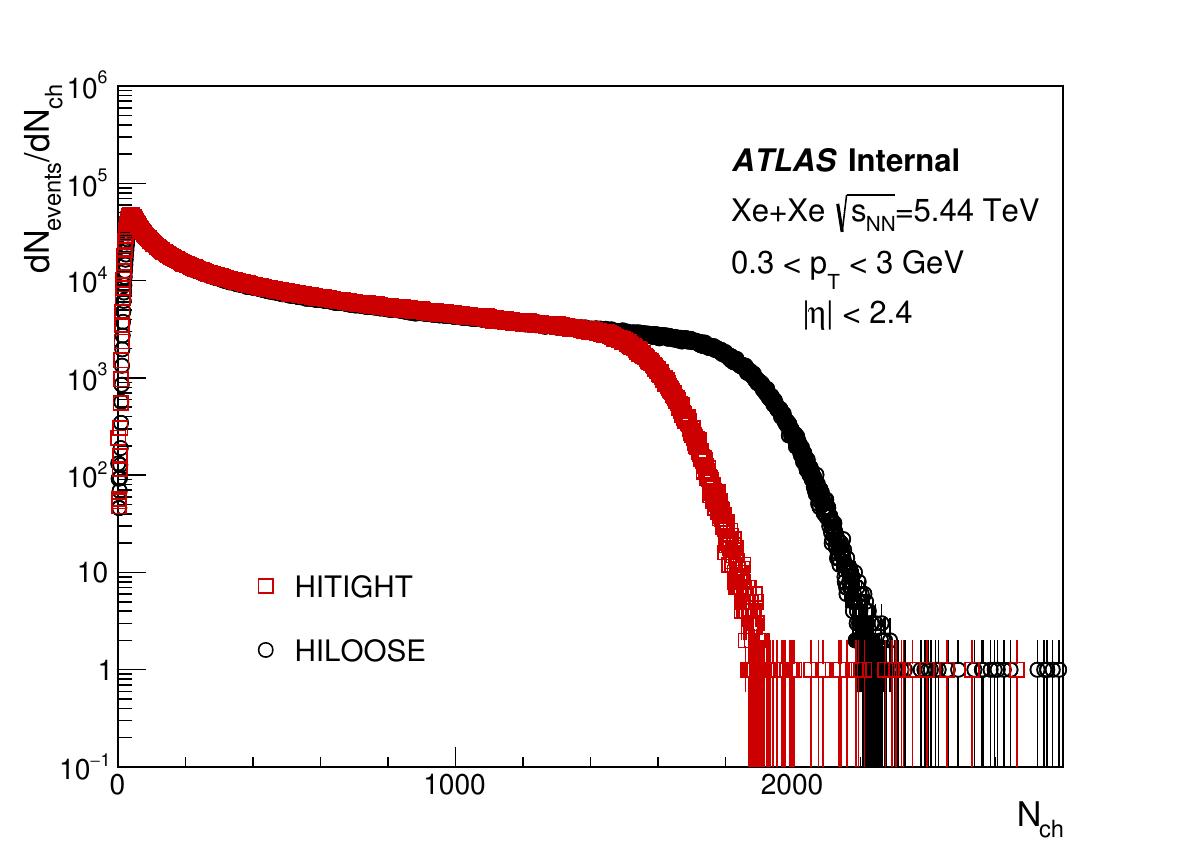}
\caption{Comparison of $N_\mathrm{ch}$ distributions for the HITIGHT and HILOOSE track selections in Xe+Xe collisions.}
\label{fig:evtSel_NCh}
\end{figure}

Figure~\ref{fig:evtSel_NCh} compares the distributions of the number of reconstructed tracks ($N_\mathrm{ch}$) using HITIGHT and HILOOSE selections for Xe+Xe collisions, illustrating that HITIGHT yields fewer tracks due to its stricter requirements. A similar behavior is observed for Pb+Pb collisions.

\newpage

\section{Monte Carlo Simulation and Corrections}
\label{sec:mc_corrections}
Monte Carlo (MC) simulated event samples are essential for correcting detector inefficiencies and accounting for the presence of fake tracks in the data. Track-by-track efficiency weights are derived from full \GEANT~\cite{AGOSTINELLI2003250, GEANT4:2002zbu} simulations of the ATLAS detector. Generated MC events undergo the same reconstruction algorithms as experimental data. 

The fidelity of the HIJING Monte Carlo simulation in reproducing detector effects and the precision of the derived tracking efficiency and fake rate corrections are critical for the accuracy of the final physics results. Residual differences between data and MC, and uncertainties in these correction procedures, are systematically evaluated and contribute to the overall uncertainty on the measurements, as documented in Appendix~\ref{sec:app_syst}.

\subsection{Monte Carlo Samples}
\subsubsection{Pb+Pb Collisions}
The MC sample for Pb+Pb collisions was generated using HIJING version 1.38b~\cite{Wang:1991hta}:\\
\texttt{mc15\_5TeV.420000.Hijing\_PbPb\_5p02TeV\_MinBias\_Flow\_JJFV6.recon.AOD.}\\
\texttt{e4962\_a868\_s2921\_r9447}\\
Anisotropic flow effects were incorporated post-generation using an afterburner procedure~\cite{Jia:2013tja}, implementing the $\NchR$, $\pT$, and $\eta$ dependence of $v_n$ coefficients measured by ATLAS in $\sqrt{s_{\mathrm{NN}}} = 2.76$ TeV Pb+Pb collisions~\cite{ATLAS:2012at}.

\subsubsection{Xe+Xe Collisions}
The MC samples for Xe+Xe collisions were generated using HIJING version 1.38b~\cite{PhysRevD.44.3501}:\\
\texttt{mc16\_5p44TeV.420221.Hijing\_XeXe\_5p44TeV\_MinBias\_Flow\_JJFV6.recon.AOD.}\\
\texttt{e6537\_s3283\_s3136\_r10409}\\
Realistic flow effects were introduced via an afterburner procedure~\cite{PhysRevC.88.014907}, applying the $\pT$, $\eta$, and centrality dependence of flow harmonics ($v_n$) as measured in $\sqrt{s_{\mathrm{NN}}} = 2.76$ TeV Pb+Pb collisions~\cite{PhysRevC.86.014907}.

\subsection{Particle Definitions in Monte Carlo}\label{sec:MCtruth}
For evaluating detector performance, generated particles (truth-level) and reconstructed tracks are used. Two categories of generated particles are considered~\cite{ATLAS:2016zkp, Wozniak:2015ycu}:
\begin{enumerate}
    \item \textbf{Primary particles}: Charged particles with a mean lifetime $\tau > 300$ ps, either directly produced or from decays of directly produced particles with $\tau < 30$ ps.
    \item \textbf{Secondary particles}: Produced from decays of particles with $\tau > 30$ ps.
\end{enumerate}
Strange baryons, due to their short lifetimes, are generally not reconstructed and are explicitly excluded from the set of truth particles for efficiency calculations.

\subsection{Calculating Tracking Efficiency and Fake Rates}
\label{sec:effFak}

The matching probability, $P_\mathrm{match}$, quantifies the association quality between reconstructed tracks and truth particles. It is calculated using a hit-based methodology. Clusters associated with a reconstructed track are linked to the truth particle depositing the most energy. Weights are assigned based on detector importance: Pixel clusters (including IBL) weight 10, SCT clusters weight 5, and TRT clusters weight 1.
\begin{equation}
P_\mathrm{match} = \frac{10 \cdot N_\mathrm{Pixel}^\mathrm{common} + 5 \cdot N_\mathrm{SCT}^\mathrm{common} + 1 \cdot N_\mathrm{TRT}^\mathrm{common}}{10 \cdot N_\mathrm{Pixel}^\mathrm{track} + 5 \cdot N_\mathrm{SCT}^\mathrm{track} + 1 \cdot N_\mathrm{TRT}^\mathrm{track}}
\label{eq:Pmatch}
\end{equation}
where $N^{\mathrm{common}}$ is the number of hits common to the track and truth particle, and $N^{\mathrm{track}}$ is the total number of hits on the track. A track is matched if $P_\mathrm{match} > 0.3$; otherwise, it is considered fake. This weighting emphasizes high-precision Pixel and SCT measurements.

The tracking performance depends on overall event activity. Corrections are derived from MC events matching data multiplicities for each centrality interval.

The tracking efficiency, $\epsilon$, is defined as:
\begin{equation}
\epsilon (\NchR,\pT,\eta)\equiv \frac{N_\mathrm{ch}^\mathrm{matched-reco}}{N_\mathrm{ch}^\mathrm{primary-truth}}
\label{eq:epsilon}
\end{equation}
where $N_\mathrm{ch}^\mathrm{matched-reco}$ is the number of reconstructed tracks matched to a primary truth particle, and $N_\mathrm{ch}^\mathrm{primary-truth}$ is the number of generated primary charged particles (excluding strange baryons).

Fake tracks are reconstructed tracks not satisfying the matching criterion ($P_\mathrm{match} \le 0.3$)~\cite{ATLAS:2016zkp, Wozniak:2015ycu}. Most fakes in the $0.5 < \pT < 10$ GeV range are associated with secondary particles. The fake track fraction, $f$, is:
\begin{equation}
f(\NchR,\pT,\eta)\equiv \frac{N_\mathrm{ch}^\mathrm{unmatched-reco}}{N_\mathrm{ch}^\mathrm{matched-reco}+N_\mathrm{ch}^\mathrm{unmatched-reco}}
\label{eq:fakefraction}
\end{equation}
where $N_\mathrm{ch}^\mathrm{unmatched-reco}$ is the number of fake reconstructed tracks.

Assuming small $\pT$ resolution, a corrected efficiency, $\epsilon'$, is defined for track-by-track corrections:
\begin{equation} \label{eq:CorrEff}
\epsilon' (\NchR,\pT,\eta)\equiv \frac{N_\mathrm{ch}^\mathrm{matched-reco}+N_\mathrm{ch}^\mathrm{unmatched-reco}}{N_\mathrm{ch}^\mathrm{primary-truth}}=\frac{\epsilon}{1-f}
\end{equation}
Each track is assigned a weight $w_\mathrm{Track} = (1 - f)/\epsilon = 1/\epsilon^{\prime}$, accounting for both efficiency and fake rates. These depend on $\NchR$, $\pT$, and $\eta$, and are calculated in fine bins: 40 HILOOSE $\NchR$ tracks, 0.2 GeV in $\pT$, and 0.1 units in $\eta$.

\subsubsection{Tracking Performance in Pb+Pb Collisions}
\begin{figure}[htbp]
    \centering
    \includegraphics[scale=0.3]{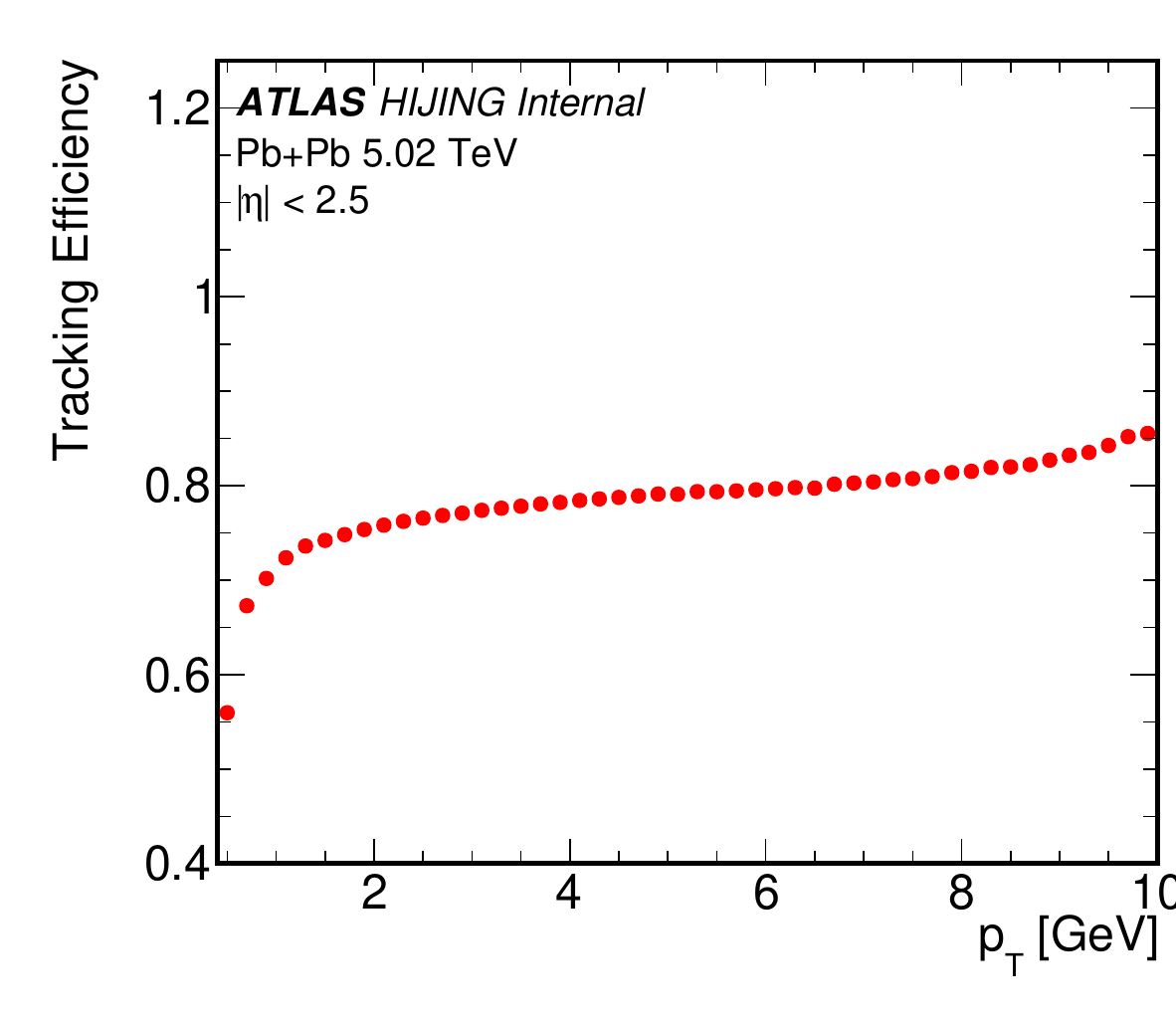}
    \caption{Tracking efficiency as a function of $\pT$ for 0-80\% centrality and $|\eta|<2.5$ in Pb+Pb collisions at $\sqrt{s_{\mathrm{NN}}} = 5.02$ TeV.}
    \label{fig:EffFull}
\end{figure}

Figure~\ref{fig:EffFull} shows tracking efficiency vs. $\pT$ for $|\eta| < 2.5$ in Pb+Pb collisions, increasing sharply up to $\pT \approx 1$ GeV, then more gradually.

\begin{figure}[htbp]
    \centering
    \includegraphics[width=0.32\linewidth]{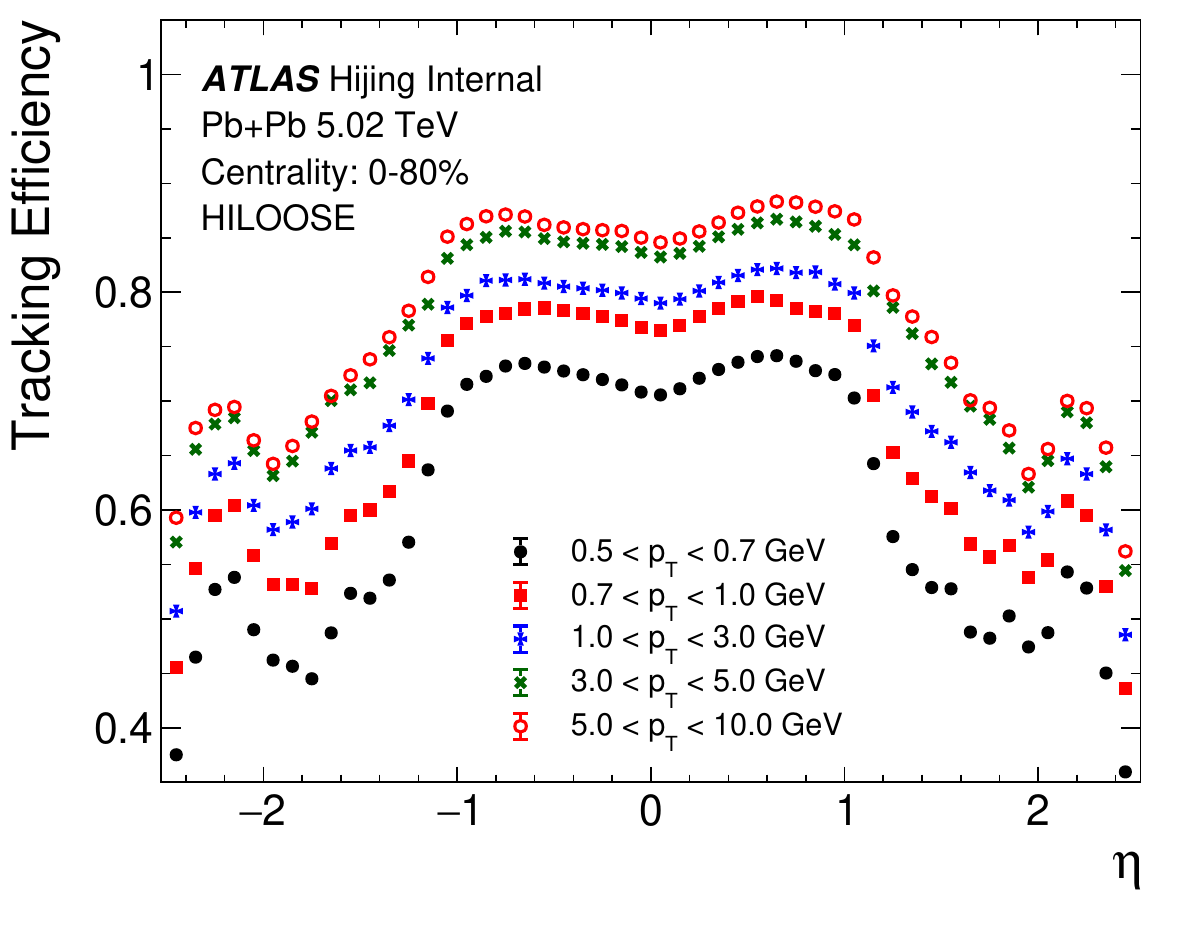}
    \includegraphics[width=0.32\linewidth]{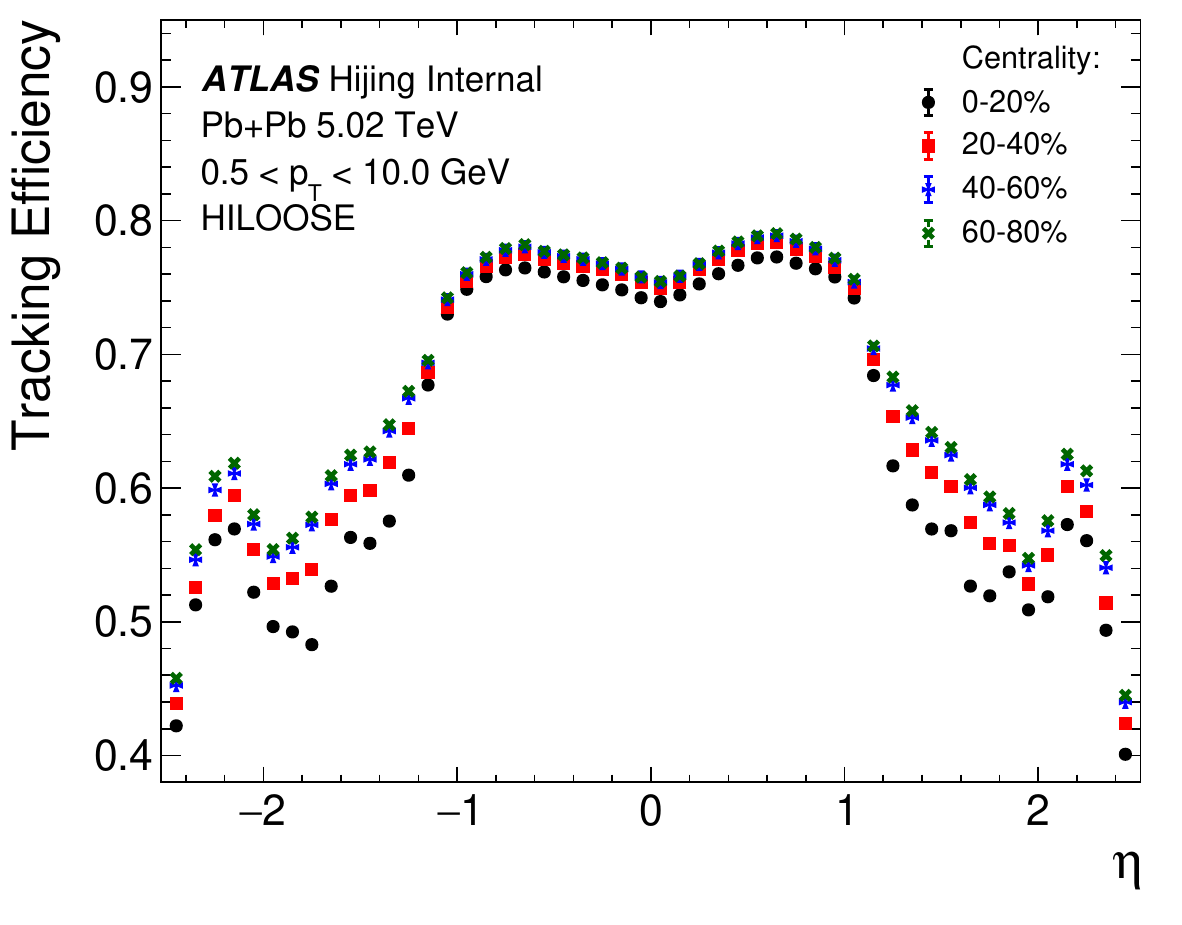}
    \includegraphics[width=0.32\linewidth]{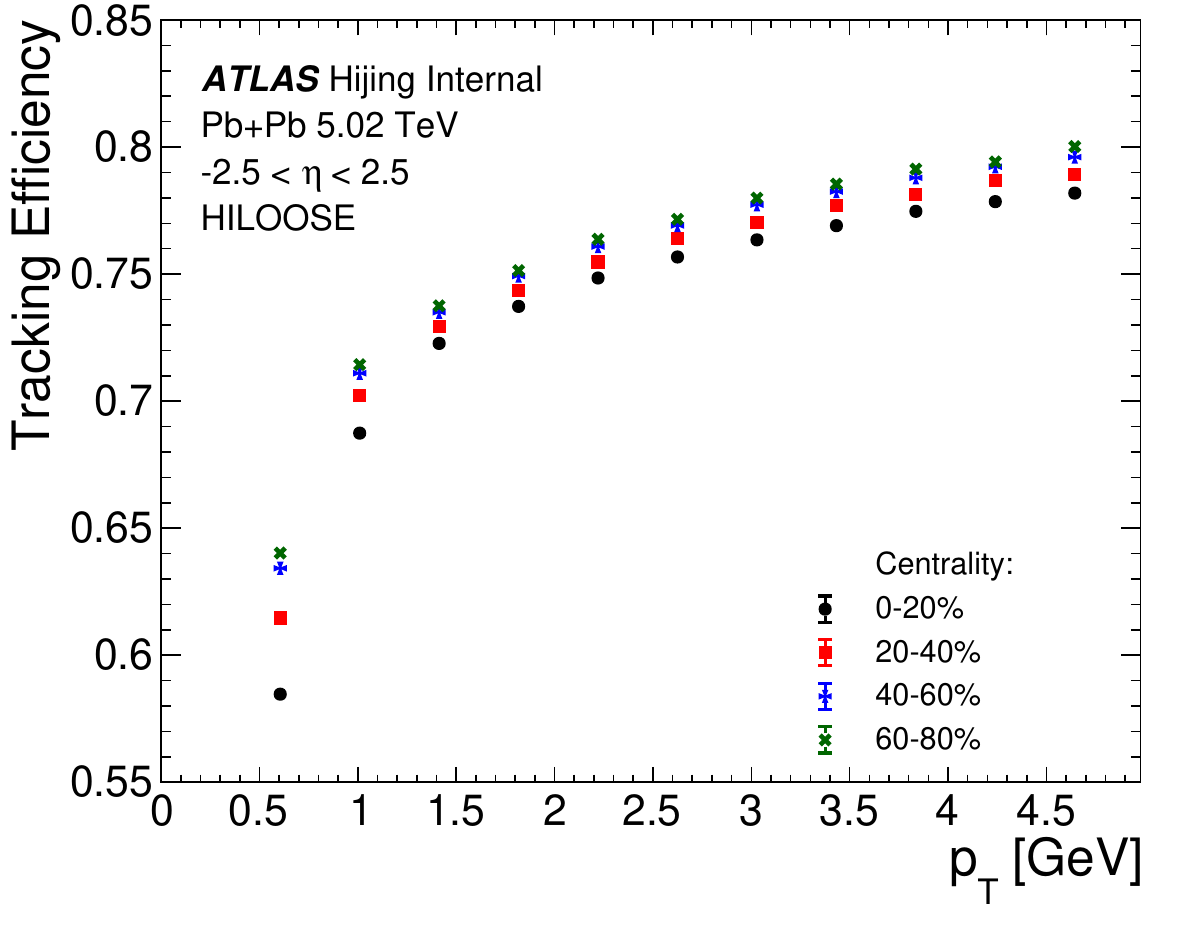}
    \caption{For Pb+Pb collisions at $\sqrt{s_{\mathrm{NN}}} = 5.02$ TeV: (Left) Tracking efficiency vs. $\eta$ for 4 $\pT$ intervals (0-80$\%$ centrality). (Center) Efficiency vs. $\eta$ for different centralities ($0.5<\pT<5.0$ GeV). (Right) Efficiency vs. $\pT$ for different centralities ($|\eta|<2.5$).}
    \label{fig:evtSel_EffFak_etaWide}
\end{figure}

Figure~\ref{fig:evtSel_EffFak_etaWide} illustrates $\epsilon$ vs. $\eta$ (left), vs. centrality (middle, for $0.5 < \pT < 5$ GeV), and vs. $\pT$ for different centralities (right, for $|\eta| < 2.5$). For HILOOSE tracks, $\epsilon$ generally increases with $|\eta|$, from central to peripheral collisions, and with $\pT$, becoming nearly constant for $\pT > 1$ GeV.

\begin{figure}[htbp]
    \centering
    \includegraphics[scale=0.3]{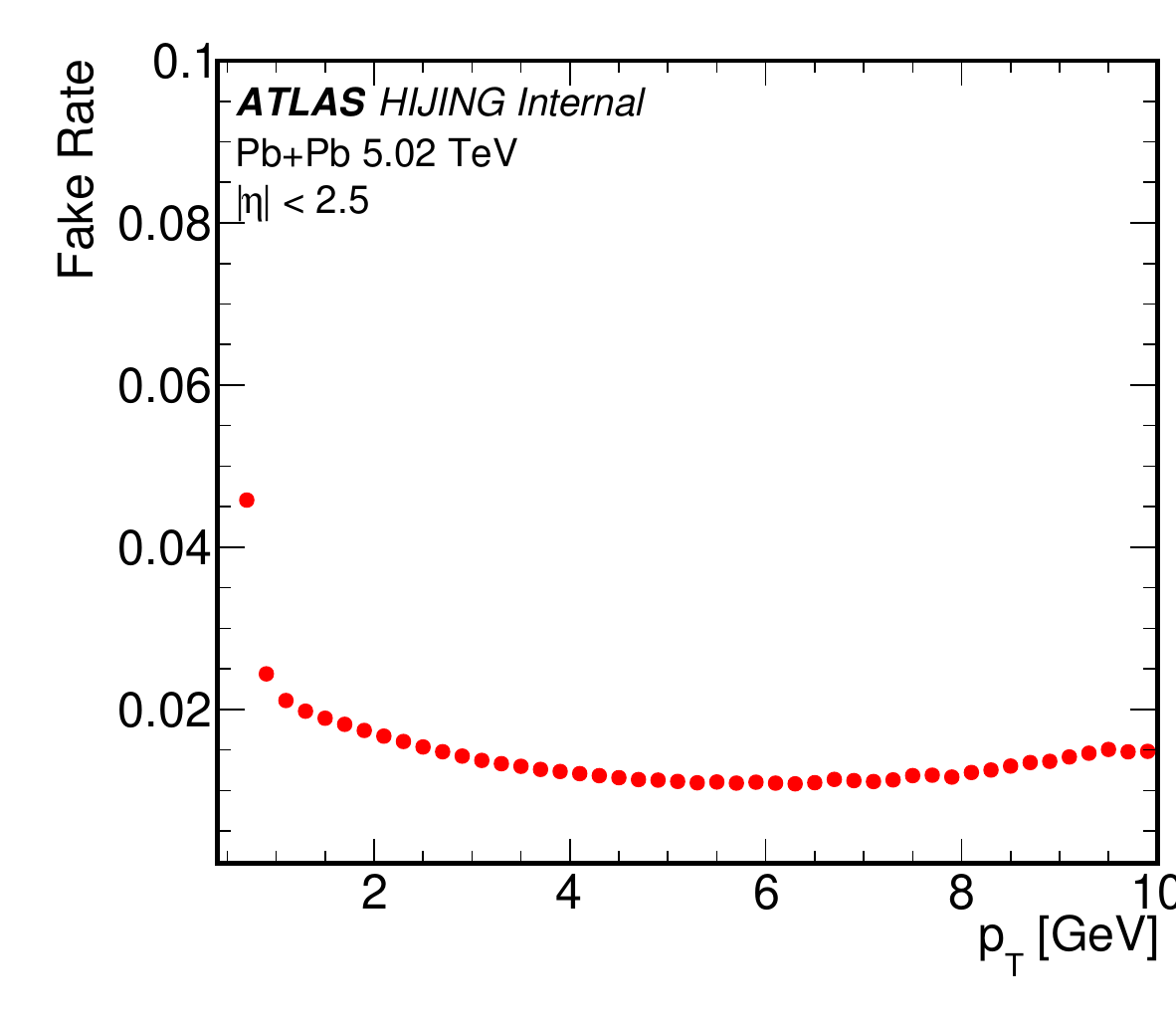}
    \caption{Fake track rates as a function of $\pT$ for 0-80\% centrality and $|\eta|<2.5$ in Pb+Pb collisions at $\sqrt{s_{\mathrm{NN}}} = 5.02$ TeV.}
    \label{fig:FakFull}
\end{figure}

Figure~\ref{fig:FakFull} shows fake track rates vs. $\pT$. The fake rate decreases sharply up to $\pT \approx 2$ GeV, then more gradually towards higher $\pT$.

\begin{figure}[htbp]
    \centering
    \includegraphics[width=0.32\linewidth]{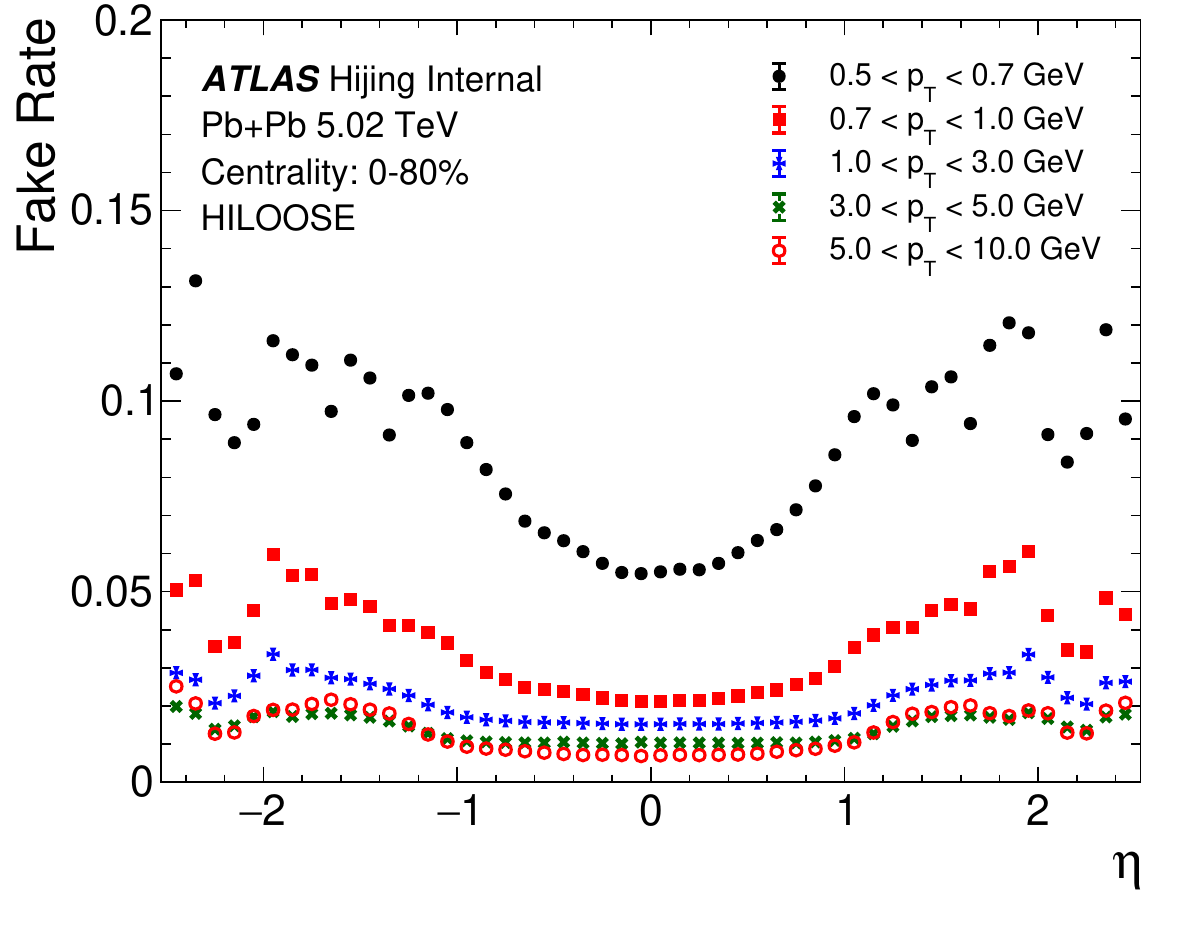}
    \includegraphics[width=0.32\linewidth]{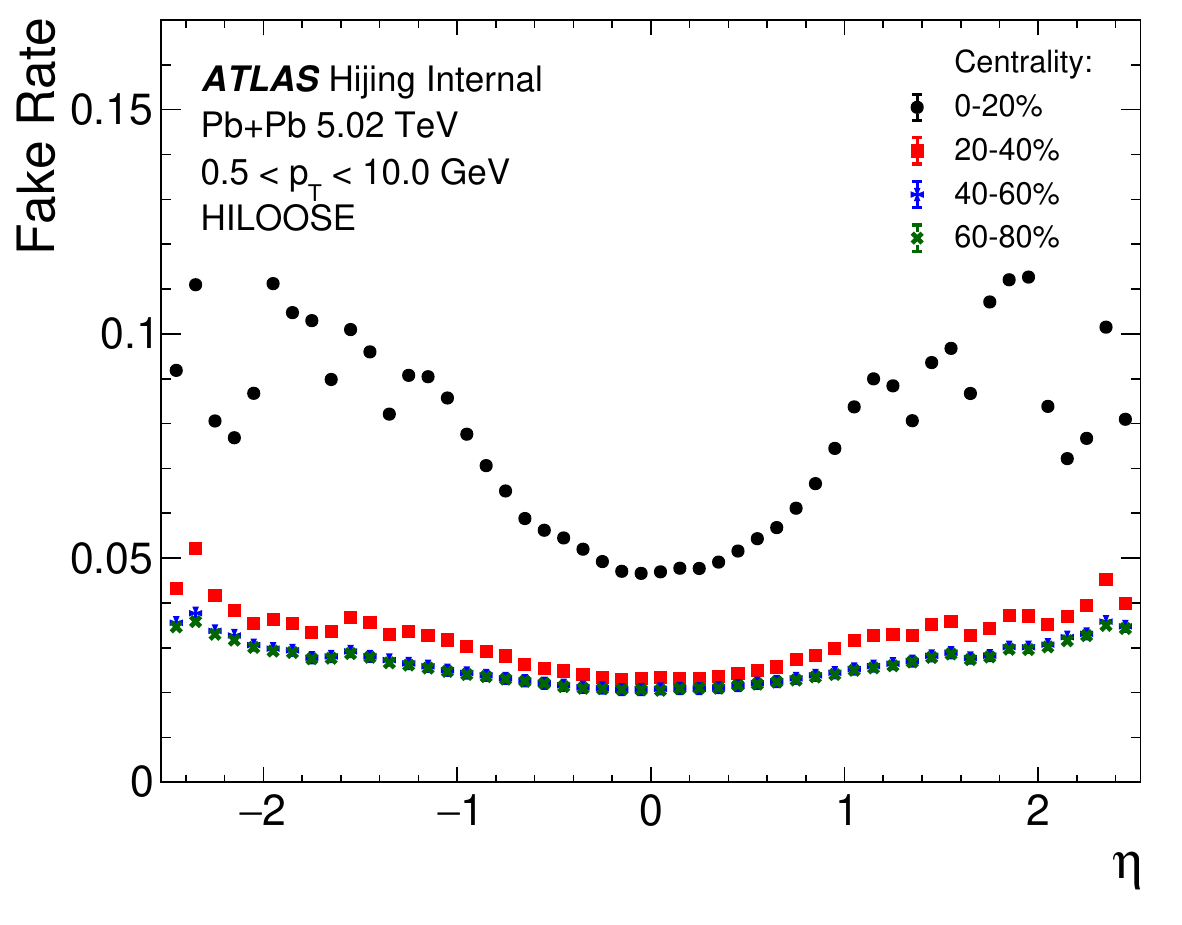}
    \includegraphics[width=0.32\linewidth]{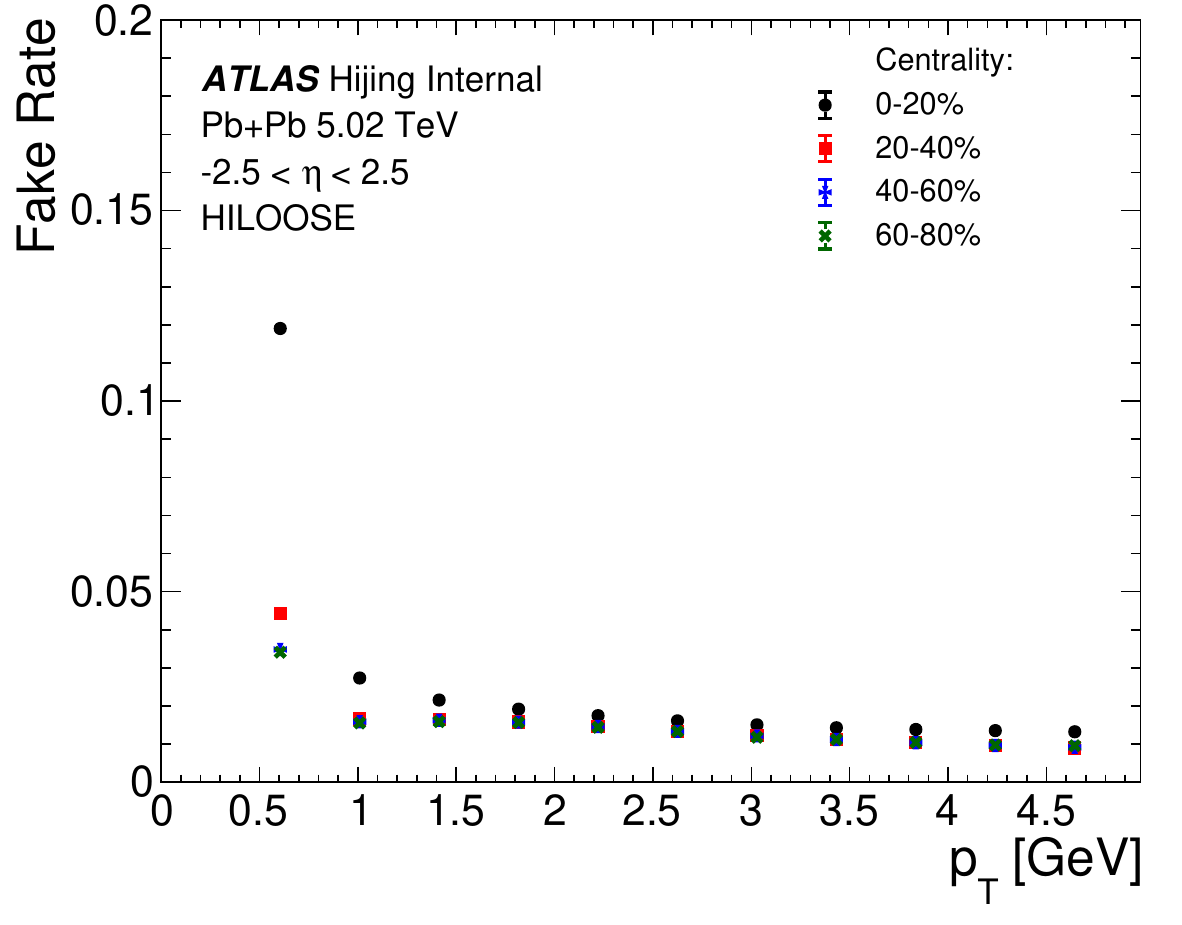}
    \caption{For Pb+Pb collisions at $\sqrt{s_{\mathrm{NN}}} = 5.02$ TeV: (Left) Fake rate vs. $\eta$ for 4 $\pT$ intervals (0-80$\%$ centrality, HILOOSE). (Center) Fake rate vs. $\eta$ for different centralities ($0.5<\pT<5.0$ GeV). (Right) Fake rate vs. $\pT$ for different centralities ($|\eta|<2.5$).}
    \label{fig:evtSel_FakTot}
\end{figure}

Figure~\ref{fig:evtSel_FakTot} illustrates fake rates vs. $\eta$ (left), vs. centrality (middle, $0.5 < \pT < 5$ GeV), and vs. $\pT$ (right, $|\eta| < 2.5$). For HILOOSE tracks, $f$ generally decreases with increasing $\pT$ and from central to peripheral collisions, stabilizing for $\pT > 1$ GeV.

\begin{figure}[htbp]
    \centering
    \includegraphics[width=0.32\linewidth]{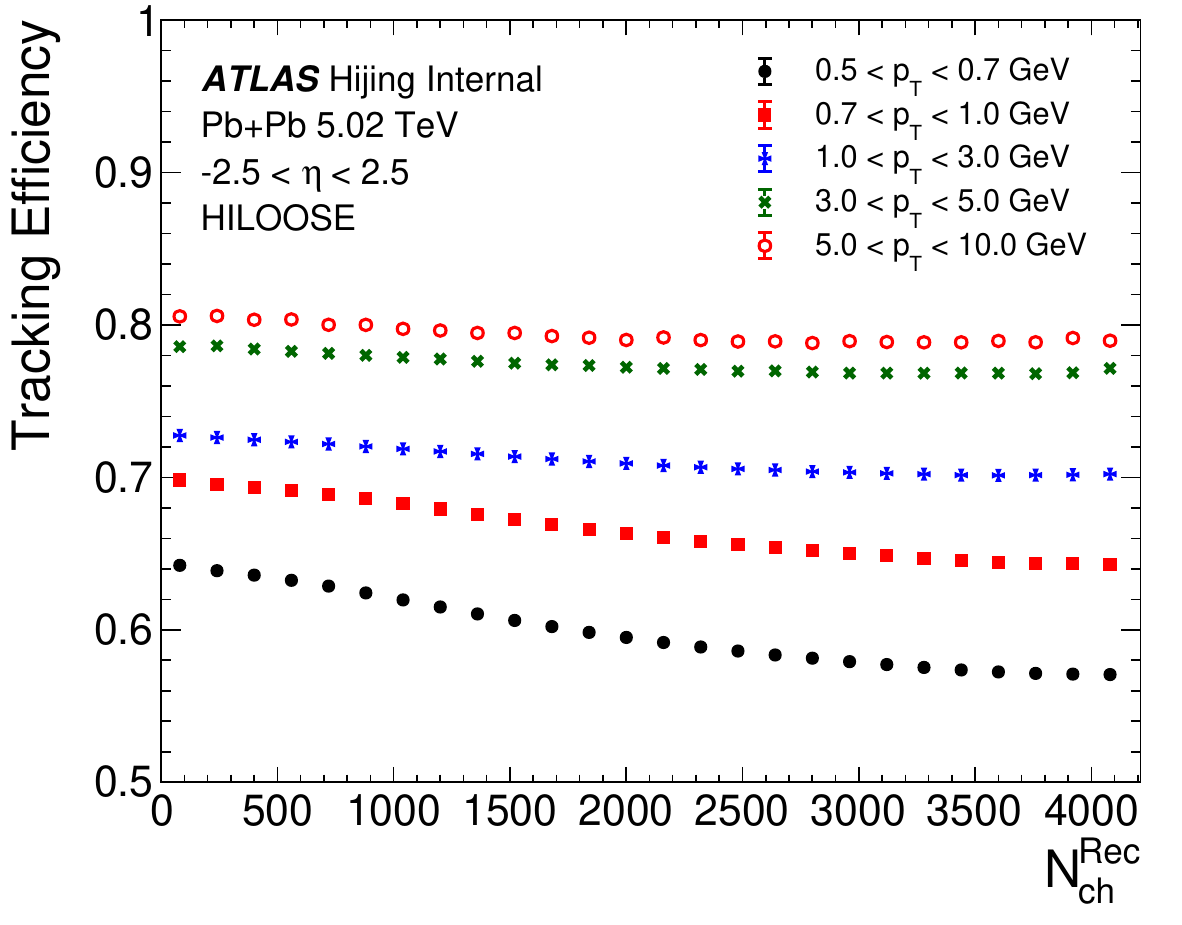}
    \includegraphics[width=0.32\linewidth]{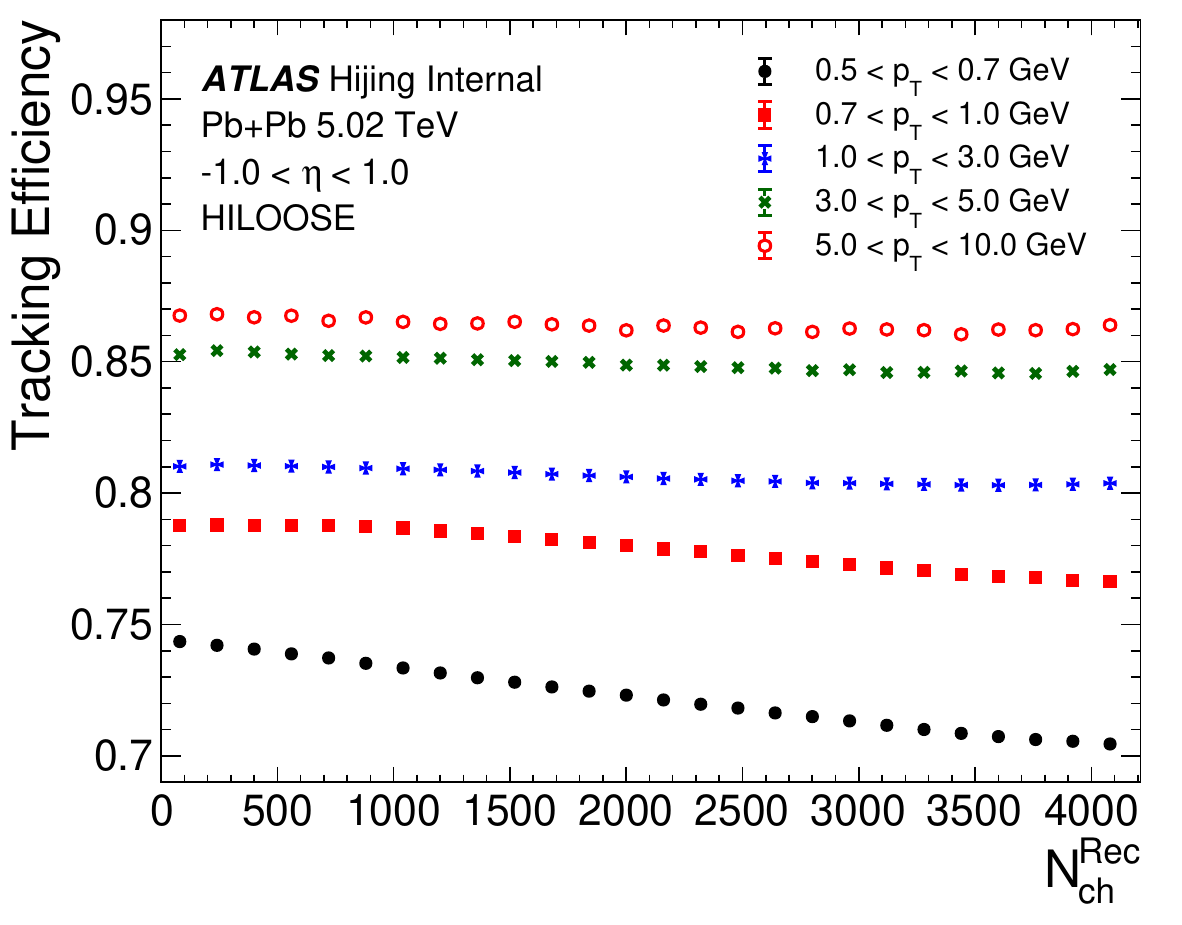}
    \includegraphics[width=0.32\linewidth]{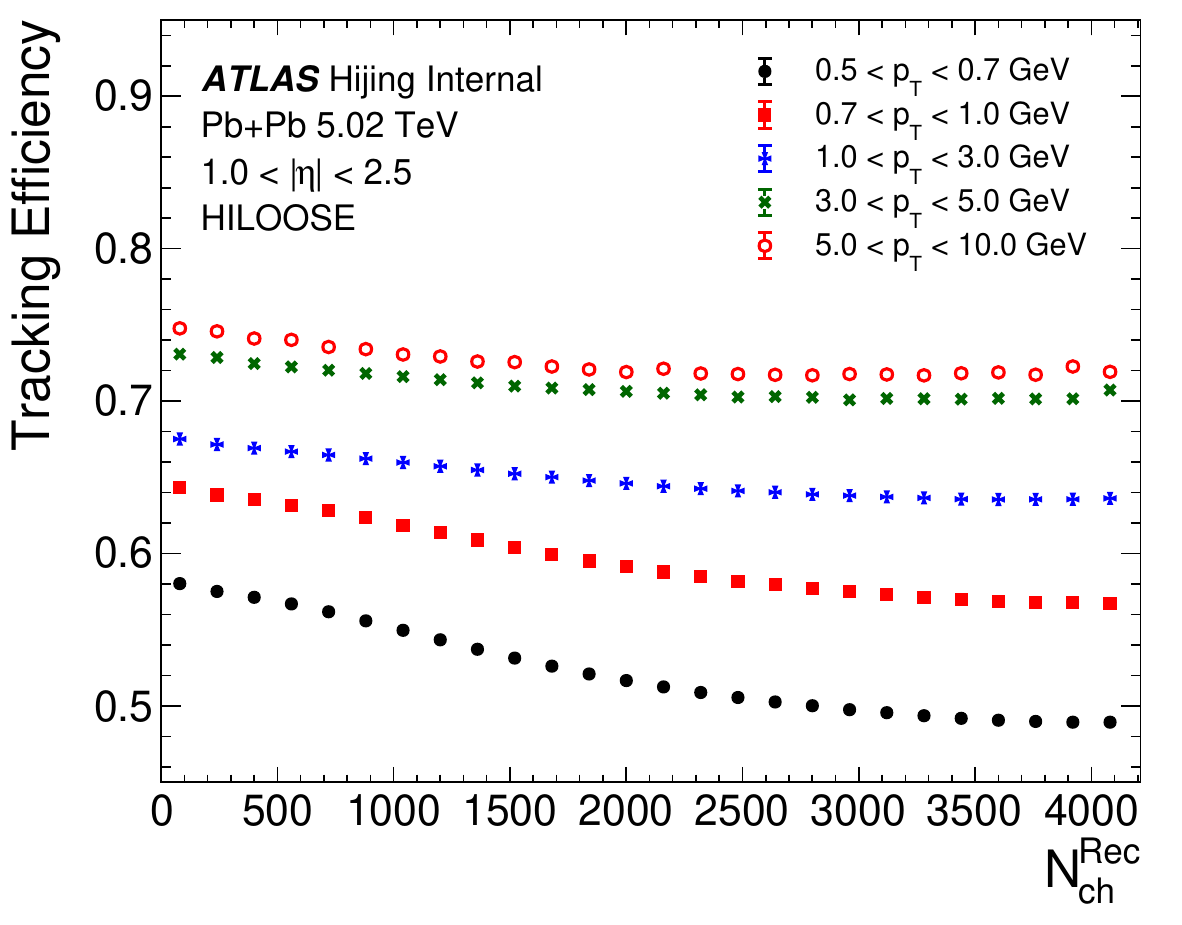}
    \includegraphics[width=0.32\linewidth]{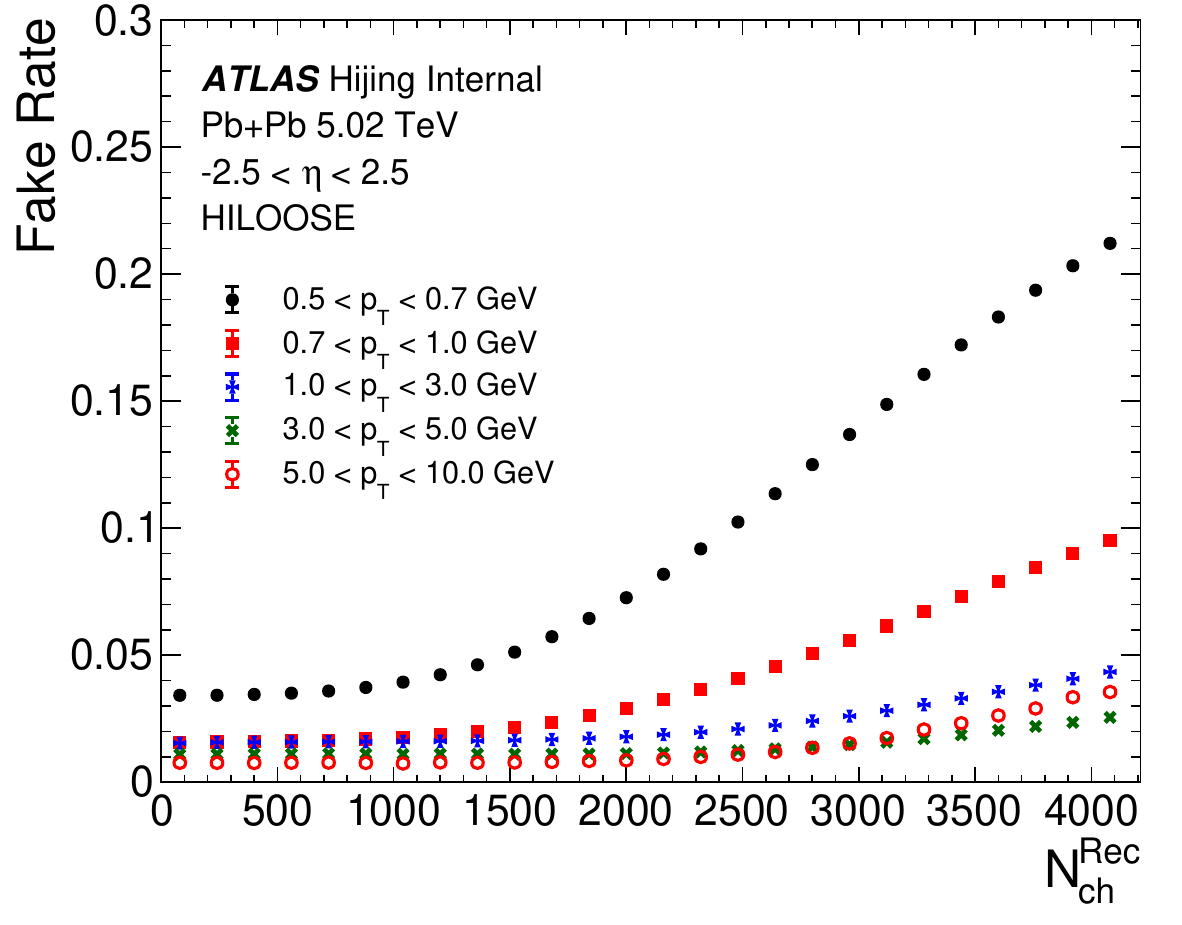}
    \includegraphics[width=0.32\linewidth]{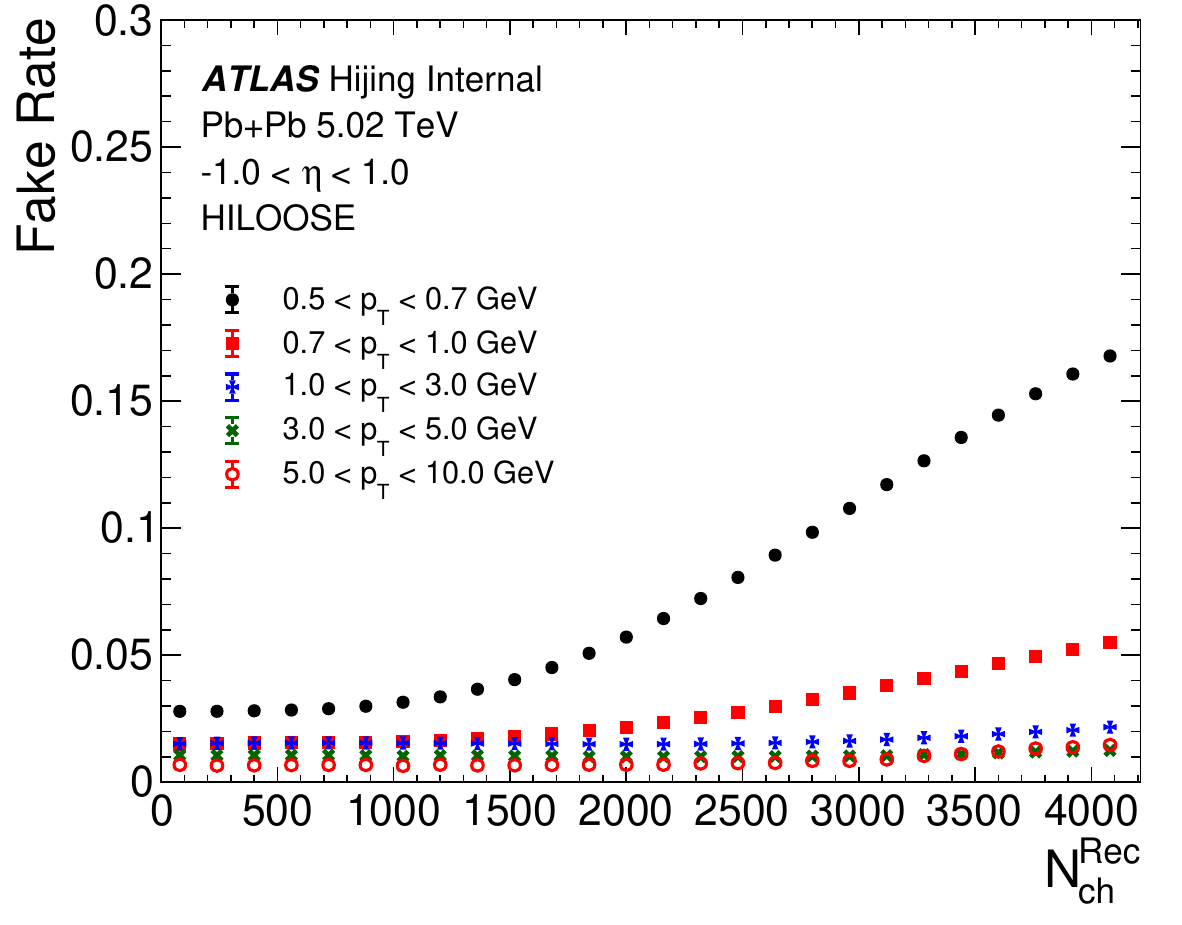}
    \includegraphics[width=0.32\linewidth]{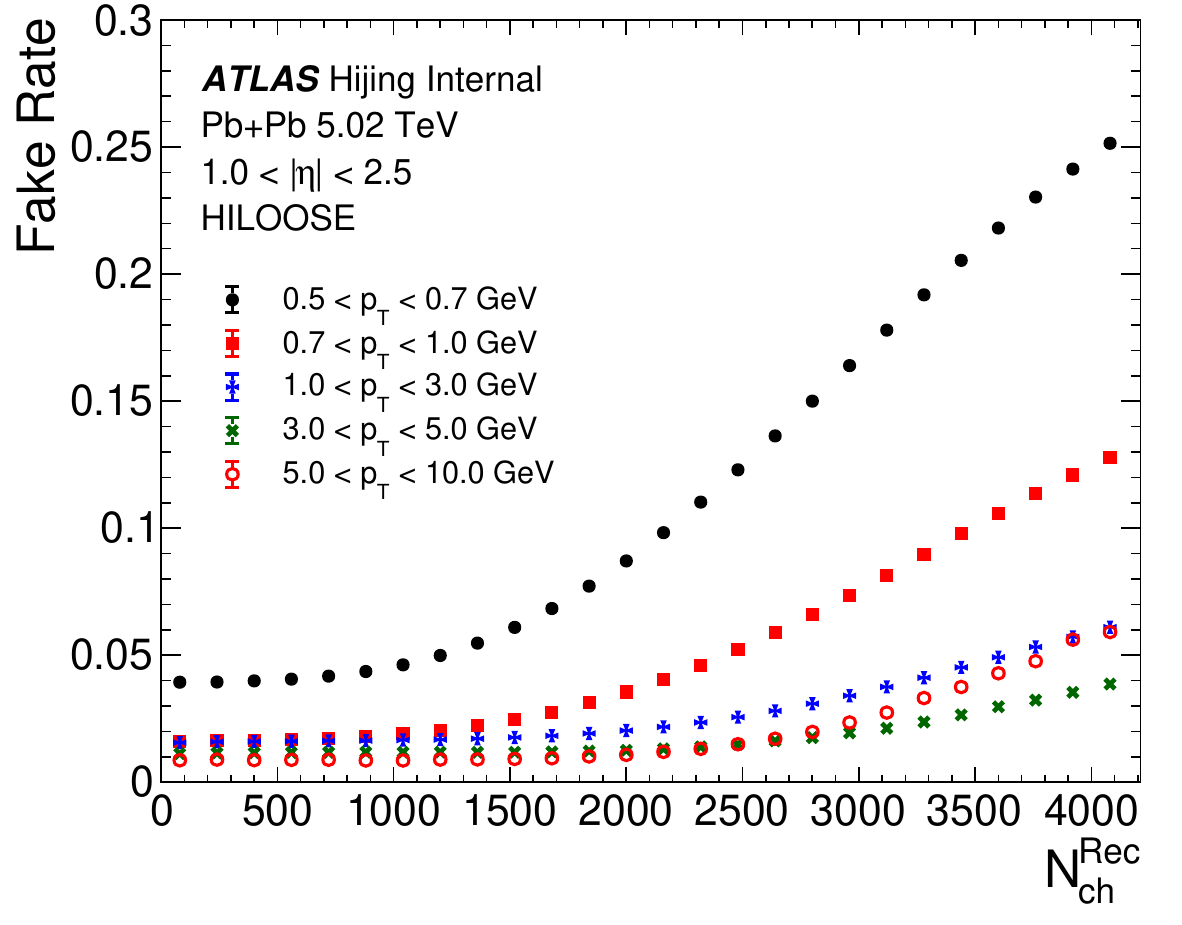}
    \caption{Tracking efficiency (top row) and Fake rate (bottom row) as functions of $\NchR$ for five $\pT$ intervals for $|\eta|<2.5$ (left), $|\eta|<1$ (center), and $1<|\eta|<2.5$ (right) in Pb+Pb collisions at $\sqrt{s_\mathrm{NN}} = 5.02$ TeV.}
    \label{fig:EffAndFake_XaxisNchRec}
\end{figure}

Figure~\ref{fig:EffAndFake_XaxisNchRec} presents $\epsilon$ (top) and $f$ (bottom) vs. HILOOSE $\NchR$ for $|\eta|<2.5$ (left), $|\eta|<1$ (center), and $1<|\eta|<2.5$ (right). Efficiency decreases with $\NchR$ and is higher at mid-rapidity. Fake rate increases with $\NchR$ and is lower at mid-rapidity.

\subsubsection{Tracking Performance in Xe+Xe Collisions}
\begin{figure}[htbp]
\centering
\includegraphics[scale=0.22]{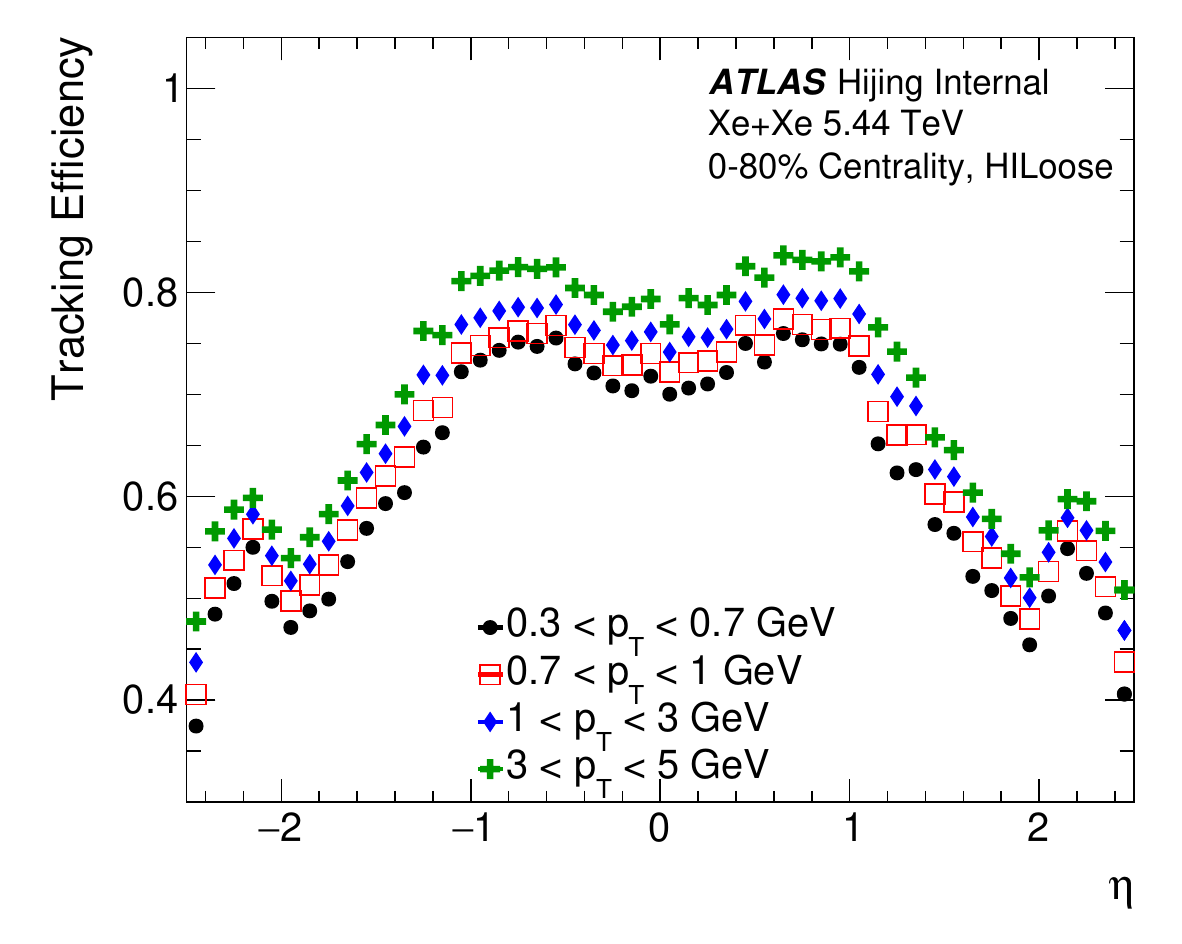}
\includegraphics[scale=0.22]{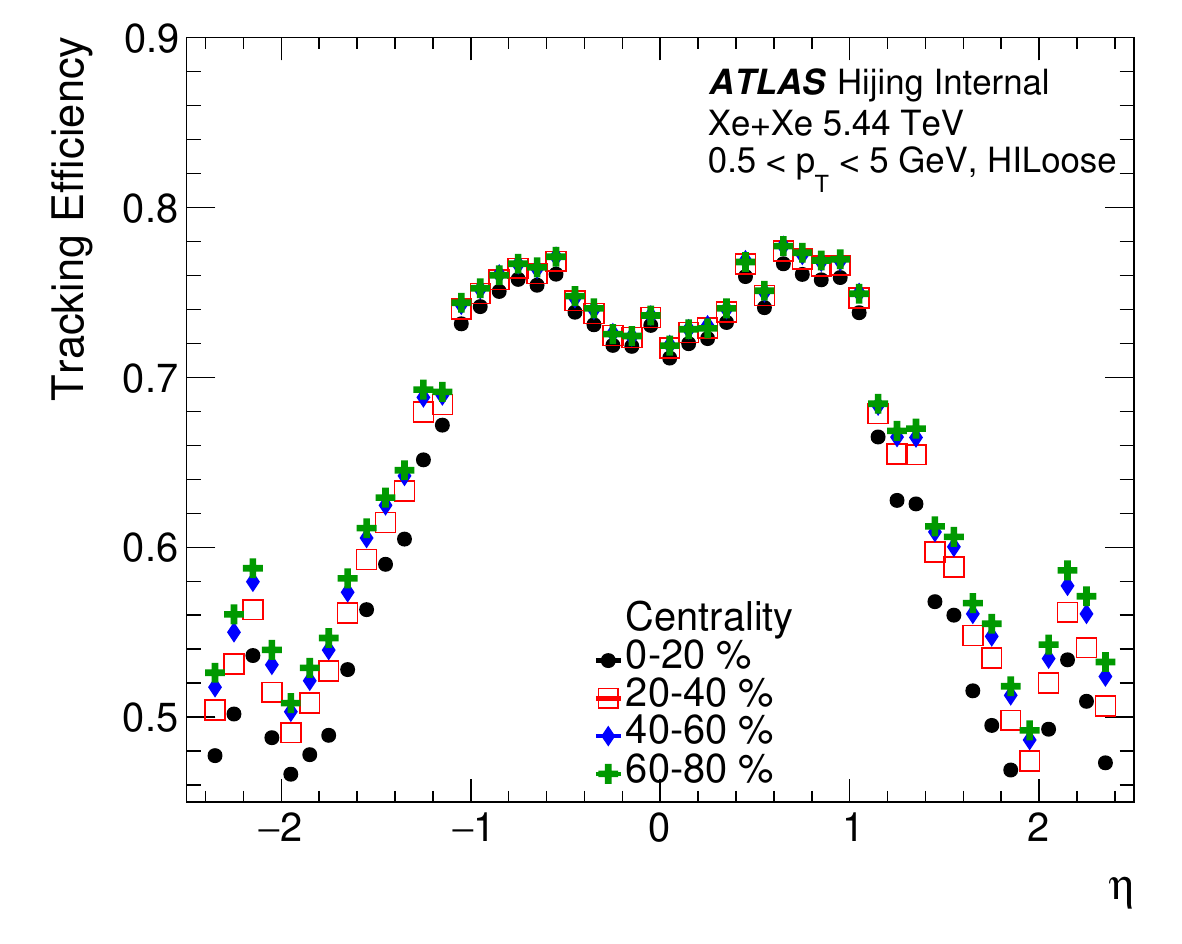}
\includegraphics[scale=0.22]{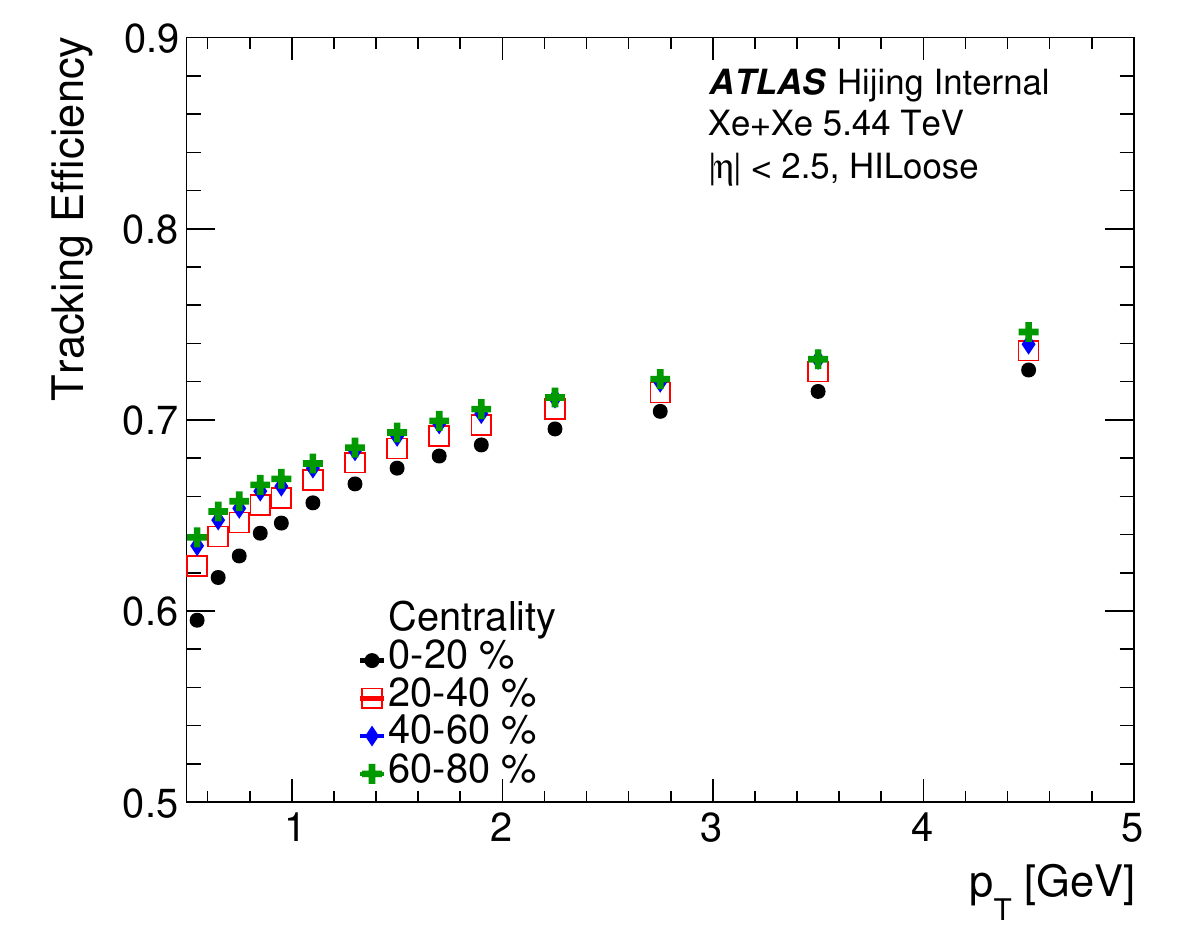}
\caption{Tracking efficiency for Xe+Xe collisions at $\sqrt{s_{\mathrm{NN}}} = 5.44$ TeV. (Left) Efficiency vs. $\eta$ for four $\pT$ intervals (0-80$\%$ centrality). (Center) Efficiency vs. $\pT$ for four centrality intervals ($|\eta|<2.4$). (Right) Efficiency vs. $\eta$ for different centrality intervals ($0.5 < \pT < 5.0$ GeV).}
\label{fig:evtSel_Eff}
\end{figure}

Figure~\ref{fig:evtSel_Eff} shows tracking efficiencies for Xe+Xe collisions: vs. $\eta$ (left, for four $\pT$ intervals, 0-80\% centrality), vs. $\pT$ (center, for four centrality intervals, $|\eta|<2.4$), and vs. $\eta$ (right, for different centralities, $0.5 < \pT < 5.0$ GeV). Efficiency increases from central to peripheral events and with $\pT$. Xe+Xe track reconstruction efficiency is generally lower than for Pb+Pb, attributed to variations in detector conditions and software versions.

\begin{figure}[htbp]
\centering
\includegraphics[scale=0.22]{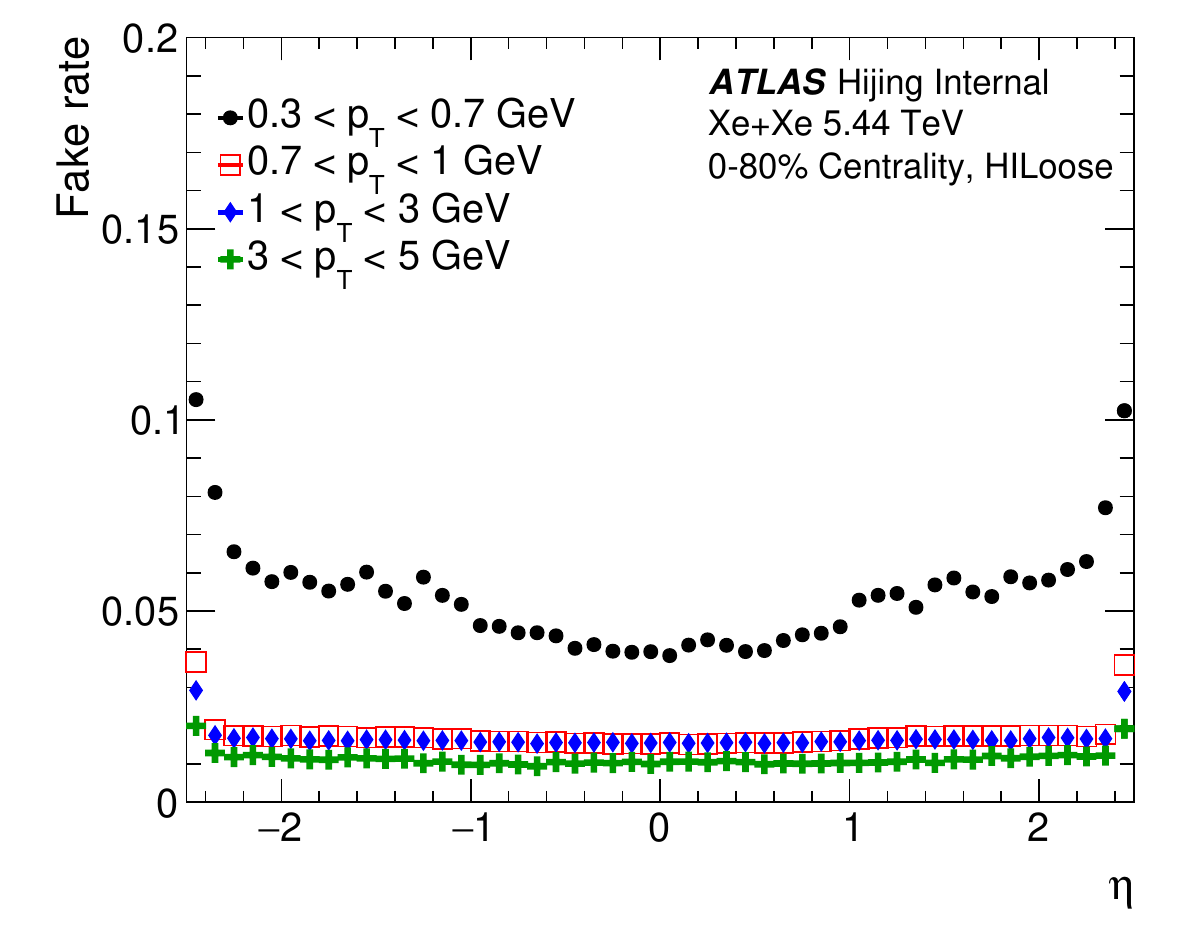}
\includegraphics[scale=0.22]{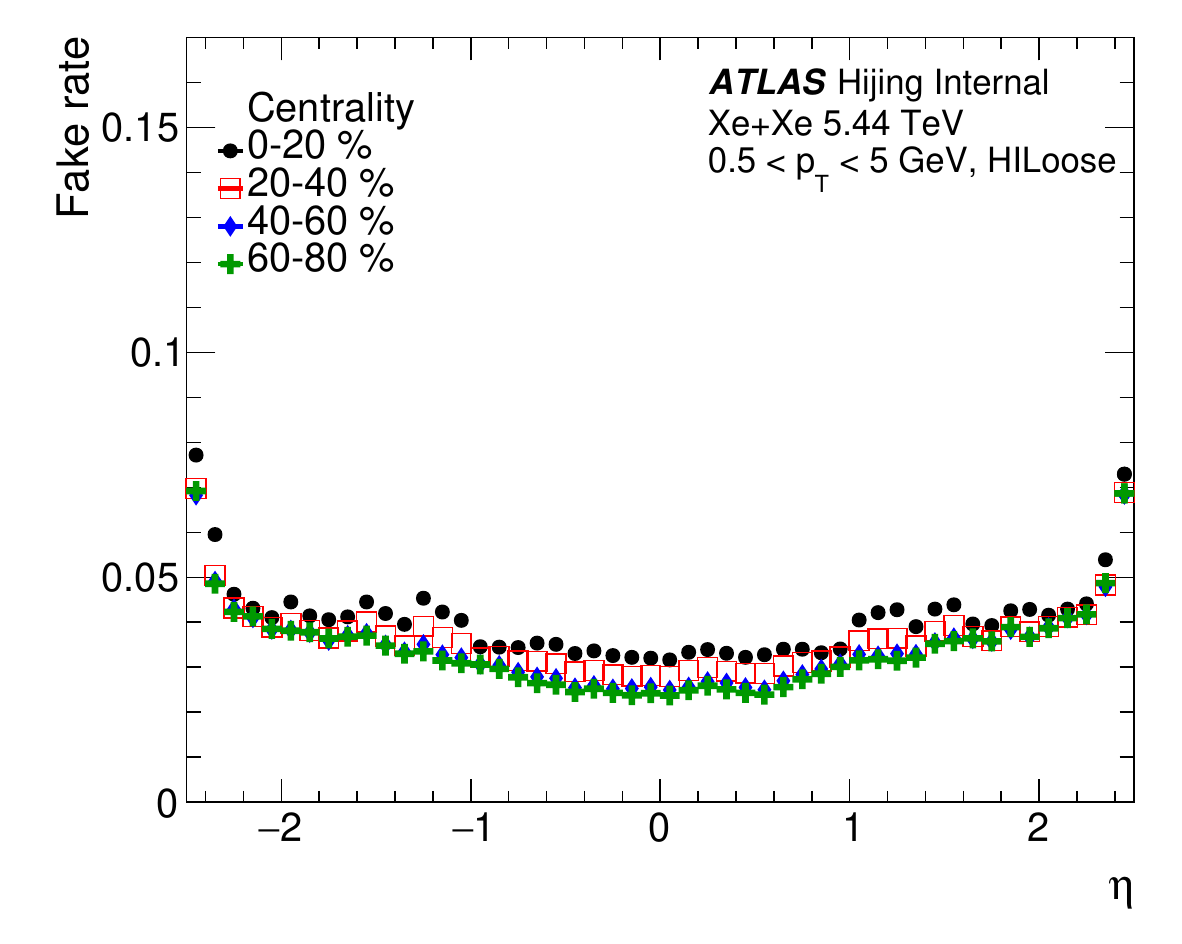}
\includegraphics[scale=0.22]{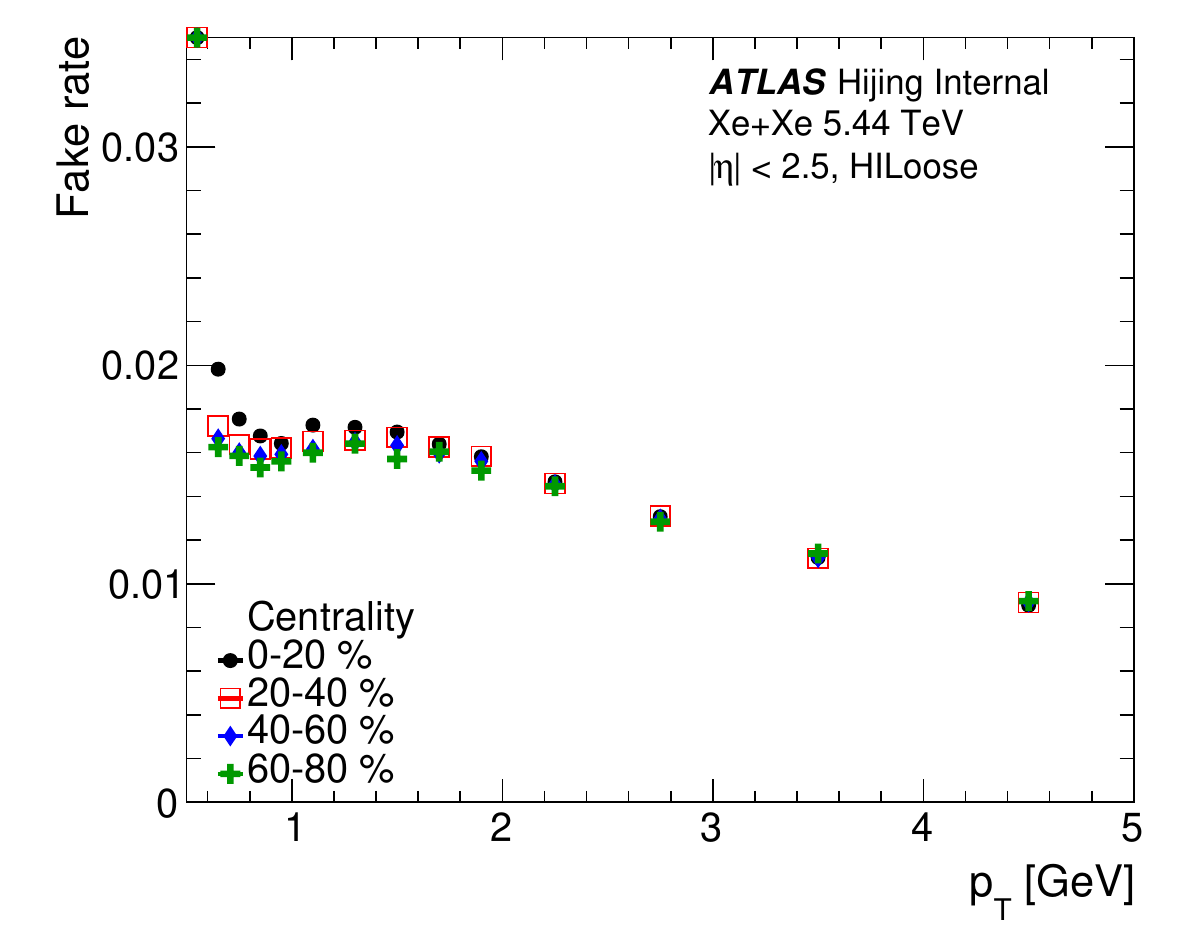}
\caption{Fake track rate for Xe+Xe collisions at $\sqrt{s_{\mathrm{NN}}} = 5.44$ TeV. (Left) Fake rate vs. $\eta$ for four $\pT$ intervals (0-80$\%$ centrality). (Center) Fake rate vs. $\pT$ for four centrality intervals ($|\eta|<2.4$). (Right) Fake rate vs. $\eta$ for different centrality intervals ($0.5 < \pT < 5.0$ GeV).}
\label{fig:evtSel_Fak}
\end{figure}

Figure~\ref{fig:evtSel_Fak} presents fake track rates for Xe+Xe MC: vs. $\eta$ (left, for four $\pT$ bins, 0-80\% centrality), vs. $\pT$ (center, for four centrality intervals, $|\eta|<2.4$), and vs. $\eta$ (right, for different centralities, $0.5 < \pT < 5.0$ GeV). Fake rates decrease from central to peripheral collisions and with increasing $\pT$.

\section{Event Activity and Centrality Definition}
\label{sec:event_activity_centrality}
Observables are measured in narrow bins of variables reflecting event activity. Centrality, a measure of the collision's geometric overlap, is determined using these activity variables.

\subsection{Event Activity Variables}
Key event activity estimators include:
\begin{itemize}
    \item $\NchR$: The number of reconstructed tracks fulfilling specific quality criteria (typically HILOOSE with $0.5 < \pT < 5.0$ GeV and $|\eta|<2.5$). Calculated as $\NchR = \sum_{i} 1$.
    \item $N_{\mathrm{ch}}$: The efficiency- and fake-corrected $\NchR$. Calculated as $N_{\mathrm{ch}} = \sum_{i} w_{i}$, where $w_{i}$ is the track correction weight (defined in Eq.~\ref{eq:CorrEff}).
    \item FCal $\sumET$: The transverse energy deposited in the Forward Calorimeter.
\end{itemize}
Observables are initially measured in fine bins (e.g., unit-width in $\NchR$/$N_{\mathrm{ch}}$, or 1 GeV in FCal $\sumET$) and then rebinned.

\subsubsection{Distributions and Correlations in Pb+Pb Collisions}
\begin{figure}[htbp]
  \centering
  \includegraphics[width=0.32\linewidth]{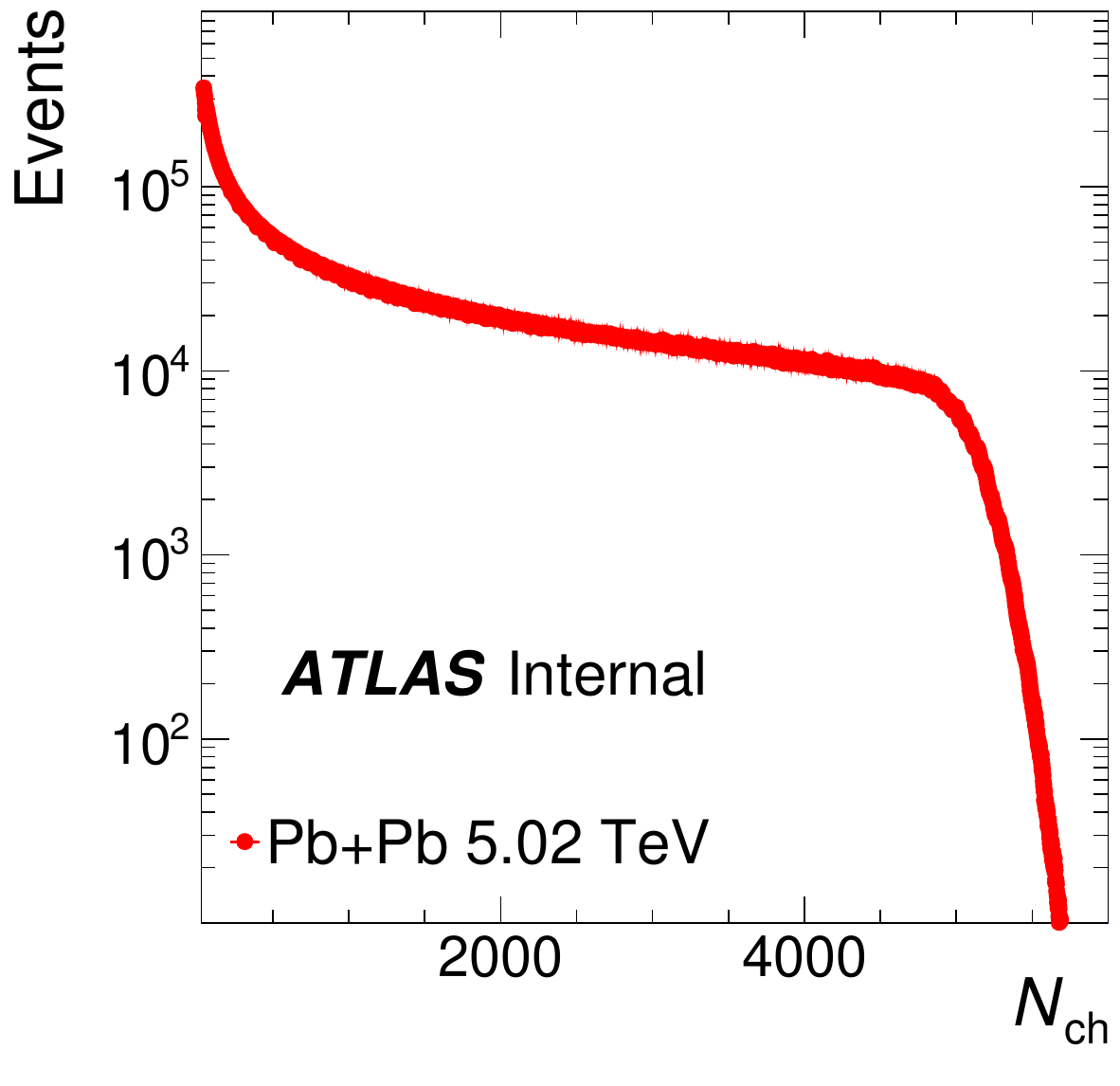}
  \includegraphics[width=0.32\linewidth]{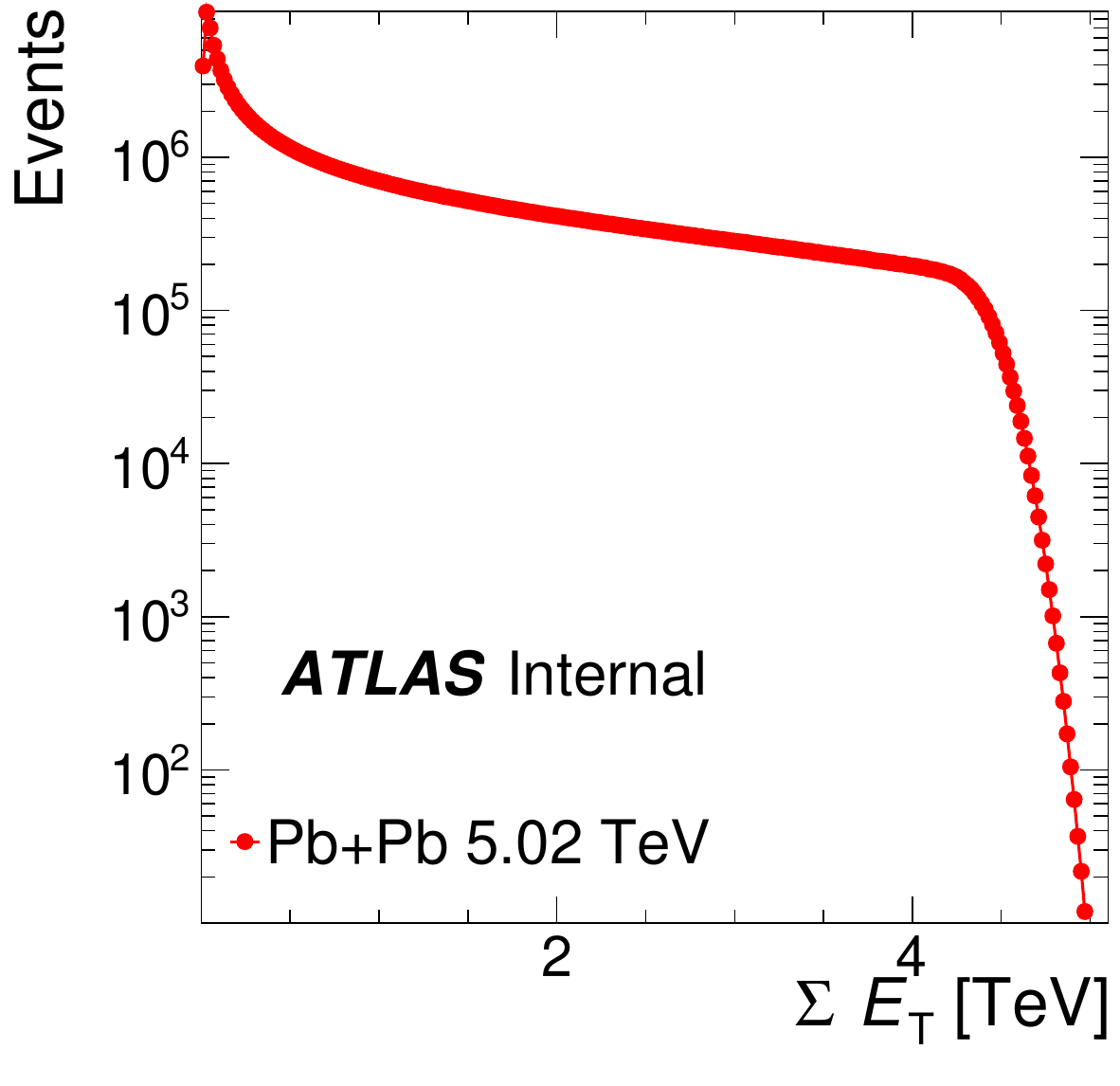}
  \includegraphics[width=0.32\linewidth]{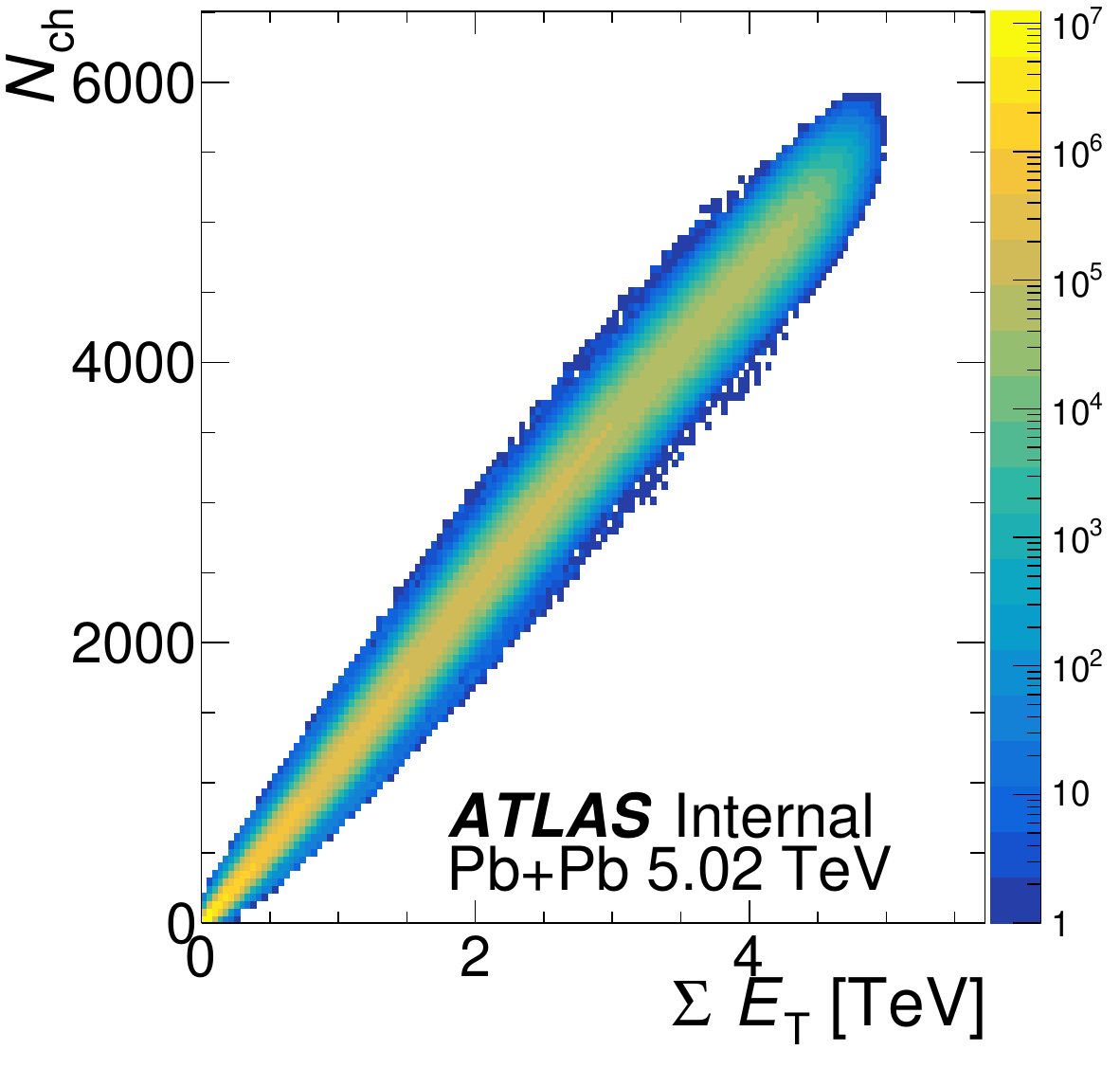}
  \caption{Distributions of FCal–$\sum E_{T}$ (left), $N_{\mathrm{ch}}$ (middle), and the correlation between $N_{\mathrm{ch}}$ and FCal–$\sum E_{T}$ (right) in Pb+Pb collisions.}
  \label{fig:evtSel_Et_cmp}
\end{figure}

Figure~\ref{fig:evtSel_Et_cmp} shows the distributions of FCal $\sumET$, $N_{\mathrm{ch}}$, and their correlation in Pb+Pb collisions.

\subsubsection{Distributions and Correlations in Xe+Xe Collisions}
\begin{figure}[htbp]
\centering
\includegraphics[width=0.3\linewidth]{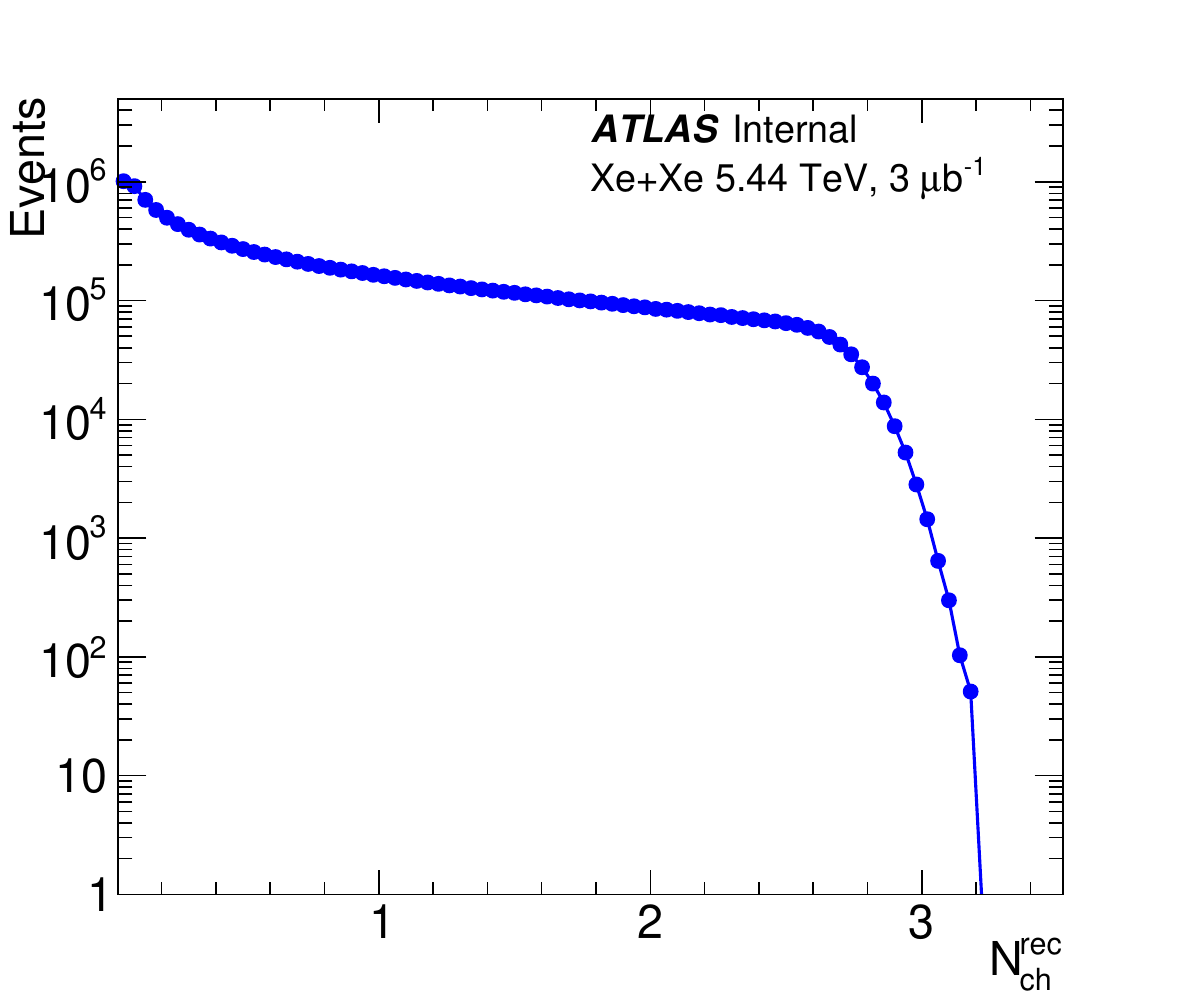}
\includegraphics[width=0.3\linewidth]{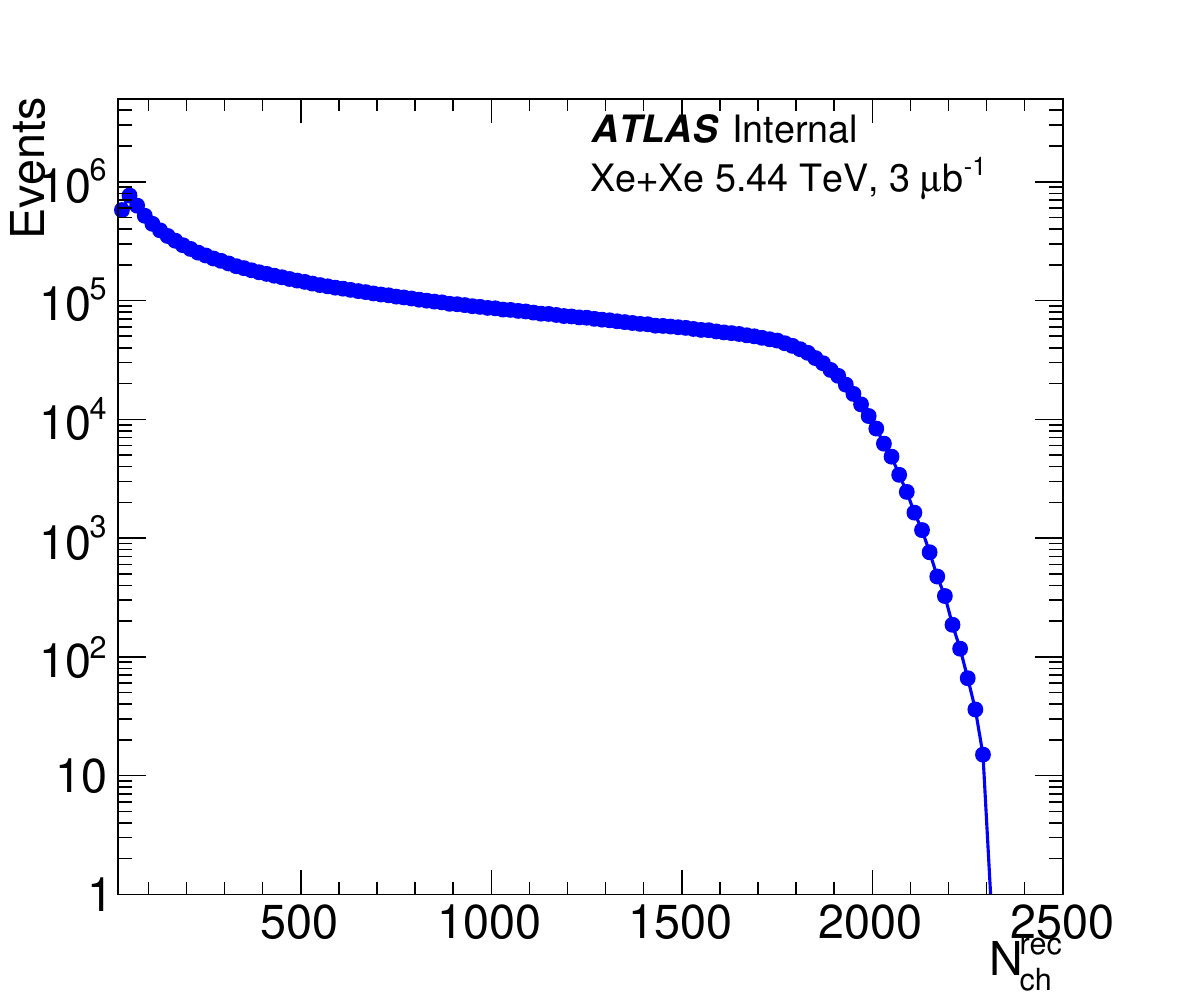}
\includegraphics[width=0.3\linewidth]{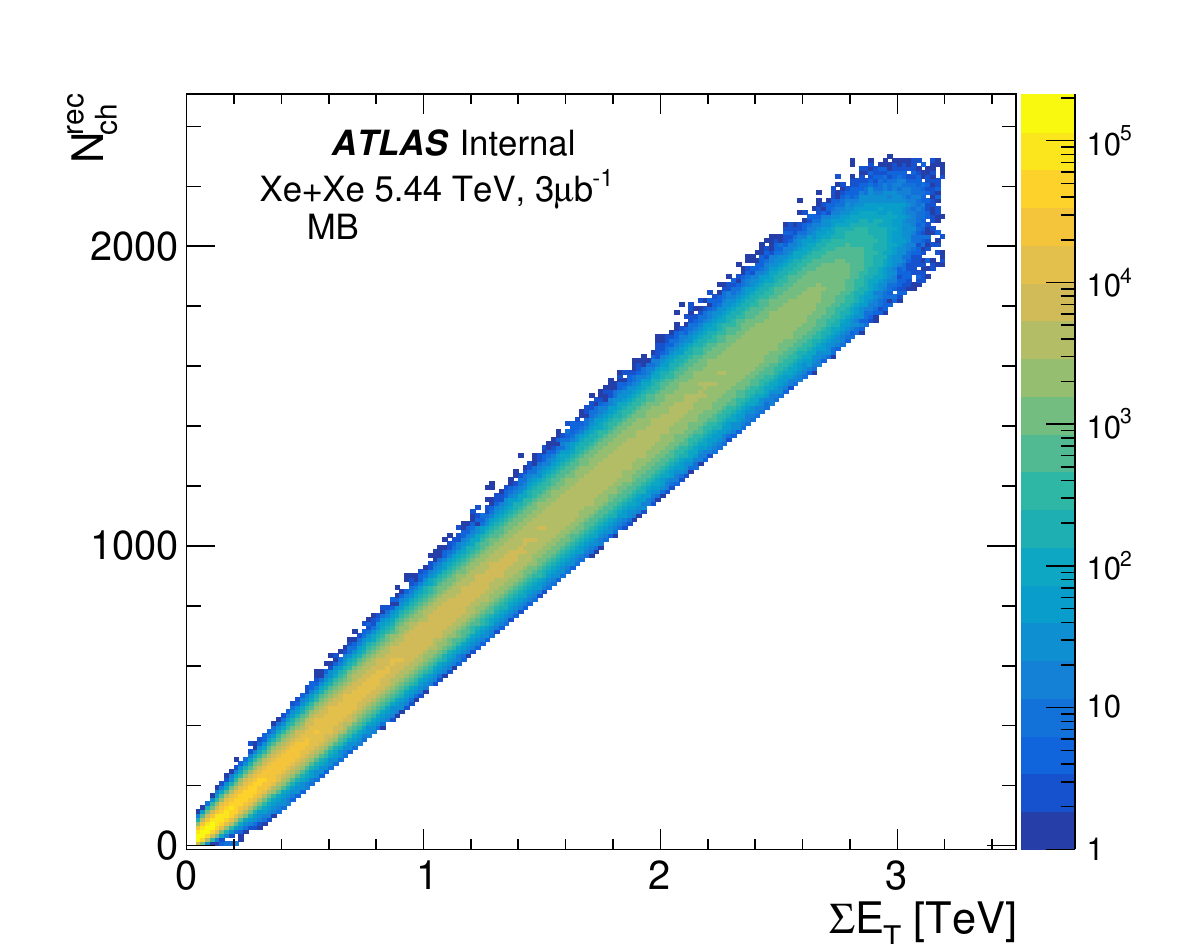}
\caption{(Left) Distribution of FCal-$\sumET$ in Xe+Xe collisions. (Middle) Distribution of $\NchR$ in Xe+Xe collisions. (Right) Correlation between FCal-$\sumET$ and $\NchR$ in Xe+Xe collisions.}
\label{fig:evtSel_Et_cmpXe}
\end{figure}

Figure~\ref{fig:evtSel_Et_cmpXe} displays the FCal-$\sumET$ distribution (left), $\NchR$ distribution (middle), and their correlation (right) for Xe+Xe collisions.

\subsection{Centrality Definition}
\label{sec:centdef}
Centrality classes are defined by fitting the FCal-$\sumET$ distribution with a Glauber model. The use of FCal $\sumET$ is motivated by the ``two-component model'' of particle production, where soft particle production (transverse energy at large $\eta$) scales with a linear combination of $N_{\mathrm{part}}$ (number of participant nucleons) and $N_{\mathrm{coll}}$ (number of binary nucleon-nucleon collisions)~\cite{Wang:2000bf,Perepelitsa:2016}.

\subsubsection{Centrality Thresholds}
The FCal-$\sumET$ thresholds for centrality definitions in Pb+Pb collisions are listed in Appendix Table~\ref{table:PbCent}~\cite{Perepelitsa:2019}. Centrality classes for Xe+Xe collisions are analogously defined based on FCal-$\sumET$ thresholds from a Glauber model fit. These thresholds, are listed in Appendix in Table~\ref{table:XeCent}.

\begin{figure}[htbp]
\centering
\includegraphics[width=0.3\linewidth]{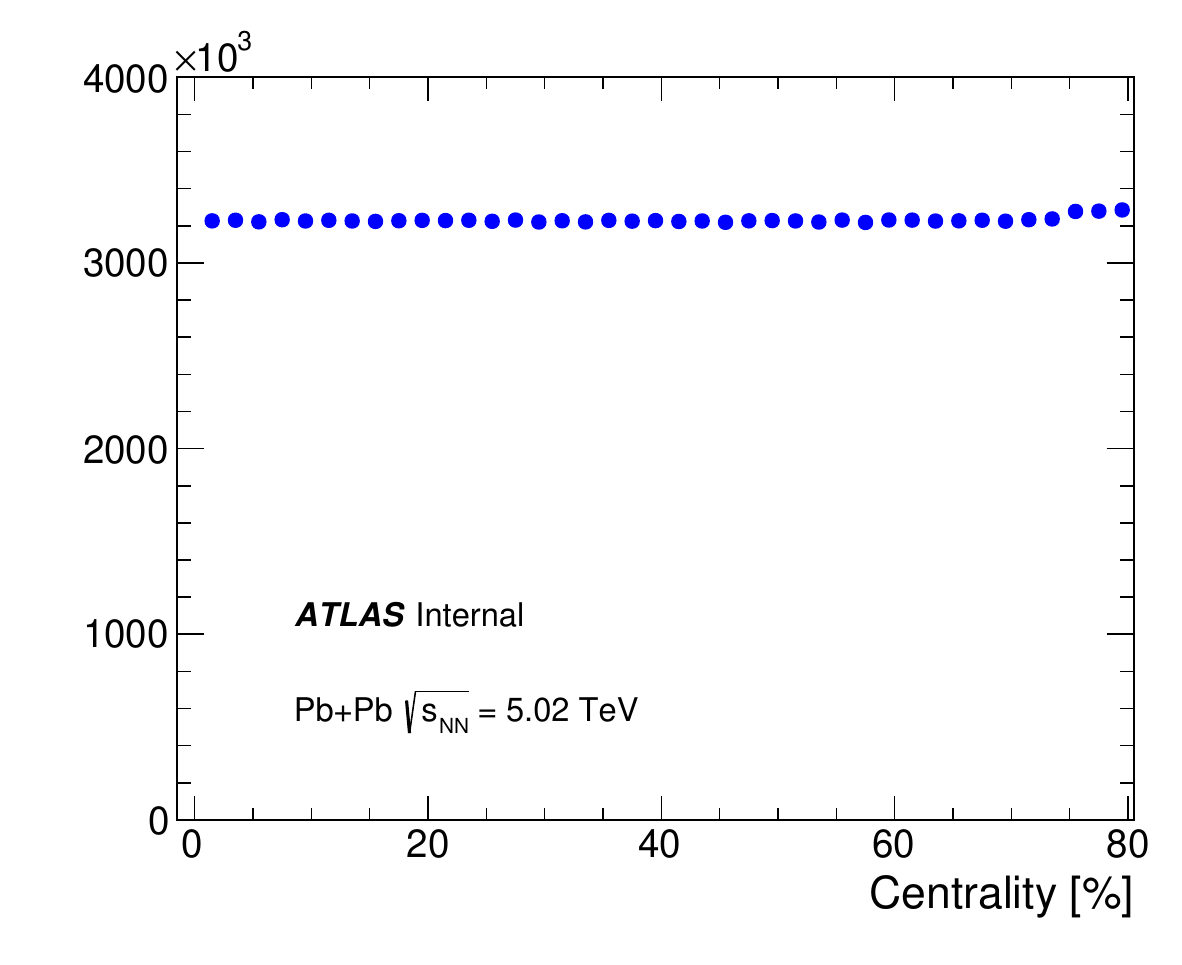}
\includegraphics[width=0.3\linewidth]{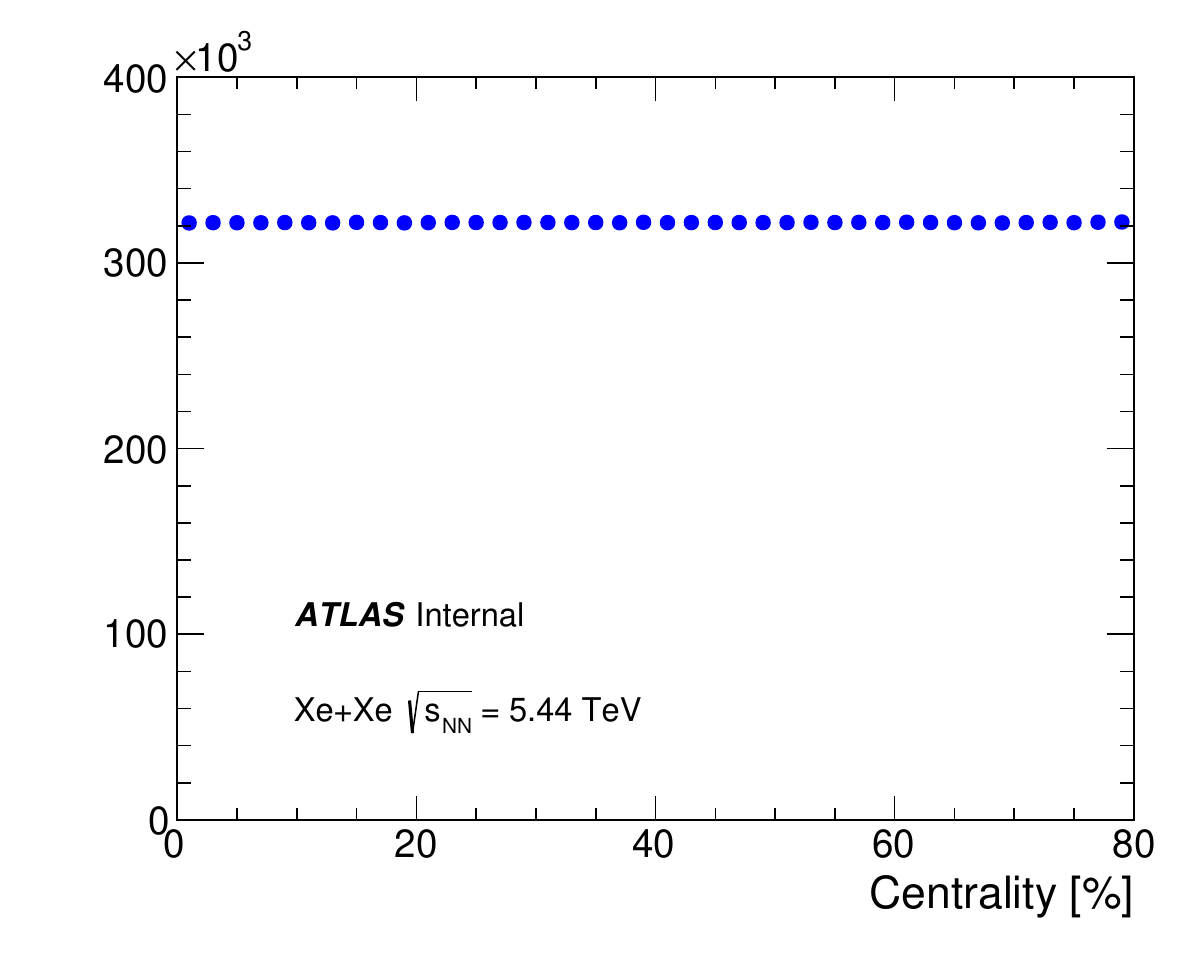}
\caption{centrality distributions in Pb+Pb collisions (left) and in Xe+Xe collisions (right). The error bars represent statistical uncertainties.}
\label{fig:Cent_cmp}
\end{figure}

The centrality distributions in Pb+Pb and Xe+Xe are shown in Fig.~\ref{fig:Cent_cmp}. The distribution is flat in the case of Xe+Xe and Pb+Pb except in peripheral Pb+Pb in 70-80$\%$ centrality where there are some deviations.

\section{Mitigation of Detector Effects}
\label{sec:further_detector_mitigation}
Additional corrections are applied to mitigate residual detector-induced effects, relevant for both Pb+Pb and Xe+Xe collision analyses.

\subsection{Track Azimuthal Distribution Flattening}
\label{subsec:phi_flattening}
In heavy-ion collisions, the particle azimuthal angle ($\phi$) distribution, averaged over many events, is expected to be uniform due to random event plane orientation. Observed non-uniformities indicate detector-induced artifacts. While MC-derived tracking efficiency corrections handle $\eta$ and $\pT$ dependencies, residual $\phi$ dependent variations, potentially arising from detector imperfections or acceptance, require an additional ``flattening'' correction.

A correction factor $w_{\phi}$ is defined as:
\begin{equation} \label{eq:flatWgt}
w_{\phi}(\mathrm{cent},\zvtx,\mathrm{charge},\pT,\eta,\phi)=\frac{\langle N(\delta\eta) \rangle_{\phi}}{N(\delta\eta,\delta\phi)}
\end{equation}
where $N(\delta\eta,\delta\phi)$ is the particle count in a differential $(\delta\eta,\delta\phi)$ window, and $\langle N(\delta\eta) \rangle_{\phi}$ is the average particle count in that $\delta\eta$ window across all $\phi$. These $w_{\phi}$ factors are calculated for both collision systems studied. Applying $w_{\phi}$ ensures a uniform $\phi$ distribution for narrow $\eta$ intervals.
Since detector effects can depend on occupancy, $w_{\phi}$ is evaluated for different centrality ranges and also depends on $\zvtx$, particle charge, $\pT$, and $\eta$.

\subsection{Recentering of Flow Vectors}
\label{subsec:recentering}
In flow analysis, ``recentering'' corrects residual detector non-uniformities affecting unit-flow vectors derived from the inner detector. This involves subtracting the run-averaged X and Y components of the event-wise q-vectors for each harmonic.

The recentered q-vector components, $q_{n,x}$ and $q_{n,y}$, are:
\begin{align} \label{recEqn}
q_{n,x} &= \frac{\sum_i w_i \cos(n\phi_i)}{\sum_i w_i} - \langle q_{n,x} \rangle \\
q_{n,y} &= \frac{\sum_i w_i \sin(n\phi_i)}{\sum_i w_i} - \langle q_{n,y} \rangle
\end{align}
where $w_i$ are particle weights (typically $w_{id}$), $\phi_i$ are azimuthal angles, $n$ is the harmonic number, and $\langle \cdot \rangle$ denotes the average over events within a specific data acquisition run. The q-vector is normalized by the sum of weights.

The recentering offsets, $\langle q_{n,x} \rangle$ and $\langle q_{n,y} \rangle$, are determined in bins of $\NchR$ and for various $\eta$ and $\pT$ bins. These offsets generally show independence from event centrality, suggesting a detector azimuthal acceptance that does not vary significantly with event activity. Their magnitudes are typically small ($< 0.001-0.005$) for all harmonics, much smaller than the expected true flow signal. Further details and plots for recentering corrections for both collision systems are provided in Appendix~\ref{sec:dAna_Pb}.

The final weight for a particle reconstructed in the inner detector, $w_{i}$, combines the flattening correction $w_{\phi}$ with the tracking efficiency $\epsilon$ and fake rate $f$ (from Section~\ref{sec:effFak}):
\begin{equation} \label{eq:totWgt}
w_{i}(\mathrm{cent},\zvtx,\mathrm{charge},\pT,\eta,\phi)=w_{\phi}(\mathrm{cent},\zvtx,\mathrm{charge},\pT,\eta,\phi)\frac{1-f(\mathrm{cent},\pT,\eta)}{\epsilon(\mathrm{cent},\pT,\eta)}
\end{equation}
This weight $w_{i}$ is used in all observables constructed from inner detector particles. Further details and plots for flattening weights for both collision systems are provided in Appendix~\ref{sec:dAna_Pb}.

With the event-selection criteria and detector corrections now established, we proceed in the next chapter to measure the novel observables in heavy-ion collision systems.
\clearpage



\chapter{Measurements and Results}
\label{chap:results}

\section{Evidence for Collective Radial Expansion of QGP}
\label{sec:chap4_v0pt}
Despite theoretical models suggesting a collective nature for the radial expansion of the QGP in HIC, direct experimental evidence has remained elusive. This chapter provides the first comprehensive experimental evidence showing that the event-by-event fluctuations in radial expansion of QGP arises as a collective response of the medium.


The collective expansion of QGP manifests itself in two complementary forms: anisotropic and radial flow. Anisotropic flow arises from the initial geometric anisotropy present in the nuclear overlap region and are quantified by the coefficients $v_n$ from the Fourier expansion of the azimuthal particle distribution. In contrast, radial flow describes the collective, isotropic outward expansion of the system. Extensive experimental measurements of $\pT$-differential studies of $v_n(\pT)$ and their event-by-event fluctuations~\cite{Adare:2011tg,ALICE:2011ab,ATLAS:2012at,Chatrchyan:2013kba,Aad:2013xma,Aad:2014fla,Aad:2015lwa,ATLAS:2019peb}, have provided stringent constraints on both the initial conditions of the collision and the transport properties of the QGP medium.

\begin{figure}[h!]
\centering
\includegraphics[width=0.5\linewidth]{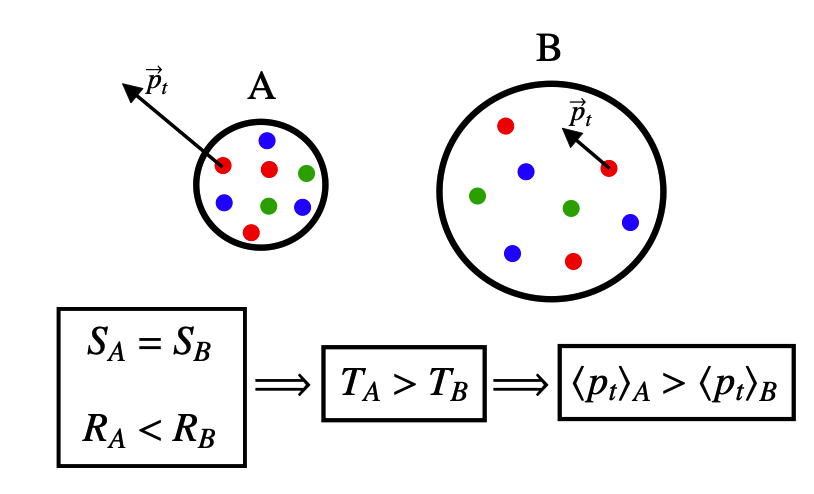}
\caption{Illustration of the relation between $R$ (system size) and $\langle p_{T}\rangle$. For two systems with the same total entropy, $S$, the one with a smaller overlap area will experience larger pressure gradients, leading to a larger outward expansion or $\langle p_{T}\rangle$~\cite{Giacalone:2020awm}.}
\label{fig:Rptcartoon}
\end{figure}

The hydrodynamic response to fluctuations in the initial overlap region's overall size leads to variations in radial flow. This phenomenon manifests as event-by-event fluctuations in the average transverse momentum, $\langle p_{T}\rangle$. Specifically, as shown in Figure~\ref{fig:Rptcartoon}, events with similar total energy but a smaller initial transverse size are expected to have larger pressure gradients in the initial state, resulting in stronger radial expansion and $\langle p_{T}\rangle$~\cite{Bozek:2012fw,Bozek:2017jog}.

\begin{figure}[h]
    \centering
    \includegraphics[width=0.8\linewidth]{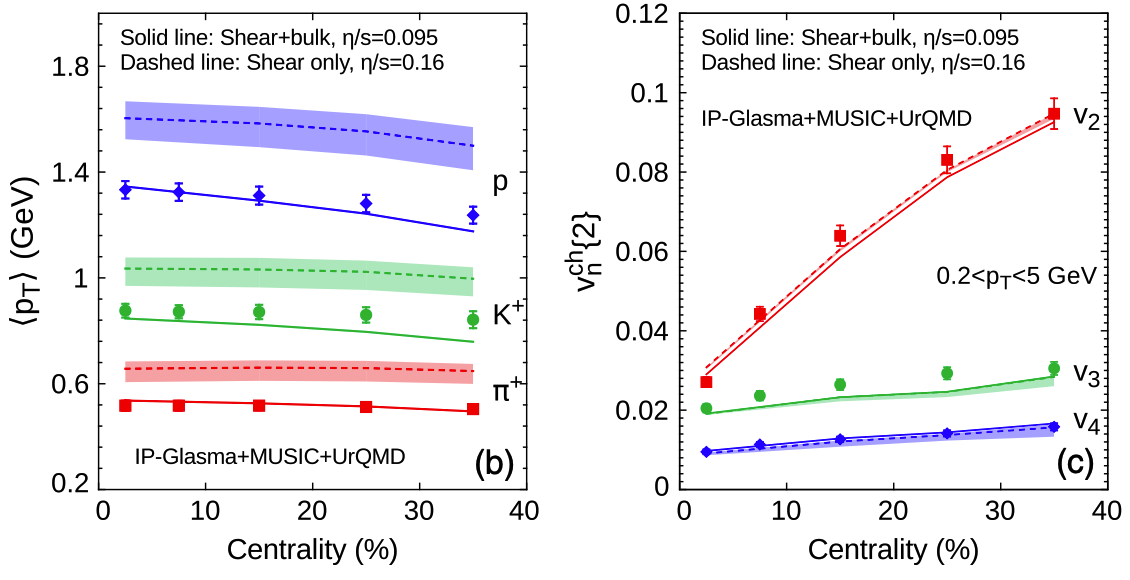}
    \caption{(Left) Ensemble-averaged event-wise transverse momentum, $\lr{[\pT]}$, and (Right) anisotropic flow coefficients $v_n\{2\}$ as functions of centrality. Bands around the dashed lines indicate the effect of varying the switching temperature for the start of hydrodynamic expansion $T_{\rm switch}$ in model calculations~\cite{Ryu:2015vwa}. Data points are from the ALICE Collaboration~\cite{ALICE:2011ab,ALICE:2013mez}.}
    \label{fig:v0pt_shearbulk}
\end{figure}

While anisotropic flow measurements primarily constrain the shear viscosity to entropy density ratio ($\eta/s$), theoretical model calculations indicate that fluctuations in radial flow are especially sensitive to the medium's bulk viscosity to entropy density ratio ($\zeta/s$). This heightened sensitivity makes measurements of radial flow fluctuations a powerful complementary probe to anisotropic flow studies for characterizing the transport properties of the QGP. As illustrated in Figure~\ref{fig:v0pt_shearbulk}, IP-Glasma+MUSIC+UrQMD simulations that exclude bulk viscosity can reproduce the centrality dependence of anisotropic flow coefficients like $v_n\{2\}$ but overestimate the event-averaged mean transverse momentum, $\lr{[\pT]}$. Conversely, the inclusion of bulk viscosity in these models effectively suppresses $\lr{[\pT]}$, leading to improved agreement with experimental data.

\subsection{Theoretical Background}\label{sec:v0pt_theory}
\subsubsection{Defining the $\pT$-differential Radial Flow Observable}
\begin{figure}[htbp]
    \centering
    \includegraphics[width=0.7\linewidth]{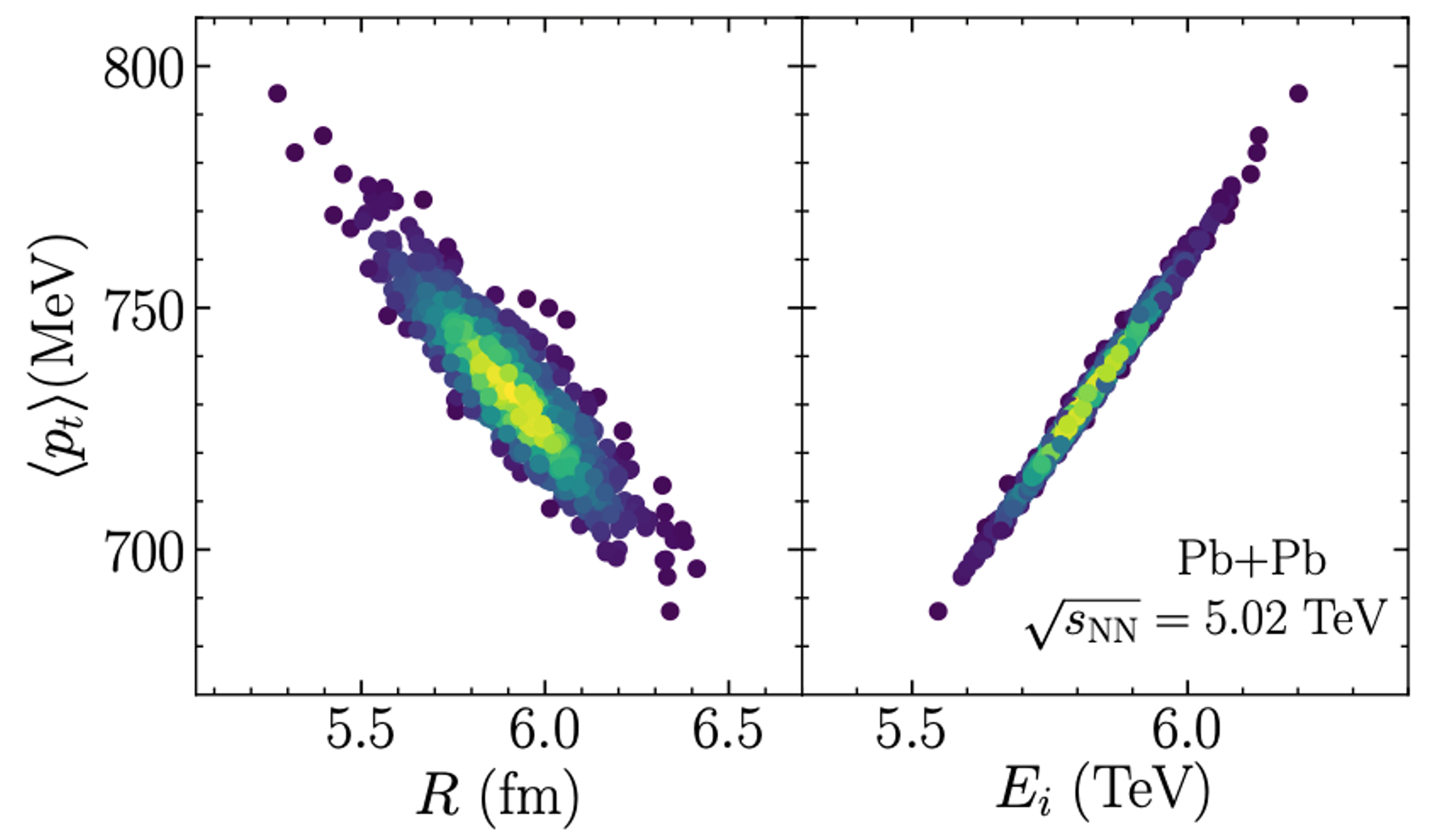}
    \caption{Ideal hydrodynamic results for Pb+Pb collisions at $\sqrt{s_{NN}}=5.02\ \mathrm{TeV}$ and impact parameter $b=2.5\ \mathrm{fm}$. Left: Event-wise average transverse momentum $[\pT]$ vs. initial size $R$. Right: $[\pT]$ vs. initial energy per unit rapidity $E_i$. Figure adapted from Ref.~\cite{Giacalone:2020byk}.}
    \label{fig:v0pt_iniFin_Giacalone}
\end{figure}

Variations in the initial overlap-region size and energy density of a heavy-ion collision leave a distinct imprint on the final-state event-wise average transverse momentum, denoted as $[\pT]$. This is evident from hydrodynamic simulations shown in Figure~\ref{fig:v0pt_iniFin_Giacalone}, demonstrating strong correlations between $[\pT]$ and both the initial transverse size $R$ and the initial energy per unit rapidity $E_i$. 

To quantify these correlations and to test whether fluctuations in radial flow indeed arise as a consequence of initial-state variations, Ref.~\cite{Schenke:2020uqq} correlated $[\pT]$ with several initial-state predictors within the IP-Glasma+MUSIC+UrQMD framework. These predictors included the average initial entropy density $[s]$, the initial elliptic area $A = \pi r^{2}\sqrt{1-\varepsilon_{2}^{2}} = \pi \sqrt{4 \langle x^{2} \rangle \langle y^{2} \rangle}$ (where $\langle x^2 \rangle, \langle y^2 \rangle$ are moments of the initial profile and $\varepsilon_2$ is the eccentricity), and the initial squared radius $r^2$. The Pearson correlation coefficient used to quantify the correlation is defined as:
\begin{equation}\label{eq:pearson}
Q_{\xi} = \frac{\langle \delta [\pT]\,\delta \hat{\xi}\rangle}
         {\sqrt{\langle (\delta [\pT])^{2}\rangle\,
                \langle (\delta \hat{\xi})^{2}\rangle}}\,,
\end{equation}
where $\delta[\pT] = [\pT] - \lr{[\pT]}$ is the fluctuation of the event-wise average $\pT$ around its ensemble mean $\lr{[\pT]}$, and $\delta \hat{\xi}$ is the fluctuation of the considered initial-state predictor $\xi$.

\begin{figure}[htbp]
    \centering
    \includegraphics[width=0.6\linewidth]{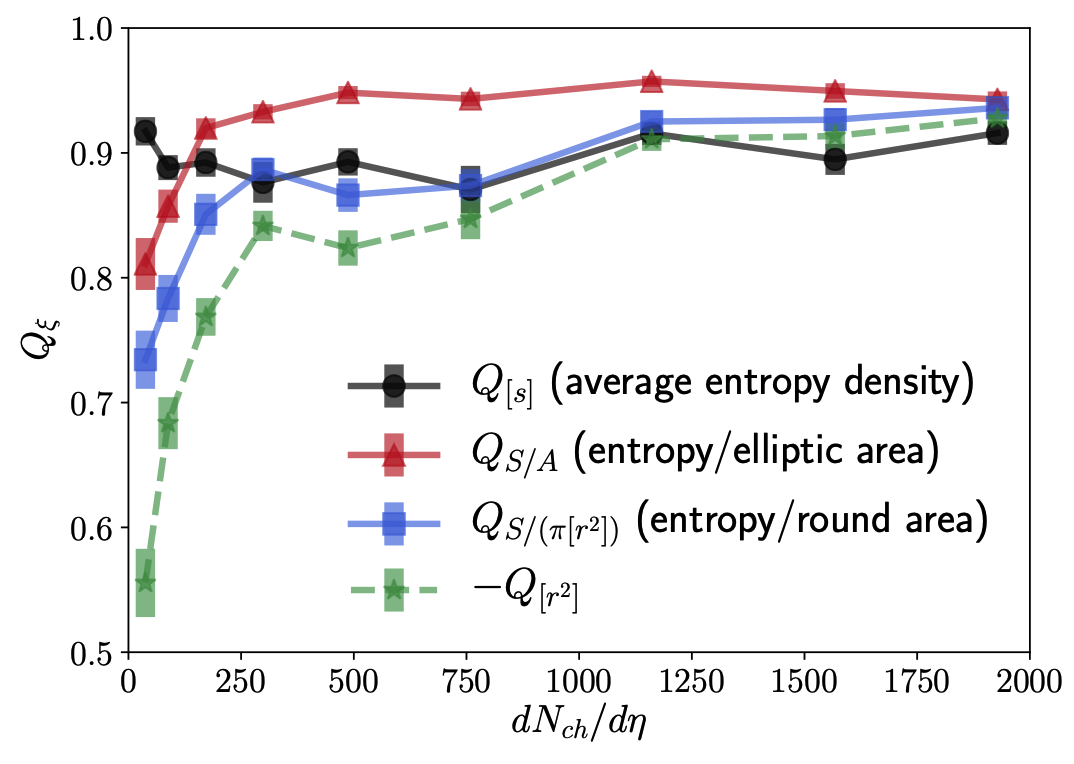}
    \caption{Pearson correlation coefficients $Q_{\xi}$ for various single-term initial-state predictors $\xi$ of event-wise average transverse momentum $[\pT]$ fluctuations in Pb+Pb collisions at $\sqrt{s_{NN}}=5.02\ \mathrm{TeV}$, from Ref.~\cite{Schenke:2020uqq}.}
    \label{fig:v0pt_pearsoninifin}
\end{figure}

Figure~\ref{fig:v0pt_pearsoninifin} shows that, in central collisions, all considered predictors correlate similarly with $[\pT]$. With decreasing multiplicity (more peripheral collisions), the overlap region becomes more elliptic; incorporating eccentricity into an entropy-per-area predictor dramatically improves its performance across centralities. In peripheral collisions where $\varepsilon_2 \sim 1$ and the region may fragment into multiple ``hot spots'', the average initial entropy density $[s]$ is found to be the most reliable predictor. The total initial energy at fixed multiplicity yields correlations comparable to $[s]$, except in the most peripheral events, and an energy-per-area predictor $E/A$ performs marginally better than an entropy-per-area predictor $S/A$. Therefore, state-of-the-art model calculations point towards the origin of final-state $[\pT]$ fluctuations as arising from initial-state fluctuations in size or energy/entropy density.

Thus, the measurement of fluctuations in $[\pT]$ using final-state particles is expected to provide a means to experimentally: (1) Determine whether radial flow fluctuations arise as a collective response to initial-state fluctuations, and, (2) Constrain initial-state fluctuations in size and energy or entropy density.

Similar to anisotropic flow coefficients $v_n$, radial flow fluctuations can be characterized in two main ways. The first is a $\pT$-integrated measure, typically defined over a specific range of $\pT$:
\begin{equation}\label{eq:v0_integrated}
v_0 = \frac{\sqrt{\langle (\delta[\pT])^2\rangle}}{\langle[\pT]\rangle}\,,
\end{equation}
which quantifies the relative standard deviation of $[\pT]$~\cite{NA49:1999inh,Bozek:2012fw,Samanta:2023amp}.

\begin{figure}[htbp]
    \centering
    \includegraphics[width=0.7\linewidth]{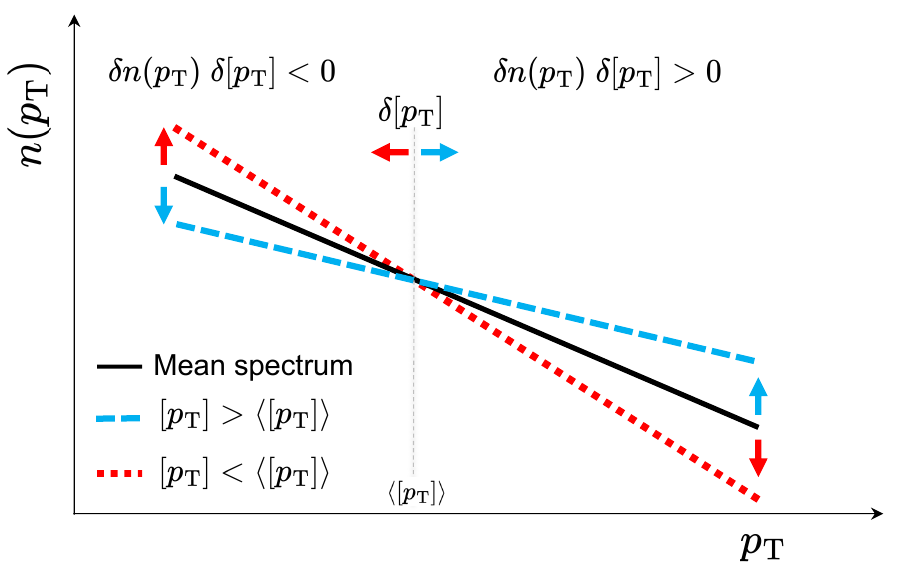}
    \caption{Schematic illustration of radial flow fluctuations. The particle yield $n(\pT)$ for an event with large radial flow (larger $[\pT]$, blue dashed line) and small radial flow (smaller $[\pT]$, red dotted line) are shown relative to the ensemble-average spectrum (black solid line). Positive covariance $\langle \delta n(\pT)\,\delta[\pT]\rangle > 0$ is expected at $\pT \gtrsim \langle[\pT]\rangle_{\text{cross}}$, and negative covariance at $\pT \lesssim \langle[\pT]\rangle_{\text{cross}}$~\cite{Parida:2024ckk}.}
    \label{fig:v0pt_1}
\end{figure}

The second way is to measure radial flow fluctuations in a $\pT$-differential manner, analogous to $v_n(\pT)$. To understand this approach, we first consider how variations in the event-wise $[\pT]$ influence the $\pT$ spectra. Assuming that an event’s $\pT$ spectrum follows a simplified Boltzmann-like distribution $f(\pT) \propto p_T e^{(-p_T/T_{\rm eff})}$, the mean $\pT$ for that event is $[\pT] = 2T_{\rm eff}$, where $T_{\rm eff}$ is the effective temperature (inverse slope parameter of the spectra). Therefore, for events with the same multiplicity, an increase in $[\pT]$ (i.e., a larger $T_{\rm eff}$) corresponds to a flatter $\pT$ spectrum. This is schematically depicted in Figure~\ref{fig:v0pt_1}: the black solid line shows the ensemble-averaged spectrum, while the blue dashed line represents an event with a larger $[\pT]$ (and thus larger $T_{\rm eff}$).

A key observation from Figure~\ref{fig:v0pt_1} is that, for events with a given multiplicity, the spectrum of an event with a higher $[\pT]$ will cross the ensemble-averaged spectrum at $\pT = \lr{[\pT]}$ of the full spectrum, denoted as $\lr{[\pT]}_{\text{cross}}$. Specifically, if $[\pT]$ is higher than the ensemble average $\lr{[\pT]}$, the particle yield $n(\pT) = \frac{dN(\pT)/dp_T}{\int N(\pT)}$ will be suppressed at low $\pT$ i.e. $\pT < \lr{[\pT]}_{\text{cross}}$ and enhanced at high $\pT$, i.e. $\pT > \lr{[\pT]}_{\text{cross}}$ relative to the average spectrum. Consequently, the covariance of the event-wise fluctuation in particle yield $\delta n(\pT) = n(\pT) - \langle n(\pT) \rangle$ with the fluctuation in the event-wise mean $\delta [\pT]$ is expected to be negative for $\pT < \lr{[\pT]}_{\text{cross}}$ and positive for $\pT > \lr{[\pT]}_{\text{cross}}$. This motivates the definition of the $\pT$-differential radial flow fluctuation observable, $v_0(\pT)$~\cite{Schenke:2020uqq,Parida:2024ckk}.

Mathematically, $v_0(\pT)$ is defined via the relation:
\begin{equation}\label{eq:v0_differential_definition}
\frac{\langle \delta n(\pT)\,\delta[\pT]\rangle}
     {\langle n(\pT)\rangle\,\langle[\pT]\rangle}
= v_0(\pT)\,v_0\,.
\end{equation}
This formulation factorizes the normalized covariance into the $\pT$-integrated fluctuation $v_0$ and the $\pT$-dependent measure $v_0(\pT)$. 


To determine if $v_0(\pT)$ is collective, we examine whether it exhibits characteristics similar to the anisotropic flow harmonics $v_n(\pT)$, whose collective nature is well-established through various experimental signatures.

\subsubsection{Hallmarks of Collectivity for Anisotropic Flow $v_n$}
To establish the collective origin of $v_n$, several experimental signatures are typically examined. These provide a template for assessing the collective nature of $v_0(\pT)$ as well. Key signatures include:
\begin{enumerate}
\item \textbf{Factorization of $v_n(\pT)$:}
A crucial signature of collective behavior is the factorization, where the two-particle correlations can be approximated by the product of single-particle anisotropy coefficients. For instance, the two-particle correlation $V_{nn}(p_{Ta}, p_{Tb})$ for particles $a$ and $b$ with transverse momenta $p_{Ta}$ and $p_{Tb}$ can be approximately expressed as:
\begin{equation}
V_{nn}(p_{Ta}, p_{Tb}) \approx v_n(p_{Ta}) v_n(p_{Tb}) + \text{non-flow terms}\,,
\end{equation}
where $v_n(\pT)$ is the $n^{\text{th}}$-order single-particle flow coefficients~\cite{PHENIX:2008osq, ATLAS:2012at}. If the anisotropy arises from a common collective response to the event geometry, $v_n(\pT)$ measured this way should be consistent across different particle pairings and kinematic regions, once non-flow effects are accounted for. 
\begin{figure}[htbp!]
    \centering
    \includegraphics[width=0.4\linewidth]{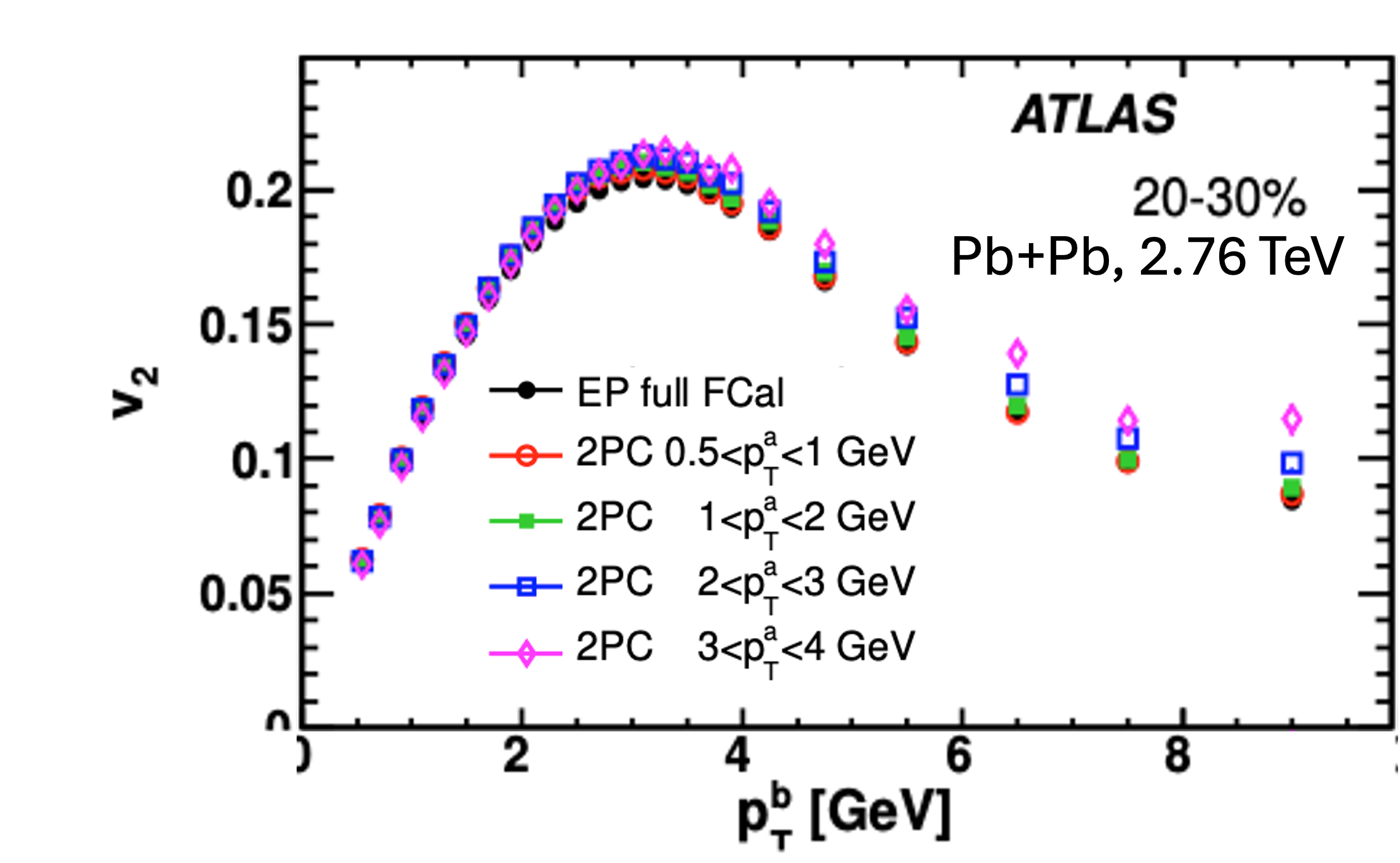}
    \caption{Comparison of $v_{2}(\pT)$ as a function of $\pT$ from mixed-$\pT$ correlations using different analysis methods for the 20–30\% centrality interval in Pb+Pb collisions at $\sqrt{s_{NN}} = 2.76\ \mathrm{TeV}$. Error bars indicate statistical uncertainties only~\cite{ATLAS:2012at}.}
    \label{fig:v0pt_factorize}
\end{figure}
Figure~\ref{fig:v0pt_factorize} illustrates this factorization for elliptic flow ($v_2$) in Pb+Pb collisions, where $v_2(\pT)$ extracted using different methods shows consistency, supporting a common collective origin.

\item \textbf{Long-range correlations in rapidity:}

Collective flow is characterized by correlations that extend over a large range in pseudorapidity. This long-range nature distinguishes it from short-range correlations (e.g., from resonance decays or jet fragmentation). The initial anisotropy of the collision zone is translated into momentum anisotropies via the medium's expansion, affecting particles across a broad rapidity span.

Figure~\ref{fig:v0pt_longrange} shows a two-particle azimuthal correlation function $C(\Delta\phi, \Delta\eta)$, where a ridge-like structure at $\Delta\phi \approx 0$ (and $\pi$ for $v_2$) extends over a large $\Delta\eta$ range, providing evidence for a long-range collective phenomena.

\begin{figure}[htbp]
    \centering
    \includegraphics[width=0.4\linewidth]{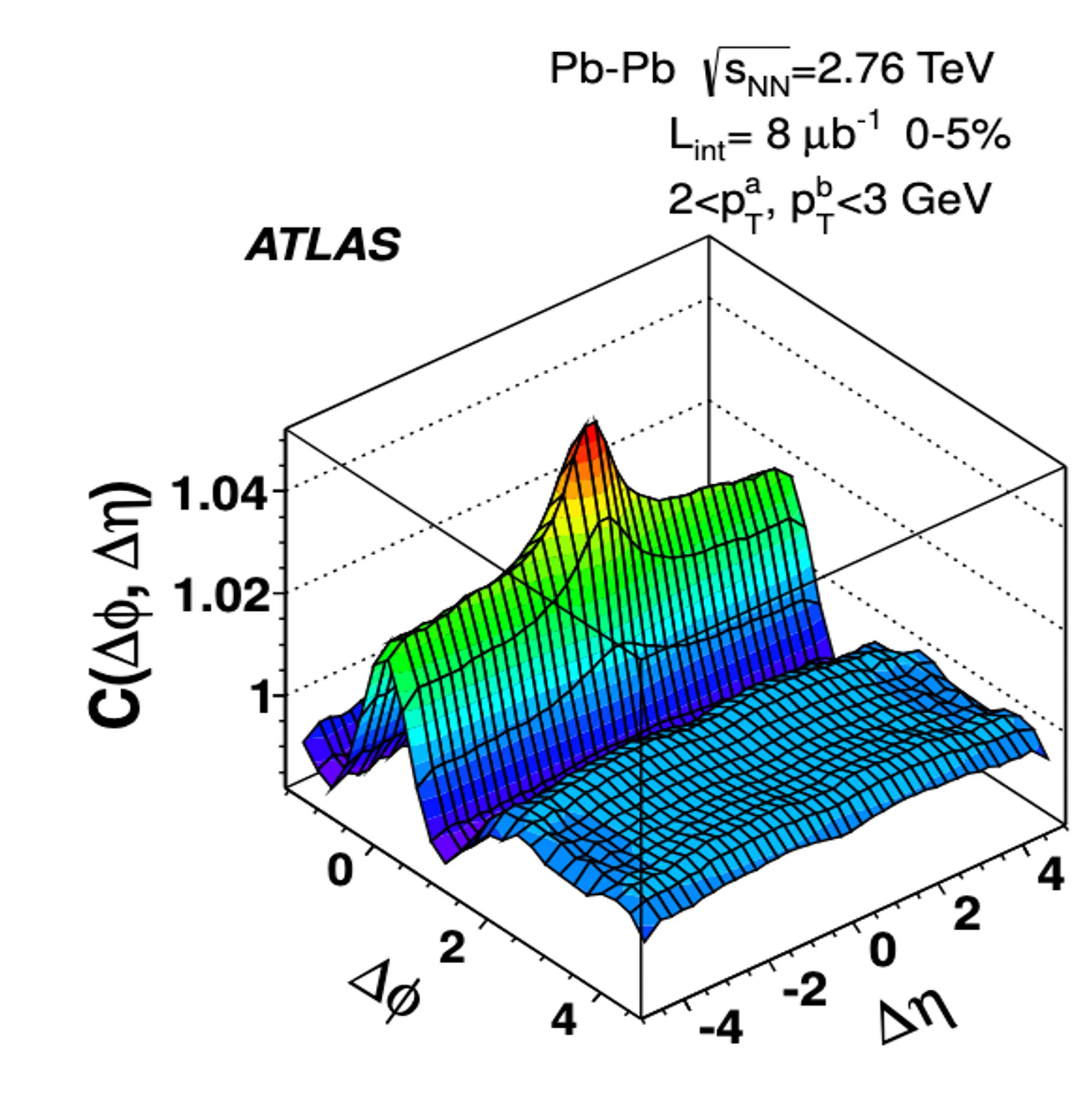}
    \caption{Two-dimensional two-particle correlation function $C(\Delta\phi,\Delta\eta)$ for Pb+Pb collisions at $\sqrt{s_{NN}} = 2.76\ \mathrm{TeV}$ in the 0-5\% centrality class, showing long-range `ridge'~\cite{ATLAS:2012at}.}
    \label{fig:v0pt_longrange}
\end{figure}

\item \textbf{Centrality independence of hydrodynamic response $k_{n}$ = $v_n(\pT)/v_{n}$:}

In a hydrodynamically driven expansion, the anisotropic flow coefficients $v_n$ are expected to be proportional to the initial spatial anisotropies $\varepsilon_n$, i.e., $v_n = \kappa_n \varepsilon_n$. While $v_n$ and $\varepsilon_n$ vary with collision centrality, their ratio $\kappa_n$ (the hydrodynamic response) is less sensitive. Therefore, the shape of $v_n(\pT)$, when scaled by the $\pT$-integrated $v_n$ (i.e., $v_n(\pT)/v_n$), is expected to show a weak dependence on centrality if the system's response is universal. Figure~\ref{fig:v0pt_scaleV2} (Right) presents $v_2(\pT)/v_2$ versus a scaled $\pT$, showing that curves for different centralities approximately overlap, suggesting centrality independence of the hydrodynamic response coefficient.

\begin{figure}[htbp]
    \centering
    \includegraphics[width=0.8\linewidth]{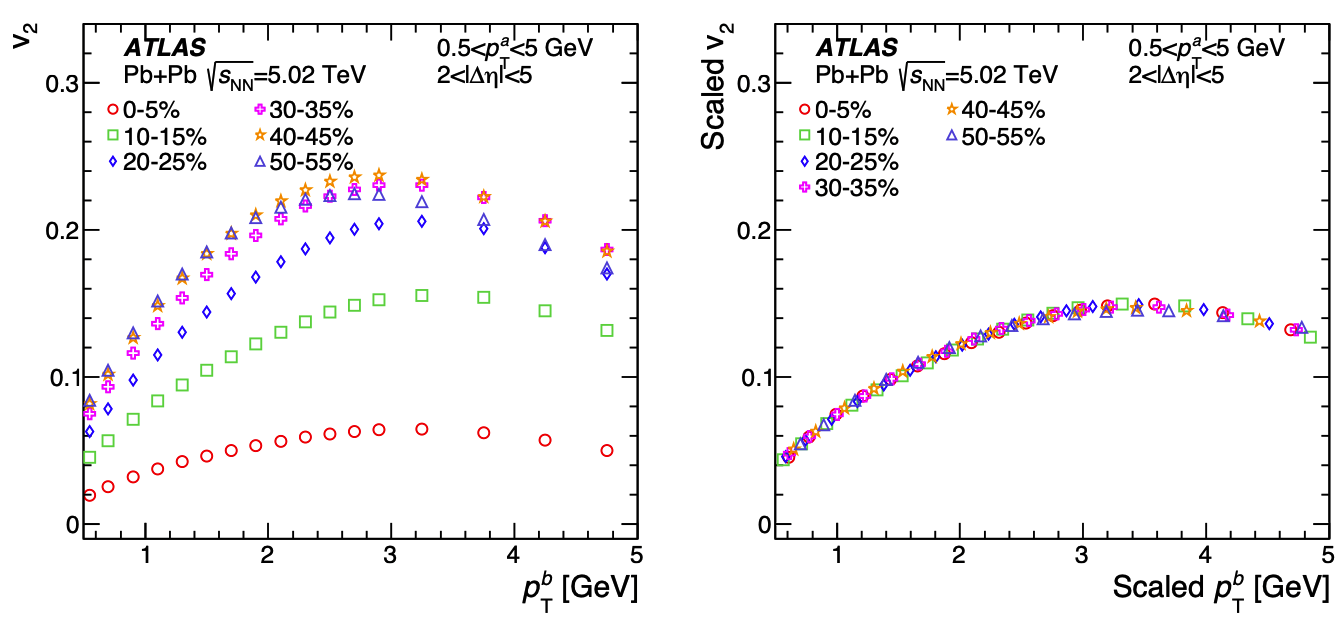}
    \caption{(Left) The elliptic flow coefficient $v_2(\pT)$ and (Right) its normalized form $v_2(\pT)/v_2$ are shown as functions of the scaled transverse momentum,  $\frac{p_T}{\overline{p}_T}\,,$  where $\overline{p}_T$ is the mean $\pT$ of all particles in the analyzed $\pT$ range. Error bars represent the combined statistical and systematic uncertainties~\cite{ATLAS:2018ezv}.}
    \label{fig:v0pt_scaleV2}
\end{figure}

\item \textbf{Particle Identity (PID) dependence of $v_n(\pT)$:}

A well-established feature of anisotropic flow is the mass ordering of $v_n(\pT)$: at low $\pT$, $v_n$ is larger for lighter particles (e.g., pions) than for heavier ones (e.g., protons). This is attributed to the effect of a common radial flow velocity field boosting particles, leading to heavier particles gaining more momentum but exhibiting smaller $v_n$ at a given low $\pT$. At higher $\pT$ (typically $> 2$ GeV), a reverse ordering or grouping by particle type (meson/baryon) can occur. This PID dependence provides important insights into the collective expansion and hadronization dynamics.
Recent predictions for $v_0(\pT)$ also suggest a distinct mass ordering~\cite{Parida:2024ckk}, where $v_0(\pT)$ (or its magnitude) is expected to decrease with increasing particle mass at a given $\pT$, reflecting how particles of different masses are affected by fluctuations in the effective temperature or radial expansion.
\end{enumerate}

To establish the collective nature of $\pT$-differential radial flow fluctuations, this work will present evidence for analogous signatures for $v_0(\pT)$. Specifically, we will investigate properties related to the first three hallmarks using data from the ATLAS experiment~\cite{ATLAS:2025ztg}. The ATLAS detector has limited particle identification capabilities for charged hadrons, so the fourth signature (PID dependence) cannot be verified in this analysis; however, a complementary study by the ALICE Collaboration has addressed this aspect~\cite{ALICE:2025iud}.

\subsection{Previous Studies}\label{sec:v0pt_prev}
Previous experimental studies on radial flow fluctuations have primarily focused on moments of the event-by-event mean transverse momentum distribution, $[\pT]$, such as the ensemble mean $\lr{[\pT]}$~\cite{Schnedermann:1993ws} and the scaled variance $v_0 = \sqrt{\lr{(\delta[\pT])^2}}/\lr{[\pT]}$~\cite{NA49:1999inh}. Higher-order cumulants, such as the skewness and kurtosis of the $[\pT]$ distribution, have also been investigated~\cite{ALICE:2023tej, ATLAS:2024jvf}. However, a dedicated observable for a differential study of radial flow fluctuations as a function of particle $\pT$, analogous to the $\pT$-differential anisotropic flow coefficients $v_n(\pT)$, has been lacking. 

To address this gap, the observable $v_0(\pT)$ was proposed and theoretically investigated~\cite{Schenke:2020uqq}, with further developments and studies in subsequent works~\cite{Samanta:2023amp, Parida:2024ckk}. By correlating event-by-event fluctuations in the shape of the particle $\pT$ spectrum with variations in the event-by-event mean transverse momentum, $[\pT]$, the $v_0(\pT)$ observable enables a $\pT$-differential study of radial flow fluctuations.

This chapter presents the first measurement of the transverse-momentum dependence of radial flow fluctuations, $v_0(\pT)$, over the range $0.5 < p_T < 10$ GeV. The measurement utilizes a two-particle correlation method and is performed using Pb+Pb collision data at $\sqn =$ 5.02 TeV recorded by the ATLAS detector. The primary objectives of this study are to: (1) identify features in the measured $v_0(\pT)$ distribution indicative of collective expansion, and (2) compare the experimental results with predictions from state-of-the-art hydrodynamic models, to assess the sensitivity of $v_0(\pT)$ to $\zeta/s$. Establishing these characteristics is important for validating the collective nature of radial flow and for leveraging its fluctuations to further constrain the transport properties of the QGP.

This chapter is organized as follows. Section~\ref{sec:method_analysis} describes the analysis methodology. Section~\ref{sec:v0pt_crossschk} presents cross-checks validating the measurement. Section~\ref{sec:v0pt_result} presents the main results, focusing on evidence for collectivity and comparisons with hydrodynamic calculations. Section~\ref{sec:v0pt_model_studies} examines model studies of event classification effects. Finally, Section~\ref{sec:v0pt_summary1} summarizes the findings and Section~\ref{sec:v0pt_summary2} offers an outlook.

\subsection{Methodology}\label{sec:method_analysis}
We start with the measurement of cumulants of the event-wise mean transverse momentum, $[\pT]$, adapting techniques well-established in anisotropic flow analyses. This approach follows the standard multi-particle cumulant method, which typically involves the calculation of Q-vectors to quantify flow harmonics~\cite{Bilandzic:2010jr, Bilandzic:2013kga, DiFrancesco:2016srj}. The underlying mathematical framework for calculating such cumulants is also well-established~\cite{Jia:2017hbm, Huo:2017nms}. A key distinction for $[\pT]$ cumulants is that the per-particle contribution to a Q-vector (e.g., $e^{in\phi}$ for anisotropic flow) is replaced by a scalar quantity representing the deviation of $[\pT]$ in an event from the event ensemble average: $\delta[\pT]$. Ensemble averages $\langle \dots \rangle$ are calculated over events within narrow bins of event activity (such as $\Nch$ or $\sumET$). The results from these narrow bins are then combined to obtain averages for wider centrality ranges.

The $n$-particle $\pT$ correlator, denoted $c_n$, is defined for a single event as:
\begin{align}\label{eq:ck}
c_{n}=\frac{\mathop{\sum}\limits_{i_1\neq ...\neq i_n } w_{i_1}...w_{i_n}(p_{T,i_1}- \MpT)...(p_{T,i_n}-\MpT)} { \mathop{\sum}\limits_{i_1\neq ...\neq i_n } w_{i_1}...w_{i_n}},
\end{align}
where $w_i$ is the weight assigned to track $i$, $p_{T,i}$ is its transverse momentum, and the sum is over all unique combinations of $n$ distinct tracks within the event. Following the framework presented in Ref.~\cite{Jia:2017hbm}, these correlators can be efficiently computed using polynomial functions of weighted single-particle sums. Further details on the exact formulae used are provided in Appendix~\ref{sec:app_method}.


Information about radial flow is encoded in the $\pT$-dependent yield in each event, $N(\pT)$, which can be factorized into the product of the global multiplicity $N = \int N(\pT)dp_T$ and the normalized (fractional) spectrum, $n(\pT)$, such that $N(\pT) = N n(\pT)$. The event-by-event fluctuations in $N(\pT)$ can be decomposed into three terms~\cite{Bhatta:2025oyp}:
\begin{equation}
\delta N(\pT) = \langle n(\pT) \rangle \delta N + \langle N \rangle \delta n(\pT) + \delta N \delta n(\pT).
\end{equation}
In most cases, the higher-order term $\delta N \delta n(\pT)$ is negligible. Variations in global multiplicity $N$ are related to centrality dependence, whereas, fluctuations in $n(\pT)$ capture genuine radial flow fluctuations related to the shape of the $\pT$ spectrum. Using the normalized yield $n(\pT)$ emphasizes this shape, a key characteristic of radial flow as discussed in Section~\ref{sec:v0pt_theory}, and suppresses trivial fluctuations related to variations in the overall event multiplicity.

To mitigate auto-correlations and non-flow, the two-subevent method is employed. Subevent `A' covers $\eta \in (-2.5, -\eta_{\mathrm{gap}}/2)$ and subevent `B' covers $\eta \in (\eta_{\mathrm{gap}}/2, 2.5)$. The analysis explores four different pseudorapidity gap sizes: $\eta_{\mathrm{gap}} = 0, 1, 2,$ and $3$. The $\delta [\pT]$ (and thus $v_0$) is calculated from subevent A, whereas the yield fluctuations $\delta n(\pT)$ are calculated from subevent B. The final covariance is obtained as a weighted average of the correlations from the two possible subevent pairings: $\langle \delta n_{\text{A}}(\pT) \delta [\pT]_{\text{B}} \rangle$ and $\langle \delta n_{\text{B}}(\pT) \delta [\pT]_{\text{A}} \rangle$. The final $v_0(\pT)$ measurement is then measured using:
\begin{equation}\label{eq:v0_differential_definition2}
\frac{\langle \delta n(\pT)\,\delta[\pT]\rangle}
     {\langle n(\pT)\rangle\,\langle[\pT]\rangle}
= v_0(\pT)\,v_0\,.
\end{equation}

The analysis is performed on data for Pb+Pb collisions at $\sqn = 5.02$ TeV, described in detail in Section~\ref{sec:datasets_systems}. Track reconstruction efficiency, $\epsilon(p_T, \eta, \Nch)$, and the rate of falsely reconstructed (``fake'') tracks, $f(p_T, \eta, \Nch)$, are determined using Monte Carlo (MC) simulations as detailed in Chapter~\ref{sec:data}. The corrected charged particle multiplicity, $\Nch$, used for event classification, is calculated as the sum of the weights used for corrections, $\sum_i w_i = \sum_i \frac{1-f_i}{\epsilon_i}$, for particles within $0.5 < p_T < 5$ GeV and $|\eta| < 2.5$. Statistical uncertainties for all measured observables are estimated using a Poisson bootstrap method~\cite{Efron:1979bxm, ATLAS:2021kho}. A further description of the Bootstrap method used can be found in Appendix~\ref{sec:app_bootstrap}.

Systematic uncertainties in the $v_0(\pT)$ measurement arise from various detector effects and analysis procedures. Key contributions include those from track selection criteria, resulting in deviations of 0.5--2.5\% for $v_0$ and, for $v_0(\pT)$, 2--7\% at low $\pT$ versus 0--1\% at high $\pT$. Tracking efficiency and fake rate corrections, linked to detector material modeling in GEANT simulations, contribute 0--1\% to $v_0$; for $v_0(\pT)$, impacts are 0.5--1\% at high $\pT$ and 2--4\% at low $\pT$, with negligible effect from azimuthal efficiency variations. Residual pileup effects lead to uncertainties below 0.5\% for all measured observables. Centrality definition uncertainties, stemming from trigger inefficiencies in peripheral event sampling, are less than 1.5\% and primarily affect centrality-binned results. An MC consistency check adds approximately 1\% uncertainty to $v_0$ and about 2\% to $v_0(\pT)$. Appendix~\ref{sec:app_syst} provides further details on the contributions from different systematic sources.

\subsection{Results} \label{sec:v0pt_result}

To establish the collective nature of the observed $p_T$-differential radial flow fluctuations, this section presents experimental evidence for three key hallmarks analogous to those observed for $v_n$: (i) the persistence of $v_0(p_T)$ correlations over large pseudorapidity gaps, indicative of a long-range phenomenon; (ii) the factorization of $v_0(p_T)$, implying its independence from the specific choice of reference transverse momentum ($\pTref$) ranges; and (iii) a universal, centrality-independent shape of the normalized response, $v_0(p_T)/v_0$. These features, if confirmed, would strongly support an early-time, bulk origin for these fluctuations, distinct from localized, non-collective effects.

\subsubsection{$v_0(\pT)$ Factorization}
\begin{figure}[htbp]
    \centering
    \includegraphics[width=1.0\linewidth]{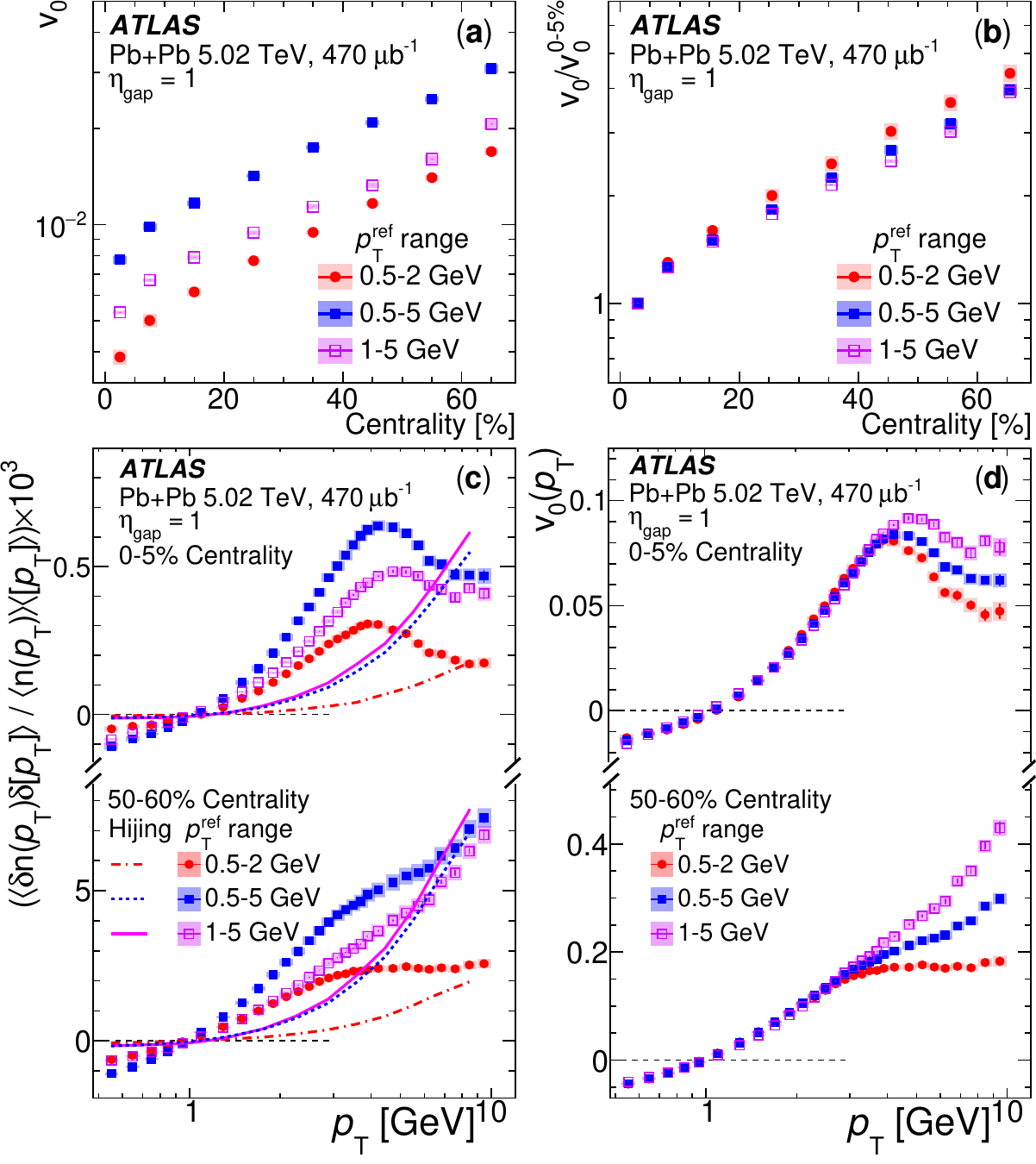}
    \caption{Centrality dependence of $v_0$ (\textbf{a}) and $v_0/v_0^{0-5\%}$ (\textbf{b}) for three $\pTref$ ranges. Panels (\textbf{c}) and (\textbf{d}) show, respectively, the normalized covariance in Eq.~\ref{eq:v0_differential_definition2}, and $v_0(\pT)$ in 0--5\% centrality (top) and 50--60\% centrality (bottom), each plotted for three $\pTref$ ranges. The lines indicate the corresponding predictions from the HIJING model containing only non-flow correlations. Bars and shaded areas indicate statistical and systematic uncertainties, respectively.}
    \label{fig:v0pt_2}
\end{figure}

Figure~\ref{fig:v0pt_2}a illustrates the centrality dependence of $v_0$ as computed in three distinct $\pTref$ ranges. While the absolute magnitude of $v_0$ varies with the chosen $\pTref$ interval, its centrality trend exhibits remarkable consistency across all selections, as depicted in Figure~\ref{fig:v0pt_2}b using the ratio $v_0/v_0^{0-5\%}$ (normalized to its value in the 0--5\% centrality range). This observation suggests that the centrality-dependent shape of $v_0$ captures the global, collective nature of radial flow fluctuations.

Next, the factorization of the $v_0(\pT)$ is further verified in Figures~\ref{fig:v0pt_2}c and \ref{fig:v0pt_2}d for two representative centrality ranges (0--5\% and 50--60\%). A key finding is that while the normalized covariance exhibits substantial variation with the chosen $\pTref$ interval, the $v_0(\pT)$ remains remarkably independent of $\pTref$ up to $\pT \approx 3$ GeV. Therefore, $v_0(\pT)$ is independent of the $\pTref$ range used for its measurement. This robust factorization behavior, analogous to that observed for anisotropic flow, strongly supports the collective origin of radial flow fluctuations. 

In central collisions, $v_0(\pT)$ exhibits a decrease above $\pT \approx 4$ GeV, mirroring similar observations for anisotropic flow, which may indicate a transition from a flow-dominated to a jet-quenching dominated regime. This decrease is not observed in relatively peripheral collisions. Figures~\ref{fig:v0pt_2}c also present a comparison of the experimental data with the non-flow expectation from the HIJING model. The HIJING model calculations generally underpredict the covariance for $\pT < 8$ GeV, implying presence of genuine correlations in data, beyond those expected from a simple superposition scenario of independent nucleon-nucleon interactions. 

However, it is important to note that when these normalized covariances are converted into $v_0(\pT)$ using Eq.~\ref{eq:v0_differential_definition2}, HIJING can artificially inflate the apparent non-flow contribution~\cite{Bhatta:2025oyp, ALICE:2025iud}. A more reliable assessment of non-flow effects from models like HIJING is reliable only at the level of 2-particle covariance.

\subsubsection{Long-Range Behavior}

Another test for the collective origin of $v_0(p_T)$ is its persistence across large pseudorapidity separations, $\eta_{\mathrm{gap}}$, which suppresses contributions from short-range non-flow correlations such as jet fragments or resonance decays.

\begin{figure}[h!]
    \centering
    \includegraphics[width=1.0\linewidth]{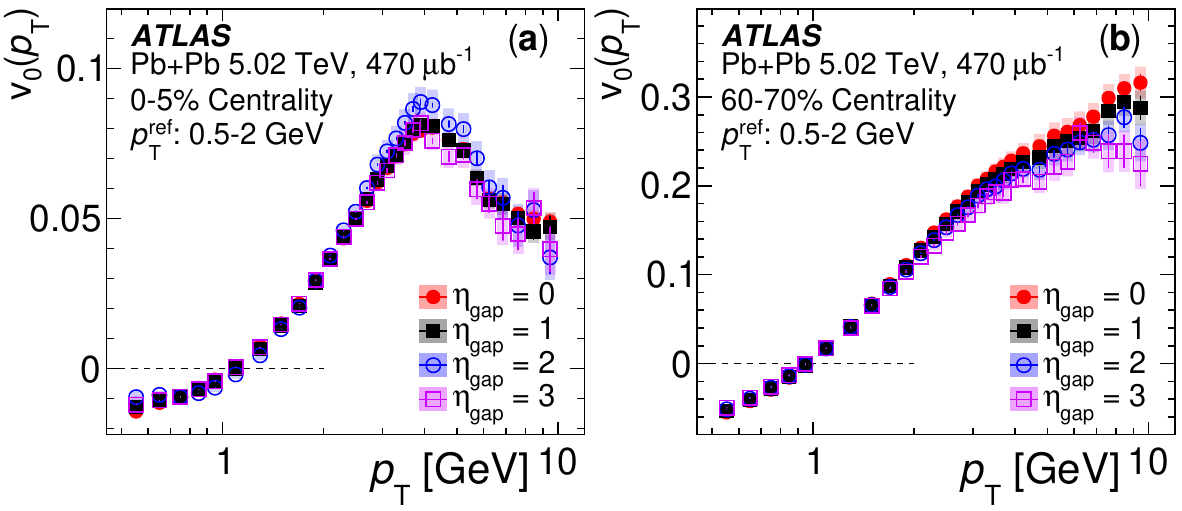}
    \caption{$v_0(\pT)$ for $\eta_{\mathrm{gap}}$=0, 1, 2, and 3 in 0--5\% (left) and 60--70\% (right) centrality. Bars and shaded areas indicate statistical and systematic uncertainties, respectively.}
    \label{fig:v0pt_3}
\end{figure}

Variations of $v_0(\pT)$ with $\eta_{\mathrm{gap}}$ are shown in Figure~\ref{fig:v0pt_3} for 0--5\% and 60--70\% centralities. In central collisions, only minor changes are observed in $v_0(\pT)$ as $\eta_{\mathrm{gap}}$ is increased from 0 to 3. In contrast, peripheral collisions (60--70\%) show a slight reduction in $v_0(\pT)$ at $\pT > 3$ GeV with increasing $\eta_{\mathrm{gap}}$, consistent with the suppression of residual non-flow contributions, potentially from away-side jet fragmentation. 

This weak dependence on $\eta_{\mathrm{gap}}$ provides strong support for the interpretation that radial flow is characterized by long-range, global correlations, akin to anisotropic flow. Consequently, an $\eta_{\mathrm{gap}} = 1$ is adopted as the default configuration for the subsequent results.

\begin{figure}[h!]
    \centering
    \includegraphics[width=1.0\linewidth]{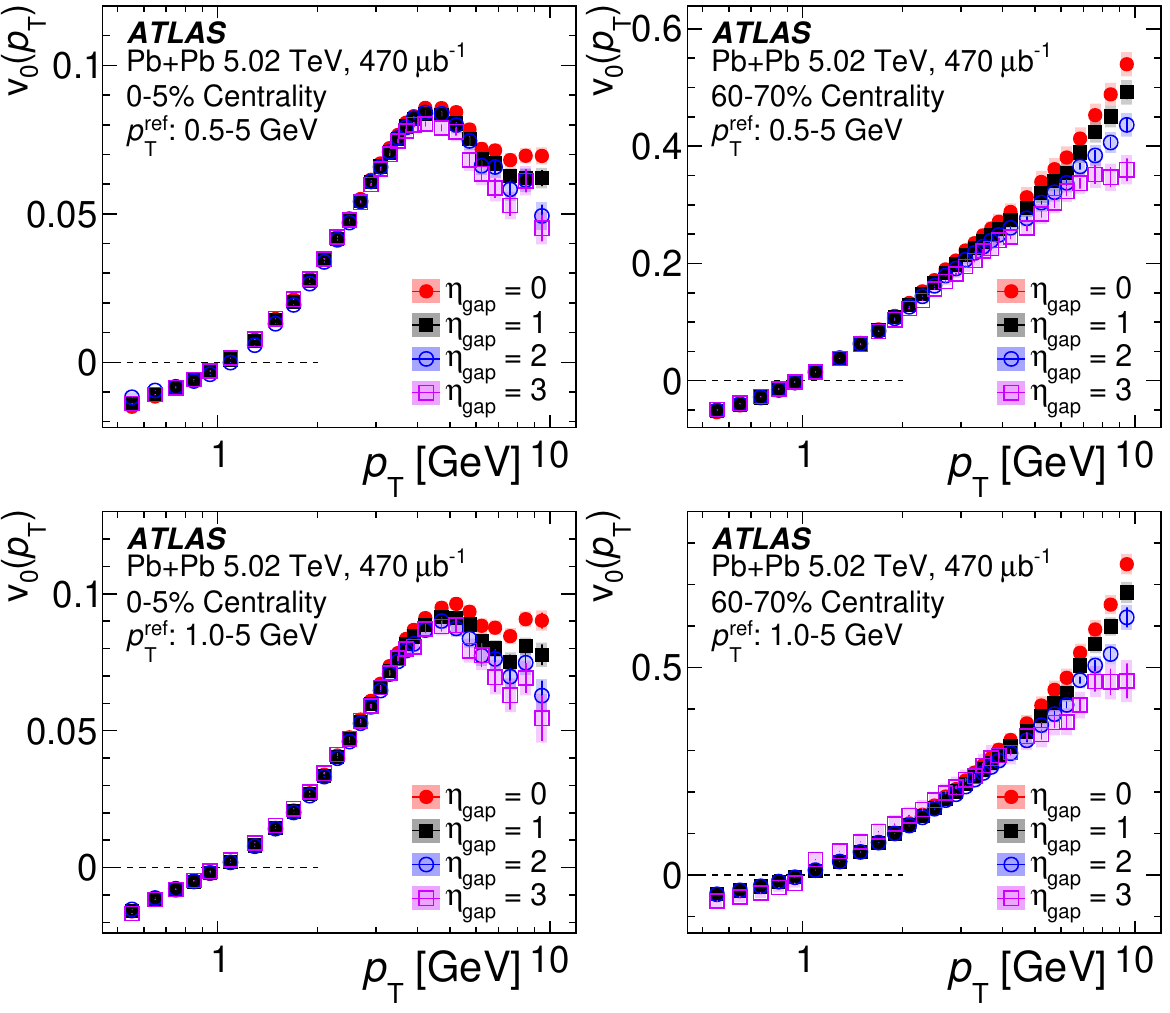}
    \caption{$v_0(\pT)$ for $\eta_{\mathrm{gap}}$=0, 1, 2, and 3 in 0--5\% (left) and 60--70\% (right) centrality and for two $\pTref$ ranges, 0.5--5 GeV (top) and 1--5 GeV (bottom). Bars and shaded areas indicate statistical and systematic uncertainties, respectively.}
    \label{fig:v0pt_3b}
\end{figure}

The study also investigated the impact of the $\pTref$ on the separation between $v_0(\pT)$ measured with varying $\eta_{\mathrm{gap}}$, as illustrated in Figure~\ref{fig:v0pt_3b}. When varying the upper or lower limits of the $\pTref$ range, the differences between results obtained with different $\eta_{\mathrm{gap}}$ values become more pronounced. This observation suggests that the choice of $\pTref$ range can influence the level of residual non-flow contributions captured by the measurement.

\subsubsection{Centrality Dependence of Hydrodynamic Response}
\begin{figure}[h!]
    \centering
    \includegraphics[width=1.0\linewidth]{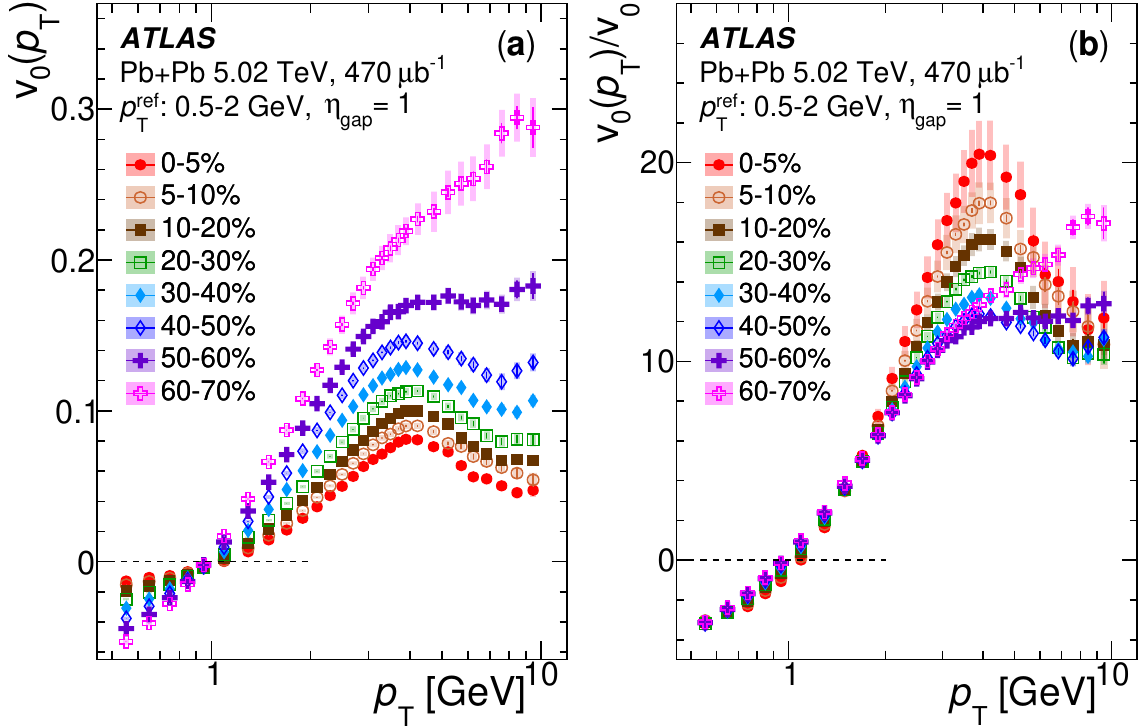}
    \caption{The $v_0(\pT)$ (left) and $v_0(\pT)/v_0$ (right) for different centrality ranges. Bars and shaded areas represent statistical and systematic uncertainties, respectively.}
    \label{fig:v0pt_4}
\end{figure}

The centrality dependence of the $v_0(\pT)$ shape is investigated next using the normalized quantity $v_0(\pT)/v_0$. Figure~\ref{fig:v0pt_4} presents a comparison of $v_0(\pT)$ and $v_0(\pT)/v_0$ across various centrality ranges. While the absolute magnitude of $v_0(\pT)$ displays a pronounced centrality dependence, the $v_0(\pT)/v_0$ exhibits a remarkable overlap for $\pT \lesssim 2.5$ GeV. This observation indicates that the radial flow fluctuations are primarily driven by event-by-event fluctuations in the initial state. This observation also implies that the shape of $v_0(\pT)$ is governed by the hydrodynamic response, $k_0(\pT) \approx v_0(\pT)/v_0$~\cite{Parida:2024ckk}, which appears to be largely independent of centrality. This scaling behavior is analogous to that observed for anisotropic flow. 

Above $\pT \approx 3$ GeV, where non-flow contamination becomes large, this shape scaling is not perfectly maintained. The characteristic rise and fall behavior of $v_0(\pT)/v_0$ is observed up to 40--50\% centrality, which could be attributed to the onset of jet quenching processes that reduce momentum fluctuations. For more peripheral collisions, however, the data show either a flatter behavior or an increasing trend with $\pT$, consistent with a larger non-flow contribution at high $\pT$ and in these peripheral events.

\begin{figure}[h!]
\centering
    \includegraphics[width=1.0\linewidth]{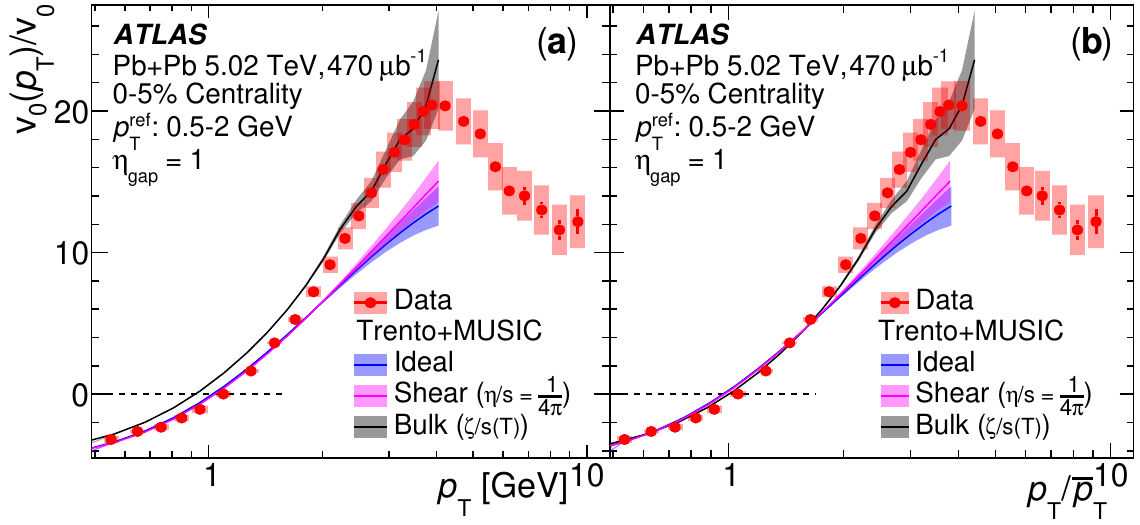}
\caption{The $v_0(\pT)/v_0$ (left) and $v_0(\pT/\pTbar)/v_0$ (right) for $0.5<\pTref<2$ GeV in 0--5\% centrality range, where $\pTbar=0.73$ GeV is the mean $\pT$ of charged particle spectrum measured by the ALICE Collaboration~\cite{ALICE:2022xip}. They are compared with hydrodynamic model predictions without viscosity (red), with constant shear viscosity (in terms of shear viscosity to entropy density ratio of $\eta/s = \frac{1}{4\pi}=0.08$, magenta), and with temperature dependent bulk viscosity (blue).}
\label{fig:v0pt_5}
\end{figure}

\subsubsection{Sensitivity to $\zeta/s$}

An advantage of $v_0(p_T)$ as a probe of QGP transport properties lies in its pronounced sensitivity to the bulk viscosity to entropy density ratio, $\zeta/s$, while exhibiting comparatively weaker dependence on shear viscosity, $\eta/s$~\cite{Parida:2024ckk}. 

Figure~\ref{fig:v0pt_5} presents a comparison of the experimental data in the 0--5\% centrality range with predictions from a hydrodynamic model described in Ref.~\cite{Parida:2024ckk}. These predictions are based on the IP-Glasma+MUSIC framework \cite{Schenke:2020mbo} and incorporate either a constant shear viscosity or a temperature-dependent bulk viscosity. 

As these calculations are not fully tuned to precisely match the overall $\pT$ spectra or event-ensemble $\pT$ fluctuations, Refs.~\cite{Parida:2024ckk,ALICE:2022xip} recommend utilizing dimensionless ratios such as $v_0(\pT)/v_0$ and $v_0(\pT/\pTbar)/v_0$. Here, $\pTbar$ represents the average transverse momentum of the inclusive charged-particle spectrum. Using these ratios helps to reduce dependencies on initial conditions and to further mitigate model dependencies on the overall spectral shape. 

Calculations performed with and without shear viscosity yield similar predictions for $v_0(\pT)/v_0$, as shown in Figure~\ref{fig:v0pt_5}. In contrast, the inclusion of a temperature-dependent bulk viscosity alters both the zero-crossing point in $\pT$ and the high-$\pT$ behavior. Without bulk viscosity, the model systematically underestimates the data at high $\pT$. However, with the addition of bulk viscosity, the predictions move closer to the experimental data, although they remain slightly above it. Rescaling the horizontal axis by $\pTbar$ further enhances the agreement in the region $\pT/\pTbar < 3$, highlighting the pronounced impact of bulk viscosity on the $\pT$-differential radial flow. While the current calculations still require further tuning, this comparison clearly demonstrates the potential of $v_0(\pT)$ data to provide valuable constraints on bulk viscosity of QGP.

\subsubsection{Cross-Checks}\label{sec:v0pt_crossschk}
To ensure the robustness and validity of the measured $v_0(\pT)$ and the interpretation of its collective nature, a series of detailed cross-checks were performed. These include the verification of sum rules for $v_0(\pT)$ and the investigation of the $v_0(\pT)$ crossing point.

\paragraph{Verification of Sum-Rules for $v_{0}(\pT)$}

There are two mathematical sum rules that $v_{0}(\pT)$ is expected to follow:
\begin{gather}
\int v_0(\pT) N_0(\pT) dp_T = 0 \label{eq:sumrule1} \\
\int (p_T - \langle p_T \rangle) v_0(\pT) N_0(\pT) dp_T = \sigma_{p_T} N_0 \label{eq:sumrule2}
\end{gather}
where $N_0(\pT)$ is the inclusive $\pT$ spectrum, $\langle p_T \rangle$ is the mean of the inclusive spectrum, and $\sigma_{p_T}$ is its standard deviation. Further discussions on these formulae can be found in Appendix~\ref{sec:app_method}.

The first sum rule in Eq.~\ref{eq:sumrule1} implies that $v_0(\pT)$ must change sign as a function of $\pT$. The measured $v_0(\pT)$ (e.g., in Figure \ref{fig:v0pt_4}) indeed exhibits a characteristic sign change, typically crossing zero near the average $\pT$ of the inclusive spectrum. The second sum rule in Eq.~\ref{eq:sumrule2} provides a quantitative relation between the $\pT$-differential $v_0(\pT)$ and the $\pT$-integrated fluctuation magnitude $v_0$, which is expected to hold if the radial flow fluctuations are collective~\cite{Schenke:2020uqq, Parida:2024ckk}.

\begin{figure}[h!]
\centering
\includegraphics[width=1.0\linewidth]{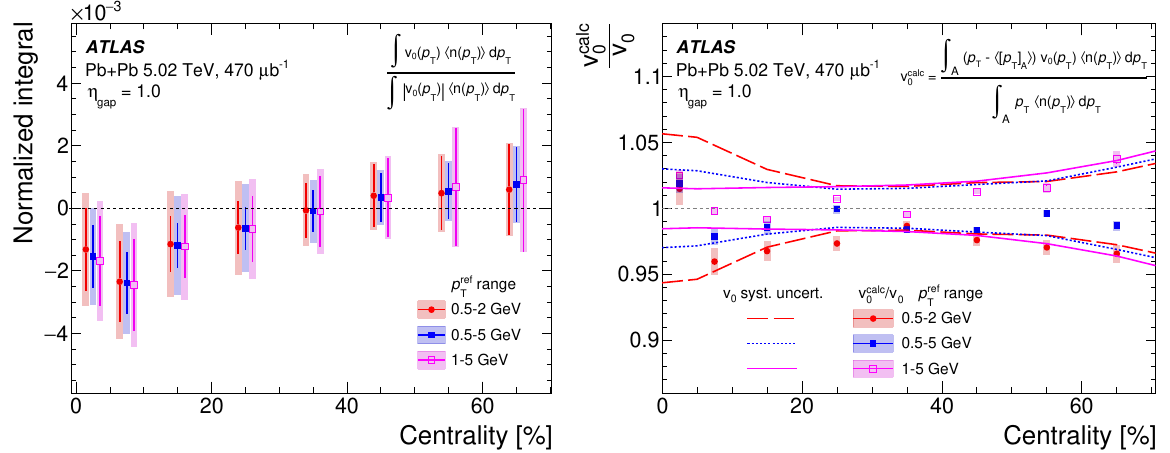}
\caption{Fractional closure for Sum Rules 1 (Eq~\ref{eq:sumrule1}) and 2 (Eq~\ref{eq:sumrule2}) is shown across different reference $\pT$ ranges in the left and middle columns, respectively. The right column compares the $v_0$ values obtained from Sum Rule 2 with those from the two-subevent method. Each row corresponds to a different $\eta_{\mathrm{gap}}$ value: $\eta_{\mathrm{gap}} = 0$ (top) and $\eta_{\mathrm{gap}} = 1$ (bottom). Error bars represent statistical uncertainties, and the shaded bands indicate systematic uncertainties.}
\label{fig:v0pt_SumRules_Consistency_Check}
\end{figure}

Figure~\ref{fig:v0pt_SumRules_Consistency_Check} presents the closure for Sum Rules 1 (Eq.~\ref{eq:sumrule1}) and 2 (Eq.~\ref{eq:sumrule2}). Sum Rule 1 exhibits excellent closure, with deviations within $0.5\%$. In contrast, Sum Rule 2, shows a small non-closure of approximately 5\%. This level of agreement is within the systematic uncertainty on the direct measurement of $v_{0}$, indicating reasonable consistency.

Minor deviations are observed in peripheral centralities, potentially due to non-flow effects. The relatively larger non-closure in Sum Rule 2 likely stems from the fact that this sum rule is approximate; it holds exactly only in the absence of non-flow effects and under the assumption of perfect factorization of $v_0(\pT)$. Given that these ideal conditions are not fully met across the entire kinematic range of $0.5 < \pT < 10$ GeV, the observed deviations are expected.

\paragraph{Crossing-Point of $v_{0}(\pT)$}
As discussed earlier, $v_0(\pT)$ crosses zero at $\MpT$ of the $\pT$ spectra~\cite{Schenke:2020uqq}. In this analysis, the $\pT$ spectrum is measured within the range $0.5 < \pT < 10$ GeV. If the origin of radial flow is collective, the crossing points of $v_0(\pT)$ for different $\pTref$ should be consistent. 

\begin{figure}[h!]
\centering
\includegraphics[width=0.7\linewidth]{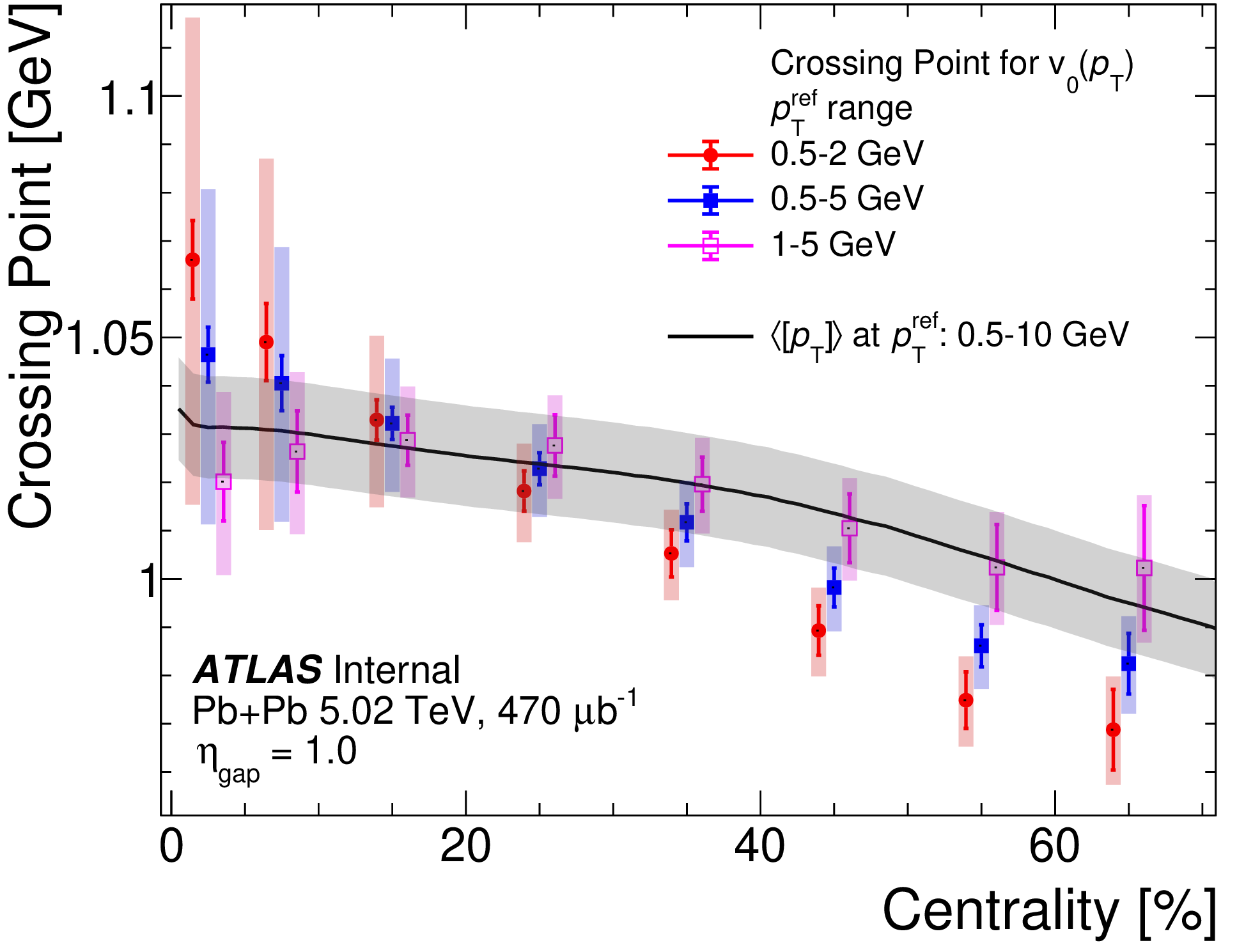}
\caption{The crossing point of $v_0(\pT)$ is plotted as a function of centrality for different reference $\pT$ ranges, with results shown for $\eta_{\mathrm{gap}} = 1$. The solid line represents $\MpT$ for tracks within $0.5 < \pT < 10$ GeV. Error bars indicate statistical uncertainties, while shaded bands represent systematic uncertainties.}
\label{fig:v0pt_Crossing_Point}
\end{figure}

Figure~\ref{fig:v0pt_Crossing_Point} examines this consistency for $\eta_{\mathrm{gap}} = 1$. For comparison, the mean transverse momentum $\MpT$ for particles within $0.5 < \pT < 10$ GeV is also shown. The results demonstrate that, within systematic uncertainties, the crossing points of $v_0(\pT)$ remain consistent across different $\pTref$ and agree with the $\MpT$ across all centralities. Minor deviations observed in peripheral centralities may be attributed to non-flow effects.

\subsection{HIJING Study on Effects of Multiplicity Fluctuations and Acceptance}
\label{sec:v0pt_model_studies}

In the data analysis presented in previous sections, events were categorized based on different measures of event activity, such as charged particle multiplicity ($\Nch$) or forward transverse energy ($\sum E_T$). The choice of this event classifier, along with specific kinematic selections, are observed to influence the measured $v_0(\pT)$. Consequently, a precise understanding of these effects is vital for an accurate interpretation of the experimental findings presented earlier.

\subsubsection{Motivation}
Certain features observed for $v_0(\pT)$ presented in Section~\ref{sec:v0pt_result} warrant a dedicated model study to unravel the underlying factors. Specifically, these are:

\begin{figure}[htbp]
\centering
\includegraphics[width=0.9\linewidth]{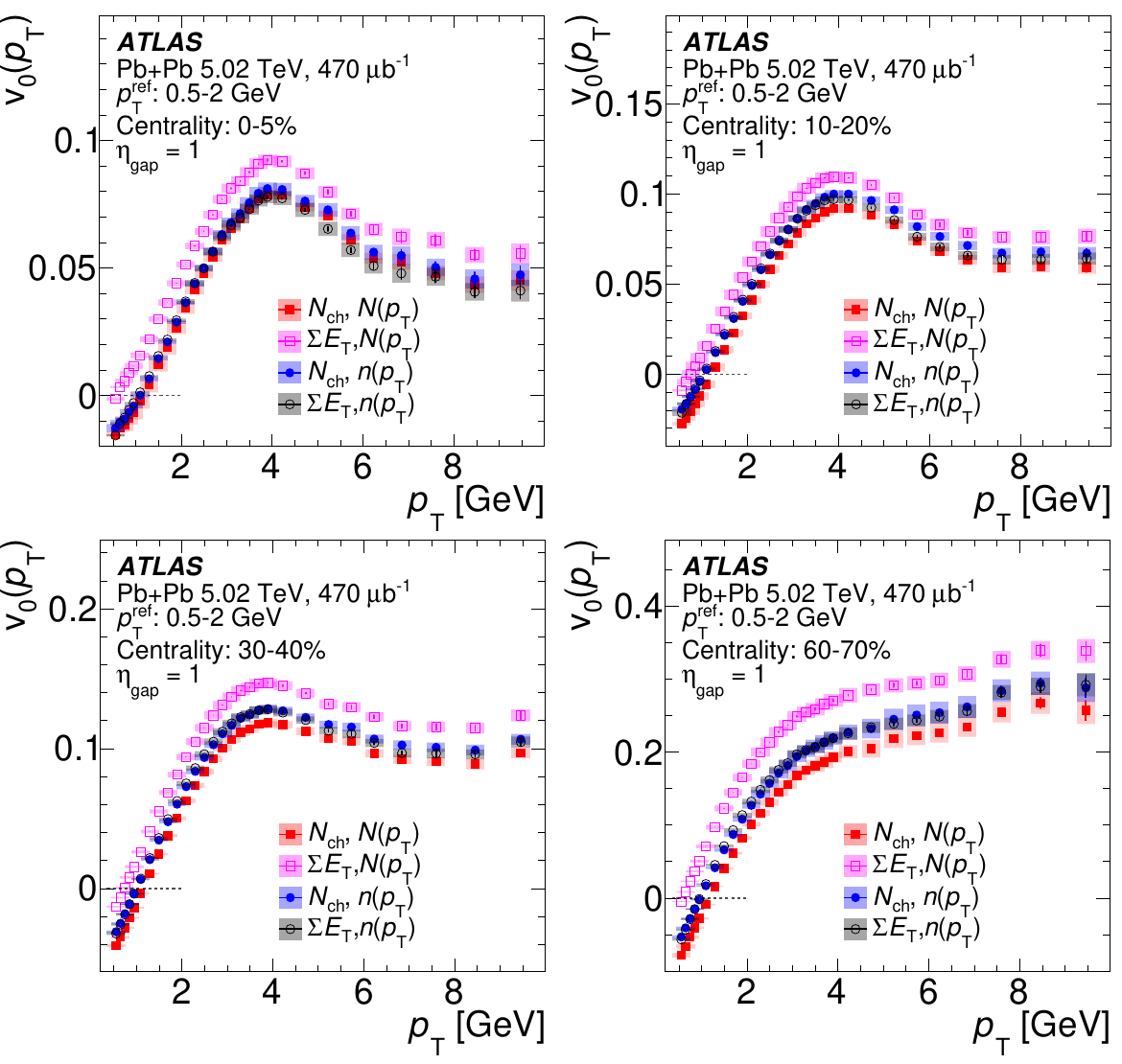}
\caption{Comparison of $v_0(\pT)$ measured using un-normalized event-by-event spectra $N(\pT)$ (open markers) and normalized spectra $n(\pT)$ (filled markers) for several centrality selections. Event averages are obtained by combining observables calculated in narrow slices of $\Nch$ (circles) or $\sum E_T$ (squares). They are obtained for $p_{T,\text{ref}}$ in $0.5$--$2\text{~GeV}$ and $\eta_{\text{gap}}=1$. Bars and shaded areas indicate statistical and total uncertainties, respectively.}
\label{fig:figaux_05}
\end{figure}

\begin{enumerate}
    \item \textbf{Dependence on Event Classifier and Spectra Normalization:}
    Experimental measurements of $v_0(p_T)$ revealed differing results depending on whether centrality was defined using $\sum E_T$ or $\Nch$. These differences were pronounced when $v_0(p_T)$ was calculated using un-normalized particle spectra, $N(p_T)$. As illustrated in Figure~\ref{fig:figaux_05}, calculations employing normalized spectra, $n(p_T) = N(p_T)/N$ (where $N$ is the total multiplicity within the considered acceptance), yield consistent $v_0(p_T)$ values for both $\sum E_T$ and $\Nch$-based centrality definitions. In contrast, when using un-normalized spectra $N(p_T)$:
    \begin{itemize}
        \item[-] $v_0(p_T)$ derived from an $\Nch$-based centrality definition is smaller than the $v_0(p_T)$ obtained using $n(p_T)$ with the same centrality definition.
        \item[-] Furthermore, $v_0(p_T)$ derived from a FCal $\sum E_T$-based centrality definition and $N(p_T)$ is larger than results obtained with $\Nch$-based centrality (for both $N(p_T)$ and $n(p_T)$ calculations).
    \end{itemize}
    These choices also led to variations in the zero-crossing point of $v_0(p_T)$.

    \item \textbf{Influence of Kinematic $p_T$ Range for Spectral Definition:}
    
    The specific transverse momentum ($p_T$) range selected to define the spectra ($N(p_T)$ or $n(p_T)$) was observed to influence the resulting $v_0(p_T)$. This is partly attributed to the relationship between the zero-crossing point of $v_0(p_T)$ and the average $p_T$ of the spectrum under study; modifying the $p_T$ range alters this average $p_T$ and consequently the behavior of $v_0(p_T)$.

    \item \textbf{Ambiguity in Existing Model Interpretations:}
    A key theoretical study discussing $v_0(p_T)$~\cite{Parida:2024ckk} did not explicitly detail the comparative effects of using un-normalized spectra $N(p_T)$ versus normalized spectra $n(p_T)$ in the calculation of $v_0(p_T)$, leaving an interpretational gap regarding experimental data.
\end{enumerate}

To investigate these observations and clarify the role of non-flow effects, multiplicity fluctuations, and analysis choices, we employ the HIJING event generator~\cite{Wang:1991hta,Gyulassy:1994ew}. HIJING lacks genuine collective radial flow, making it suitable for studying non-collective baseline effects.

\subsubsection{Methodology}
\label{ssec:hijing_theory_defs}

The $p_T$-differential radial flow $v_0(\pT)$ is fundamentally defined via the covariance of spectral fluctuations with fluctuations in the event-wise mean transverse momentum $[\pT]$:
\begin{equation}
\label{eq:v0pt_definition_hijing}
\frac{\langle \delta X(\pT) \delta [\pT] \rangle}{\langle X(\pT) \rangle \langle [\pT] \rangle_A} = v_0^{(X)}(\pT) v_{0,[\pT]},
\end{equation}
where $X(\pT)$ can be either the normalized spectrum $n(\pT)$ or the un-normalized spectrum $N(\pT)$, leading to $v_0(\pT)$ or $v_0'(\pT)$ respectively. The term $v_{0,[\pT]} = \sqrt{\langle (\delta [\pT])^2 \rangle_A} / \langle [\pT] \rangle_A$ quantifies the relative fluctuation of $[\pT]$ calculated within a specific reference $p_T$ range $A$ (denoted $p_{T,\text{ref}}$). 

The $v_0(\pT)$ can also be expressed using the Pearson correlation coefficient $\rho$:
\begin{equation}
\label{eq:v0pt_pearson_hijing}
v_0(\pT) = \rho(n(\pT), [\pT]) \frac{\sqrt{\langle (\delta n(\pT))^2 \rangle}}{\langle n(\pT) \rangle}.
\end{equation}
This highlights that $v_0(\pT)$ isolates the part of the spectral shape fluctuation in $\delta n(\pT)$ that is linearly correlated with $\delta [\pT]$. The magnitude $|v_0(\pT)| \leq \sqrt{\langle (\delta n(\pT))^2 \rangle}/\langle n(\pT) \rangle$. This correlation can be reduced by decorrelation effects, e.g., if $[\pT]$ is measured at low $p_T$ and $n(\pT)$ fluctuations are probed at much higher $p_T$.

\paragraph{Effect of Using un-normalized Spectra $N(\pT)$}
If $v_0'(\pT)$ is defined using the un-normalized spectrum $N(\pT)$ as per Eq.~\ref{eq:v0pt_definition_hijing}, its difference from $v_0(\pT)$ (defined with $n(\pT)$) can be approximated as follows. Assuming $N(\pT) = N \cdot n(\pT)$ and that fluctuations $\delta N$ and $\delta n(\pT)$ are not strongly correlated (i.e., $\langle \delta N \delta n(\pT) \rangle \approx 0$, making $\langle N(\pT) \rangle \approx \langle N \rangle \langle n(\pT) \rangle$), the difference $\Delta v_0 \equiv v_0'(\pT) - v_0(\pT)$ is given by:
\begin{equation}
\label{eq:delta_v0_N_vs_n}
\Delta v_0 \approx \frac{\langle \delta N \delta [\pT] \rangle}{\langle N \rangle \sqrt{\langle (\delta [\pT])^2 \rangle}} = \rho(N, [\pT]) \frac{\sqrt{\langle (\delta N)^2 \rangle}}{\langle N \rangle}.
\end{equation}
This offset $\Delta v_0$ depends on the relative fluctuation of the total multiplicity $N$ within the considered acceptance and its correlation $\rho(N, [\pT])$ with $[\pT]$. The sign of $\Delta v_0$ is determined by $\rho(N, [\pT])$, which is sensitive to the event activity classifier used.

\paragraph{Effect of Restricted $p_T$ Acceptance for Normalization}
\label{sssec:hijing_theory_acceptance}
Experiments typically measure particles within a restricted $p_T$ range, say $R$. If the fractional spectrum is defined only within this range, $n_R(\pT) = N(\pT)/N_R$ (where $N_R = \int_R N(\pT) dp_T$), this introduces further effects. Even if the multiplicity $N_{\text{full}}$ over a very wide range is fixed, $N_R$ fluctuates due to the stochastic nature of particle population in the subrange $R$. For fixed $N_{\text{full}}$, $N_R$ follows a binomial distribution with sampling probability $\epsilon = \langle N_R \rangle / \langle N_{\text{full}} \rangle$.

\begin{figure}[htbp]
\centering
\includegraphics[width=0.5\textwidth]{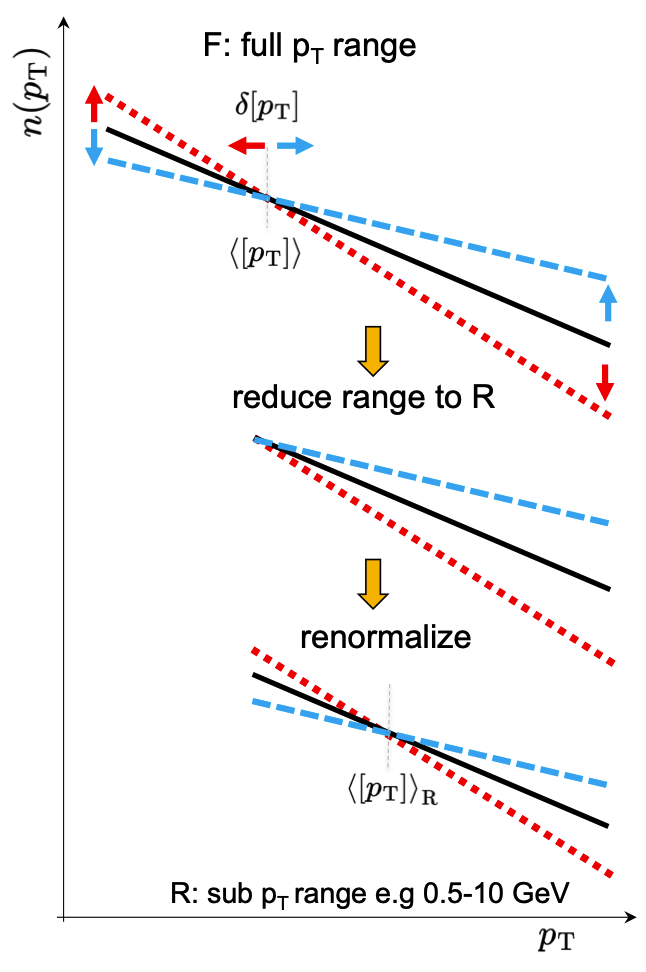}
\caption{Top: Schematic illustration of how event-by-event ($EbE$) radial flow fluctuations can create correlations between the $p_T$-differential yield $n(\pT)$ and the $EbE$ average transverse momentum $[\pT]$. The blue (red) curve represents an event with larger (smaller) than average radial flow, and the black curve is the ensemble-averaged spectrum. Bottom: Fractional spectra defined over a wide range $F$ (e.g., $n_F(\pT)=N(\pT)/N_F$) versus a subrange $R$ (e.g., $n_R(\pT)=N(\pT)/N_R$) are normalized differently, affecting $v_0(\pT)$ calculations.}
\label{fig:ptrange}
\end{figure}

Let $v_{0,\text{full}}(\pT)$ be the observable defined using $n_{\text{full}}(\pT)$ (normalized by $N_{\text{full}}$) and $v_{0,R}(\pT)$ be defined using $n_R(\pT)$ (normalized by $N_R$). The difference $\Delta v_{0,R} = v_{0,R}(\pT) - v_{0,\text{full}}(\pT)$ arises. If $N_{\text{full}}$ is fixed, this offset is primarily due to $N_R$ fluctuations:
\begin{equation}
\label{eq:delta_v0R_fixedNfull_approx_hijing}
\Delta v_{0,R} \approx - \left( \rho(N_R, [\pT]) \sqrt{\frac{\langle (\delta N_R)^2 \rangle_N}{\langle N_R \rangle^2}} \right)_{N_{\text{full}}},
\end{equation}
where the subscript $N_{\text{full}}$ indicates evaluation at fixed total multiplicity. Using the binomial property $\langle (\delta N_R)^2 \rangle_{N_{\text{full}}} = \langle N_R \rangle (1-\epsilon)$:
\begin{equation}
\label{eq:delta_v0R_binomial_hijing}
\Delta v_{0,R} \approx - \left( \rho(N_R, [\pT]) \sqrt{\frac{1-\epsilon}{\langle N_R \rangle}} \right)_{N_{\text{full}}}.
\end{equation}
This shift can alter the zero-crossing point of $v_{0,R}(\pT)$, typically moving it towards $\langle [\pT] \rangle_R$, the average $p_T$ within the range $R$. This is illustrated schematically in Figure~\ref{fig:ptrange} (bottom panel). It is important that this renormalization primarily affects the crossing point, while the fundamental factorization property of $v_0(\pT)$ (its independence from $p_{T,\text{ref}}$ for $v_{0,[\pT]}$) is expected to hold.

To study these effects, Pb+Pb collisions were generated using HIJING (version with quenching, HIPR1(10)=2 GeV for the hard scale cutoff) at $\sqrt{s_{NN}} = 5.02$ TeV. Charged particles with $|\eta| < 2.5$ and $p_T < 10$ GeV were used. For two-subevent calculations, particles were divided into subevent A ($-2.5 < \eta_a < -\eta_{\text{gap}}/2$) and subevent B ($\eta_{\text{gap}}/2 < \eta_b < 2.5$), with $\eta_{\text{gap}} = 0$ and $3$ explored.
The event-wise $[\pT]$ was calculated using reference particles in three $p_{T,\text{ref}}$ ranges: $0.5$--$2$ GeV, $0.5$--$5$ GeV, and $1$--$5$ GeV. The spectra $N(\pT)$ or $n(\pT)$, and subsequently $v_0(\pT)$ or $v_0'(\pT)$, were determined in three overall $p_T$ analysis ranges: $0$--$10$ GeV (full range), $0.5$--$10$ GeV (subrange $R_1$), and $1$--$10$ GeV (subrange $R_2$).
Events were classified based on various definitions of charged particle multiplicity ($\Nch$) in different $\eta$ windows, and by the number of participant nucleons ($N_{\text{part}}$), as detailed in Table~\ref{tab:hijing_selections}.

\begin{table}[h!]
\centering
\caption{Kinematic selections for particles used in defining $n(\pT)$ or $N(\pT)$, for calculating $[\pT]$ (via $p_{T,\text{ref}}$), and for event activity classification in the HIJING study.}
\begin{tabular}{|c|c|c|}
\hline
$p_T$ range for $N(\pT)$, $n(\pT)$ & $p_{T,\text{ref}}$ for $[\pT]$ (GeV) & Event Activity Classifier \\
\hline
$0$--$10$ GeV, $|\eta| < 2.5$ & $0.5$--$2$ & $\Nch$ in $|\eta| < 2.5$ \\
$0.5$--$10$ GeV, $|\eta| < 2.5$ & $0.5$--$5$ & $\Nch$ in $2.5 < |\eta| < 3.2$ \\
$1$--$10$ GeV, $|\eta| < 2.5$ & $1$--$5$ & $\Nch$ in $3.2 < |\eta| < 4$ \\
& & $\Nch$ in $4 < |\eta| < 5$ \\
& & $N_{\text{part}}$ \\
\hline
\end{tabular}
\label{tab:hijing_selections}
\end{table}

For each event, $[\pT]$, $N(\pT)$, and $n(\pT)$ were calculated for each subevent. Two-particle cumulants were used for ensemble averages: $\langle (\delta[\pT])^2 \rangle = \langle \delta[\pT]_a\delta[\pT]_b \rangle$, and $\langle \delta X(\pT)\delta[\pT] \rangle = \frac{1}{2}\langle \delta X_a(\pT)\delta[\pT]_b + \delta X_b(\pT)\delta[\pT]_a \rangle$.
The offset term $\Delta v_0$ (Eq.~\ref{eq:delta_v0_N_vs_n}) was estimated using the two-subevent method as:
\begin{equation}
\label{eq:delta_v0_hijing_calc}
\Delta v_0^{\text{calc}} = \frac{\langle \delta N_a \delta [\pT]_b \rangle}{\langle N_a \rangle \sqrt{\langle \delta[\pT]_a \delta[\pT]_b \rangle}}.
\end{equation}
This analysis procedure mirrors that used in the ATLAS measurements~\cite{ATLAS:2025ztg}.

Since HIJING is based on an independent particle production source model, various quantities are expected to follow simple scaling laws with the number of sources $N_s$. For instance, if particles for event activity selection are independent of those in the analysis:
\begin{align}
\langle \delta N_a \delta N_b \rangle &\propto N_s & \langle \delta [\pT]_a \delta [\pT]_b \rangle &\propto 1/N_s \nonumber \\
\sqrt{\langle (\delta N_a \delta N_b \rangle} / \langle N_a \rangle &\propto 1/\sqrt{N_s} & \Delta v_0^{\text{calc}} &\propto 1/\sqrt{N_s} \nonumber \\
\langle \delta N_a \delta [\pT]_b \rangle &\sim \text{const} & \rho(N_a, [\pT]_b) &\sim \text{const} \nonumber \\
\sqrt{N_s} \Delta v_0^{\text{calc}} &\sim \text{const} &&
\label{eq:hijing_scaling_laws}
\end{align}
where ``const'' implies independence from $N_s$.

\subsubsection{Results}

\begin{figure}[htbp]
\centering
\includegraphics[width=0.5\textwidth]{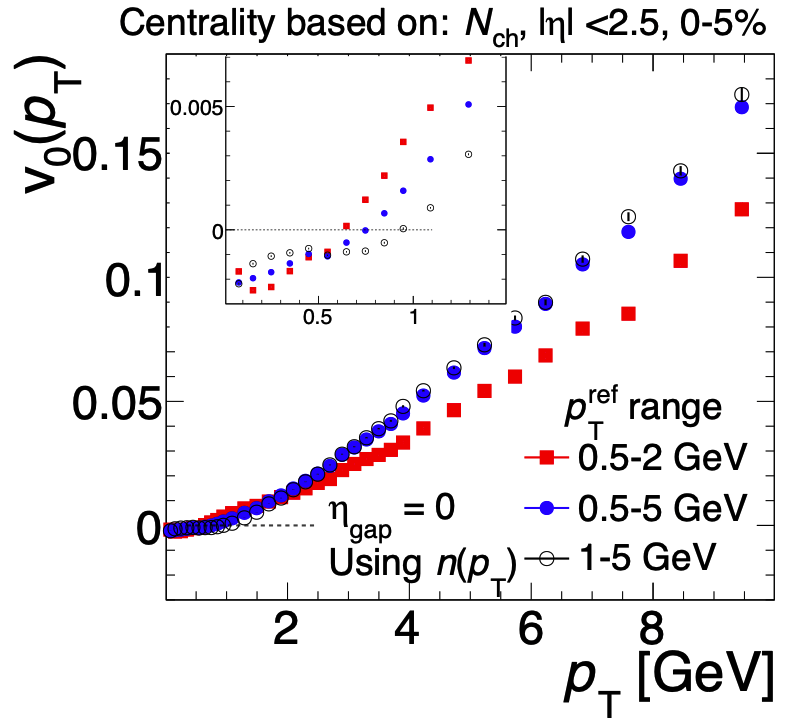}
\caption{The $v_0(\pT)$ calculated from HIJING using fractional spectra $n(\pT)$ (normalized in $0$--$10$ GeV range) for three $p_{T,\text{ref}}$ selections, in $0$--$5\%$ most central Pb+Pb collisions. The inset provides a zoomed-in view of the low $p_T$ region.}
\label{fig:v0pthij_fig2}
\end{figure}

Figure~\ref{fig:v0pthij_fig2} shows $v_0(\pT)$ from HIJING using $n(\pT)$ for three $p_{T,\text{ref}}$ selections in central collisions. Non-flow correlations aare oberved to extend to low $p_T$ and also cause variations in the zero-crossing point (see inset). The overall magnitude of $v_0(\pT)$ in HIJING is much smaller than in experimental data, supporting the interpretation that collective radial flow, absent in HIJING, dominates the experimental measurements.

\begin{figure}[h!]
\centering
\includegraphics[width=1.0\textwidth]{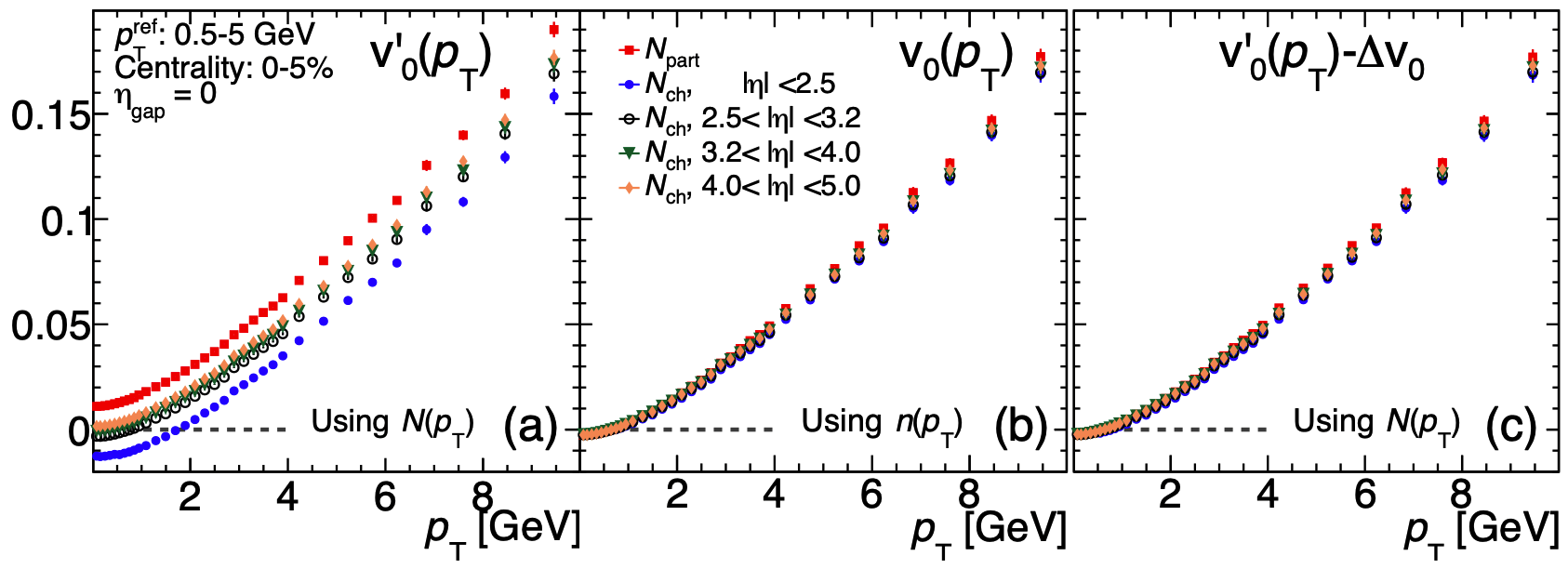}
\caption{HIJING results for $v_0(\pT)$ in $0$--$5\%$ most central Pb+Pb collisions: (a) $v_0'(\pT)$ obtained using $N(\pT)$; (b) $v_0(\pT)$ obtained using $n(\pT)$; (c) $v_0'(\pT)$ after applying the correction $\Delta v_0^{\text{calc}}$ from Eq.~\ref{eq:delta_v0_hijing_calc}. Different markers correspond to different event activity selections.}
\label{fig:v0pthij_fig3}
\end{figure}

\paragraph{Influence of Event Classification and Spectral Normalization}
Figure~\ref{fig:v0pthij_fig3}(a) presents $v_0'(\pT)$, calculated using un-normalized spectra $N(\pT)$, for various event activity selections. The results exhibit similar increasing trends with $p_T$ but show substantial vertical offsets relative to each other. These offsets, dependent on the event classifier, drastically alter the zero-crossing point. For event classes based on $N_{\text{part}}$ or forward $\Nch$, $v_0'(\pT)$ may not cross zero at all, indicating a strong influence of local spectral fluctuations with global multiplicity fluctuations.

In contrast, Figure~\ref{fig:v0pthij_fig3}(b) shows $v_0(\pT)$ (from normalized spectra $n(\pT)$). These results cluster closely together for all event classifiers, sharing a common zero-crossing point near the average $p_T$ of the inclusive spectrum. This demonstrates that $v_0(\pT)$ from $n(\pT)$ is less sensitive to the choice of event classifier.
Figure~\ref{fig:v0pthij_fig3}(c) shows that the differences between $v_0'(\pT)$ in panel (a) and $v_0(\pT)$ in panel (b) are largely accounted for by the calculated offset $\Delta v_0^{\text{calc}}$ from Eq.~\ref{eq:delta_v0_hijing_calc}, effectively reconciling the two definitions.

\begin{figure}[h!]
\centering
\includegraphics[width=0.5\textwidth]{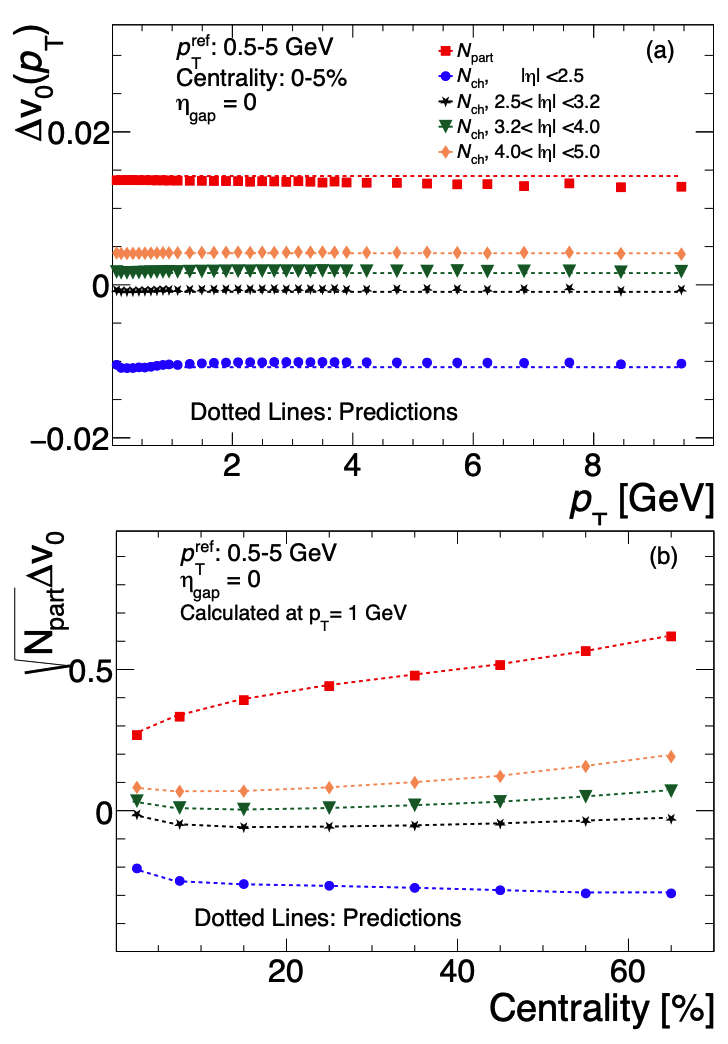}
\caption{Top (a): Comparison of $v_0'(\pT)-v_0(\pT)$ (points) with the estimated $\Delta v_0^{\text{calc}}$ (dotted lines) in $0$--$5\%$ central HIJING Pb+Pb collisions. Bottom (b): Centrality dependence of $\sqrt{N_{\text{part}}}(v_0'(\pT) - v_0(\pT))$ at $p_T = 1$ GeV (points) compared with $\sqrt{N_{\text{part}}}\Delta v_0^{\text{calc}}$ (dotted lines).}
\label{fig:v0pthij_fig4}
\end{figure}

The validity of the $\Delta v_0$ correction is further tested in Figure~\ref{fig:v0pthij_fig4}. Panel (a) compares the measured difference $v_0'(\pT) - v_0(\pT)$ (points) with the predicted $\Delta v_0^{\text{calc}}$ (dotted lines) versus $p_T$. The difference is nearly $p_T$-independent and well described by $\Delta v_0^{\text{calc}}$. Panel (b) shows the centrality dependence of this difference at $p_T = 1$ GeV, scaled by $\sqrt{N_{\text{part}}}$. Again, remarkable agreement is observed.

These findings confirm that $v_0(\pT)$ calculated with $n(\pT)$ effectively separates global multiplicity fluctuations from $p_T$-dependent spectral shape fluctuations. The correction term $\Delta v_0$ increases with the $\eta$ gap between analyzed particles and those used for event classification, consistent with known multiplicity decorrelation effects~\cite{Bzdak:2012tp, Jia:2015jga, ATLAS:2016rbh}, which can be linked to initial-state longitudinal fluctuations~\cite{Jia:2020tvb}.

\begin{figure}[h]
\centering
\includegraphics[width=0.7\textwidth]{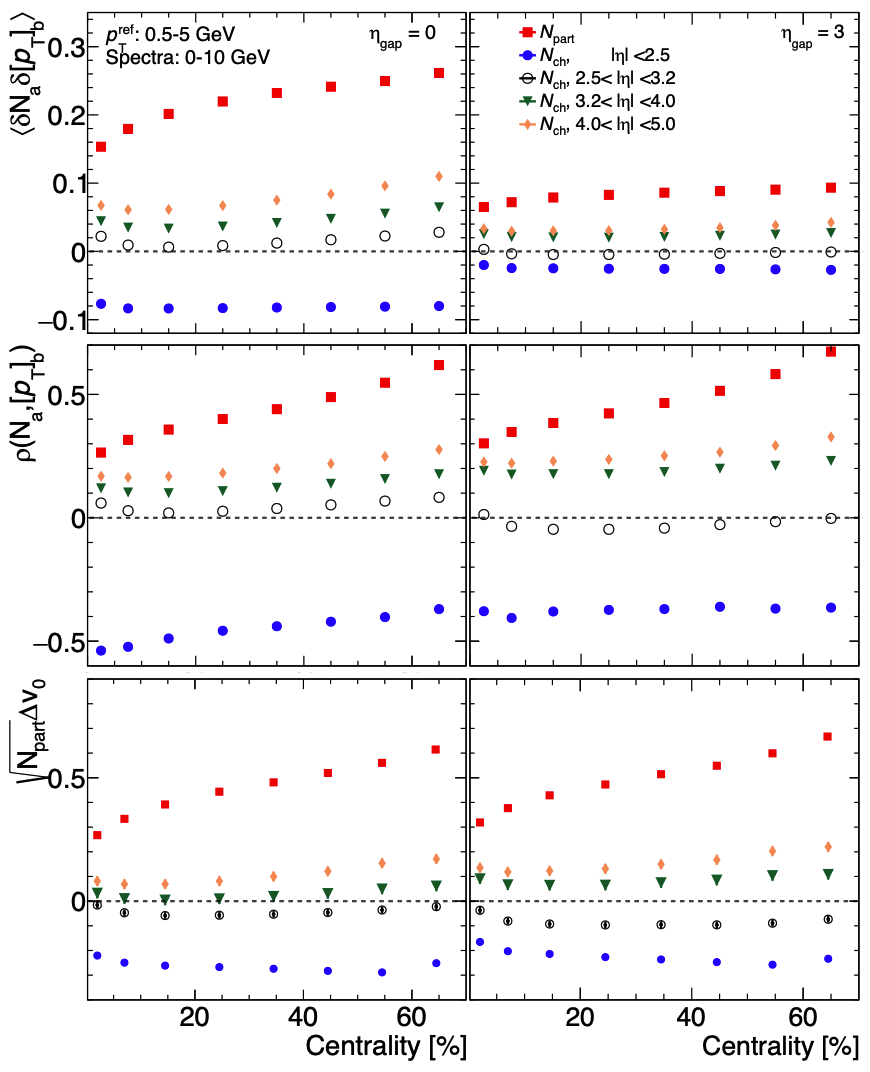}
\caption{Components of $\Delta v_0^{\text{calc}}$ from HIJING: $\langle \delta N_a\delta[\pT]_b \rangle$ (top), $\rho(N_a, [\pT]_b)$ (middle), and $\sqrt{N_s}\Delta v_0^{\text{calc}}$ (here shown as $\sqrt{N_{\text{part}}}\Delta v_0^{\text{calc}}$, bottom) vs. centrality for $\eta_{\text{gap}} = 0$ (left) and $\eta_{\text{gap}} = 3$ (right), for various event activity classifiers.}
\label{fig:v0pthij_fig5}
\end{figure}

The expected scaling laws in Eq.~\ref{eq:hijing_scaling_laws} from the independent source picture in HIJING were also tested. Figure~\ref{fig:v0pthij_fig5} shows that quantities like $\langle \delta N_a\delta[\pT]_b \rangle$ and $\sqrt{N_{\text{part}}}\Delta v_0^{\text{calc}}$ are approximately constant or vary slowly with centrality, supporting the theoretical expectations when event selection and analysis regions are sufficiently separated.

However, when the particles for event activity definition strongly overlap with those for analysis (e.g., $\Nch$ in $|\eta| < 2.5$, blue symbols in Fig.~\ref{fig:v0pthij_fig5}), observables like $\langle \delta N_a\delta[\pT]_b \rangle$ become small. This is due to kinematic constraints: fixing total $\Nch = N_a + N_b + N_c$ (where $N_c$ is in the gap) induces anticorrelations between $N_a, N_b, N_c$. This anticorrelation is strongest for $\eta_{\text{gap}}=0$ ($N_c=0$) and weakens as $\eta_{\text{gap}}$ increases. For well-separated regions or $N_{\text{part}}$-based selection, correlations are small or slightly positive, with magnitudes depending on the $\eta$ range and increasing towards peripheral collisions for $N_{\text{part}}$ selection. This behavior helps explain the ordering of $v_0(\pT)$ observed experimentally for different centrality definitions as observed in ATLAS measurements and shown in Figure~\ref{fig:figaux_05}.

\paragraph{Consequence of Varying $p_T$ Acceptance for Normalization}
As discussed in Section~\ref{sssec:hijing_theory_acceptance}, normalizing spectra within a restricted $p_T$ acceptance $R$ (i.e., using $n_R(\pT)=N(\pT)/N_R$) can lead to an offset $\Delta v_{0,R}$ (Eqs.~\ref{eq:delta_v0R_fixedNfull_approx_hijing}-\eqref{eq:delta_v0R_binomial_hijing}) compared to using a wider normalization range. This primarily affects the zero-crossing point of $v_0(\pT)$.

\begin{figure}[h]
\centering
\includegraphics[width=1.0\textwidth]{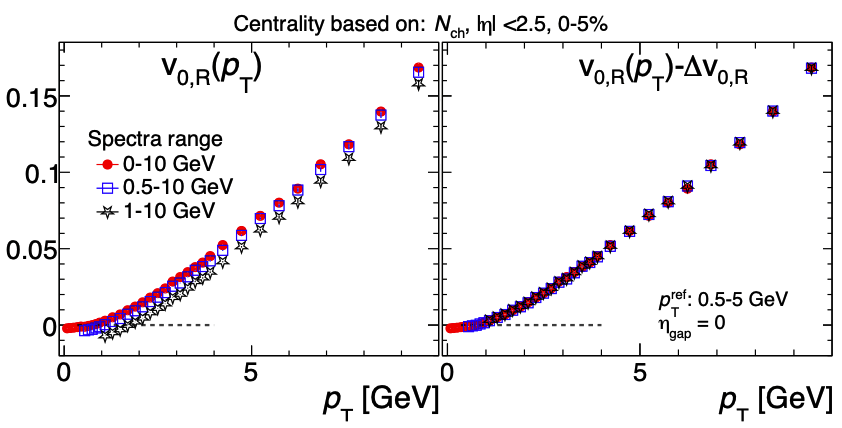}
\caption{(Left) HIJING $v_{0}(p_{T})$ calculated using spectra $n_R(\pT)$ normalized by $N_{R} = \int_{A}^{B} N(p_{T})\,dp_{T}$ where $R$ corresponds to three different $p_{T}$ ranges ($0$--$10$, $0.5$--$10$, and $1$--$10$ GeV) in $0$--$5\%$ most central collisions. (Right) The results after correcting for the acceptance-induced offset, showing improved consistency.}
\label{fig:v0pthij_fig6}
\end{figure}

Figure~\ref{fig:v0pthij_fig6}(a) shows $v_0(\pT)$ from HIJING where $n_R(\pT)$ was calculated using three different $p_T$ ranges $R$ for $N_R$. The results clearly depend on the chosen range $R$. Figure~\ref{fig:v0pthij_fig6}(b) shows that these differences are largely reconciled after applying a correction for this acceptance-induced offset shown in Eq.~\ref{eq:delta_v0R_fixedNfull_approx_hijing}. This confirms that the precise zero-crossing location of $v_0(\pT)$ in such cases is strongly influenced by the kinematic properties of the spectrum within the chosen acceptance $R$, rather than being a pure indicator of dynamical feature across the full spectrum.

\subsection{Conclusion}\label{sec:v0pt_summary1}

This chapter presents the first measurement of the transverse momentum dependence of radial flow fluctuations, $v_0(\pT)$, in Pb+Pb collisions at $\sqn = 5.02$ TeV, performed by the ATLAS Collaboration~\cite{Bhatta:2025oyp}. The experimental data reveal strong evidence for the collective nature of radial flow, highlighted by three key features. 
Firstly, $v_0(\pT)$ exhibits minimal dependence on the pseudorapidity gap ($\eta_{\mathrm{gap}}$) indicating that radial flow is a global phenomenon extending over large longitudinal distances. Secondly, for $\pT \lesssim 3$ GeV, the $v_0(\pT)$ factrorizes, i.e it is independent of the specific $p^{ref}_\mathrm{T}$ range used to define $v_0$, consistent with the factorization behavior observed for $v_{n}$ with $n\geq2$. Thirdly, when normalized by the integrated fluctuation $v_0$, the shape of $v_0(\pT)$ (i.e., $v_0(\pT)/v_0$) is almost independent of collision centrality for $\pT \lesssim 2.5$ GeV. This striking feature suggests a universal hydrodynamic response to radial expansion, with the magnitude of fluctuations primarily driven by initial-state geometry.

Comparisons with hydrodynamic model calculations show that the inclusion of a temperature-dependent bulk viscosity, parameterized according to lattice QCD expectations, significantly improves the description of the $v_0(\pT)/v_0$ data, especially concerning its zero-crossing point and high-$\pT$ behavior. This provides a novel experimental avenue to quantitatively constrain $\zeta/s$, a transport coefficient that has proven more challenging to access than shear viscosity but is important for a complete understanding of QGP properties.

A detailed HIJING model study~\cite{Bhatta:2025oyp} demonstrated that $v_0(\pT)$ measured using fractional spectra agree better between different centrality definitions, unlike $N(\pT)$ which exhibits shifts dependent on centrality definition. This difference is attributed to the trivial correlation between total multiplicity and average transverse momentum. 
Furthermore, the study demonstrates that the chosen $\pT$ acceptance for defining the normalized spectrum introduces a horizontal shift along the $\pT$ axis for the zero-crossing point of $v_0(\pT)$. These results explains the influence of event classification schemes, spectral normalization procedures, and experimental acceptance on $v_0(\pT)$ to better interpret its features in experimental measurements and further assist probing the genuine collective dynamics in heavy-ion collisions.

\subsection{Outlook}\label{sec:v0pt_summary2}

The results presented in this chapter open up several exciting avenues for future research, leveraging $v_0(\pT)$ to gain deeper insights into the QGP. The observed decrease of $v_0(\pT)$ at $\pT > 3$--$4$ GeV in central collisions, similar to trends seen in anisotropic flow, warrants further investigation. This region may mark a transition where the interplay between collective hydrodynamic flow and energy loss mechanisms for hard-scattered partons (jet quenching) becomes important, and studying $v_0(\pT)$ at higher $\pT$ could offer new perspectives on this interplay. 

Concurrently, while the two-subevent method with an $\eta_{\mathrm{gap}}$ effectively suppresses short-range non-flow correlations, their potential contributions at higher $\pT$ and in peripheral collisions necessitate continued scrutiny. Model studies, such as those employing HIJING, and advanced techniques for non-flow subtraction or more sophisticated modeling of non-flow sources could help to further isolate the collective signal.

Beyond these refinements, extending $v_0(\pT)$ measurements to smaller collision systems (e.g., p+Pb, O+O), different heavy-ion species (e.g., Xe+Xe), and a range of collision energies is crucial. This would allow for systematic studies of the system-size and energy dependence of radial flow fluctuations, addressing fundamental questions about the onset and evolution of collectivity. 

Finally, building on the current analysis demonstrating the long-range nature of $v_0(\pT)$ by varying $\eta_{\mathrm{gap}}$, more differential studies of the decorrelation of $v_0(\pT)$ as a function of the pseudorapidity separation between the measured particle and the region defining $\MpT$ could provide insights into the longitudinal structure and dynamics of radial flow fluctuations, similar to rapidity decorrelation studies performed for anisotropic flow \cite{Chatterjee:2017mhc, Jia:2014ysa}.
\clearpage 

\section{Disentangling Initial State Fluctuations Using Radial Flow}
\label{sec:chap5_ptfluc}
The previous chapter established radial flow as a collective phenomenon arising from the hydrodynamic expansion of the QGP and demonstrated its sensitivity to initial-state fluctuations. This collective outward boost influences the $\pT$ spectra of particle produced in the final-state. Specifically, the slope of the $\pT$ spectra is a direct reflection of the event-wise average transverse momentum, $[\pT]$. 

Historically, the degree of radial flow in events has been inferred from fitting particle spectra using the blast-wave parametrization~\cite{Braun-Munzinger:2018hat, ALICE:2013mez, STAR:2005gfr}. This approach is built on the physical picture of a thermally equilibrated source expanding with a collective velocity. In the blast-wave model, the final-state $\pT$ of a particle is understood as a superposition of two components: a random, thermal motion characterized by the kinetic freeze-out temperature, $T_{\text{kin}}$, and an ordered, collective motion described by the radial flow velocity, $\beta$. In a simplified non-relativistic picture, a particle's total energy is the sum of its thermal energy and the kinetic energy from the collective expansion, $E_{\text{tot}} = E_{\text{th}}(T_{\text{kin}}) + \frac{1}{2}m\beta^2$. By simultaneously fitting this model to the measured $p_T$ spectra of identified hadrons, these two parameters, $T_{\text{kin}}$ and $\beta$, can be extracted, effectively constraining these parameters.

\begin{figure}[htbp!]
\centering
\includegraphics[width=0.6\linewidth]{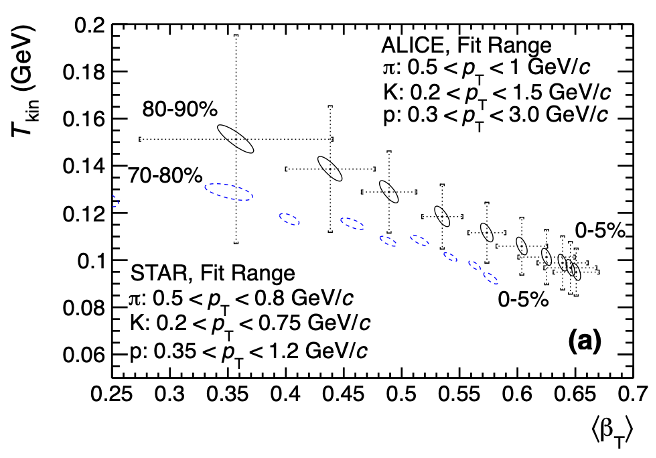}
\caption{Kinetic freezeout temperature ($T_{\text{kin}}$) vs. average radial flow velocity ($\langle \beta \rangle$) extracted from blast-wave model fits to identified particle spectra by the ALICE and STAR collaborations for different centrality classes. The ellipses represent statistical and systematic uncertainties~\cite{Braun-Munzinger:2018hat, ALICE:2013mez, STAR:2005gfr}.}
\label{fig:Tbeta}
\end{figure}

Figure~\ref{fig:Tbeta} presents a historical overview of results from the STAR and ALICE collaborations, showing $T_{\text{kin}}$ versus $\langle \beta \rangle$ extracted from these ensemble-averaged fits for different centrality bins and particle species~\cite{Braun-Munzinger:2018hat, ALICE:2013mez, STAR:2005gfr}. The plot illustrates a characteristic anticorrelation: more central collisions (e.g., 0--5\%) exhibit higher $\langle \beta \rangle$ and lower $T_{\text{kin}}$ compared to peripheral collisions (e.g., 80--90\%). However, these extracted parameters are ensemble averages and do not directly capture the event-by-event fluctuations in the system's collective expansion. To probe the dynamical nature of radial flow fluctuations and their connection to the QGP properties, it is necessary to study the distribution of event-by-event fluctuations in $[\pT]$.

\subsection{Theoretical Background}

The distribution of the $[\pT]$, $P([\pT])$, is proposed to be a sensitive probe of the initial-state geometry, its fluctuations, and the thermodynamic properties, including the equation of state (EoS) and the speed of sound squared ($c_s^2$), of the QGP~\cite{Gardim:2019xjs,Gardim:2019brr,Samanta:2023amp,Samanta:2023kfk,Nijs:2023bzv, Gardim:2024zvi,SoaresRocha:2024drz}. The shape of the $P([\pT])$ distribution can be characterized by its moments: the mean $\lr{[\pT]}$, variance $\lr{(\delta \pT)^2}$, and skewness $\lr{(\delta \pT)^3}$, where $\delta \pT = [\pT]-\lr{[\pT]}$ represents the event-by-event fluctuation of $[\pT]$ around its ensemble average $\lr{[\pT]}$.

In a simple independent superposition scenario, where nuclear collisions are viewed as a collection of independent particle production from participating nucleons followed by final-state interactions, both geometrical and intrinsic fluctuations are expected to scale with the charged particle multiplicity, $\Nch$, approximately as $\lr{(\delta \pT)^2}\propto 1/\Nch$ and $\lr{(\delta \pT)^3} \propto 1/\Nch^{\;2}$~\cite{Gavin:2011gr, Cody:2021cja}. However, experimental measurements of $[\pT]$ correlations have revealed fluctuations in excess of this independent superposition baseline, providing evidence for the existence of additional dynamical correlations in heavy-ion collisions~\cite{STAR:2005vxr, ALICE:2014gvd}.

\subsubsection{Sources of $[\pT]$ Fluctuations}
Drawing from the understanding that individual particle transverse momentum reflects both the system's collective expansion and its thermal properties, the overall event-by-event fluctuation in average transverse momentum, $\delta[\pT]$, can be considered to originate from broadly two distinct categories of effects:

\begin{itemize}
    \item \textbf{Geometrical fluctuations:} Variations in the initial size and shape of the nuclear overlap region from event to event. The hydrodynamic response translates fluctuations in the transverse size $R$ into fluctuations in the collective radial flow, approximated as $\delta\pT/\lr{[\pT]} \approx -\delta R/\lr{R}$~\cite{Bozek:2012fw}.
    \item \textbf{Intrinsic fluctuations:} Other stochastic sources at a fixed geometry, including fluctuations in the positions of nucleons and partons, initial energy deposition, and the temperature profile of the QGP fluid~\cite{Samanta:2023amp}.
\end{itemize}
Recent phenomenological studies suggest that these two components of fluctuations can be effectively investigated using moments of $P([\pT])$ in ultra-central collisions (UCC)~\cite{Gazdzicki:1992ri,Samanta:2023amp, Samanta:2023kfk} as is described below. 

\subsubsection{Disentangling Components of $[\pT]$ Fluctuations}
\begin{figure}[htbp]
  \centering
  \includegraphics[width=0.6\linewidth]{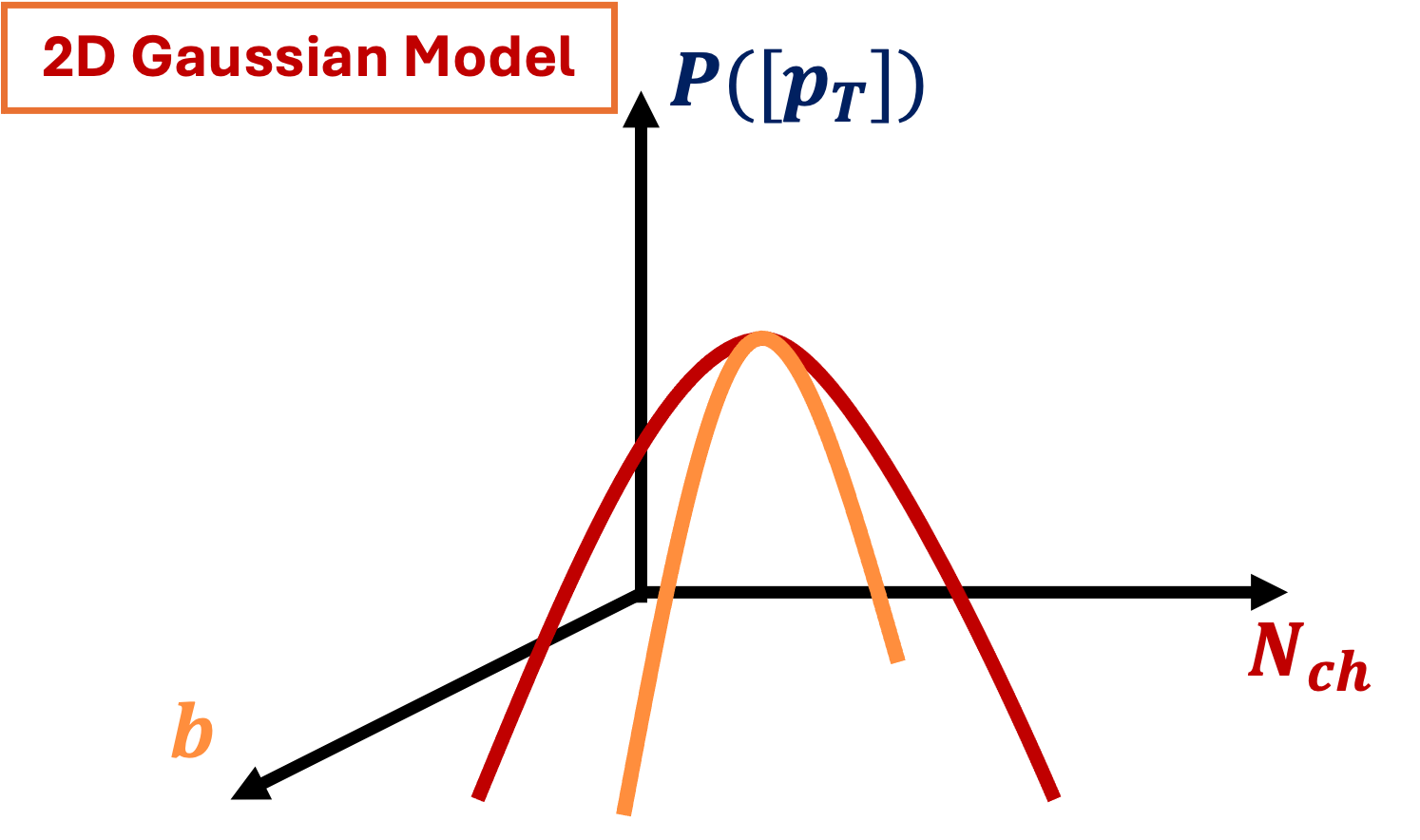}
  \caption{Conceptual illustration of the two-dimensional Gaussian model for $[\pT]$ fluctuations. This model is defined using two axes representing $b$ and $\Nch$. Event-by-event variations in~$b$ and~$\Nch$ are described by the 2D-Gaussian distribution, and the~$[\pT]$ for each event is treated as a single-valued function of~$b$ and~$\Nch$.}
  \label{fig:2dg}
\end{figure}

Figure~\ref{fig:2dg} shows a schematic of the two-dimensional Gaussian model, which describes two components of fluctuations along two axes: one representing variations in the impact parameter~$b$ and the other representing variations in charged-particle multiplicity~$\Nch$. Within this model framework, at a fixed~$b$, fluctuations in the initial energy or entropy lead to variations in~$\Nch$, capturing the intrinsic component of the fluctuations. Conversely, at a fixed~$\Nch$, event-by-event changes in~$b$ capture the geometric component of the fluctuations.

\begin{figure}[htbp]
  \centering
  \includegraphics[width=0.5\linewidth]{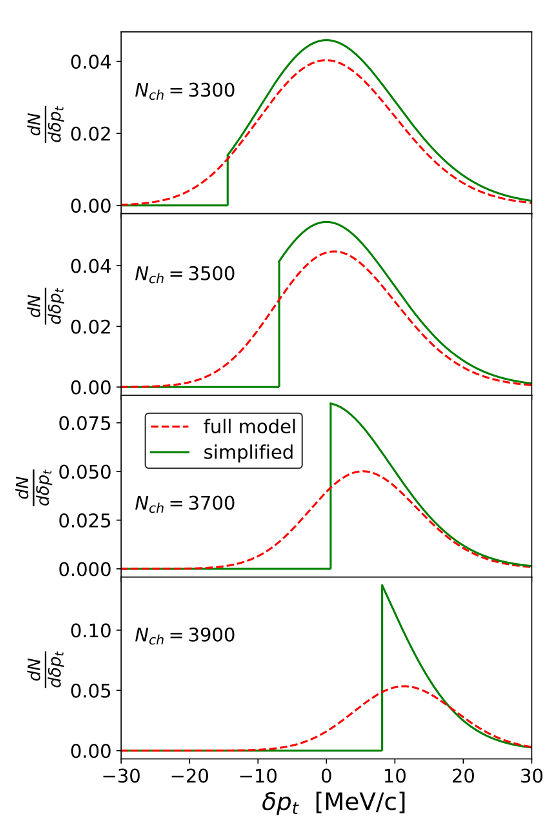}
  \caption{Model calculations showing the distribution of $\delta[\pT]$ for different values of charged‐particle multiplicity $N_{\mathrm{ch}}$. The solid green lines represent a simplified model, while the dashed red lines show a more complete calculation. The shape and skewness of these distributions are sensitive to the underlying fluctuation sources.}
  \label{fig:ptskew}
\end{figure}

In UCC, as $b \to 0$, the transverse overlap radius~$R$ approaches its maximum. This inherently suppresses geometric fluctuations. Given that~$\delta R$ and~$\delta[\pT]$ are anticorrelated, this suppression of~$\delta R$ truncates the lower tail of the~$\delta[\pT]$ distribution. This effect is clearly depicted in Figure~\ref{fig:ptskew}, which shows~$\delta[\pT]$ distributions for increasing~$\Nch$ in UCC from the 2D Gaussian model simulations~\cite{Samanta:2023kfk}. For more central collision event selections by selecting on $\Nch$, the allowed range for fluctuations in ~$b$ within the model framework gets further suppressed resulting in the $\delta[\pT]$ distribution being increasingly cut off from the left. This truncation leads to a sharp decrease in the variance and a corresponding increase in the skewness of the~$\delta[\pT]$ distribution.

These pronounced changes in variance and skewness specifically arise from the constraint imposed on geometric fluctuations in UCC, while intrinsic fluctuations are largely unaffected. Therefore, measuring the moments of the~$[\pT]$ distribution in UCC offers a valuable experimental handle to disentangle contributions from geometric fluctuations from those arising from intrinsic sources.

\subsubsection{Constraining $c_s^2$}
In addition to constraining sources of initial-state fluctuations, the measurement of $\langle [\pT] \rangle$ in UCC is also suggested to be sensitive to the speed of sound, $c_s^2$, in the QGP. Thermodynamically, $c_s^2$ is defined as the derivative of pressure, $P$, with respect to energy density, $\epsilon$:
\begin{equation}
c_s^2 = \frac{dP}{d\epsilon}.
\end{equation}
This can be related to derivatives involving temperature, $T$, and entropy density, $s$. It has been argued that the effective temperature of the system, $T_{\text{eff}}$, is approximately proportional to the average transverse momentum, $\langle \pT \rangle$, and that the entropy density, $s$, is approximately proportional to the charged-particle multiplicity, $\Nch$~\cite{Gardim:2019xjs, Gardim:2019brr}. These proportionalities imply the following approximate relations for their logarithmic derivatives:
\begin{equation}
\begin{aligned}
\frac{dT_{\text{eff}}}{T_{\text{eff}}} &\approx \frac{d\langle \pT \rangle}{\langle \pT \rangle} \quad \Rightarrow \quad d(\ln T_{\text{eff}}) \approx d(\ln \langle \pT \rangle), \\
\frac{ds(T_{\text{eff}})}{s(T_{\text{eff}})} &\approx \frac{d\Nch}{\Nch} \quad \Rightarrow \quad d(\ln s) \approx d(\ln \Nch).
\end{aligned}
\end{equation}
Using these approximations, and the thermodynamic identity $c_s^2 = d(\ln T)/d(\ln s)$ (for a system at negligible chemical potential and fixed volume), $c_s^2$ can be proposed as an experimentally accessible quantity:
\begin{equation}
\label{eq:cs2_experimental}
c_s^2 = \frac{d(\ln T)}{d(\ln s)} \approx \frac{d(\ln \langle \pT \rangle)}{d(\ln \Nch)}.
\end{equation}
In UCC, the transverse overlap area is approximately fixed, approximately satisfying the criteria of a fixed volume. Within this fixed volume, variations in the initial entropy is expected to lead to variations in $\Nch$. At the same time, a larger initial entropy density is expected to drive a stronger collective radial expansion, resulting in a larger $\langle [\pT] \rangle$. Consequently, the slope of $\ln \langle [\pT] \rangle$ versus $\ln \Nch$ in UCC, i.e., $\frac{d(\ln \langle [\pT] \rangle)}{d(\ln \Nch)}$, is predicted to provide a direct experimental constraint on $c_s^2$.

\subsection{Previous Studies}
The mean $\lr{[\pT]}$ and variance $\lr{(\delta \pT)^2}$ of the $P([\pT])$ distribution have been measured across various collision energies and system sizes over the past two decades~\cite{NA49:1999inh, PHENIX:2002aqz, PHENIX:2003ccl, STAR:2003cbv, NA49:2003hxt, CERES:2003sap, STAR:2005vxr, NA49:2008fag,STAR:2013sov, ALICE:2014gvd, STAR:2019dow}. These studies consistently show an increase in $\lr{[\pT]}$ towards more central collisions and confirm that the variance approximately follows the expected $1/\Nch$ power-law scaling predicted by the independent superposition scenario. 

More recently, ALICE reported initial measurements of the skewness $\lr{(\delta \pT)^3}$ and kurtosis $\lr{(\delta \pT)^4}$ in broad multiplicity ranges for Xe+Xe and Pb+Pb collisions~\cite{ALICE:2023tej}. A detailed study by CMS collaboration focused on the behavior of $\lr{[\pT]}$ in Pb+Pb UCC and attempted to extract $c_s^2$~\cite{CMS:2024sgx}, although with noted caveats regarding the interpretation~\cite{Nijs:2023bzv, Gardim:2024zvi,SoaresRocha:2024drz,Gavassino:2025bts}. 

A common limitation of these previous studies is that they could not disentangle the contributions of geometrical and intrinsic fluctuations to the $P([\pT])$ distribution and its moments. Achieving this disentanglement requires precise measurements of higher-order moments, in narrow centrality intervals, especially in UCC.

This chapter presents the measurement of the mean, variance, and skewness of $P([\pT])$ as a function of $\Nch$ in $^{208}$Pb+$^{208}$Pb collisions at $\sqn = 5.02$~TeV and $^{129}$Xe+$^{129}$Xe collisions at $\sqn = 5.44$~TeV. Comparing results from the smaller Xe+Xe system with Pb+Pb provides a unique opportunity to study the role of system size in the scaling behavior of these moments and the underlying fluctuation components.

\subsection{Methodology}
The moments of the $P([\pT])$ distribution are calculated using multiparticle correlation methods, similar to techniques developed for anisotropic flow analysis~\cite{Bilandzic:2010jr, Bhatta:2021qfk}. For a single event, the event-wise average transverse momentum and the unnormalized 2- and 3-particle correlators are computed as follows:
\begin{align}
[\pT] &= \frac{\sum_{i} w_{i}p_{\mathrm{T},i}}{\sum_{i} w_{i}} \\
c_{2} &= \frac{\sum_{i\neq j} w_{i}w_{j}\delta p_{i}\delta p_{j}}{\sum_{i\neq j } w_{i}w_{j}} \\
c_{3} &= \frac{\sum_{i\neq j \neq k } w_{i}w_{j}w_{k}\delta p_{i}\delta p_{j}\delta p_{k}}{\sum_{i\neq j\neq k } w_{i}w_{j}w_{k}}
\end{align}
Here, $p_{\mathrm{T},i}$ is the transverse momentum of track $i$, $\delta p_{i} \equiv p_{\mathrm{T},i}- \lr{[\pT]}$ is the deviation of the track's $\pT$ from the ensemble-averaged mean transverse momentum for the given event class, and $w_{i} \equiv (1-f_i)/\epsilon_i$ are weights applied to track $i$ to correct for reconstruction efficiency, $\epsilon_i$, and fake rate, $f_i$~\cite{Aad:2019fgl}. The ensemble averages of these correlators, denoted by $\lr{c_2}$ and $\lr{c_3}$, are then used to define dimensionless, normalized moments:
\begin{equation}
k_{2}=\frac{\lr{c_{2}}}{\lr{[\pT]}^{2}},\quad k_{3}=\frac{\lr{c_{3}}}{\lr{[\pT]}^{3}},\quad \gamma = \frac{\lr{c_{3}}}{\lr{c_{2}}^{3/2}},\quad \Gamma = \frac{\lr{c_{3}}\lr{[\pT]}}{\lr{c_{2}}^{2}}. \label{eq:ck2}
\end{equation}
In this context, $\gamma$ is referred to as the ``standard skewness'' as it represents the skewness for a distribution with unit variance, while $\Gamma$ is termed the ``intensive skewness''~\cite{Giacalone:2020lbm}. Under the independent superposition scenario, these normalized moments are expected to scale approximately as $k_{2} \propto 1/\Nch$, $k_{3}\propto 1/(\Nch)^2$, and $\gamma \propto 1/\sqrt{\Nch}$, while $\Gamma$ should be roughly independent of $\Nch$. Further discussions on these formulae can be found in Appendix~\ref{sec:app_method}.

Statistical uncertainties for these observables are determined using a Poisson bootstrap method~\cite{Efron:1979bxm, ATLAS:2021kho}. Further description on the Bootstrap method used can be found in Appendix~\ref{sec:app_bootstrap}. Systematic uncertainties arise from various experimental factors, including track selection criteria, reconstruction efficiency corrections, residual pileup effects, centrality definition, and potential biases assessed via Monte Carlo (MC) consistency checks.

Systematic uncertainties from track selection, evaluated by varying selection criteria, are generally small, below 0.5\% for $\lr{[\pT]}$ and $k_3$, and ranging from 0.5\% to 4\% for $k_2$, $\gamma$, and $\Gamma$. Efficiency uncertainties, primarily from detector material simulation in \textsc{Geant}, can be up to 4\% for individual track efficiency~\cite{Aad:2019fgl}, propagating to uncertainties around 1\% for $\lr{[\pT]}$, 0.5\% for $k_2$, 2--2.5\% for $k_3$, 1--1.5\% for $\gamma$, and 1.5--2.5\% for $\Gamma$. Azimuthal efficiency variations are accounted for via track weights and found to have negligible impact. Residual pileup contributes uncertainties less than 0.5\% for all moments. Centrality definition uncertainties, assessed by varying Glauber model parameters, are relevant when binning results by centrality and are below 0.5\% in UCC.

Consistency checks using HIJING MC samples involve calculating moments using truth particles and comparing them to results obtained from reconstructed tracks with the same analysis procedure and corrections applied to real data~\cite{ATLAS:2019peb,ATLAS:2019dct}\footnote{While absolute $[\pT]$-moments in HIJING may differ from data, the MC check validates the analysis procedure's ability to reproduce moments from reconstructed tracks relative to truth level particles discussed in Section~\ref{sec:MCtruth}, accounting for effects like smearing of $[\pT]$ and $\Nch$.}. These checks show differences less than 0.25\% for $\lr{[\pT]}$ and $k_2$, and about 1.2\% for $k_3$, $\gamma$, and $\Gamma$.

Total systematic uncertainties are obtained by adding individual contributions in quadrature. For most observables, total systematic uncertainties are smaller than statistical uncertainties, except for $\lr{[\pT]}$ where they are comparable. The largest systematic contributions generally come from track selection. Across both Pb+Pb and Xe+Xe systems, total systematic uncertainties are typically below 1\% for $\lr{[\pT]}$, 2--4\% for $k_2$, 2--5\% for $k_3$, and 2--4\% for $\gamma$ and $\Gamma$. The uncertainty in the measured $\Nch$ itself, mainly due to tracking efficiency and fake track corrections, reaches up to 3\% in Pb+Pb UCC.  More details on the contribution of different systematic sources can be found in Appendix~\ref{sec:app_syst}.

\subsection{Results}
\subsubsection{Dependence on $\Nch$}
\begin{figure}[htbp]
\centering
\includegraphics[width=0.8\linewidth]{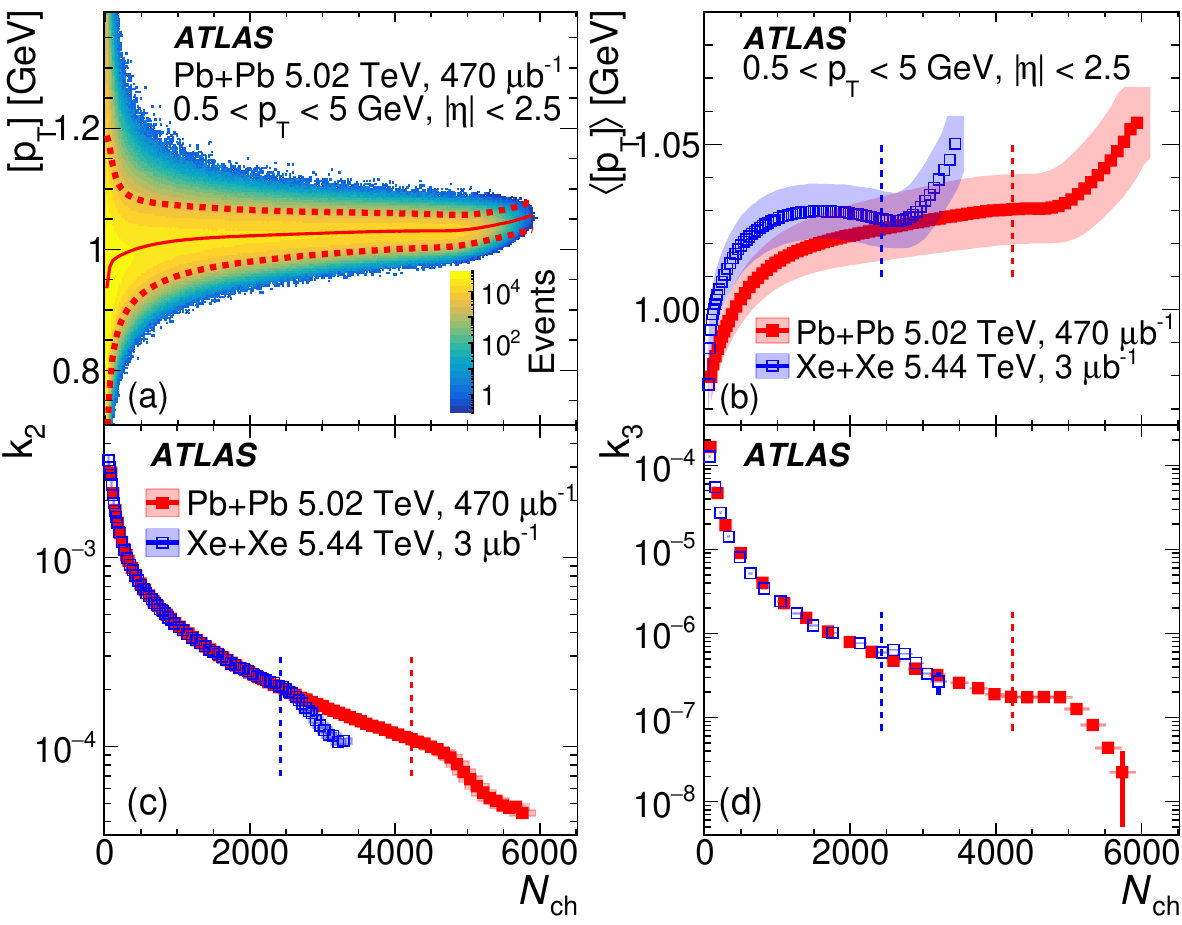}
\caption{Panel (a) depicts the 2D distribution of $[\pT]$ vs. $\Nch$ in Pb+Pb collisions, with the solid and dashed lines indicating the mean and two standard deviations, respectively. Panels (b), (c), and (d) show the $\Nch$ dependence of $\lr{[\pT]}$, $k_2$, and $k_3$, respectively, for both Pb+Pb and Xe+Xe collisions. Error bars represent statistical uncertainties and shaded boxes denote systematic uncertainties. Vertical dashed lines mark $\Nch$ values at 5\% centrality, 4230 for Pb+Pb, 2425 for Xe+Xe.}
\label{fig:fig1}
\end{figure}

Figure~\ref{fig:fig1}(a) shows the two-dimensional distribution of event-wise $[\pT]$ as a function of $\Nch$ for Pb+Pb collisions. The solid line traces the mean of $[\pT]$ at a given $\Nch$, while the dashed lines indicate the $\pm 2$ standard deviation contours, illustrating the width of the $P([\pT])$ distribution. With increasing $\Nch$, a clear increase in the mean $[\pT]$ and a decrease in the width of the distribution is observed. However, the width of the $P([\pT])$ shown in this panel has statistical contributions from a limited number of tracks in an event, which is taken care of by using moment definitions.

The moments of $P([\pT])$ are presented as a function of $\Nch$ for both Pb+Pb and Xe+Xe collisions in Figure~\ref{fig:fig1}(b)-(d). The mean transverse momentum $\lr{[\pT]}$ increases with $\Nch$, with the rise being steeper in peripheral collisions and becoming milder towards mid-centrality. The normalized variance $k_2$ and skewness $k_3$ both exhibit a power-law-like decrease with increasing $\Nch$ across most centrality ranges, broadly consistent with expectations from the independent superposition scenario. However, for UCCs, beyond $\Nch$ values marking 5\% centrality, all three observables show distinct deviations from these general trends: $\lr{[\pT]}$ displays a notable sharp increase, while $k_2$ and $k_3$ undergo a pronounced decrease.

\subsubsection{Scaling Behavior and UCC Non-monotonicity}

\begin{figure}[htbp]
\centering
\includegraphics[width=0.55\linewidth]{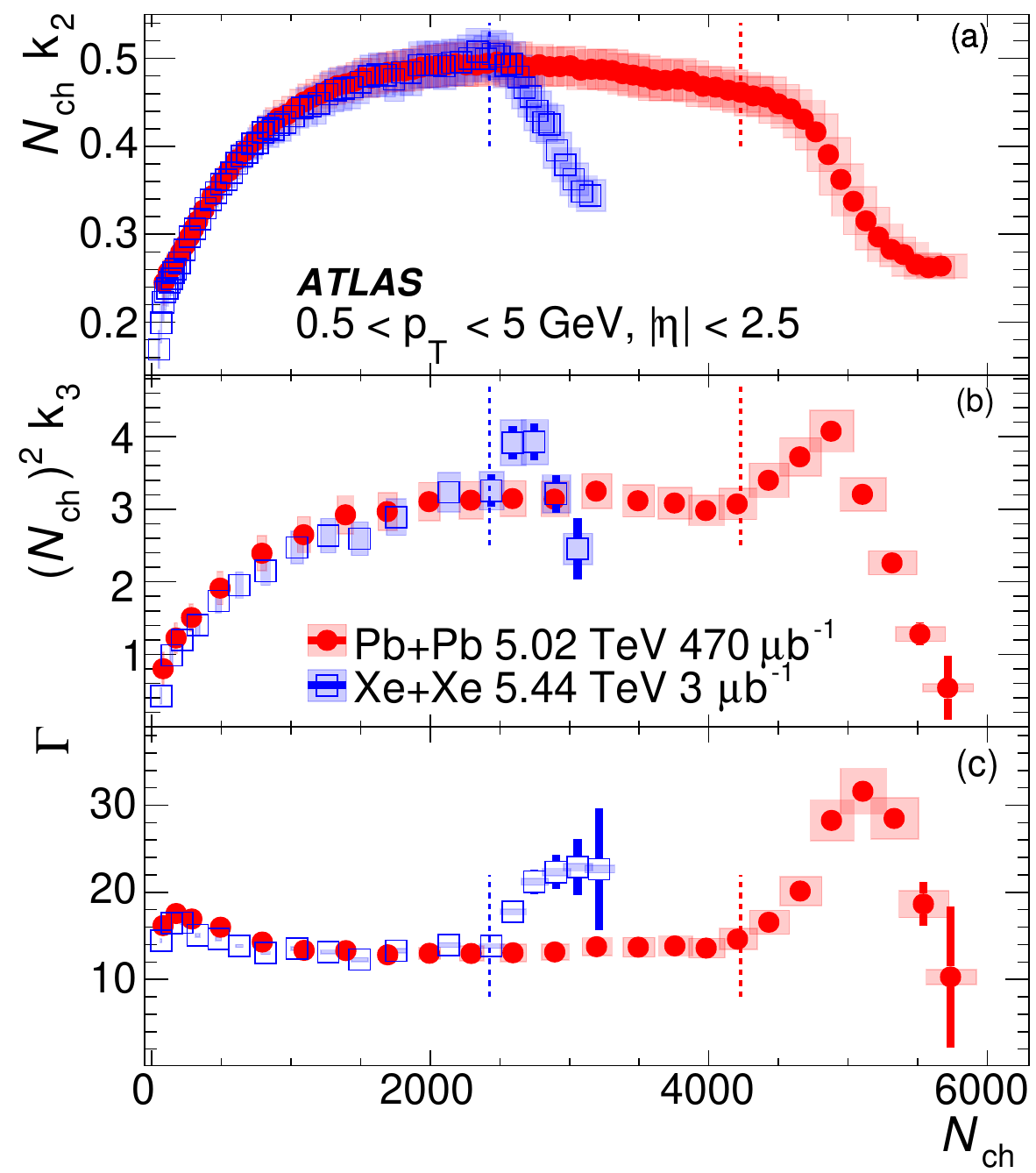}
\caption{The $\Nch$ dependence of (a) $\Nch k_{2}$, (b) $(\Nch)^{2} k_{3}$, and (c) $\Gamma$ in Pb+Pb and Xe+Xe collisions. These quantities are normalized to highlight deviations from simple power-law scaling. Error bars show statistical uncertainties; shaded boxes represent systematic uncertainties. Vertical dashed lines mark $\Nch$ values at 5\% centrality.}
\label{fig:fig2}
\end{figure}

To better visualize deviations from simple power-law scaling, Figure~\ref{fig:fig2} presents $\Nch k_2$ in panel (a), $(\Nch)^2 k_3$ in panel (b), and $\Gamma$ in panel (c) as a function of $\Nch$. In both systems, $\Nch k_2$ and $(\Nch)^2 k_3$ show an initial increase up to $\Nch \approx 1500$, consistent with the development of radial flow~\cite{Voloshin:2003ud}. Beyond this point, their variation becomes more gradual until the UCC region. The intensive skewness $\Gamma$ exhibits a slight decrease from peripheral to mid-central collisions before becoming relatively flat until the UCC region.

Within the UCC, a pronounced non-monotonic behavior is observed in all these scaled quantities. $\Nch k_2$ decreases, while $(\Nch)^2 k_3$ and $\Gamma$ show abrupt increases followed by sharp decreases. These non-monotonic features are interpreted as a consequence of the suppression of the initial nuclear overlap transverse size distribution, $P(R)$, as the impact parameter approaches zero and $R$ reaches its maximum value~\cite{Samanta:2023amp}. This suppression at large $R$ values initially introduces a positive skewness in the $P([\pT])$ distribution, which then diminishes as the variance of $P(R)$ is further reduced in the most central events.

\subsubsection{Extraction of Geometrical and Intrinsic Components}
\begin{figure}[htbp]
\centering
\includegraphics[width=0.9\linewidth]{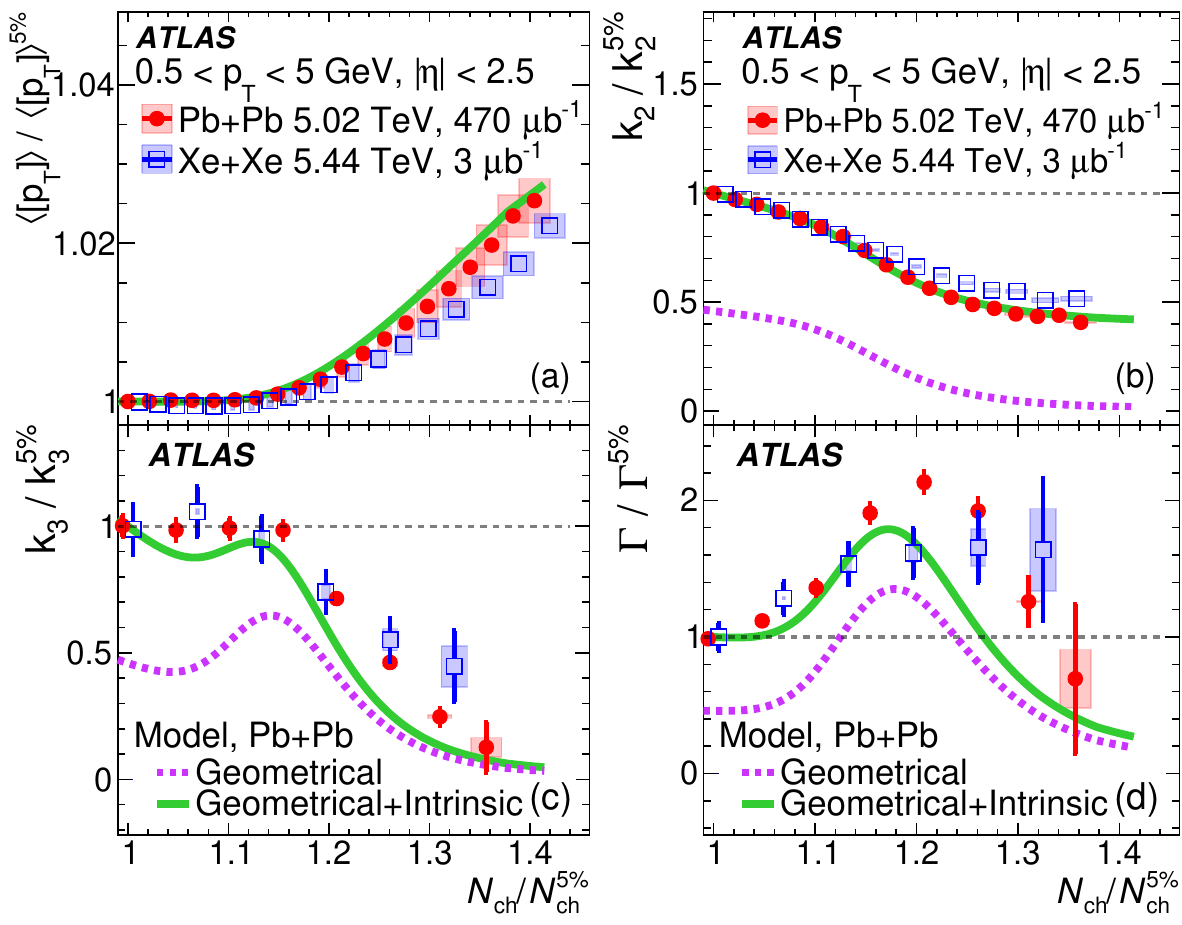}
\caption{The (a) $\lr{[\pT]}$, (b) $k_2$, (c) $k_3$, and (d) $\Gamma$, scaled by their respective values at 5\% centrality, as a function of $\Nch/N_\mathrm{ch}^\mathrm{5\%}$ for Pb+Pb and Xe+Xe collisions. Data are compared to predictions from Ref.~\cite{Samanta:2023kfk} and its estimated geometrical component (dashed lines). Error bars show statistical uncertainties; shaded boxes represent systematic uncertainties.}
\label{fig:fig3}
\end{figure}
\begin{figure}[htbp]
\centering
\includegraphics[width=0.85\linewidth]{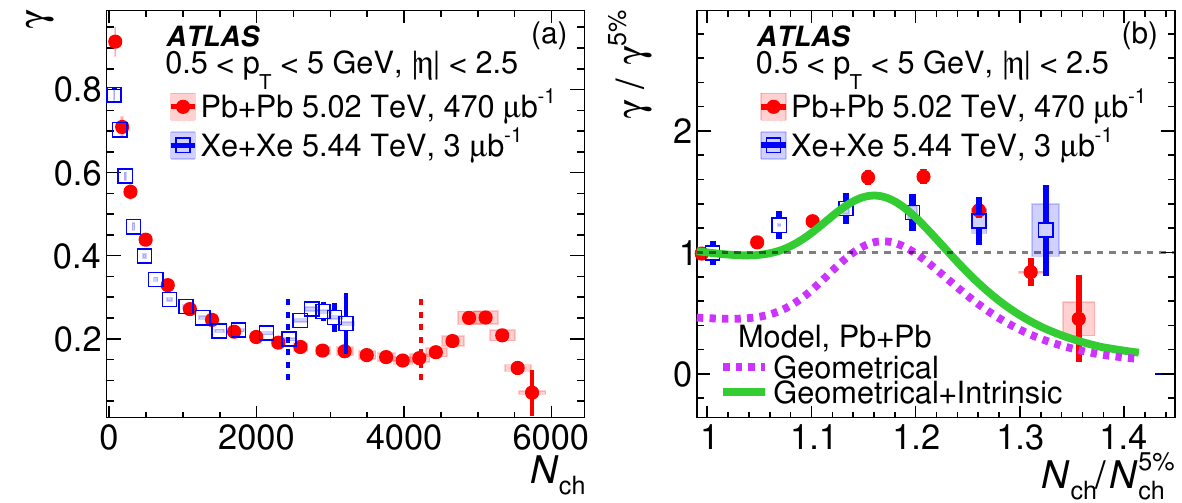}
\caption{(a) $\gamma$ as a function of $\Nch$ and (b) scaled-$\gamma$ as a function of $\Nch / N_\mathrm{ch}^\mathrm{5\%}$ in Pb+Pb and Xe+Xe collisions compared to predictions from Ref.~\cite{Samanta:2023kfk}. Error bars and shaded areas represent statistical and systematic uncertainties.}
\label{fig:app4}
\end{figure}

To highlight the non-monotonic trends in UCC and facilitate a direct comparison between collision systems, Figure~\ref{fig:fig3} shows $\lr{[\pT]}$, $k_2$, $k_3$, and $\Gamma$, each scaled by their values at 5\% centrality, plotted against normalized multiplicity $\Nch/N_\mathrm{ch}^\mathrm{5\%}$. Figure~\ref{fig:app4} presents the standard skewness $\gamma$ and scaled-$\gamma$ similarly. 

Qualitatively similar trends are observed across all scaled observables in both Pb+Pb and Xe+Xe systems. However, the magnitude of the variations in UCC is somewhat attenuated in the smaller Xe+Xe system. This difference is consistent with expectations, as the broader $\Nch$ distribution in Xe+Xe (due to its smaller size) results in a less effective selection of events with minimal geometric fluctuations at the highest multiplicities compared to Pb+Pb. This underscores the importance of system-size comparisons for understanding the nature of these fluctuations.

These results are compared to a recent model that describes $P([\pT])$ fluctuations using a 2D Gaussian distribution in $\Nch$ and impact parameter, attributing fluctuations at fixed $\Nch$ primarily to variations in impact parameter and transverse size $R$~\cite{Samanta:2023amp,Samanta:2023kfk}. Within this framework, the scaled observables are decomposed into geometrical and intrinsic components. 

The model broadly reproduces the increasing trend of scaled $\lr{[\pT]}$ and the decreasing trend of scaled $k_2$ observed in Pb+Pb data. However, it predicts a steeper decrease for scaled $k_3$, $\Gamma$, and $\gamma$ in Figures~\ref{fig:fig3} and~\ref{fig:app4}, with increasing $\Nch/N_\mathrm{ch}^\mathrm{5\%}$ compared to the data. The larger measured values for the scaled skewness observables suggest the presence of additional sources of skewness in $P([\pT])$ that are not fully captured by the current model implementation. The dashed lines in Figure~\ref{fig:fig3} illustrate the estimated contribution from the geometrical component according to the model, indicating that the UCC variation in $k_2$, $k_3$, $\Gamma$ and $\gamma$ can be attributed to geometrical fluctuations. This comparison highlights the model's potential for disentangling components while also pointing to areas for further theoretical refinement.

\subsubsection{Constraining $c^{2}_{s}$ using UCC}
\begin{figure}[htbp]
\centering
\includegraphics[width=0.75\linewidth]{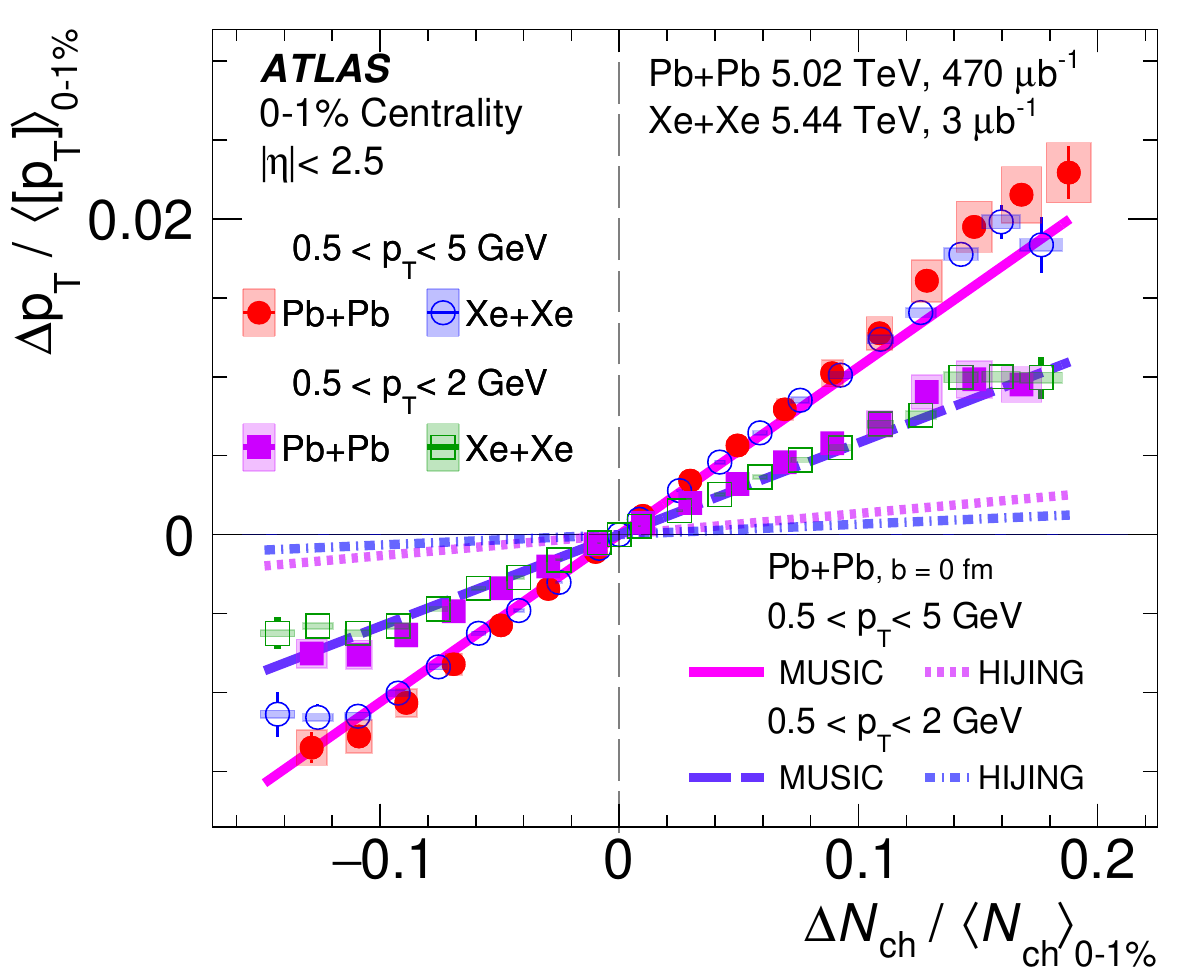}
\caption{Correlation between $\Delta\pT/\lr{[\pT]}_{0-1\%}$ and $\Delta\Nch/\lr{\Nch}_{0-1\%}$ in the 0--1\% most central Pb+Pb and Xe+Xe collisions for two $\pT$ ranges. Error bars show statistical uncertainties; shaded boxes represent systematic uncertainties. Data are compared to the MUSIC hydrodynamic model ($c_s^2\approx0.23$ at $T_{\mathrm{eff}}\approx 222$ MeV) and HIJING model at $b=0$~fm~\cite{Samanta:2023amp}.}
\label{fig:fig4}
\end{figure}

The observed non-monotonic behavior of the skewness measures, notably their sharp `bump'-like shape in the ultra-central collisions, provides evidence for the suppression of geometric fluctuation sources. These geometric sources become constrained as $b \to 0$, allowing for enhanced sensitivity to intrinsic fluctuations, which may originate from quantum variations in the initial energy deposition profiles or from thermal fluctuations within the QGP medium at a nearly fixed collision geometry.

To specifically investigate the correlation between $[\pT]$ and $\Nch$ in the 0--1\% most central events, we analyze $\Delta \pT/\lr{[\pT]}_{0-1\%}$ as a function of $\Delta\Nch/\lr{\Nch}_{0-1\%}$, where average values are taken over the 0--1\% centrality bin. Figure~\ref{fig:fig4} shows this correlation for two different $\pT$ ranges for both Pb+Pb and Xe+Xe collisions. A nearly linear relationship is observed in both systems, with consistent slopes between Pb+Pb and Xe+Xe. The improved consistency between the systems in this plot compared to Figure~\ref{fig:fig3} is attributed to the more stringent centrality selection on both axes for normalization. The slope of this correlation, however, exhibits a dependence on the selected $\pT$ range, indicating a kinematic sensitivity of the radial flow effects.

The data are compared to predictions from the HIJING model~\cite{Gyulassy:1994ew}, which lacks collective final-state interactions, and the state-of-the-art MUSIC hydrodynamic model~\cite{Schenke:2010nt}, which incorporates the full hydrodynamic evolution. HIJING underpredicts the measured slopes. In contrast, the MUSIC model successfully reproduces the slopes for Pb+Pb data across both $\pT$ ranges. This strong agreement with a hydrodynamic model in UCC where geometrical fluctuations are minimized, suggests that these measured slopes are a direct manifestation of the QGP's hydrodynamic response to fluctuations in energy density at a nearly fixed transverse size.

 The measured slopes in Figure~\ref{fig:fig4} provide an experimental estimate for $c_s^2(T_{\mathrm{eff}})$. The MUSIC model's description of the Pb+Pb data is achieved with $c_s^2 \approx 0.23$ at an effective temperature $T_{\mathrm{eff}} \approx 222$ MeV, consistent with some previous estimations~\cite{CMS:2024sgx}. However, the extraction of $c_s^2$ is known to be sensitive to various factors, including the kinematic selection criteria used for particles~\cite{Nijs:2023bzv,SoaresRocha:2024drz}. Crucially, since the slopes in Figure~\ref{fig:fig4} are primarily driven by the intrinsic component of $P([\pT])$ fluctuations in UCC, which is linked to the higher-order moments shown in Figure~\ref{fig:fig3}, any model aiming for a quantitative extraction of $c_s^2$ must simultaneously describe all these observables.

\subsubsection{Caveats in extraction of $c^2_s$}
The extraction of the sound speed squared ($c_s^2$) from $[\pT]$ fluctuations is subject to several dependencies. Experimentally, the inferred $c_s^2$ value can be sensitive to the event centrality definition (e.g., using charged particle multiplicity or forward energy) and the kinematic cuts applied to particles included in the $[\pT]$ calculation, such as the lower kinematic cut on $\pT$ of the particles \cite{Nijs:2023bzv, Gardim:2024zvi, SoaresRocha:2024drz}.  An alternate method to calculate the dependence of the extracted $c^{2}_{s}$ on the kinematic cuts on the $\pT$-range is further described in Appenidx~\ref{sec:chap5b_cs2}.

Furthermore, the relationship between measured observables and $c_s^2$ is inherently model-dependent, relying on the specific hydrodynamic framework used for interpretation. Different models or parameter choices can yield varying conclusions about the $c_s^2$ value that best fits the data \cite{Nijs:2023bzv, Gardim:2024zvi, SoaresRocha:2024drz}. Moreover, the assumption that $[\pT]$ fluctuations solely probe the thermodynamic $c_s^2$ may be an oversimplification.

\subsection{Non-Collective Baseline from HIJING}
\subsubsection{Motivation}
he HIJING model serves as a baseline for studying particle production and correlations in the absence of QGP-induced collective effects. This study~\cite{Bhatta:2021qfk} established a comprehensive framework for calculating $[\pT]$ fluctuations using HIJING simulations across various systems ($pp$, $p$+Pb, Xe+Xe, Pb+Pb) at $\sqn \sim$ 5 TeV.
\subsubsection{Methodology}
The methodology involved calculating $\pT$ cumulants as a function of $\Nch$. The formulae are described in details, previously in Eq.~\ref{eq:ck2} and in Appendix~\ref{sec:app_method}. Two methods were employed: the standard method considering all particles within $|\eta|<2.5$, and a two-subevent method dividing the event into rapidity-separated regions (e.g., $-2.5<\eta<-0.75$ and $0.75<\eta<2.5$). Comparing these methods allows for disentangling short-range correlations from longer-range ones. 

\subsubsection{Results}
\paragraph{$\Nch$ Scaling of Cumulants}
In large collision systems like Pb+Pb and Xe+Xe, $\pT$ cumulants $\lr{c_n}$ exhibit a clear power-law dependence on $\Nch$, approximately scaling as $\propto 1/\Nch^{(n-1)}$ for $n=2,3,4$. This is consistent with the independent superposition picture inherent in HIJING, where $\Nch$ serves as a proxy for the number of independent particle-producing sources. 

\begin{figure}[htbp]
    \centering
    \includegraphics[width=0.7\linewidth]{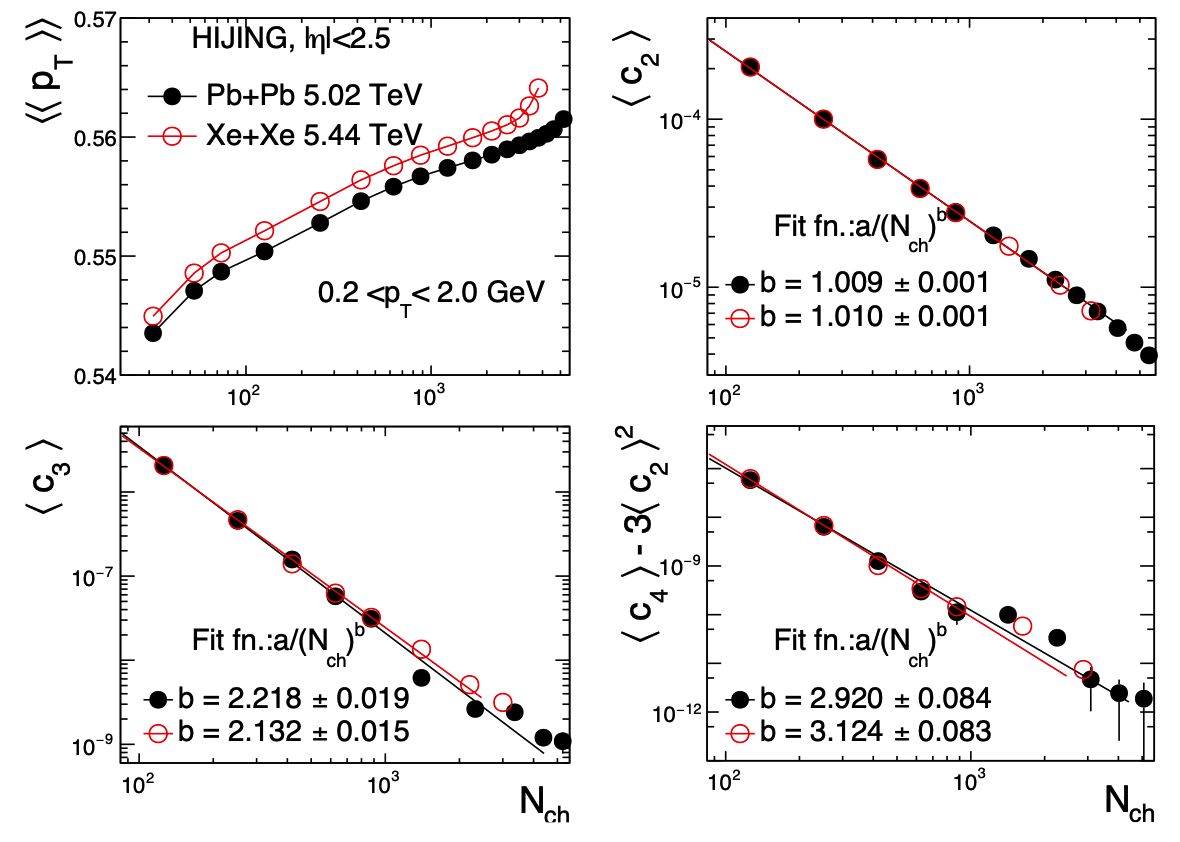}
    \caption{The $[\pT]$ cumulants (without normalization) for particles in $0.2 < \pT < 2 \text{ GeV}$ vs $\Nch$ in Pb+Pb and Xe+Xe collisions. Solid lines show fits to $a/(\Nch)^b$~\cite{Bhatta:2021qfk}. Here, $\lr{\lr{\pT}} = \lr{[\pT]}$.}
    \label{fig:ptfluchij_fig1}
\end{figure}

To quantify the dependence of the cumulants on $\Nch$, they were fitted with a power-law expression of the form: $a/(\Nch)^b$. Figure~\ref{fig:ptfluchij_fig1} illustrates this scaling, showing that the cumulants decrease in magnitude with increasing $\Nch$, well-described by the power-law fits. The fitted exponent $b$ is close to $n-1$ for $\lr{c_{n}}$ supporting the understanding that the $[\pT]$ fluctuations within HIJING originate from a simple independent source scenario.

\begin{figure}[htbp]
    \centering
    \includegraphics[width=1.0\linewidth]{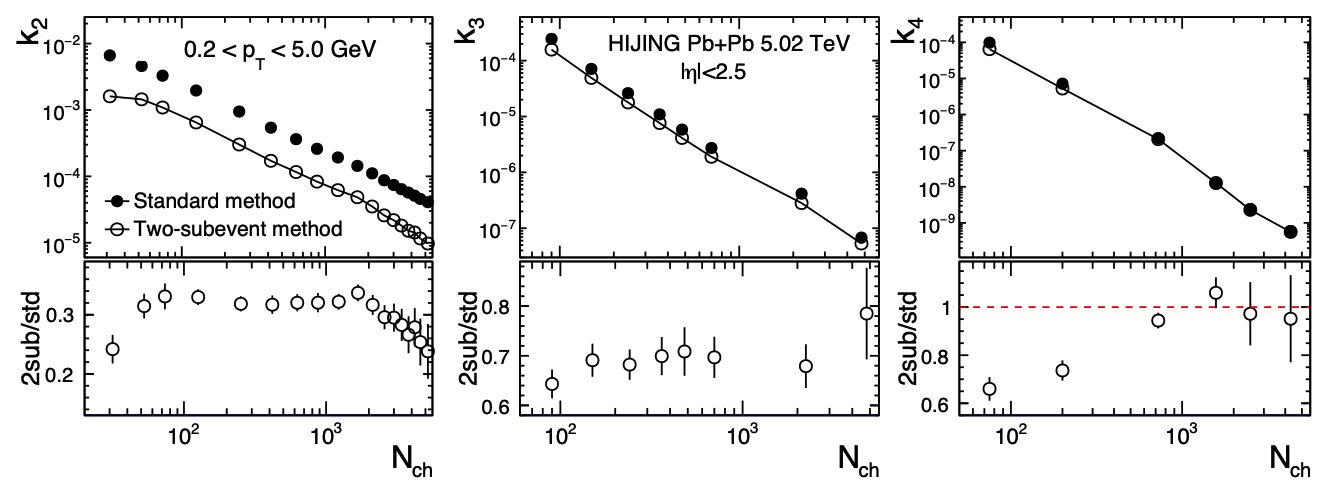}
    \caption{Variance (left), skewness ($k_3$, middle) and kurtosis ($k_4$, right) $\pT$ cumulants in Pb+Pb using the standard (solid points) and two-subevent method (open points) for particles in $0.2 < \pT < 5.0 \text{ GeV}$ vs $\Nch$~\cite{Bhatta:2021qfk}.}
    \label{fig:ptfluchij_fig2}
\end{figure}

\paragraph{Subevent Analysis and Correlation Length}
The two-subevent method is intended to suppress short-range correlations in multi-particle correlation measurements. Figure~\ref{fig:ptfluchij_fig2} compares the normalized cumulants $k_2$, $k_3$, and $k_4$ obtained using the standard and two-subevent methods as a function of the charged-particle multiplicity, $\Nch$. The variance ($k_2$) calculated using the two-subevent method is approximately three times smaller than that obtained with the standard method in Pb+Pb collisions. This suppression indicates that short-range correlations, likely originating from jet fragmentation and resonance decays, predominantly contribute to the total $p_T$ variance in the HIJING model.

In contrast, higher-order cumulants, $k_3$ and $k_4$, exhibit progressively smaller differences between the two methods as the order of correlation increases. This trend aligns with the expectation that short-range non-flow correlations primarily influence correlations involving fewer particles, and their contributions diminish with an increasing number of correlated particles.

\begin{figure}[htbp]
    \centering
    \includegraphics[width=0.8\linewidth]{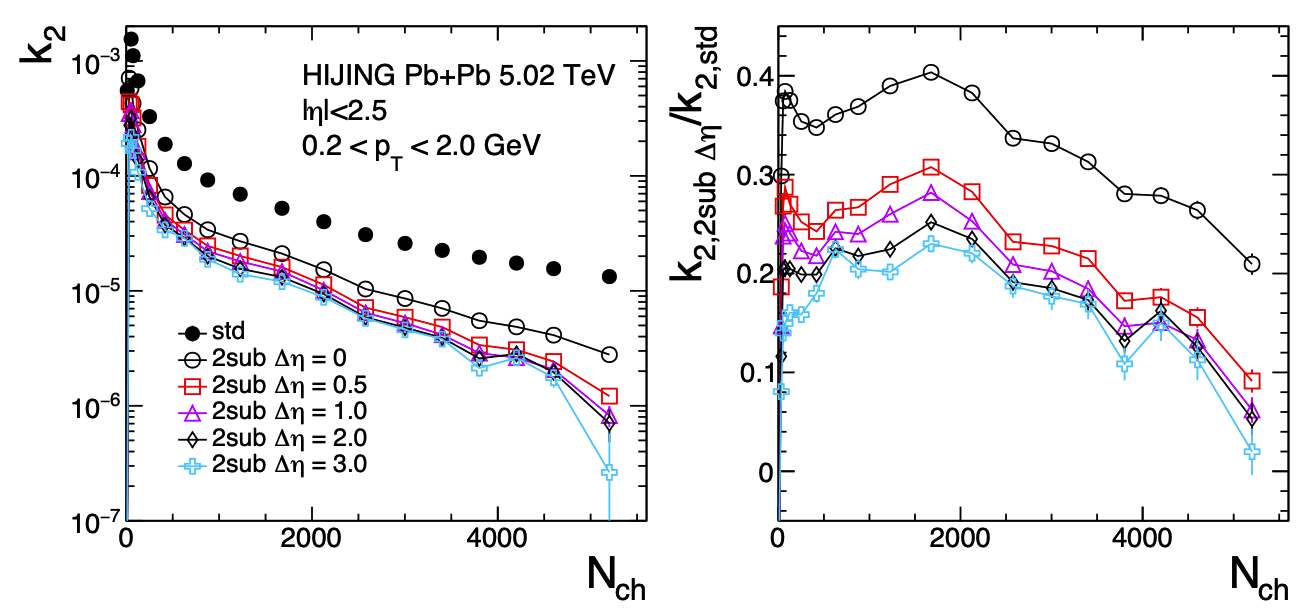}
    \caption{(Left) Variance obtained for standard and two-subevent methods with various rapidity separations vs $\Nch$ in Pb+Pb for $0.2 < \pT < 2 \text{ GeV}$. (Right) Ratio of results from two-subevent method to standard method~\cite{Bhatta:2021qfk}.}
    \label{fig:ptfluchij_fig3}
\end{figure}

Furthermore, the dependence of the variance ($k_2$) on the rapidity separation ($\Delta\eta$) between particles is analyzed in Figure~\ref{fig:ptfluchij_fig3}. As $\Delta\eta$ increases, $k_2$ decreases and saturates for rapidity separations greater than approximately $0.5$ units. This saturation implies that genuinely long-range correlations present in HIJING, which are mainly attributed to string dynamics, have a characteristic pseudorapidity correlation length of around $0.5$ units.

\paragraph{System Size Dependence}

\begin{figure}[htbp]
    \centering
    \includegraphics[width=0.8\linewidth]{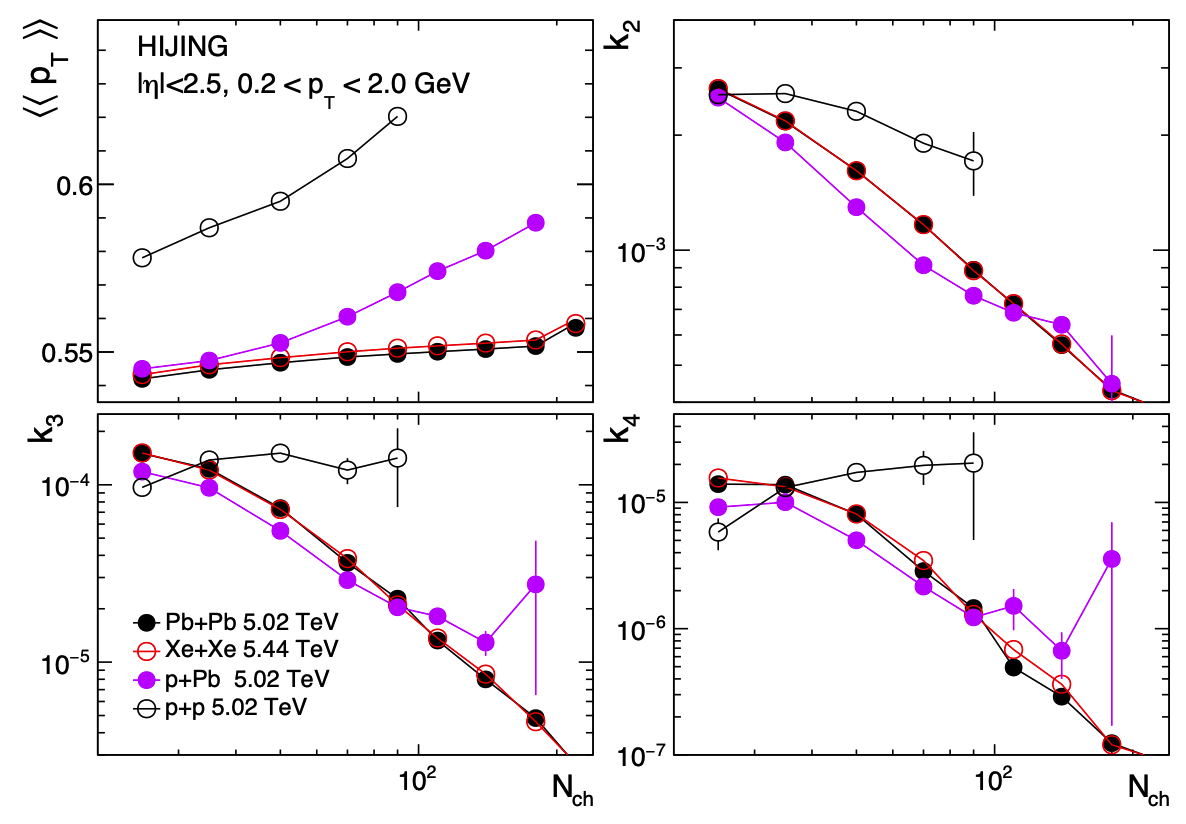}
    \caption{Comparison of $\pT$ cumulants between different collision systems as a function of $\Nch$~\cite{Bhatta:2021qfk}.}
    \label{fig:ptfluchij_fig4}
\end{figure}

Figure~\ref{fig:ptfluchij_fig4} compares cumulants across collision systems of varying sizes. The expectation from such a comparison is that fluctuations will decrease with increasing system size due to the larger number of contributing sources. It is observed that, although the average transverse momentum ($\langle [p_T] \rangle$) increases with system size, the higher-order cumulants, $k_2$, $k_3$, and $k_4$, at comparable charged-particle multiplicities ($\Nch$) are larger in $pp$ collisions compared to A+A collisions. 

This behavior arises from different dominant sources of fluctuations: per-source fluctuations, such as those from minijets, play a significant role in high-multiplicity $pp$ events. In contrast, fluctuations in the number of sources dominate in A+A collisions. Thus, distinct fluctuation behaviors are observed depending on the collision system type at a given multiplicity.

\subsection{Conclusion}
This chapter presented a comprehensive study of the event-by-event average transverse momentum distribution, $P([\pT])$, as a sensitive probe of initial-state geometry, its fluctuations, and the subsequent hydrodynamic evolution of the QGP. A key objective was to experimentally disentangle the contributions from geometrical and intrinsic fluctuations by measuring the mean, variance, and skewness of $P([\pT])$ across a broad $\Nch$ range in $^{208}$Pb+$^{208}$Pb and $^{129}$Xe+$^{129}$Xe collisions.

The variance and skewness are observed to follow an approximate power-law scaling with $\Nch$ consistent with the independent superposition scenario in most centrality ranges, but marked deviations are observed in ultra-central collisions. These behaviors are consistent with the suppression of geometrical fluctuations as the overlap region reaches its maximum size near-zero impact parameter.

The HIJING study provides a quantitative baseline for $\pT$ fluctuations in the absence of strong collective final-state effects. The observed $\Nch$ scaling in A+A collisions supports an independent source picture. The subevent method reveals that short-range correlations dominate the $\pT$ variance in HIJING, with a characteristic rapidity length scale around 0.5 units. System size comparisons highlight different fluctuation mechanisms in $pp$ vs A+A collisions at high $\Nch$. Deviations of experimental data from these HIJING predictions can serve as signatures for collective phenomena and guide the development of more complete theoretical models.

The measured correlation between $\Delta[\pT]/\lr{[\pT]}$ and $\Delta\Nch/\lr{\Nch}$ in UCC is well-reproduced by state-of-the-art hydrodynamic models, suggesting that this correlation is a direct manifestation of the QGP's hydrodynamic response to intrinsic fluctuations in UCC. The agreement is achieved with a $c_s^2 \approx 0.23$ at an effective temperature $T_{\mathrm{eff}} \approx 222$ MeV. The comparison between Pb+Pb and Xe+Xe collisions revealed that the centrality dependence of the moments in UCC is slightly less pronounced in the smaller Xe+Xe system, highlighting the sensitivity of these measurements to system size.

In conclusion, the investigation of $[\pT]$ fluctuations provides a robust and complementary approach for constraining initial-state fluctuations and refining the understanding of the hydrodynamic response. The observed non-monotonic behavior of higher moments and the correlation between $[\pT]$ and $\Nch$ in UCC, offer powerful constraints for theoretical models aiming to describe the QGP EoS. Future work should focus on developing models capable of simultaneously reproducing all measured moments across the full centrality range and their kinematic dependencies for a more precise extraction of QGP properties like the speed of sound.

\subsection{Outlook}
The results presented in this chapter open several avenues for future research. Further experimental measurements of higher-order moments of the P($[\pT]$) distribution with increased precision, esspecially in ultra-central collisions and across a wider range of system sizes and collision energies, will provide more stringent constraints for theoretical models. Detailed comparisons of these measurements with predictions from various hydrodynamic models, employing different initial-state descriptions and equations of state, are essential for a more robust extraction of QGP properties, including the speed of sound. Ultimately, the continued investigation of event-by-event fluctuations in observables like $[\pT]$ promises to refine understanding of the initial conditions of heavy-ion collisions and the fundamental properties of the Quark-Gluon Plasma.

\clearpage

\section{Constraining Nuclear Structure Using Elliptic and Radial Flow Correlation}
\label{sec:chap6_vnpt}

Building upon the established collective nature of radial flow fluctuations discussed in Chapter~\ref{sec:chap4_v0pt} and their sensitivity to initial geometrical fluctuations discussed in Chapter~\ref{sec:chap5_ptfluc}, this chapter focuses on leveraging multi-particle correlations between anisotropic flow ($v_n$) and the event-wise mean transverse momentum ($[\pT]$) in the final state to constrain the geometry of colliding nuclei and initial-state conditions. 

The initial-state correlations between eccentricity, $\varepsilon_{n}^{2}$, and the inverse of the system size, $1/R_{\perp}$, survive the collective expansion of the system and, as a collective response, leave their imprints on final-state $v_n$-$[\pT]$ correlations. As will be discussed later in this chapter, these $v_n$-$[\pT]$ correlations are highly sensitive to: (1) Initial-state conditions, such as the nucleon width, and (2) Initial momentum anisotropy. Thus, experimental measurement of $v_n$-$[\pT]$ correlations can provide strong constraints on the initial conditions of HIC.

Beyond initial conditions, relativistic HIC also offers a unique femtoscopic lens onto the structure of the colliding nuclei themselves~\cite{STAR:2024wgy}. This chapter leverages this potential by employing event-by-event correlations between $v_n$ and $[\pT]$, as a sensitive probe of nuclear deformation parameters. By comparing such correlations in collisions of approximately spherical $^{208}\text{Pb}$ nuclei with those involving $^{129}\text{Xe}$, this work aims to extract quantitative information about the shape of $^{129}\text{Xe}$. Of particular interest is the constraint on its quadrupole deformation parameter $\beta_2$ and, notably, its triaxiality parameter $\gamma_{Xe}$, the latter being especially challenging to determine for odd-mass nuclei via traditional low-energy spectroscopic methods.

\subsection{Theoretical Background}

\subsubsection{Nuclear Structure}

Atomic nuclei are not always spherical~\cite{Moller:2015fba,Bohr:1957ht}. Nuclei with partially filled nucleon shells often deviate from a spherical shape due to intrinsic quadrupole moment. The low-lying energy structure of such deformed nuclei can be described by collective rotational bands, where energy levels follow a rotational pattern:
$$E(I)\approx \frac{\hbar^{2}}{2\mathcal{I}} I(I+1),$$
where $I$ represents the angular momentum and $\mathcal{I}$ the nuclear moment of inertia. This pattern is particularly pronounced in even–even nuclei, where rotational levels typically appear with $I=0,2,4,\dots$, corresponding to collective rotational excitations. A key experimental indicator of this behavior is the ratio of the energy of the first $4^{+}$ excited state to that of the first $2^{+}$ state, $E(4^{+})/E(2^{+})$. For an ideal rigid rotor, this ratio is $10/3\approx3.33$, considerably higher than the value of approximately $2.0$ expected for vibrational excitations in spherical nuclei.

\begin{figure}[htbp]
\centering
\includegraphics[width=0.48\linewidth]{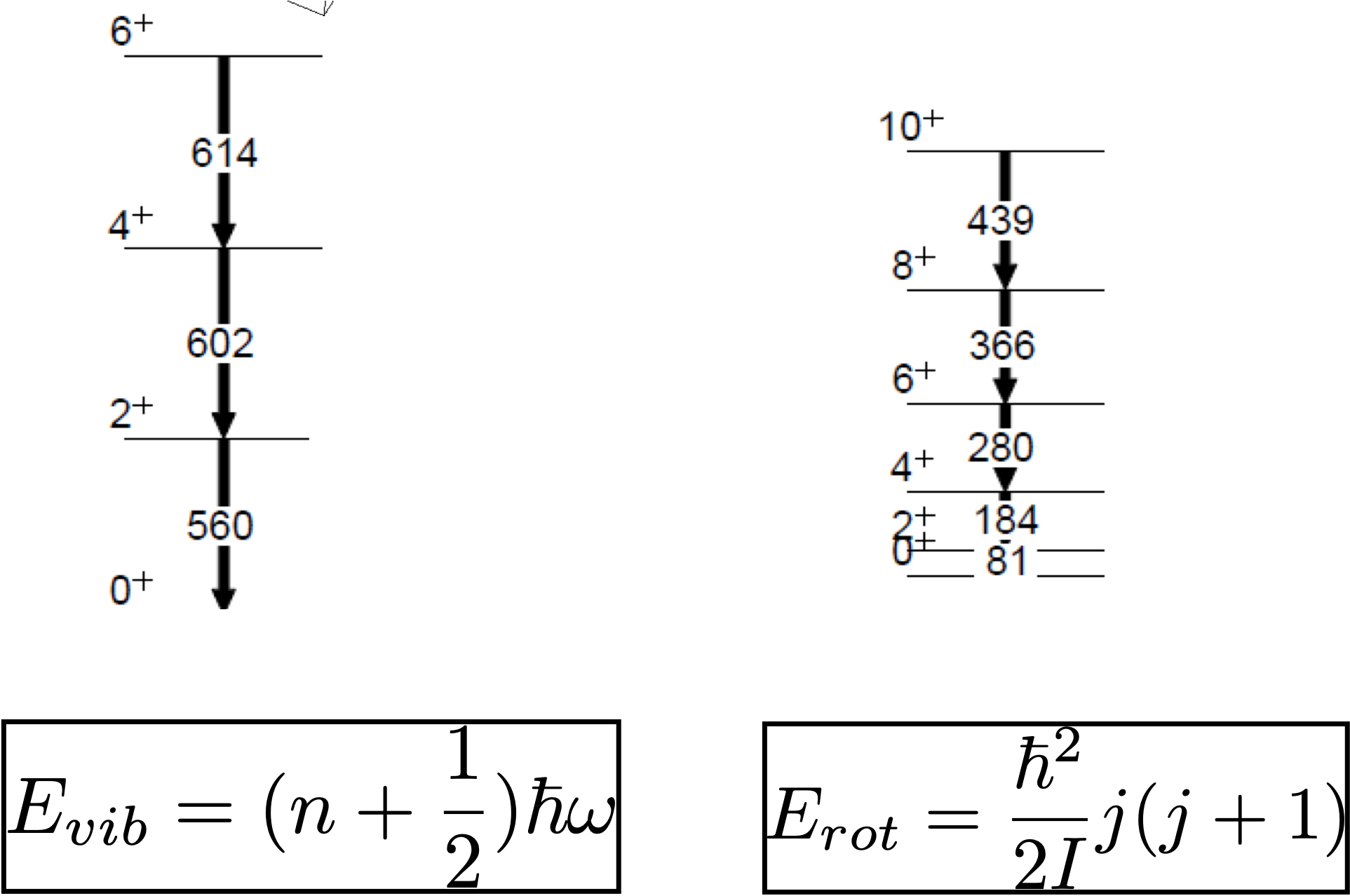}
\includegraphics[width=0.48\linewidth]{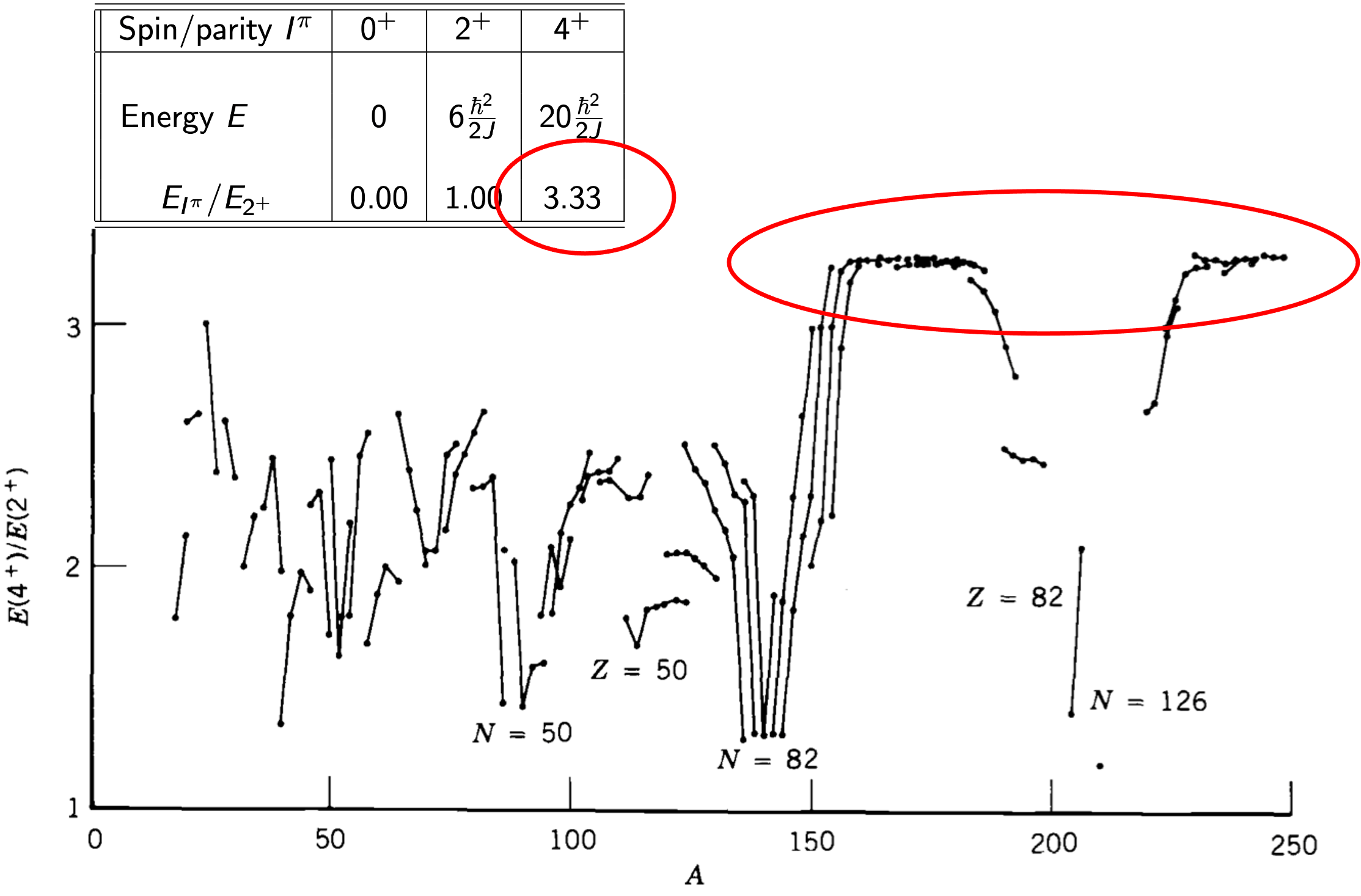}
\caption{(Left) Typical energy level schemes for rotational and vibrational states. (Right) The ratio $E(4^{+})/E(2^{+})$ vs. the mass number of nuclei. Red circles denote nuclei with a high $E(4^{+})/E(2^{+})$ ratio, indicating strong rotational behavior and deformation.}
\label{fig:enter-label}
\end{figure}

The degree of non-sphericity is quantified by the quadrupole deformation parameter, $\beta$, which is directly related to the intrinsic quadrupole moment $Q_{0}$, a measure of the nuclear density distribution. For an axially symmetric nucleus, the nuclear radius $R$ as a function of the angle $\theta'$ (relative to the symmetry axis) can be expressed using spherical harmonics $Y_{\ell m}$~\cite{Jia:2021qyu}:
\begin{equation}
R(\theta') = R_{0}\bigl(1 + \beta\,Y_{20}(\theta')\bigr).
\end{equation}
A positive $\beta$ indicates a prolate (football-like) deformation, while a negative value signifies an oblate (disc-like) deformation. This inherent deformation, probed through energy level structures and spectroscopic signatures, is important for understanding nuclear structure and influences high-energy nuclear collisions by affecting the initial geometric anisotropies.

\begin{figure}[htbp]
\centering
\includegraphics[width=1.0\linewidth]{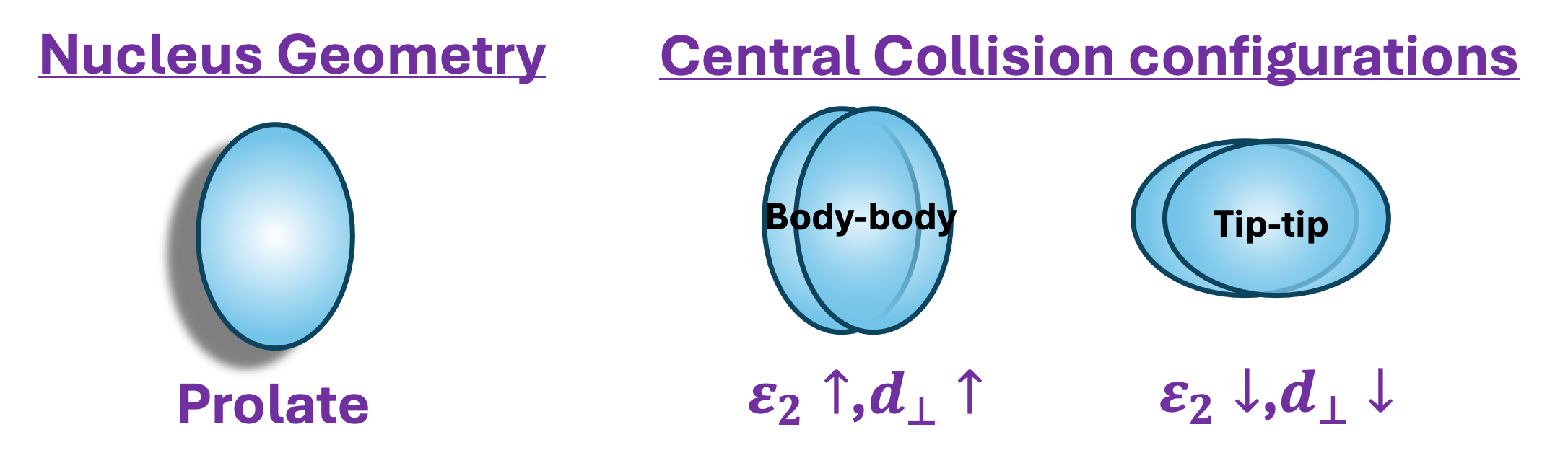}
\caption{Cartoon of two extreme orientations for collisions of two prolately deformed nuclei. $\varepsilon_{2}$ is ellipticity whereas $d_{\perp}=1/R_{\perp}$ is the inverse transverse size.}
\label{fig:prolatecoll}
\end{figure}

In collisions involving deformed nuclei, fluctuations in the overlap area size and eccentricity are considerably larger than in collisions of spherical nuclei. Consider, for instance, a prolately deformed nucleus. While a central collision ($b=0$) of spherical nuclei results in a circular overlap, the overlap characteristics for a prolately deformed nucleus vary with its orientation. A ``tip–tip'' configuration (collision along the deformation axis) yields a smaller, more circular overlap compared to spherical nuclei, leading to a smaller effective overlap radius $R_{\mathrm{overlap}}$ and ellipticity $\varepsilon_{2}$. Conversely, a ``body–body'' configuration (collision perpendicular to the deformation axis) results in a larger overlap area, yielding increased $R_{\mathrm{overlap}}$ and $\varepsilon_{2}$ compared to a central collision of spherical nuclei, as depicted in Figure~\ref{fig:prolatecoll}.

\subsubsection{Effect of Nuclear Deformation on $\rho(v^2_{2},[\pT])$}
The geometry of collisions involving deformed nuclei is orientation-dependent, leading to specific correlations that are not typically observed in collisions of spherical nuclei. For a given centrality, a “tip–tip” orientation (resulting in a smaller, rounder overlap) leads to a smaller initial eccentricity $\varepsilon_{2}$ and consequently a smaller anisotropic flow $v_{2}$. However, this configuration also results in a larger event-wise average transverse momentum, $\langle p_{T}\rangle_{\text{evt}}$, due to stronger pressure gradients arising from the smaller system size.

Conversely, a “body–body” orientation (resulting in a larger, more elongated overlap) leads to a larger $\varepsilon_{2}$ and thus a larger $v_{2}$, but a smaller $\langle p_{T}\rangle_{\text{evt}}$. This interplay between the collision geometry and the resulting flow and momentum is expected to generate a negative correlation between $v_{2}$ (or $v_2^2$) and $\langle p_{T}\rangle_{\text{evt}}$ in the final state for such deformed systems.

To quantify these correlations, the Pearson correlation coefficient, $\rho_{n}$, has been proposed~\cite{Bozek:2016yoj}:
\begin{equation} \label{eq:rho}
\rho_n
= \frac{\bigl\langle\bigl\langle v_n^2\,\delta \pT \bigr\rangle\bigr\rangle}
{\sqrt{\bigl(\langle v_n^4\rangle - \langle v_n^2\rangle^2\bigr)}\,
\sqrt{\bigl\langle\langle (\delta \pT)^2\rangle\bigr\rangle}},
\end{equation}
where $\delta p_{T}=p_{T}-\langle p_{T}\rangle_{\rm evt}$ and $\langle p_{T}\rangle_{\rm evt}$ is the average transverse momentum of particles in a given event. The notation $\langle\langle\cdot\rangle\rangle$ denotes averaging over all particle pairs (or triplets, depending on the specific observable definition) within events of comparable particle multiplicity, while $\langle\cdot\rangle$ denotes averaging over all such events.

\begin{figure}[htbp]
\centering
\includegraphics[width=0.45\linewidth]{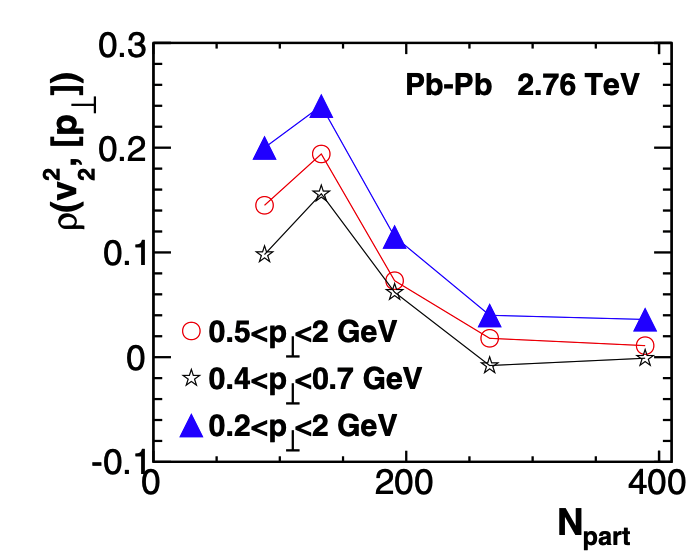}
\includegraphics[width=0.45\linewidth]{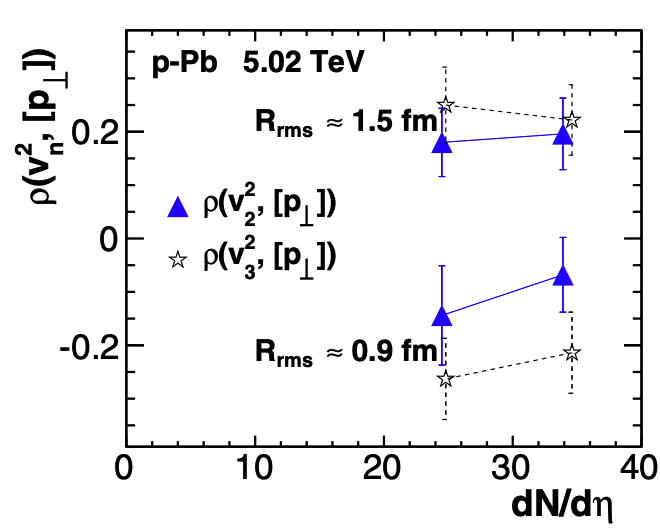}
\caption{$v_{2}$-$\langle p_{T}\rangle_{\text{evt}}$ correlation (denoted $\rho_2$) for Pb+Pb collisions. The stars denote the Pearson coefficient, the circles denote the correlation coefficient without self-correlations, and the triangles denote calculations with oversampled events~\cite{Bozek:2016yoj}.}
\label{fig:Bozek}
\end{figure}

Initial calculations for collisions of spherical nuclei, such as Pb+Pb, predicted that $\rho_{n}$ (specifically for $n=2$) would be large and positive in mid-central collisions, approaching zero in peripheral collisions, as shown in Figure~\ref{fig:Bozek}. This positive correlation was attributed to overall system size fluctuations linked to centrality variations.

Recent studies have further highlighted the sensitivity of $\rho_{n}$ in central collisions to the detailed shape of atomic nuclei, including triaxiality~\cite{Giacalone:2019pca,Giacalone:2020awm,Jia:2021tzt,Jia:2021wbq,Bally:2021qys,Jia:2021qyu}. Most nuclei exhibit some degree of deformation into an ellipsoidal shape, which can be described by the nuclear surface equation~\cite{Bohr:1957ht}:
\begin{equation}\label{eq:R_triaxial}
R(\theta,\phi)
= R_{0}\Bigl(1 + \beta\bigl[\cos\gamma\,Y_{2,0}(\theta,\phi)
+ \tfrac{\sin\gamma}{\sqrt{2}}\bigl(Y_{2,2}(\theta,\phi)+Y_{2,-2}(\theta,\phi)\bigr)\bigr]\Bigr),
\end{equation}
where $R_{0}$ is the average nuclear radius, $Y_{\ell m}$ are spherical harmonics, and $\beta$ and $\gamma$ are the quadrupole deformation parameters. The parameter $\beta$ (often denoted $\beta_2$ for quadrupole deformation) quantifies the magnitude of the deformation (typically 0.1–0.4~\cite{Moller:2015fba}), while the angle $\gamma$ (ranging from $0^\circ$ to $60^\circ$) describes the triaxiality, representing the relative lengths of the three semi-axes ($r_{1},r_{2},r_{3}$) of the ellipsoid. Specifically, $\gamma=0^\circ$ corresponds to a prolate shape ($r_{1}=r_{2}<r_{3}$, with $r_3$ as the symmetry axis), $\gamma=60^\circ$ to an oblate shape ($r_{1}<r_{2}=r_{3}$, with $r_1$ as the symmetry axis), and intermediate values like $\gamma=30^\circ$ to a maximally triaxial shape ($2r_{2}\approx r_{1}+r_{3}$).

Traditionally, nuclear shapes are inferred from low-energy spectroscopic measurements, which determine $(\beta,\gamma)$ for even–even nuclei (e.g., deformed isotopes of Samarium or Uranium, while $^{208}$Pb is spherical)~\cite{Raman:2001nnq}. The shapes of odd-mass nuclei, such as $^{129}$Xe, are typically calculated using nuclear structure models tuned to data from even–even nuclei. Therefore, flow measurements in high-energy HIC offer a novel tool to probe nuclear shapes, particularly for odd-mass nuclei.

\begin{figure}[htbp]
\centering
\includegraphics[width=1.0\linewidth]{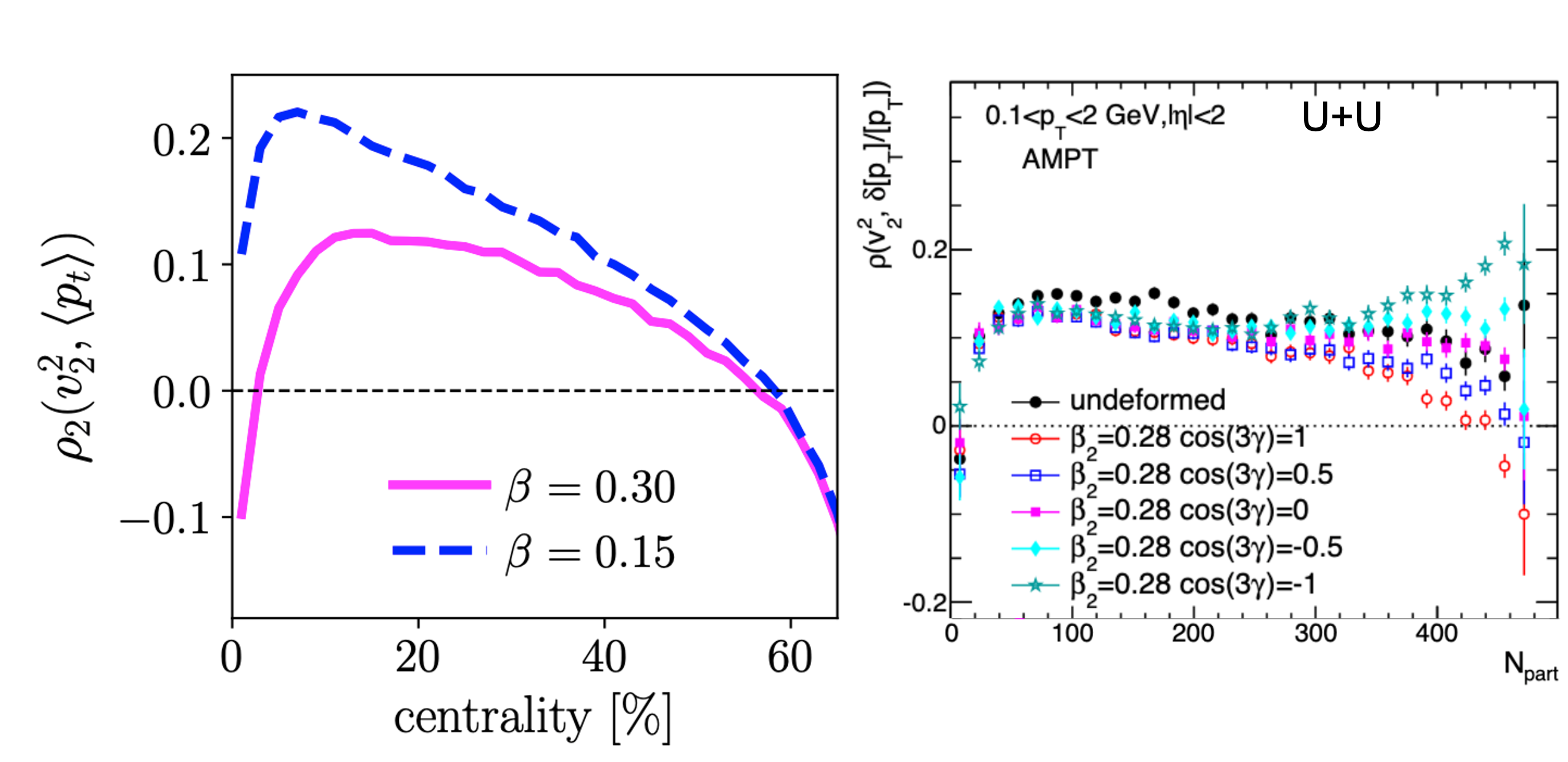}
\caption{(Left) Influence of increasing quadrupole deformation $\beta_{2}$ on $\rho_{2}$~\cite{PhysRevC.102.024901}. (Right) Influence of triaxiality $\gamma$ on $\rho_{2}$~\cite{Jia:2021qyu} in model simulated U+U collisions at $\sqn$=200 GeV.}
\label{fig:rhogamma}
\end{figure}

Furthermore, both the magnitude of quadrupole deformation $\beta_{2}$ and the triaxiality parameter $\gamma$ are expected to influence $\rho_{2}$ (the correlation coefficient for $n=2$). This provides a valuable means to constrain even these higher-order deformations in nuclei like $^{129}$Xe, which are otherwise challenging to extract using traditional spectroscopic methods. Figure~\ref{fig:rhogamma} presents model studies demonstrating the sensitivity of $\rho_{2}$ to $\beta_{2}$ and $\gamma$. The left panel shows that increasing $\beta_{2}$ of the colliding deformed nucleus leads to an increase in $\rho_{2}$, consistent with the parametric form discussed later in the text. The right panel illustrates that transitioning from a prolately deformed nucleus ($\gamma<30^\circ$) to an oblate one ($\gamma>30^\circ$) decreases $\rho_{2}$ in the most central collisions.

Therefore, precise measurements of $\rho_{n}$ are crucial for constraining:
\begin{itemize}
    \item The nuclear structure of deformed nuclei, including odd-mass nuclei like $^{129}$Xe.
    \item Refining initial conditions for HIC models, such as the nucleon width.
    \item Addressing the long-standing question of experimental evidence for initial momentum anisotropy.
\end{itemize}

\begin{figure}[htbp]
\centering
\includegraphics[width=0.32\linewidth]{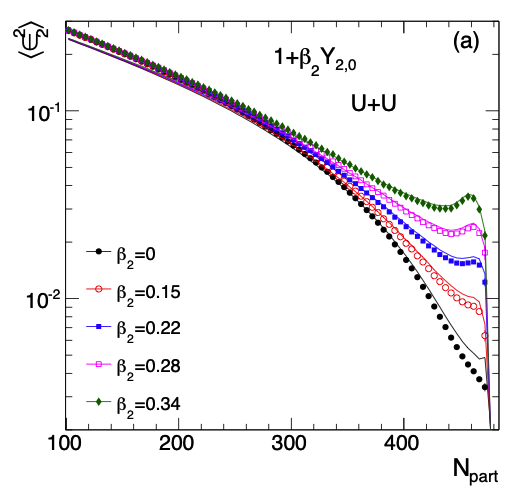}
\includegraphics[width=0.32\linewidth]{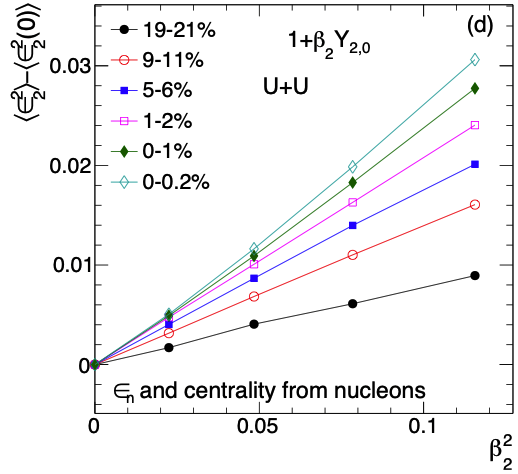}
\includegraphics[width=0.32\linewidth]{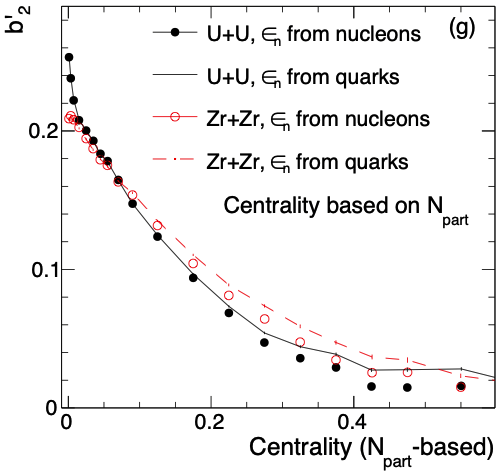}
\caption{(Left) $\langle\varepsilon_{2}^{2}\rangle$ in collisions of U+U nuclei with different quadrupole deformation parameters $\beta_{2}$. (Center) Excess $\langle\varepsilon_{2}^{2}\rangle$ with respect to the spherical baseline, $\langle\varepsilon_{2}^{2}(0)\rangle$, vs. $\beta_{2}^2$, showing linear correlation and validating the form in Eq.~\ref{eq:form} for $v_2^2$. (Right) Value of extracted coefficient $b_{2}'$ vs. centrality for the same system, using a form similar to Eq.~\ref{eq:form} for $\rho_2$.}
\label{fig:jiaGlauber}
\end{figure}

Recent model studies suggest that $v_{2}^{2}$ and $\rho_{2}$ follow simple parametric forms~\cite{Giacalone:2018apa,Jia:2021tzt,Jia:2021qyu}:
\begin{equation}\label{eq:form}
v_2^2 \approx a + b\,\beta^2,
\quad
\rho_2 \approx a' + b'\,\cos(3\gamma)\,\beta^3.
\end{equation}
The parameters $a$ and $a'$ represent the baseline values for collisions of spherical nuclei ($\beta=0$) and are smallest in central collisions. The parameters $b$ and $b'$ are nearly independent of centrality. Consequently, the impact of nuclear deformation (characterized by $\beta$ and $\gamma$) on $v_{2}^{2}$ and $\rho_{2}$ is expected to be most pronounced in central collisions. This is further illustrated by simulations of deformed U+U collisions using a Glauber model, as shown in Figure~\ref{fig:jiaGlauber}. These simulations demonstrate that as the quadrupole deformation $\beta$ (or $\beta_{2}$) increases, $\langle\varepsilon_{2}^{2}\rangle$ clearly increases, with the largest increase observed in the most central collisions. 

The middle panel of Figure~\ref{fig:jiaGlauber} shows that the excess $\varepsilon_{2}^{2}$ relative to the spherical baseline, $\varepsilon_{2}^{2}(0)$, is linearly correlated with $\beta_{2}^2$, consistent with Eq.~\ref{eq:form}. Lastly, the extracted coefficient $b'$ for $n=2$ as a function of centrality shows a slow rise towards the most central collisions. This indicates that nuclear deformation is expected to maximally influence the measured flow coefficients and their correlations in the most central collisions, providing an experimental avenue for its extraction.

\subsubsection{Effect of initial conditions}
Subsequent model investigations have revealed the sensitivity of $\rho_{n}$ to several factors beyond just system size and deformation. In central and mid-central collisions, $\rho_{n}$ is generally expected to be positive due to centrality fluctuations: larger systems (corresponding to higher centrality, i.e., more participating nucleons) tend to exhibit both larger $v_{n}$ and larger $\langle p_{T}\rangle_{\text{evt}}$~\cite{Schenke:2020uqq}.

\begin{figure}[htbp]
\centering
\includegraphics[width=0.6\linewidth]{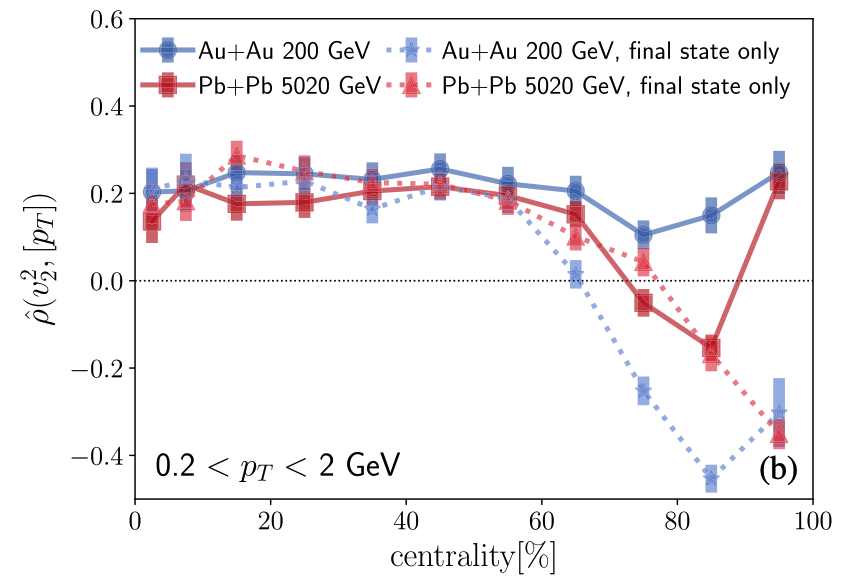}
\caption{Effect of initial momentum anisotropy on $v_{2}$-$\langle p_{T}\rangle_{\text{evt}}$ correlations within IP-Glasma+MUSIC+URQMD model simulations for Pb+Pb collisions at 5.02~TeV and Au+Au collisions at 200~GeV~\cite{Giacalone:2020byk}.}
\label{fig:imaschenke}
\end{figure}

\begin{figure}[htbp]
\centering
\includegraphics[width=0.8\linewidth]{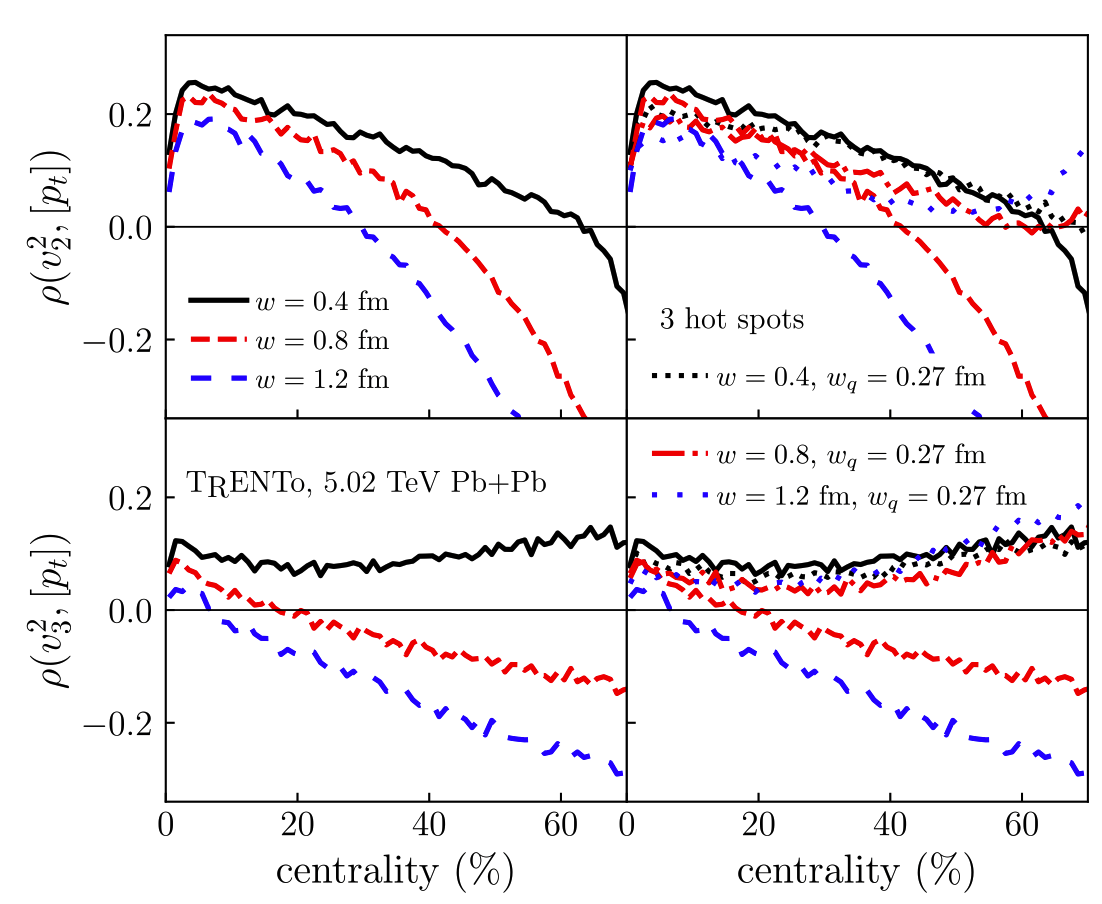}
\caption{$v_{2}$-$\langle p_{T}\rangle_{\text{evt}}$ correlation for Pb+Pb collisions in the TRENTo model with varying nucleon width parameter, $w$, for smooth nucleons and nucleons with 3 hotspots with constituent quark width $w_{q}$~\cite{Giacalone:2021clp}.}
\label{fig:nucleonsize}
\end{figure}

Furthermore, the presence of initial momentum anisotropy can cause $\rho_{n}$ to become positive in the most peripheral collisions, as illustrated in Figure~\ref{fig:imaschenke}~\cite{Giacalone:2020byk}. The root-mean-square size of the nucleon also influences the behavior of $\rho_{n}$ in peripheral collisions; for instance, a larger nucleon size tends to decrease the value of $\rho_{n}$, as shown in Figure~\ref{fig:nucleonsize}~\cite{Giacalone:2021clp}.

\subsection{Previous Studies}
\begin{figure}[htbp]
\centering
\includegraphics[width=1.0\linewidth]{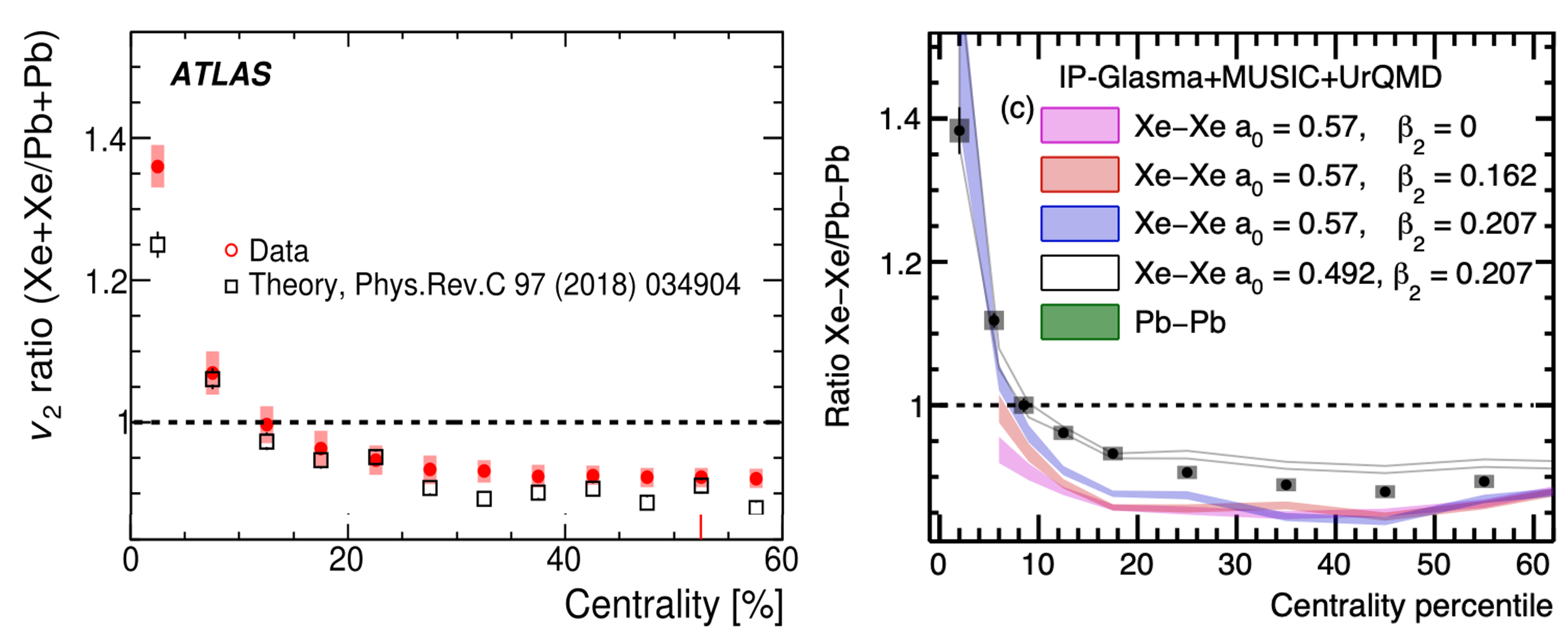}
\caption{$v_{2}\{2\}$ ratio between Xe+Xe collisions at 5.44 TeV and Pb+Pb collisions at 5.02 TeV as a function of centrality from ATLAS (Left) and ALICE (Right) collaborations~\cite{ATLAS:2019dct,ALICE:2024nqd}. While the ATLAS measurement is compared to Trento model calculations~\cite{Giacalone:2017dud}, the ALICE measurement is compared to IP-Glasma+MUSIC+URQMD model calculations~\cite{Schenke:2020mbo}.}
\label{fig:aliceatlasv2}
\end{figure}

A quadrupole deformation for $^{129}$Xe, $\beta_{2,\rm Xe}\approx0.16\!-\!0.2$, was inferred from the enhanced ratio $v_{2,\rm Xe}/v_{2,\rm Pb}$ observed in central collisions by both ALICE and ATLAS~\cite{ALICE:2024nqd,ATLAS:2019dct}. Figure~\ref{fig:aliceatlasv2} illustrates that incorporating this large $\beta_{2,\rm Xe}$ in HIC models is essential for describing the $v_{2}$ ratio between deformed $^{129}$Xe and spherical $^{208}$Pb nuclei.

ATLAS previously published a measurement of the $v_n$-$[\pT]$ correlation coefficient, $\rho_{n}$, for $n=2,3,$ and 4 in Pb+Pb collisions at $\sqrt{s_{NN}}=5.02$~TeV~\cite{Aad:2019fgl}. This was followed by similar measurements from ALICE in Pb+Pb and Xe+Xe collisions~\cite{ALICE:2021gxt}. These results generally show positive correlations for all harmonics, except in the peripheral region where $\rho_{2}$ becomes negative. These behaviors have been qualitatively reproduced by recent initial state and hydrodynamic model calculations~\cite{Schenke:2020uqq,Giacalone:2020dln}.

This chapter details a study of the centrality and system-size dependence of $\rho_{n}$ in $^{129}\mathrm{Xe}+^{129}\mathrm{Xe}$ and $^{208}\mathrm{Pb}+^{208}\mathrm{Pb}$ collisions. The goal is to elucidate the effects of initial-state conditions and nuclear deformation. Measurements are performed across several ranges of transverse momentum ($p_{T}$) and pseudorapidity ($\eta$) to quantify the influence of final-state effects~\cite{Schenke:2020uqq}. The $\rho_{n}$ values are also affected by nonflow contributions, primarily from resonance decays and jets. These contributions can be mitigated using the "subevent method," as elucidated in Section~\ref{sec:challenge}.

\subsection{Methodology}
\label{sec:4}

In the experimental analysis, the measurement of $\rho_n$ using Eq.~\ref{eq:rho} requires the calculation of its constituent terms. For this purpose, we define:
\begin{align} \label{eq:rho1}
\rho_n &= \frac{\cov{n}}{\sqrt{\var{n}}\sqrt{c_{k}}},\quad \text{where} \\
\cov{n} &= \llrr{v_n^2\,\delta \pT}, \nonumber \\
\var{n} &= \lr{v_n^4}-\lr{v_n^2}^2, \nonumber \\
c_k &= \llrr{\delta \pT\,\delta \pT}. \nonumber
\end{align}

The covariance $\cov{n}$ is a three-particle correlator, obtained by first calculating the average product for unique particle triplets within each event and subsequently averaging this quantity over all events in a specified event-activity class based on the number of reconstructed charged particles, $\NchR$, and the total forward transverse energy, $\SumET$:
\begin{align}
\cov{n} &= \lr{\frac{\sum\nolimits_{i,j,k \,\atop i\neq j\neq k} w_i w_j w_k \,e^{i n(\phi_i-\phi_j)}\bigl(p_{T,k}-[\pT]\bigr)}
{\sum\nolimits_{i,j,k \,\atop i\neq j\neq k} w_i w_j w_k}}.
\end{align}
In this formula, the indices $i,j,k$ iterate over distinct charged particles to include all unique triplets. The particle weight $w_i$, detailed in Section~\ref{sec:data}, corrects for detector non-uniformities and tracking inefficiencies. The expression for $\cov{n}$ can be expanded algebraically within the cumulant framework~\cite{Bilandzic:2010jr,Bilandzic:2013kga,Jia:2017hbm,Huo:2017nms} into a polynomial function of flow vectors and momentum-scalar quantities. These are constructed from event-wise sums:
\begin{align}\label{eq:2}
\bq_{n;k} &= \frac{\sum_i w_i^k e^{i n\phi_i}}{\sum_i w_i^k},\quad
p_{m;k} = \frac{\sum_i w_i^k\bigl(p_{T,i}-[\pT]\bigr)^m}{\sum_i w_i^k},\quad
[\pT] = \frac{\sum_i w_i\,p_{T,i}}{\sum_i w_i},
\end{align}
where $k$ and $m$ are integer powers. Further details of this expansion are available in Ref.~\cite{Zhang:2021phk}.

The statistical uncertainty of the measurements is determined using a standard Poisson bootstrap method~\cite{Efron:1979bxm,ATLAS:2021kho}. A further description of the Bootstrap method used can be found in Appendix~\ref{sec:app_bootstrap}.

To compute the Pearson coefficient in Eq.~\ref{eq:rho1}, the variances $c_k$ and $\var{n}$ must also be calculated. The term $c_k$ represents the event-averaged two-particle $p_T$ covariance:
\[
c_k = \lr{\frac{\sum\nolimits_{i,j \,\atop i\neq j} w_i w_j\,(p_{T,i}-[\pT])\,(p_{T,j}-[\pT])}
{\sum\nolimits_{i,j \,\atop i\neq j} w_i w_j}}.
\]
This is obtained using all distinct particle pairs in the full event (i.e., within $|\eta|<2.5$). 

The term $\var{n}$ (the variance of $v_n^2$) is calculated using a hybrid approach that combines cumulants from different methods to balance nonflow reduction and statistical precision:
\begin{align} \label{eq:4}
\mathrm{Var}(v_n\{2\}^2)_{\mathrm{dyn}} &=  c_n\{4\}_{\mathrm{standard}} + c_n\{2\}^2_{\mathrm{two-subevent}}
\end{align}
Here, $c_n\{2\}\equiv\lr{v_n^2}$ is the two-particle cumulant and $c_n\{4\}\equiv\lr{v_n^4}-2(\lr{v_n^2})^2$ is the four-particle cumulant~\cite{Borghini:2000sa}; these are described further in an Appendix.

The $c_n\{4\}$ term, being relatively insensitive to nonflow correlations but having poorer statistical precision~\cite{Bilandzic:2010jr}, is obtained from the standard method using the full event. The $c_n\{2\}$ term, more susceptible to nonflow but with better precision, is calculated using the two-subevent method. The calculation of $c_n\{2\}$ and $c_n\{4\}$ follows established procedures~\cite{Bilandzic:2010jr,Jia:2017hbm}, expressing them in terms of the flow vectors $\bq_{n;k}$ (defined in Eq.~\ref{eq:2}). Further discussions on these formulas can be found in Appendix~\ref{sec:app_method}.

Charged particles in this analysis are selected from predefined $p_T$ ranges listed in Table~\ref{tab:1}. For Pb+Pb data, two $p_T$ ranges are used: $0.5<p_T<5\,$GeV and $0.5<p_T<2\,$GeV. For Xe+Xe data, an additional range, $0.3<p_T<2\,$GeV, is included. The primary results are based on the $0.5<p_T<5\,$GeV range due to its superior statistical precision.

\begin{table}[!h]
\centering
\caption{\label{tab:1} The $\eta$ and $p_T$ ranges chosen for the standard and subevent methods.}
\scalebox{0.8}{%
\begin{tabular}{|l|c|c|}
\hline
Method & Default $\eta$ selection & Alternative $\eta$ selection \\ \hline
Standard & $|\eta|<2.5$ & $|\eta|<1$ \\
Two-subevent & $-2.5<\eta_a<-0.75$, $0.75<\eta_c<2.5$ & $-1<\eta_a< -0.35$, $0.35<\eta_c<1$ \\
Three-subevent & $-2.5<\eta_a<-0.75$, $|\eta_b|<0.5$, $0.75<\eta_c<2.5$ & $-1<\eta_a< -0.35$, $|\eta_b|<0.3$, $0.35<\eta_c<1$ \\ \hline\hline
\multicolumn{2}{|c|}{$p_T$ selection for Xe+Xe} & $p_T$ selection for Pb+Pb \\ \hline
\multicolumn{2}{|c|}{$0.3<p_T<2\,$GeV, $0.5<p_T<5\,$GeV, $0.5<p_T<2\,$GeV} & $0.5<p_T<5\,$GeV, $0.5<p_T<2\,$GeV \\ \hline
\end{tabular}
}
\end{table}

The measurement of $\cov{n}$, $\var{n}$, and $c_k$ follows a three-step procedure~\cite{Aaboud:2017blb,Aad:2019fgl}. First, these correlators are calculated event-wise. Second, their values are averaged over events with comparable multiplicity (e.g., similar $\SumET$ or the same $\NchR$ bin) and then combined in broader multiplicity ranges. The Pearson coefficients $\rho_n$ are then obtained using Eq.~\ref{eq:rho1}. This averaging procedure helps to reduce centrality fluctuation effects~\cite{Zhou:2018fxx,Schenke:2020uqq,Bozek:2020drh}. Third, the $\NchR$ dependence is converted to a centrality percentile dependence for comparability with other results.

Particle weights $w_i$, accounting for detector inefficiencies $\epsilon(\eta,p_T)$ and azimuthal non-uniformities, are applied. Residual $\phi$-dependencies in flow vectors $\bq_{n;k}$ are corrected by subtracting an event-averaged offset, as discussed in Section~\ref{sec:dAna_Pb}.

\textbf{Comparison and Selection of Subevent Methods}

To suppress short‐range nonflow contributions (e.g., from resonance decays and jets), an $\eta$ gap is applied between particles, and three primary methods for calculating the correlators are compared~\cite{Jia:2017hbm,Zhang:2021phk}:
\begin{itemize}
  \item \textbf{Standard method:} All charged particles with $|\eta|<2.5$ are used.
  \item \textbf{Two‐subevent method:} Particles are taken from two separated subevents, $a\:(-2.5<\eta_a<-0.75)$ and $c\:(0.75<\eta_c<2.5)$. For $\cov{n}$, one particle each contributing to $v_n^2$ (one from $a$, one from $c$) is used, and the third particle (for the $p_T$ term) is drawn from either subevent $a$ or $c$, ensuring it is distinct from the first two.
  \item \textbf{Three‐subevent method:} Particles are taken from three subevents: $a\:(-2.5<\eta_a<-0.75)$, $b\:(|\eta_b|<0.5)$, and $c\:(0.75<\eta_c<2.5)$. For $\cov{n}$, the particles contributing to $v_n^2$ are from subevents $a$ and $c$, and the particle for the $p_T$ term is from subevent $b$.
  \item \textbf{Combined-subevent method:} The three-particle correlators (like $\cov{n}$) are calculated by averaging the results obtained from triplets formed using both the two-subevent and three-subevent configurations.
\end{itemize}

Table~\ref{tab:1} lists the $\eta$ and $p_T$ selections used in this analysis, which are relevant for studying non-flow and longitudinal effects, along with the corresponding results presented in the following section. Furthermore, the justification for using certain subevent methods for specific results is provided in detail in Appendix~\ref{sec:subevent_vnpt}.

\subsection{Results}\label{sec:7}
The values of the constituent correlators $c_k$, $\var{n}$ (variance of $v_n^2$), and $\cov{n}$ (covariance term) are calculated and combined to obtain the Pearson correlation coefficient $\rho_n$ for each choice of $p_T$ and $\eta$ ranges in Pb+Pb and Xe+Xe collisions, as defined in Table~\ref{tab:1}. In each case, these quantities can be obtained with the event-averaging procedure performed in intervals of either $\SumET$ or $\NchR$; these intervals are then translated to average centrality values. The default event-averaging procedure is based on $\SumET$. 

As described in Section~\ref{sec:4}, the primary results shown are calculated for charged particles in the range $0.5<p_T<5$~GeV, using the three-subevent method for $\rho_2$ and the combined-subevent method for $\rho_3$ and $\rho_4$.

\subsubsection{Dependence on method and collision systems}\label{sec:7.1}
\begin{figure}[htbp]
\centering
\includegraphics[width=0.97\linewidth]{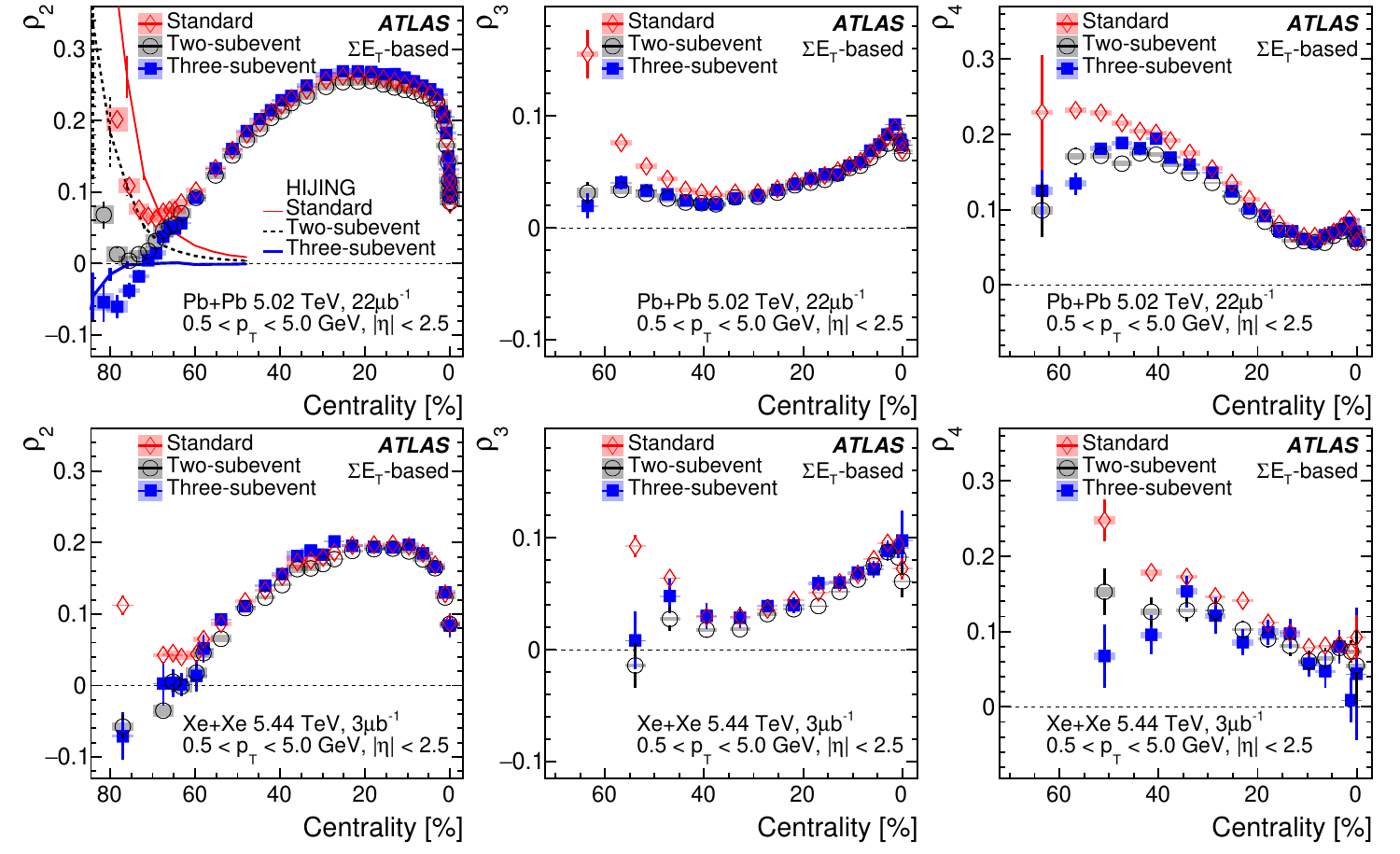}
\caption{The centrality dependence of $\rho_n$ for $n=$2 (left), 3 (middle), and 4 (right) in Pb+Pb (top) and Xe+Xe (bottom) collisions, calculated for the standard, two-subevent, and three-subevent methods. They are calculated using the event-averaging procedure based on $\SumET$. The error bars and shaded boxes represent statistical and systematic uncertainties, respectively. To reduce statistical fluctuations in the Xe+Xe data, wider centrality binning is used in the bottom row. The Pb+Pb $\rho_2$ data are also compared with HIJING calculations from Ref.~\cite{Zhang:2021phk}, which include only nonflow correlations.}
\label{fig:1}
\end{figure}

Figure~\ref{fig:1} shows the $\rho_n$ values obtained from the standard, two-subevent, and three-subevent methods for charged particles with $0.5<p_T<5$~GeV in Pb+Pb and Xe+Xe collisions. These are obtained using the event-averaging procedure based on $\SumET$ and plotted as a function of centrality. 

The results from the different methods are close to each other in central and mid-central collisions. In peripheral collisions beyond 60\% centrality, the values from the standard method are larger than those obtained from the subevent methods. This is consistent with the presence of nonflow correlations arising from resonance decays and jets, which typically give positive contributions to both $v_n$ and $[\pT]$ in the standard method, thereby enhancing the correlation. 

The nonflow effects in the two-subevent method, reflected by the difference from the three-subevent method, are also visible beyond 70\% centrality. Smaller differences, albeit weakly dependent on centrality, are also observed between the two-subevent method and the three-subevent method in mid-central and central collisions. These differences are expected because the $v_n$ signal, as well as the decorrelations of $v_n$ and $[\pT]$, depend on the chosen $\eta$ intervals and the $\eta$-gap ($\Deta$), which differ between the two methods~\cite{Jia:2014ysa,Pang:2015zrq,Bozek:2016yoj}. 

The influence of nonflow effects was investigated recently in models~\cite{Zhang:2021phk,Lim:2021auv}. The $\rho_2$ values obtained from the HIJING model, which generates only nonflow correlations, show a similar ordering between the three methods, as seen in Figure~\ref{fig:1}. In particular, the values of $\rho_2$ from the three-subevent method are closer to zero in the multiplicity range corresponding to the centrality range shown. The non-zero $\rho_2$ signal observed in the data in the peripheral region cannot be reproduced by the HIJING model, indicating contributions beyond these nonflow correlations.

The results from the subevent methods show similar centrality dependencies between the Pb+Pb and Xe+Xe systems: the $\rho_2$ values reach a minimum in peripheral collisions, increase to a positive maximum value, and then decrease in the most central collisions; the $\rho_3$ values show a mild increase towards central collisions; the $\rho_4$ values show an increase then a gradual decrease towards central collisions. 

In the ultracentral collision region, all the $\rho_n$ values show a sharp decrease towards the most central collisions. This decrease is much clearer in the Pb+Pb system due to its better statistical precision and better centrality resolution than in the Xe+Xe system. This sharp decrease starts at around 1.6\% centrality in Pb+Pb, which corresponds approximately to the location of the "knee" in the minimum-bias $\SumET$ distribution~\cite{Aaboud:2019sma}. For events having $\SumET$ values beyond the knee, essentially all nucleons participate in the collision. As a result, geometric fluctuations that typically enhance the $\rho_n$ values are suppressed. A similar suppression of fluctuations has also been observed for other flow observables~\cite{Aaboud:2019sma}.

\begin{figure}[htbp!]
\vspace*{-0.2cm}\centering
\includegraphics[width=0.97\linewidth]{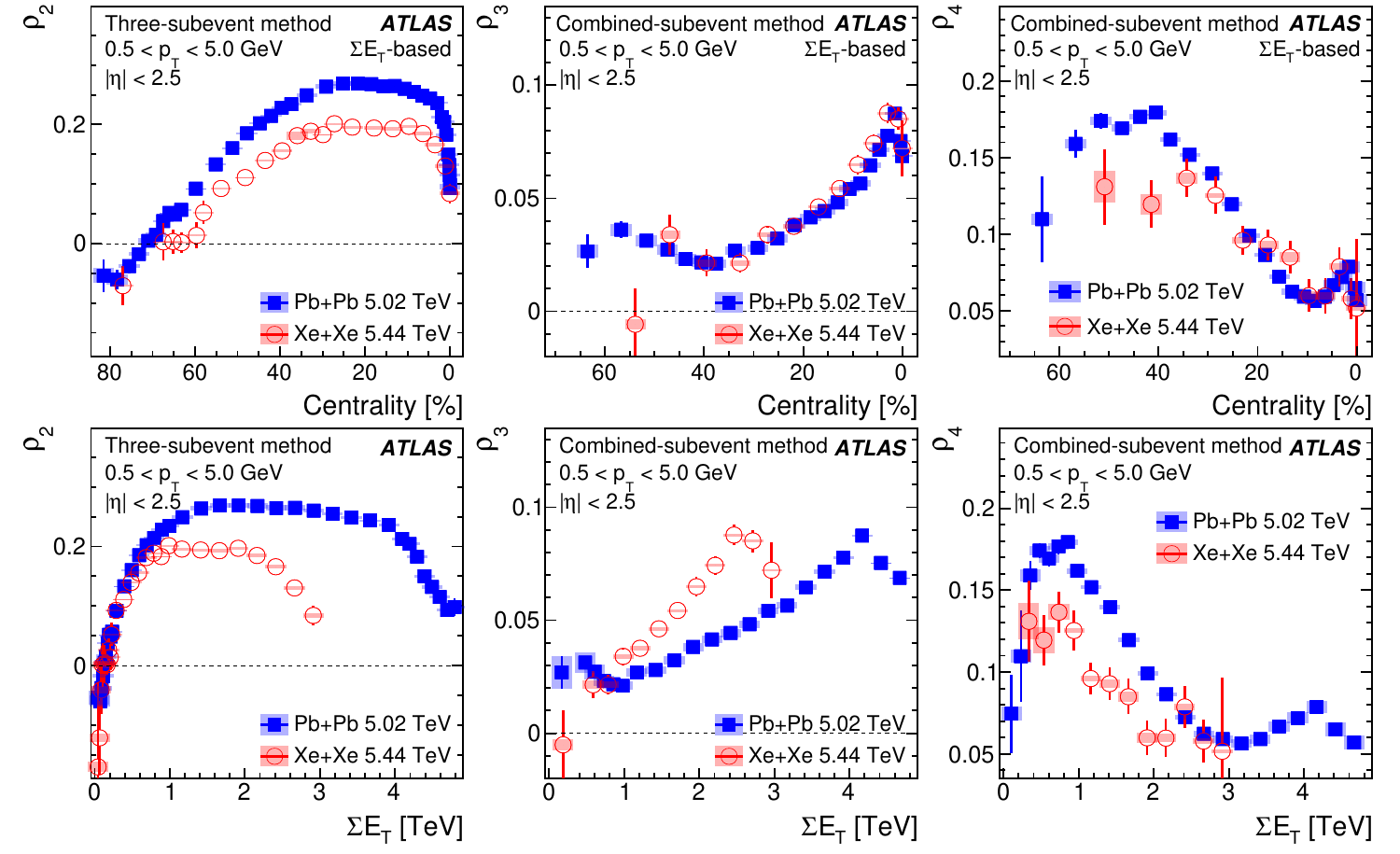}
\vspace*{-0.2cm}\caption{The centrality (top) and $\SumET$ (bottom) dependences of $\rho_n$ for $n=2$ (left), 3 (middle) and 4 (right) in Pb+Pb and Xe+Xe collisions. They are calculated using the event-averaging procedure based on $\SumET$. The error bars and shaded boxes represent statistical and systematic uncertainties, respectively.}
\label{fig:2}
\end{figure}

Figure~\ref{fig:2} provides a direct comparison of the Pb+Pb and Xe+Xe $\rho_n$ values as a function of centrality (top) and $\SumET$ (bottom). These two different choices for the $x$-axis test whether the system-size dependence of $\rho_n$ scales with centrality or $\SumET$. When compared at the same centralities, the Xe+Xe $\rho_2$ values are everywhere smaller than the Pb+Pb values. However, when compared using $\SumET$, the Pb+Pb and Xe+Xe $\rho_2$ values agree for small $\SumET$ values ($\SumET<0.5$~TeV) but differ for larger $\SumET$. When plotted as a function of $\SumET$, the $\rho_3$ values in Pb+Pb and Xe+Xe collisions are similar only at low $\SumET$, while they are similar over the full range when plotted as a function of centrality. The $\rho_4$ values for the two systems are similar when plotted as a function of centrality in the 0--40\% centrality range, but not when plotted as a function of $\SumET$.

\subsubsection{Dependence on the $p_T$ and $\eta$ ranges}\label{sec:7.2}
\begin{figure}[htbp]
\centering
\includegraphics[width=0.97\linewidth]{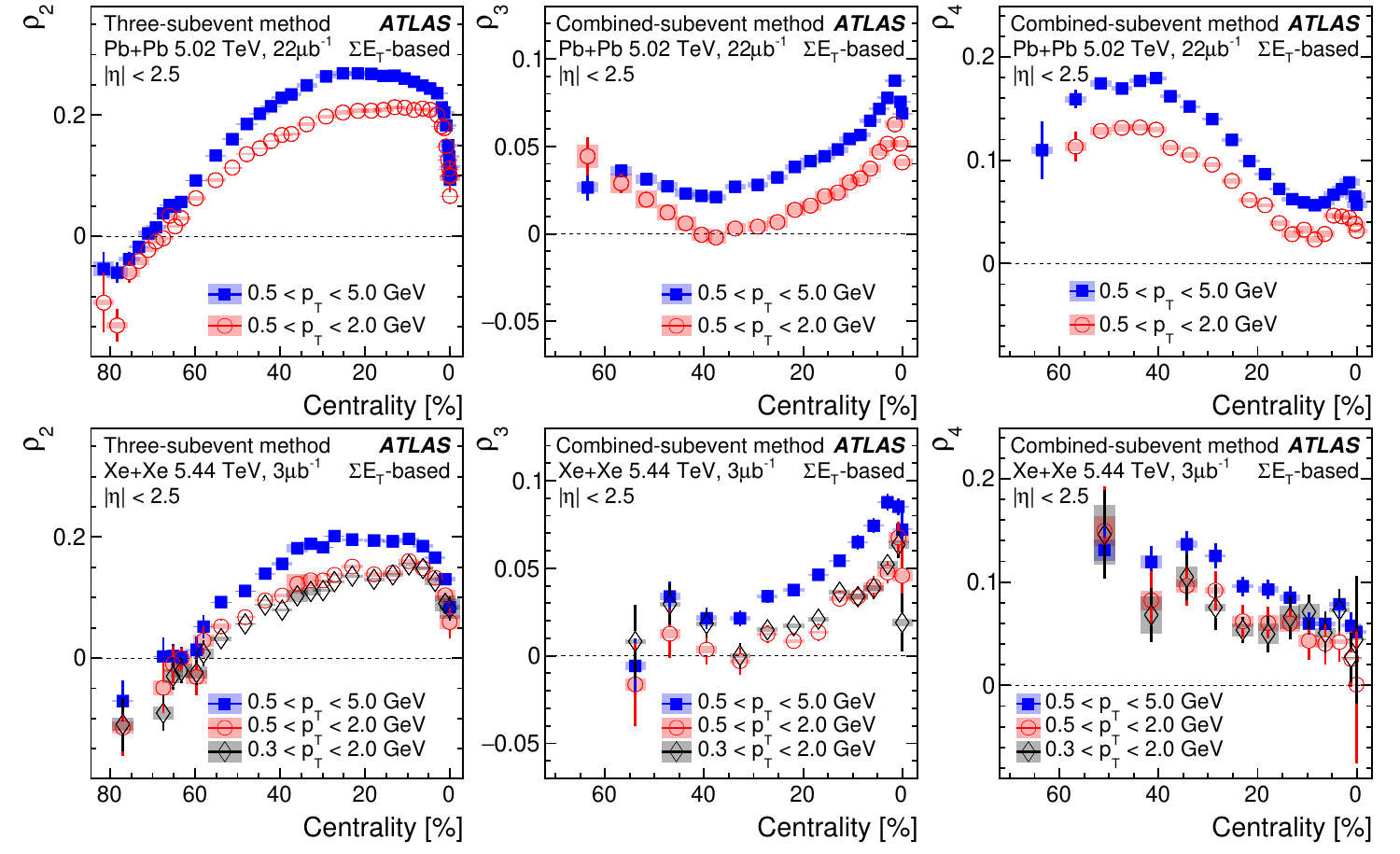}
\caption{The centrality dependence of $\rho_n$ for two $p_T$ ranges in Pb+Pb collisions (top) and three $p_T$ ranges in Xe+Xe collisions (bottom) for $n=$2 (left), 3 (middle) and 4 (right). They are obtained via the event-averaging procedure based on $\SumET$. The error bars and shaded boxes represent statistical and systematic uncertainties, respectively.}
\label{fig:3}
\end{figure}

Figure~\ref{fig:3} shows the centrality dependence of $\rho_n$ in two $p_T$ ranges for Pb+Pb collisions and three $p_T$ ranges for Xe+Xe collisions. It is observed that the $\rho_n$ values for $0.5<p_T<2$~GeV are smaller than those for $0.5<p_T<5$~GeV in both systems, but the overall centrality dependence remains similar. In Xe+Xe collisions, the $\rho_n$ values obtained for a lower $p_T$ range of $0.3<p_T<2$~GeV are found to be close to those obtained for $0.5<p_T<2$~GeV. This is expected since the collective behavior of the bulk particles in the $0.5<p_T<2$~GeV range primarily reflects the hydrodynamic response, so including more particles by further lowering the $p_T$ threshold does not significantly change $\rho_n$. This is an important observation for comparison with other experiments or model calculations, where different $p_T$ ranges are often used.

\begin{figure}[htbp]
\centering
\includegraphics[width=0.97\linewidth]{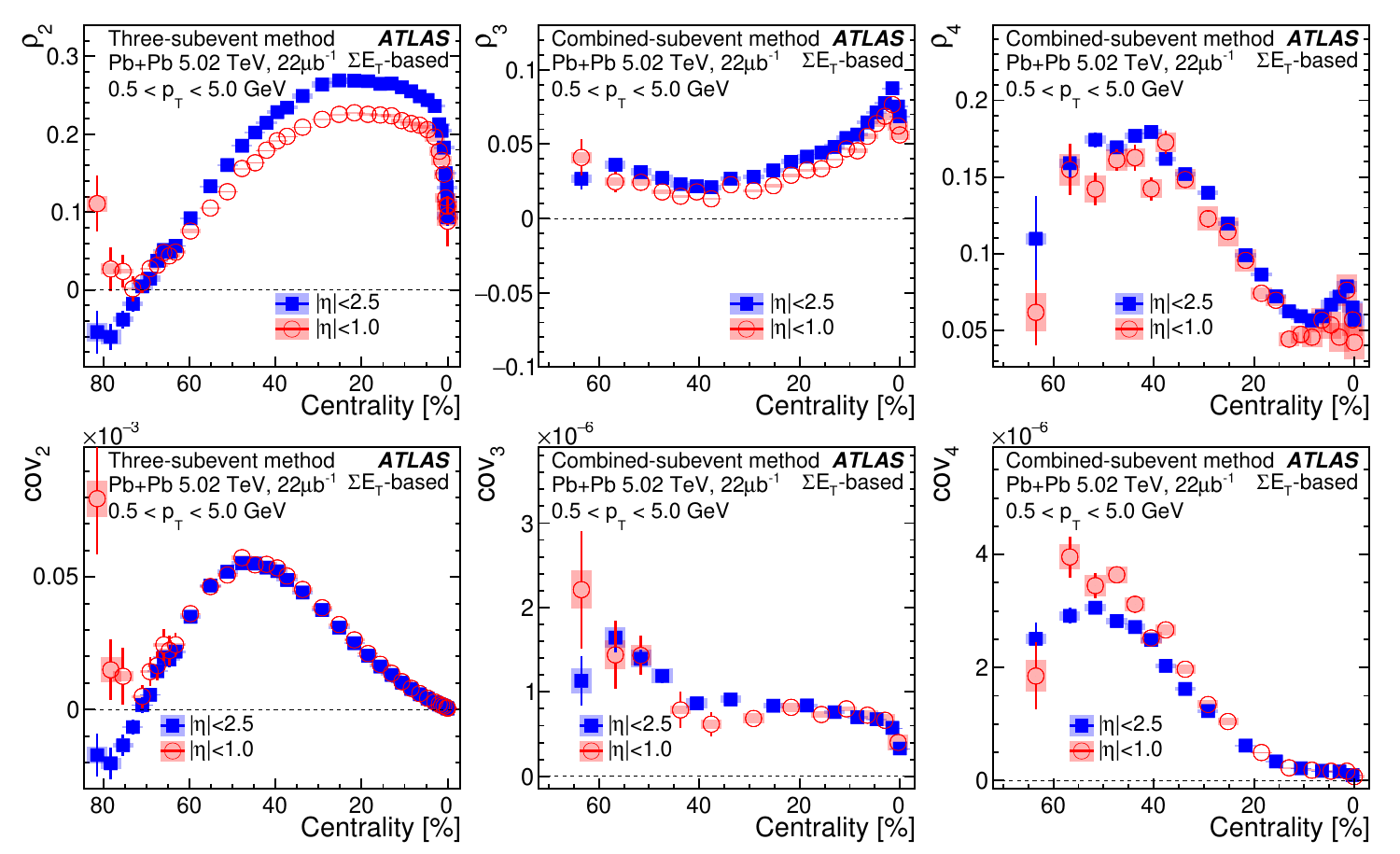}
\vspace*{-0.0cm}
\caption{The centrality dependence of $\rho_n$ (top) and $\cov{n}$ (bottom) for $n=$2 (left), 3 (middle), and 4 (right) in Pb+Pb collisions compared between the two choices for the $\eta$ ranges from Table~\ref{tab:1}. They are calculated using the event-averaging procedure based on $\SumET$. The error bars and shaded boxes represent statistical and systematic uncertainties, respectively.}
\label{fig:4}
\end{figure}

The analysis is also repeated for the $\eta$ range closer to mid-rapidity, $|\eta|<1$, as listed in Table~\ref{tab:1}. Figure~\ref{fig:4} compares the centrality dependence of $\rho_n$ and its covariance term $\cov{n}$ for the two $\eta$ ranges. The results for $\cov{n}$ are almost in agreement with each other, except for $n=2$ and 4 in peripheral collisions. In contrast, the results for $\rho_n$ are systematically lower for $|\eta|<1$ than for $|\eta|<2.5$. This implies that the difference in $\rho_n$ arises mainly from the $\eta$ dependence of the variance terms $\var{n}$ and $c_k$ used in its calculation. The values of $\cov{2}$ and $\rho_2$ for centrality above 70\% are larger for the $|\eta|<1$ selection, likely due to larger residual nonflow effects associated with selecting particles from a smaller $\eta$ range, which can reduce the effectiveness of the $\eta$-gap in subevent methods.

\subsubsection{Effects of centrality fluctuations}\label{sec:7.3}
\begin{figure}[htbp]
\centering
\includegraphics[width=0.97\linewidth]{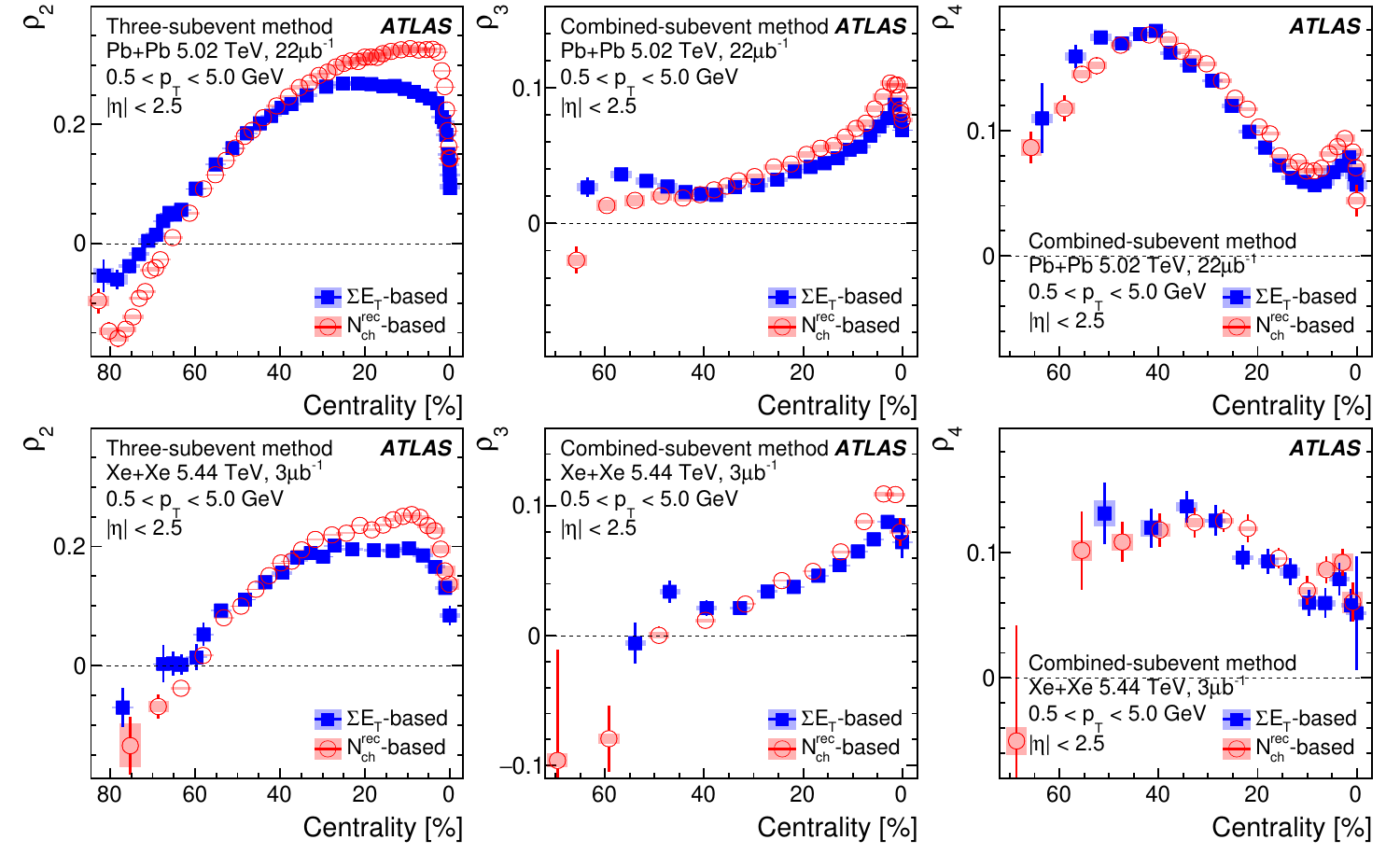}
\caption{The centrality dependence of $\rho_n$ in Pb+Pb (top) and Xe+Xe (bottom) collisions for $n=$2 (left), 3 (middle), and 4 (right), compared between the $\NchR$-based event-averaging procedure (solid squares) and the $\SumET$-based event-averaging procedure (solid circles). The results are calculated using charged particles with $0.5<p_T<5$~GeV. The error bars and shaded boxes represent statistical and systematic uncertainties, respectively.}
\label{fig:5}
\end{figure}

As discussed in the introduction, due to the finite resolution of any event-activity estimator used to characterize event centrality, multiparticle cumulants for flow and $[\pT]$ fluctuations are sensitive to the multiplicity observable used in the event-averaging procedure. The results for $\rho_n$ as a function of centrality in Pb+Pb and Xe+Xe collisions are shown in Figure~\ref{fig:5}, comparing event averaging based on $\SumET$ versus $\NchR$. Large differences between the $\rho_2$ values are observed in central collisions and in peripheral collisions: compared to results based on $\SumET$, the results based on $\NchR$ are larger in central collisions and smaller in peripheral collisions. Differences between the two event-activity estimators are also observed for $\rho_3$ and $\rho_4$.

The influence of centrality fluctuations on $\rho_n$ was recently studied in a transport model framework~\cite{Jia:2021wbq}, albeit at RHIC energies of $\sqn=0.2$~TeV. That study found that the $\rho_2$ values based on particle multiplicity at mid-rapidity are different from those based on particle multiplicity at forward rapidity. These differences are qualitatively similar to those observed in Figure~\ref{fig:5}. The $\rho_2$ values obtained using event activity at forward rapidities were also found to be more consistent with results obtained using the number of participating nucleons~\cite{Jia:2021wbq}. That finding reinforces the notion that the event-activity estimator in ATLAS based on $\SumET$ may have better centrality resolution than the estimator based on $\NchR$. 

Recently, it was argued that $\rho_2$ is a sensitive probe of the nature of collectivity in small collision systems and peripheral HIC, in particular for isolating the contribution from initial momentum anisotropy in a gluon saturation picture~\cite{Giacalone:2020dln}. The hydrodynamic expansion in the final state produces a negative (positive) $\rho_2$ in peripheral (non-peripheral) collisions~\cite{Bozek:2020drh,Giacalone:2020dln,Schenke:2020uqq}, while initial momentum anisotropy is expected to give a large positive contribution in the most peripheral collisions~\cite{Giacalone:2020byk}. Therefore, the centrality dependence of $\rho_2$, after considering both initial state and final-state effects, is predicted to exhibit an increasing trend toward the most peripheral centralities~\cite{Giacalone:2020byk}. However, Figure~\ref{fig:5} shows that the trends of $\rho_2$ in peripheral collisions could still be significantly modified by centrality fluctuations.

\subsubsection{Searching for a Signature of Initial Momentum Anisotropy}\label{sec:results_search_ima}
\begin{figure}[htbp]
\centering
\includegraphics[width=1\linewidth]{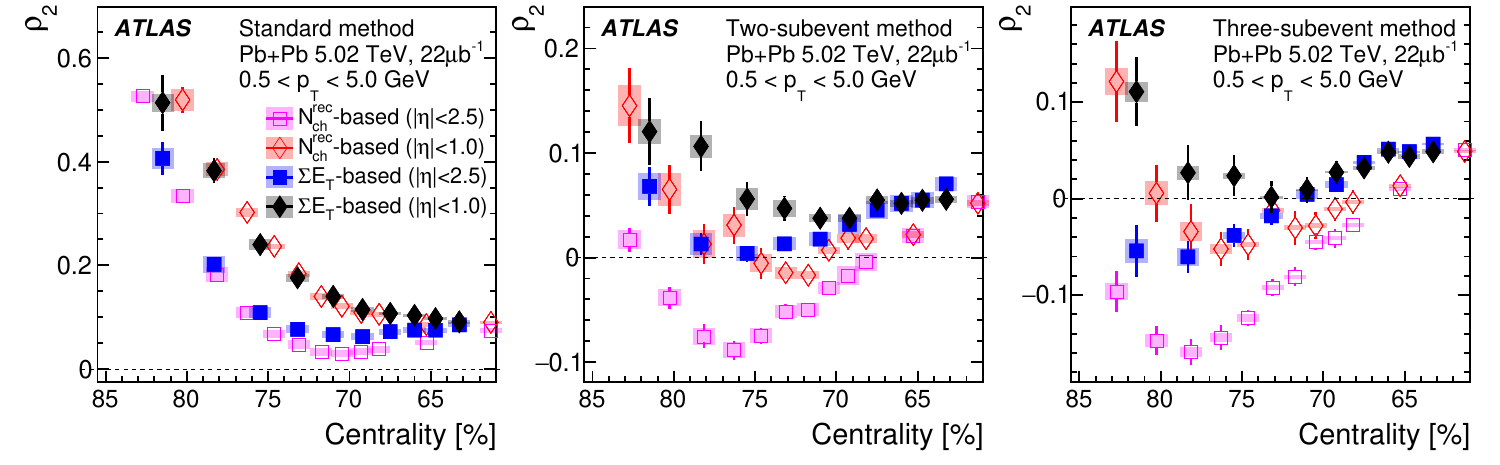}
\caption{ The centrality dependence of $\rho_2$ in Pb+Pb collisions in the peripheral region of 60--84\% for the standard method (left), two-subevent method (middle) and three-subevent method (right), compared between the $\NchR$-based and $\SumET$-based event-averaging procedures and two $\eta$ ranges. The error bars and shaded boxes represent statistical and systematic uncertainties, respectively.}
\label{fig:6}
\end{figure}

Figure~\ref{fig:6} compares the centrality dependence of $\rho_2$ in the $|\eta|<2.5$ and $|\eta|<1$ selections, based on both $\SumET$ and $\NchR$, in more detail over the 60--84\% centrality range for Pb+Pb collisions. It is shown separately for the standard method and subevent methods in order to better separate the influence of nonflow effects from other effects. The successive reduction of $\rho_2$ from the standard method (left panel), to the two-subevent method (middle panel), and to the three-subevent method (right panel) is a robust feature indicating the suppression of nonflow correlations~\cite{Zhang:2021phk}. 

In the right panel, where the residual nonflow is expected to be the smallest, two interesting features can be observed: 1) the $\rho_2$ values obtained for the narrower $|\eta|<1$ range are much larger than those for $|\eta|<2.5$, suggesting that the results obtained in $|\eta|<1$ might still have large nonflow contributions despite the subevent technique; 2) the differences between the $\rho_2$ values from the two event-activity estimators are large for both $\eta$ ranges, reflecting the persistent impact of centrality fluctuations. 

The results from this measurement do not show clear evidence for initial-state momentum anisotropy through an upturn in $\rho_2$ in very peripheral collisions. Future, more detailed studies of the behavior of $\rho_2$ in very peripheral collisions, including those in smaller $pp$ and $p$+Pb collision systems, will be useful to disentangle the effects of nonflow, centrality fluctuations, and initial momentum anisotropy.

\subsubsection{Constraining Xe Triaxiality}\label{sec:results_constraining_xe_triaxiality}

Various models predict different behaviors for $\rho_n$ depending on the assumptions made regarding initial conditions and final-state dynamics. Initial condition models like Glauber or Trento~\cite{Moreland:2014oya} typically rely on linear response relations: $v_n \propto \varepsilon_n$ and $[\pT] \propto E/S$ (where $E/S$ is the initial energy to entropy ratio)~\cite{Schenke:2020uqq,Giacalone:2020byk}. These models estimate $\rho_n$ from event-by-event distributions of $\varepsilon_n$ and $E/S$. More realistic hydrodynamic models evolve systems from Glauber or Trento initial states using 2D boost-invariant (e.g., v-USPhydro~\cite{Giacalone:2020dln}, Trajectum~\cite{Nijs:2021clz}) or 3D (e.g., IP-Glasma+MUSIC~\cite{Schenke:2020uqq,Giacalone:2020byk}) hydrodynamic equations. IP-Glasma+MUSIC uses a 3D initial condition from gluon saturation models and includes an option for initial momentum anisotropy ($\epsilon_p$). Predictions from IP-Glasma+MUSIC with and without $\epsilon_p$ are compared to the data. Most models are tuned to bulk observables such as $v_2$, $v_3$, and $p_T$ spectra. Many incorporate nuclear quadrupole deformation with specific $(\beta, \gamma)$ values, such as $\beta_{\mathrm{Xe}}=0.2$ in Trento, 0.16 in Trajectum, 0.18 in IP-Glasma+MUSIC, and $\beta_{\mathrm{Pb}}=0.06$ in Trento (and zero in other models for Pb). Default triaxiality parameters like $\gamma_{\mathrm{Xe}}=30^{\circ}$ and $\gamma_{\mathrm{Pb}}=27^{\circ}$ are implemented, though sometimes only for specific models like Trento in this comparison set. These deformation parameters are motivated by nuclear energy-density functional calculations~\cite{Bally:2021qys,Budaca:2020ynk} which predict most probable values around $(\beta_{\mathrm{Xe}},\gamma_{\mathrm{Xe}})\approx(0.2,27^{\circ})$ and $(\beta_{\mathrm{Pb}},\gamma_{\mathrm{Pb}})\approx(0.06,27^{\circ})$, although these calculations also indicate the possibility of shape fluctuations over a broad $\gamma$ range.

\begin{figure}[htbp]
\centering
\includegraphics[width=1\linewidth]{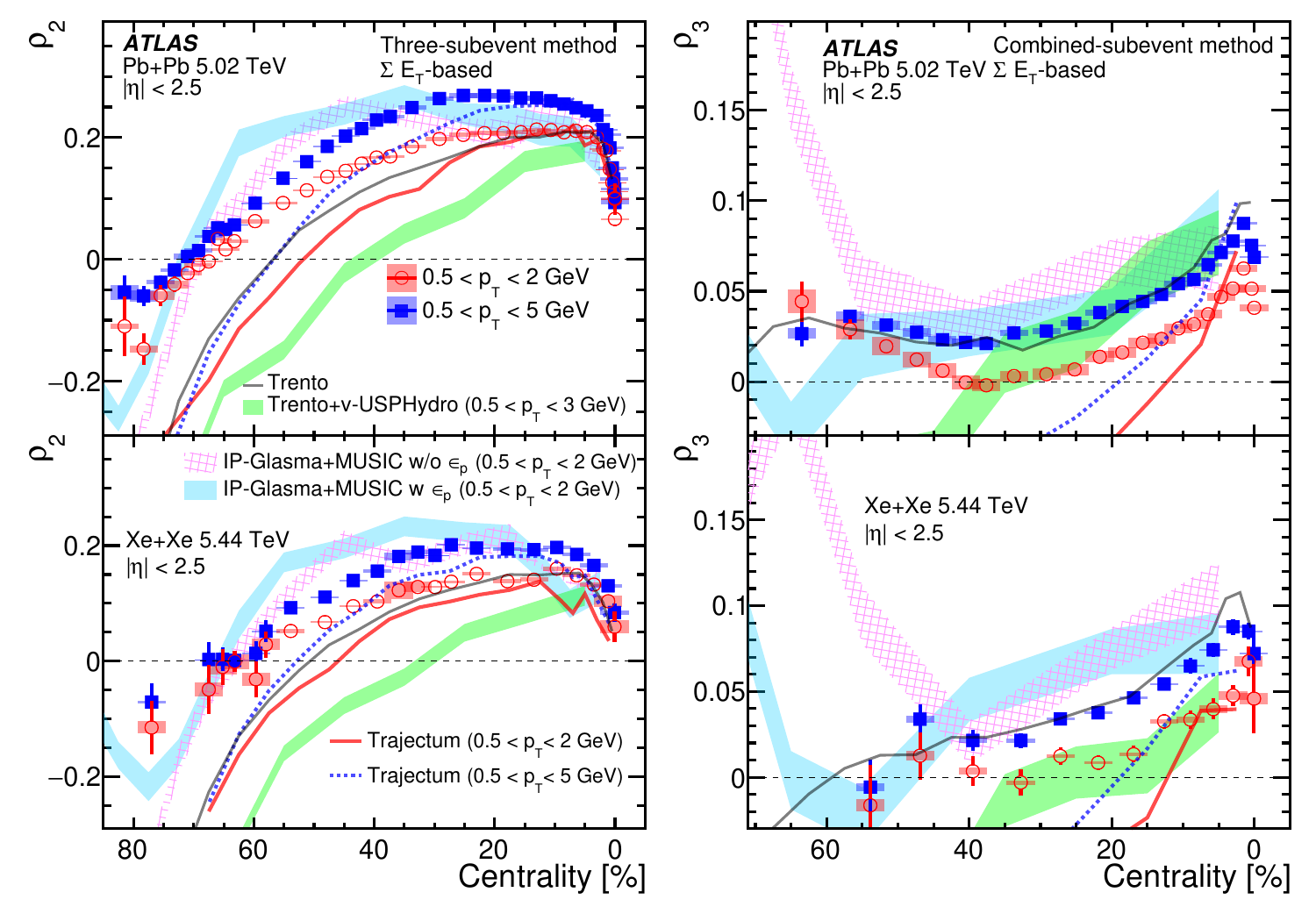}
\caption{The $\rho_2$ (left) and $\rho_3$ (right) values in Pb+Pb (top) and Xe+Xe (bottom) collisions in two $p_T$ ranges and $|\eta|<2.5$, compared with various models: Trento~\cite{Giacalone:2020ymy} and Trajectum~\cite{Nijs:2021clz} models (solid lines), and v-USPhydro~\cite{Giacalone:2020dln} and IP-Glasma+MUSIC~\cite{Giacalone:2020byk} hydrodynamic models (shaded bands, representing statistical uncertainties of the model calculations).}
\label{fig:8}
\end{figure}

Figure~\ref{fig:8} shows the $\rho_2$ and $\rho_3$ values for two $p_T$ ranges in Pb+Pb (top) and Xe+Xe (bottom) collisions. They are compared with the various models described above. 

In the 0--10\% centrality range, where nuclear deformation effects are expected to be important, models generally agree well with each other and with the data. The Trajectum model accurately reproduces the ordering between the $0.5<p_{T}<2$~GeV and $0.5<p_{T}<5$~GeV ranges for $\rho_2$, although it tends to underestimate $\rho_2$ in central Xe+Xe collisions, possibly due to the assumed value of $\beta_{\mathrm{Xe}}$. 

In non-central collisions, models show notable differences, primarily reflecting variations in initial condition parameters like nucleon size~\cite{Giacalone:2021clp}. For peripheral collisions, all model predictions for $\rho_2$ qualitatively match the ATLAS data's sharp decrease and sign change. The initial-state-only Trento model generally underestimates $\rho_2$ across all $p_{T}$ ranges. It describes $\rho_3$ for $0.5<p_{T}<5$~GeV reasonably well but overestimates it elsewhere; this model inherently lacks $p_{T}$ or $\eta$ dependence by construction. 

Both v-USPhydro and Trajectum underestimate $\rho_2$ and $\rho_3$ in non-central collisions. The IP-Glasma+MUSIC model, with and without initial momentum anisotropy ($\epsilon_p$), tends to overestimate the data in mid-central collisions (30--60\%) but underestimates it in more peripheral ones. It predicts the location of the sign change for $\rho_2$ but overestimates its magnitude in mid-central collisions. Differences due to the inclusion of $\epsilon_p$ appear mainly in peripheral collisions beyond 70\% centrality, where current data precision is limited. This highlights the need for more detailed measurements in this region and in smaller systems like $pp$ and $p$+Pb to clarify the role of initial momentum anisotropy.

\begin{figure}[htbp]
\centering
\includegraphics[width=0.97\linewidth]{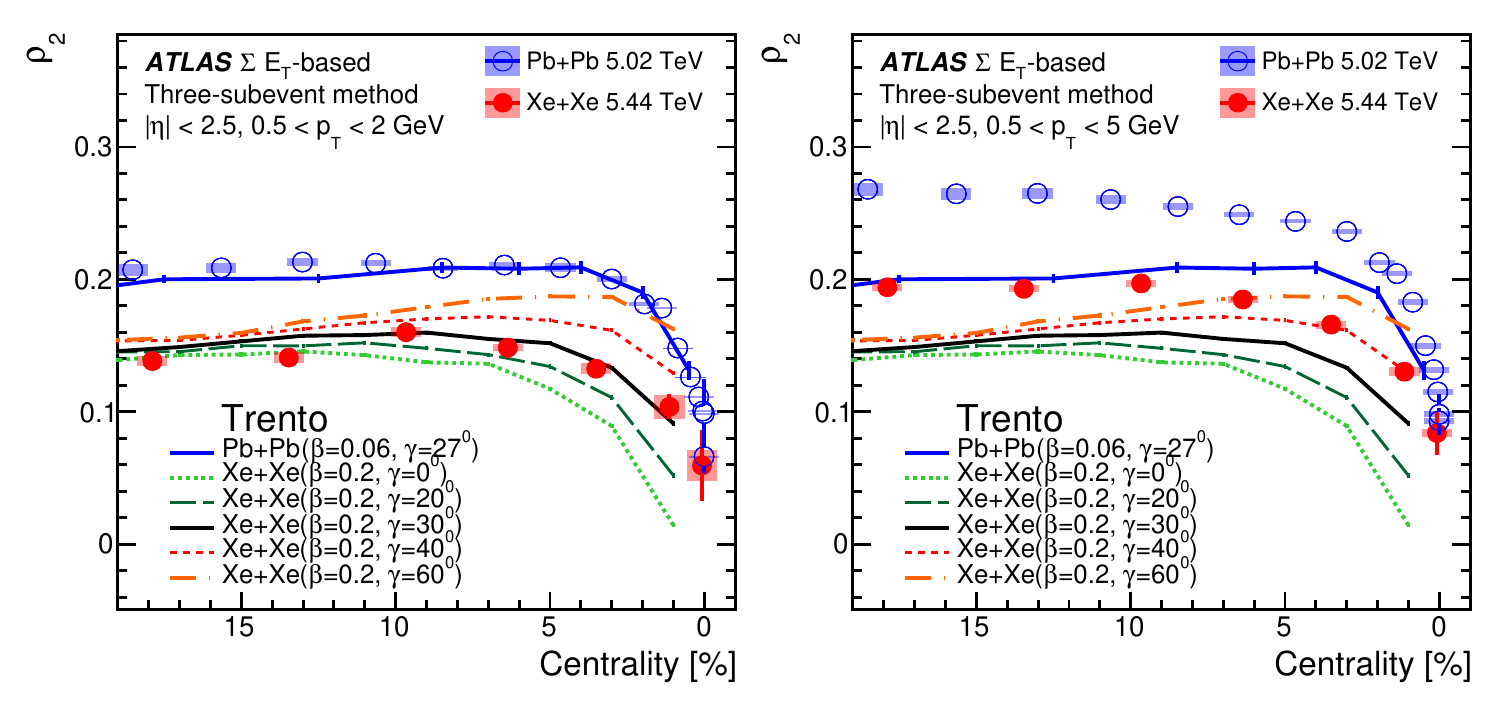}
\caption{Comparison of $\rho_2$ in Xe+Xe and Pb+Pb collisions with the Trento model for various quadrupole deformation parameter assumptions~\cite{Bally:2021qys} in $0.5<p_T<2$~GeV (left) and $0.5<p_T<5$~GeV (right) as a function of centrality. The same Trento model results are used in both panels, and they are connected by lines for better visualization.} 
\label{fig:9}
\end{figure}

Figure~\ref{fig:9} compares $\rho_2$ data in the 0--20\% centrality range with Trento model calculations to investigate the influence of triaxiality~\cite{Bally:2021qys}. Because of the large quadrupole deformation of the $^{129}$Xe nucleus ($\beta_{\mathrm{Xe}}\approx 0.2$), $\rho_2$ should be sensitive to the triaxiality parameter $\gamma_{\mathrm{Xe}}$. This expectation is confirmed in the Trento model, which produces very different trends for $\rho_2$ as a function of centrality for different $\gamma_{\mathrm{Xe}}$ values. However, direct comparisons between the Trento model and data require care, as the $p_T$ dependence of $\rho_n$ is absent in the Trento model by construction.

\begin{figure}[htbp]
\centering
\includegraphics[width=1.0\linewidth]{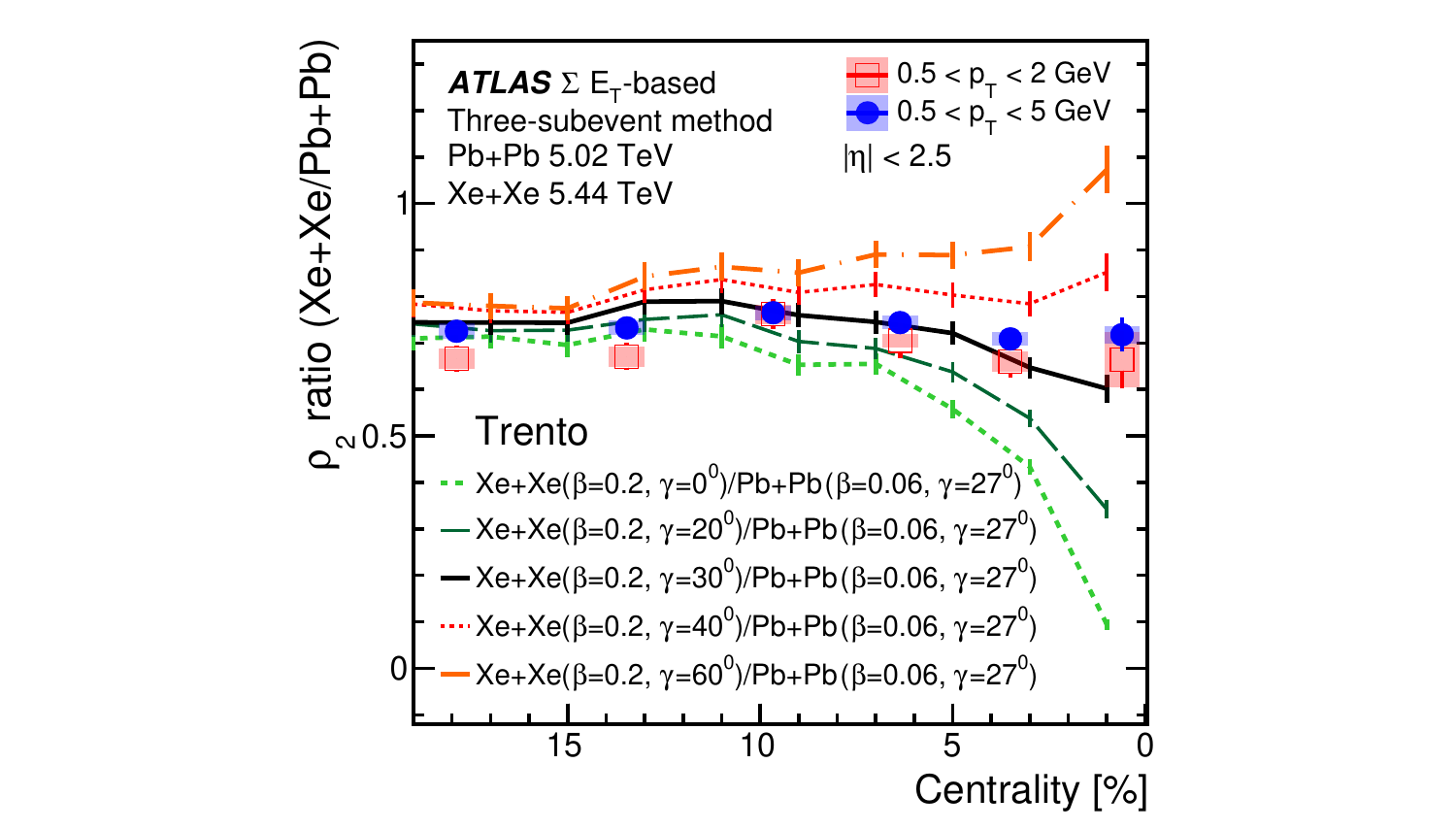}
\caption{Comparison of $\rho_{2}$ ratios, $\rho_{2,{\mathrm{Xe+Xe}}}/\rho_{2,{\mathrm{Pb+Pb}}}$, with the Trento model for various triaxiality parameter ($\gamma_{\mathrm{Xe}}$) assumptions, keeping $\beta_{\mathrm{Xe}}$ and $\beta_{\mathrm{Pb}}$ fixed as per Ref.~\cite{Bally:2021qys}, in two $p_T$ ranges. The Trento model results are connected by lines for better visualization.}
\label{fig:10}
\end{figure}

To mitigate the impact of the differing $p_{T}$ dependencies between data and the Trento model (which lacks $p_T$ dependence), ratios of $\rho_2$ values between Xe+Xe and Pb+Pb collisions are calculated for the two $p_{T}$ ranges. These data ratios are then compared with the corresponding ratios obtained from the Trento model in Figure~\ref{fig:10}. The experimental ratio of $\rho_2$ values is found to be approximately 0.7, and it is slightly lower in the $0.5<p_{T}<2$~GeV range than in the $0.5<p_{T}<5$~GeV range. In the 10--20\% centrality range, where triaxiality is predicted to play a minor role, the model calculation for the ratio is very close to the data ratio. In the 0--10\% centrality range, where the predicted $\rho_2$ values show large dependence on triaxiality, the comparison between the model and data ratios favors a $\gamma_{\mathrm{Xe}}\approx 30^{\circ}$. 


The comparison of the measured ratio $\rho_{2,Xe+Xe}/\rho_{2,Pb+Pb}$ with TRENTo model calculations incorporating varying $^{129}\text{Xe}$ deformation parameters~\cite{Bally:2021qys}, strongly favors a significantly triaxial shape for $^{129}\text{Xe}$, with $\gamma_{Xe} \approx 30^\circ$. This finding provides strong experimental evidence for a non-axial ground state deformation in $^{129}\text{Xe}$. 

While the present analysis robustly points towards triaxiality in $^{129}\text{Xe}$ nucleus, it is acknowledged that distinguishing definitively between a rigid triaxial shape and a $\gamma$-soft potential in nuclei like $^{129}\text{Xe}$ may require complementary observables, as discussed in recent Ref.~\cite{Dimri:2023wup}. Nevertheless, this work established $\rho(v_2^2,)$ as a potent new tool for such nuclear structure investigations, providing the first experimental evidence for $^{129}\text{Xe}$ triaxiality using Heavy-Ion Collisions.

\subsection{Non-Flow Baseline from HIJING and Pythia}
\subsubsection{Motivation}
Model studies using event generators such as \textsc{HIJING} and \textsc{PYTHIA} help establish crucial baseline estimates of non-flow contributions to $v_n$--$[p_{\mathrm{T}}]$ covariances by modeling particle production primarily without collective (flow) effects~\cite{Zhang:2021phk}. The analysis method applied to these simulated data is identical to that described for experimental data in Section~\ref{sec:4}.

\subsubsection{Results}
\begin{figure}[htbp]
\centering
\includegraphics[width=1\linewidth]{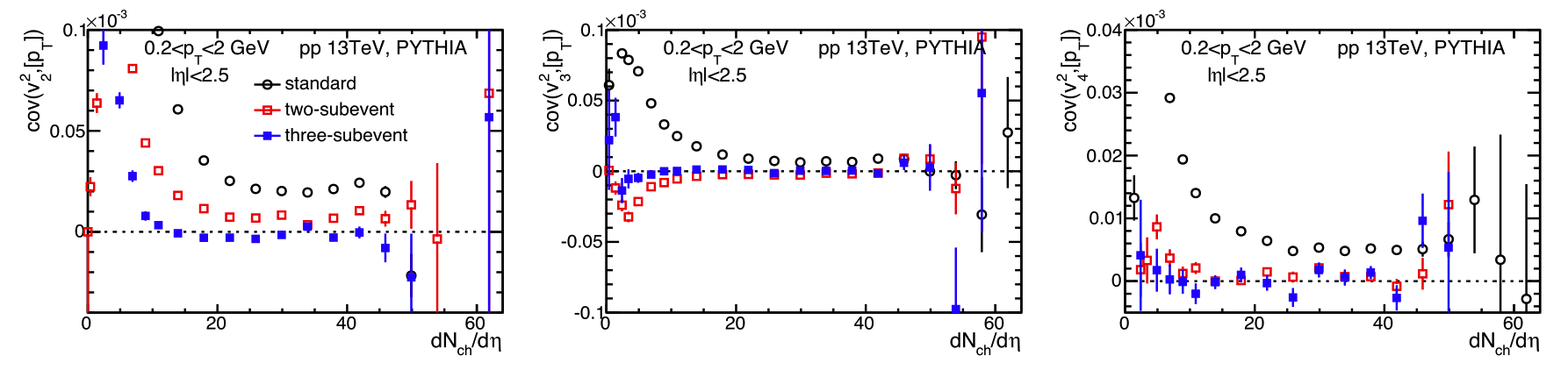}
\caption{$\mathrm{cov}(v_n^2, [\pT])$ as a function of $dN_{\mathrm{ch}}/d\eta$ for $n = 2$ (left), 3 (middle), and 4 (right), compared between the standard, two-subevent, and three-subevent methods for charged particles in $0.2 < \pT < 2$ GeV, obtained from 13 TeV $pp$ Pythia8 simulations.}
\label{fig:vnpthij1}
\end{figure}

Figure~\ref{fig:vnpthij1} illustrates the $\mathrm{cov}(v_n^2, [\pT])$ results obtained from Pythia8 $pp$ simulations, comparing the standard, two-subevent, and three-subevent methods. The standard method produces positive values for all harmonics ($n=2,3,4$). This is attributed to the dominance of short-range jet fragmentation, which creates particle clusters characterized by higher average $p_T$ and enhanced azimuthal correlations near $\Delta\phi \sim 0$ (relative to the jet axis). These features lead to a simultaneous increase in both the apparent $v_n^2$ and the event-wise mean transverse momentum, $[\pT]$. 

With the two-subevent method, the covariances are generally positive for even harmonics ($v_2^2, v_4^2$) and negative for odd harmonics ($v_3^2$). This pattern is consistent with the influence of away-side jet fragments. For instance, a di-jet event can lead to positive $v_2^2$ (from the back-to-back structure) and increased $[\pT]$, while for $v_3^2$, the azimuthal structure might lead to a more negative apparent $v_3^2$ while still increasing $[\pT]$, resulting in a negative $\mathrm{cov}(v_3^2, [\pT])$. For the three-subevent method, $\mathrm{cov}(v_2^2, [\pT])$ remains positive up to $dN_{\mathrm{ch}}/d\eta \lesssim 10$ before becoming slightly negative at higher multiplicities. Across all methods, the magnitudes of $\mathrm{cov}(v_n^2, [\pT])$ are largest using the standard method and smallest using the three-subevent method. This observation strongly suggests that the three-subevent method is the most effective among those tested at suppressing these non-flow contributions.

\begin{figure}[htbp]
    \centering
    \includegraphics[width=1\linewidth]{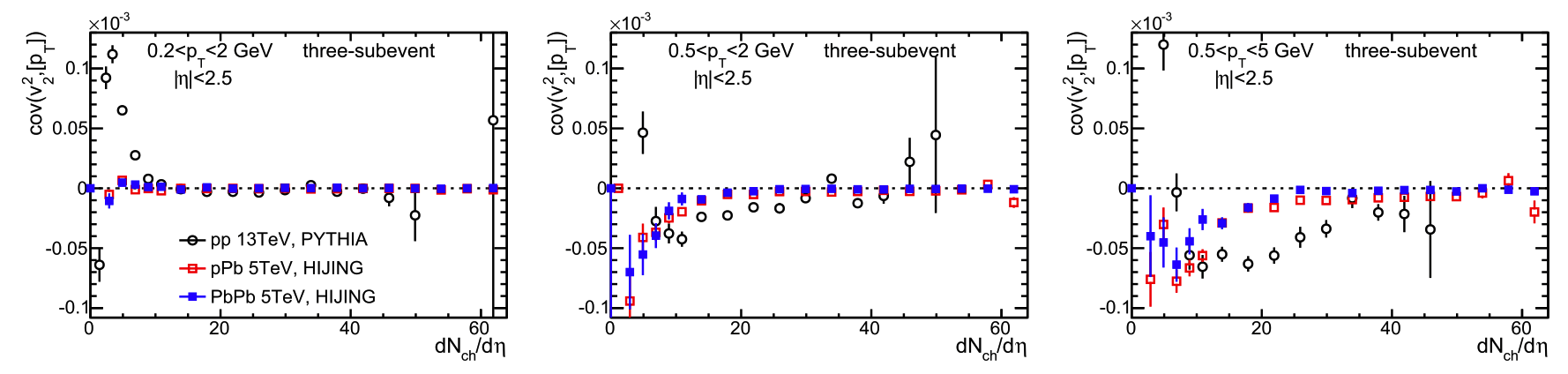}
    \caption{$\mathrm{cov}(v_2^2, [p_{\mathrm{T}}])$ as a function of $dN_{\mathrm{ch}}/d\eta$ from the three-subevent method, compared between $pp$ (Pythia8), $p$+Pb (HIJING), and Pb+Pb (HIJING) collision systems for three $p_{\mathrm{T}}$ ranges: $0.2 < \pT < 2$ GeV (left), $0.5 < \pT < 2$ GeV (middle), and $0.5 < \pT < 5$ GeV (right).}
    \label{fig:vnpthij2}
\end{figure}

Figure~\ref{fig:vnpthij2} compares the three-subevent $\mathrm{cov}(v_2^2, [p_{\mathrm{T}}])$ across $pp$ collisions from Pythia8 and both $p$+Pb and Pb+Pb collisions from HIJING in three different $p_T$ ranges. At low $dN_{\mathrm{ch}}/d\eta$, $p$+Pb and Pb+Pb simulations show negative covariance values, the magnitudes of which tend to increase with the $p_T$ range considered. This contrasts with the positive values observed in $pp$ simulations at similar low multiplicities. For $dN_{\mathrm{ch}}/d\eta > 10$, the $pp$ values also become negative and are generally lower (more negative) than those for $p$+Pb and Pb+Pb. The consistently lower values for $p$+Pb compared to Pb+Pb in these HIJING simulations might suggest slightly different characteristics or magnitudes of residual non-flow in $p$+Pb collisions within this model framework.

To estimate non-flow effects on the Pearson coefficient $\rho(v_n^2, [p_{\mathrm{T}}])$, a suitable normalization for the covariance is required. The $\mathrm{var}(v_n^2)$ term obtained directly from these non-flow models cannot be used for data, as the model $\mathrm{var}(v_n^2)$ would only contain non-flow contributions, whereas data contains both flow and non-flow. Instead, for estimating the non-flow impact on $\rho(v_n^2, [p_{\mathrm{T}}])$ when comparing with data, the denominator term $\mathrm{var}(v_n^2)$ is estimated from published experimental $v_n\{2\}$ and $v_n\{4\}$ measurements, which use methods that aim to subtract non-flow correlations in these collision systems~\cite{CMS:2015yux,ATLAS:2017rtr, ATLAS:2018ngv}. A formula used for this estimation is:
\begin{equation}\label{eq:hijdatvar_vn2}
\mathrm{var}(v_n^2)
= \langle v_n^4\rangle - \langle v_n^2\rangle^2
\approx v_{n,\mathrm{data}}\{2\}^4
\Bigl(1 - \bigl[\tfrac{v_n\{4\}_{\mathrm{data}}}{v_n\{2\}_{\mathrm{data}}}\bigr]^4\Bigr).
\end{equation}
Here, $v_{n,\mathrm{data}}\{2\}$ and $v_n\{4\}_{\mathrm{data}}$ are the experimental cumulants.

\begin{figure}[htbp]
        \centering
        \includegraphics[width=1\linewidth]{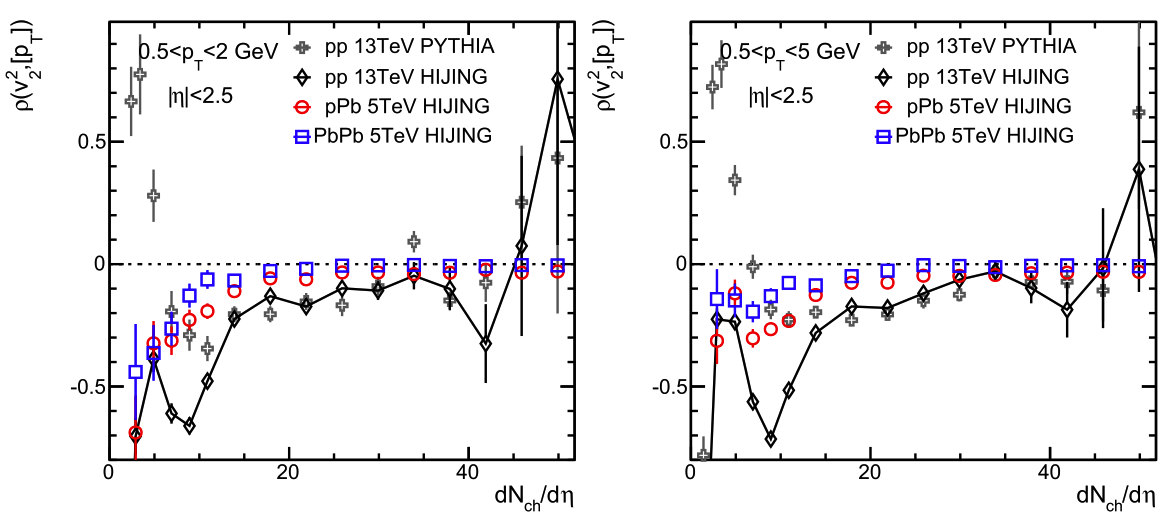}
        \caption{Estimated non-flow contribution to $\rho(v_2^2, [p_{\mathrm{T}}])$ from the three-subevent method as a function of $dN_{\mathrm{ch}}/d\eta$. The covariance term is from non-flow simulations (Pythia8 for $pp$, HIJING for $p$+Pb, Pb+Pb), while $\mathrm{var}(v_2^2)$ is estimated using Eq.~\ref{eq:hijdatvar_vn2} with experimental data inputs. Comparisons are shown between the three collision systems for $0.5 < p_{\mathrm{T}} < 2$ GeV (left) and $0.5 < p_{\mathrm{T}} < 5$ GeV (right).}
        \label{fig:vnpthij3}
    \end{figure}

Figure~\ref{fig:vnpthij3} shows the estimated non-flow contributions to $\rho(v_2^2, [p_{\mathrm{T}}])$ across the three collision systems, using the simulated covariances and the data-driven variance normalization. In the $0.2 < p_{\mathrm{T}} < 2$ GeV range, for $dN_{\mathrm{ch}}/d\eta > 12$, the magnitude of this non-flow estimate for $\rho(v_2^2, [p_{\mathrm{T}}])$ is typically less than $0.02$ in $p$+Pb and Pb+Pb collisions and less than $0.05$ in $pp$ collisions. For $dN_{\mathrm{ch}}/d\eta > 20$ across both $p_T$ ranges shown ($0.5 < p_T < 2$ GeV and $0.5 < p_T < 5$ GeV), the magnitude of the non-flow $\rho(v_2^2, [p_{\mathrm{T}}])$ is approximately $0.02$ in Pb+Pb, $0.06$ in $p$+Pb, and larger in $pp$ collisions (around $0.1$--$0.2$). An experimental measurement of $\rho(v_2^2, [p_{\mathrm{T}}])$ that exceeds these baseline values would suggest the presence of correlations unrelated to these modeled non-flow sources, potentially stemming from initial- or final-state collective effects.

\begin{figure}[htbp]
        \centering
        \includegraphics[width=0.5\linewidth]{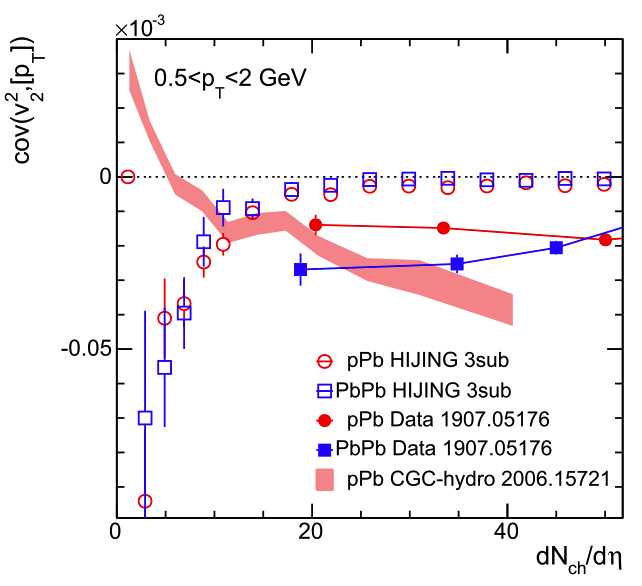}
        \caption{$\mathrm{cov}(v_2^2, [p_{\mathrm{T}}])$ from the three-subevent method (HIJING for $p$+Pb and Pb+Pb) versus $dN_{\mathrm{ch}}/d\eta$ for collisions in the $0.5 < p_{\mathrm{T}} < 2$ GeV range, compared with ATLAS data~\cite{ATLAS:2022dov} and CGC-hydro calculations that include initial-state momentum correlations but no explicit non-flow from fragmentation~\cite{Giacalone:2020byk}.}
        \label{fig:vnpthij4}
    \end{figure}

Figure~\ref{fig:vnpthij4} compares the estimated non-flow contribution to $\mathrm{cov}(v_2^2, [p_{\mathrm{T}}])$ from HIJING with ATLAS experimental data~\cite{ATLAS:2022dov} and calculations from a CGC-hydro model~\cite{Giacalone:2020byk} (which models collectivity but does not include explicit jet/fragmentation-based non-flow). The non-flow estimate is much smaller than the experimental data in high-$dN_{\mathrm{ch}}/d\eta$ Pb+Pb collisions, indicating that the data are dominated by collective phenomena. In $p$+Pb collisions for $0.5 < p_T < 2$ GeV, the non-flow estimate is significant, reaching approximately 30--40\% of the experimental values at $dN_{\mathrm{ch}}/d\eta \sim 20$. The $p$+Pb CGC-hydro model over-predicts the data at $dN_{\mathrm{ch}}/d\eta > 20$, where final-state collective effects are expected to dominate the model calculation. At $dN_{\mathrm{ch}}/d\eta < 10$, the CGC-hydro model is dominated by a positive initial-state momentum correlation contribution, which appears smaller in magnitude than the estimated negative non-flow baseline in this region. In small systems like $p$+Pb, the combination of the negative non-flow baseline and the positive initial-state contribution for $0.5 < p_T < 2$ GeV might result in a complex interplay, potentially making the unambiguous extraction of initial momentum anisotropy effects challenging in experimental measurements without careful separation of all contributions.

\subsection{Conclusion}
\label{sec:chap6_conclusion} 

This study presented experimental measurements of the correlation between harmonic flow $v_n^2$ and the event-by-event mean transverse momentum $[p_{\mathrm{T}}]$, quantified by the Pearson correlation coefficient $\rho(v_n^2, [p_{\mathrm{T}}])$, in Pb+Pb and Xe+Xe collisions using the ATLAS detector. This correlation is recognized as a potential probe sensitive to the initial collision geometry, nuclear deformation, and initial momentum anisotropy. Results were presented as a function of centrality, evaluated using two distinct event-activity estimators: the number of reconstructed charged particles ($\NchR$) and the total forward transverse energy ($\SumET$). Comparing these two estimators revealed the effects of centrality fluctuations, leading to the selection of $\SumET$-based results as the default for model comparisons due to its better centrality resolution.

Comparisons of the measured $\rho(v_n^2, [p_{\mathrm{T}}])$ with predictions from the Trento initial-state model and several hydrodynamic models (v-USPhydro, Trajectum, IP-Glasma+MUSIC), including one with an optional initial momentum anisotropy component (IP-Glasma+MUSIC), show that these models qualitatively describe the overall centrality- and system-dependent trends. However, they generally fail to quantitatively reproduce all features of the data. In peripheral collisions, the interpretation of $\rho(v_2^2, [p_{\mathrm{T}}])$ in terms of initial momentum anisotropy is complicated by possible residual non-flow effects and centrality fluctuations. In mid-central collisions, models tend to over- or under-predict the magnitudes of $\rho(v_2^2, [p_{\mathrm{T}}])$ and $\rho(v_3^2, [p_{\mathrm{T}}])$. Most models exhibit good agreement with the measured $\rho(v_2^2, [p_{\mathrm{T}}])$ values in central collisions. Notably, a comparison of the ratio $\rho_{2,\mathrm{Xe+Xe}}/\rho_{2,\mathrm{Pb+Pb}}$ with the Trento model strongly supports a highly deformed triaxial $^{129}$Xe nucleus, favoring a triaxiality parameter value of $\gamma_{\mathrm{Xe}}\approx 30^{\circ}$.

To better understand and quantify non-flow effects, their influence on $\mathrm{cov}(v_2^2, [p_{\mathrm{T}}])$ was studied in $pp$, $p$+Pb, and peripheral Pb+Pb collisions using Pythia8 and HIJING event generators. These models simulate correlations from processes like jet fragmentation and resonance decays, which are sources of non-flow, but do not include correlations arising from the collective, flow-driven evolution of the system.

The efficacy of the rapidity-separated three-subevent method for non-flow suppression was investigated. This method was found to yield the smallest $|\mathrm{cov}(v_2^2, [p_{\mathrm{T}}])|$ values compared to standard and two-subevent methods, indicating its superior capability in reducing non-flow contributions. Using the three-subevent method, the estimated non-flow component of $\mathrm{cov}(v_2^2, [p_{\mathrm{T}}])$ is negative at high $dN_{\mathrm{ch}}/d\eta (> 20)$ in $p$+Pb and Pb+Pb HIJING simulations, approaching zero at higher multiplicities. The magnitudes of these non-flow estimates are much smaller than the experimental values in $p$+Pb and Pb+Pb collisions in this high-multiplicity region, suggesting that the measured $\mathrm{cov}(v_2^2, [p_{\mathrm{T}}])$ there primarily reflects genuine collectivity-driven correlations.

At lower $dN_{\mathrm{ch}}/d\eta (< 20)$, the behavior of the non-flow estimate for $\mathrm{cov}(v_2^2, [p_{\mathrm{T}}])$ is complex and exhibits strong model dependence. In HIJING simulations of $p$+Pb and Pb+Pb, these values decrease towards more negative values at lower $dN_{\mathrm{ch}}/d\eta$. Conversely, in simulations of $pp$ collisions, the behavior is different between HIJING and Pythia8, with values reaching a maximum around $dN_{\mathrm{ch}}/d\eta \sim 20$ before decreasing again at lower multiplicities, and even sign differences are observed between the two generators. These discrepancies highlight the considerable model dependence of non-flow estimates in the low-multiplicity region.

The potential sign change of $\mathrm{cov}(v_2^2, [p_{\mathrm{T}}])$ predicted from initial-state momentum correlations could be made ambiguous depending on the sign and magnitude of these model-dependent non-flow contributions, especially in smaller systems or peripheral collisions.

\subsection{Outlook}
\label{sec:chap6_outlook}

Looking ahead, further experimental measurements of $v_n$--$[p_{\mathrm{T}}]$ correlations in different collision systems, such as $pp$ and $p$+A collisions at higher statistics and potentially with varying beam energies, would be highly valuable. More differential measurements, for instance, as a function of particle species or finer bins in pseudorapidity, could provide additional constraints on theoretical models and help to disentangle initial-state and final-state effects more effectively. Exploring other correlation observables that are sensitive to the initial state and its fluctuations would also be beneficial for a comprehensive understanding.

On the theoretical front, continued development of event generators and hydrodynamic models is crucial. Future models should aim to incorporate increasingly realistic initial-state descriptions, including not only nucleon substructure and deformations but also momentum correlations. Furthermore, a more sophisticated and consistent treatment of non-flow contributions across all multiplicity ranges and collision systems is needed. 

Detailed quantitative studies are required to precisely quantify the relative contributions of initial-state effects, final-state dynamics, and non-flow to the measured correlations, especially in the complex low-multiplicity regime. Such concerted efforts will be essential for unambiguously interpreting experimental data and extracting fundamental properties of the QGP and the initial state of heavy-ion collisions. Ultimately, this work contributes to the ongoing program of using flow-momentum correlations as a sensitive probe of the initial conditions and the subsequent evolution of the system created in these energetic collisions.

\clearpage

\newpage
\addtocontents{toc}{\protect\setcounter{tocdepth}{0}}%
\setcounter{tocdepth}{0}%
\chapter{Summary}
\label{sec:summary}

This thesis addresses key open questions regarding the initial state and subsequent evolution of the Quark-Gluon Plasma (QGP) created in heavy-ion collisions, leveraging detailed analyses of event-by-event fluctuations and multi-particle correlations using ATLAS data. The overarching goal is to gain detailed insight into the earliest, experimentally inaccessible stages of the collision and the ensuing evolution of the medium, complementing traditional inclusive measurements.

\section{Establishing Collective Nature of Radial Flow Fluctuations}

The goal of the first part of this thesis was to provide experimental evidence establishing radial flow fluctuations as a long-range property of the medium arising from single-particle signals. Furthermore, while radial flow and its fluctuations were predicted to be strongly sensitive to bulk viscosity, $\zeta/s$, strong experimental constraints on this transport coefficient beyond those from $\langle p_{\mathrm{T}}\rangle$ measurements were lacking.

Chapter~\ref{sec:chap4_v0pt} addressed these points by first measuring the $p_{\mathrm{T}}$-differential radial flow fluctuation observable, $v_0(p_{\mathrm{T}})$. By demonstrating that $v_0(p_{\mathrm{T}})$ exhibits genuine collective behavior that persists over large pseudorapidity gaps and shows $p_{\mathrm{T}}$-factorization, the study confirmed its origin in the bulk, early-time dynamics of the system. The near-universal shape of $v_0(p_{\mathrm{T}})$ across different centralities points to the initial state as the origin of these radial flow fluctuations. This work provided the first experimental evidence to explicitly establish the collective nature of radial flow fluctuations, filling a significant gap in the experimental understanding of QGP expansion.

Future directions could include measurements of $v_0(p_{\mathrm{T}})$ in small systems such as $pp$, $p$+Pb, and O+O collisions to search for signals of collective medium formation, the evidence for which is ambiguous using anisotropic flow alone. Additionally, the decreasing behavior of $v_0(p_{\mathrm{T}})$ for $p_{\mathrm{T}} > 3$ GeV might stem from energy loss fluctuations where jets are quenched while traversing the QGP medium, providing an additional handle to constrain jet-medium interactions. Finally, detailed comparisons with hydrodynamic models that explicitly incorporate the effects of bulk viscosity on this observable are needed to extract quantitative constraints on $\zeta/s$.

\section{Disentangling Sources of Initial-State Fluctuations and extraction of $c^2_s$}

Once radial flow was shown to originate as a collective response to initial-state fluctuations in Chapter~\ref{sec:chap4_v0pt}, the next goal was to experimentally disentangle the various sources of these fluctuations. The $p_{\mathrm{T}}$-differential $v_0(p_{\mathrm{T}})$ observable, introduced in Chapter~\ref{sec:chap4_v0pt}, has contributions from many different sources of initial-state fluctuations. Experimentally disentangling these contributions provides strong constraints on these sources, such as separating those arising from event-by-event variations in overlap geometry from non-geometric ones, like quantum mechanical fluctuations in the initial-state or thermal fluctuations during medium evolution. Event-by-event fluctuations in the mean transverse momentum, $[\pT]$, measured in ultra-central collisions were used as a tool to achieve this. 

Chapter~\ref{sec:chap5_ptfluc} addressed this by analyzing higher moments of the $[\pT] $ distribution, such as the variance, $k_2$, and skewnes, $k_3$. It was observed that in UCC, these moments show peculiar non-monotonic behavior. This is attributed to the natural upper limit on the overlap geometry as the impact parameter approaches zero, which truncates the available phase space for geometric fluctuations. By comparison with models, it was demonstrated that this UCC feature can be excellently explained as arising from these constrained geometric fluctuations, whereas in other centralities, the contributions from geometric and non-geometric or intrinsic sources were found to be comparable. This analysis provided a new experimental method for disentangling different sources of initial-state fluctuations, which have been difficult to constrain using previously existing observables.

 Using hydrodynamic model comparisons, this study also related the increase of $\langle [\pT] \rangle$ with charged particle multiplicity in UCC to the speed of sound squared, $c_s^2$, constraining it to  $\approx 0.23$ at an effective temperature $T_{\mathrm{eff}} \approx 222$ MeV.




\section{Constraining Triaxial Deformation in Odd-mass $^{129}$Xe Nucleus}

Next leveraging the findings presented in Chapter~\ref{sec:chap4_v0pt} and Chapter~\ref{sec:chap5_ptfluc} regarding collective nature of radial flow fluctuations and its sensitivity to initial-state geometrical variations, a novel correlator using harmonic flow coefficients and radial flow coefficients, $v_n$--$[p_{\mathrm{T}}]$, was used to constrain the structure of colliding nuclei. This is especially relevant for odd-mass nuclei like $^{129}$Xe, where extracting higher-order deformation parameters using traditional spectroscopic methods is complicated.

Chapter~\ref{sec:chap6_vnpt} addressed this by presenting measurements of the Pearson correlator $\rho(v_n^2, [p_{\mathrm{T}}])$ in spherical $^{208}$Pb+$^{208}$Pb collisions and comparing them to those from $^{129}$Xe+$^{129}$Xe. A key achievement of this work is providing the first experimental evidence for the triaxial deformation of $^{129}$Xe, which consrained its triaxiality parameter ($\gamma_{\mathrm{Xe}} \approx 30^{\circ}$). This analysis demonstrates that heavy-ion collisions can serve as a novel probe of nuclear geometry, overcoming the challenges that traditional spectroscopic methods face in extracting deformation parameters, especially for odd-mass nuclei.

Furthermore, this study also searched for signals of the elusive initial momentum anisotropy in $\rho(v_2^2, [p_{\mathrm{T}}])$ in peripheral collision events. By critically assessing the impact of non-flow effects using event generators such as Pythia8 and HIJING models, as well as different subevent methods, and by studying the effect of centrality fluctuations, it was found that the presence of model-dependent non-flow in peripheral collision events complicates the interpretation of the double sign-change behavior observed in data.

Future work should focus on more precise experimental measurements in peripheral collisions and smaller systems ($pp$, $p$+Pb) to clarify the role of initial momentum anisotropy. Continued theoretical development is required to accurately model non-flow across all multiplicities and to better constrain the interplay between initial-state effects, non-flow, and final-state dynamics. Moreover, probing the existence of nuclear shape fluctuations using heavy-ion collision events would be an interesting direction in future.





\clearpage



\begingroup
\let\chapter\section  
\bibliographystyle{IEEEtran}
\addcontentsline{toc}{chapter}{Bibliography}  
\bibliography{Thesis_Som}
\endgroup

\newpage
\chapter*{Appendix A: Kinematic Dependence of Extracted $c_s^2$} \label{sec:chap5b_cs2}
\addcontentsline{toc}{chapter}{Appendix A: Kinematic Dependence of Extracted $c_s^2$}

One of the key findings discussed in the previous chapter is that the ultra-central rise of $\lr{[\pT]}$ as a function of $\Nch$ is sensitive to the speed of sound in the medium, $c_{s}^{2}$. The CMS collaboration reported the first such measurement of $c_{s}^{2}$ using Pb+Pb collisions at $\sqrt{s_{NN}} = 5.02$ TeV~\cite{CMS:2024sgx}. A key aspect of this extraction is the requirement to include particles across the entire $p_T$ range, i.e., $p_T \geq 0$ GeV. To address this challenge, the CMS collaboration, followed by the ALICE collaboration, fitted particle spectra using the Hagedorn function to extrapolate the spectra to $p_T = 0$ GeV and calculate $\lr{[\pT]}$~\cite{ALICE:2024sqmcs2}. Therefore, it becomes important to measure and understand how the slope of $\lr{[\pT]}$ versus $\Nch$ in UCC approaches the value related to $c_{s}^{2}$ as the minimum considered $p_T$ tends towards $0$ GeV.

This chapter introduces an independent, novel method for studying the dependence of the $\lr{[\pT]}$ versus $\Nch$ relationship in UCC on varying $p_T$ selections of the tracks used in the analysis. Based on simplified assumptions regarding the nature of radial flow fluctuations and utilizing published $p_T$ spectra from Pb+Pb collisions at $\sqrt{s_{NN}} = 5.02$ TeV in the 0--5\% centrality class, this approach aims to provide a more straightforward and complementary alternative to the methods utilized by previous analyses from CMS and ALICE~\cite{CMS:2024sgx, ALICE:2024sqmcs2}.

\section{Theoretical Background}\label{sec:theory_pt_scaling}
Before proceeding to the explicit extraction of the $p_T$-range dependence of the extracted speed of sound, $c_{s}^{2}$, it is necessary to first examine how the moments of event-wise mean transverse momentum, $[\pT]$, fluctuations vary with changing kinematic selections on $p_T$. This understanding will be leveraged later to motivate the formulas central to the methodology employed in this study.

Experimental measurements of $[\pT]$ fluctuations are often constrained to a specific kinematic acceptance, typically a limited range in transverse momentum, $[p_{T,\text{min}}, p_{T,\text{max}}]$. These acceptance cuts can modify the observed fluctuation strength relative to the true, underlying fluctuation that would be observed over the full, unconstrained kinematic range. The dependence of the normalized variance of $\lr{[p_T]}$, denoted $k_2 = \sigma^2_{\lr{[p_T]}}/\lr{[\pT]}^2$, on the $p_T$-range is related to Sum-Rule 2 (Eq.~\eqref{eq:sumrule2b_revised_again}, reproduced below for clarity). Specifically, $k_2$ is influenced by the integration range chosen for $v_0(\pT)$ when applying this sum rule:
\begin{gather}
\int (p_T - \langle p_T \rangle) v_0(\pT) N_0(\pT) dp_T = \sigma_{p_T} N_0 \label{eq:sumrule2b_revised_again}
\end{gather}
where the integration range on the left-hand side (LHS) can be varied to extract $\sigma_{p_T}$, which in turn can be used to calculate $k_2$ within a limited $p_T$ range.

The relationship between $\sigma_{p_T}$ measured within a given acceptance $A$ and its counterpart in the full $p_T$ range can be described by a dimensionless correction factor, $C_A$. This factor, $C_A$, can be derived from Sum-Rule 2 (Eq.~\eqref{eq:sumrule2b_revised_again}) under the assumption of a collective evolution of the system, which relates fluctuations in the particle spectra to fluctuations in the event-wise quantity $[\pT]$. This collective assumption can be expressed numerically as:
\begin{equation}\label{eq:CA_hydro_revised}
\frac{\delta n(p_{\mathrm{T}})}{\langle n(p_{\mathrm{T}}) \rangle}  
=  
\frac{v_0(p_{\mathrm{T}})}{v_0}\frac{\delta [p_{\mathrm{T}}]}{\langle p_{\mathrm{T}} \rangle}
\end{equation}
Here, $v_0(\pT)$ represents the $p_T$-dependent radial flow velocity, and $v_0$ is the $p_T$-integrated radial flow fluctuation measure.

Following the derivation in Refs.~\cite{Parida:2024ckk, Bhatta:2025oyp}, $C_A$ can be written as: 
\begin{align} \label{eq:CA_definition_revised}
C_A \equiv \frac{1}{N_{0A}\lr{[p_T]}_A} \int_{\pT \in A} (\pT - \lr{[p_T]}_A) \frac{v_0(\pT)}{v_0} N_0(\pT) d\pT
\end{align}
Here, $N_{0A} = \int_{\pT \in A} N_0(\pT) d\pT$ is the particle yield within the acceptance $A$. If the experimental acceptance covers the full $p_T$ range, then $C_A = 1$. Note that the definition of $C_A$ explicitly assumes that the radial flow fluctuation is collective in nature~\cite{Parida:2024ckk, Bhatta:2025oyp}.

This factor $C_A$ relates measurements of $k_2$ in a limited $p_T$ acceptance to those in the full $p_T$ range as follows:
\begin{align} \label{eq:k2_scaling_CA_normalized_revised}
\frac{k_{2,A}}{\lr{[p_T]}_A^2} = C_A^2 \frac{k_2}{\lr{[p_T]}^2}
\end{align}
Thus, the variance measured in acceptance $A$, $k_{2,A}$, can be related to the true variance $k_2$ by:
\begin{align} \label{eq:k2_scaling_CA_revised}
k_{2,A} = C_A^2 \left(\frac{\lr{[p_T]}_A}{\lr{[p_T]}}\right)^2 k_2
\end{align}

Equation~\eqref{eq:k2_scaling_CA_revised} further implies that $k_2$ measured as a function of $N_{\mathrm{ch}}$ in two different $p_T$ acceptances should be related by a constant scale factor. Conversely, observing such scaling in experimental measurements would indicate a collective evolution of the system. 

Finally, the acceptance factor $C_{A}$ also relates the slope of $\lr{[\pT]}$ versus $N_{\mathrm{ch}}$ in UCC measured with different $p_T$ selection cuts. This is particularly relevant in the context of determining the speed of sound, $c_{s}^{2}$. A fundamental definition relates $c_{s}^{2}$ to the logarithmic derivatives of mean transverse momentum and multiplicity:
\begin{equation}
c_{s}^{2} \;\equiv\; \frac{d \ln \langle \pT \rangle}{d \ln N_{\mathrm{ch}}} \,. \label{eq:cs2_fundamental_def_final}
\end{equation}
When momentum cuts are applied, limiting the acceptance to $A$, the change in the logarithm of the mean transverse momentum measured in this acceptance, $d \ln \langle \pT \rangle_{A}$, is related to the true change (over the full kinematic range) by $C_A$:
\begin{equation}
d \ln \langle \pT \rangle_{A} \;=\; C_{A}\; d \ln \langle \pT \rangle \,. \label{eq:dlnpt_correction_final}
\end{equation}
This implies that $C_A$ is important for connecting measurements made in limited acceptance to the actual value of $c_{s}^{2}$.

The methodology to measure the mean and variances of the $[\pT]$ fluctuations are identical to those used in Chapter~\ref{sec:chap5_ptfluc}.

\section{Results}

\subsection{$p_T$-Range Sensitivity of $[p_T]$ Cumulants} \label{sec:pt_range_sensitivity}
\begin{figure}[htbp]
\centering
\includegraphics[width=0.48\linewidth]{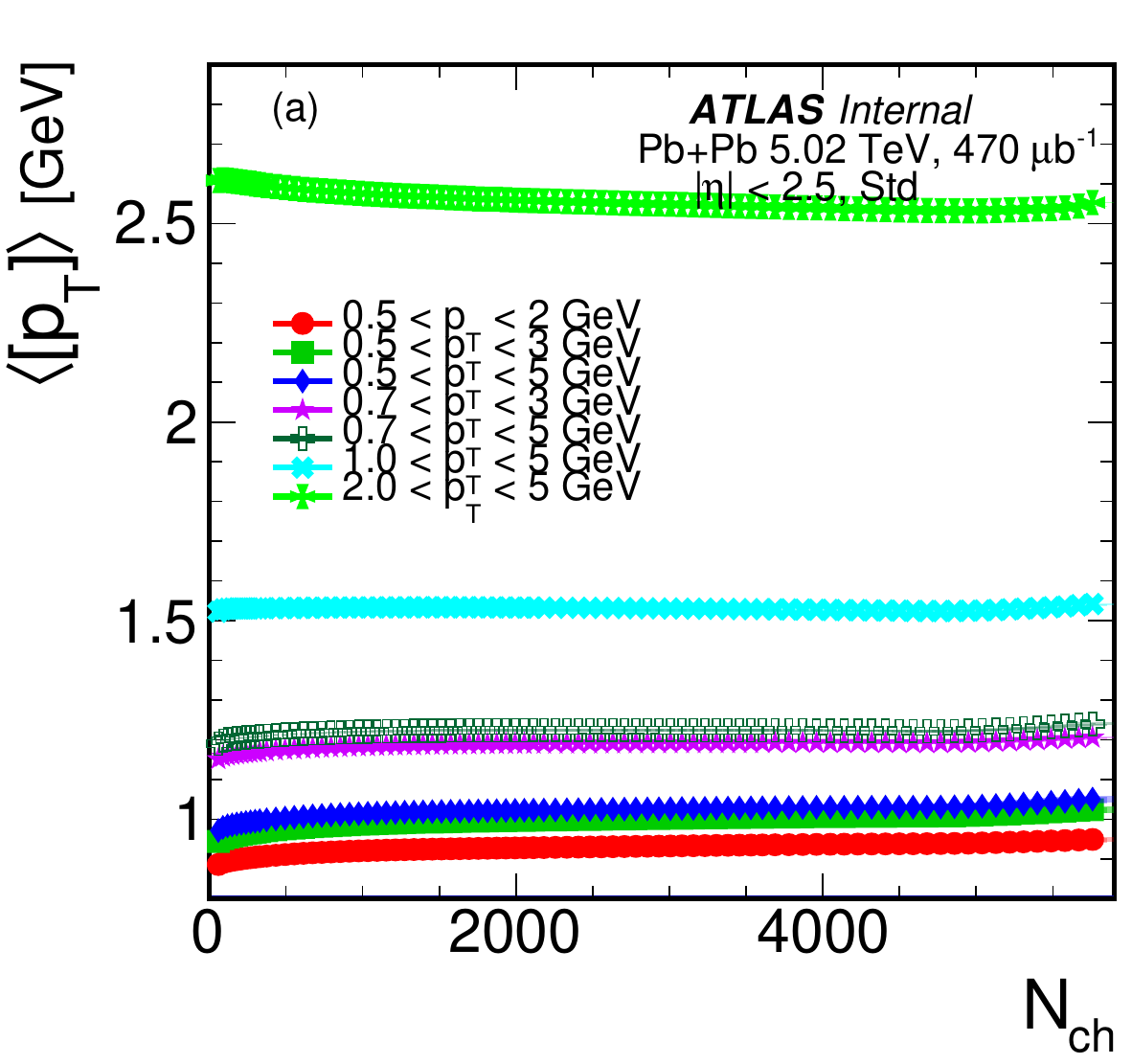}
\includegraphics[width=0.48\linewidth]{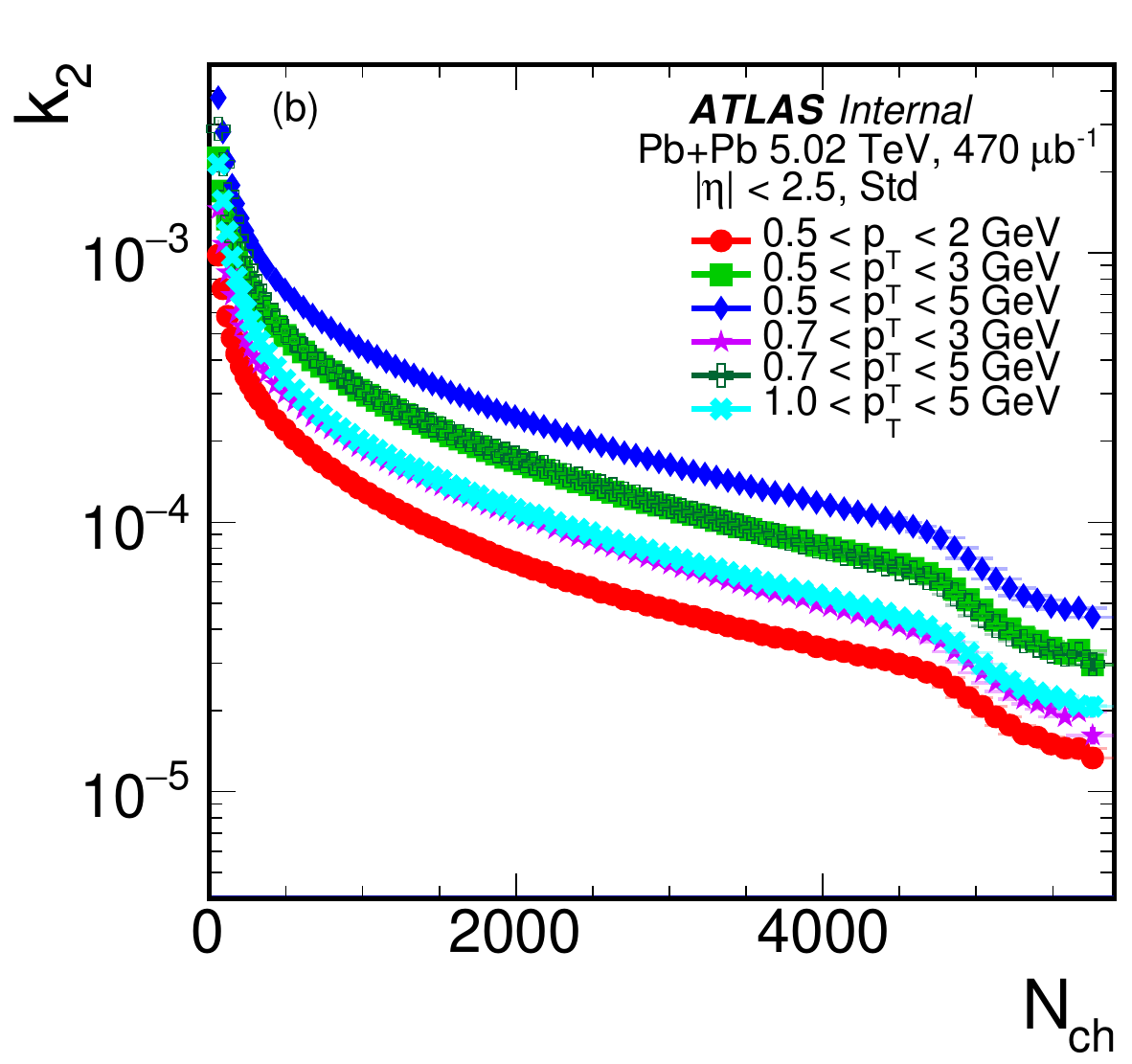}
\caption[Comparison of $\MpT$ and $k_2$ for different $p_T$-ranges]{Comparison of $\MpT$ (left) and $k_2$ (right) between different $p_T$-ranges as a function of $\Nch$. The error bars and shaded areas would indicate statistical and systematic uncertainties, respectively.}
\label{fig:Obs_pTRangeDep}
\end{figure}

This section investigates the kinematic sensitivity of normalized $\lr{[p_T]}$ cumulants to variations in the $p_T$ range of particles used in their calculation, as illustrated in Figure~\ref{fig:Obs_pTRangeDep}. The mean transverse momentum, denoted $\MpT$ in Figure~\ref{fig:Obs_pTRangeDep} (and representing $\lr{[p_T]}$), is observed to increase from peripheral to mid-central collisions for most $p_T$ ranges. Conversely, the second cumulant, $k_2$, exhibits an approximate power-law dependence on $\Nch$ across all $p_T$ ranges, following the relationship $k_n \propto 1/\Nch^{n-1}$~\cite{Bhatta:2021qfk}.

Figure~\ref{fig:Obs_pTRangeDep} also highlights the $p_T$-range dependence of these fluctuation measurements. It is observed that $\MpT$ increases with an increase in either the lower or upper limit of the $p_T$ interval. For $k_2$, expanding the $p_T$ range generally results in an increased variance. However, an exception is noted for the $p_T$ range $1 < p_T < 5$ GeV, which yields a $k_2$ value close to that for the $0.7 < p_T < 3$ GeV range.

Beyond $\Nch \sim 4229$, corresponding to 5\% centrality (denoted as $N_{\mathrm{ch}}^{5\%}$), all observables in Figure~\ref{fig:Obs_pTRangeDep} exhibit deviations from their overall $\Nch$ dependence across all $p_T$ ranges. Notably, $\MpT$ shows a sharp increase, whereas $k_2$ displays a sharp drop at the highest $\Nch$ values. These deviations in the overall behavior of $\lr{[p_T]}$ fluctuations, observed in ultra-central collisions (UCC), are expected due to the gradual suppression of geometrical fluctuations. This suppression allows for the extraction of the relative contributions from both geometrical and intrinsic fluctuations~\cite{Samanta:2023kfk}.

\begin{figure}[htbp]
\centering
\includegraphics[width=0.48\linewidth]{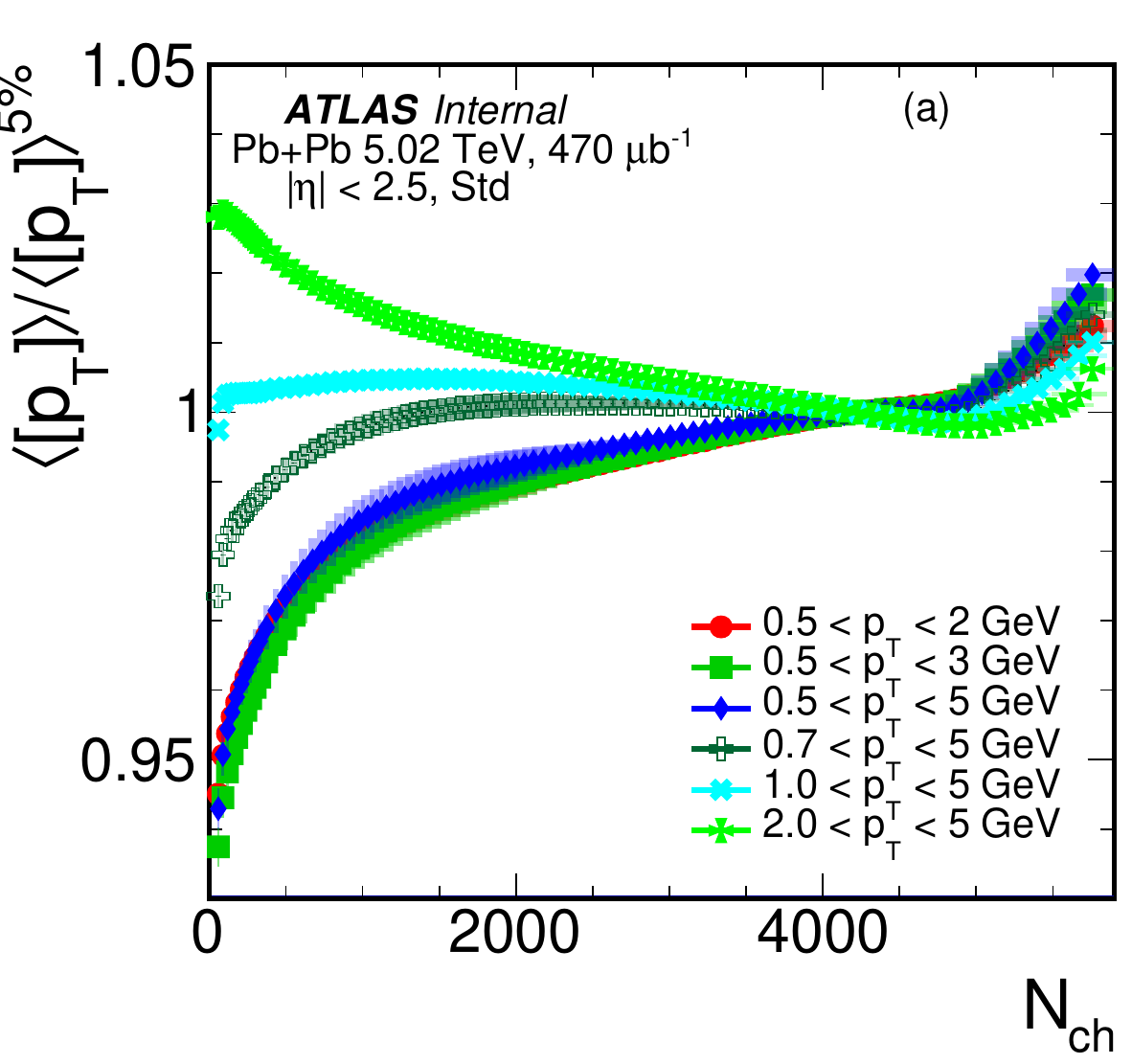}
\includegraphics[width=0.48\linewidth]{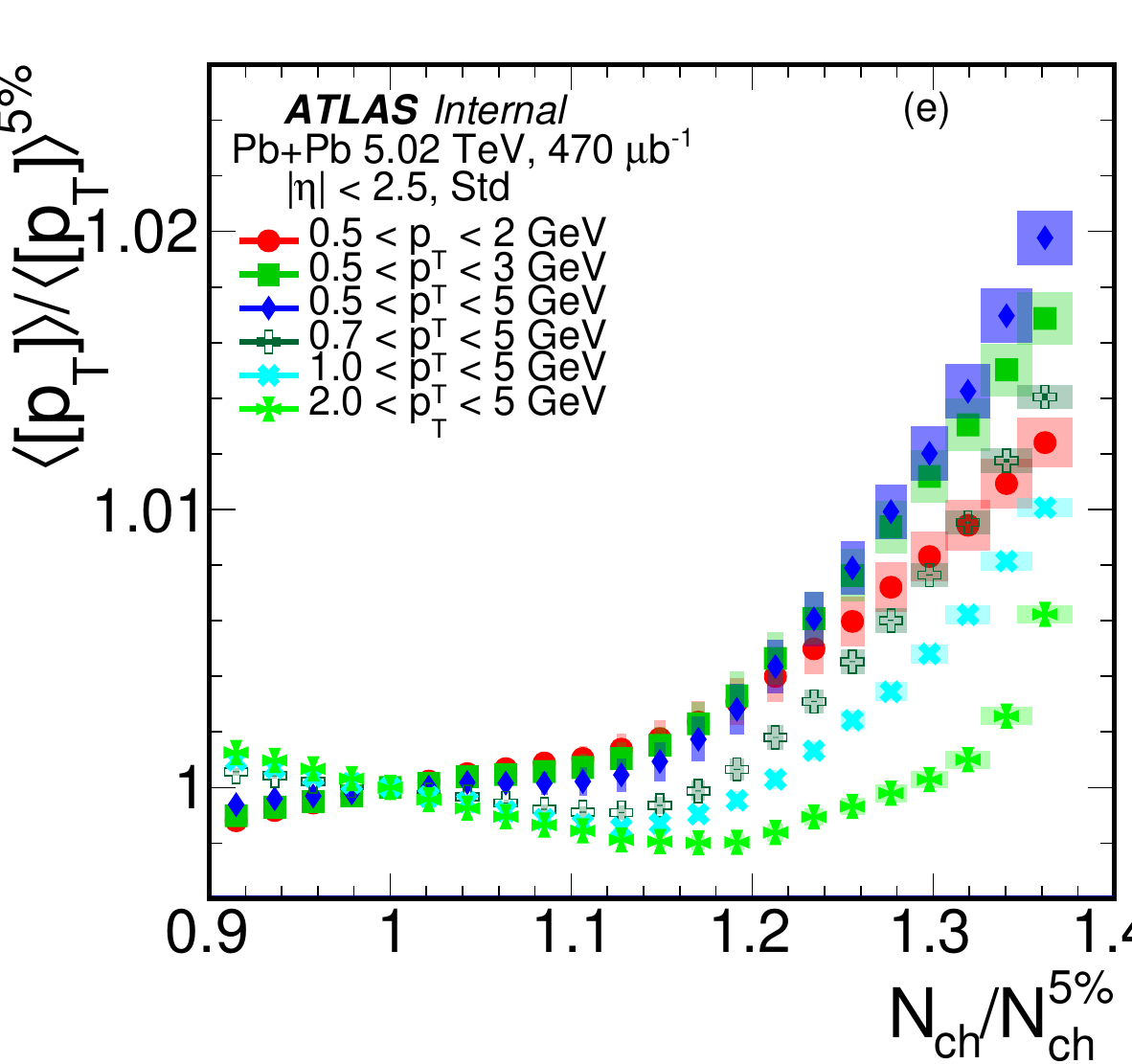}
\includegraphics[width=0.48\linewidth]{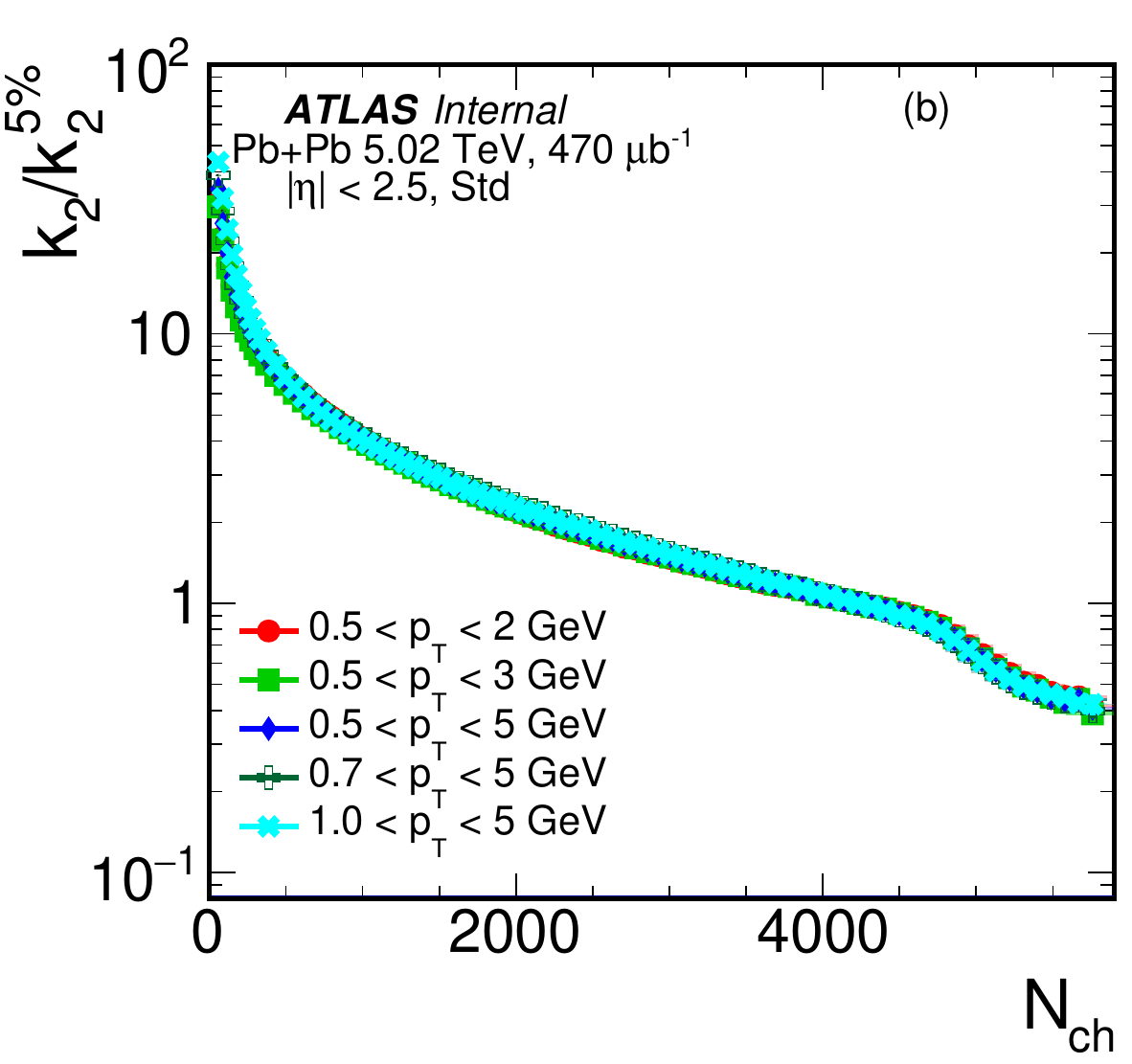}
\includegraphics[width=0.48\linewidth]{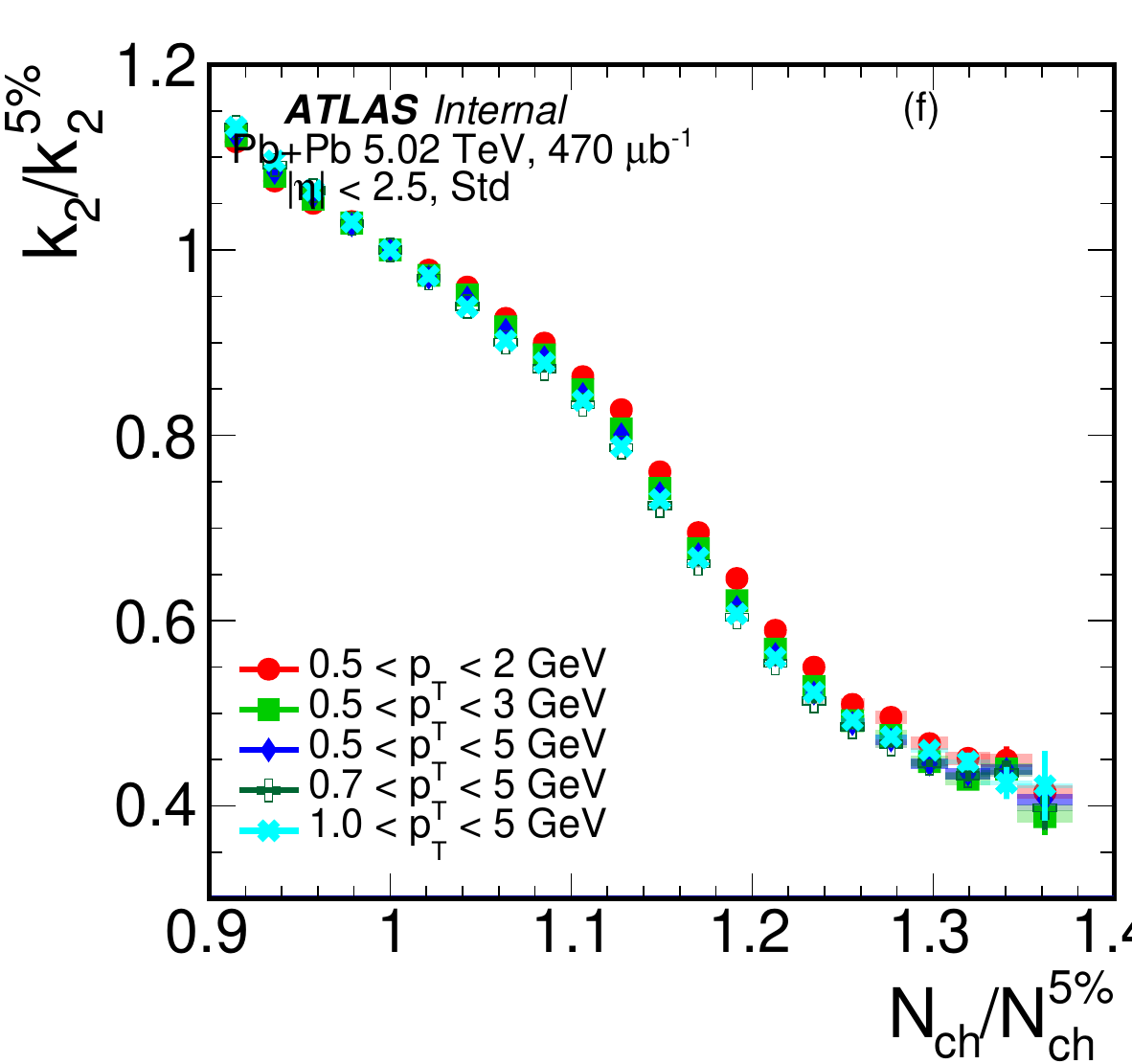}
\caption[Normalized $\MpT$ and $k_2$ for different $p_T$-ranges]{Comparison of $\MpT/\MpT^{5\%}$ (top row) and $k_2/k_2^{5\%}$ (bottom row) for different $p_T$-ranges, normalized by their respective values at 5\% $\Nch$, presented as a function of $\Nch$ (left column) and $\Nch/N^{5\%}_{\mathrm{ch}}$ (right column). The error bars and shaded areas would indicate statistical and systematic uncertainties, respectively.}
\label{fig:Obs_pTRangeDep_Norm5PercentY}
\end{figure}

Furthermore, it is important to study whether the contributions from geometric and intrinsic components exhibit any final-state effects, i.e., $p_T$-range dependence. Any kinematic dependence in the intrinsic component, arising from the medium's evolution, would alter the UCC behavior of the cumulants. To address this, the cumulants are normalized by their values at $N_{\mathrm{ch}}^{5\%}$ for a closer examination. Figure~\ref{fig:Obs_pTRangeDep_Norm5PercentY} illustrates these normalized $\lr{[p_T]}$ cumulants, shown as a function of $\Nch$ as well as $\Nch/N_{\mathrm{ch}}^{5\%}$ (right column).

It is evident that for $\MpT/\MpT^{5\%}$, different $p_T$ ranges display different $\Nch$ dependencies. Specifically focusing on UCC (e.g., panel (b) of Figure~\ref{fig:Obs_pTRangeDep_Norm5PercentY} top-right, showing $\MpT/\MpT^{5\%}$ vs $\Nch/N_{\mathrm{ch}}^{5\%}$), the slope of $\MpT/\MpT^{5\%}$ increases when the upper $p_T$ limit is increased while keeping the lower limit fixed. In UCC, increasing $\Nch$ selects events with larger entropy while keeping the overlap geometry approximately fixed. This increase in entropy density drives radial flow, thereby increasing the slope of $\MpT/\MpT^{5\%}$ versus $\Nch$ in UCC. Therefore, either increasing the upper $p_T$ limit (with a fixed lower $p_T$ limit) or decreasing the lower $p_T$ limit (with a fixed upper $p_T$ limit) allows more particles from different regions of the spectrum to contribute to $\MpT$. This reflects a larger entropy density in UCC and thus increases the slope of $\MpT/\MpT^{5\%}$ versus $\Nch$.

For $k_2$, measurements across different $p_T$ ranges, when normalized by their values at 5\% centrality, show strong agreement. This scaling behavior for $k_2$ was predicted in Ref.~\cite{Parida:2024ckk} as new evidence supporting the hydrodynamic nature of radial flow fluctuations (as discussed in Section~\ref{sec:theory_pt_scaling}). This leads to one of the key findings of this study: $k_2$ indeed exhibits the predicted scaling behavior, providing evidence of hydrodynamic evolution in the system. This suggests that the scaling behavior of $\lr{[p_T]}$ cumulants offers a novel approach to demonstrate the collective evolution of QGP.

The insensitivity of these scaled $\lr{[p_T]}$ cumulants to the $p_T$ range also implies that the fluctuations within any selected $p_T$ range are self-similar to those of the full $P(\lr{[p_T]})$ distribution in the experimentally accessible $p_T$ range. Specifically in UCC, where the peculiar behavior of these cumulants arises from the interplay between unhindered intrinsic fluctuations and diminished geometric fluctuations, the observed vanishing of $p_T$-range dependence in the scaled $k_2$ indicates that the relative contributions from geometrical and intrinsic fluctuations are largely independent of the $p_T$ selection.

The dependence of $v_0(p_T)$ on $p_T$ (discussed in Section~\ref{sec:theory_pt_scaling}) leads to variations in the acceptance correction factor $C_A$. This factor, in turn, affects how the fluctuations scale with the chosen $p_T$ range, and consequently influences the extracted speed of sound. Building upon these observations regarding $p_T$-range sensitivity and the role of $C_A$, this chapter now transitions to a novel method for studying how the extraction of the speed of sound, $c_s^2$, itself depends on the $p_T$ range.

\subsection{Extracting Speed of Sound from ATLAS Data} \label{sec:extract_cs2_atlas}

This work introduces a novel method to determine the speed of sound, $c_s^2$, that aims to be less dependent on extrapolations across the full, experimentally unobserved, kinematic range—a challenge highlighted in the Introduction. A key advantage of our approach, detailed in the subsequent sections, is its reliance on readily available data: the $p_T$ spectra from central collisions (0-5\% centrality in this study) and the observed slope of $\lr{[p_T]}$ versus $\Nch$ in UCC within a limited, experimentally accessible $p_T$-range. This provides an independent route for measuring $c_s^2$, enhancing the robustness of such direct probes of QGP properties.

\subsection{Methodology for $c_s^2$ Extraction} \label{sec:extract_cs2_theory}
The speed of sound squared, $c_s^2$, is related to the logarithmic derivative of the mean transverse momentum, $\MpT$, with respect to the charged particle multiplicity, $\Nch$, following:
\begin{align}
c_s^2 = \frac{d\,\ln{\MpT}}{d\,\ln{\Nch}} \approx \frac{\Delta \MpT / \MpT}{\Delta \Nch / \lr{\Nch}} \label{eq:cs2_redefined_main_chap5b}
\end{align}

Our framework builds upon the assumption, following Ref.~\cite{Gardim:2019iah}, that the relative change in the single-particle $p_T$ spectrum due to small variations in fluid velocity is linear:
\begin{align} \label{eq:SoS_Base0_redefined_main_chap5b}
\frac{N(\pT) - \lr{N(\pT)}}{\lr{N(\pT)}} &= \chi(\pT - \lr{[p_T]}) \\
\Rightarrow \quad N(\pT) &= \lr{N(\pT)} \left[ 1 + \chi(\pT - \lr{[p_T]}) \right] \nonumber
\end{align}
Here, $N(\pT)$ is the event's $p_T$ spectrum, normalized such that $\int_{p_{\min}}^{p_{\max}} N(\pT) d\pT = 1$ within the acceptance $p_{\min} < \pT < p_{\max}$. The term $\lr{N(\pT)}$ represents the $p_T$ spectrum averaged over events in a specific centrality class (e.g., 0-5\%). The quantity $\lr{[p_T]}$ is the event-averaged mean $p_T$, defined for this context as $\lr{[p_T]} \equiv \int_{p_{\min}}^{p_{\max}} \pT \lr{N(\pT)} d\pT$. The parameter $\chi$ represents the proportionality constant linking spectral changes to fluid velocity variations (in general, $\chi$ could depend on $p_T$).

The event-wise mean transverse momentum, which can be written as $\MpT_{\text{event}} = \lr{[p_T]} + \Delta[p_T]$, can be expressed using Eq.~\eqref{eq:SoS_Base0_redefined_main_chap5b}:
\begin{align} \label{eq:SoS_Integration_redefined_main_chap5b}
\lr{[p_T]} + \Delta[p_T] &= \int_{p_{\min}}^{p_{\max}} \pT N(\pT) d\pT \nonumber \\
&= \int_{p_{\min}}^{p_{\max}} \pT \lr{N(\pT)} \left[ 1 + \chi(\pT - \lr{[p_T]}) \right] d\pT \nonumber \\
&= \lr{[p_T]} + \chi \int_{p_{\min}}^{p_{\max}} \pT (\pT - \lr{[p_T]}) \lr{N(\pT)} d\pT \nonumber \\
&= \lr{[p_T]} + \chi \left( \lr{[p_T^2]} - \lr{[p_T]}^2 \right)
\end{align}
where $\Delta[p_T]$ is the event-by-event deviation from the centrality-averaged $\lr{[p_T]}$. The averages $\lr{[p_T^n]}$ are moments of the centrality-averaged spectrum $\lr{N(\pT)}$\footnote{Calculating these moments by integrating $\lr{N(\pT)}$ is equivalent to first averaging $p_T^n$ over tracks within an event and then averaging over all events in the centrality class.}. From Eq.~\eqref{eq:SoS_Integration_redefined_main_chap5b}, the relative deviation in mean $p_T$ is:
\begin{align} \label{eq:SoS_DefineF_redefined_main_chap5b}
\frac{\Delta[p_T]}{\lr{[p_T]}} = \left( \frac{\lr{[p_T^2]}}{\lr{[p_T]}} - \lr{[p_T]} \right) \chi \equiv F \chi
\end{align}
This defines the factor $F$ for a given $p_T$ range $[p_{\min}, p_{\max}]$:
\begin{align} \label{eq:SoS_ExplainF_redefined_main_chap5b}
F_{p_{\min} - p_{\max}} = \frac{\lr{[p_T^2]}}{\lr{[p_T]}} - \lr{[p_T]}, \quad \text{where } \lr{[p_T^n]} = \int_{p_{\min}}^{p_{\max}} \pT^n \lr{N(\pT)} d\pT
\end{align}
In this analysis, the experimental $p_T$ acceptance is typically $0.5 \text{ GeV} < p_T < 5 \text{ GeV}$. However, as established in the Introduction, an accurate determination of $c_s^2$ requires considering contributions from all particles, including those down to $p_T = 0 \text{ GeV}$. We address this by extrapolating the measured $\lr{N(\pT)}$ from $0.5 \text{ GeV}$ down to $0 \text{ GeV}$ using functions discussed briefly in the Introduction and detailed in the Methodology section (Section~\ref{sec:extract_cs2_methodology}). While an ideal upper limit $p_{\max}$ would be $\infty$, using $p_{\max} = 5 \text{ GeV}$ is a reasonable approximation. Particles with $p_T > 5 \text{ GeV}$ are mainly products of hard scatterings, are less sensitive to collective medium effects, and constitute a small fraction ($<1\%$) of the total yield due to the steeply falling $p_T$ spectrum. Their impact on $\lr{[p_T]}$ and $\lr{[p_T^2]}$ is thus considered minimal.

Combining these elements, the experimentally measured slope in UCC for a given $p_T$ range is:
\begin{align}\label{eq:SoS_Main_redefined_main_chap5b}
\text{slope}_{p_{\min}-p_{\max}} = \left(\frac{\Delta [\pT]/\lr{[p_T]}}{\Delta \Nch/\Nch}\right)_{p_{\min}-p_{\max}} = (F \chi)_{p_{\min}-p_{\max}} \left(\frac{\Delta N_{\mathrm{ch}\,event}}{\Delta \Nch}\right) \approx (F \chi)_{p_{\min}-p_{\max}}
\end{align}
Here, we assume that the proportionality $\chi$ implicitly contains the response to changes in entropy density, which scale with $\Nch$. Theoretically, this measured slope approaches $c_s^2$ following Eq.~\ref{eq:cs2_redefined_main_chap5b} when the full kinematic range of $0 \leq p_T < \infty$ is considered. Our methodology, detailed next, demonstrates that considering particles within $0 \leq p_T \leq 5 \text{ GeV}$ is sufficient for a robust extraction of $c_s^2$ using this framework, where $\alpha \equiv ( \Delta \Nch / \Nch )^{-1}$. The term $(\Delta \Nch / \Nch )^{-1}$  absorbed into how $\chi$ or the overall response is interpreted. For simplicity in notation, we write the response as $(F \chi)_{p_{\min}-p_{\max}}$. 


\subsection{Steps for extracting $c^{2}_{s}$} \label{sec:extract_cs2_methodology}
The extraction of $c_s^2$ involves several steps, beginning with the treatment of the experimental $p_T$ spectra and culminating in the calculation of the slope parameter that approximates $c_s^2$.

\subsubsection{Step 1: Extrapolation of $p_T$ Spectra} \label{sec:method_step1_chap5b}
The analysis utilizes the ATLAS $p_T$ spectra for 0-5\% central Pb-Pb collisions at $\sqn = 5.02 \text{ TeV}$ from Ref.~\cite{ATLAS:2022kqu}, shown in the left panel of Fig.~\ref{fig:SoSMethodology_Step1_2_chap5b}. Since the published spectra have a lower $p_T$ limit of $0.5 \text{ GeV}$, an extrapolation to $0 \text{ GeV}$ is necessary. This low-$p_T$ region, dominated by soft particle production, significantly contributes to the factor $F$ (Eq.~\eqref{eq:SoS_ExplainF_redefined_main_chap5b}), which is crucial for the $c_s^2$ extraction via Eq.~\eqref{eq:SoS_Main_redefined_main_chap5b}. Following a similar approach by CMS~\cite{CMS:2024sgx}, the Hagedorn function (Eq.~\eqref{eq:Hagedron_redefined_main_chap5b}) is the default choice for this extrapolation.
\begin{align} \label{eq:Hagedron_redefined_main_chap5b}
\left(\frac{1}{N_{\text{evt}}} \frac{d\Nch}{d\pT}\right)_{\text{Hagedorn}} = A \cdot \pT \left(1+\frac{1}{\sqrt{1-\lr{\beta_{T}}^{2}}} \frac{\left(\sqrt{\pT^{2}+m^{2}}-\lr{\beta_{T}}\pT\right)}{nT}\right)^{-n}
\end{align}
For systematic uncertainty estimation, two commonly used Levy-Tsallis functional forms~\cite{Cleymans:2012ya,STAR:2006nmo,Yang:2021bab} are also employed:
\begin{align} \label{eq:Levy-Tsallis_redefined_main_chap5b}
\left( \frac{1}{N_{\text{evt}}} \frac{d\Nch}{d\pT d\eta} \right)_{\text{Levy-Tsallis}} = A \cdot \pT\frac{(n-1)(n-2)}{n C\left[n C+m_0(n-2)\right]}\left(1+\frac{\sqrt{\pT^2+m_0^2}-m_0}{n C}\right)^{-n}
\end{align}
\begin{align} \label{eq:Levy-Tsallis2_redefined_main_chap5b}
\left( \frac{1}{N_{\text{evt}}} \frac{d\Nch}{d\pT d\eta} \right)_{\text{Levy-Tsallis Form 2}} = A \cdot \pT \sqrt{\pT^{2}+m^{2}_{0}}\cosh(y)\left(1+\frac{\sqrt{\pT^{2}+m^{2}_{0}} \cosh(y)-m_{0}}{nT}\right)^{-(n+1)}
\end{align}
In these functions, $A$ is an overall normalization factor, while $m_0$ (or $m$), $\lr{\beta_{T}}$, $n$, $T$, and $C$ are free fit parameters. These parameters are determined by fitting the functions to the experimental data in the $0.5 < p_T < 5 \text{ GeV}$ range.

\subsubsection{Step 2: Computation of the Factor $F_{x-5}$} \label{sec:method_step2_chap5b}
\begin{figure}[H]
\centering
\includegraphics[width=0.49\linewidth]{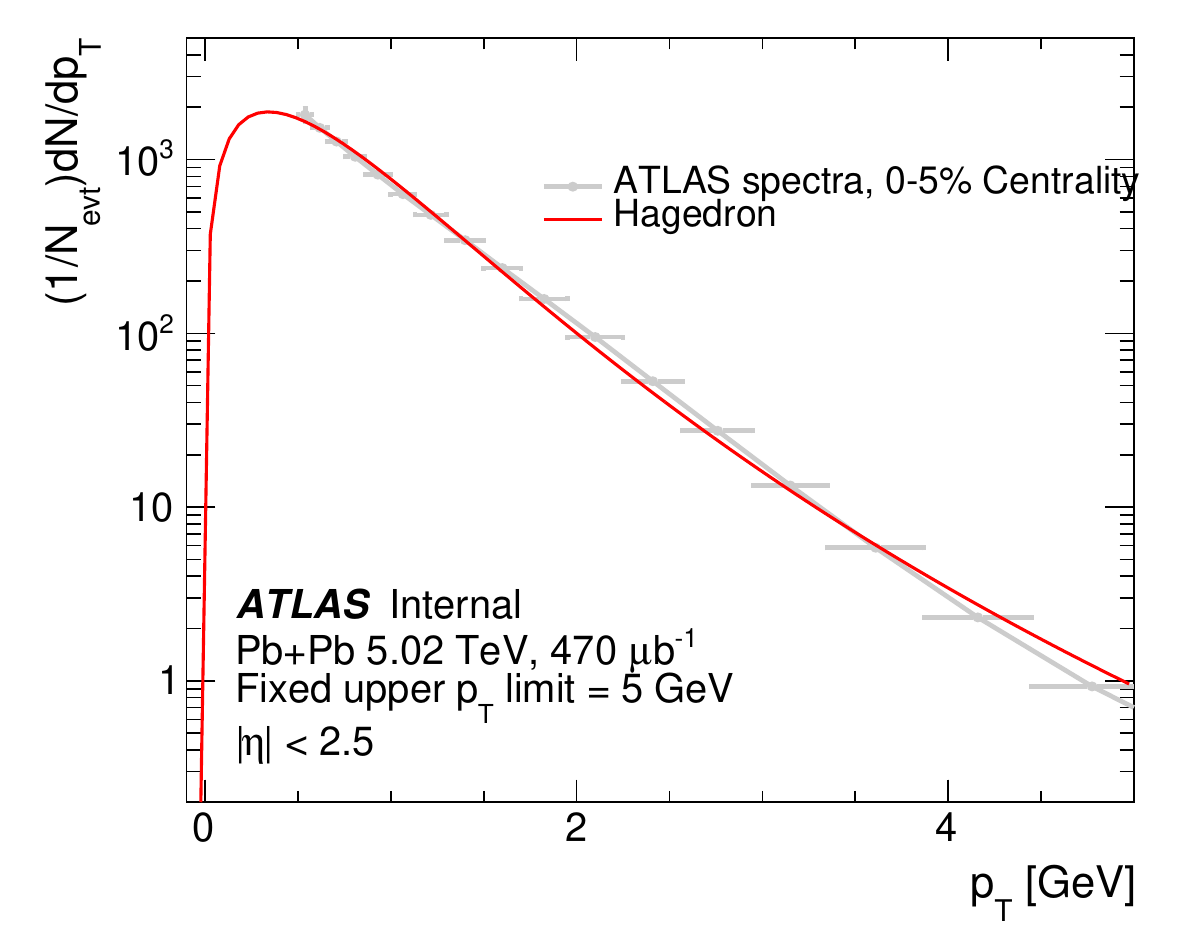}
\includegraphics[width=0.49\linewidth]{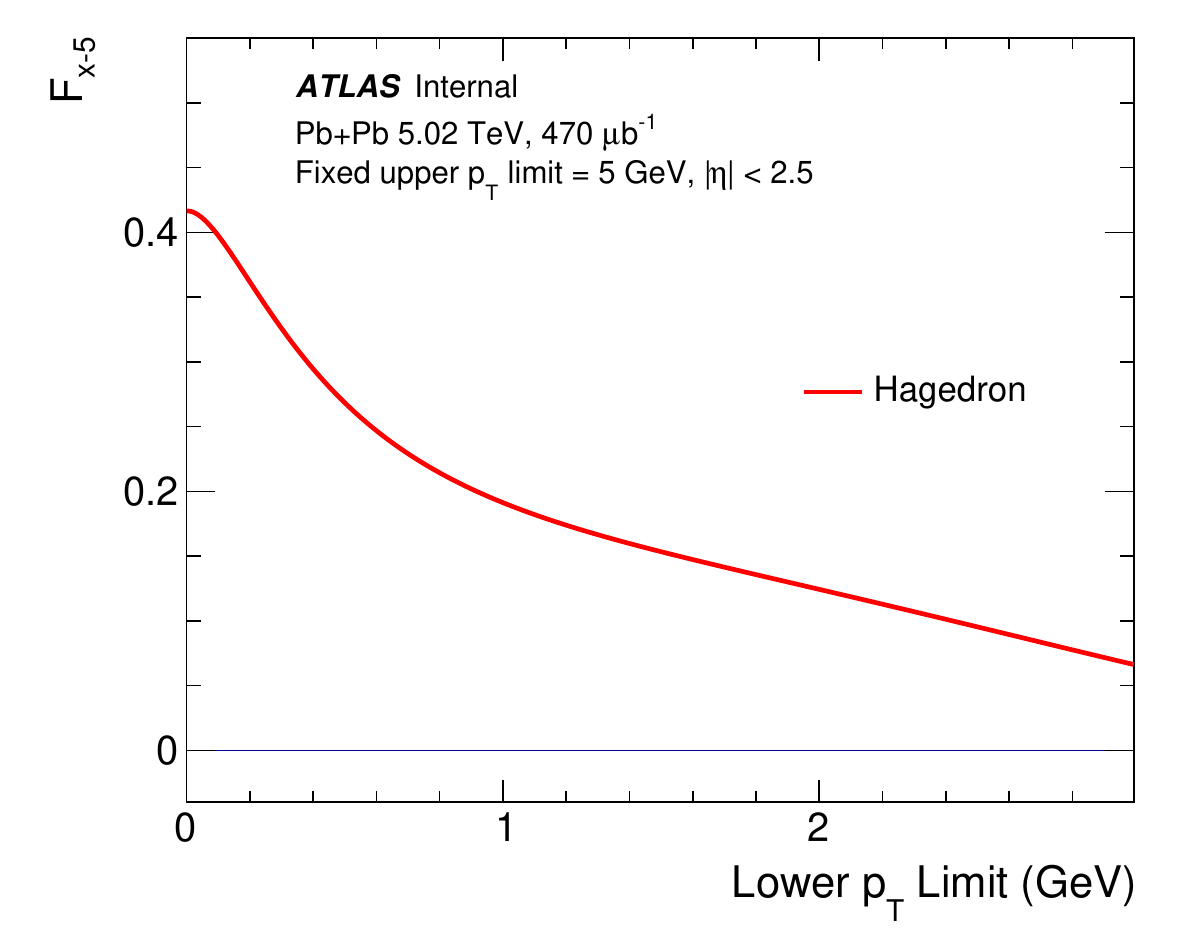}
\caption{(Left) Extrapolation of the $p_T$ spectra (0-5\% centrality) to $0 \text{ GeV}$ using the Hagedorn function (Eq.~\eqref{eq:Hagedron_redefined_main_chap5b}). (Right) The factor $F_{x-5}$ computed using the Hagedorn-extrapolated $p_T$ spectra as a function of the lower $p_T$ limit $x$, with the upper limit fixed at $5 \text{ GeV}$.}
\label{fig:SoSMethodology_Step1_2_chap5b}
\end{figure}

Using the extrapolated $p_T$ spectra obtained in Step 1, the function $F_{x-5}$ is computed according to Eq.~\eqref{eq:SoS_ExplainF_redefined_main_chap5b}. The subscript $x-5$ indicates that the calculation is performed for a $p_T$ range from a variable lower limit $x$ (GeV) to an upper limit fixed at $5 \text{ GeV}$. The right panel of Fig.~\ref{fig:SoSMethodology_Step1_2_chap5b} shows $F_{x-5}$ as a function of the lower $p_T$ limit $x$, calculated using the Hagedorn-extrapolated spectra.

\subsubsection{Step 3: Determination of $\alpha\chi_{0.5-5}$} \label{sec:method_step3_chap5b}
The term $\alpha\chi$ effectively relates relative changes in $\lr{[p_T]}$ to relative changes in $\Nch$ via the factor $F$. It is first determined for the experimentally accessible range $0.5 < p_T < 5 \text{ GeV}$, denoted here as $\alpha\chi_{0.5-5}$. This is achieved by fitting the slope of $\lr{[p_T]}$ versus $\Nch$ in the UCC region, as shown in the left panel of Fig.~\ref{fig:SoSMethodology_Step3_4_chap5b}.

The data are plotted as $\lr{[p_T]}/\lr{[p_T]}^{5\%}$ versus $\Nch/N^{5\%}_{\mathrm{ch}}$ and fitted with the functional form $\lr{[p_T]} = A'(\Nch/B')^{s}$ (where $A'$ and $B'$ are normalization constants and $s$ is the slope), following Refs.~\cite{Gardim:2019brr, CMS:2024sgx}. The extracted slope parameter $s$ corresponds to $\left(\frac{\Delta [\pT]/\lr{[p_T]}}{\Delta \Nch/\Nch}\right)_{0.5-5}$. The fitting range is optimized to capture the UCC behavior reliably. The measured slope is then combined with $F_{0.5-5}$ (calculated in Step 2 using $x=0.5 \text{ GeV}$) to obtain $\alpha\chi_{0.5-5}$:
\begin{align}\label{eq:alphachi1_redefined_main_chap5b}
\alpha\chi_{0.5-5} = \frac{1}{F_{0.5-5}}\left(\frac{\Delta [\pT]/\lr{[p_T]}}{\Delta \Nch/\Nch}\right)_{0.5-5}
\end{align}

\subsubsection{Step 4: Extrapolation of $(\alpha\chi)$ and Calculation of $c_s^2$} \label{sec:method_step4_chap5b}
\begin{figure}[H]
\centering
\includegraphics[width=0.49\linewidth]{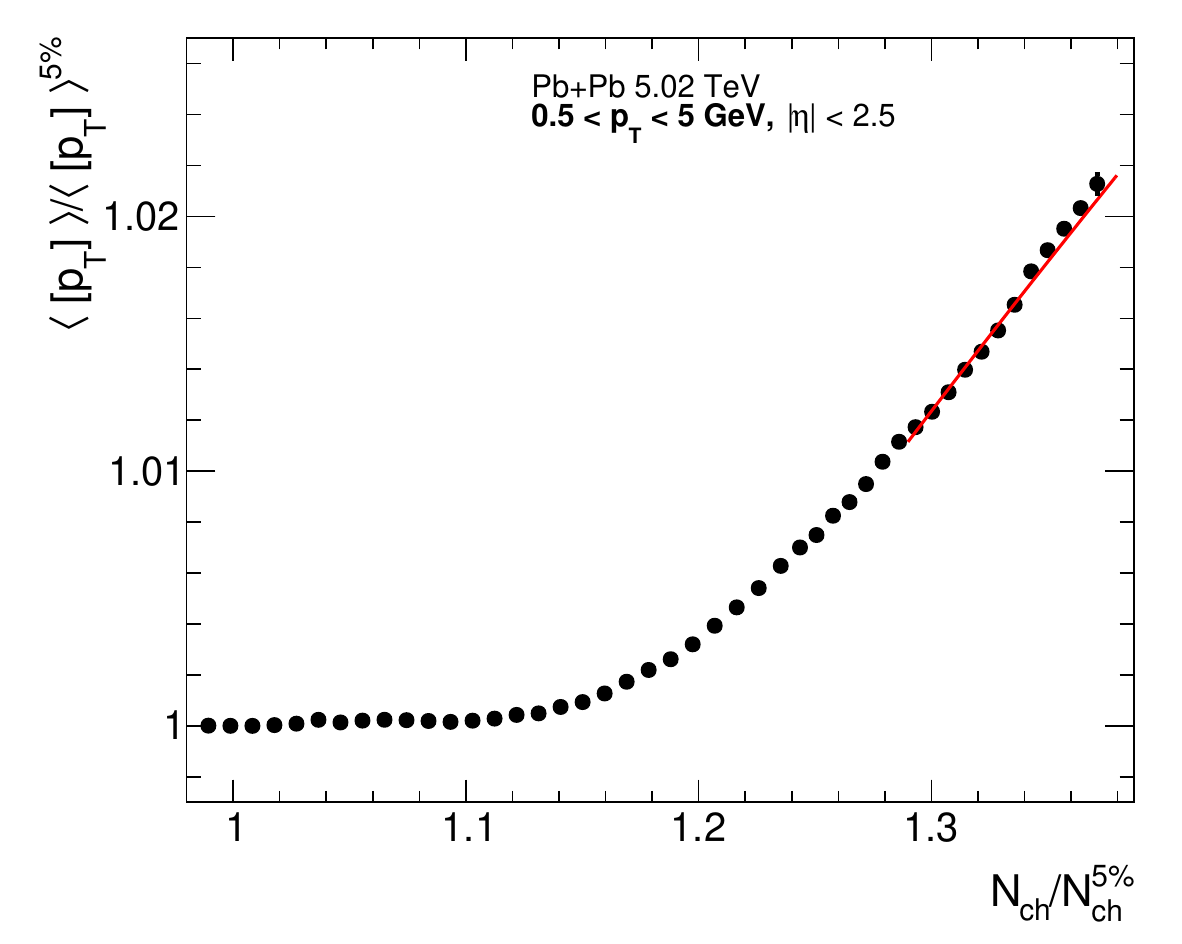}
\includegraphics[width=0.49\linewidth]{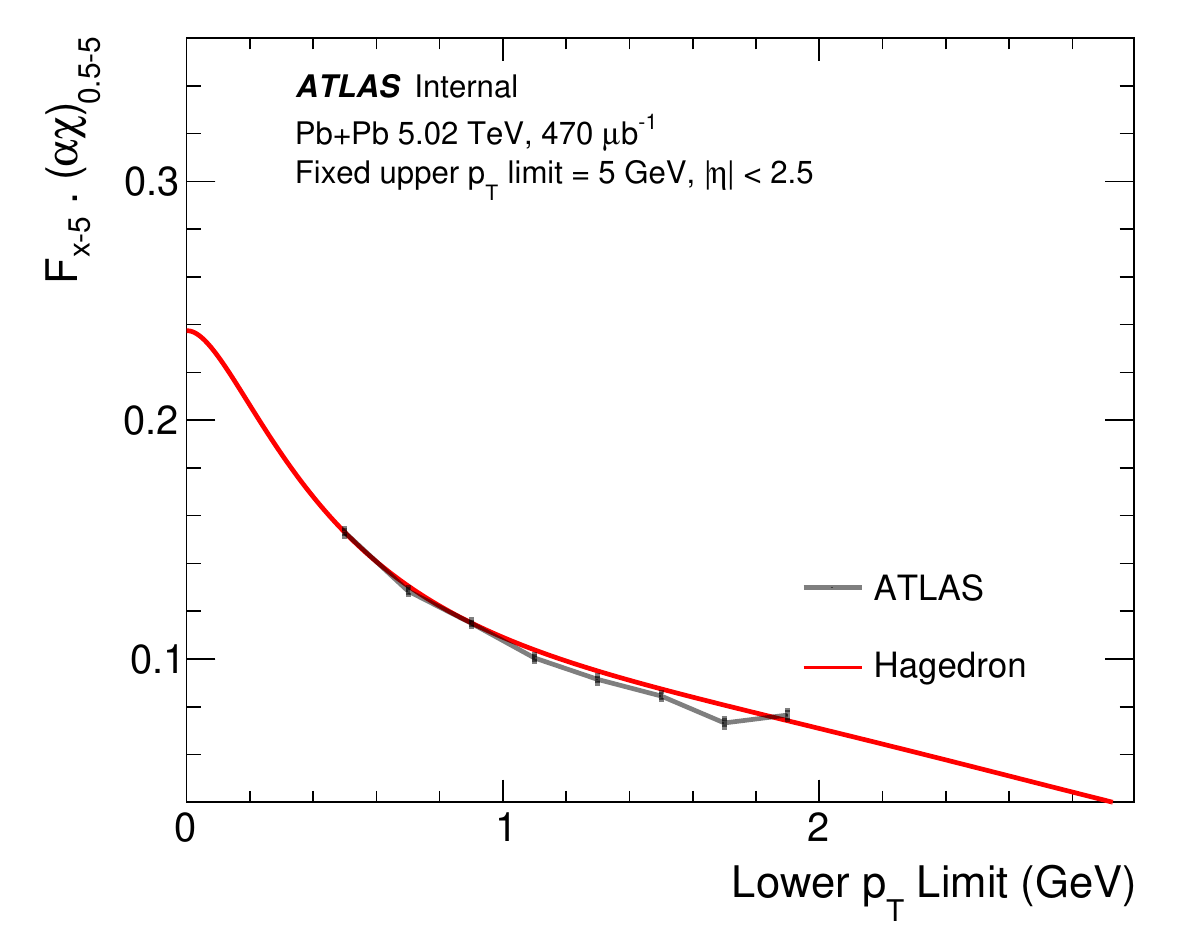}
\caption{(Left) Fit to $\lr{[p_T]}$ vs $\Nch$ in UCC (normalized axes) for the $0.5 < \pT < 5 \text{ GeV}$ range to extract the slope $\left(\frac{\Delta [\pT]/\lr{[p_T]}}{\Delta \Nch/\Nch}\right)_{0.5-5}$. This slope is then used with $F_{0.5-5}$ to calculate $\alpha\chi_{0.5-5}$. Statistical errors are shown. (Right) Extraction of the speed of sound, $c_s^2$. The points show $\text{slope}_{x-5}$ (directly measured for $x \ge 0.5 \text{ GeV}$) and the curve represents $F_{x-5} \alpha\chi_{0.5-5}$. The value of $F_{x-5} \alpha\chi_{0.5-5}$ at $p_{T,\min} (\text{or } x) = 0 \text{ GeV}$ is taken as $c_s^2$.}
\label{fig:SoSMethodology_Step3_4_chap5b}
\end{figure}

To determine $c_s^2$, which corresponds to $\text{slope}_{0-5}$ (i.e., the slope for the $p_T$ range $0 \text{ GeV}$ to $5 \text{ GeV}$), we require the value of $\alpha\chi_{0-5}$. A key assumption of this methodology is that $\alpha\chi_{x-5}$ is approximately constant, or varies slowly, with the lower $p_T$ limit $x$, for a fixed upper limit $p_{\text{max}} = 5 \text{ GeV}$. Thus, we assume $\alpha\chi_{0-5} \approx \alpha\chi_{0.5-5}$. This assumption allows us to write:
\begin{align}\label{eq:cs2Formula_redefined_main_chap5b}
c_s^2 = \text{slope}_{0-5} = F_{0-5} \alpha\chi_{0-5} \approx F_{0-5} \alpha\chi_{0.5-5}
\end{align}
The validity of the slow variation of $\alpha\chi_{x-5}$ with $x$ is tested by comparing the directly measured slopes, $\text{slope}_{x-5}$ (from data within various $p_T$ windows $x-5 \text{ GeV}$, where $x \ge 0.5 \text{ GeV}$), with the curve $F_{x-5} \alpha\chi_{0.5-5}$. As shown in the right panel of Fig.~\ref{fig:SoSMethodology_Step3_4_chap5b}, the measured slopes for different $p_{T,\min}$ values (with $p_{T,\max}=5 \text{ GeV}$) agree well with the curve generated under this assumption. Finally, $c_s^2$ is obtained by evaluating $F_{x-5} \alpha\chi_{0.5-5}$ at $x=0 \text{ GeV}$ (i.e., using $F_{0-5}$). Using this method, the value converges to $c_s^2 = 0.237$ at $p_{T,\min} = 0 \text{ GeV}$, as illustrated in the right panel of Fig.~\ref{fig:SoSMethodology_Step3_4_chap5b}.

\subsection{Systematic Uncertainty in Extraction of $c_s^2$} \label{sec:extract_cs2_syst_chap5b}
Several sources of systematic uncertainty can impact the extracted value of $c_s^2$. These are discussed and quantified in the following subsections.
\begin{enumerate}
    \item \textbf{Choice of Extrapolation Function:}\\
    Uncertainty in $c_s^2$ arises from the method used to extrapolate the $p_T$ spectra to $0 \text{ GeV}$. While this analysis defaults to the Hagedorn function (Eq.~\eqref{eq:Hagedron_redefined_main_chap5b}), alternative extrapolations using two Levy-Tsallis variants (Eqs.~\eqref{eq:Levy-Tsallis_redefined_main_chap5b},~\eqref{eq:Levy-Tsallis2_redefined_main_chap5b}) are performed to assess this uncertainty. Figure~\ref{fig:Sys_spectraFitFunc_chap5b} illustrates these functions and their impact on the extracted $c_s^2$. The first Levy-Tsallis form yields a $c_s^2$ value close to the Hagedorn result, whereas the second form introduces a systematic variation of less than 7\%.
    \begin{figure}[H]
    \centering
    \includegraphics[width=0.49\linewidth]{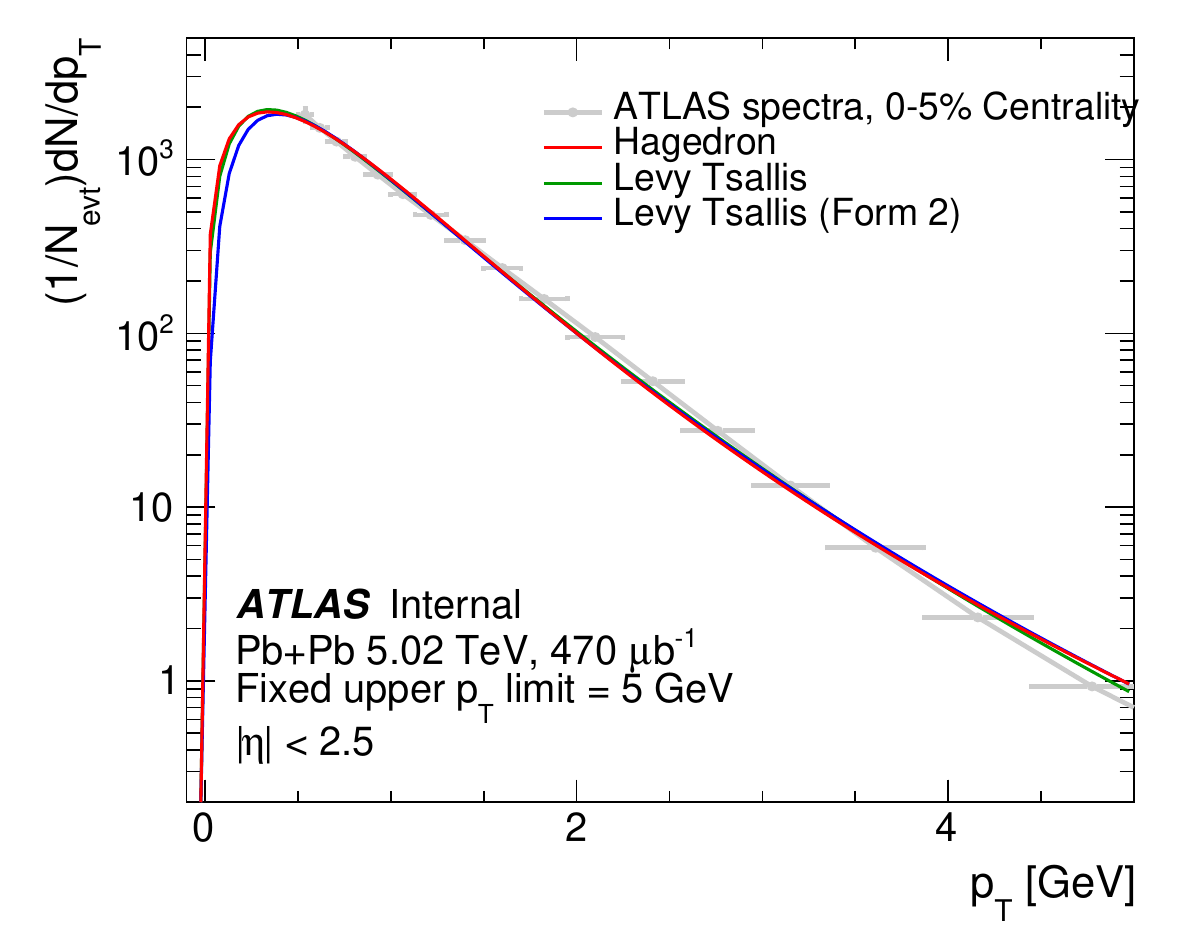}
    \includegraphics[width=0.49\linewidth]{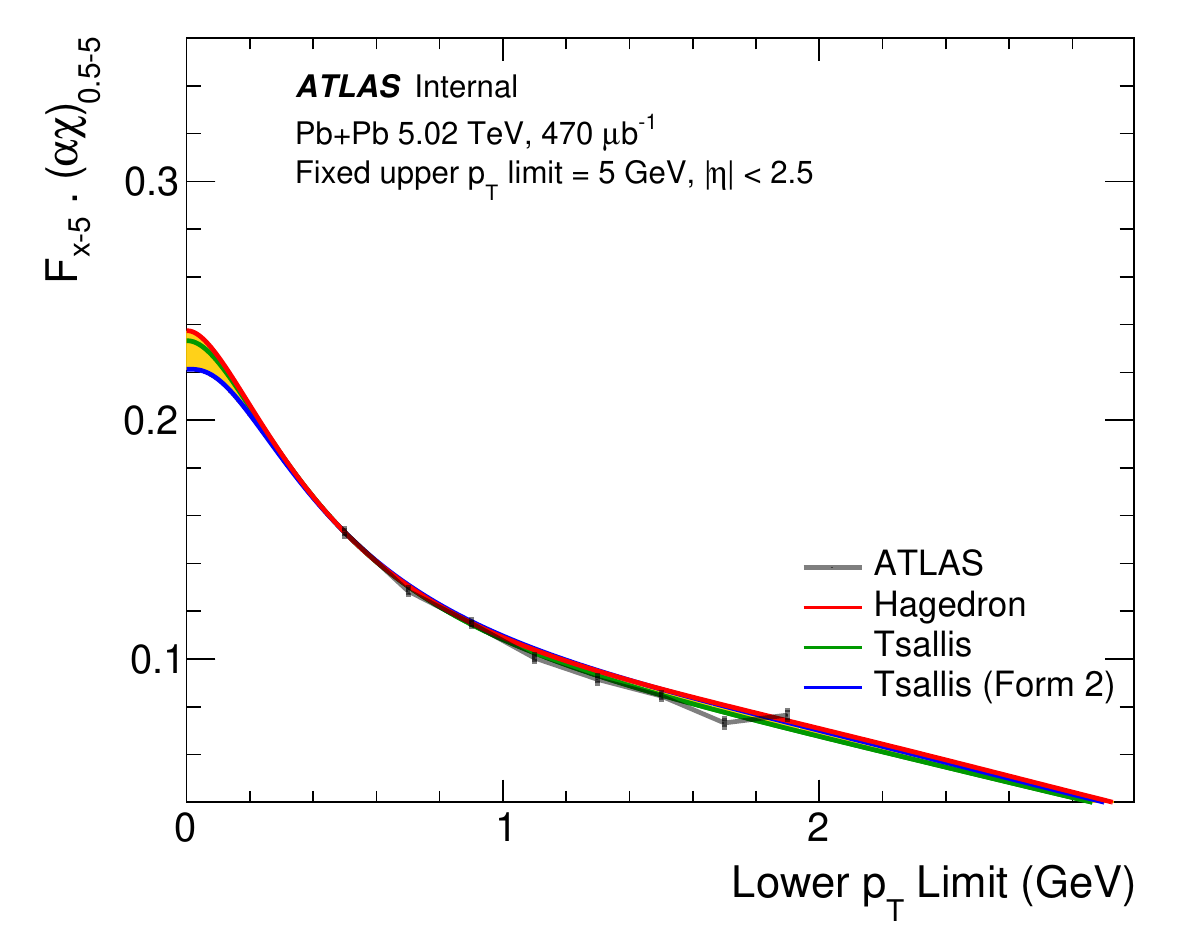}
    \caption{(Left) Extrapolation of $p_T$ spectra to $0 \text{ GeV}$ using Hagedorn and two Levy-Tsallis parameterizations (Eqs.~\eqref{eq:Hagedron_redefined_main_chap5b},~\eqref{eq:Levy-Tsallis_redefined_main_chap5b}, and \eqref{eq:Levy-Tsallis2_redefined_main_chap5b}). (Right) Computation of the curve $F_{x-5} \alpha\chi_{0.5-5}$ using $p_T$ spectra extrapolated with the different functions, illustrating the impact on the $c_s^2$ extraction path.}
    \label{fig:Sys_spectraFitFunc_chap5b}
    \end{figure}

    \item \textbf{Fitting Range for Spectra Extrapolation:}\\
    The $p_T$-range used to fit the extrapolation functions can influence the shape of the extrapolated spectra at low $p_T$ and, consequently, the value of $c_s^2$. To quantify this, the upper limit of the fitting range for the Hagedorn function was varied (e.g., $0.5 < p_T < 2 \text{ GeV}$, $0.5 < p_T < 3 \text{ GeV}$, etc., up to the default $0.5 < p_T < 5 \text{ GeV}$). Figure~\ref{fig:Sys_spectraFitRang_chap5b} demonstrates that this variation contributes approximately 8\% to the systematic uncertainty on $c_s^2$.
    \begin{figure}[H]
    \centering
    \includegraphics[width=0.49\linewidth]{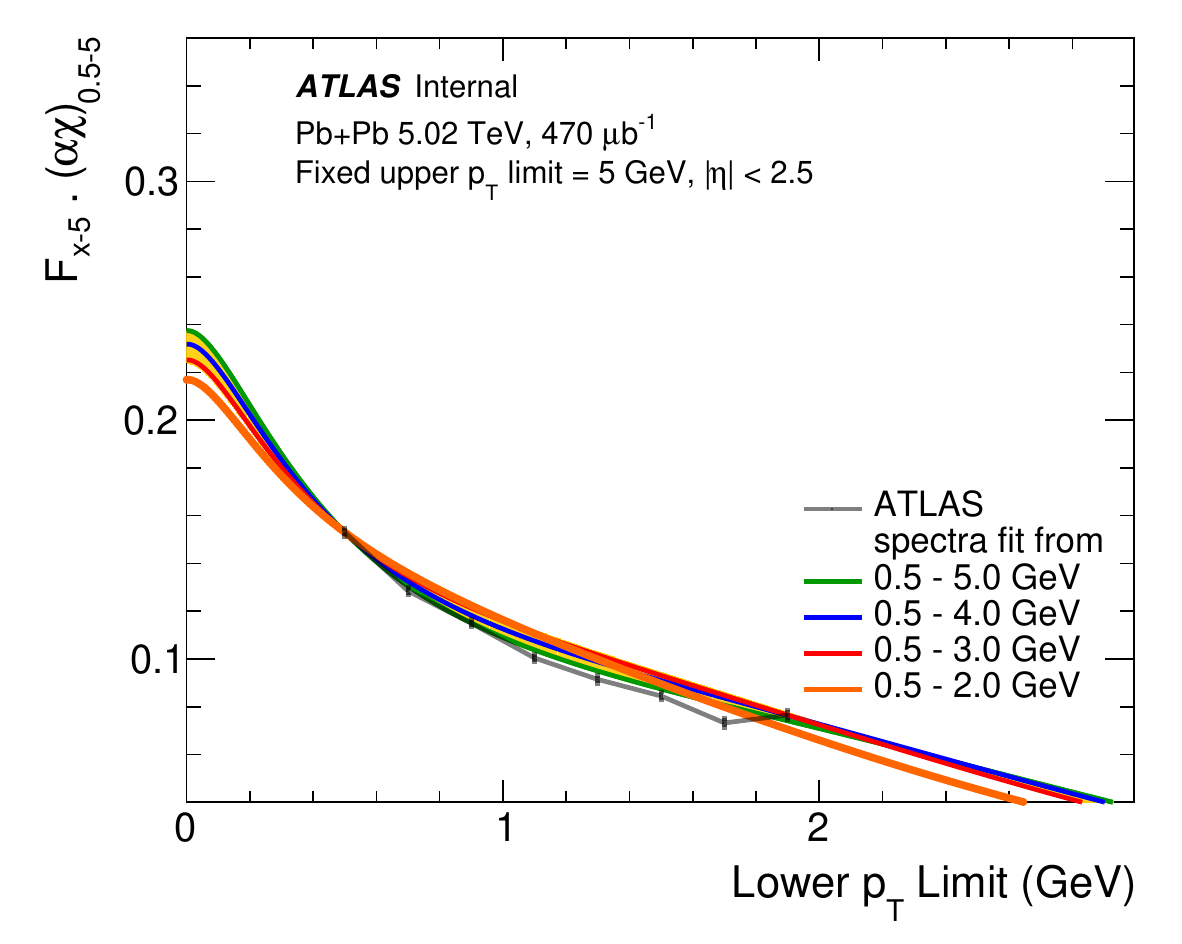}
    \caption{Effect of varying the $p_T$-range selected for fitting the $p_T$ spectra (0-5\% centrality) with the Hagedorn function on the extracted $F_{x-5} \alpha\chi_{0.5-5}$ curve used for $c_s^2$ determination.}
    \label{fig:Sys_spectraFitRang_chap5b}
    \end{figure}

    \item \textbf{Fitting Range for UCC Slope ($\lr{[p_T]}$ vs $\Nch$):}\\
    The estimation of $\alpha\chi_{0.5-5}$ from the slope of $\lr{[p_T]}$ versus $\Nch$ in UCC is sensitive to the chosen $\Nch$ fitting range. The default range in this study is $\Nch > \Nch^{\text{knee}} + 2.5\sigma_{\Nch}$, corresponding to $\Nch/N^{5\%}_{\mathrm{ch}} > 1.286$. This choice is based on Ref.~\cite{Gardim:2019brr}, which suggests fitting for $\Nch > \Nch^{\text{knee}}$ (the "knee" of the multiplicity distribution), and differs from the fixed range used by CMS~\cite{CMS:2024sgx}. To estimate the systematic uncertainty, this lower limit of the fitting range was varied by $\pm \delta_{(\Nch/N^{5\%}_{\mathrm{ch}})}$. This variation results in an uncertainty of about 6\% on $c_s^2$, as depicted in Fig.~\ref{fig:Sys_mptnchFitRang_chap5b}. Variations by $\pm 2\delta_{(\Nch/N^{5\%}_{\mathrm{ch}})}$ showed minimal additional impact.
    \begin{figure}[H]
    \centering
    \includegraphics[width=0.49\linewidth]{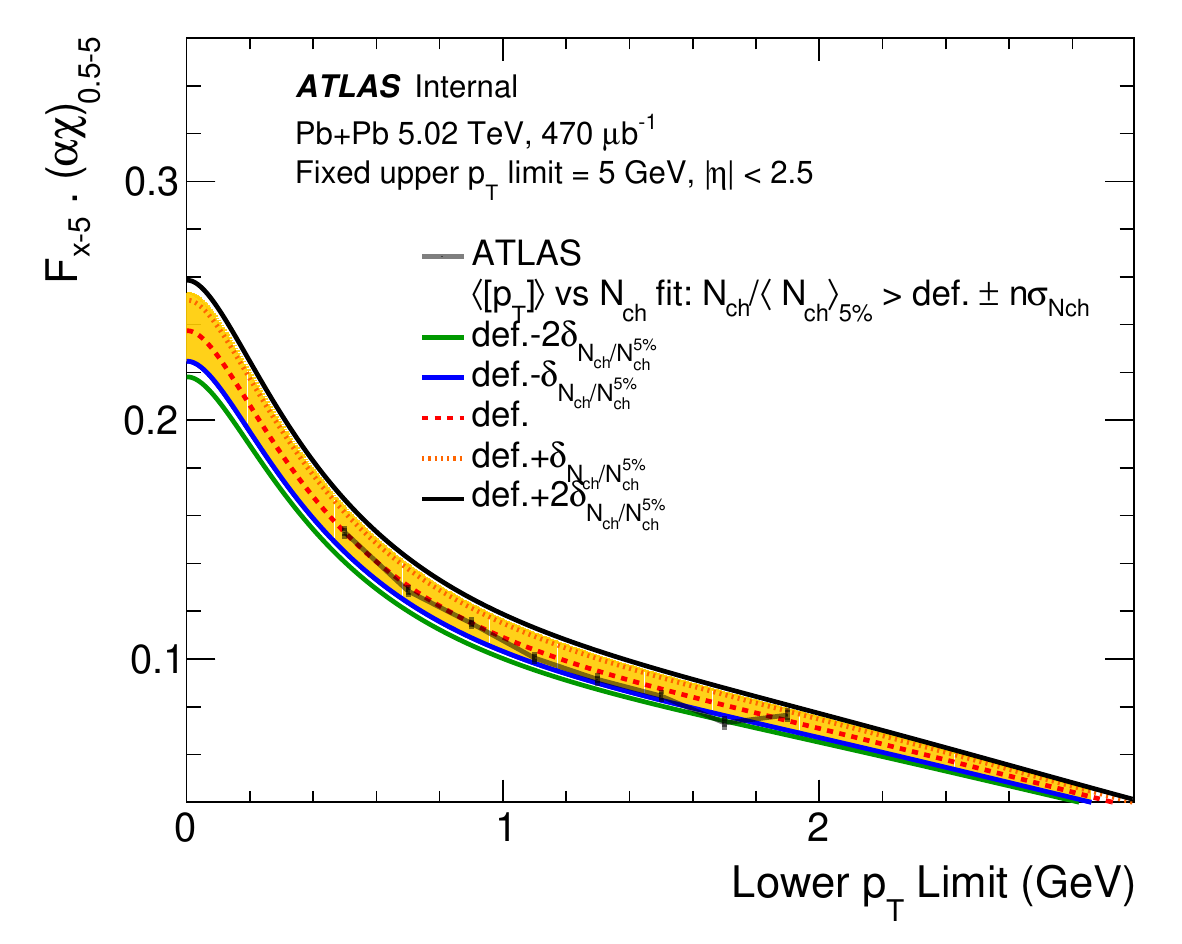}
    \caption{Effect of varying the $\Nch$-range for fitting $\lr{[p_T]}$ vs $\Nch$ on the determination of the $F_{x-5} \alpha\chi_{0.5-5}$ curve used for $c_s^2$ extraction.}
    \label{fig:Sys_mptnchFitRang_chap5b}
    \end{figure}

    \item \textbf{Estimation of $\alpha\chi_{0-5}$:}\\
    The primary assumption for estimating $\alpha\chi_{0-5}$ is that $\alpha\chi_{x-5} \approx \alpha\chi_{0.5-5}$, implying that $(\alpha\chi)$ is a slowly varying function of the lower $p_T$ limit, $x$, when the upper limit is fixed. To test this, $\alpha\chi_{x-5}$ was calculated by extracting the UCC slope for various $p_T$ windows (e.g., $x$ between 0.5 and 2 GeV, with $p_{T,\text{max}}=5$ GeV) and using the corresponding $F_{x-5}$ values. The resulting variations in the $c_s^2$ value obtained by extrapolating these different $\alpha\chi_{x-5}$ values (assuming each is constant down to $x=0$) were found to be negligible, as shown in Fig.~\ref{fig:Sys_matching_chap5b}. This validates the assumption and indicates a minimal systematic contribution from this source.
    \begin{figure}[H]
    \centering
    \includegraphics[width=0.49\linewidth]{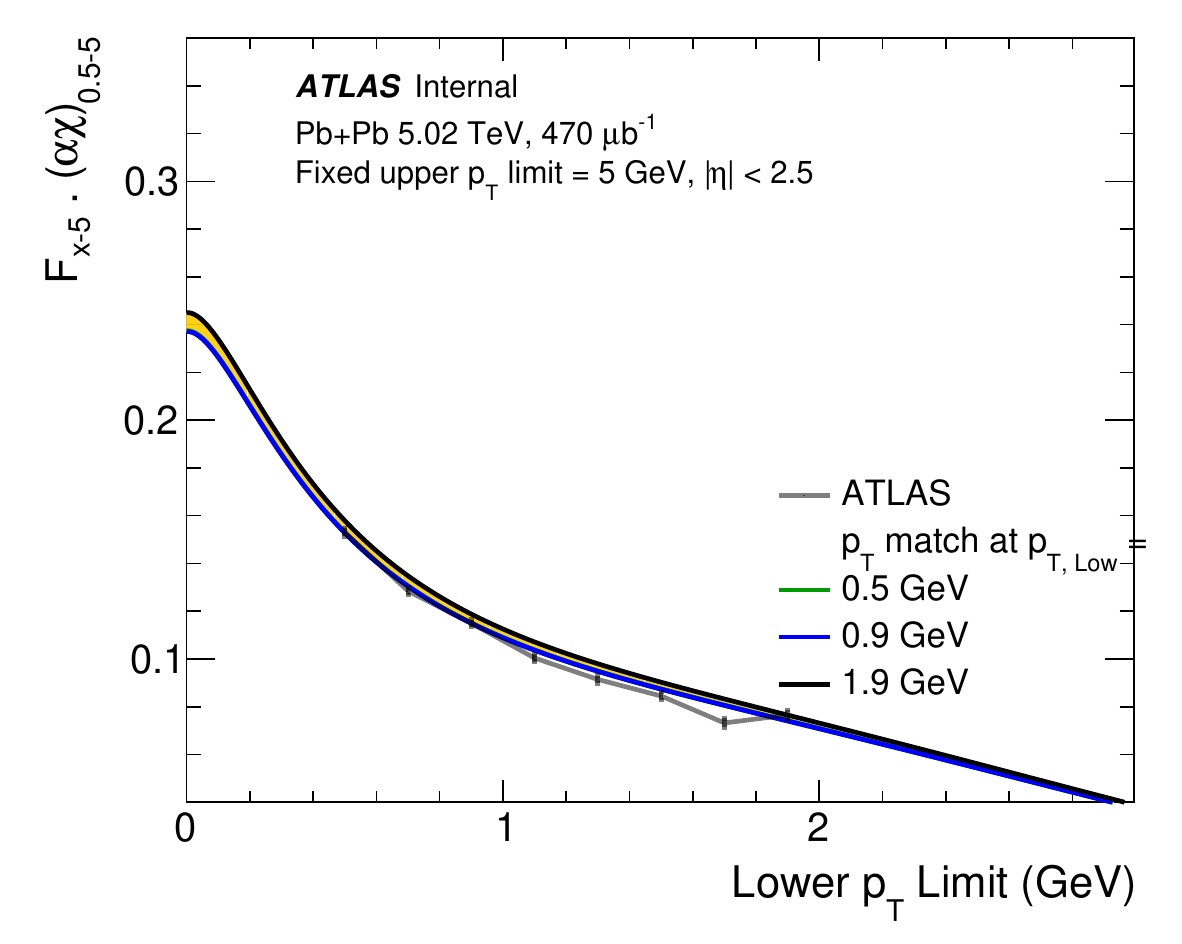}
    \caption{Effect on the $c_s^2$ determination (value of $F_{x-5} \alpha\chi_{\text{ref}}$ at $x=0$) from varying the $p_{T,\min}$ (denoted as $x_{\text{ref}}$) used for the reference estimation of $\alpha\chi_{\text{ref}}$, assuming $\alpha\chi_{x-5}$ is constant for $x < x_{\text{ref}}$.}
    \label{fig:Sys_matching_chap5b}
    \end{figure}

    \item \textbf{Standard Experimental Uncertainties Affecting Slope:}\\
Standard experimental systematic uncertainties, which are detailed in Appendix~\label{sec:sysptfluc}, primarily affect the magnitude of $\lr{[p_T]}$ and thus the measured slope of $\lr{[p_T]}$ versus $\Nch$. Their impact on the extracted $c_s^2$ is evaluated by propagating these uncertainties. Figure~\ref{fig:Sys_def_chap5b} shows that the combined systematic uncertainty on $c_s^2$ from these sources exhibits no strong dependence on the lower $p_T$-limit of the analysis range ($x$). An average of this uncertainty across the considered data points is extrapolated to $p_T = 0 \text{ GeV}$, yielding an estimated systematic contribution of approximately 1\% to $c_s^2$.
    \begin{figure}[H]
    \centering
    \includegraphics[width=0.49\linewidth]{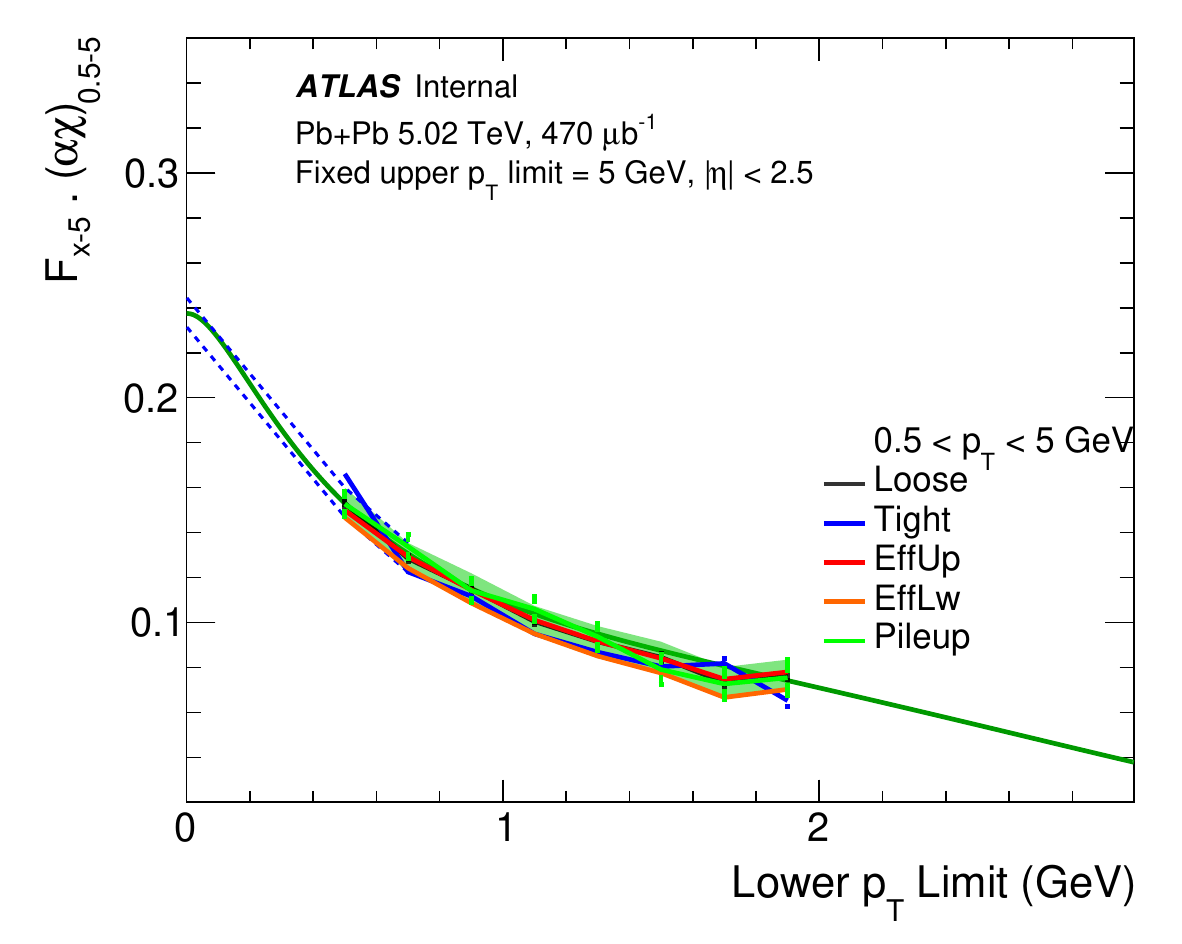}
    \caption{Effect of standard experimental systematic variations (detailed in a previous chapter/section) on the extracted value of $F_{x-5} \alpha\chi_{0.5-5}$ as a function of $p_{T,\min}$ (denoted as $x$).}
    \label{fig:Sys_def_chap5b}
    \end{figure}
\end{enumerate}

\begin{figure}[htbp]
\centering
\includegraphics[width=0.9\linewidth]{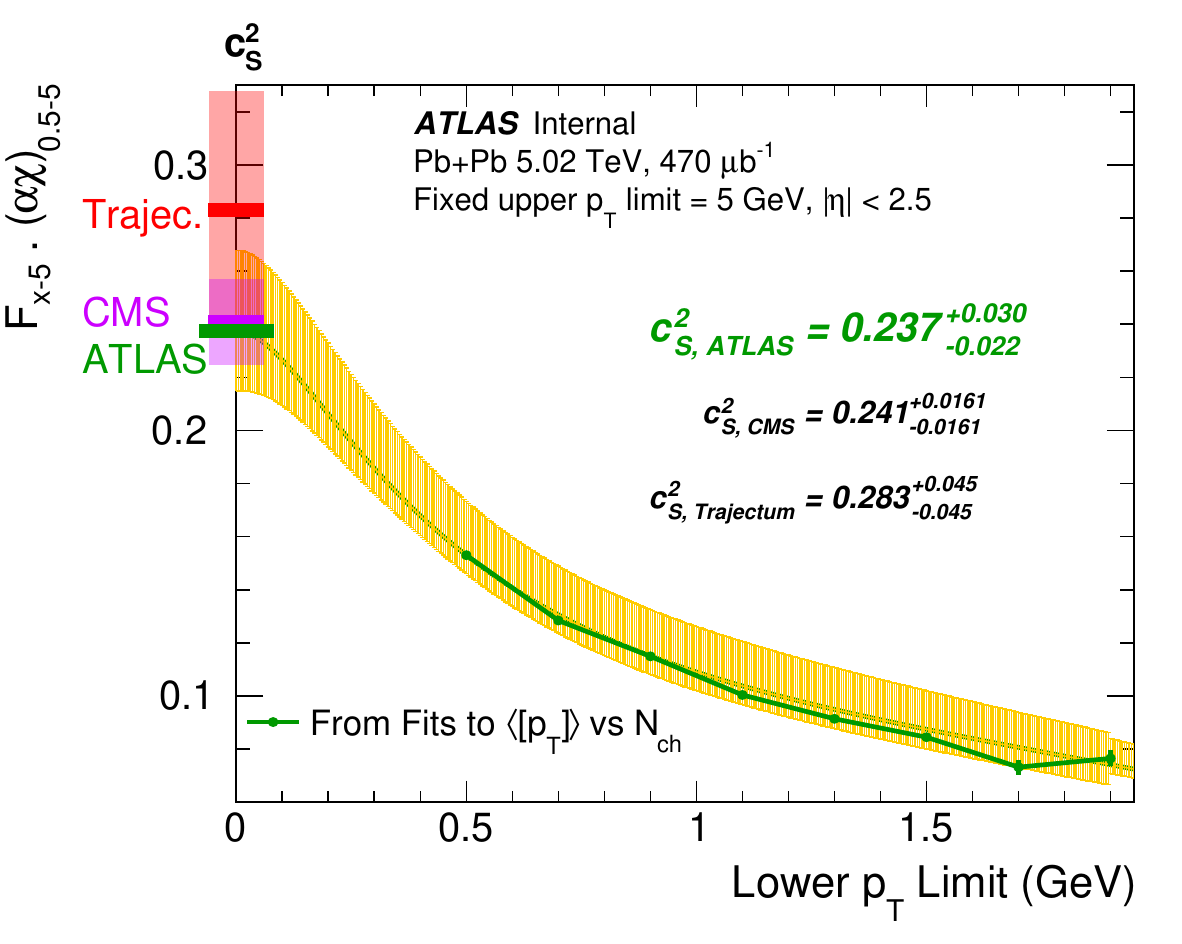}
\caption{$F_{x-5} \alpha\chi_{0.5-5}$ calculated from ATLAS data with varying $p_{\mathrm{T,min}}$ (denoted as $x$). The value of this curve at $p_{\mathrm{T,min}} = 0 \text{ GeV}$ is the extracted $c_s^2$. The value of $c_s^2$ reported by CMS and predicted by Trajectum are given by the magenta and red colored boxes, respectively. The shaded boxes and bands represent the total errors on these values of $c_s^2$.}
\label{fig:totcS2_chap5b}
\end{figure}

The contributions from these different sources of systematic uncertainties are added in quadrature to obtain the total systematic uncertainty on the extracted value of $c_s^2$. Figure~\ref{fig:totcS2_chap5b} shows the derived quantity $F_{x-5} \alpha\chi_{0.5-5}$ (which corresponds to the slope of $\lr{[p_T]}$ vs $\Nch$ under the assumption of constant $\chi$) for varying lower $p_T$ limits $x$, with a fixed upper $p_T$-limit of 5 GeV. This curve converges to $c_s^2$ at $p_T = 0 \text{ GeV}$. 

From the current analysis, the extracted value is {\bf $c_s^2 = 0.237^{+0.031}_{-0.024}$}. For comparison, the value of $c_s^2$ reported by CMS~\cite{CMS:2024sgx} is shown as a magenta box, and the prediction from the Trajectum hydrodynamic model~\cite{Nijs:2021clz} is shown as a red box, both evaluated at $p_T=0 \text{ GeV}$. The corresponding shaded bands represent their respective combined errors reported on the extracted value of $c_s^2$. As can be observed, our analysis, using an independent and novel extraction method, reports a value of $c_s^2$ consistent with that reported by CMS.

\section{Conclusion} \label{sec:chap5b_conclusion}

This chapter has detailed a novel method for the direct extraction of the speed of sound squared, $c_s^2$, in the Quark-Gluon Plasma. The method leverages experimentally accessible $p_T$ ranges and published $p_T$ spectra from central Pb+Pb collisions, focusing on the relationship between $\lr{[p_T]}$ and $\Nch$ in ultra-central collisions. A key aspect of this approach is its systematic treatment of extrapolating observed quantities to the full kinematic phase space relevant for an accurate $c_s^2$ determination. From this analysis, we report a value of $c_s^2 = 0.237^{+0.031}_{-0.024}$. This result is consistent with values previously reported by the CMS collaboration, thereby validating our independent approach.

Alongside the $c_s^2$ extraction, this work has also highlighted findings regarding the general kinematic sensitivities of $\lr{[p_T]}$ cumulants. The observed self-similarity of the second cumulant, $k_2$, when appropriately scaled and considered across different kinematic ranges in UCC suggests that the relative contributions from geometrical and intrinsic fluctuations remain constant, even with varying $p_T$ cuts. This scaling behavior is further elucidated by the theoretical framework discussed in Section~\ref{sec:theory_pt_scaling}, through the role of the acceptance correction factor $C_A$.

This $k_2$ scaling indicates that the initial-state contributions are not disproportionately affected by the dynamics of the later stages of QGP evolution, which typically govern the kinematic dependence of many observables in heavy-ion collisions. The consistency of these higher-order scaled cumulants across various $p_T$ ranges further implies that these initial-state contributions are independent of kinematic selection beyond an overall scaling factor, a phenomenon explained by the factor $C_A$~\cite{Parida:2024ckk}.

The proposed method for $c_s^2$ extraction is distinct from other approaches, such as those employed by CMS and ALICE, offering an independent experimental determination of this fundamental QGP property. This work not only provides a new value for $c_s^2$ but also illuminates how the slope of $\lr{[p_T]}$ versus $\Nch$ in UCC evolves with varying lower $p_T$ cuts. This provides deeper insights into the interplay between soft and hard physics in these extreme conditions. Ultimately, this independent measurement contributes to a more robust understanding of QGP properties and aids in constraining theoretical models.

\clearpage

\newpage
\chapter*{Appendix B}\label{sec:Appendix}
\addcontentsline{toc}{chapter}{Appendix B}

\section{Methodology}\label{sec:app_method}
We outline the methodology for measurement of observables used in this dissertation.
\subsection{For $[\pT]$ Cumulants}

The analysis of $\lr{\pT}$ fluctuations begins with defining the $n$-particle $\pT$ correlator for a single event:
\begin{equation}\label{eq:cn_def}
c_{n}=\frac{\sum_{i\neq \dots \neq n } w_i \dots w_n(p_{T,i}- \MpT) \dots (p_{T,n}-\MpT)} { \sum_{i\neq \dots \neq n } w_i \dots w_n}
\end{equation}
where $w_i$ is the weight and $p_{T,i}$ is the transverse momentum for track $i$. The sums are over unique track combinations within the event, and $\MpT \equiv \lr{p_{11}}$ denotes the event-averaged mean $\pT$. Following established methods, $c_n$ are expanded algebraically using auxiliary quantities $p_{m k} = \sum_{i}w_{i}^{k}p^{m}_{T,i} / \sum_{i}w_{i}^{k}$ and their centered counterparts:
\begin{align}\label{eq:pmk_centered}
& \bar{p}_{1 k} \equiv p_{1 k}-\MpT \\
& \bar{p}_{2 k} \equiv p_{2 k}-2p_{1 k}\MpT+\MpT^{2} \notag \\
& \bar{p}_{3 k} \equiv p_{3 k}-3p_{2 k}\MpT+3p_{1 k}\MpT^{2}-\MpT^{3} \\
& \bar{p}_{4 k} \equiv p_{4 k}-4p_{3 k}\MpT+6p_{2 k}\MpT^{2}-4p_{1 k}\MpT^{3}+\MpT^{4} \quad \text{etc.} \notag
\end{align}
For example, the 2- and 3-particle correlators ($c_2$ and $c_3$) can be expressed in terms of $\bar{p}_{mk}$ and $\tau_{k}=\sum_{i} w_{i}^{k+1} / (\sum_{i} w_{i})^{k+1}$.

Event-averaged $\lr{\pT}$ cumulants are then obtained by averaging the event correlators $c_n$ over an ensemble of events, $\lr{c_{n}}$:

To minimize artificial contributions, the cumulants of $[\pT]$ distribution are normalized into dimensionless quantities:
\begin{align}\label{eq:kn_normalized_def}
& \mathrm{Normalized\;Variance} = k_{2} = \frac{\lr{c_{2}}}{\lr{[\pT]}^2}, \\
& \mathrm{Normalized\;Skewness} =\gamma = \frac{\lr{c_{3}}}{\lr{c_2}^{3/2}}, \\
& \mathrm{Intensive\;Skewness} =\Gamma = \frac{\lr{c_{3}}\lr{[\pT]}}{\lr{c_2}^{2}}, \\
& \mathrm{Normalized\;Kurtosis} =\kappa = \frac{\lr{c_{4}}-3\lr{c_{2}}^2}{\lr{c_2}^{2}}.
\end{align}

\subsection{For $v_0(\pT)$ Observable}

The $\pT$-differential observable $v_0(\pT)$ probes event-by-event fluctuations in the shape of the $\pT$ spectrum. To avoid mixing correlated tracks, a two-subevent method is employed. The fluctuations in the yield and normalized yield of the $\pT$ spectra in subevent 'a' are given by:
\begin{equation}
\begin{aligned}
\delta N_{\text{a}}(\pT) &= N_{\text{a}}(\pT) - \lr{N_{\text{a}}(\pT)} \\
\delta n_{\text{a}}(\pT) &= n_{\text{a}}(\pT) - \lr{n_{\text{a}}(\pT)}
\end{aligned}
\end{equation}
where
\begin{equation*}
N_{\text{a}}(\pT) = \sum_{i \in a}^{\pT} w_i
\end{equation*}
and
\begin{equation*}
n_{\text{a}}(\pT) = \frac{N_{\text{a}}(\pT)}{S_{\text{a}1}}
\end{equation*}
with
\begin{equation*}
S_{\text{a}1} = \sum_{i \in a} w_i.
\end{equation*}
The fluctuation in the event-averaged $\pT$ for subevent 'b' is:
\begin{equation}
\delta [\pT]_{\text{b}} = \frac{\sum_{i \in b} w_i(p_{T,i} - \MpT_{\text{b}})}{S_{\text{b}1}}.
\end{equation}
The calculation of $v_0(\pT)$ and $\vOpTF$ uses cross-correlations between disjoint subevents 'a' and 'b'. The event-averaged $\pT$ and the ensemble-averaged 2-subevent correlator are:
\begin{align*}
[\pT] &= \frac{S_{\text{a}1}p_{a11} + S_{\text{b}1}p_{b11}}{S_{\text{a}1} + S_{\text{b}1}} \\
\lr{c_{2,\text{2sub}}} &= \lr{\bar{p}_{\text{a}11}\bar{p}_{\text{b}11}}
\end{align*}
For $v_0(\pT)$ (normalized spectra):
\begin{align}\label{eq:v0pT_2sub_norm}
& \nCov_{\text{ab}} = \lr{n_{\text{a}}(\pT)\bar{p}_{\text{b}11}} - \lr{n_{\text{a}}(\pT)}\lr{\bar{p}_{\text{b}11}} \\
& v_{0,\text{ab}}(\pT) = \frac{\nCov_{\text{ab}}}{\lr{n_{\text{a}}(\pT)}\sqrt{\lr{c_{2,\text{2sub}}}}}
\end{align}
For $\vOpTF$ (non-normalized spectra):
\begin{align}\label{eq:v0pT_2sub_nonorm}
& \Cov_{\text{ab}} = \lr{N_{\text{a}}(\pT)\bar{p}_{\text{b}11}} - \lr{N_{\text{a}}(\pT)}\lr{\bar{p}_{\text{b}11}} \\
& v^{\prime}_{0,\text{ab}}(\pT) = \frac{\Cov_{\text{ab}}}{\lr{N_{\text{a}}(\pT)}\sqrt{\lr{c_{2,\text{2sub}}}}}
\end{align}
The final $v_0(\pT)$ and $\vOpTF$ are weighted averages of 'ab' and 'ba' pairs. The integrated $v_0$ is defined as:
\begin{equation}\label{eq:V0_integrated_def}
v_{0} = \frac{\sqrt{\lr{c_{2,\text{2sub}}}}}{\lr{[\pT]}}
\end{equation}
The scaled observable $v_0(\pT)/v_0$ is also used.

\subsection{For $v_n - \pT$ Correlation ($\rho(v_n^2, [\pT])$)}\label{sec:method_vnpt}

We outline the methodology for calculating the Pearson Correlation Correlator (PCC), $\rho(v_n^2, [\pT])$, which quantifies the correlation between the $n^{\mathrm{th}}$-order anisotropic flow $v_n$ and the event-averaged transverse momentum $[\pT]$. 

The correlation coefficient is defined as :
\begin{equation}
    \rho(v_n\{2\}^2,[\pT])=\frac{\mathrm{cov}(v_n\{2\}^2,[\pT])}{\mathrm{Var}(v_n\{2\}^2)_{\mathrm{dyn}}c_k}
\end{equation}
We first define the following quantities :
\begin{align}
    Q_{nk}=&\sum_i w_i^k e^{in\phi_i}\\
    S_{mk}=&(\sum_i w_i^k)^m\\
    O_{nk}=&\sum_i w_i^k e^{in\phi_i}p_i\\
    P_{mk}=&\sum_i w_i^kp_i^m\\
    \llangle p\rrangle=&\frac{1}{N_{ev}}\sum_{evts}\frac{\sum_i w_ip_i}{\sum_iw_i}\\
    \bar{p}_i=&p_i-\llangle p\rrangle\\
    \bar{O}_{nk}=&\sum_i w_i^k e^{in\phi_i}\bar{p}_i=O_{nk}-Q_{nk}\llangle p\rrangle\\
    \bar{P}_{mk}=&\sum_i w_i^k\bar{p}_i^m
\end{align}
Note $P_{0k}\equiv S_{1k}$, and there is the following simple relation that are handy for later discussions:
\begin{align}\nonumber
&\bar{P}_{1k}=P_{1k}-S_{1k}\llangle p\rrangle\\\nonumber
&\bar{P}_{2k}=P_{2k}-2P_{1k}\llangle p\rrangle+S_{1k}\llangle p\rrangle^2\\\label{eq:sh}
&\bar{P}_{3k}=P_{3k}-3P_{2k}\llangle p\rrangle+3P_{1k}\llangle p\rrangle^2-S_{1k}\llangle p\rrangle^3
\end{align}

\subsubsection{Formula for Covariance}
\paragraph{standard (full-event) method}
Let us derive the formula for the covariance in case of 1 subevent. 
Now the three particle correlation of the $\mathrm{cov}(v_n\{2\}^2,[\pT])$ is given as :
\begin{align}
    \mathrm{cov}(v_n\{2\}^2,[\pT])=\frac{\sum_i\sum_j'\sum_h' w_iw_jw_he^{in(\phi_i-\phi_j)}(p_h-\llangle p\rrangle)} { \sum_i\sum_j'\sum_h' w_iw_jw_h}
\end{align}
First lets expand the denominator :
\begin{align*}
    D&=\sum_i\sum_j\sum_h w_iw_jw_h - 3\sum_i\sum_j' w_i^2w_j - \sum_i w_i^3\\
    &=S_{31}-3[\sum_i\sum_j w_i^2w_j-\sum_i w_i^3]-S_{13}\\
    &=S_{31}-3S_{12}S_{11}+3S_{13}-S_{13}\\
    &=S_{31}-3S_{12}S_{11}+2S_{13}
\end{align*}
Next lets expand the numerator :
\begin{align*}
    N=&\sum_i\sum_j\sum_h w_iw_jw_he^{in(\phi_i-\phi_j)}(p_h-\llangle p\rrangle) - \sum_i\sum_h' w_i^2w_h(p_h-\llangle p\rrangle) - 2 \sum_i\sum_j' w_i^2w_je^{in(\phi_i-\phi_j)}(p_i-\llangle p\rrangle)- \sum_i w_i^3 (p_i-\llangle p\rrangle)\\
    =&|Q_{n1}|^2P_{11}-|Q_{n1}|^2S_{11}\llangle p\rrangle-\{\sum_i\sum_h w_i^2w_h(p_h-\llangle p\rrangle)-\sum_i w_i^3(p_i-\llangle p\rrangle) \}\\
    &-2\{ \sum_i\sum_j w_i^2w_je^{in(\phi_i-\phi_j)}(p_i-\llangle p\rrangle)-\sum_i w_i^3(p_i-\llangle p\rrangle) \}-P_{13}+S_{13}\llangle p\rrangle\\
    =&|Q_{n1}|^2P_{11}-|Q_{n1}|^2S_{11}\llangle p\rrangle-S_{12}P_{11}+S_{12}S_{11}\llangle p\rrangle+P_{13}-S_{13}\llangle p\rrangle\\
    &-2\Re(O_{n2}Q_{n1}^*)+2\Re(Q_{n2}Q_{n1}^*)\llangle p\rrangle+2P_{13}-2S_{13}\llangle p\rrangle-P_{13}+S_{13}\llangle p\rrangle\\
    =&|Q_{n1}|^2P_{11}-S_{12}P_{11}-2\Re(O_{n2}Q_{n1}^*)+2P_{13}\\
    &-\llangle p\rrangle\{ |Q_{n1}|^2S_{11}-S_{12}S_{11}-2\Re(Q_{n2}Q_{n1}^*)+2S_{13} \}
\end{align*}
So the covariance for the standard method becomes :
\begin{align}
Cov[FSE]=&\frac{|Q_{n1}|^2P_{11}-S_{12}P_{11}-2\Re(O_{n2}Q_{n1}^*)+2P_{13}}{S_{31}-3S_{12}S_{11}+2S_{13}}\\
    &-\llangle p\rrangle\frac{ |Q_{n1}|^2S_{11}-S_{12}S_{11}-2\Re(Q_{n2}Q_{n1}^*)+2S_{13}}{S_{31}-3S_{12}S_{11}+2S_{13}}
\end{align}
or in terms of $\bar{p}$ :
\begin{align}
Cov[FSE]=&\frac{|Q_{n1}|^2\bar{P}_{11}-S_{12}\bar{P}_{11}-2\Re(\bar{O}_{n2}Q_{n1}^*)+2\bar{P}_{13}}{S_{31}-3S_{12}S_{11}+2S_{13}}
\end{align}

\paragraph{Two-subevent}
The three particle correlation of the $\mathrm{cov}(v_n\{2\}^2,[\pT])$ is given as :
\begin{align}
    \mathrm{cov}(v_n\{2\}^2,[\pT])=\frac{\sum_i\sum_j\sum_h' w_iw_jw_he^{in(\phi_i-\phi_j)}(p_h-\llangle p\rrangle)} { \sum_i\sum_j\sum_h' w_iw_jw_h}
\end{align}
In 2-subevent case the particle i and j belong to two different subevents - A and C and are never the same.\\
The expansion of denominator :
\begin{align*}
    D&=\sum_i\sum_j\sum_h w_iw_jw_h - \sum_i\sum_j w_i^2w_j- \sum_i\sum_j w_iw_j^2\\
    &=S_{A11}S_{C11}S_{11}-S_{A12}S_{C11}-S_{A11}S_{C12}
\end{align*}
The expansion of numerator :
\begin{align*}
    N=&\sum_i\sum_j\sum_h w_iw_jw_he^{in(\phi_i-\phi_j)}(p_h-\llangle p\rrangle) - \sum_i\sum_j w_i^2w_je^{in(\phi_i-\phi_j)}(p_i-\llangle p\rrangle) - \sum_i\sum_j w_iw_j^2e^{in(\phi_i-\phi_j)}(p_j-\llangle p\rrangle)\\
    =&\Re(Q_{An1}Q_{Cn1}^*)P_{11}-\Re(Q_{An1}Q_{Cn1}^*)S_{11}\llangle p\rrangle-\Re(O_{An2}Q_{Cn1}^*)+\llangle p\rrangle\Re(Q_{An2}Q_{Cn1}^*)\\
    &-\Re(Q_{An1}O_{Cn2}^*)+\llangle p\rrangle\Re(Q_{An1}Q_{Cn2}^*)\\
    =&\Re(Q_{An1}Q_{Cn1}^*)P_{11}-\Re(O_{An2}Q_{Cn1}^*)-\Re(Q_{An1}O_{Cn2}^*)\\
    &-\llangle p\rrangle\{\Re(Q_{An1}Q_{Cn1}^*)S_{11}-\Re(Q_{An2}Q_{Cn1}^*)-\Re(Q_{An1}Q_{Cn2}^*) \}
\end{align*}
So the covariance for the 2-subevent becomes :
\begin{align}
Cov[2SE]=&\frac{\Re(Q_{An1}Q_{Cn1}^*)P_{11}-\Re(O_{An2}Q_{Cn1}^*)-\Re(Q_{An1}O_{Cn2}^*)}{S_{A11}S_{C11}S_{11}-S_{A12}S_{C11}-S_{A11}S_{C12}}\\
    &-\llangle p\rrangle\frac{\Re(Q_{An1}Q_{Cn1}^*)S_{11}-\Re(Q_{An2}Q_{Cn1}^*)-\Re(Q_{An1}Q_{Cn2}^*)}{S_{A11}S_{C11}S_{11}-S_{A12}S_{C11}-S_{A11}S_{C12}}
\end{align}
or in terms of $\bar{p}$ : 
\begin{align}
Cov[2SE]=&\frac{\Re(Q_{An1}Q_{Cn1}^*)\bar{P}_{11}-\Re(\bar{O}_{An2}Q_{Cn1}^*)-\Re(Q_{An1}\bar{O}_{Cn2}^*)}{S_{A11}S_{C11}S_{11}-S_{A12}S_{C11}-S_{A11}S_{C12}}
\end{align}

\paragraph{Three-subevent}
The three particle correlation of the $\mathrm{cov}(v_n\{2\}^2,[\pT])$ is given as :
\begin{align}
    \mathrm{cov}(v_n\{2\}^2,[\pT])=\frac{\sum_i\sum_j\sum_h w_iw_jw_he^{in(\phi_i-\phi_j)}(p_h-\llangle p\rrangle)} { \sum_i\sum_j\sum_h w_iw_jw_h}
\end{align}
In three-subevent case the particle i and j belong to two different subevents - A and C and particle k is from the subevent B. The i,j and k particle are never the same.\\
The expansion of denominator :
\begin{align*}
    D&=\sum_i\sum_j\sum_h w_iw_jw_h\\
    &=S_{A11}S_{C11}S_{B11}
\end{align*}
The expansion of numerator :
\begin{align*}
    N=&\sum_i\sum_j\sum_h w_iw_jw_he^{in(\phi_i-\phi_j)}(p_h-\llangle p\rrangle)\\
    =&\Re(Q_{An1}Q_{Cn1}^*)P_{B11}-\Re(Q_{An1}Q_{Cn1}^*)S_{B11}\llangle p\rrangle
\end{align*}
So the covariance for the three-subevent becomes :
\begin{align}
Cov[3SE]=&\frac{\Re(Q_{An1}Q_{Cn1}^*)P_{B11}-\llangle p\rrangle \Re(Q_{An1}Q_{Cn1}^*)S_{B11}}{S_{A11}S_{C11}S_{B11}}
\end{align}
or in terms of $\bar{p}$ : 
\begin{align}
Cov[3SE]=&\frac{\Re(Q_{An1}Q_{Cn1}^*)\bar{P}_{B11}}{S_{A11}S_{C11}S_{B11}}
\end{align}

\subsubsection{Formula for $c_k$}
\paragraph{standard}
Let us derive the formula for the $c_k$ in case of 1 subevent. 
The two particle correlation is given as :
\begin{align}
    c_k=\frac{\sum_i\sum_j' w_iw_j(p_i-\llangle p\rrangle)(p_j-\llangle p\rrangle)} { \sum_i\sum_j' w_iw_j}
\end{align}
The expansion of the denominator :
\begin{align*}
    D=&\sum_i\sum_j w_iw_j - \sum_i w_i^2\\
    =&S_{21}-S_{12}
\end{align*}
The expansion of numerator :
\begin{align*}
    N=&\sum_i\sum_j w_iw_j(p_i-\llangle p\rrangle)(p_j-\llangle p\rrangle)- \sum_i w_i^2 (p_i-\llangle p\rrangle)^2\\
    =&\sum_i\sum_jw_iw_j\{ p_ip_j-\llangle p\rrangle p_i-\llangle p\rrangle p_j +\llangle p\rrangle^2 \} - \sum_i w_i^2 \{ p_i^2+\llangle p\rrangle^2-2p_i\llangle p\rrangle \}\\
    =&P_{11}^2-2\llangle p\rrangle P_{11}S_{11}+\llangle p\rrangle^2S_{21}-P_{22}-\llangle p\rrangle^2S_{12}+2\llangle p\rrangle P_{12}\\
    =& P_{11}^2-P_{22}-2\llangle p\rrangle (P_{11}S_{11}-P_{12})+\llangle p\rrangle^2 (S_{21}-S_{12})
\end{align*}
So the $c_k$ is :
\begin{align}
    c_k=\frac{P_{11}^2-P_{22}-2\llangle p\rrangle (P_{11}S_{11}-P_{12})+\llangle p\rrangle^2 (S_{21}-S_{12})}{S_{21}-S_{12}}
\end{align}
or in terms of $\bar{p}$ (We should simply use this form since we can use Eq.~\eqref{eq:sh}):
\begin{align}
    c_k=\frac{\bar{P}_{11}^2-\bar{P}_{22}}{S_{21}-S_{12}}
\end{align}

\paragraph{Two-subevent}
The two particle correlation is given as :
\begin{align}
    c_k=\frac{\sum_i\sum_j w_iw_j(p_i-\llangle p\rrangle_A)(p_j-\llangle p\rrangle_C)} { \sum_i\sum_j w_iw_j}
\end{align}
In 2-subevent case the particle i and j belong to two different subevents - A and C and are never the same.\\
The expansion of the denominator :
\begin{align*}
    D=&\sum_i\sum_j w_iw_j\\
    =&S_{A11}*S_{C11}
\end{align*}
The expansion of numerator :
\begin{align*}
    N=&\sum_i\sum_j w_iw_j(p_i-\llangle p\rrangle_A)(p_j-\llangle p\rrangle_C)\\
    =&\sum_i\sum_jw_iw_j\{ p_ip_j-\llangle p\rrangle_C p_i-\llangle p\rrangle_A p_j +\llangle p\rrangle_A\llangle p\rrangle_C \}\\
    =&P_{A11}P_{C11}-\llangle p\rrangle_C P_{A11}S_{A11}-\llangle p\rrangle_A P_{C11}S_{C11}+\llangle p\rrangle_A\llangle p\rrangle_C S_{A11}S_{C11}
\end{align*}
So the $c_k$ is :
\begin{align}
    c_k=\frac{P_{A11}P_{C11}-\llangle p\rrangle_C P_{A11}S_{A11}-\llangle p\rrangle_A P_{C11}S_{C11}+\llangle p\rrangle_A\llangle p\rrangle_C S_{A11}S_{C11}}{S_{A11}*S_{C11}}
\end{align}
or in terms of $\bar{p}$ :
\begin{align}
    c_k=\frac{\bar{P}_{A11}*\bar{P}_{C11}}{S_{A11}*S_{C11}}
\end{align}

\subsubsection{Formula for $\mathrm{Var}(v_{n}\{2\}^2$)}
The variance of $v_{n}\{2\}^2$ is given by :
\begin{align}
    \mathrm{Var}(v_n\{2\}^2)=\llangle 4 \rrangle - \llangle 2 \rrangle^2
\end{align}
where $\langle 2 \rangle$ is the 2-particle correlation and $\langle 4 \rangle$ is the 4-particle correlation in an event.
\paragraph{standard}
Two-particle correlation :
\begin{align}
    \langle 2 \rangle&=\frac{\sum_i\sum_j' w_iw_je^{in(\phi_i-\phi_{j})}} { \sum_i\sum_j' w_iw_j}\\
    &=\frac{\sum_i\sum_j w_iw_je^{in(\phi_i-\phi_j)}-\sum_iw_i^2} {\sum_i\sum_j w_iw_j-\sum_iw_i^2}\\
    &=\frac{|Q_{11}|^2-S_{12}}{S_{21}-S_{12}}
\end{align}
Four-particle correlation :
\begin{align}
    \langle 4 \rangle&=\frac{\sum_i\sum_j'\sum_h'\sum_l' w_iw_jw_hw_le^{in(\phi_i+\phi_{j}-\phi_{h}-\phi_{l})}} { \sum_i\sum_j'\sum_h'\sum_l' w_iw_jw_hw_l}
\end{align}
The denominator can be expanded as :
\begin{align*}
    D&=\sum_i\sum_j'\sum_h'\sum_l' w_iw_jw_hw_l\\
    &=\sum_i\sum_j\sum_h\sum_l w_iw_jw_hw_l - 6\sum_i\sum_j'\sum_h' w_i^2w_jw_h-4\sum_i\sum_j' w_i^3w_j-3\sum_i\sum_j'w_i^2w_j^2-\sum_iw_i^4\\
    &=S_{41}-6(S_{12}S_{11}S_{11}-2\sum_i\sum_j' w_i^3w_j-\sum_i\sum_j' w_i^2w_j^2-S_{14})-4(S_{13}S_{11}-S_{14})-3(S_{22}-S_{14})-S_{14}\\
    &=S_{41}-4S_{13}S_{11}-3S_{22}-6S_{21}S_{12}+12S_{14}+12\sum_i\sum_j' w_i^3w_j+6\sum_i\sum_j' w_i^2w_j^2\\
    &=S_{41}-4S_{13}S_{11}-3S_{22}-6S_{21}S_{12}+12S_{14}+12(S_{13}S_{11}-S_{14})+6(S_{22}-S_{14})\\
    &=S_{41}+8S_{13}S_{11}+3S_{22}-6S_{21}S_{12}-6S_{14}
\end{align*}
The numerator can be expanded as :
\begin{align*}
    N&=\sum_i\sum_j'\sum_h'\sum_l' w_iw_jw_hw_le^{in(\phi_i+\phi_{j}-\phi_{h}-\phi_{l})}\\
    &=\sum_i\sum_j\sum_h\sum_l w_iw_jw_hw_le^{in(\phi_i+\phi_{j}-\phi_{h}-\phi_{l})} 
    -4\sum_i\sum_j'\sum_h' w_i^2w_jw_he^{in(\phi_{j}-\phi_{h})}\\
    &-2\sum_i\sum_j'\sum_h' w_i^2w_jw_he^{in(2\phi_{i}-\phi_{j}-\phi_{h})}
    -4\sum_i\sum_j' w_i^3w_je^{in(\phi_{i}-\phi_{j})}\\
    &-\sum_i\sum_j' w_i^2w_j^2e^{i2n(\phi_{i}-\phi_{j})}-2\sum_i\sum_j' w_i^2w_j^2-\sum_iw_i^4\\
    &=|Q_{11}|^4-S_{14}\\
    &-4( S_{12}|Q_{11}|^2-2\sum_i\sum_j'w_i^3w_je^{in(\phi_{i}-\phi_{j})}-\sum_i\sum_j'w_i^2w_j^2-S_{14} )\\
    &-2( Q_{22}Q_{11}^*Q_{11}^*-2\sum_i\sum_j'w_i^3w_je^{in(\phi_{i}-\phi_{j})}-\sum_i\sum_j'w_i^2w_j^2e^{i2n(\phi_{i}-\phi_{j})}-S_{14} )\\
    &-4( Q_{13}Q_{11}^*-S_{14} ) - (|Q_{22}|^2-S_{14}) - 2(S_{22}-S_{14})\\
    &=|Q_{11}|^4-|Q_{22}|^2-2Q_{22}Q_{11}^*Q_{11}^*-4Q_{13}Q_{11}^*-4S_{12}|Q_{11}|^2-2S_{22}+12S_{14}\\
    &+12\sum_i\sum_j'w_i^3w_je^{in(\phi_{i}-\phi_{j})}+2\sum_i\sum_j'w_i^2w_j^2e^{i2n(\phi_{i}-\phi_{j})}+4\sum_i\sum_j'w_i^2w_j^2\\
    &=|Q_{11}|^4-|Q_{22}|^2-2Q_{22}Q_{11}^*Q_{11}^*-4Q_{13}Q_{11}^*-4S_{12}|Q_{11}|^2-2S_{22}+12S_{14}\\
    &+12(Q_{13}Q_{11}^*-S_{14})+2(|Q_{22}|^2-S_{14})+4(S_{22}-S_{14})\\
    &=|Q_{11}|^4+|Q_{22}|^2-2Q_{22}Q_{11}^*Q_{11}^*+8Q_{13}Q_{11}^*-4S_{12}|Q_{11}|^2+2S_{22}-6S_{14}
\end{align*}
So the final form of $\langle 4 \rangle$ is :
\begin{align}
    \langle 4 \rangle&=\frac{|Q_{11}|^4+|Q_{22}|^2-2\Re(Q_{22}Q_{11}^*Q_{11}^*)+8\Re(Q_{13}Q_{11}^*)-4S_{12}|Q_{11}|^2+2S_{22}-6S_{14}} {S_{41}+8S_{13}S_{11}+3S_{22}-6S_{21}S_{12}-6S_{14}}
\end{align}

\paragraph{Two-subevent}
Two-particle correlation :
\begin{align}
    \langle 2 \rangle_{a|c}&=\frac{\sum_i\sum_j w_iw_je^{in(\phi_i-\phi_{j})}} { \sum_i\sum_j w_iw_j}\\
    &=\frac{\Re(Q_{A11}Q_{C11}^*)}{S_{A11}S_{C11}}
\end{align}
Particle i is from subevent A and particle j is from subevent C.
Four-particle correlation :
\begin{align}
    \langle 4 \rangle_{aa|cc}&=\frac{\sum_i\sum_j'\sum_h\sum_l' w_iw_jw_hw_le^{in(\phi_i+\phi_{j}-\phi_{h}-\phi_{l})}} { \sum_i\sum_j'\sum_h\sum_l' w_iw_jw_hw_l}
\end{align}
Particle i and j are from subevent A and particle h and l are from subevent C.
The denominator can be expanded as :
\begin{align*}
    D&=\sum_i\sum_j'\sum_h\sum_l' w_iw_jw_hw_l\\
    &=\sum_i\sum_j\sum_h\sum_l w_iw_jw_hw_l - \sum_i\sum_h\sum_l' w_i^2w_hw_l-\sum_i\sum_j'\sum_h w_iw_jw_h^2-\sum_i\sum_h w_i^2w_h^2\\
    &=S_{A21}S_{C21}-(S_{A12}S_{C21}-S_{A12}S_{C12})-(S_{A21}S_{C12}-S_{A12}S_{C12})-S_{A12}S_{C12}\\
    &=S_{A21}S_{C21}-S_{A12}S_{C21}-S_{A21}S_{C12}+S_{A12}S_{C12}\\
    &=(S_{A21}-S_{A12})(S_{C21}-S_{C12})
\end{align*}
The numerator can be expanded as :
\begin{align*}
    N&=\sum_i\sum_j'\sum_h\sum_l' w_iw_jw_hw_le^{in(\phi_i+\phi_{j}-\phi_{h}-\phi_{l})}\\
    &=\sum_i\sum_j\sum_h\sum_l w_iw_jw_hw_le^{in(\phi_i+\phi_{j}-\phi_{h}-\phi_{l})} \\
    &-\sum_i\sum_h\sum_l' w_i^2w_jw_he^{in(2\phi_{i}-\phi_{h}-\phi_l)}\\
    &-\sum_i\sum_j'\sum_h w_iw_jw_h^2e^{in(\phi_{i}+\phi_{j}-2\phi_{h})}\\
    &-\sum_i\sum_h w_i^2w_h^2e^{in2(\phi_{i}-\phi_{h})}\\
    &=|Q_{A11}Q_{C11}^*|^2-(Q_{A22}Q_{C11}^*Q_{C11}^*-Q_{A22}Q_{C22}^*)\\
    &-(Q_{A11}Q_{A11}Q_{C22}^*-Q_{A22}Q_{C22}^*)-Q_{A22}Q_{C22}^*\\
    &=|Q_{A11}Q_{C11}^*|^2-Q_{A22}Q_{C11}^*Q_{C11}^*-Q_{A11}Q_{A11}Q_{C22}^*+Q_{A22}Q_{C22}^*
    &=(Q_{A11}^2-Q_{A22})(Q_{C11}^2-Q_{C22})^*
\end{align*}
So the final form of $\langle 4 \rangle_{aa|cc}$ is :
\begin{align}
    \langle 4 \rangle_{aa|cc}&=\frac{\Re(Q_{A11}^2-Q_{A22})\Re(Q_{C11}^{*2}-Q_{C22}^*)} {(S_{A21}-S_{A12})(S_{C21}-S_{C12})}
\end{align}

\subsubsection{Formulae using $q$-vectors}
Recall the definitions :
\begin{align}
    Q_{nk}=&\sum_i w_i^k e^{in\phi_i}\\
    S_{mk}=&(\sum_i w_i^k)^m\\
    O_{nk}=&\sum_i w_i^k e^{in\phi_i}p_i\\
    P_{mk}=&\sum_i w_i^kp_i^m\\
    \llangle p\rrangle=&\frac{1}{N_{ev}}\sum_{evts}\frac{\sum_i w_ip_i}{\sum_iw_i}\\
\end{align}
Lets define some normalised quantities :
\begin{align}
    \hat{q}_{nk}&=Q_{nk}/S_{1k}\\
    \hat{o}_{nk}&=O_{nk}/S_{1k}\\
    \hat{p}_{nk}&=P_{mk}/S_{1k}\\
    \tau_{k}&=\frac{\sum_i w_i^{k+1}}{(\sum_i w_i)^{k+1}}\\
    &=\frac{S_{1,k+1}}{S_{k+1,1}} 
\end{align}
1. FSE 2-particle correlation :
\begin{align}
    \langle 2 \rangle&=\frac{|Q_{11}|^2-S_{12}}{S_{21}-S_{12}}\\
    &=\frac{|\hat{q}_{11}|^2-S_{12}/S_{21}}{1-S_{12}/S_{21}}\\
    &=\frac{|\hat{q}_{11}|^2-\tau_1}{1-\tau_1}
\end{align}
2. 2-SE 2-particle correlation :
\begin{align}
    \langle 2 \rangle_{a|c}&=\frac{\Re(Q_{A11}Q_{C11}^*)}{S_{A11}S_{C11}}\\
    &=\Re(\hat{q}_{A11}\hat{q}_{C11}^*)
\end{align}
3. FSE 4-particle correlation :
\begin{align}
    \langle 4 \rangle&=\frac{|Q_{11}|^4+|Q_{22}|^2-2\Re(Q_{22}Q_{11}^*Q_{11}^*)+8\Re(Q_{13}Q_{11}^*)-4S_{12}|Q_{11}|^2+2S_{22}-6S_{14}} {S_{41}+8S_{13}S_{11}+3S_{22}-6S_{21}S_{12}-6S_{14}}\\
    &=\frac{|\hat{q}_{11}|^4+\tau_1^2|\hat{q}_{22}|^2-2\tau_1\hat{q}_{22}\hat{q}_{11}^*\hat{q}_{11}^*+8\tau_2\hat{q}_{13}\hat{q}_{11}^*-4\tau_1|\hat{q}_{11}|^2+2\tau_1^2-6\tau_3} {1+8\tau_2+3\tau_1^2-6\tau_1-6\tau_3}\\
    &=\frac{|\hat{q}_{11}|^4+\tau_1^2(2+|\hat{q}_{22}|^2)-2\tau_1(\hat{q}_{22}\hat{q}_{11}^*\hat{q}_{11}^*+2|\hat{q}_{11}|^2)+8\tau_2\hat{q}_{13}\hat{q}_{11}^*-6\tau_3} {1-6\tau_1+8\tau_2+3\tau_1^2-6\tau_3}
\end{align}
4. 2-SE 4-particle correlation :
\begin{align}
    \langle 4 \rangle_{aa|cc}&=\frac{\Re(Q_{A11}^2-Q_{A22})\Re(Q_{C11}^{*2}-Q_{C22}^*)} {(S_{A21}-S_{A12})(S_{C21}-S_{C12})}\\
    &=\frac{\Re(\hat{q}_{A11}^2-\tau_{A1}\hat{q}_{A22})\Re(\hat{q}_{C11}^{*2}-\tau_{C1}\hat{q}_{C22}^*)} {(1-\tau_{A1})(1-\tau_{C1})}
\end{align}
5. 3-SE Covariance :
\begin{align}
Cov[3SE]&=\frac{\Re(Q_{An1}Q_{Cn1}^*)P_{B11}-\llangle p\rrangle \Re(Q_{An1}Q_{Cn1}^*)S_{B11}}{S_{A11}S_{C11}S_{B11}}\\
&=\Re(\hat{q}_{An1}\hat{q}_{Cn1}^*)\hat{p}_{B11}-\llangle p\rrangle \Re(\hat{q}_{An1}\hat{q}_{Cn1}^*)\\
&=\Re(\hat{q}_{An1}\hat{q}_{Cn1}^*)(\hat{p}_{B11}-\llangle p\rrangle)
\end{align}
6. FSE Covariance :
\begin{align}
Cov[FSE]&=\frac{|Q_{n1}|^2P_{11}-S_{12}P_{11}-2\Re(O_{n2}Q_{n1}^*)+2P_{13}}{S_{31}-3S_{12}S_{11}+2S_{13}}\\
    &-\llangle p\rrangle\frac{ |Q_{n1}|^2S_{11}-S_{12}S_{11}-2\Re(Q_{n2}Q_{n1}^*)+2S_{13}}{S_{31}-3S_{12}S_{11}+2S_{13}}\\
&=\frac{|\hat{q}_{n1}|^2\hat{p}_{11}-\tau_1\hat{p}_{11}-2\tau_1\Re(\hat{o}_{n2}\hat{q}_{n1}^*)+2\tau_2\hat{p}_{13}}{1-3\tau_1+2\tau_2}\\
    &-\llangle p\rrangle\frac{ |\hat{q}_{n1}|^2-\tau_1-2\tau_1\Re(\hat{q}_{n2}\hat{q}_{n1}^*)+2\tau_2}{1-3\tau_1+2\tau_2}\\
    &=\frac{|\hat{q}_{n1}|^2(\hat{p}_{11}-\llangle p\rrangle)-\tau_1(\hat{p}_{11}-\llangle p\rrangle)-2\tau_1(\Re(\hat{o}_{n2}\hat{q}_{n1}^*)-\llangle p\rrangle\Re(\hat{q}_{n2}\hat{q}_{n1}^*))+2\tau_2(\hat{p}_{13}-\llangle p\rrangle)}{1-3\tau_1+2\tau_2}
\end{align}
7. 2-SE Covariance :
\begin{align}
Cov[2SE]=&\frac{\Re(Q_{An1}Q_{Cn1}^*)P_{11}-\Re(O_{An2}Q_{Cn1}^*)-\Re(Q_{An1}O_{Cn2}^*)}{S_{A11}S_{C11}S_{11}-S_{A12}S_{C11}-S_{A11}S_{C12}}\\
    &-\llangle p\rrangle\frac{\Re(Q_{An1}Q_{Cn1}^*)S_{11}-\Re(Q_{An2}Q_{Cn1}^*)-\Re(Q_{An1}Q_{Cn2}^*)}{S_{A11}S_{C11}S_{11}-S_{A12}S_{C11}-S_{A11}S_{C12}}\\
    &=\frac{\Re(\hat{q}_{An1}\hat{q}_{Cn1}^*)\hat{p}_{11}-\tau_{A1}(S_{A11}/S_{11})\Re(\hat{o}_{An2}\hat{q}_{Cn1}^*)-\tau_{C1}(S_{C11}/S_{11})\Re(\hat{q}_{An1}\hat{o}_{Cn2}^*)}{1-\tau_{A1}(S_{A11}/S_{11})-\tau_{C1}(S_{C11}/S_{11})}\\
    &-\llangle p\rrangle\frac{\Re(\hat{q}_{An1}\hat{q}_{Cn1}^*)-\tau_{A1}(S_{A11}/S_{11})\Re(\hat{q}_{An2}\hat{q}_{Cn1}^*)-\tau_{C1}(S_{C11}/S_{11})\Re(\hat{q}_{An1}\hat{q}_{Cn2}^*)}{1-\tau_{A1}(S_{A11}/S_{11})-\tau_{C1}(S_{C11}/S_{11})}\\
    &=\frac{\Re(\hat{q}_{An1}\hat{q}_{Cn1}^*)\hat{p}_{11}-\tau_{A1}\alpha_A\Re(\hat{o}_{An2}\hat{q}_{Cn1}^*)-\tau_{C1}\alpha_C\Re(\hat{q}_{An1}\hat{o}_{Cn2}^*)}{1-\tau_{A1}\alpha_A-\tau_{C1}\alpha_C}\\
    &-\llangle p\rrangle\frac{\Re(\hat{q}_{An1}\hat{q}_{Cn1}^*)-\tau_{A1}\alpha_A\Re(\hat{q}_{An2}\hat{q}_{Cn1}^*)-\tau_{C1}\alpha_C\Re(\hat{q}_{An1}\hat{q}_{Cn2}^*)}{1-\tau_{A1}\alpha_A-\tau_{C1}\alpha_C}\\
    &=\{ \Re(\hat{q}_{An1}\hat{q}_{Cn1}^*)(\hat{p}_{11}-\llangle p\rrangle)-\tau_{A1}\alpha_A(\Re(\hat{o}_{An2}\hat{q}_{Cn1}^*)-\llangle p\rrangle\Re(\hat{q}_{An2}\hat{q}_{Cn1}^*))\\
    &-\tau_{C1}\alpha_C(\Re(\hat{q}_{An1}\hat{o}_{Cn2}^*)-\llangle p\rrangle\Re(\hat{q}_{An1}\hat{q}_{Cn2}^*)) \} / \{ 1-\tau_{A1}\alpha_A-\tau_{C1}\alpha_C \}
\end{align}
where $\alpha_A=S_{A11}/S_{11}$ and $\alpha_C=S_{C11}/S_{11}$.

\clearpage

\section{Comparison between different subevent methods for $v_n - \pT$ Correlation ($\rho(v_n^2, [\pT])$)}\label{sec:subevent_vnpt}

\subsection{Covariance comparison}
\begin{figure}[htbp]
\centering
\includegraphics[width=0.32\linewidth]{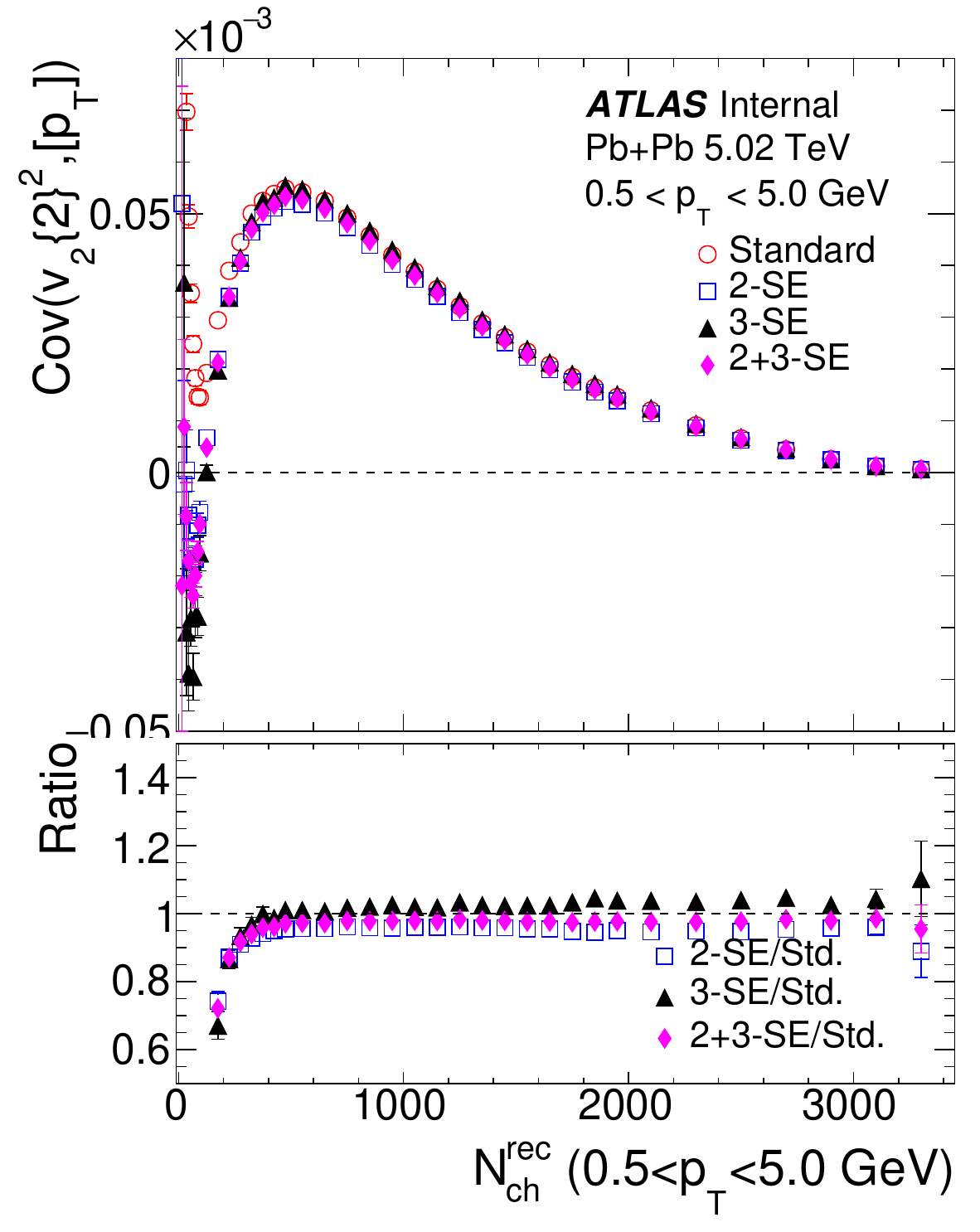}
\includegraphics[width=0.32\linewidth]{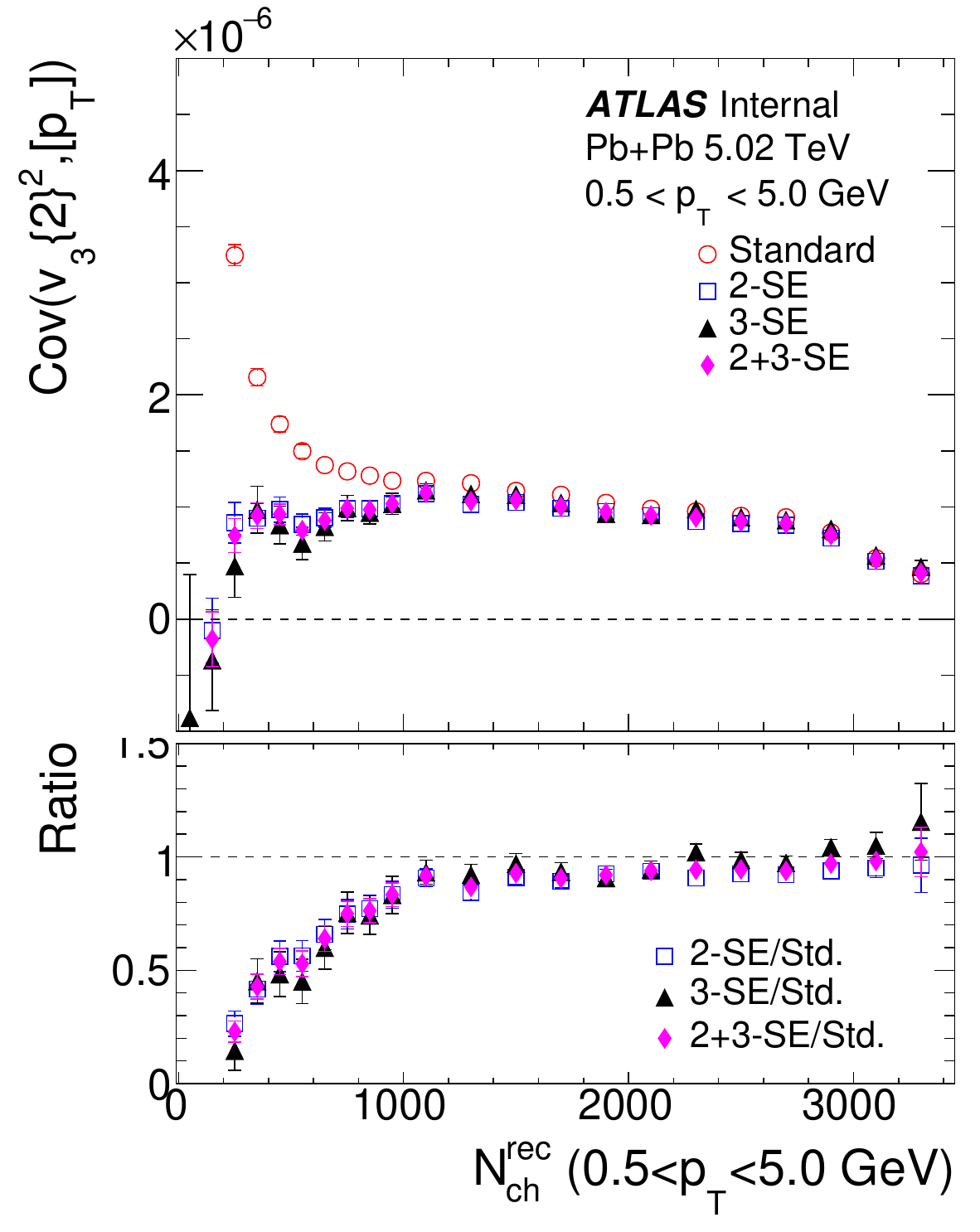}
\includegraphics[width=0.32\linewidth]{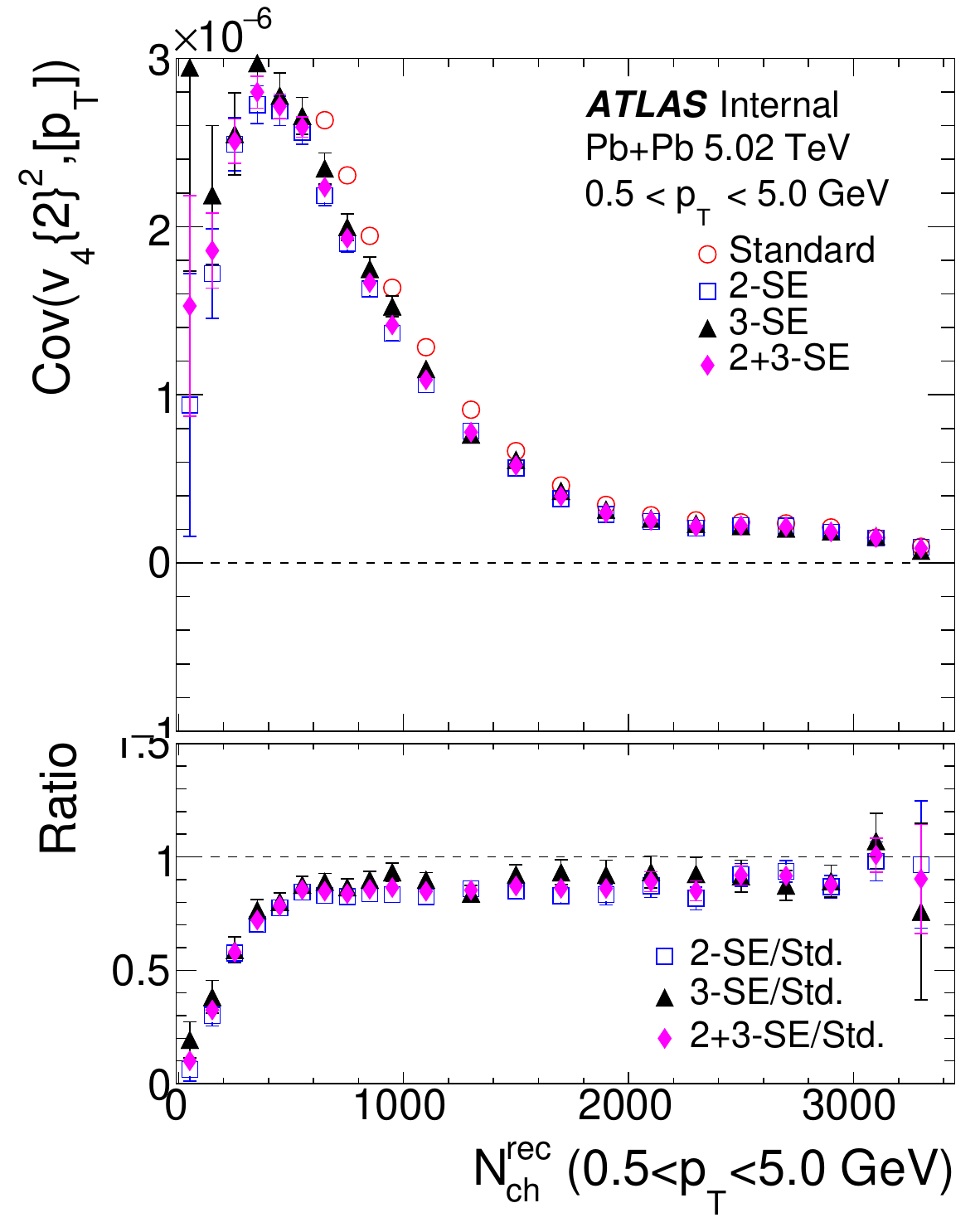}
\caption{Comparison of $\mathrm{cov}(v_n\{2\}^2,[\pT])$ using the standard, two-subevent and three-subevent methods in Pb+Pb for $n=2$, 3 and 4. The result is obtained for the $|\eta|<2.5$. The error bars represent statistical uncertainties.}
\label{fig:CovS_Pb}
\end{figure}

\begin{figure}[htbp]
\centering
\includegraphics[width=0.32\linewidth]{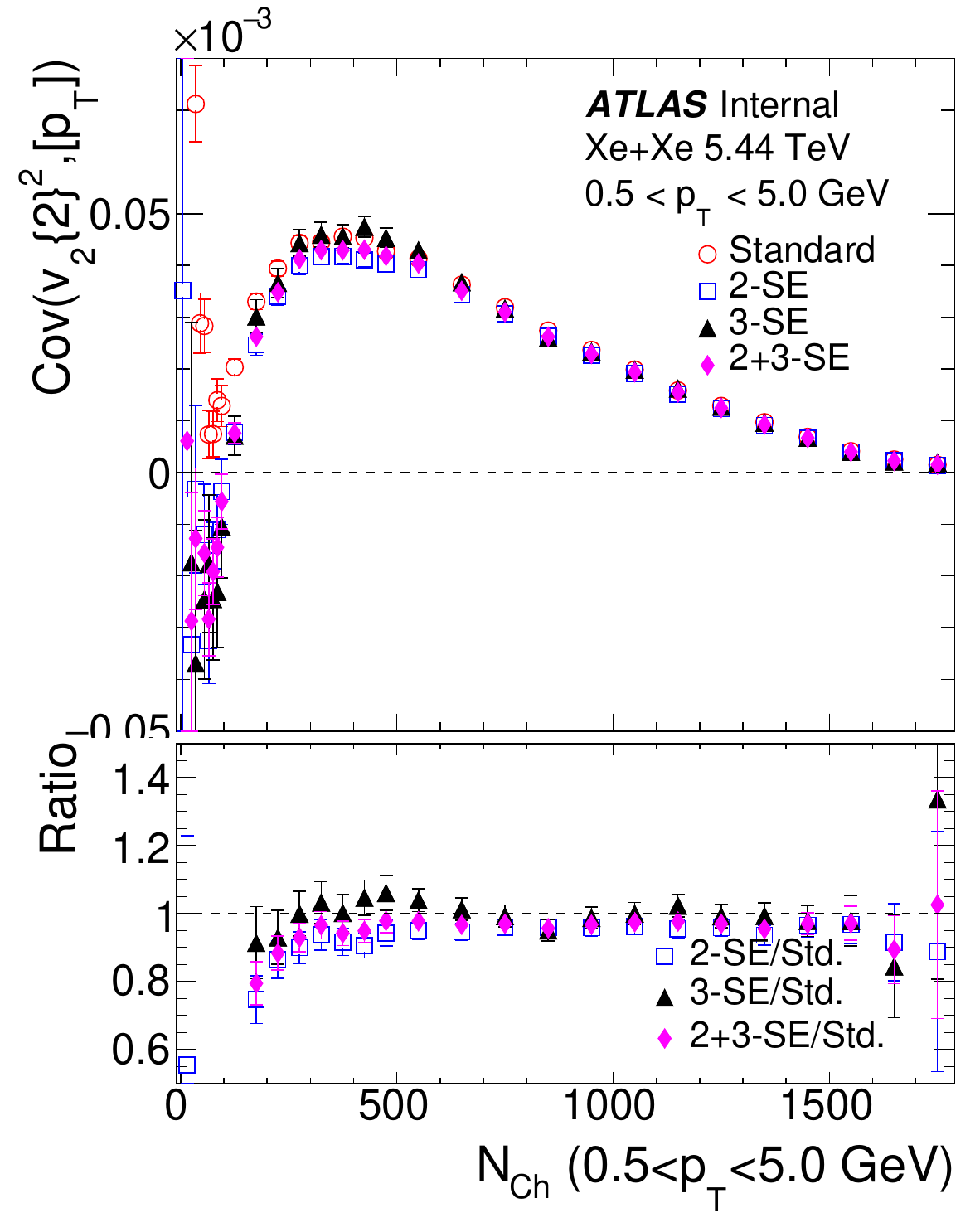}
\includegraphics[width=0.32\linewidth]{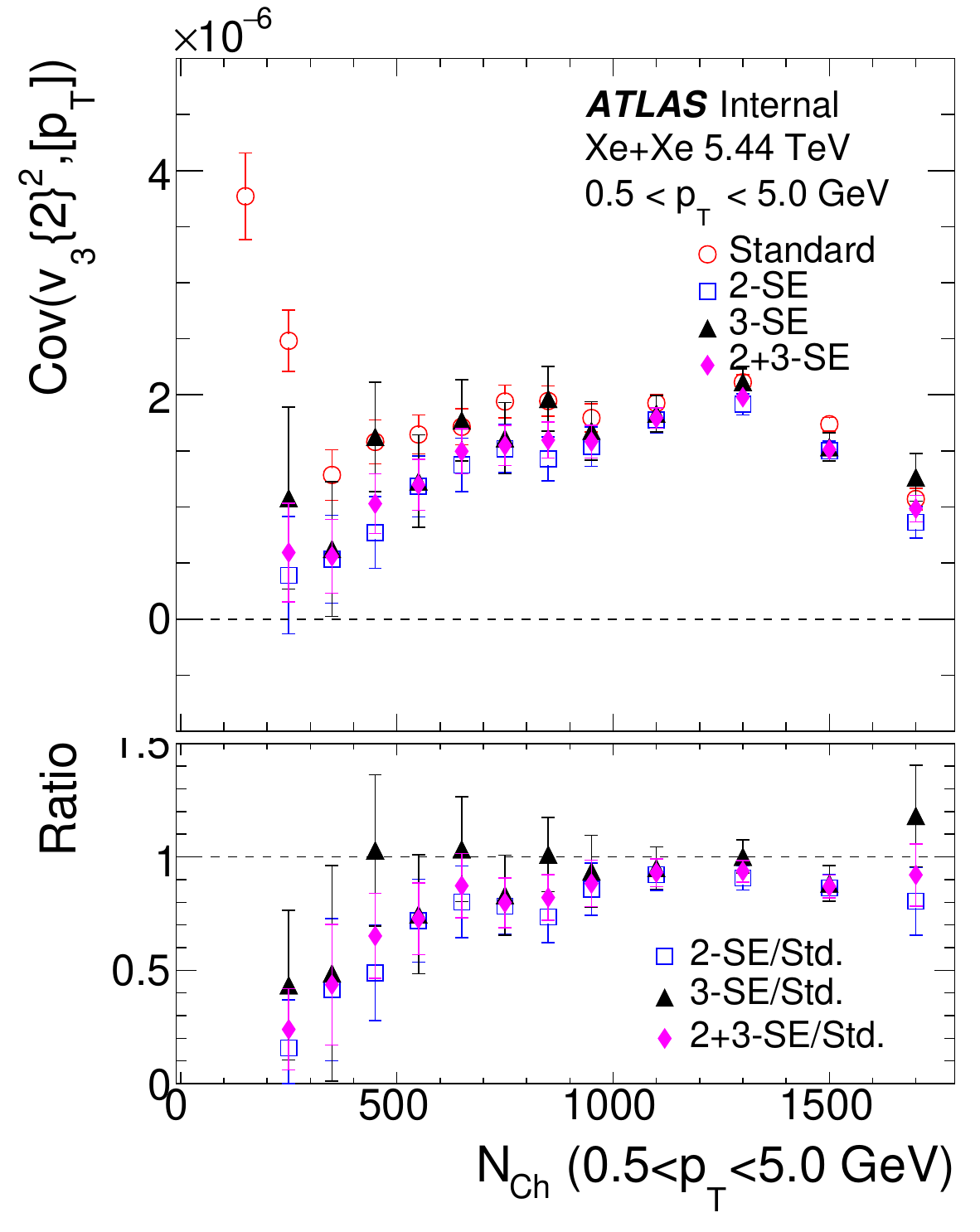}
\includegraphics[width=0.32\linewidth]{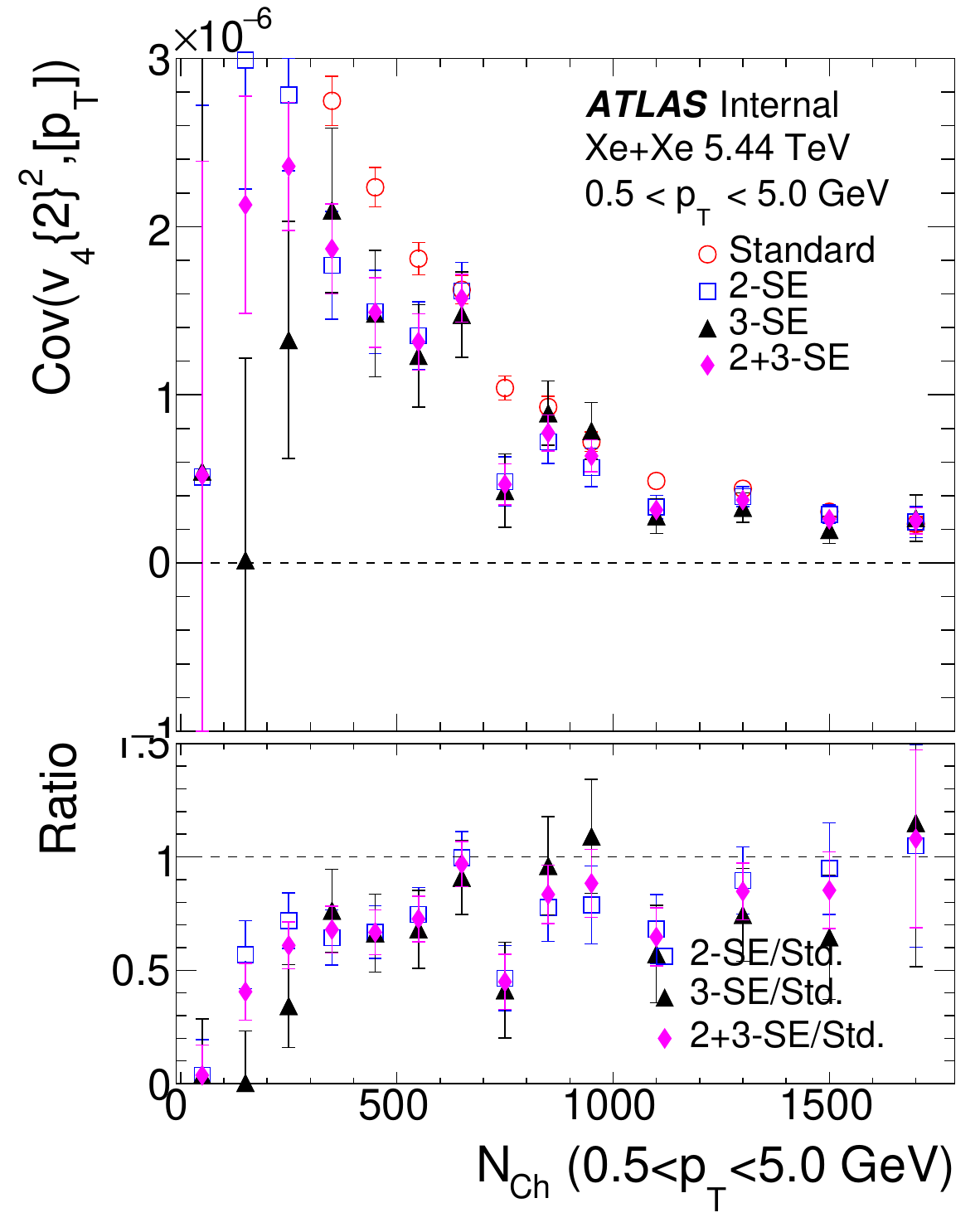}
\caption{Comparison of $\mathrm{cov}(v_n\{2\}^2,[\pT])$ using the standard, two-subevent and three-subevent methods in Xe+Xe for $n=2$, 3 and 4. The result is obtained for the $|\eta|<2.5$. The error bars represent statistical uncertainties.}
\label{fig:CovS_Xe}
\end{figure}

The covariance $\cov{n}=\langle\langle v_n^2\,\delta \pT\rangle\rangle$ as a function of $\NchR$ is shown in Figures~\ref{fig:CovS_Pb} and \ref{fig:CovS_Xe}. The standard method exhibits elevated values in peripheral bins, indicating nonflow contamination. The two- and three-subevent methods converge at higher multiplicities; for $n=3,4$ they agree within errors, while some deviation persists for $n=2$. Averaging the two- and three-subevent results (“2+3‐SE”) improves precision, and this average is used for $\cov{3,4}$. However, for $\cov{2}$, the three‐subevent result alone is adopted.

\subsection{Variance comparison}
\begin{figure}[htbp]
\centering
\includegraphics[width=0.32\linewidth]{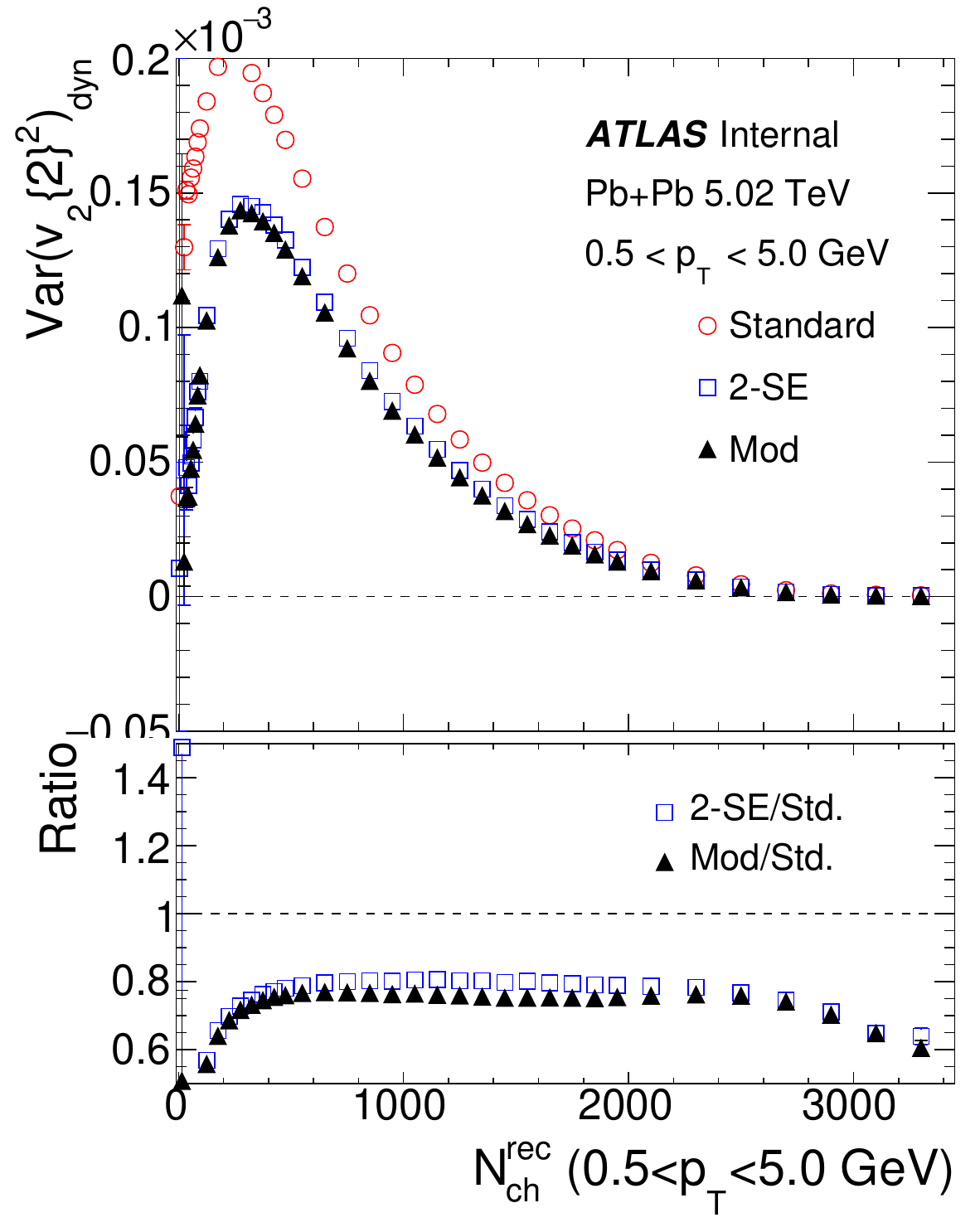}
\includegraphics[width=0.32\linewidth]{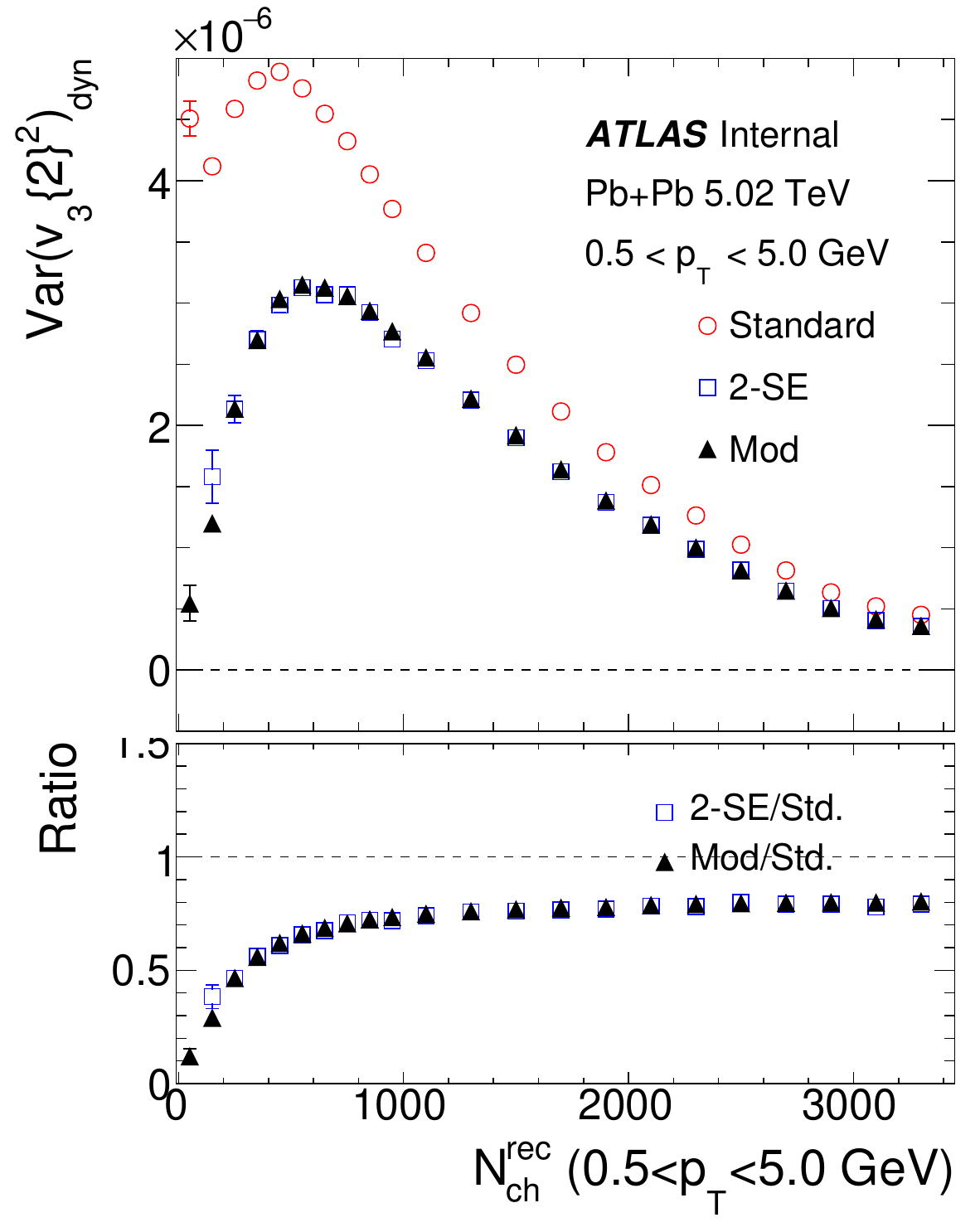}
\includegraphics[width=0.32\linewidth]{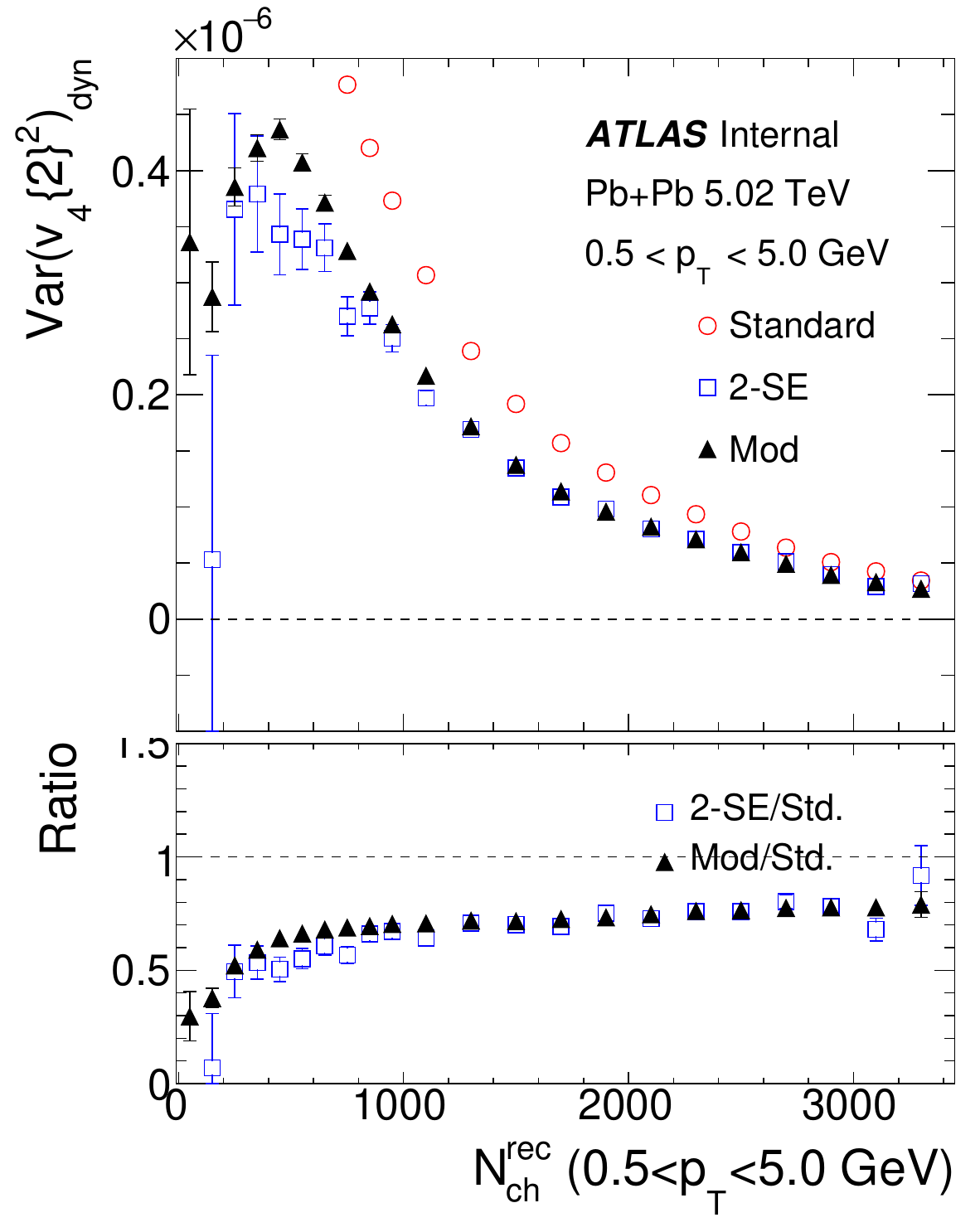}
\caption{Comparison of $\mathrm{Var}(v_n\{2\}^2)_{\mathrm{dyn}}$ using the standard and two-subevent methods as well as method based on Eq.~\ref{eq:4} in Pb+Pb for $n=2$, 3 and 4. The result is obtained for the $|\eta|<2.5$. The error bars represent statistical uncertainties.}
\label{fig:VarS_Pb}
\end{figure}

\begin{figure}[htbp]
\centering
\includegraphics[width=0.32\linewidth]{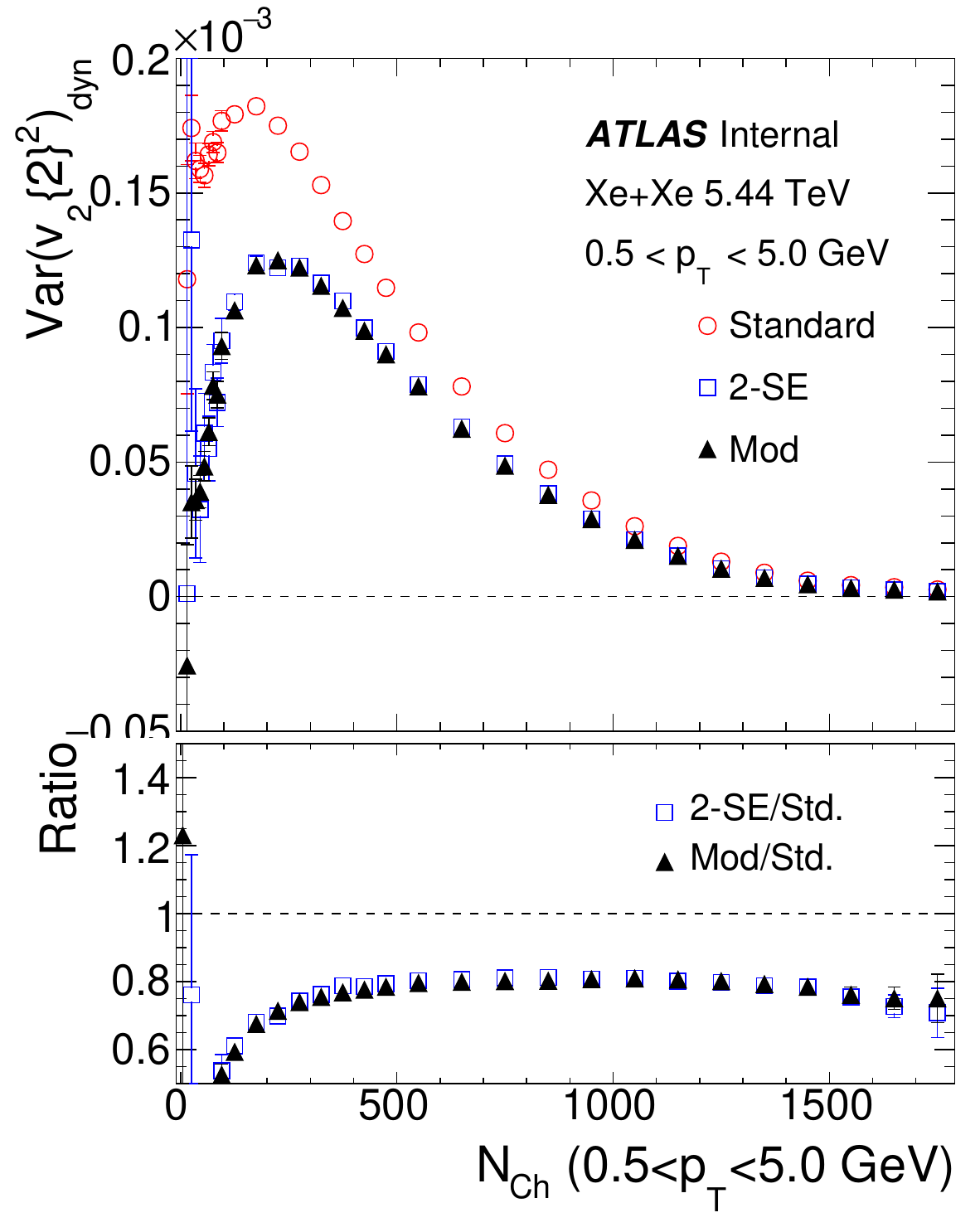}
\includegraphics[width=0.32\linewidth]{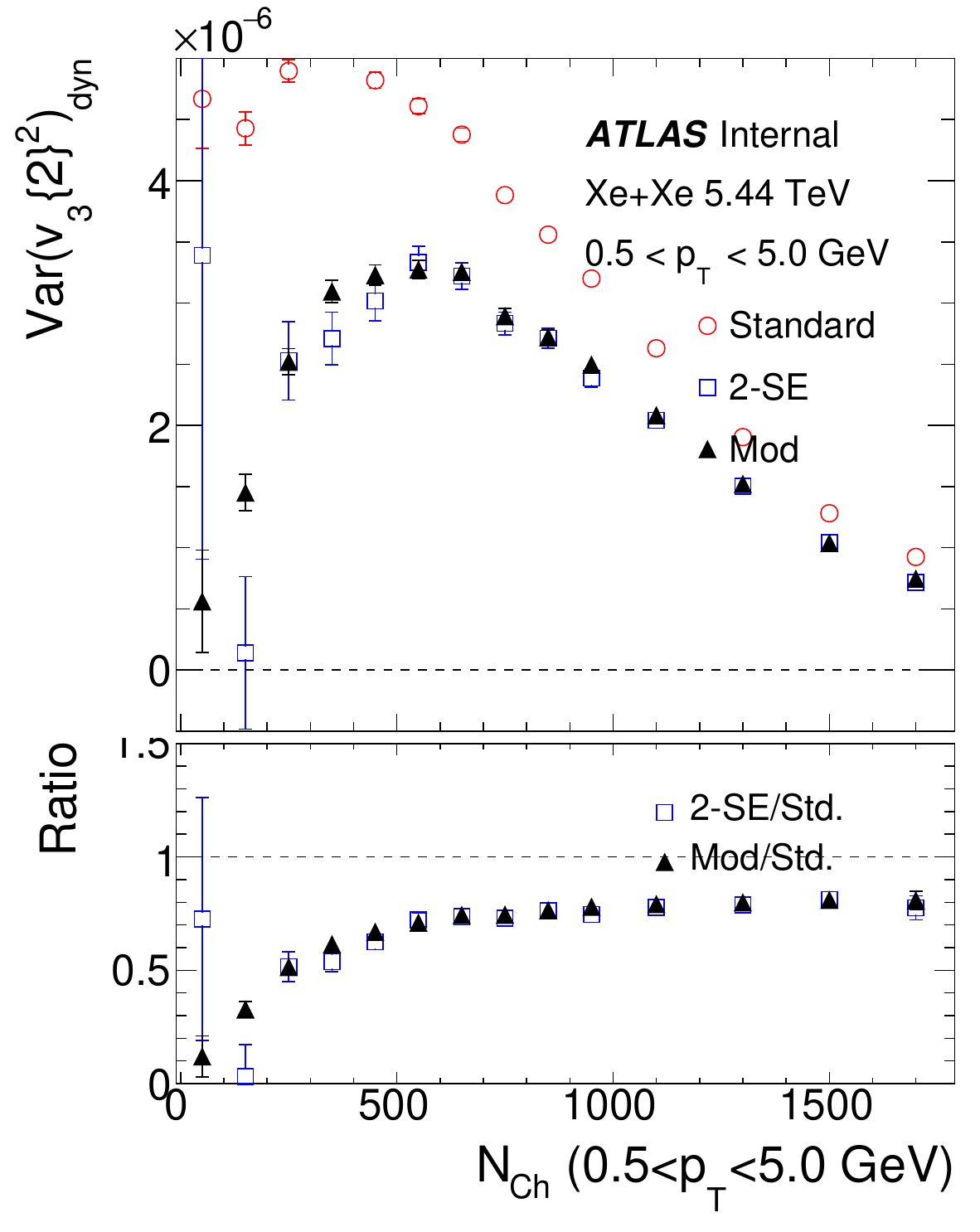}
\includegraphics[width=0.32\linewidth]{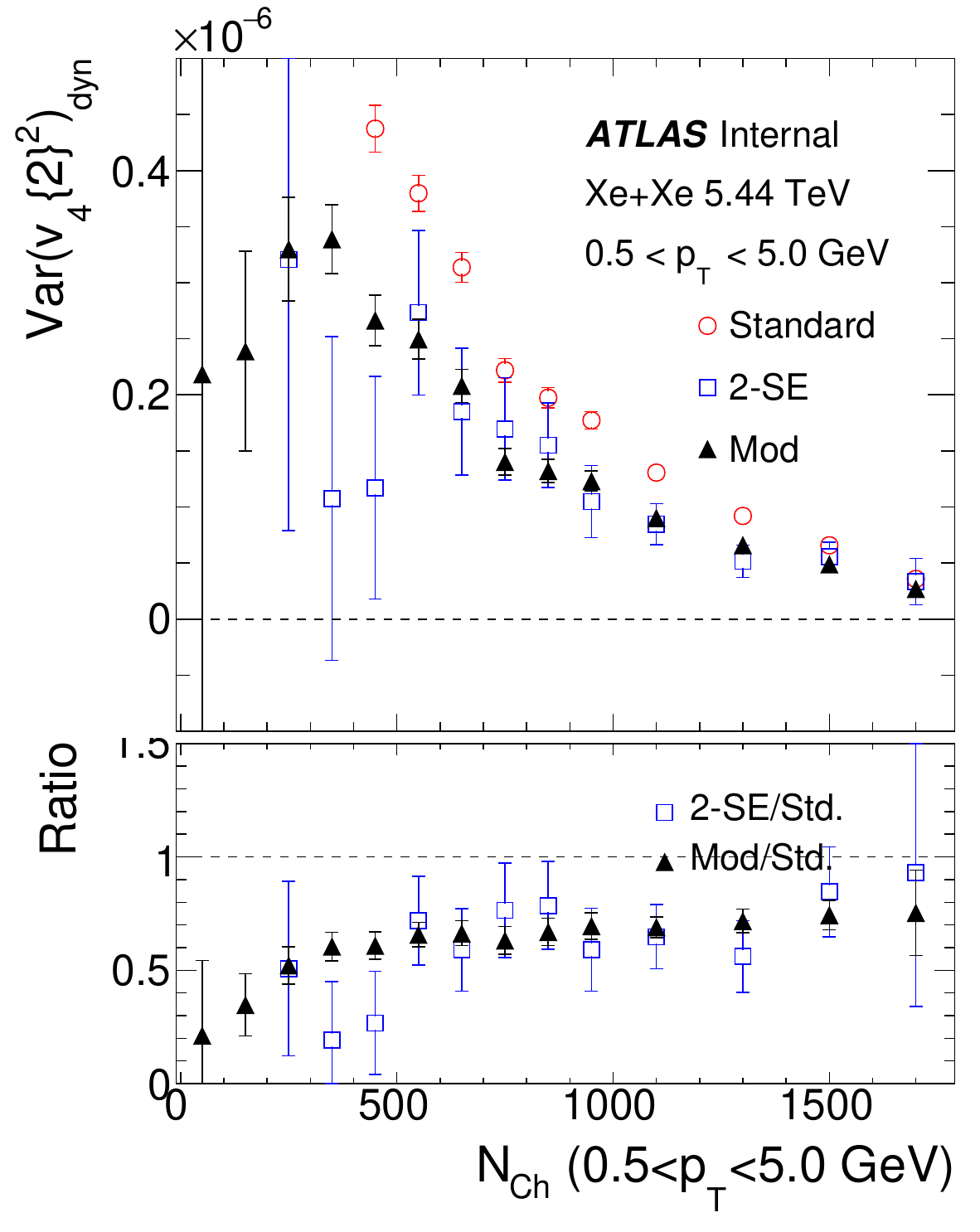}
\caption{Comparison of $\mathrm{Var}(v_n\{2\}^2)_{\mathrm{dyn}}$ using the standard and two-subevent methods as well as method based on Eq.~\ref{eq:4} in Xe+Xe for $n=2$, 3 and 4. The result is obtained for the $|\eta|<2.5$. The error bars represent statistical uncertainties.}
\label{fig:VarS_Xe}
\end{figure}

Figures~\ref{fig:VarS_Pb} and \ref{fig:VarS_Xe} compare the variance $\mathrm{Var}\bigl((v_n\{2\})^2\bigr)_{\mathrm{dyn}} = c_n\{4\} - [c_n\{2\}]^2$ evaluated using standard, two‐subevent, and hybrid methods [Eq.~\eqref{eq:4}]. The standard approach overestimates the variance in peripheral events. The hybrid prescription, which retains nonflow suppression while improving statistical precision, is chosen for $\var{n}$ in the final $\rho_n$.

\subsection{Pearson coefficient}
\begin{figure}[htbp]
\centering
\includegraphics[width=0.32\linewidth]{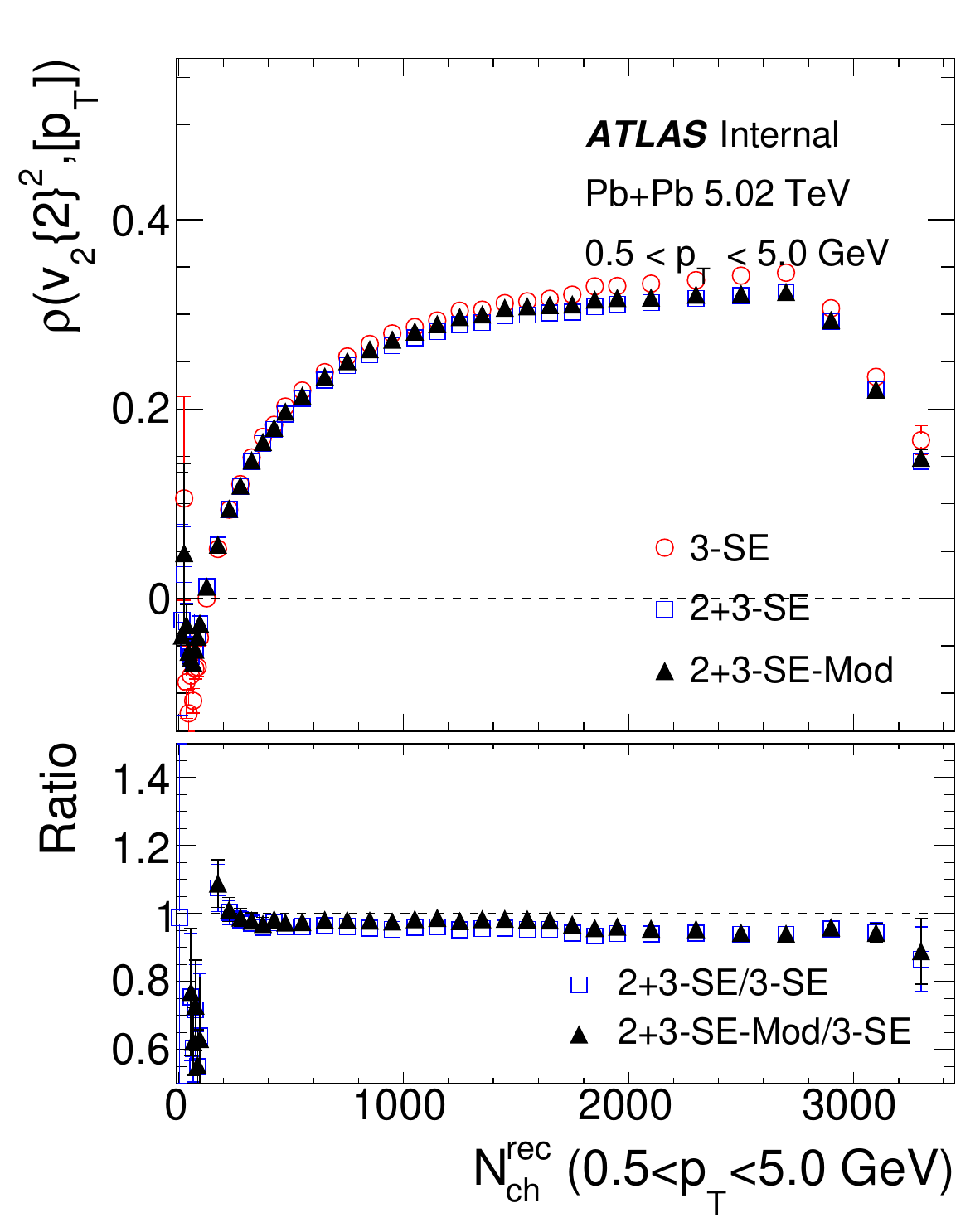}
\includegraphics[width=0.32\linewidth]{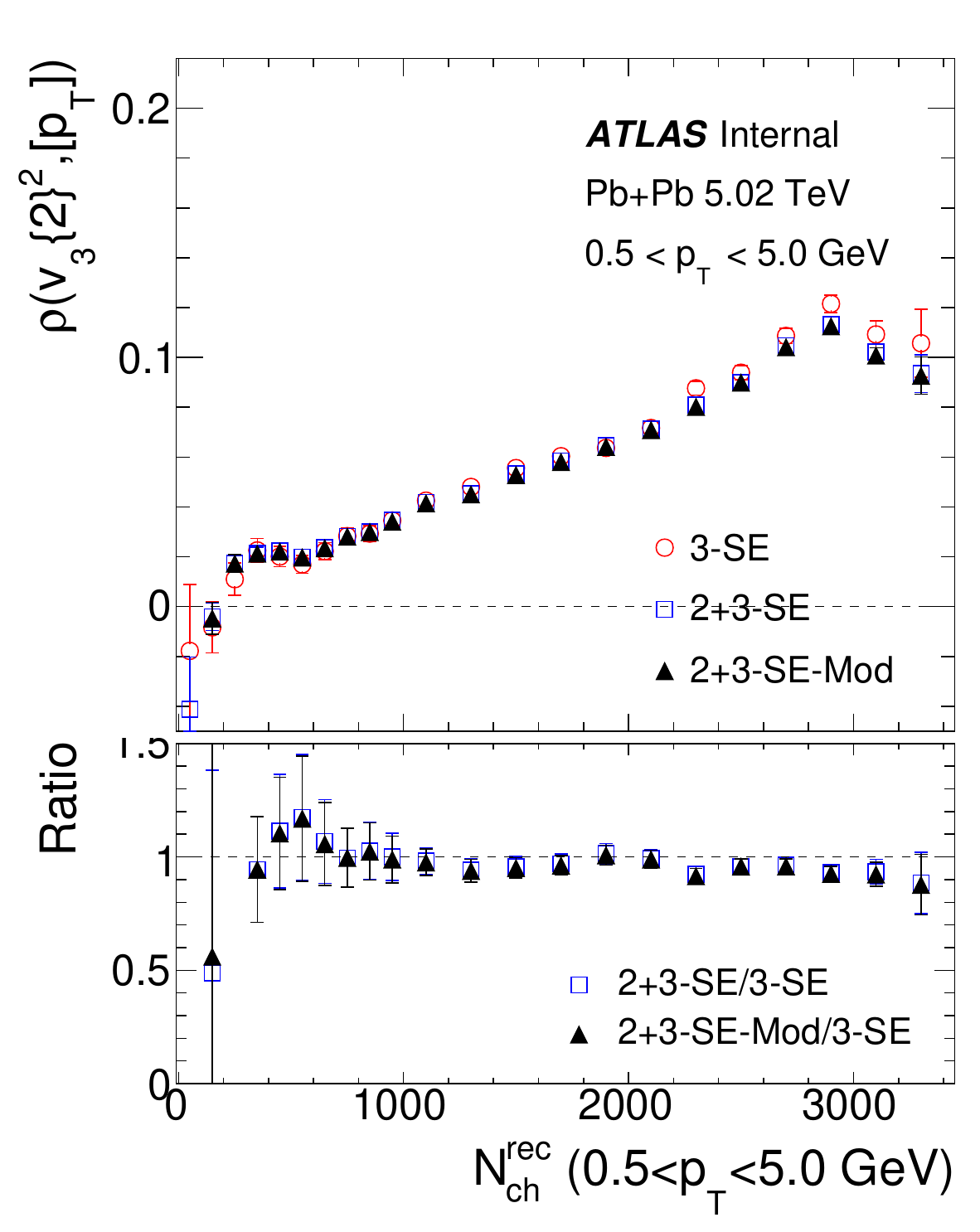}
\includegraphics[width=0.32\linewidth]{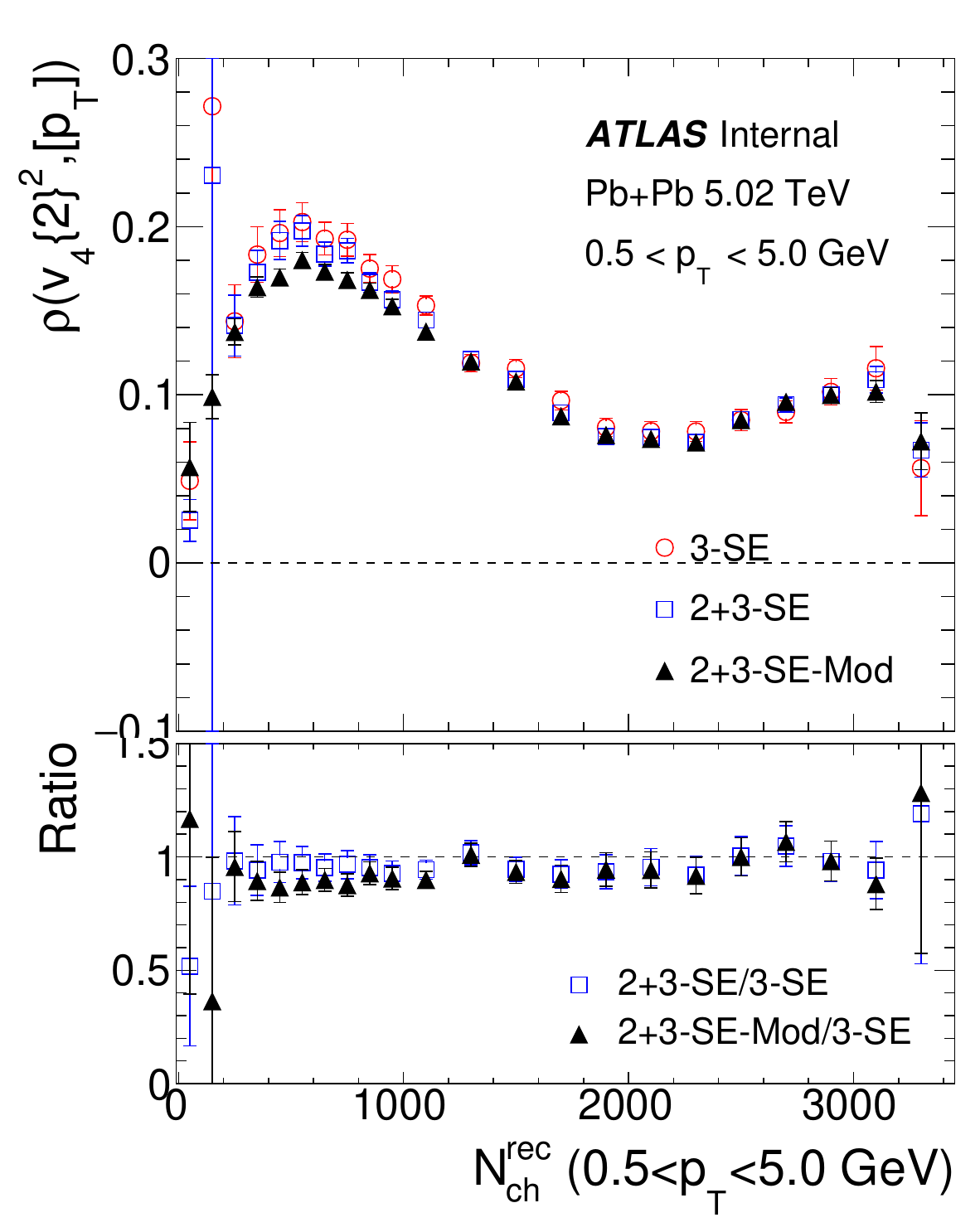}
\caption{Comparison of $\rho(v_n\{2\}^2,[\pT])$ using different $\eta$-subevents in Pb+Pb collisions for $n=2$, 3 and 4. The result is obtained for the $|\eta|<2.5$. The error bars represent statistical uncertainties.}
\label{fig:RhoS_Pb}
\end{figure}

\begin{figure}[htbp]
\centering
\includegraphics[width=0.32\linewidth]{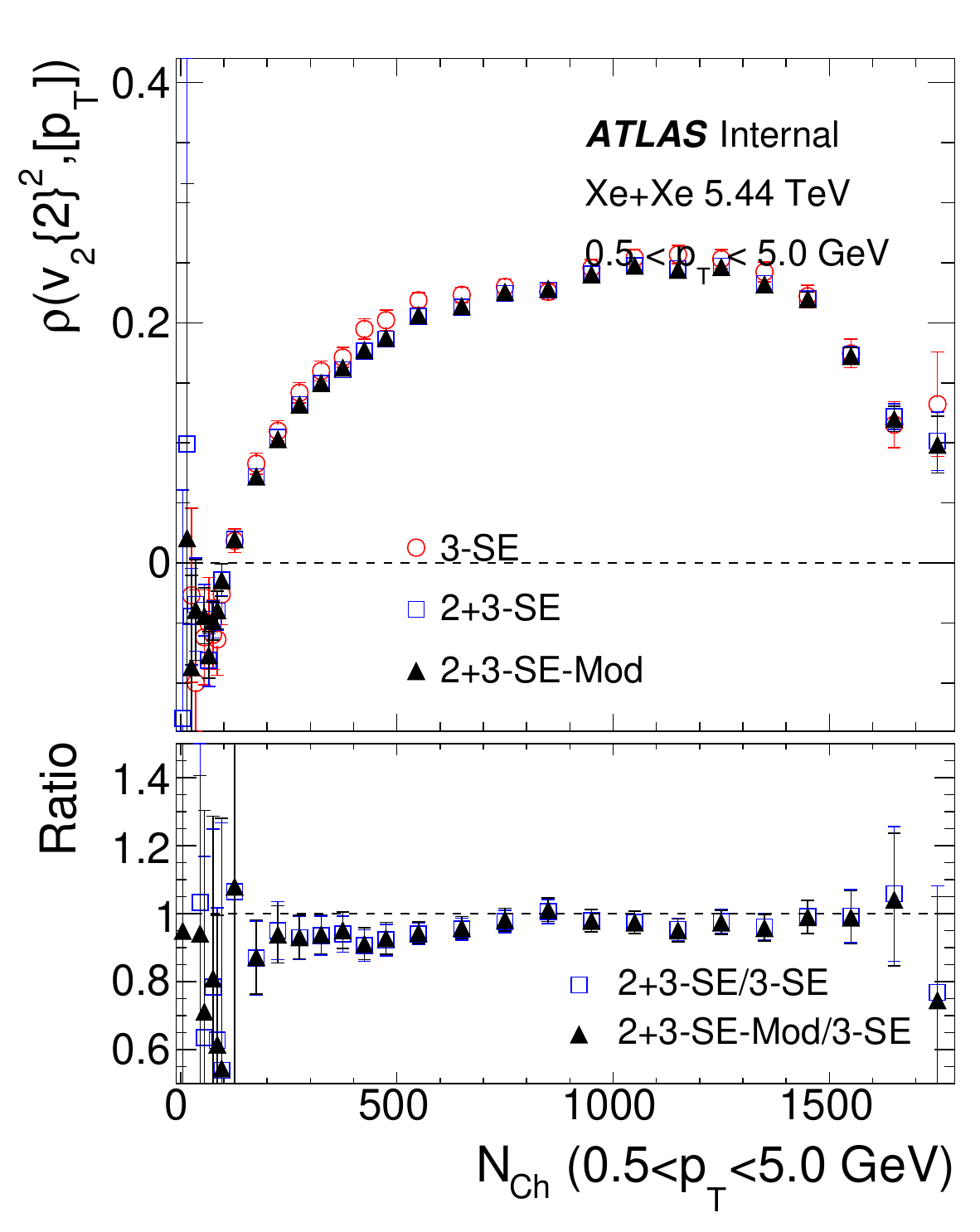}
\includegraphics[width=0.32\linewidth]{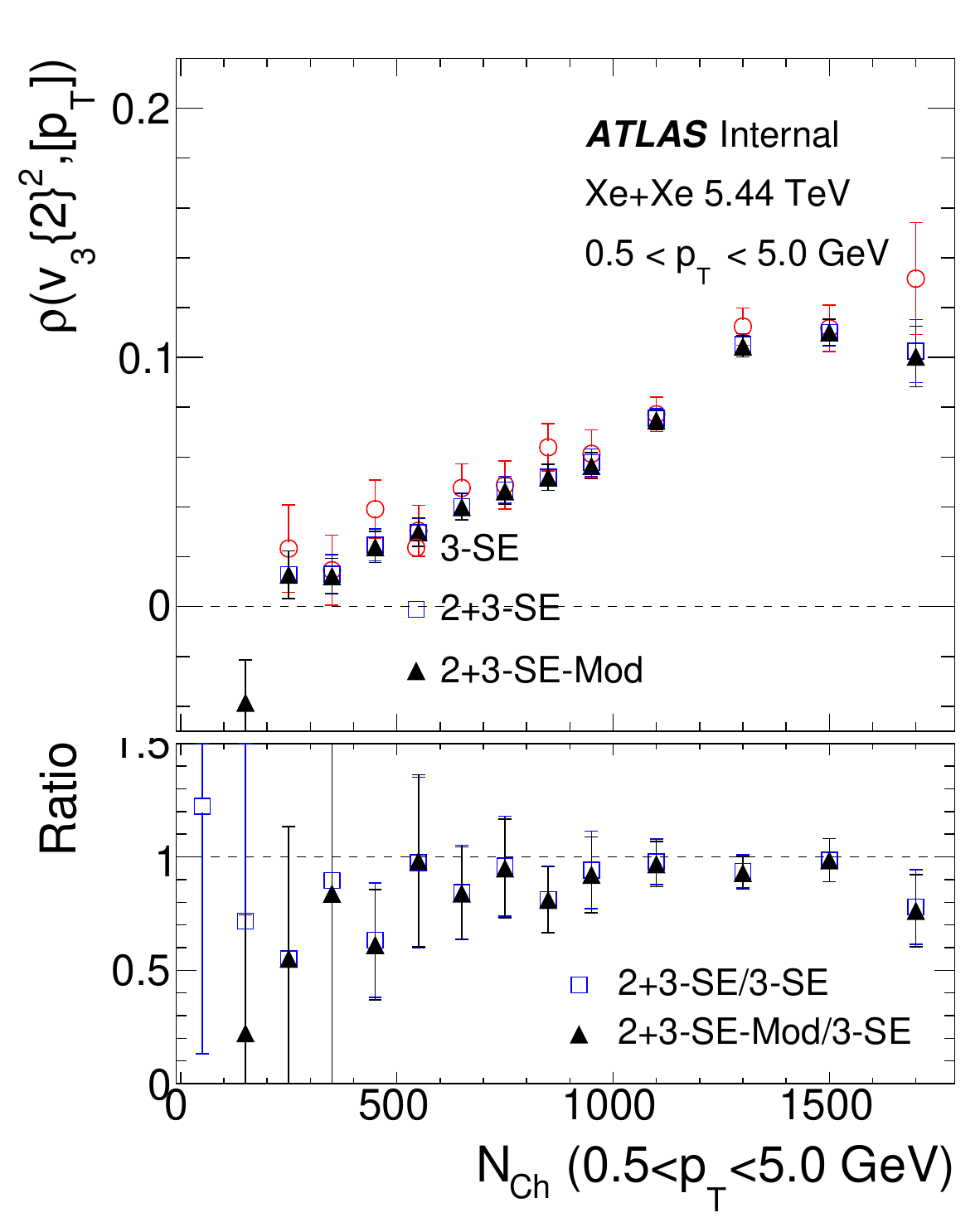}
\includegraphics[width=0.32\linewidth]{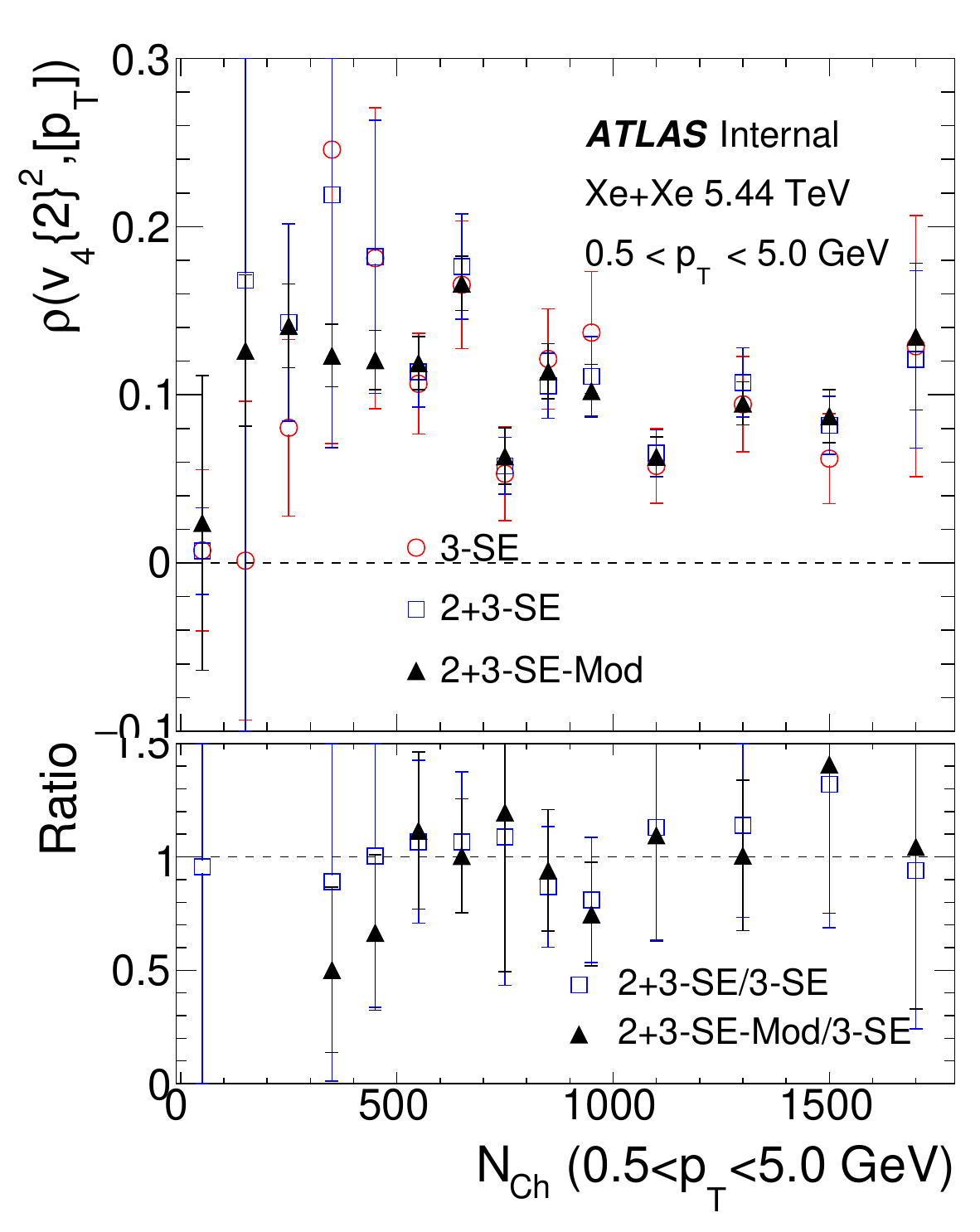}
\caption{Comparison of $\rho(v_n\{2\}^2,[\pT])$ using different $\eta$-subevents in Xe+Xe collisions for $n=2$, 3 and 4. The result is obtained for the $|\eta|<2.5$. The error bars represent statistical uncertainties.}
\label{fig:RhoS_Xe}
\end{figure}

Finally, three combinations for the Pearson coefficient $\rho_n$ are tested: \emph{3‐SE} ($\cov{n}$ from three‐subevent, $\var{n}$ from two‐subevent), \emph{2+3‐SE} ($\cov{n}$ averaged two‐ and three‐subevent, $\var{n}$ two‐subevent), and \emph{2+3‐SE‐Mod} (same $\cov{n}$, hybrid $\var{n}$). These are shown in Figures~\ref{fig:RhoS_Pb} and \ref{fig:RhoS_Xe} for $n=2,3,4$. All methods agree for $n=3,4$, with “2+3‐SE‐Mod” offering the best precision, and thus is adopted for the final $\rho_n$ results.

\clearpage

\begin{figure}[htbp]
\centering
\includegraphics[width=0.8\linewidth]{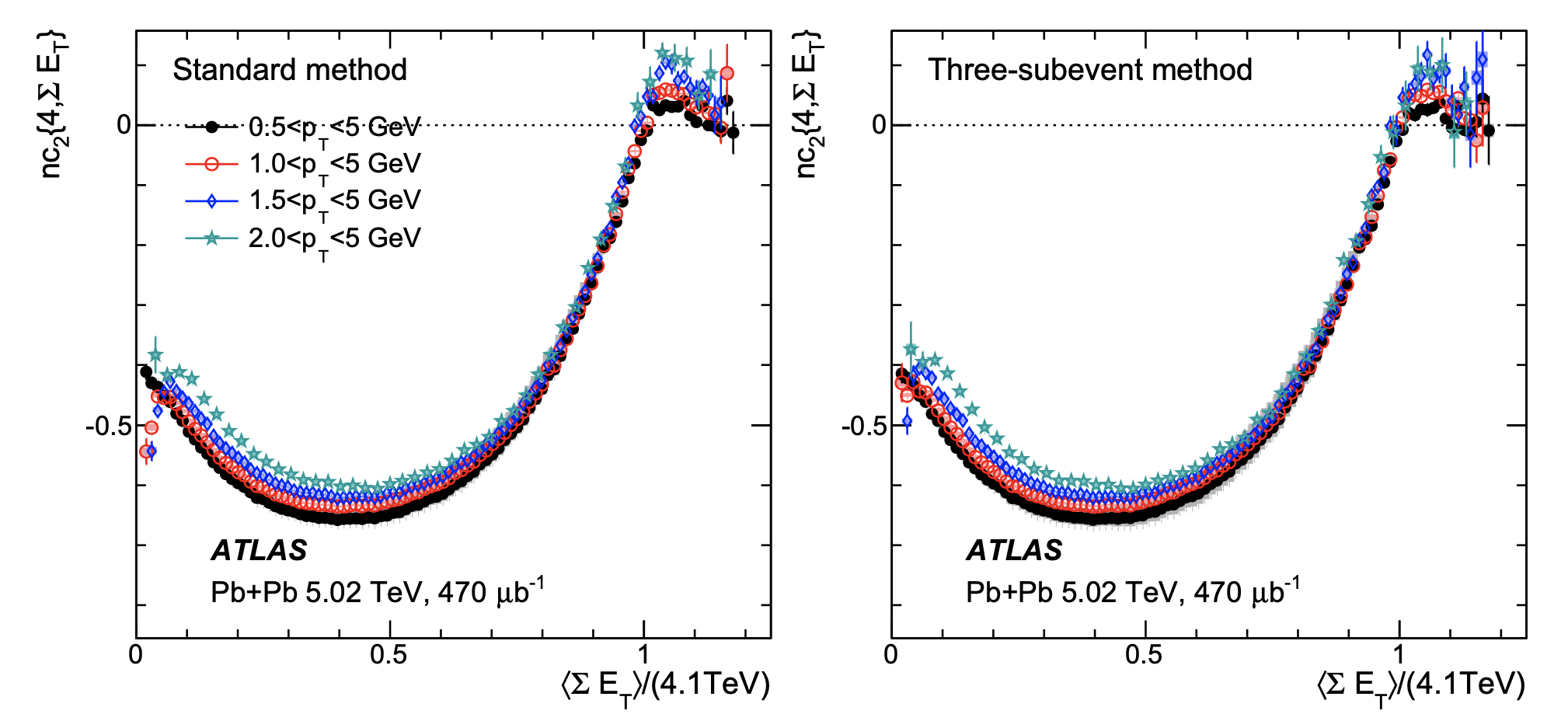}
\caption{The $nc_2\{4\}$ values calculated for charged particles in several $\pT$-ranges with the standard method (left) and three-subevent method (right). The error bars and shaded boxes represent the statistical and systematic uncertainties, respectively.}
\label{fig:Ming_decor}
\end{figure}

Some of the differences between the different subevent methods were previously observed in other analysis. Figure~\ref{fig:Ming_decor} is taken from the ATLAS publication on higher order cumulants in Pb+Pb~\cite{Aaboud:2019sma}. It shows the comparison of the normalized four-particle cumulant $nc_2\{4\}$ from standard and three-subevent methods in different $\pT$-ranges in Pb+Pb collisions. Significant differences were observed between the two methods over a wide range in centrality. These differences in $nc_2\{4\}$ and also similar kind of differences in ATLAS publication of mixed-harmonic symmetric cumulants~\cite{Aaboud:2018syf} were attributed to longitudinal flow decorrelations. Following the same logic, the differences in the observables we observe in this analysis between different methods based on subevents can also be attributed to flow decorrelations.

\begin{figure}[htbp]
\centering
\includegraphics[width=0.8\linewidth]{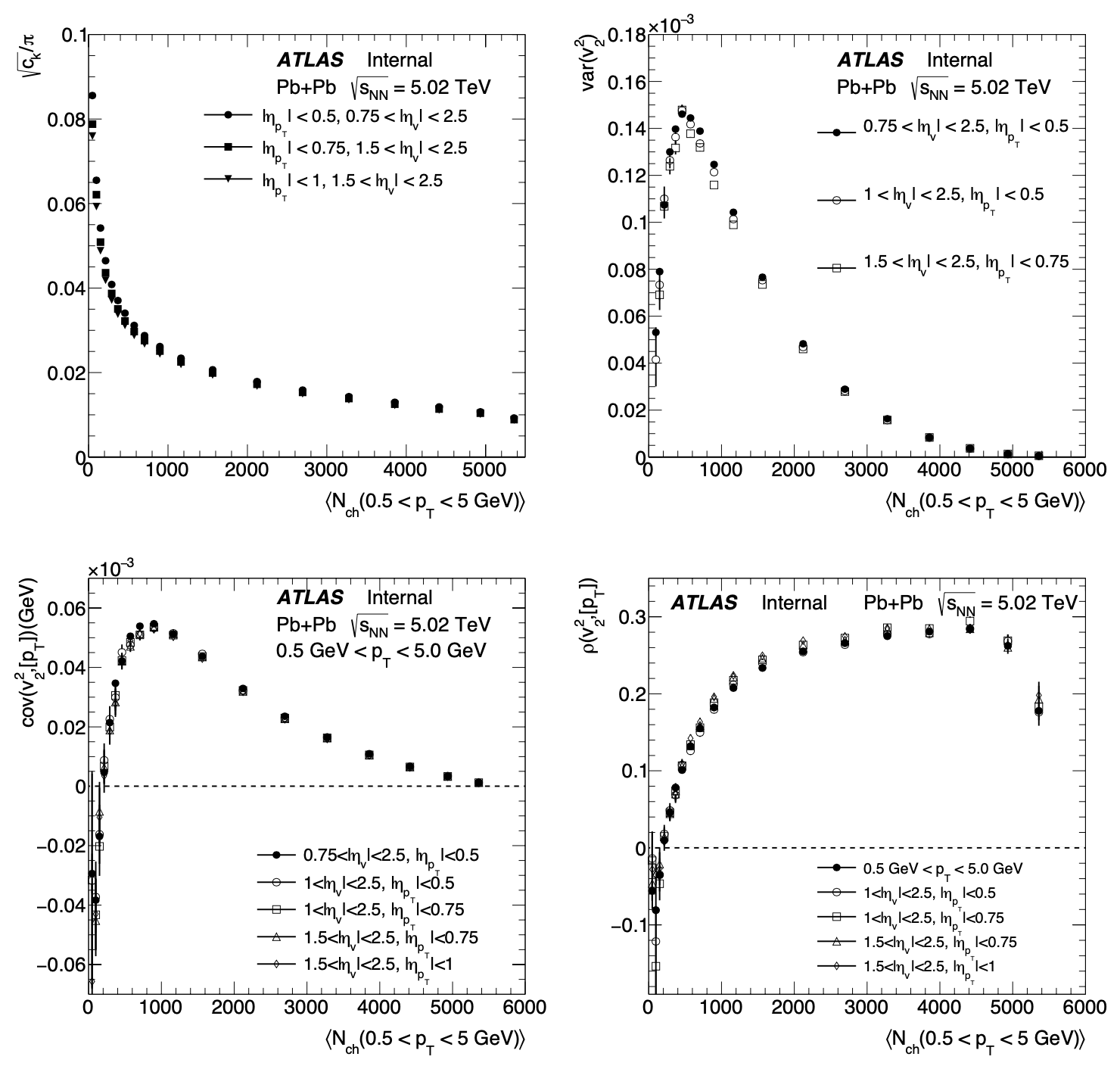}
\caption{$\sqrt{c_k}/\langle [\pT] \rangle$, $\mathrm{Var}(v_2\{2\}^2)_{\mathrm{dyn}}$, $\mathrm{cov}(v_2\{2\}^2,[\pT])$ and $\rho(v_n\{2\}^2,[\pT])$ for different $\eta-gaps$ selections between A, B and C in Pb+Pb collisions.}
\label{fig:Adam_etagap}
\end{figure}

Due to applied changes in $\eta$-ranges in subevents A, B and C, not only the track multiplicity changed significantly but also changed the gap widths between sub-events A and C as well as the gap between sub-events A$+$C and B. This variation in $\eta$-gaps allowed the study of impact of “non-flow” correlations on the obsevables. Figure~\ref{fig:Adam_etagap} shows the comparison of the observables - $\sqrt{c_k}/\langle [\pT] \rangle$, $\mathrm{Var}(v_2\{2\}^2)_{\mathrm{dyn}}$, $\mathrm{cov}(v_2\{2\}^2,[\pT])$ and $\rho(v_n\{2\}^2,[\pT])$ in Pb+Pb collisions for different $\eta$-gap selections between A, B and C. The observables were seen to depend very weakly on the $\eta$-gap choices.

The primary systematic uncertainty in $\rho(v_n\{2\}^2,[\pT])$ stems from $\mathrm{cov}(v_n\{2\}^2,[\pT])$, with partial cancellation from tracking efficiency and track selection cuts. For $\mathrm{cov}(v_2\{2\}^2,[\pT])$, systematics are $\lesssim 5\%$ across all centralities. For $\mathrm{cov}(v_3\{2\}^2,[\pT])$ and $\mathrm{cov}(v_4\{2\}^2,[\pT])$, systematics are $\lesssim 5\%$ for central/mid-central events; relative uncertainties are not quoted for peripheral collisions due to large statistical errors or sign changes, but full systematic uncertainties will be shown. Pileup contamination contributes $<0.5\%$ systematic uncertainty across all centralities. Flattening has minimal systematic impact for all observables and centralities. For $n=2$, track quality variation and efficiency correlation dominate systematics for all observables. For $n=3, 4$, efficiency variations contribute comparably or more to systematic uncertainty. Centrality percentile uncertainty is most significant in peripheral regions.
\clearpage

\section{Flattening and Recentering Corrections for Flow Vectors}\label{sec:dAna_Pb}
\subsection{Flattening}
\begin{figure}[htbp]
\centering
\includegraphics[width=1.\linewidth]{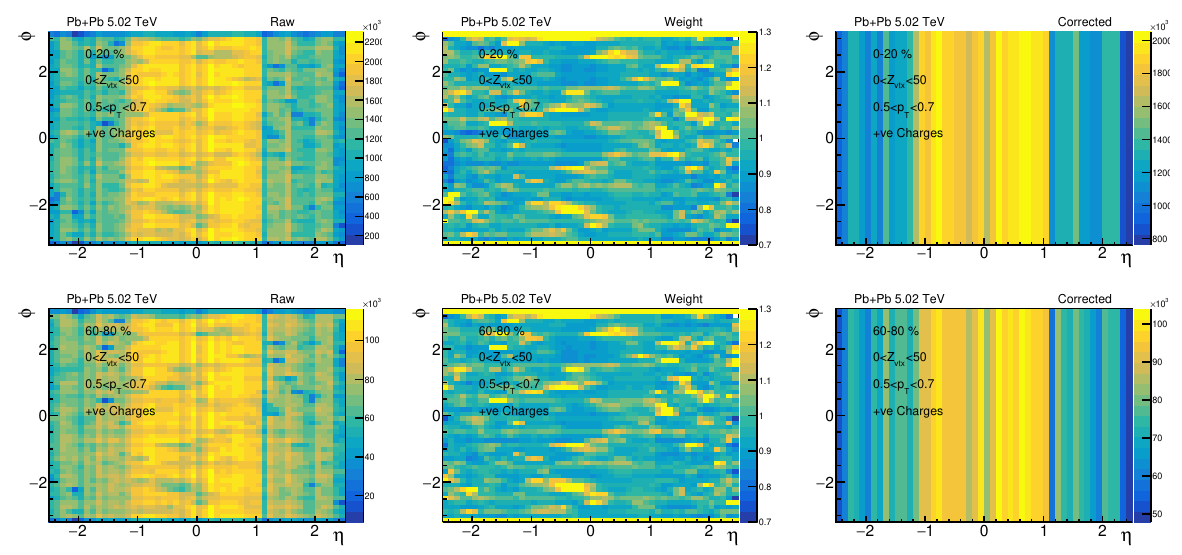}
\caption{Flattening procedure in Pb+Pb for two centrality ranges. Top row is for 0-20$\%$ centrality and bottom row is for 60-80$\%$ centrality. Left column shows the raw ($\eta$,$\phi$) distributions, middle column shows the estimated correction factors $w_{\phi}$ and right column shows the corrected ($\eta$,$\phi$) distributions of the selected tracks.}
\label{fig:flatPb_1}
\end{figure}

Raw $(\eta, \phi)$ distributions of selected tracks, as shown in the left column of Figure \ref{fig:flatPb_1} for central (0-20$\%$) and peripheral (60-80$\%$) events, exhibit significant non-uniformities and voids along $\phi$ for various $\eta$ intervals. The middle column of Figure \ref{fig:flatPb_1} displays this correction factor for the specified centralities. The right column demonstrates the $(\eta, \phi)$ distribution subsequent to applying the $w_{\phi}$ correction.

\begin{figure}[htbp]
\centering
\includegraphics[width=1.\linewidth]{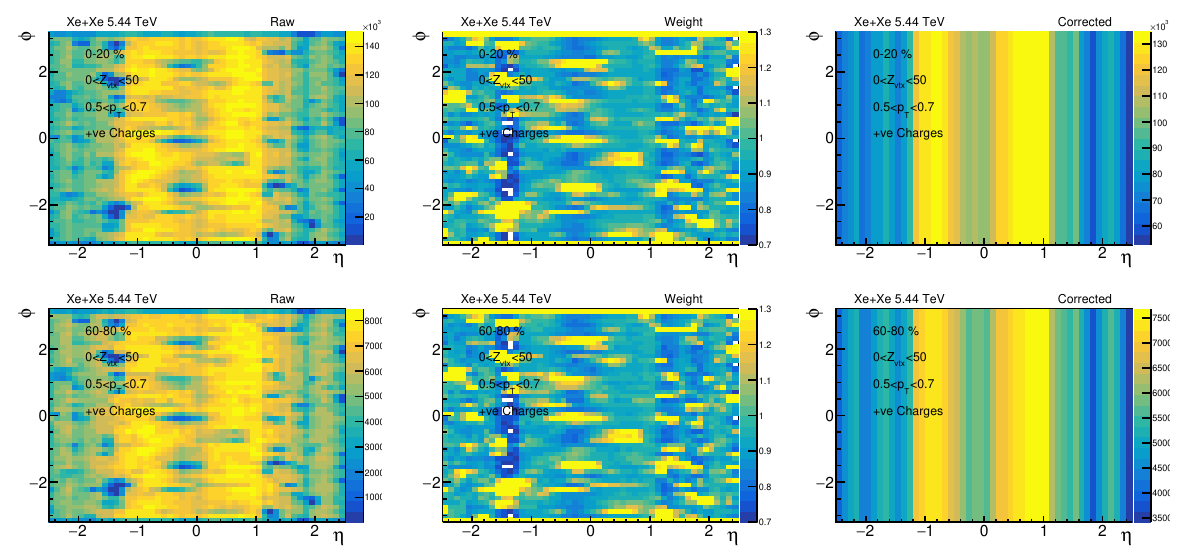}
\caption{Flattening procedure applied to Xe+Xe data for two centrality ranges. The top row corresponds to 0-20$\%$ centrality, and the bottom row to 60-80$\%$. The left column shows the raw ($\eta$,$\phi$) distributions of selected tracks. The middle column displays the estimated correction factors $w_{\phi}$. The right column presents the ($\eta$,$\phi$) distributions after applying the $w_{\phi}$ corrections, showing improved uniformity in $\phi$.}
\label{fig:flatXe_1}
\end{figure}

The top row and bottom row of Figure \ref{fig:flatPb_1} shows such acceptance effects for central 0-20$\%$ and peripheral 60-80$\%$ events. The structures are similar but not identical between the two event classes.
Similar plots for the Xe+Xe collisions are shown in Figure~\ref{fig:flatXe_1}.

\clearpage

\subsection{Recentering of Flow Vectors}\label{q-vecRec}
$q_n,x$ and $q_n,y$ for $n=2, 3, 4$ in Pb+Pb and Xe+Xe collisions for different $\eta$ regions are shown before and after recentering in this subsection.

\begin{figure}[htbp]
\centering
\includegraphics[width=0.32\linewidth]{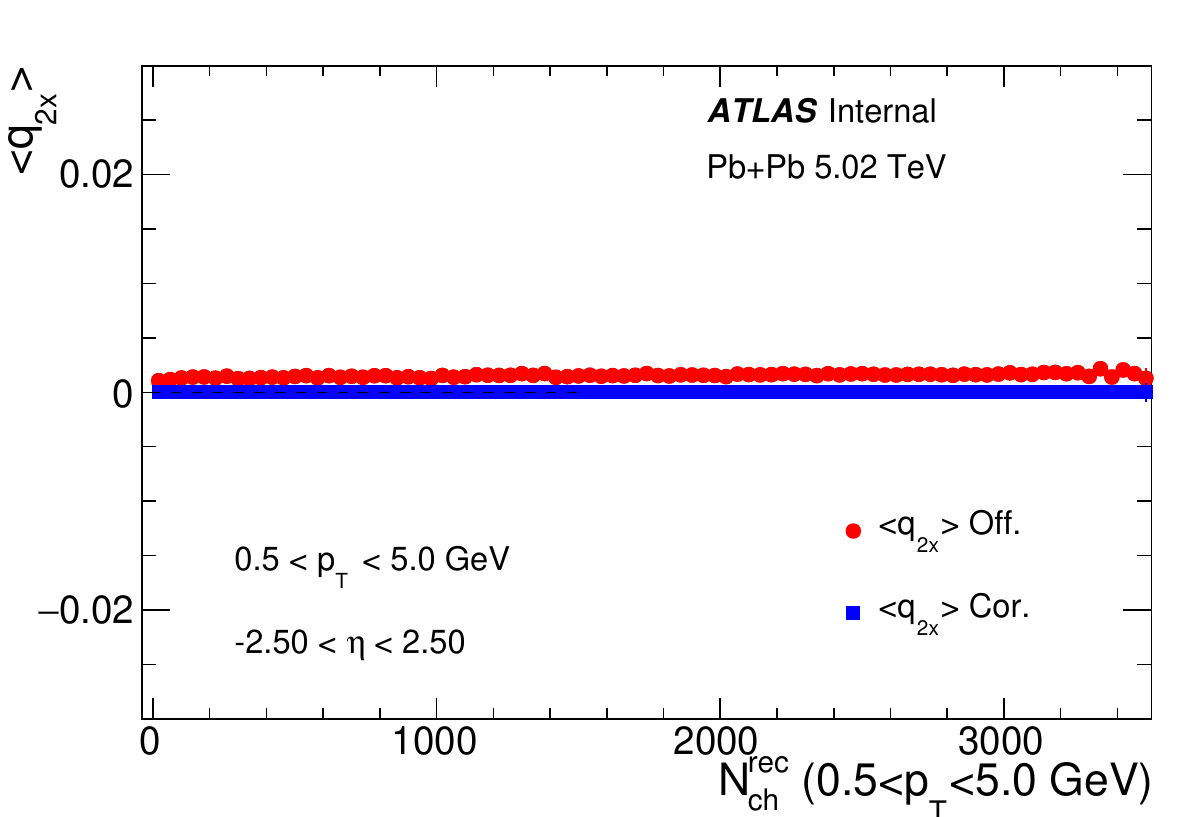}
\includegraphics[width=0.32\linewidth]{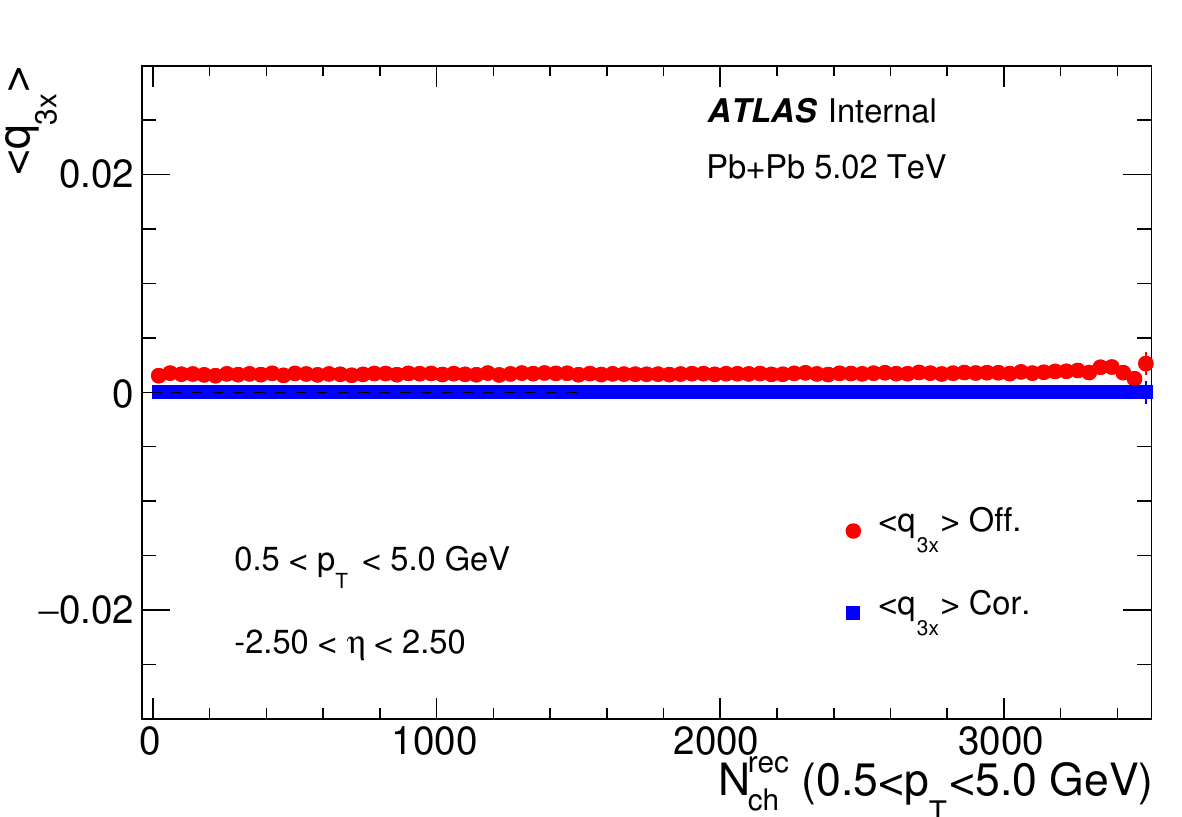}
\includegraphics[width=0.32\linewidth]{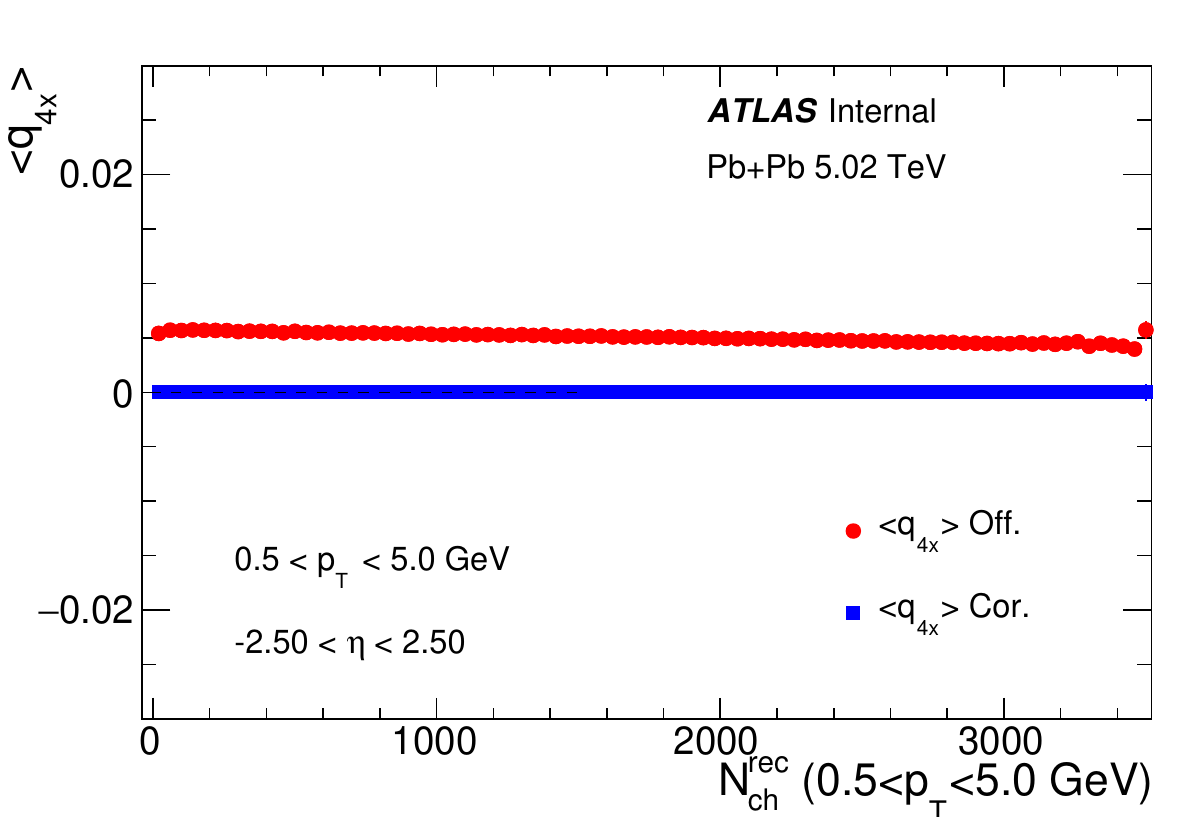}
\includegraphics[width=0.32\linewidth]{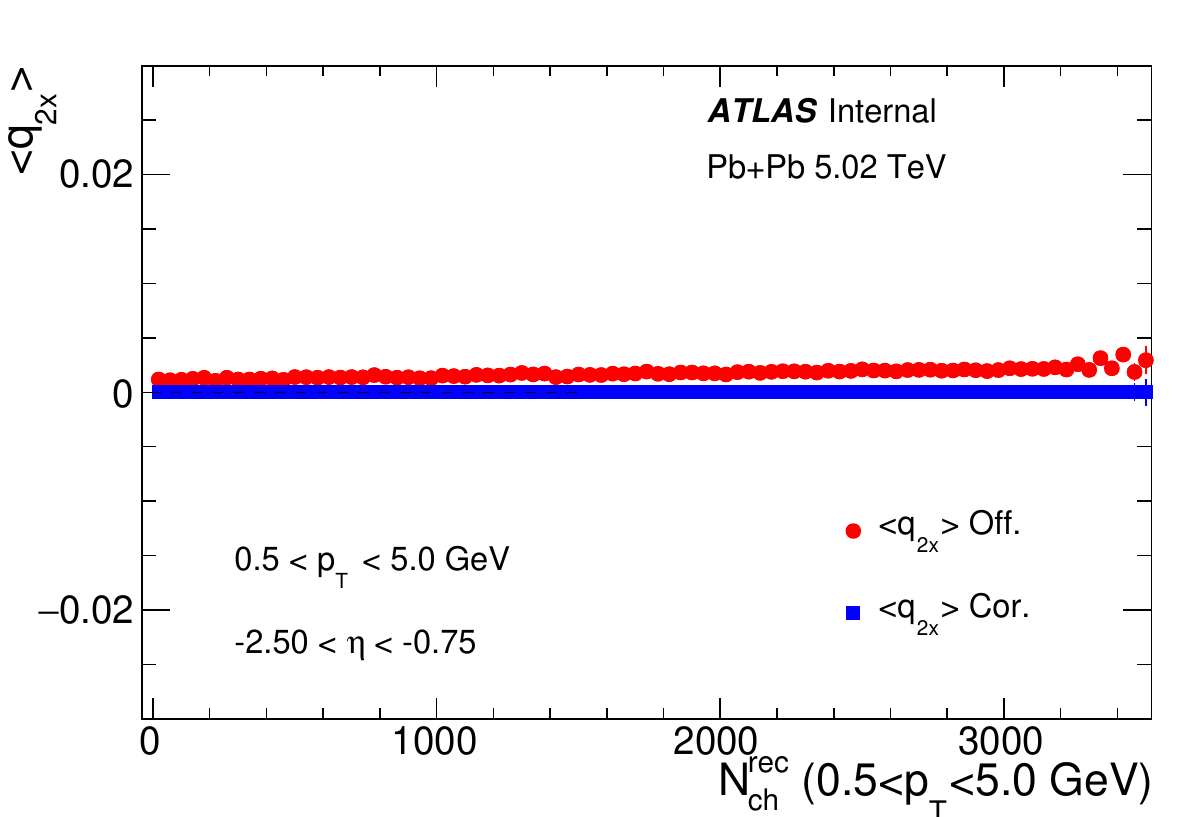}
\includegraphics[width=0.32\linewidth]{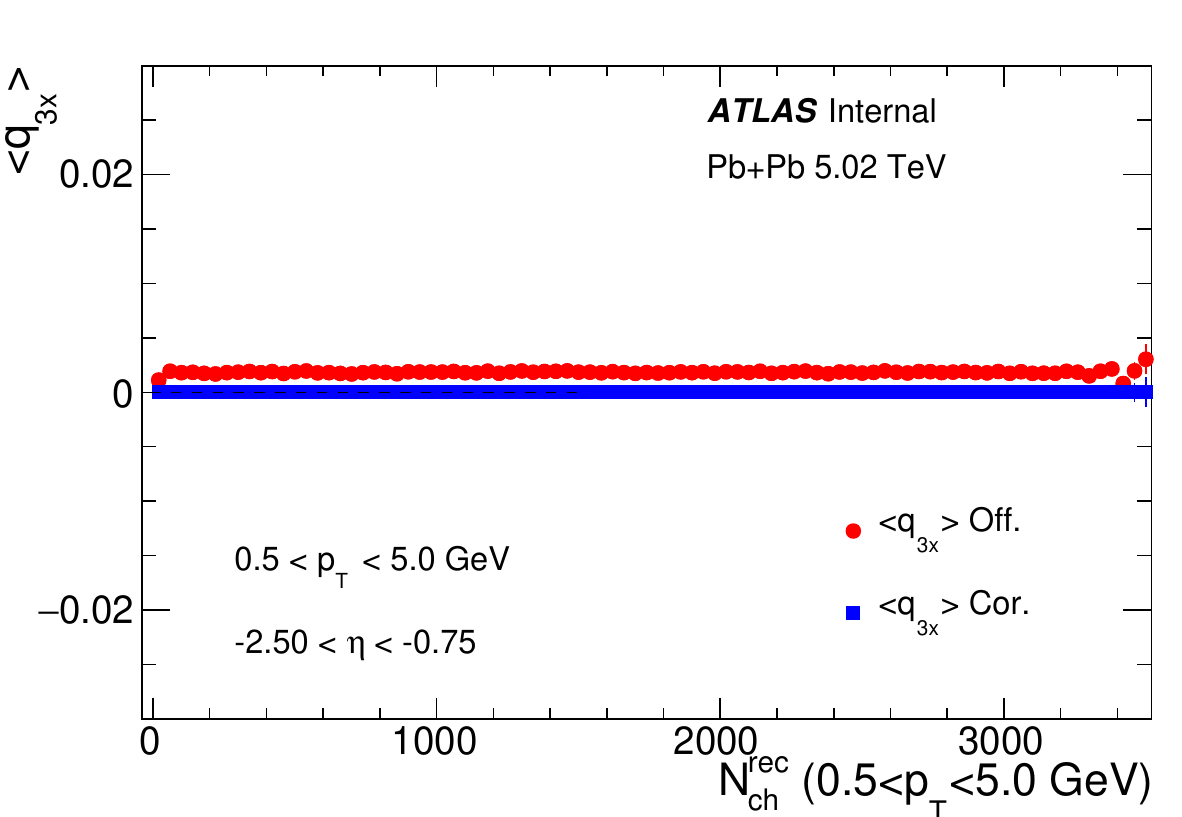}
\includegraphics[width=0.32\linewidth]{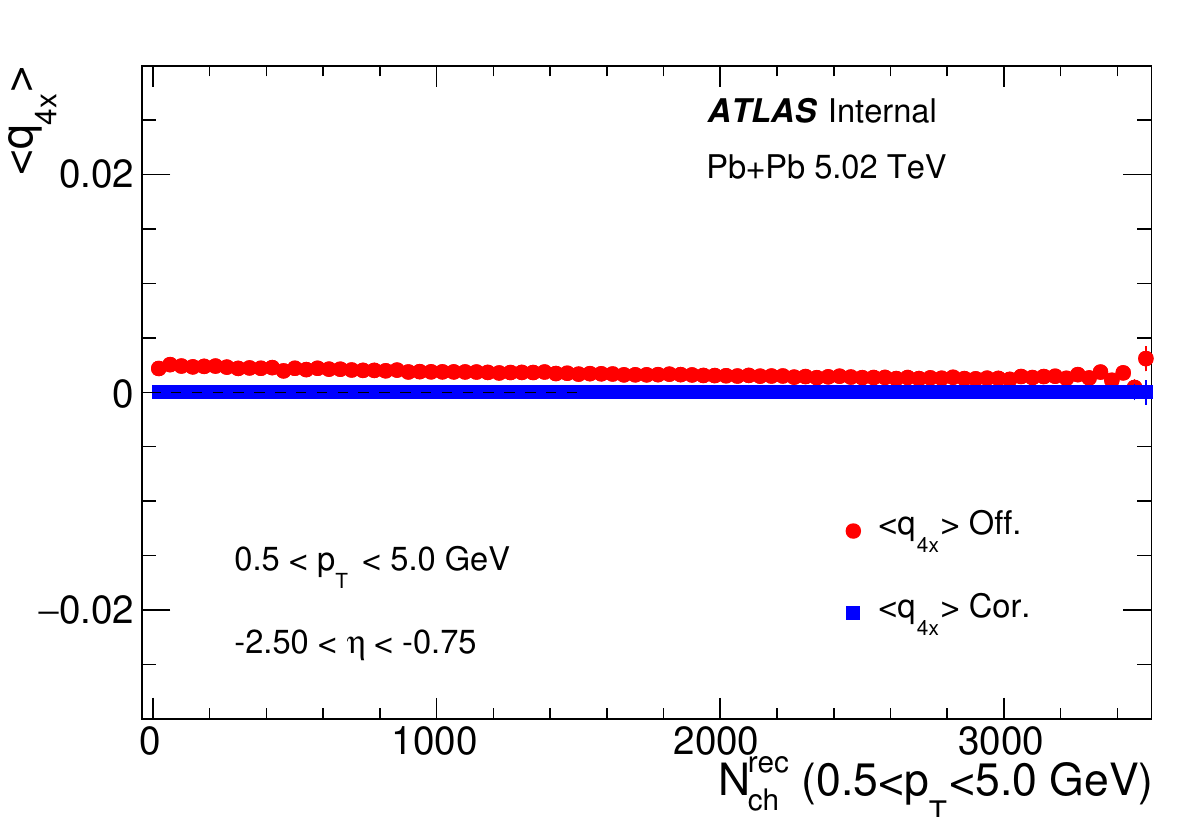}
\includegraphics[width=0.32\linewidth]{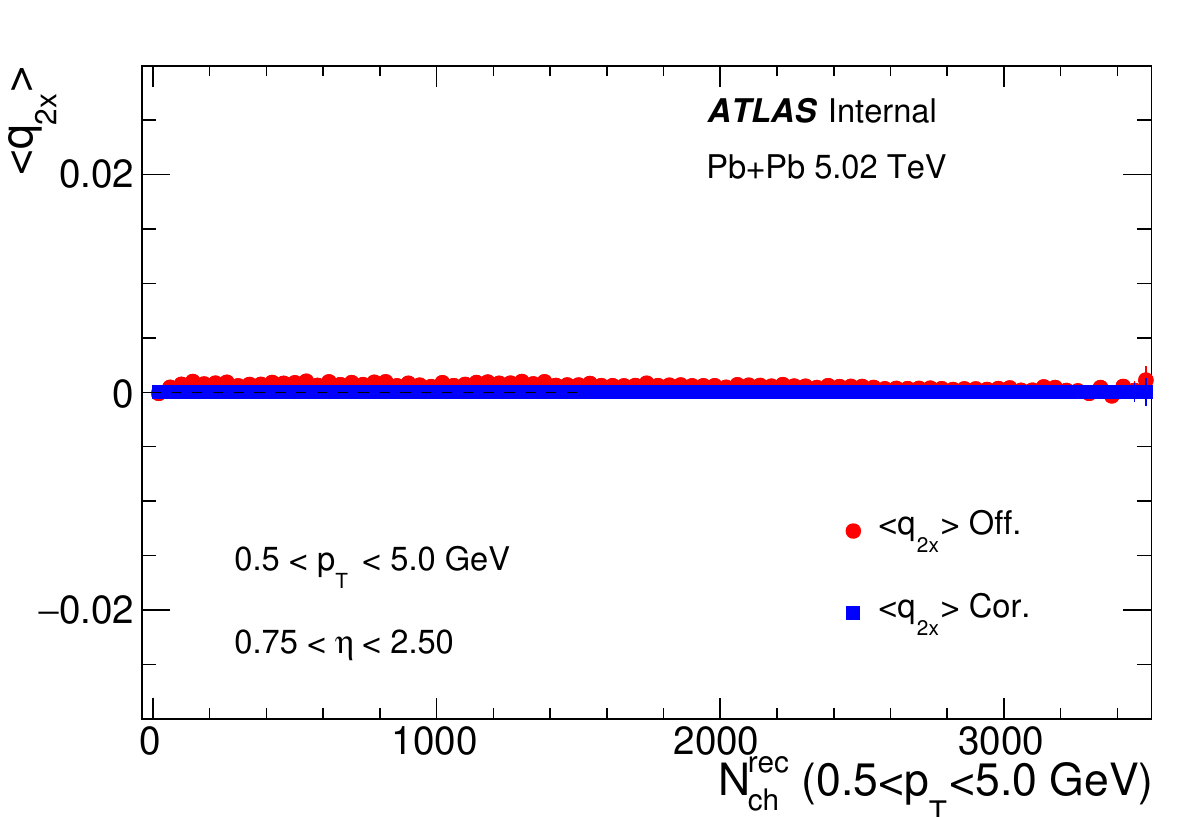}
\includegraphics[width=0.32\linewidth]{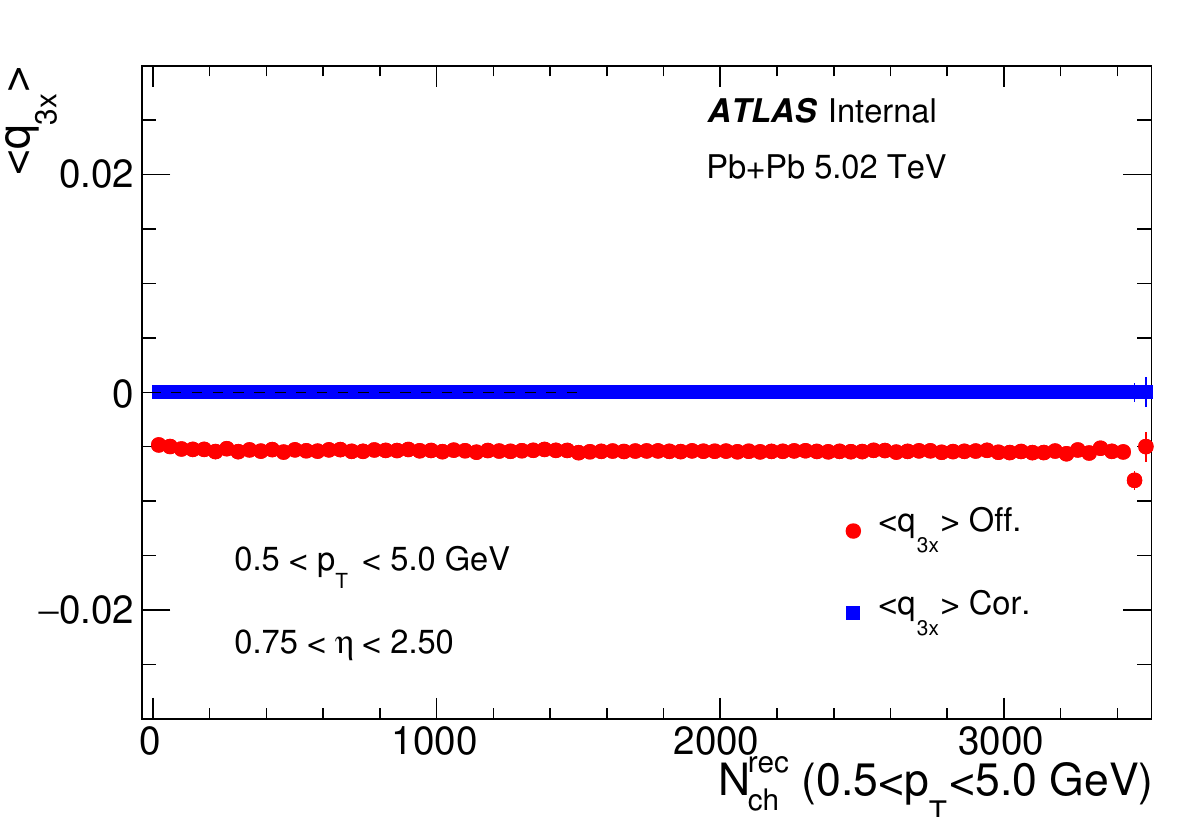}
\includegraphics[width=0.32\linewidth]{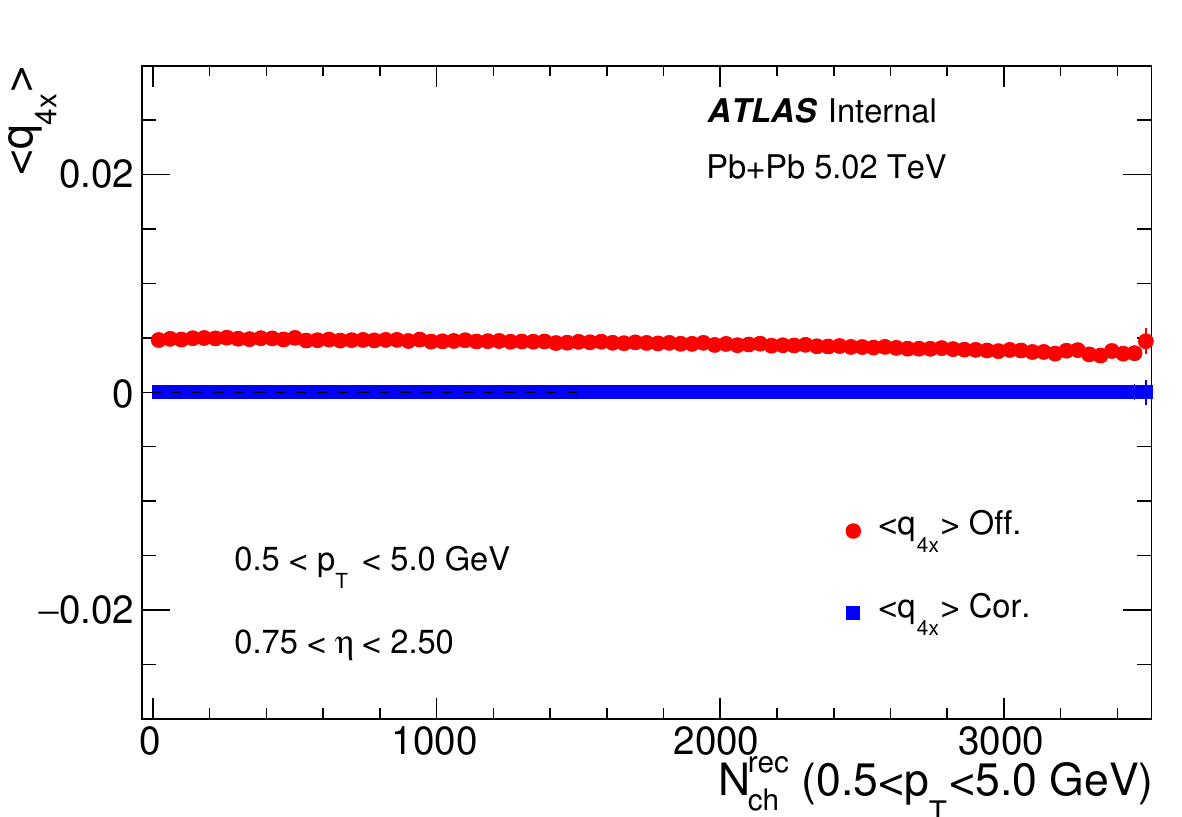}
\caption{Comparison of raw and the recentered $\langle q_{n,x}\rangle$ in Pb+Pb for $n=2$, 3 and 4 and for $-2.50<\eta<-0.75$ (top) and $0.75<\eta<2.50$ (bottom). The error bars represent statistical uncertainties.}
\label{fig:RecX_Pb}
\end{figure}

\begin{figure}[htbp]
\centering
\includegraphics[width=0.32\linewidth]{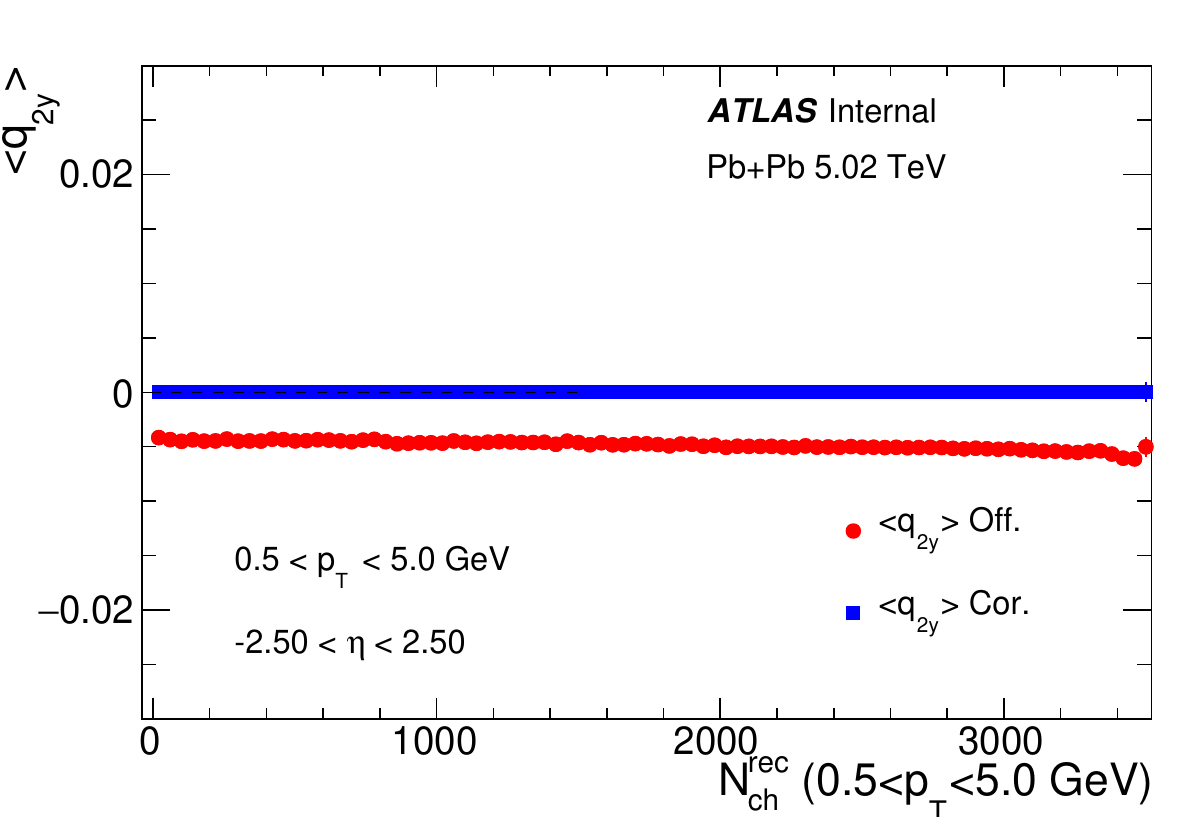}
\includegraphics[width=0.32\linewidth]{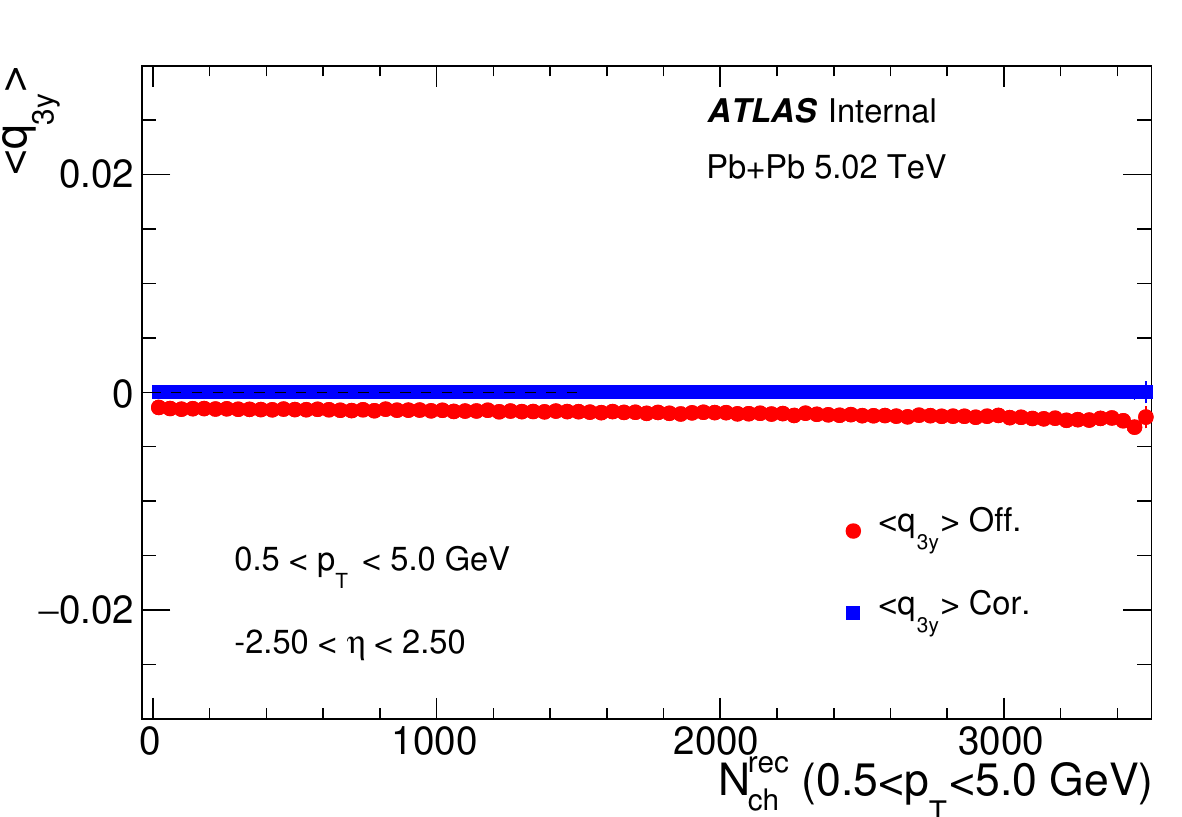}
\includegraphics[width=0.32\linewidth]{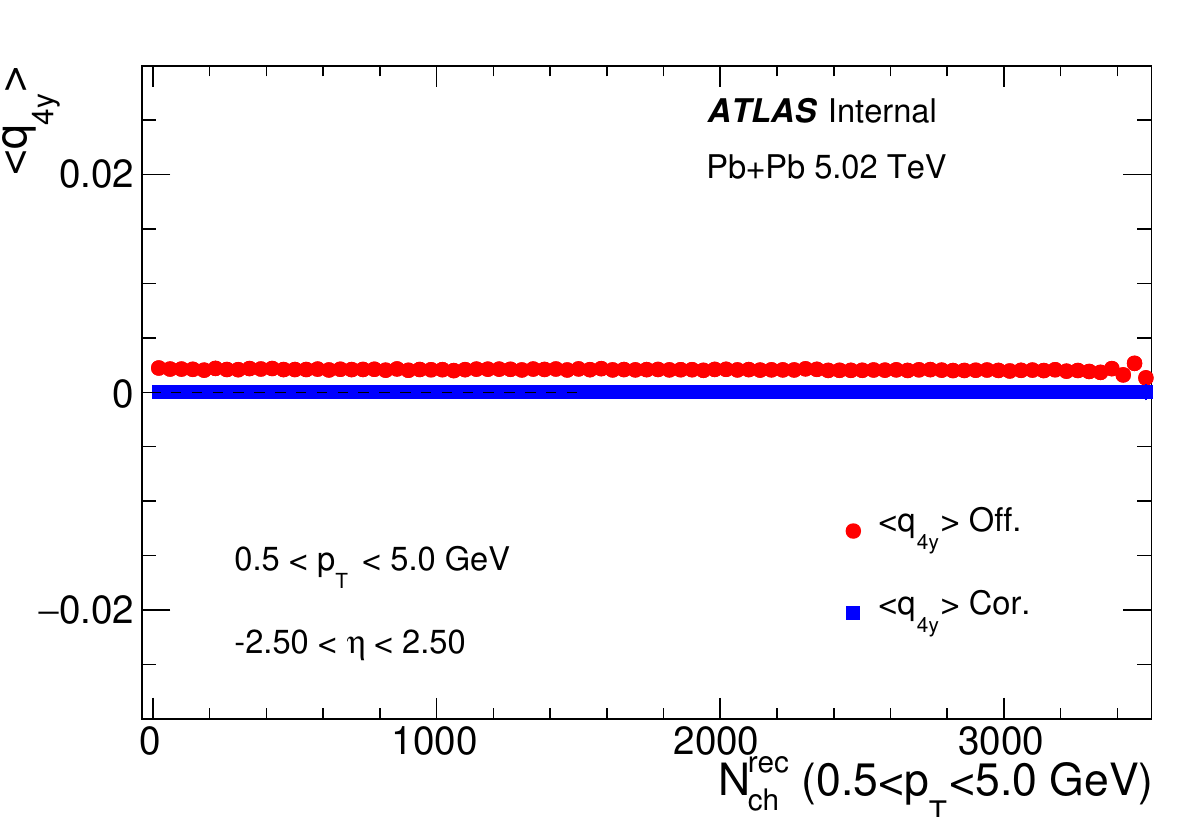}
\includegraphics[width=0.32\linewidth]{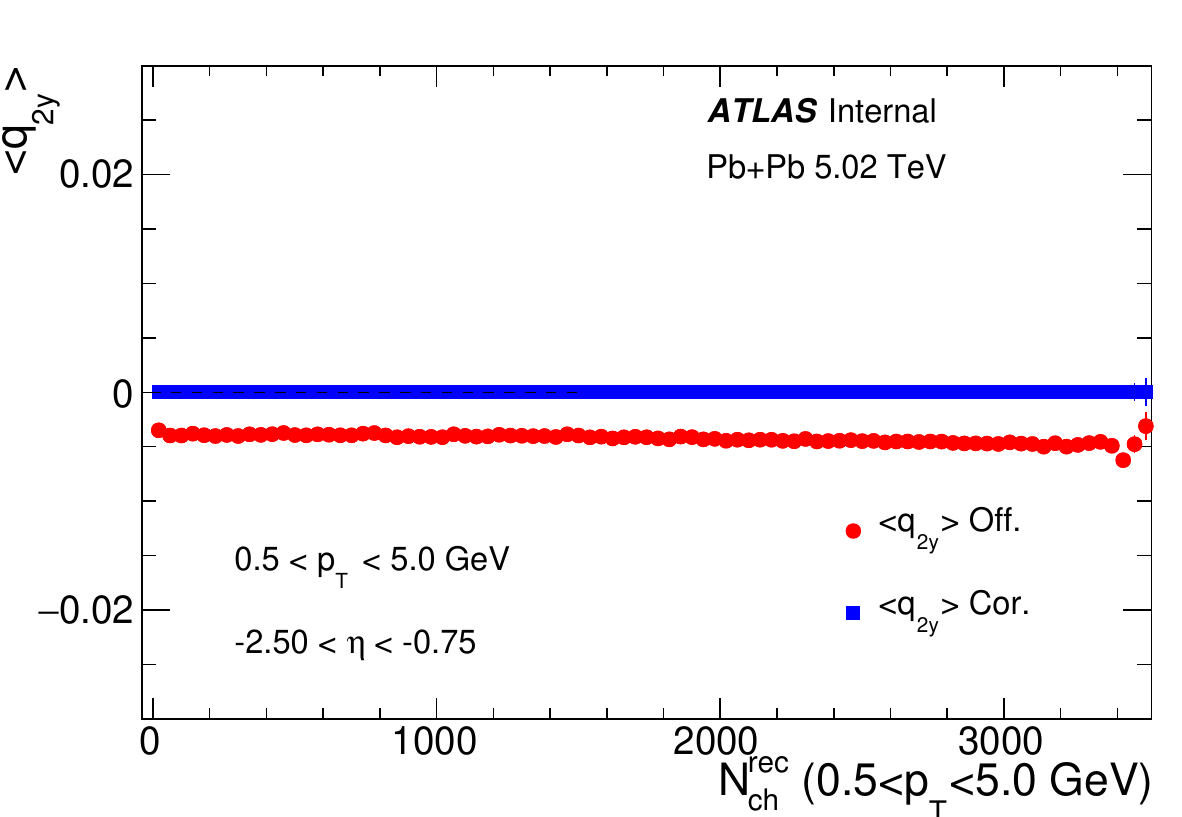}
\includegraphics[width=0.32\linewidth]{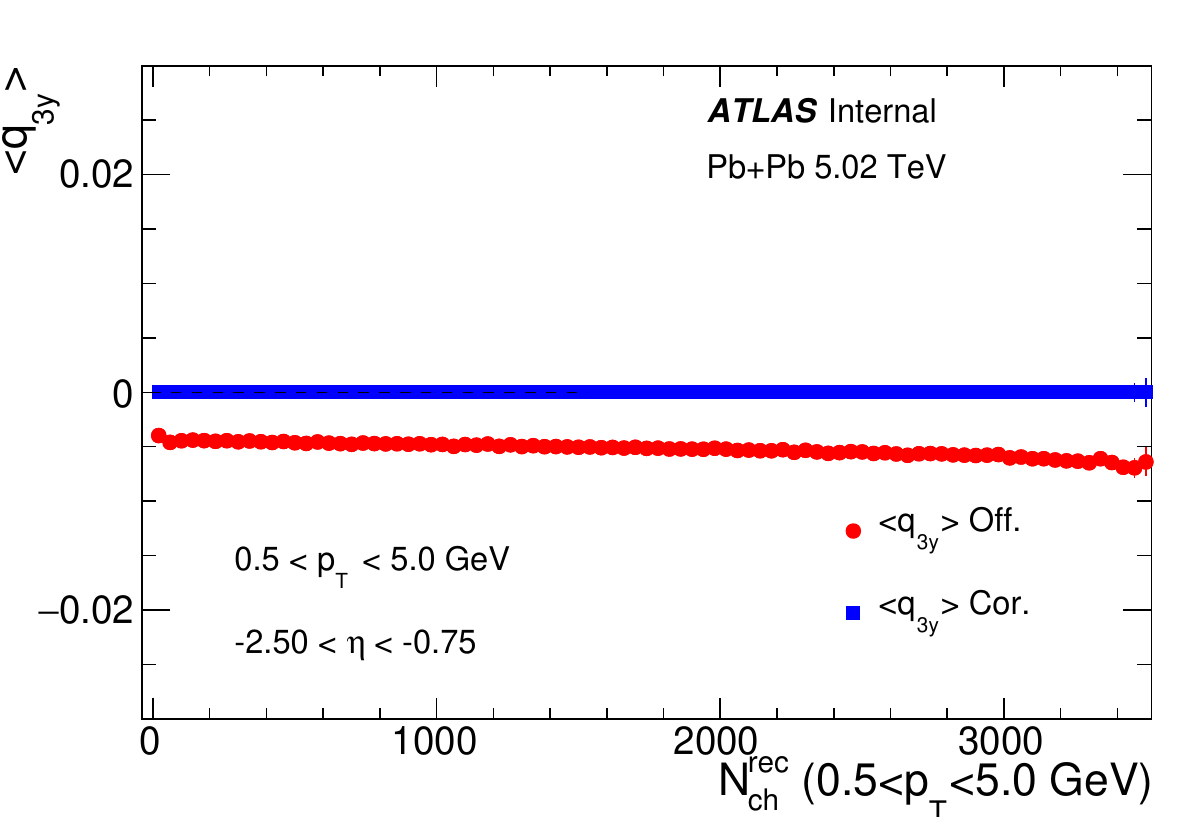}
\includegraphics[width=0.32\linewidth]{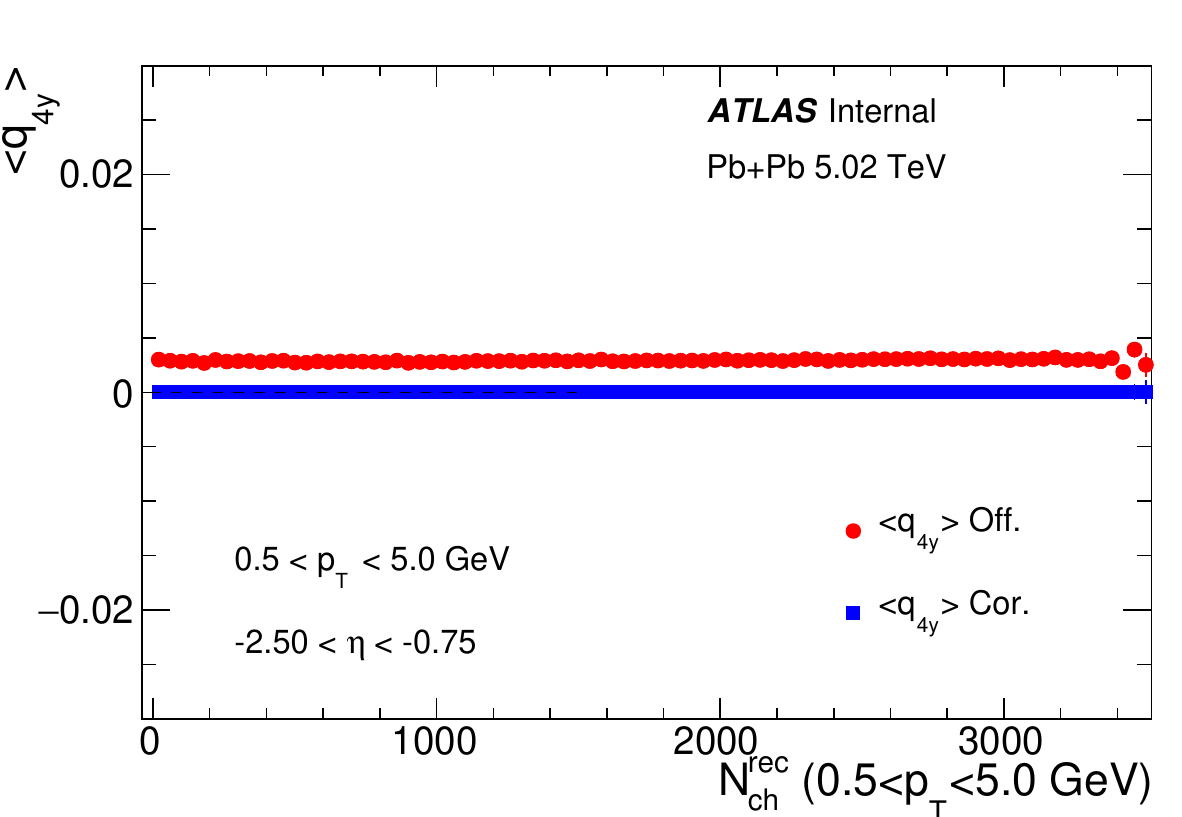}
\includegraphics[width=0.32\linewidth]{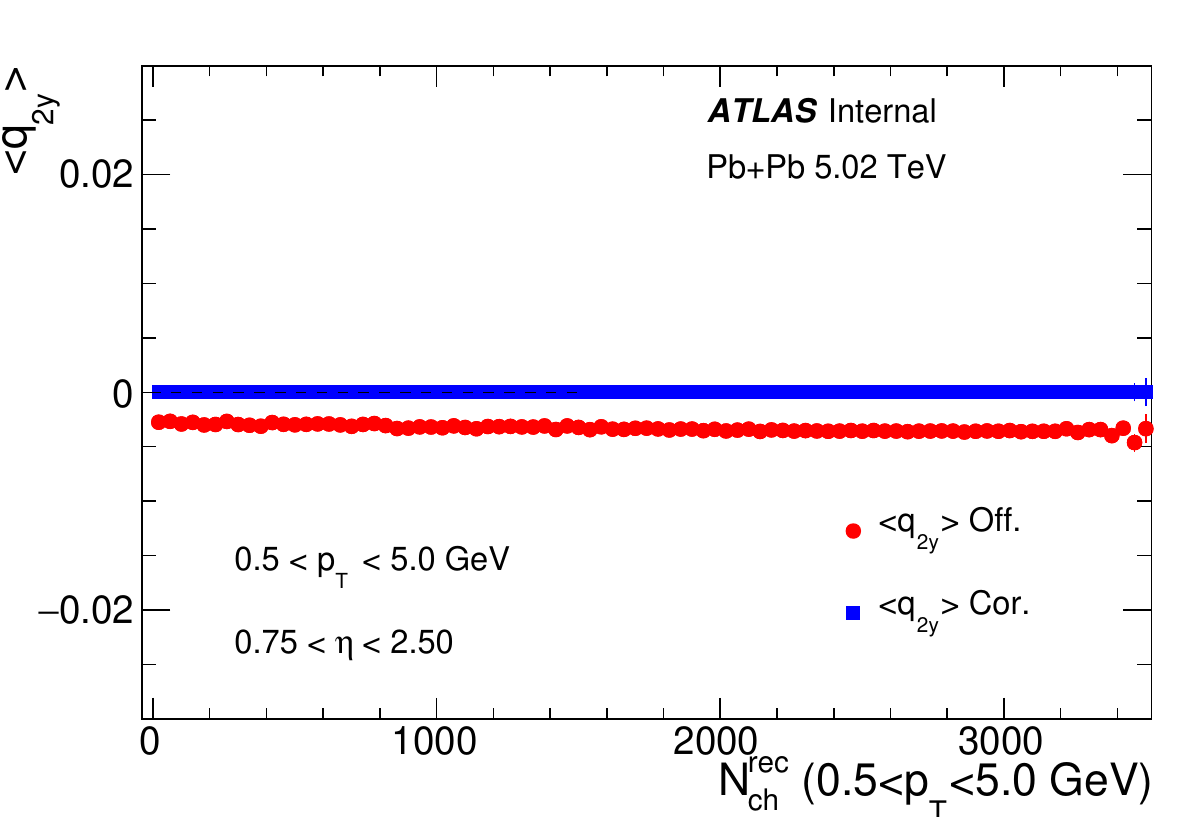}
\includegraphics[width=0.32\linewidth]{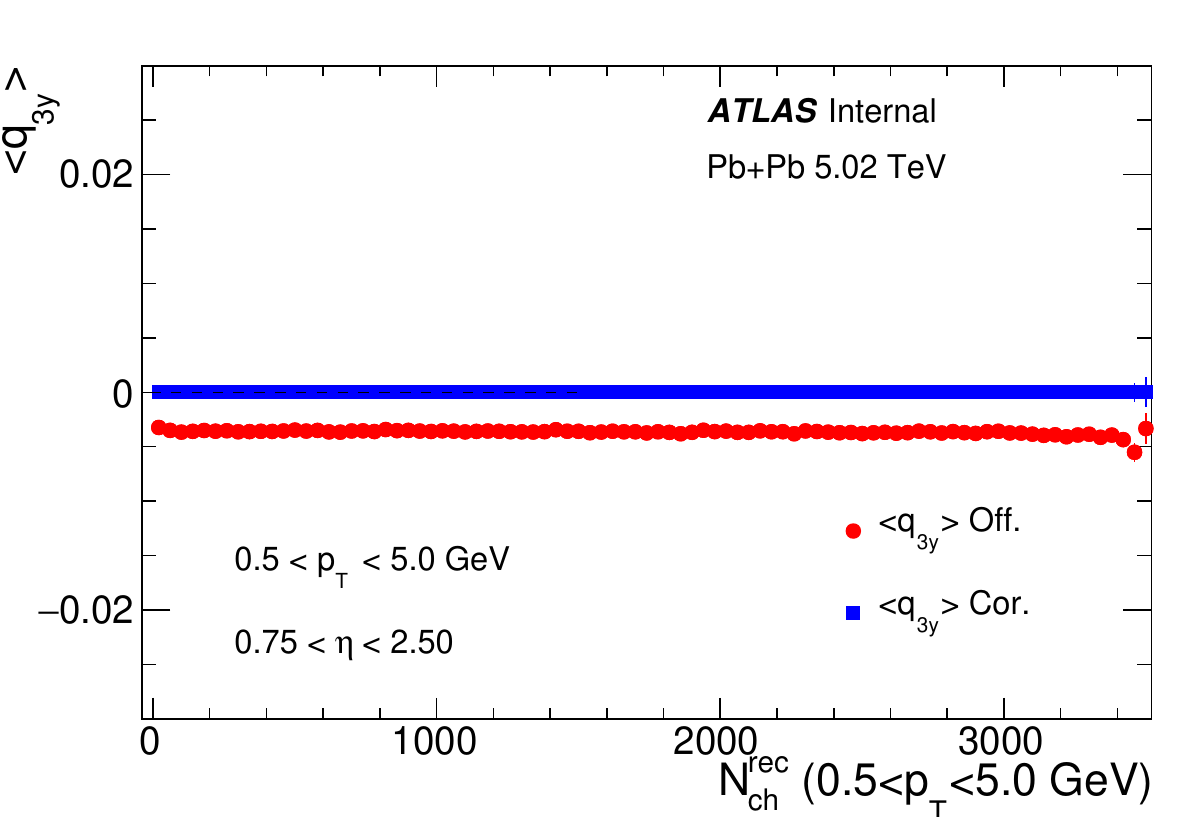}
\includegraphics[width=0.32\linewidth]{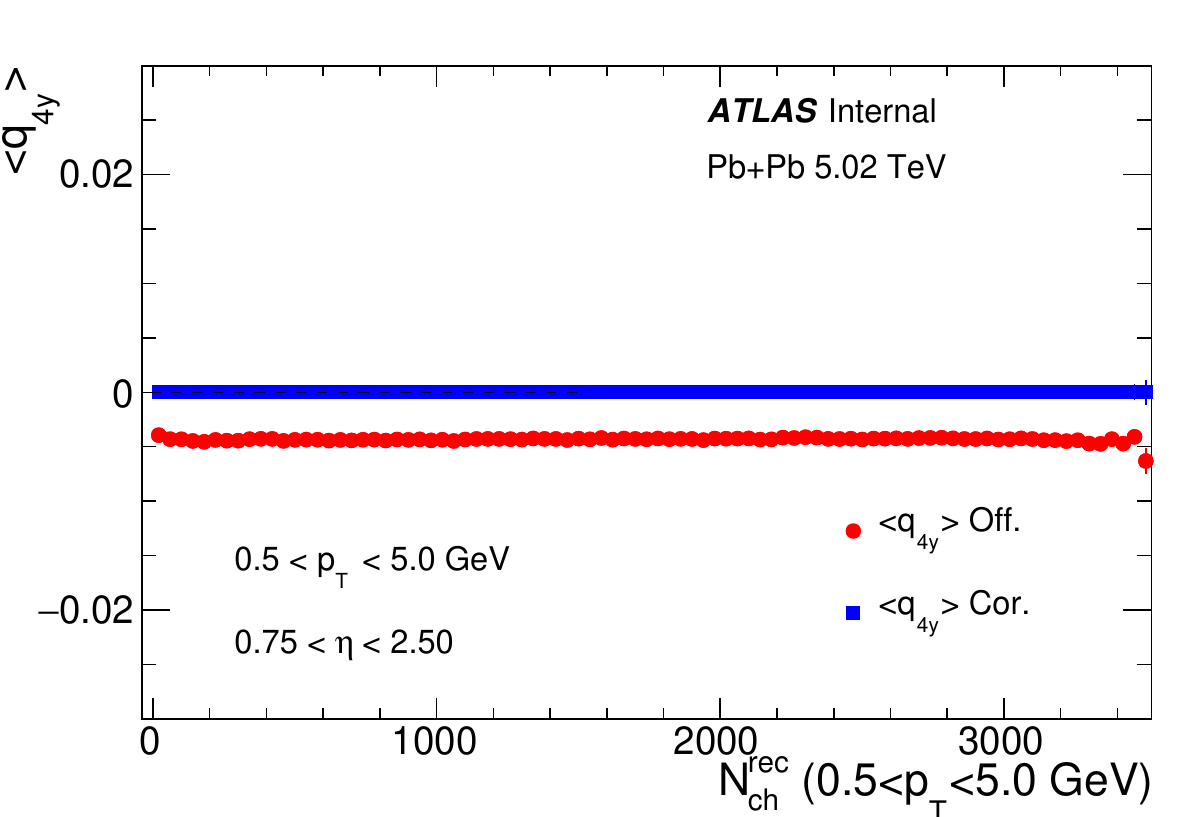}
\caption{Comparison of raw and the recentered $\langle q_{n,y}\rangle$ in Pb+Pb for $n=2$, 3 and 4 and for $-2.50<\eta<-0.75$ (top) and $0.75<\eta<2.50$ (bottom). The error bars represent statistical uncertainties.}
\label{fig:RecY_Pb}
\end{figure}

\begin{figure}[htbp]
\centering
\includegraphics[width=0.32\linewidth]{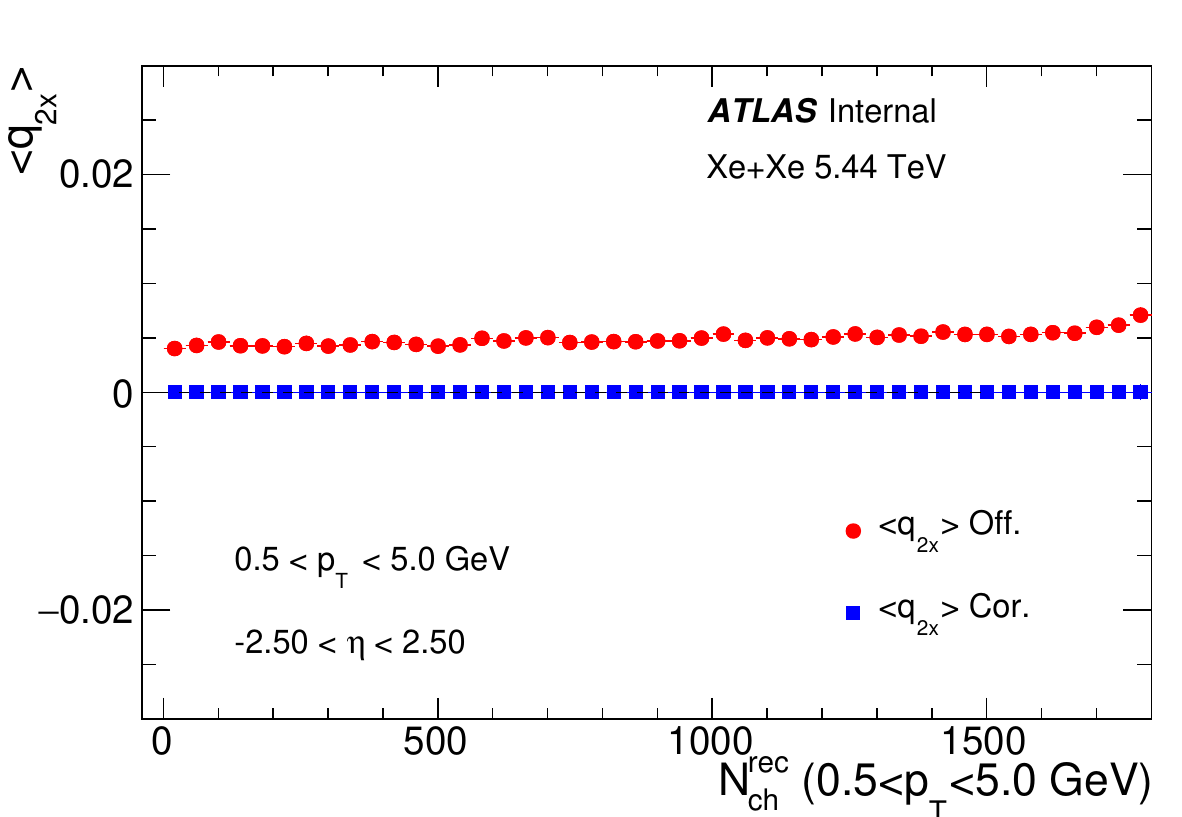}
\includegraphics[width=0.32\linewidth]{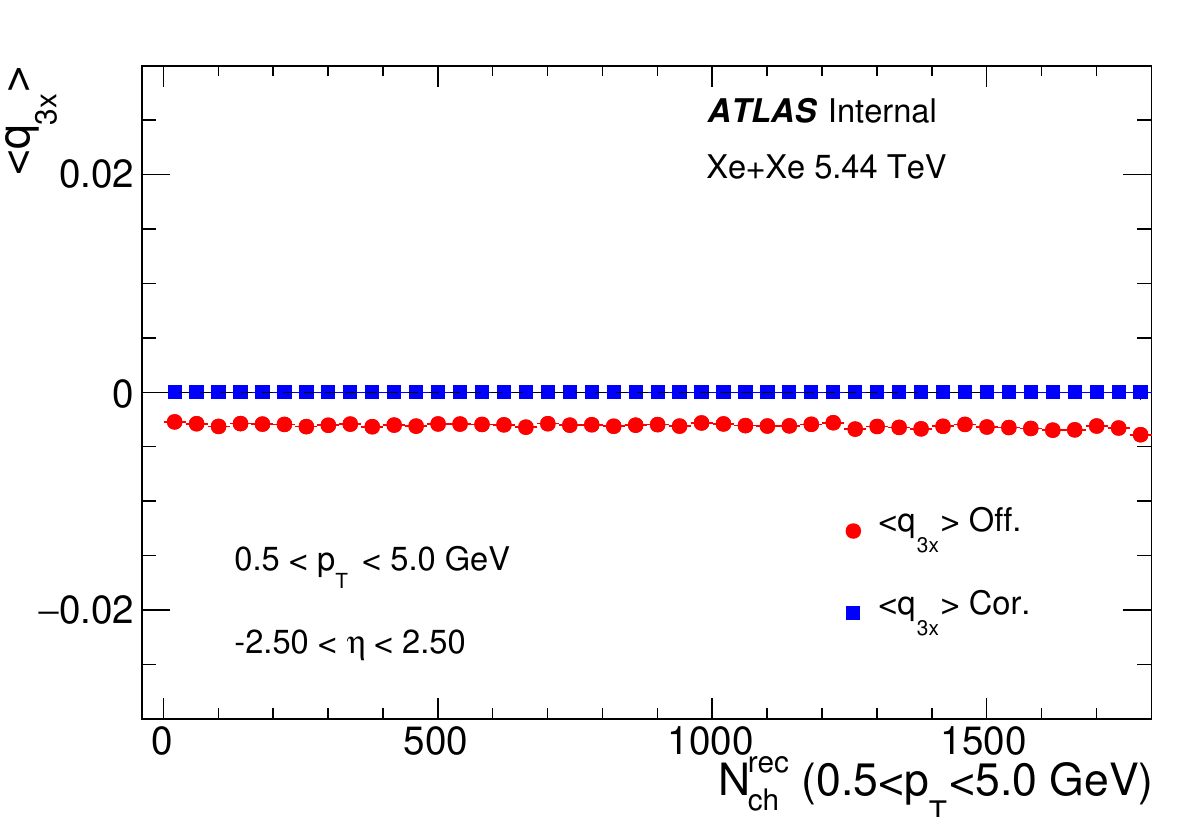}
\includegraphics[width=0.32\linewidth]{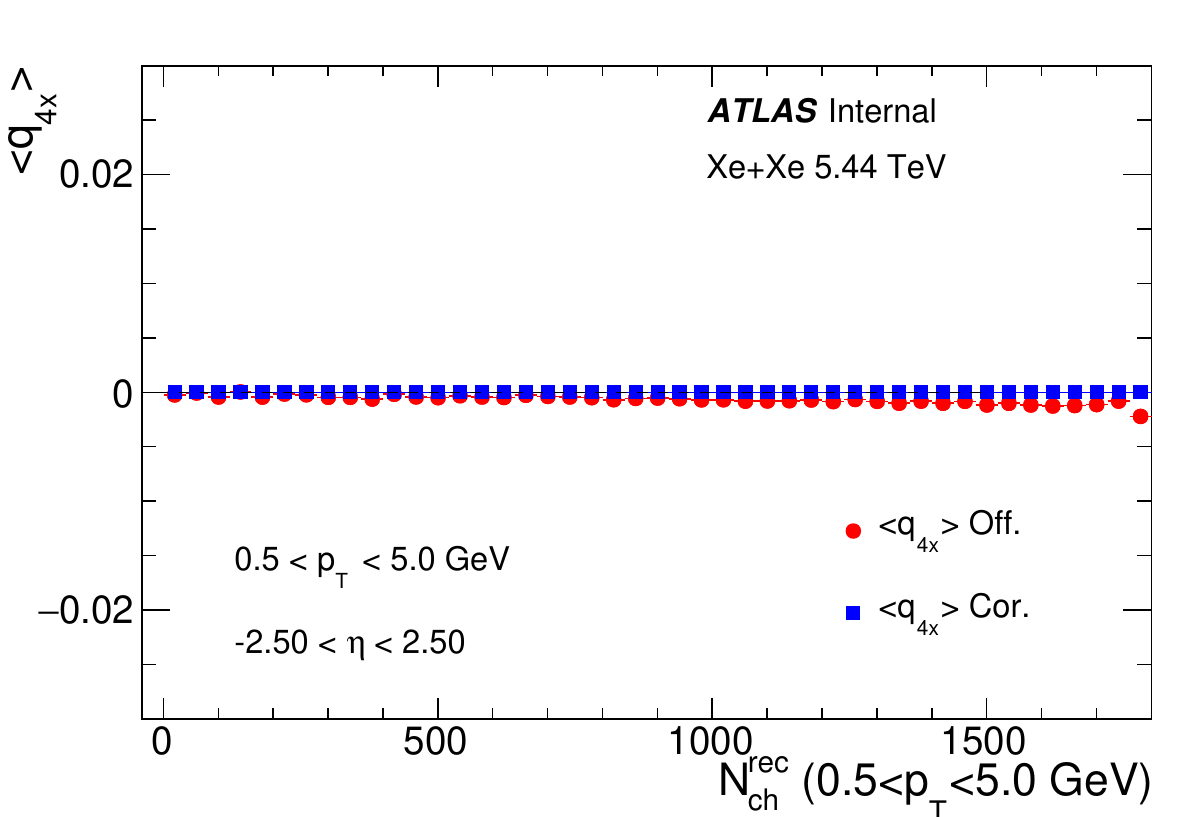}
\includegraphics[width=0.32\linewidth]{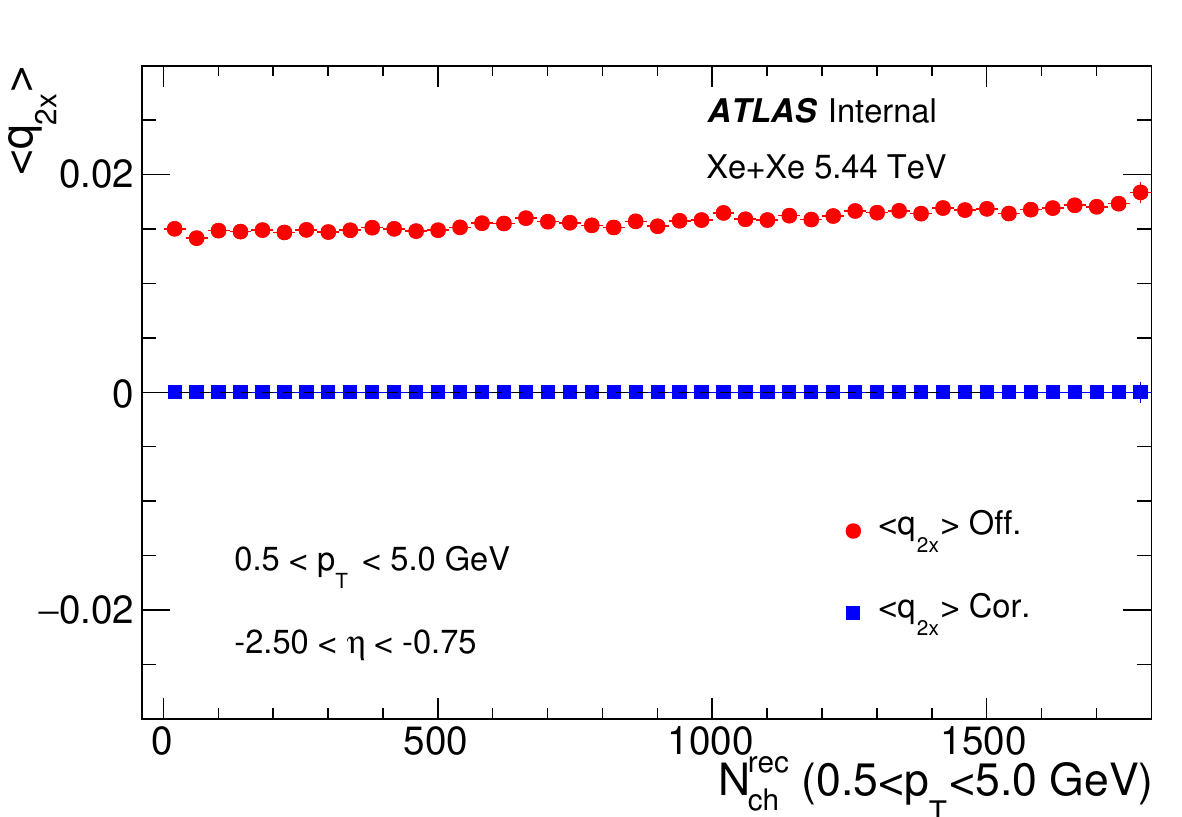}
\includegraphics[width=0.32\linewidth]{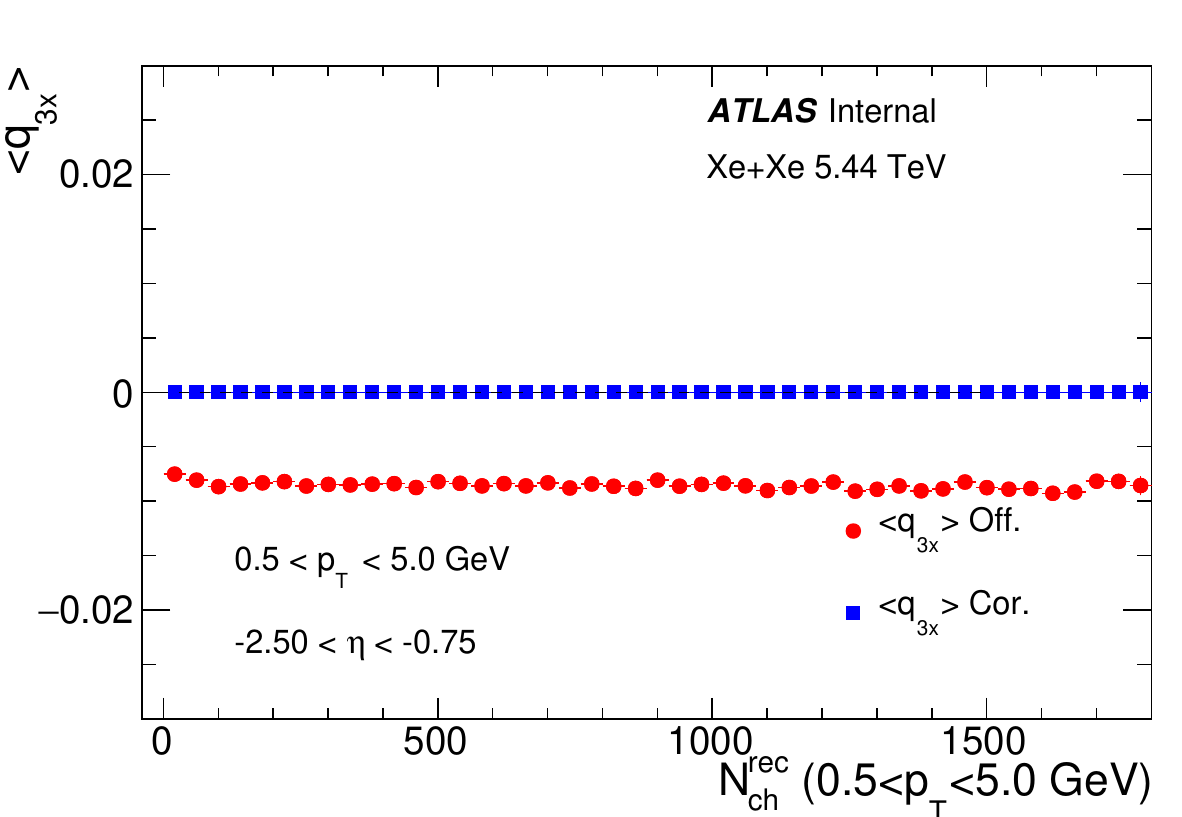}
\includegraphics[width=0.32\linewidth]{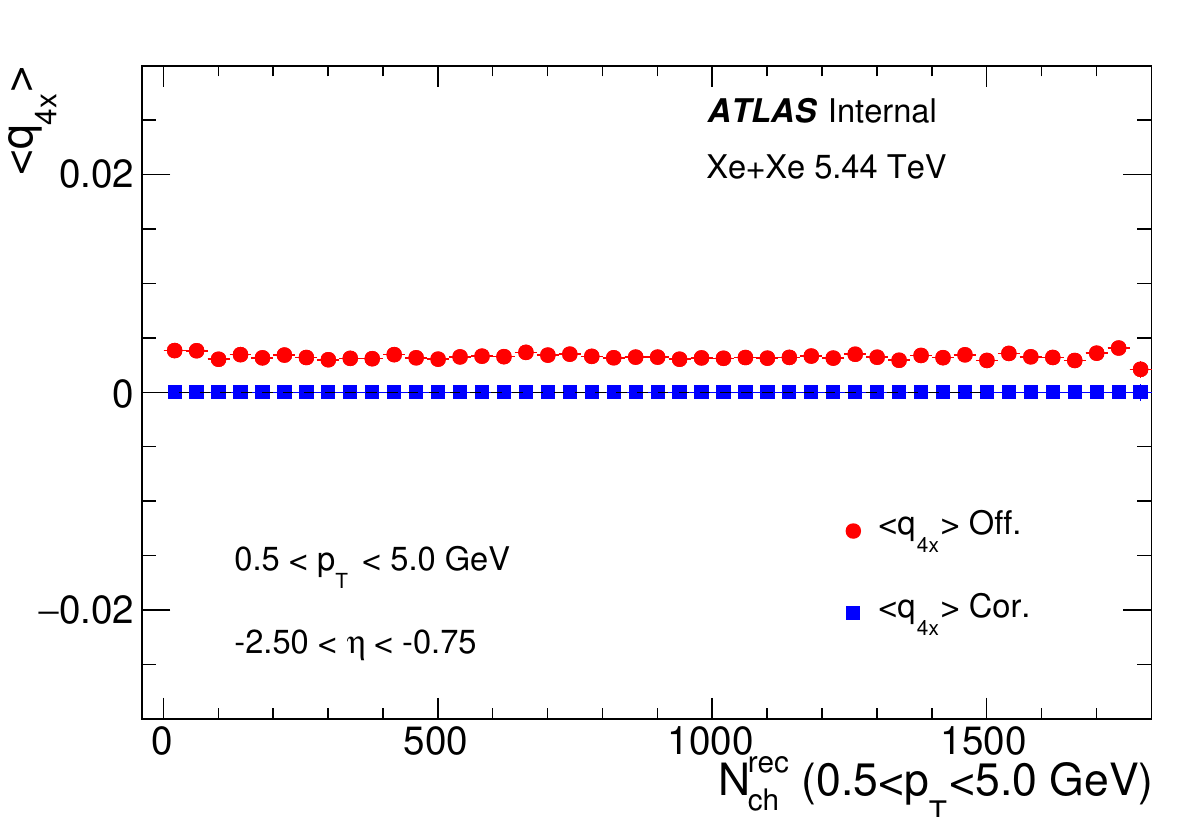}
\includegraphics[width=0.32\linewidth]{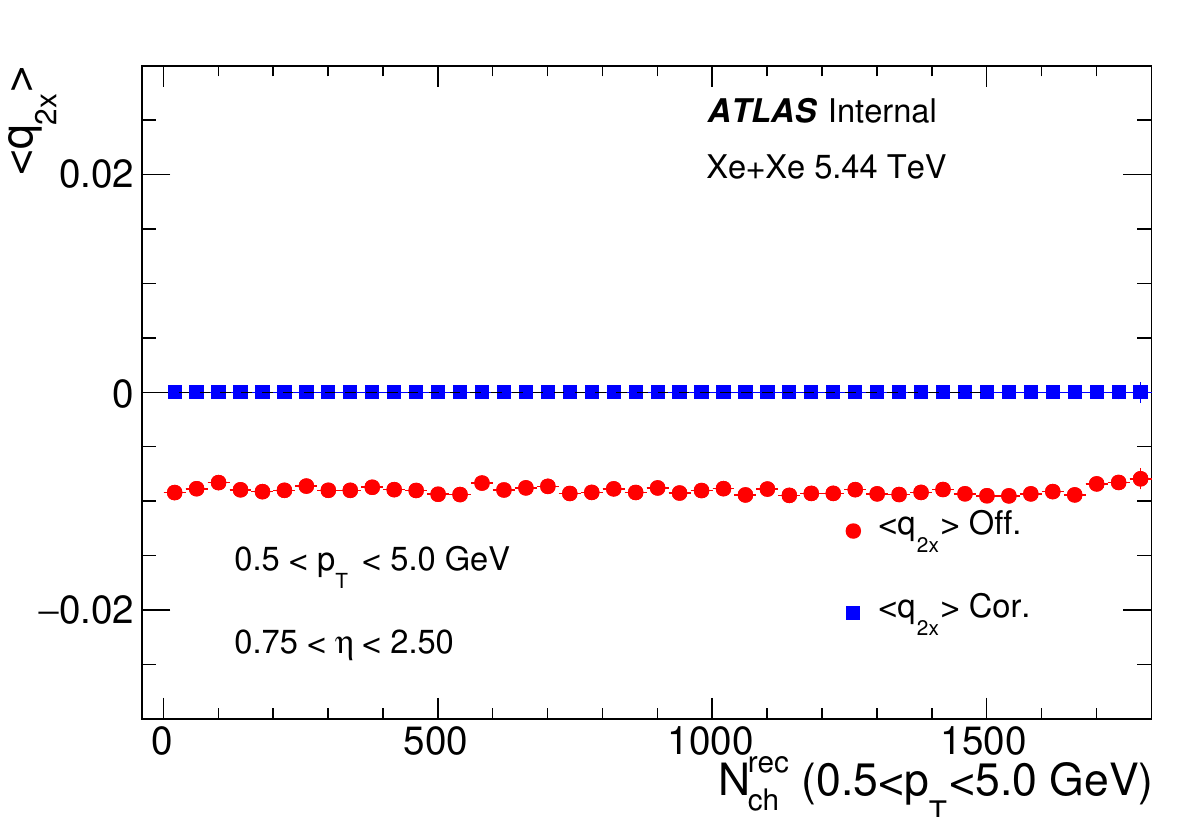}
\includegraphics[width=0.32\linewidth]{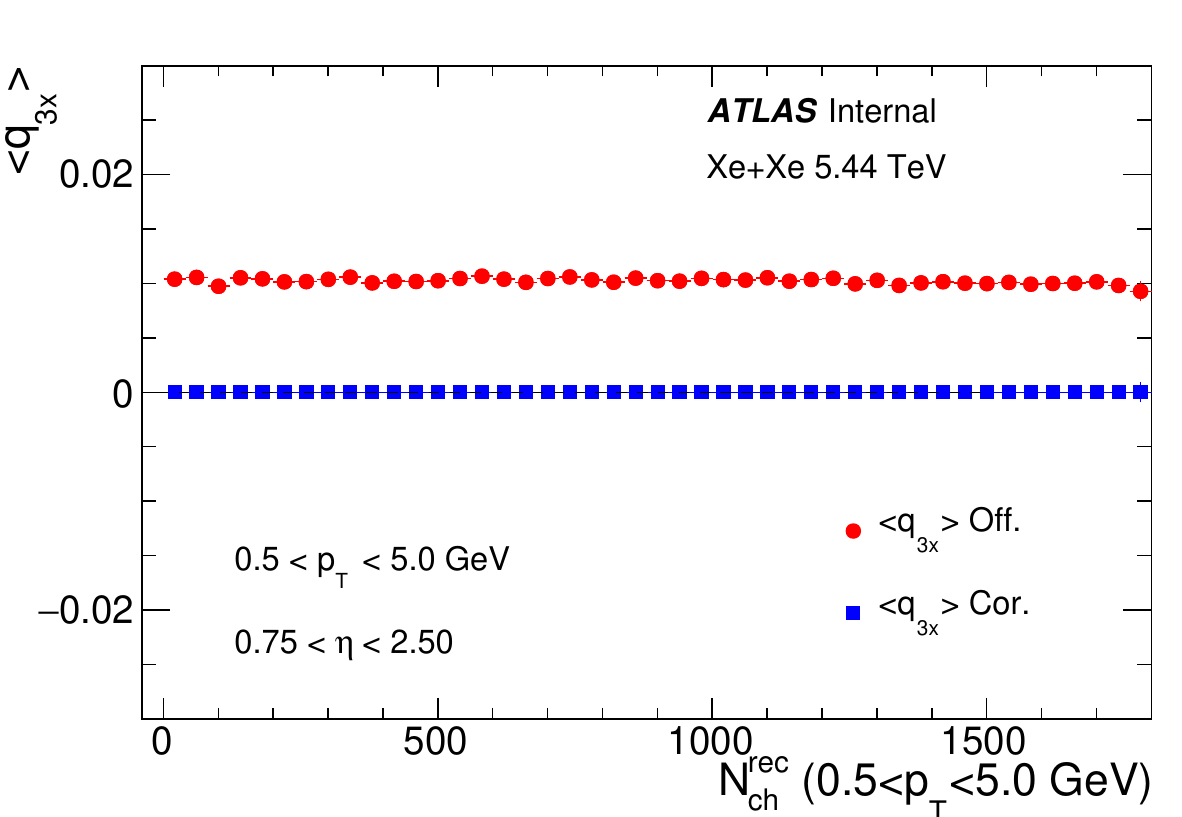}
\includegraphics[width=0.32\linewidth]{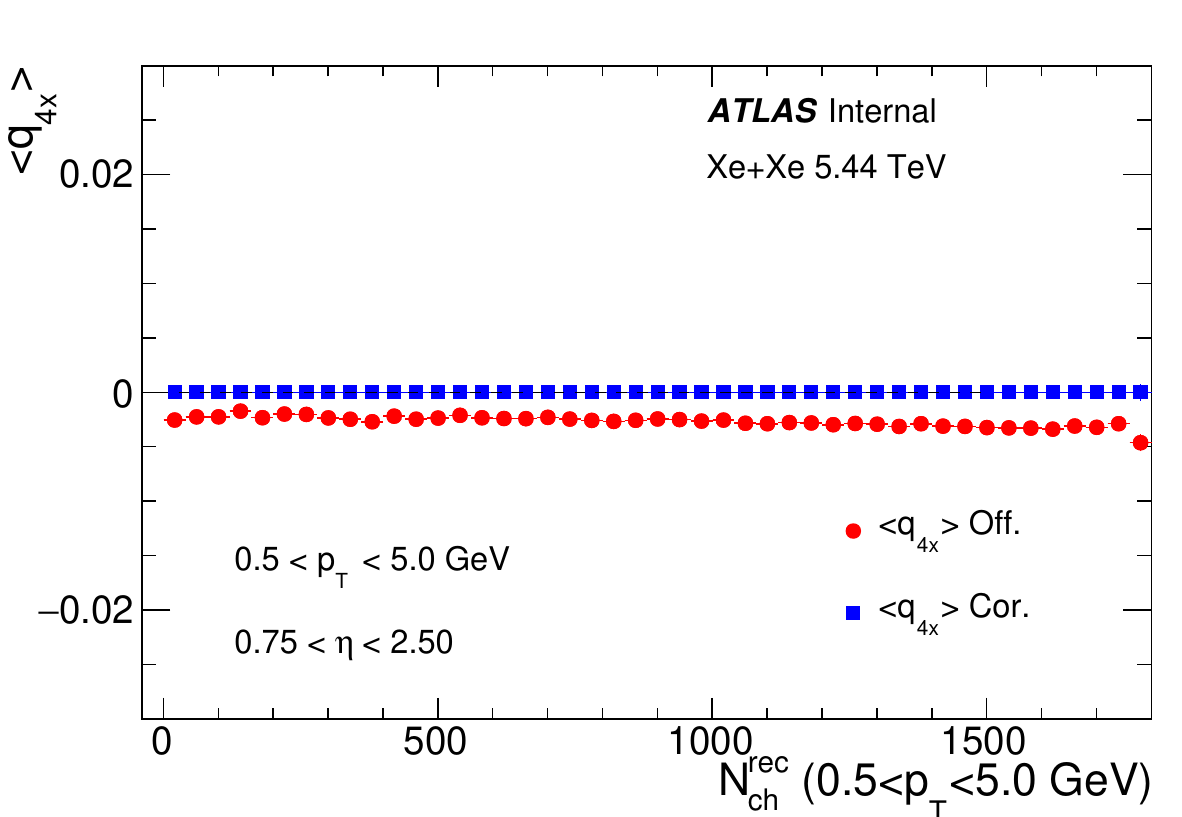}
\caption{Comparison of raw (points) and recentered (lines) average $q_{n,x}$ components in Xe+Xe collisions for $n=2$ (left column), $n=3$ (middle column), and $n=4$ (right column). Results are shown for $|\eta| < 2.5$ (top row), $-2.50<\eta<-0.75$ (middle row), and $0.75<\eta<2.50$ (bottom row). The error bars represent statistical uncertainties.}
\label{fig:RecX_Xe}
\end{figure}
\begin{figure}[htbp]
\centering
\includegraphics[width=0.32\linewidth]{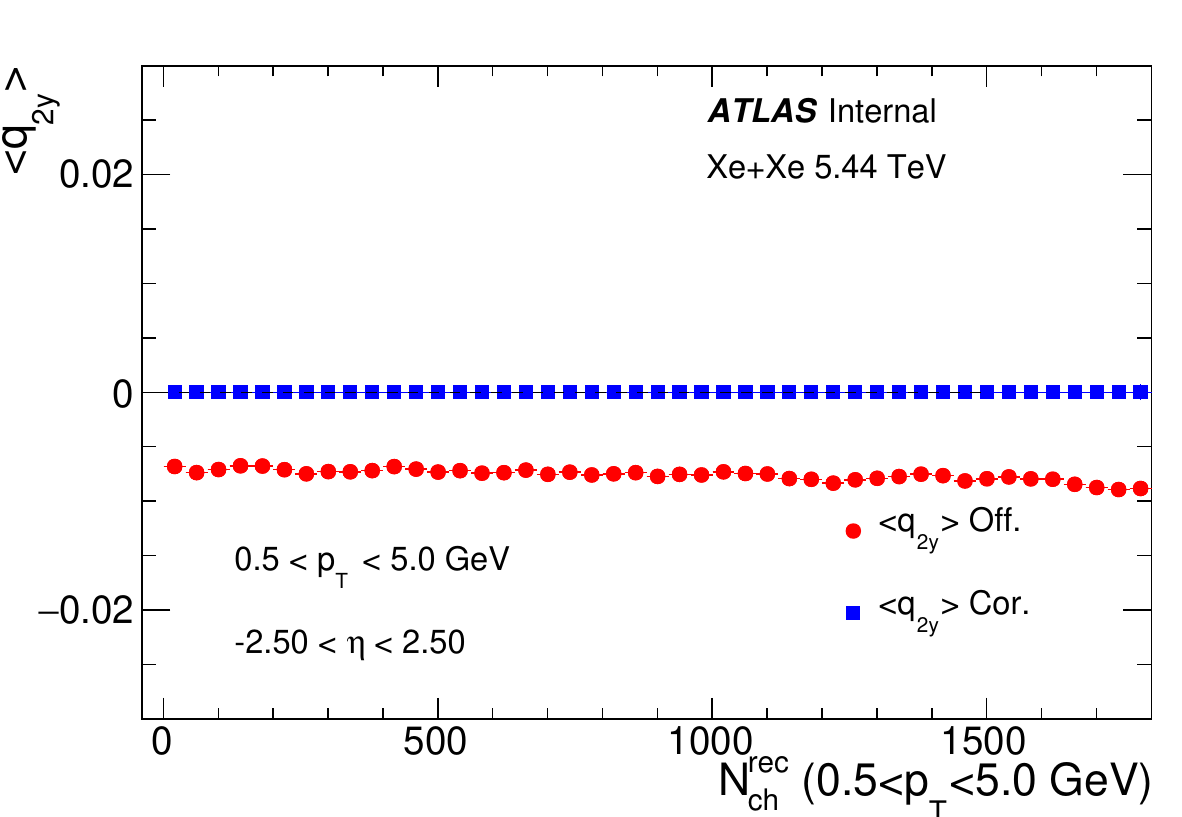}
\includegraphics[width=0.32\linewidth]{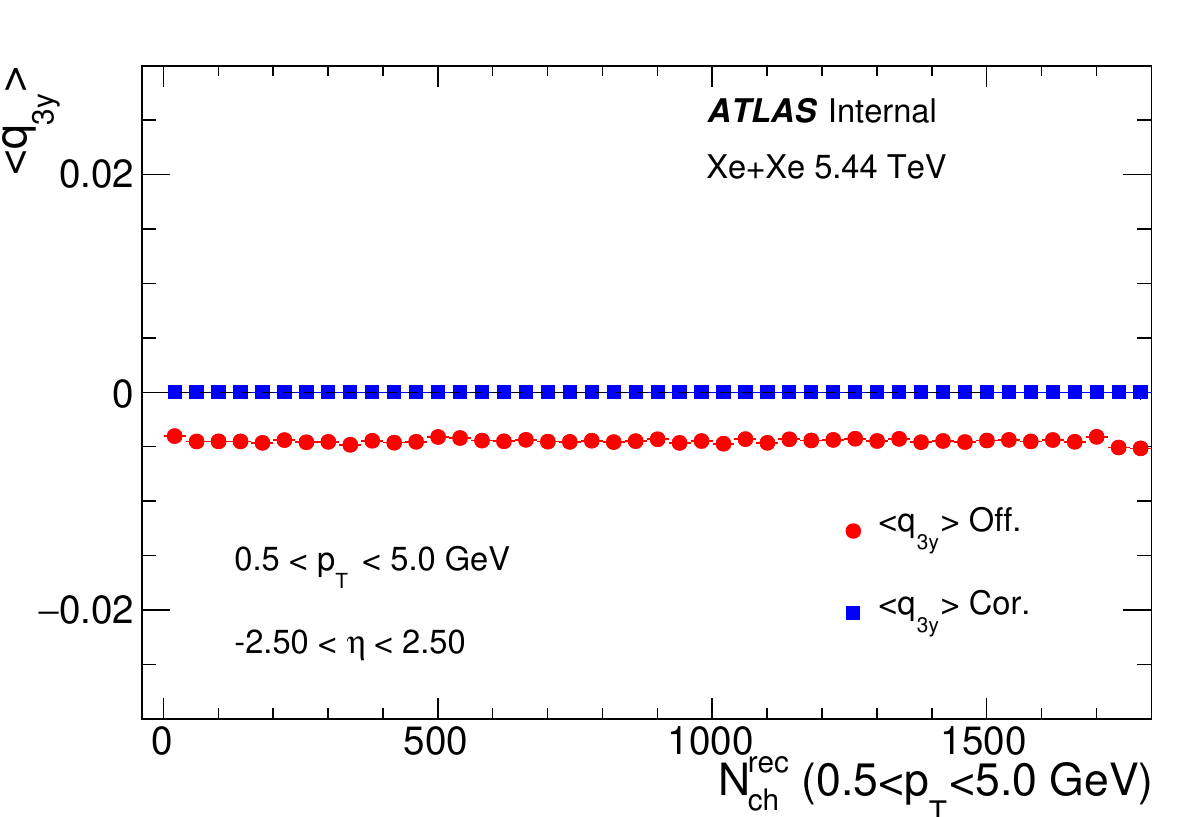}
\includegraphics[width=0.32\linewidth]{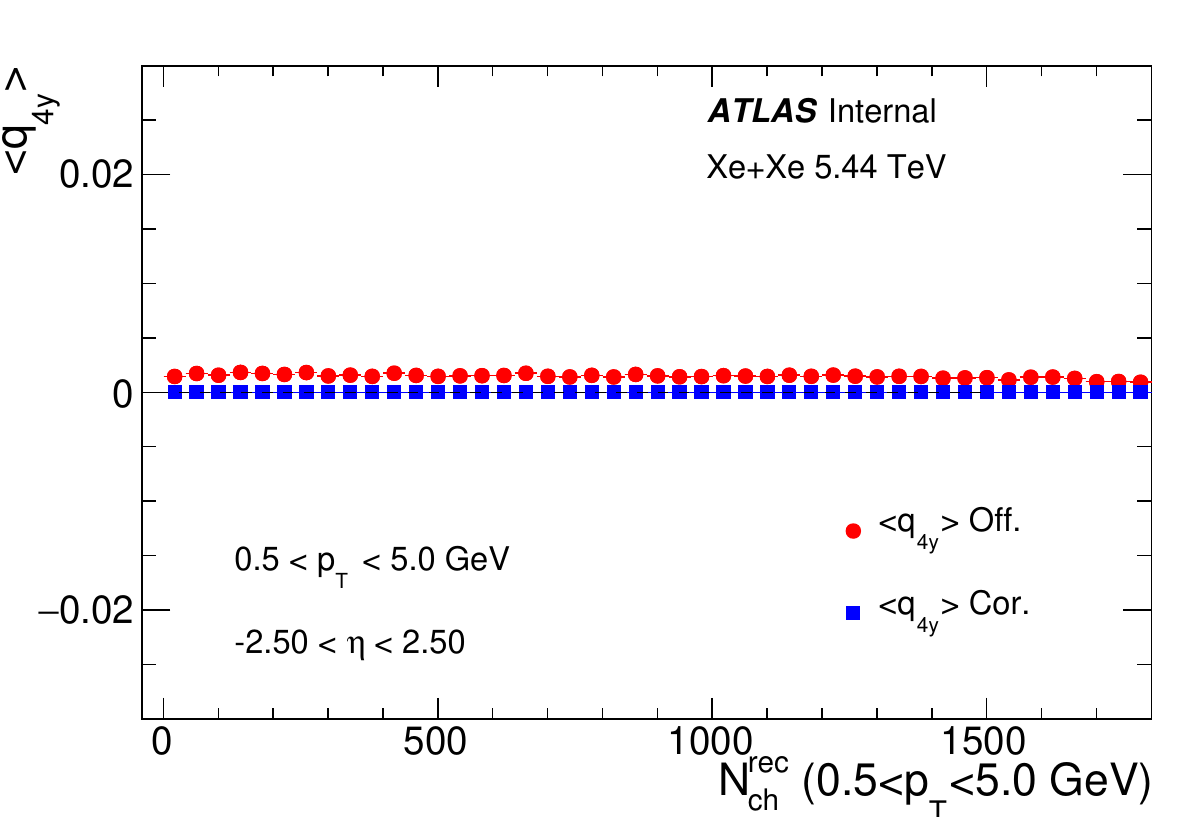}
\includegraphics[width=0.32\linewidth]{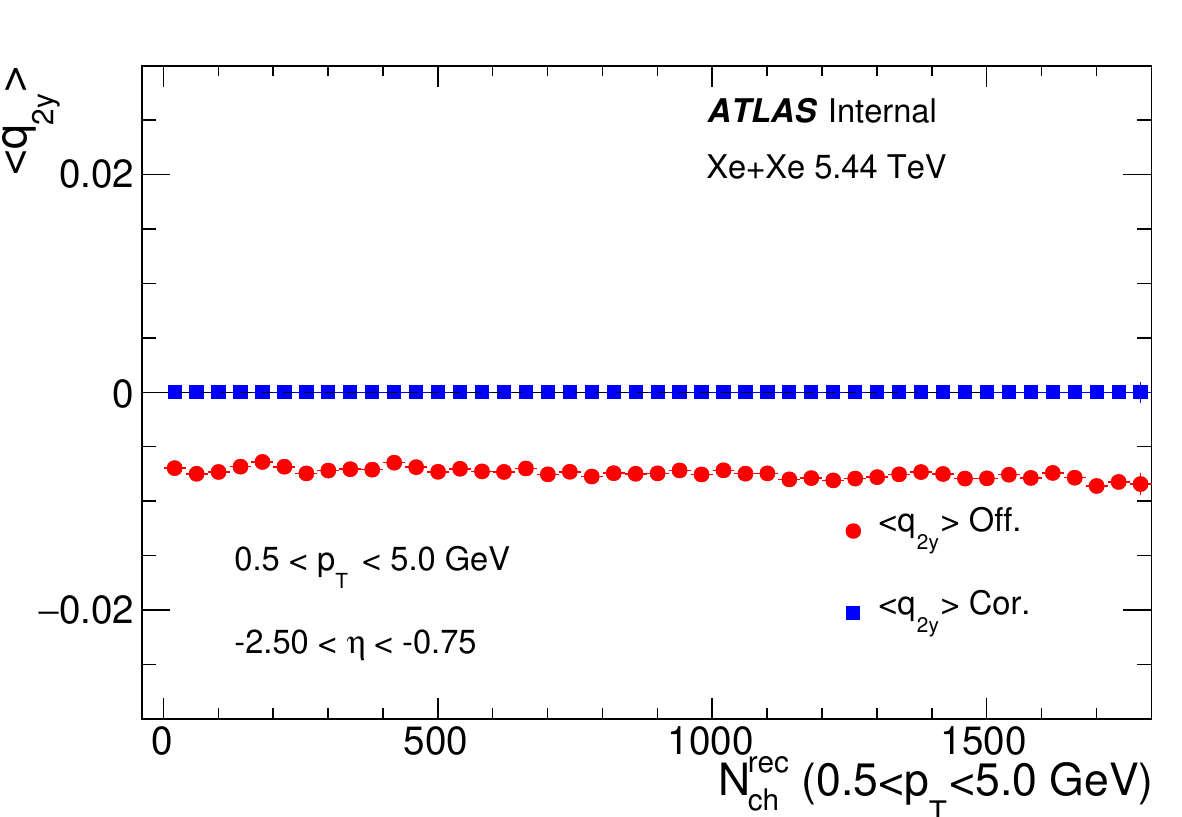}
\includegraphics[width=0.32\linewidth]{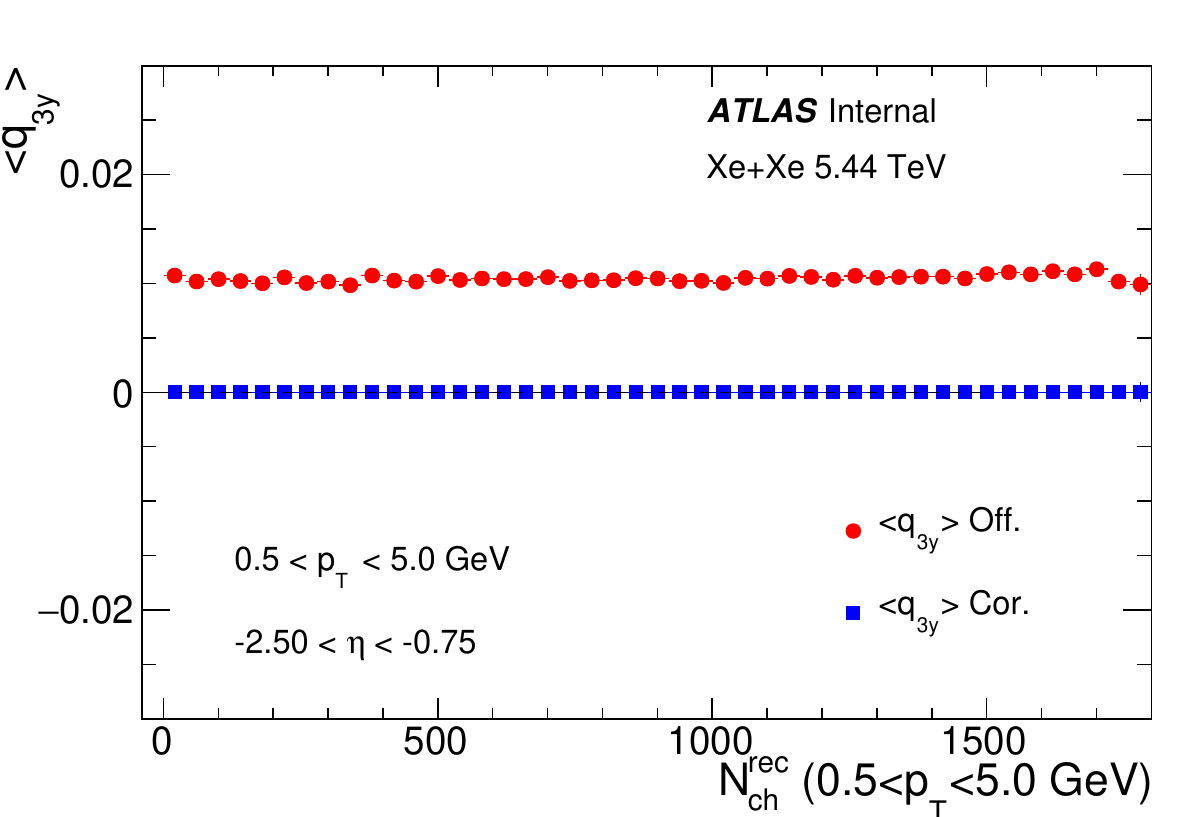}
\includegraphics[width=0.32\linewidth]{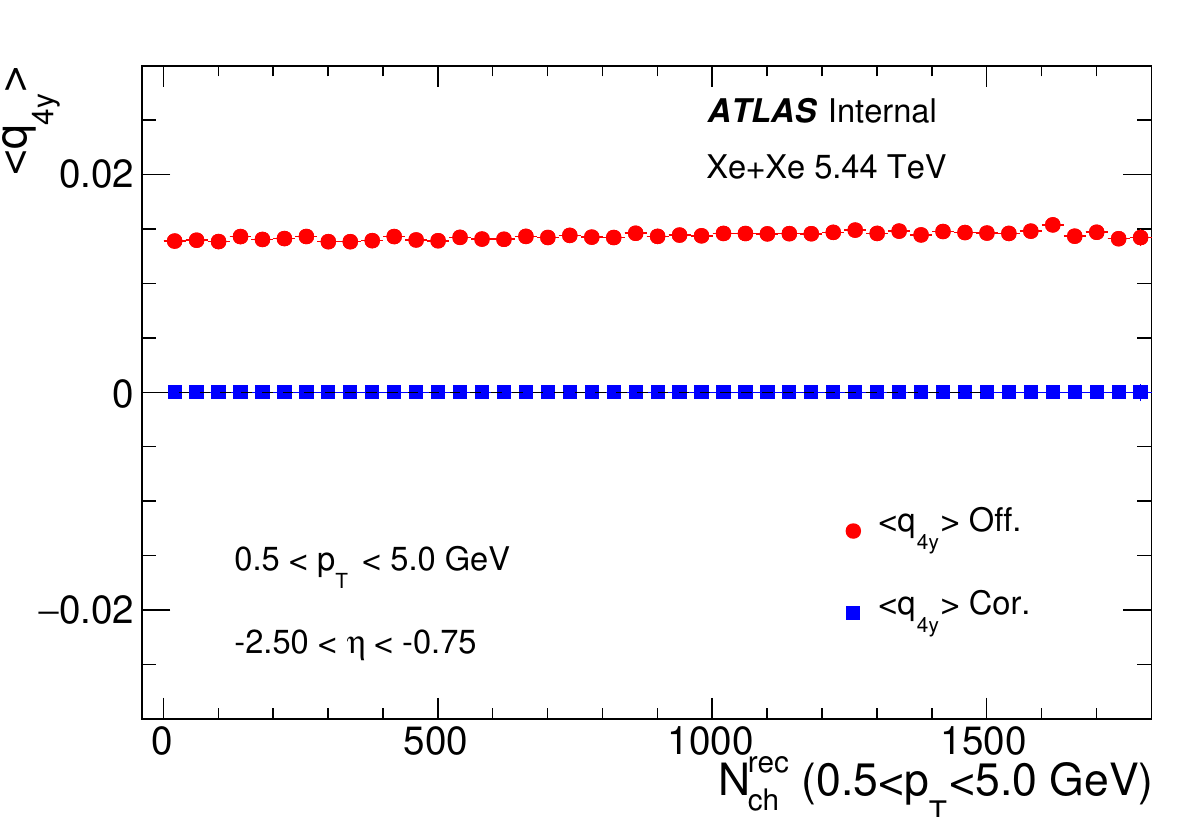}
\includegraphics[width=0.32\linewidth]{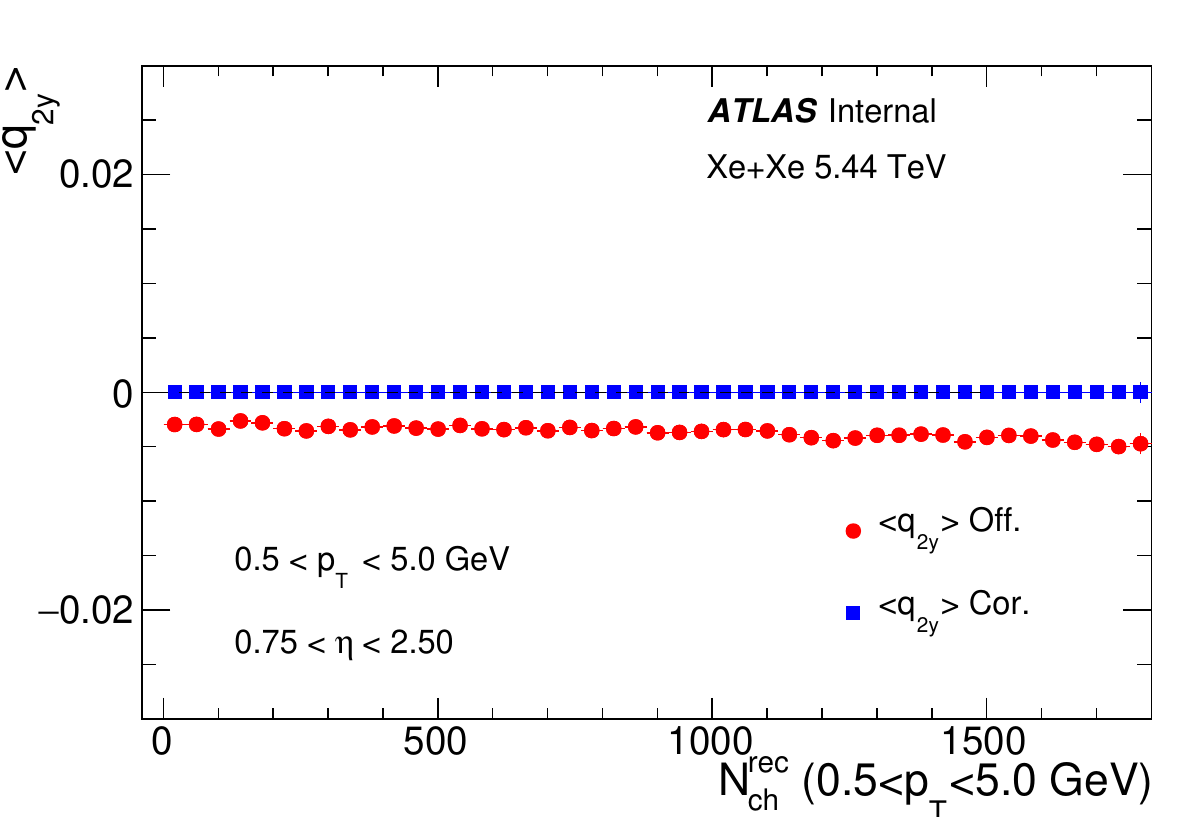}
\includegraphics[width=0.32\linewidth]{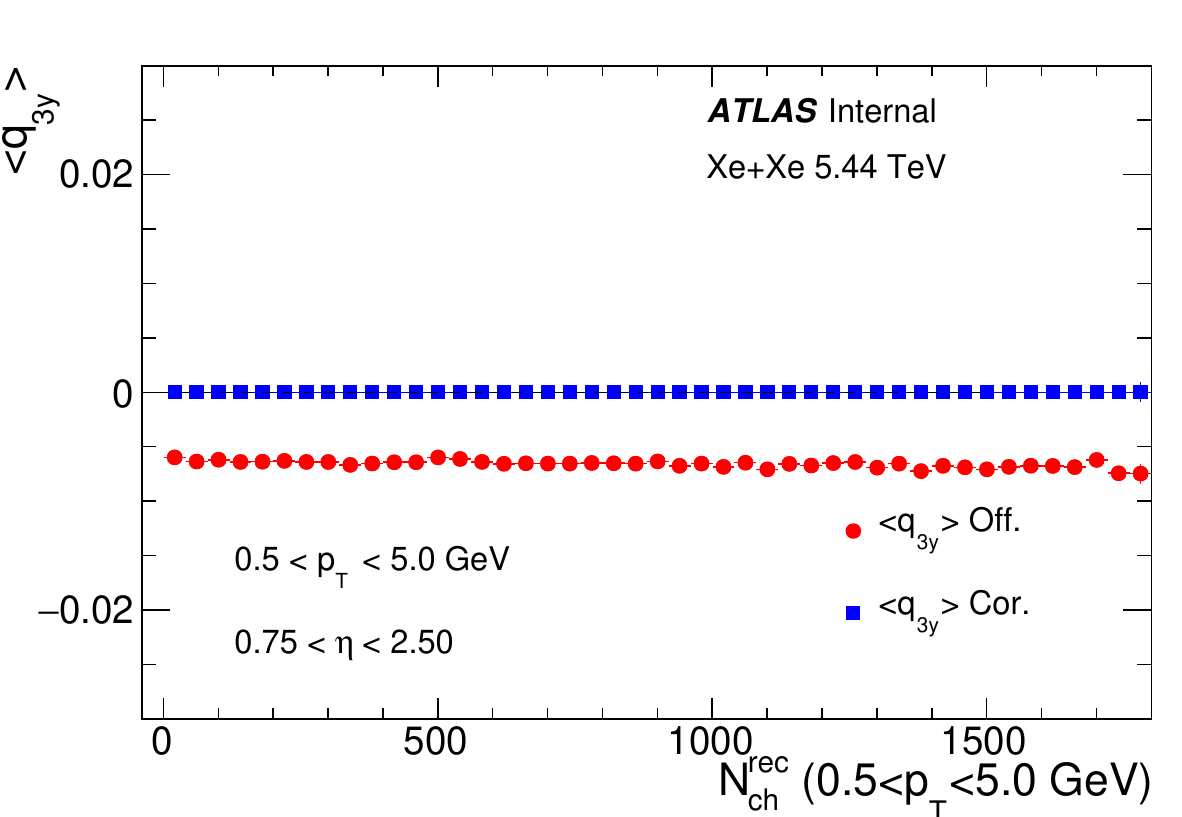}
\includegraphics[width=0.32\linewidth]{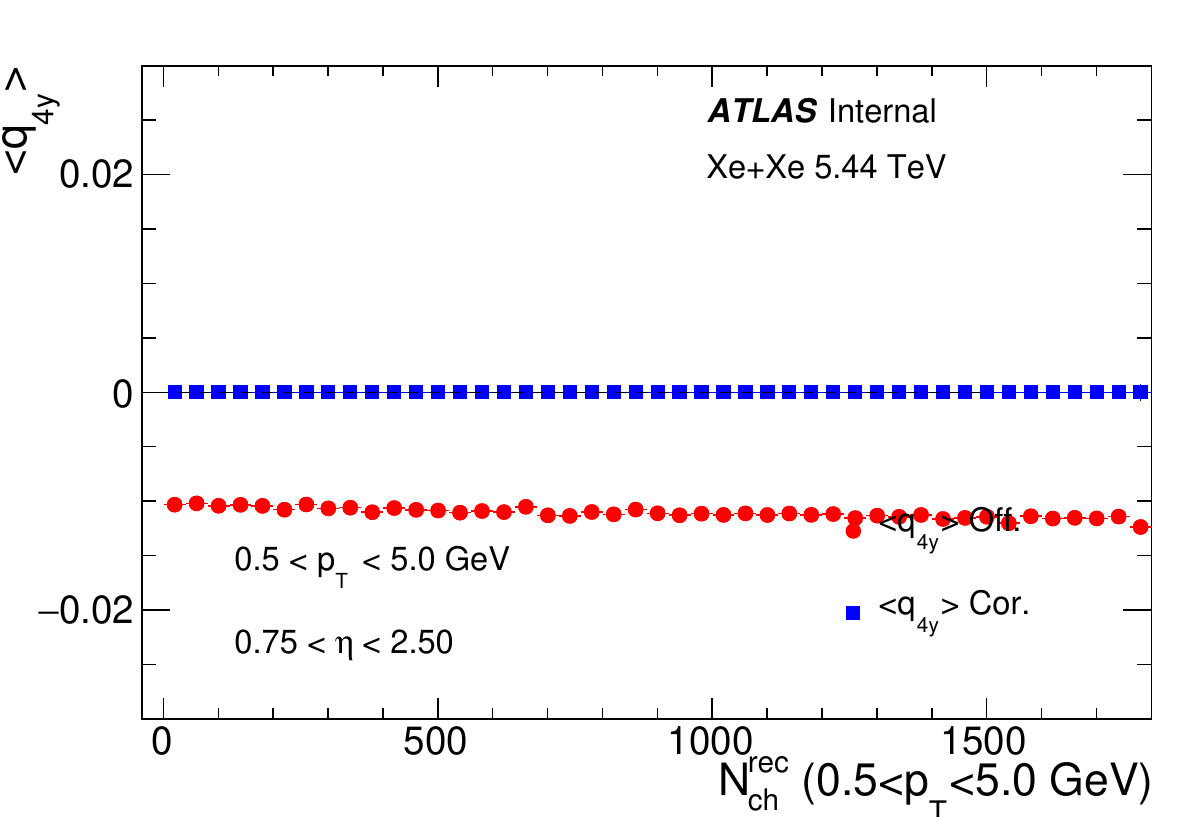}
\caption{Comparison of raw (points) and recentered (lines) average $q_{n,y}$ components in Xe+Xe collisions for $n=2$ (left column), $n=3$ (middle column), and $n=4$ (right column). Results are shown for $|\eta| < 2.5$ (top row), $-2.50<\eta<-0.75$ (middle row), and $0.75<\eta<2.50$ (bottom row). The error bars represent statistical uncertainties.}
\label{fig:RecY_Xe}
\end{figure}

Figures \ref{fig:RecX_Xe} and \ref{fig:RecY_Xe} illustrate the raw and recentered average $q_{n,x}$ and $q_{n,y}$ values, respectively, for $n=2, 3, 4$ in Xe+Xe collisions for different $\eta$ regions. While these offsets are still considerably smaller than the expected magnitude of the actual flow signal (which is approximately 0.05--0.12 for $v_2$ and 0.03--0.04 for $v_3$~\cite{Aad:2019xmh}), this correction is particularly important for $v_4$, which has a smaller $\pT$-integrated signal typically around 0.01--0.02. Earlier analyses have shown that without proper flattening and recentering corrections, the measured $v_4$ values can be significantly biased in analyses~\cite{Mohapatra:2298472}.

\clearpage

\section{Centrality Thresholds}

\begin{table}[htbp]
\centering
\begin{tabular}{cc|cc}
\hline
Centrality & FCal–$\sum E_{T}$ & Centrality & FCal–$\sum E_{T}$ \\
(\%)       & (TeV)              & (\%)       & (TeV)              \\
\hline
1   & 4.2637  & 41 & 0.8442 \\
2   & 4.0834  & 42 & 0.8046 \\
3   & 3.9204  & 43 & 0.7663 \\
4   & 3.7670  & 44 & 0.7293 \\
5   & 3.6226  & 45 & 0.6936 \\
6   & 3.4856  & 46 & 0.6593 \\
7   & 3.3547  & 47 & 0.6260 \\
8   & 3.2297  & 48 & 0.5941 \\
9   & 3.1103  & 49 & 0.5633 \\
10  & 2.9959  & 50 & 0.5336 \\
11  & 2.8857  & 51 & 0.5051 \\
12  & 2.7797  & 52 & 0.4777 \\
13  & 2.6776  & 53 & 0.4515 \\
14  & 2.5796  & 54 & 0.4264 \\
15  & 2.4848  & 55 & 0.4021 \\
16  & 2.3931  & 56 & 0.3790 \\
17  & 2.3045  & 57 & 0.3569 \\
18  & 2.2189  & 58 & 0.3357 \\
19  & 2.1362  & 59 & 0.3155 \\
20  & 2.0558  & 60 & 0.2962 \\
21  & 1.9780  & 61 & 0.2777 \\
22  & 1.9027  & 62 & 0.2602 \\
23  & 1.8297  & 63 & 0.2436 \\
24  & 1.7591  & 64 & 0.2278 \\
25  & 1.6905  & 65 & 0.2127 \\
26  & 1.6243  & 66 & 0.1984 \\
27  & 1.5601  & 67 & 0.1849 \\
28  & 1.4975  & 68 & 0.1721 \\
29  & 1.4373  & 69 & 0.1600 \\
30  & 1.3789  & 70 & 0.1486 \\
31  & 1.3221  & 71 & 0.1379 \\
32  & 1.2671  & 72 & 0.1277 \\
33  & 1.2139  & 73 & 0.1183 \\
34  & 1.1623  & 74 & 0.1093 \\
35  & 1.1122  & 75 & 0.1009 \\
36  & 1.0637  & 76 & 0.0931 \\
37  & 1.0169  & 77 & 0.0857 \\
38  & 0.9715  & 78 & 0.0788 \\
39  & 0.9276  & 79 & 0.0724 \\
40  & 0.8852  & 80 & 0.0664 \\
\hline
\end{tabular}
\caption{Centrality thresholds for 5.02 TeV Pb+Pb collisions determined from the Glauber model fit to the FCal-$\sumET$ distribution.}
\label{table:PbCent}
\end{table}

\begin{table}[h!]
\begin{center}
 \begin{tabular}{||c c || c c||}
 \hline
 Centrality & FCal-$\sum E_{T}$ & Centrality & FCal-$\sum E_{T}$ \\ [0.5ex]
 \hline\hline
 1 & 2.67806 & 41 & 0.54619 \\
 \hline
 2 & 2.5599 & 42 & 0.52125 \\
 \hline
 3 & 2.45935 & 43 & 0.49721 \\
 \hline
 4 & 2.3653 & 44 & 0.47396 \\
 \hline
 5 & 2.27602 & 45 & 0.45155 \\
 \hline
 6 & 2.19142 & 46 & 0.43001 \\
 \hline
 7 & 2.11048 & 47 & 0.40917 \\
 \hline
 8 & 2.03308 & 48 & 0.38904 \\
 \hline
 9 & 1.95875 & 49 & 0.36972 \\
 \hline
 10 & 1.88743 & 50 & 0.35113 \\
 \hline
 11 & 1.8191 & 51 & 0.3331 \\
 \hline
 12 & 1.75344 & 52 & 0.31578 \\
 \hline
 13 & 1.68991 & 53 & 0.29915 \\
 \hline
 14 & 1.62866 & 54 & 0.28314 \\
 \hline
 15 & 1.56988 & 55 & 0.26785 \\
 \hline
 16 & 1.51282 & 56 & 0.25316 \\
 \hline
 17 & 1.45772 & 57 & 0.23913 \\
 \hline
 18 & 1.40429 & 58 & 0.22567\\
 \hline
 19 & 1.35255 & 59 & 0.2127 \\
 \hline
 20 & 1.30248 & 60 & 0.20038 \\
 \hline
 21 & 1.25404 & 61 & 0.18854\\
 \hline
 22 & 1.20703 & 62 & 0.17722 \\
 \hline
 23 & 1.16155 & 63 & 0.16646 \\
 \hline
 24 & 1.11755 & 64 & 0.15627 \\
 \hline
 25 & 1.07498 & 65 & 0.14653 \\
 \hline
 26 & 1.03359 & 66 & 0.13725 \\
 \hline
 27 & 0.99358 & 67 & 0.12843 \\
 \hline
 28 & 0.95471 & 68 & 0.12006 \\
 \hline
 29 & 0.91707 & 69 & 0.11213 \\
 \hline
 30 & 0.88055 & 70 & 0.10461 \\
 \hline
 31 & 0.84512 & 71 & 0.09749 \\
 \hline
 32 & 0.81063 & 72 & 0.09074 \\
 \hline
 33 & 0.77738 & 73 & 0.08435 \\
 \hline
 34 & 0.74505 & 74 & 0.07833 \\
 \hline
 35 & 0.71376 & 75 & 0.07266 \\
 \hline
 36 & 0.68344 & 76 & 0.06734 \\
 \hline
 37 & 0.65421 & 77 & 0.06233 \\
 \hline
 38 & 0.62591 & 78 & 0.0576 \\
 \hline
 39 & 0.59847 & 79 & 0.05317 \\
 \hline
 40 & 0.57184 & 80 & 0.049 \\ [1ex]
 \hline
 \end{tabular}
\end{center}
\caption{FCal-$\sumET$ thresholds defining centrality intervals for Xe+Xe collisions at $\sqrt{s_{\mathrm{NN}}}$ = 5.44 TeV.}
\label{table:XeCent}
\end{table}
\clearpage

\clearpage
\section{Statistical Error Estimation using Poisson Bootstrap Method}\label{sec:app_bootstrap}

Estimating statistical errors for multi-particle cumulants and correlation observables, such as $\lr{\pT}$ cumulants, $v_0(\pT)$, and $\rho(v_n^2, [\pT])$ components, presents challenges for standard error propagation. This is because standard methods can lead to an overestimation of the statistical uncertainty, particularly for higher-order correlations and combinations involving statistically correlated quantities (e.g., $\lr{\mathrm{corr}_n\{4\}}$ and $\lr{\mathrm{corr}_n\{2\}}^2$ in $\mathrm{Var}(v_n^2)_{\mathrm{dyn}}$).

Therefore, robust resampling techniques like bootstrapping are employed for proper error estimation in this dissertation. Specifically, a **Poisson bootstrap** method is used to estimate the statistical errors for all observables. The choice of Poisson bootstrap is motivated by its computational efficiency compared to standard resampling; instead of randomly selecting events, it applies a Poisson weight (drawn from a distribution with a mean of one) to the observable calculated for each event. This method is mathematically justified when the number of events in a given event class is sufficiently large, such that the multinomial probability of event observables can be approximated by a Poisson probability.

Compared to straightforward analytical error calculations, the bootstrap errors are generally smaller, especially for higher-order correlations. For this analysis, Poisson bootstrap with 80 iterations for the Xe+Xe dataset and 60 iterations for the Pb+Pb dataset was performed to ensure reliable statistical error estimation for all observables presented.

\section{Systematic Uncertainties}\label{sec:app_syst}
This section discusses the sources for systematic uncertainties and cross-checks. 
The systematic uncertainties in these analyses arise from several common sources, which are detailed below:

\begin{enumerate}
    \item \textbf{Track Quality Selection (HILOOSE vs HITIGHT):} The selection of tracks based on cuts applied to parameters such as $d_0$, $z_0\sin\theta$, pixel hits, and SCT hits leads to two primary categories: HILOOSE (looser cuts) and HITIGHT (tighter cuts). The choice between these selections affects the tracking efficiency and the balance between real and fake tracks. In some analyses, the average of measurements obtained with HILOOSE and HITIGHT tracks is taken as the default value to symmetrize the systematic variation. The variation due to track quality is often treated as a one-sided uncertainty, and in such cases, half of the difference between the HILOOSE and HITIGHT results is added symmetrically to the total systematic uncertainty.
    \item \textbf{Tracking Efficiency Uncertainty:} The efficiency of track reconstruction depends on the number of charged particles ($\NchR$), transverse momentum ($p_T$), and pseudorapidity ($\eta$). Uncertainties in the tracking efficiency, typically ranging from 1\% to 4\%, stem from discrepancies in the detector material budget between the experimental data and the simulations (GEANT). These uncertainties are often evaluated by varying the efficiency according to a specific functional form, such as
   
    \begin{equation}
         \epsilon_{\pm}(\pT) \equiv \epsilon(\pT) \pm 0.04 \frac{\epsilon(\pT)-\epsilon(\pT^{\mathrm{high}}=5 \mathrm{GeV})}{\epsilon(\pT^{\mathrm{low}}=0.5 \mathrm{GeV})-\epsilon(\pT^{\mathrm{high}}=5 \mathrm{GeV})}.
    \end{equation}
   
    \item \textbf{Centrality Definition Uncertainty (Lower vs Higher FCal-$\sumET$ Cuts):} The determination of centrality intervals is subject to uncertainties related to the min-bias trigger efficiency. This can lead to a $\pm 1\%$ uncertainty in the nominal centrality range. The impact of this uncertainty is assessed by re-evaluating observables with slightly shifted centrality ranges (e.g., for a nominal range of 0-85.5\%, variations to 0-84.5\% and 0-86.5\% are considered). This source of uncertainty is relevant when results are presented as a function of centrality percentiles.
    \item \textbf{$\phi$-Flattening Procedure:} Detector effects can cause non-uniformities in the azimuthal angle ($\phi$) distribution of tracks. A $\phi$-flattening procedure, involving track weights, is applied to correct for these effects. The systematic uncertainty associated with this correction is estimated by repeating the analysis without applying the $\phi$-flattening.
    \item \textbf{Pileup Contamination:} To estimate the systematic effect from pileup contamination, the significance cut applied on the ZDC-$\sumET$ correlation plot is varied from 6-$\sigma$ to 7-$\sigma$. The resulting systematic effect on the observables is found to be less than $0.5\%$ for all observables and all centralities.
\end{enumerate}

The relative systematic uncertainty for an individual source is generally calculated as 
\begin{equation}
    \delta_{\mathrm{sys}} = \frac{{O}_{\mathrm{check}}-O_{\mathrm{default}}}{O_{\mathrm{default}}},
\end{equation}
where $O$ is the observable. For observables that can take values close to zero, the absolute uncertainty,
\begin{equation}
    \delta_{\mathrm{sys,Abs}} = {O}_{\mathrm{check}}-O_{\mathrm{default}}
\end{equation}

\subsection{For $\Nch$} \label{sec:sysNch}
Using efficiency corrected $\NchR$, $\textit{N}_{\mathrm{ch}}$ as event class introduces systematic uncertainties arising from the efficiency and Fake rates used to calculate it. A described in the previous section, the efficiency and Fake rates have their own associated uncertainties, which, lead to an uncertainty on the calculated $\textit{N}_{\mathrm{ch}}$. 

\begin{figure}[htbp]
\centering
\includegraphics[width=0.4\linewidth]{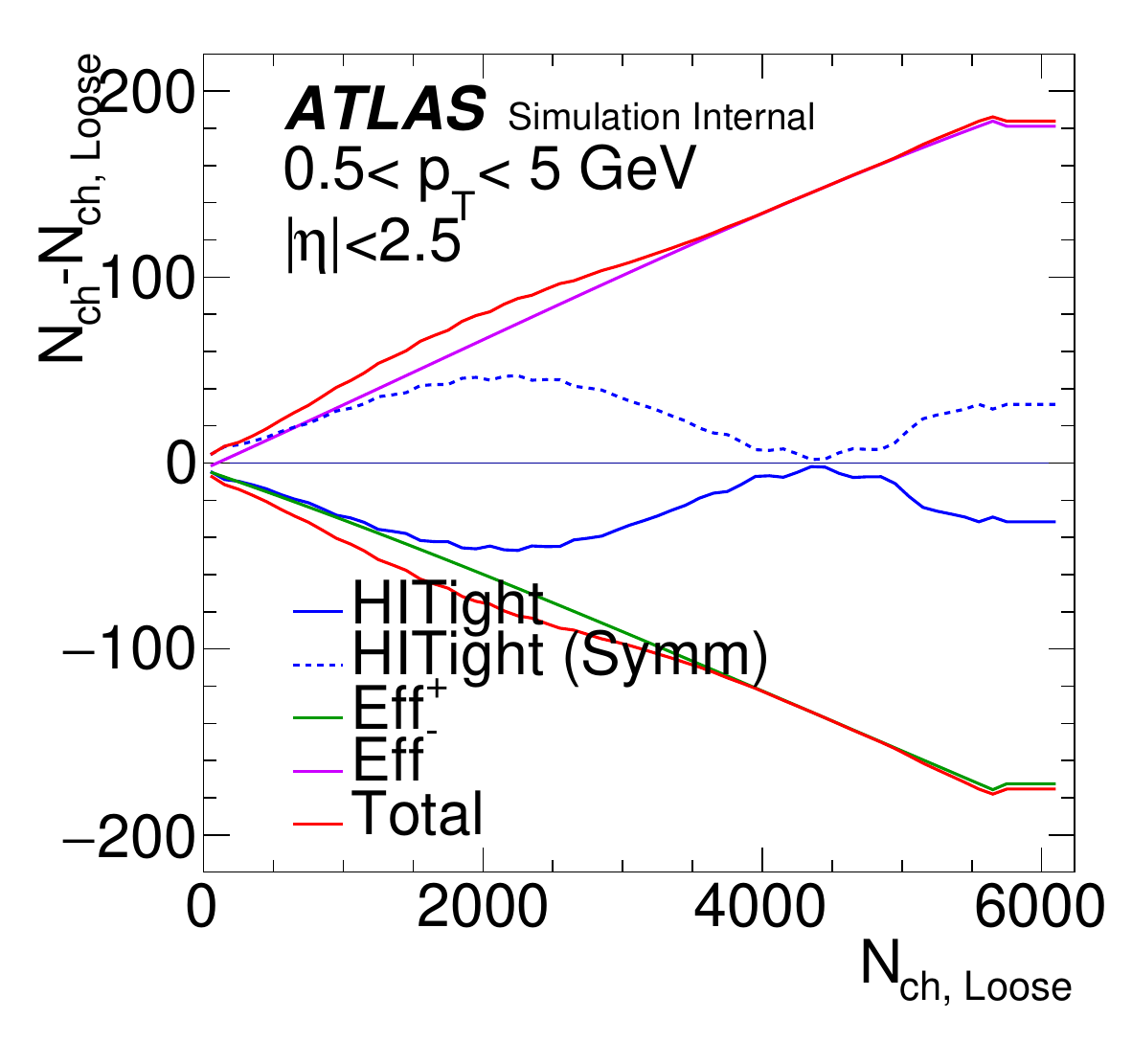}
\includegraphics[width=0.4\linewidth]{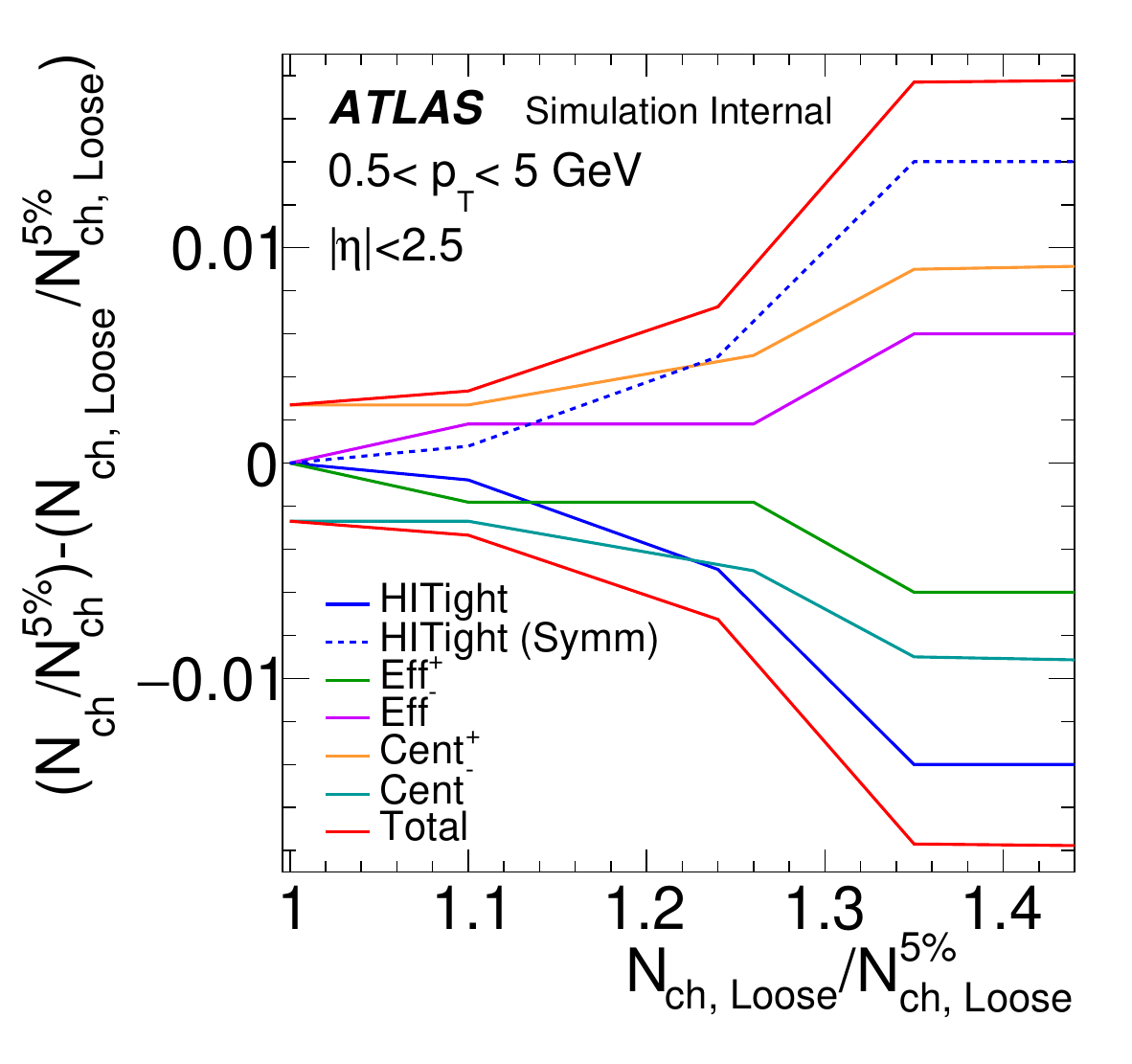}
\caption{(Left) Different Systematic contributions for $\Nch$ as a function of the default choice of $N_{\mathrm{ch, Loose}}$, (Right) Different Systematic contributions for $\Nch/N^{5\%}_{ch}$ as a function of the default choice of $N_{\mathrm{ch, Loose}}/N^{5\%}_{\mathrm{ch, Loose}}$,  in Pb+Pb collisions at $\sqn=$ 5.02 TeV. The red line shows the total systematics obtained by taking a square-root of the quadrature-sum of different systematic contributions.}
\label{fig:sysnch}
\end{figure}

In this analysis, $\textit{N}_{\mathrm{ch, Loose}}$ is used as the default choice. The difference between $\textit{N}_{\mathrm{ch}}$ calculated for different systematic variations and $\textit{N}_{\mathrm{ch, Loose}}$  are shown in left panel of fig.~\ref{fig:sysnch} as a function of $\textit{N}_{\mathrm{ch, Loose}}$. The systematic contribution from track selection criteria (HITight) is symmetrized to account for systematics from a more relaxed tack selection criteria than HILoose. The maximum contribution to the total systematics on $\textit{N}_{\mathrm{ch, Loose}}$ arises from the $\pm 4\%$ efficiency variation. The magnitude of total systematics for a given value of $\textit{N}_{\mathrm{ch, Loose}}$ is obtained from taking a square-root of the quadrature-sum of different contributions for a given value of $\textit{N}_{\mathrm{ch, Loose}}$. In most central collisions, the total systematic uncertainty on the $\textit{N}_{\mathrm{ch, Loose}}$ is around 3\%.

Similarly, this analysis also uses $\Nch/N^{5\%}_{ch}$ as event class to plot certain results, which has its own associated systematic uncertainty. The different systematic contributions for $\Nch/N^{5\%}_{ch}$ are shown in the right panel of Fig.~\ref{fig:sysnch}. The systematic variations are smoothed and conservative estimates are shown for each contribution. After normalization by $N^{5\%}_{ch}$, the largest systematic contribution arises from track quality variation from HILoose to HITight. For largest value of $N_{\mathrm{ch, Loose}}/N^{5\%}_{\mathrm{ch, Loose}}$, the total systematic uncertainty is around 1.5\%.

\clearpage

\subsection{For $[\pT]$ Cumulants}\label{sec:sysptfluc}

The systematic uncertainties for the cumulants $k_n$ ($n=2, 3$) and their normalized counterparts ($nk_2$, $\gamma$, $\Gamma$) are evaluated for particles within $|\eta|<2.5$ and $0.5<p_T<5.0$ GeV for Pb+Pb collisions and $0.3<p_T<5.0$ GeV for Xe+Xe collisions. Similar to the $v_0(p_T)$ analysis, the track quality selection, tracking efficiency, and centrality definition contribute to the systematic uncertainties. The contribution from the track quality variation (HILOOSE vs HITIGHT) is typically symmetrized by dividing the difference by two and adding it symmetrically to the total systematic uncertainty. For higher-order measurements, statistical fluctuations in the systematic contributions are smoothed out by averaging neighboring points. The HIJING closure check, performed for Pb+Pb collisions, provides an estimate of the non-closure, which is then added in quadrature to the total systematic uncertainty for both Pb+Pb and Xe+Xe systems. The systematic effect from pileup contamination is estimated to be less than $0.5\%$ for these observables across all centralities.

There are several observables in the analysis, each with several methods and $\pT$ and $\eta$ selections. Instead of showing the systematic checks for all of them, we focus on the three most important physics quantities: $k_{n}$ for $n=2$, 3 and their normalized counterparts, Normalized variance ($nk_{2}$), Normalized skewness ($\gamma$) and intensive skewness ($\Gamma$) using the standard method for particles within $|\eta|<$2.5 and 0.5$<\pT$<5.0 GeV for Pb+Pb collisions and 0.3$<\pT$<5.0 GeV for Xe+Xe collisions. In the following, we summarize them separately. The systematics for Pb+Pb collisions are shown in Section~\ref{sec:sys:Pb} and for Xe+Xe collisions are shown in Section~\ref{sec:sys:Xe}. 

Please note that all sources of systematics have an upward and downward variation about the default measurement whereas, the track quality variation is one-sided (i.e varied between HILOOSE and HITIGHT). Therefore the contribution from track quality variation is divided by 2 and is added symmetrically between the upper and lower limits of the total systematic variations in both systems (denoted as `TightSym' on the bottom panel of the systematics plots). In addition, there are statistical jumps going from one bin to the next for the systematic contributions for higher-order measurements. In this case, the points are averaged taking the neighborhood points into account to smooth out the jumps.

\subsubsection{In Pb+Pb Collisions}\label{sec:sys:Pb}
\begin{figure}[htbp]
\centering
\includegraphics[width=0.325\linewidth]{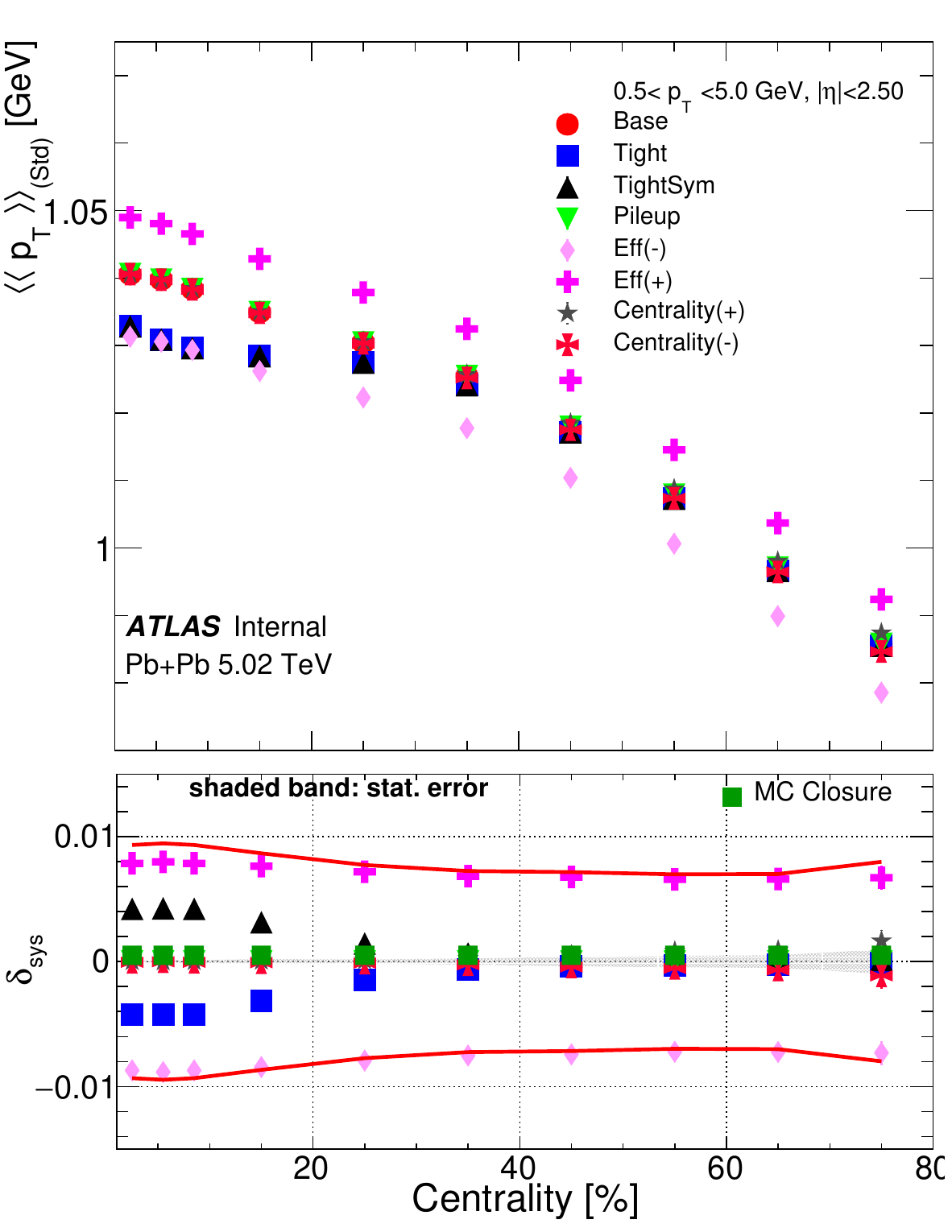}
\includegraphics[width=0.325\linewidth]{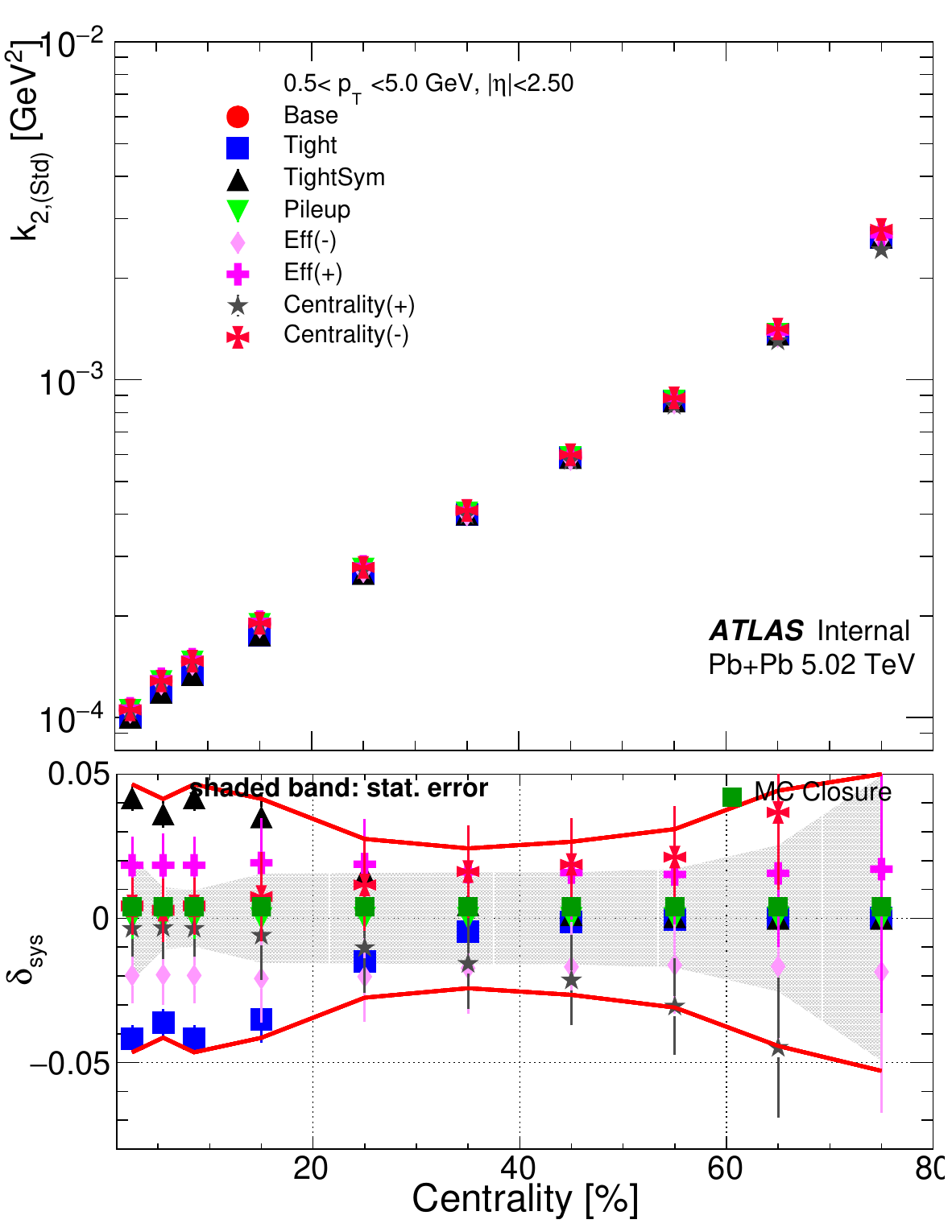}
\includegraphics[width=0.325\linewidth]{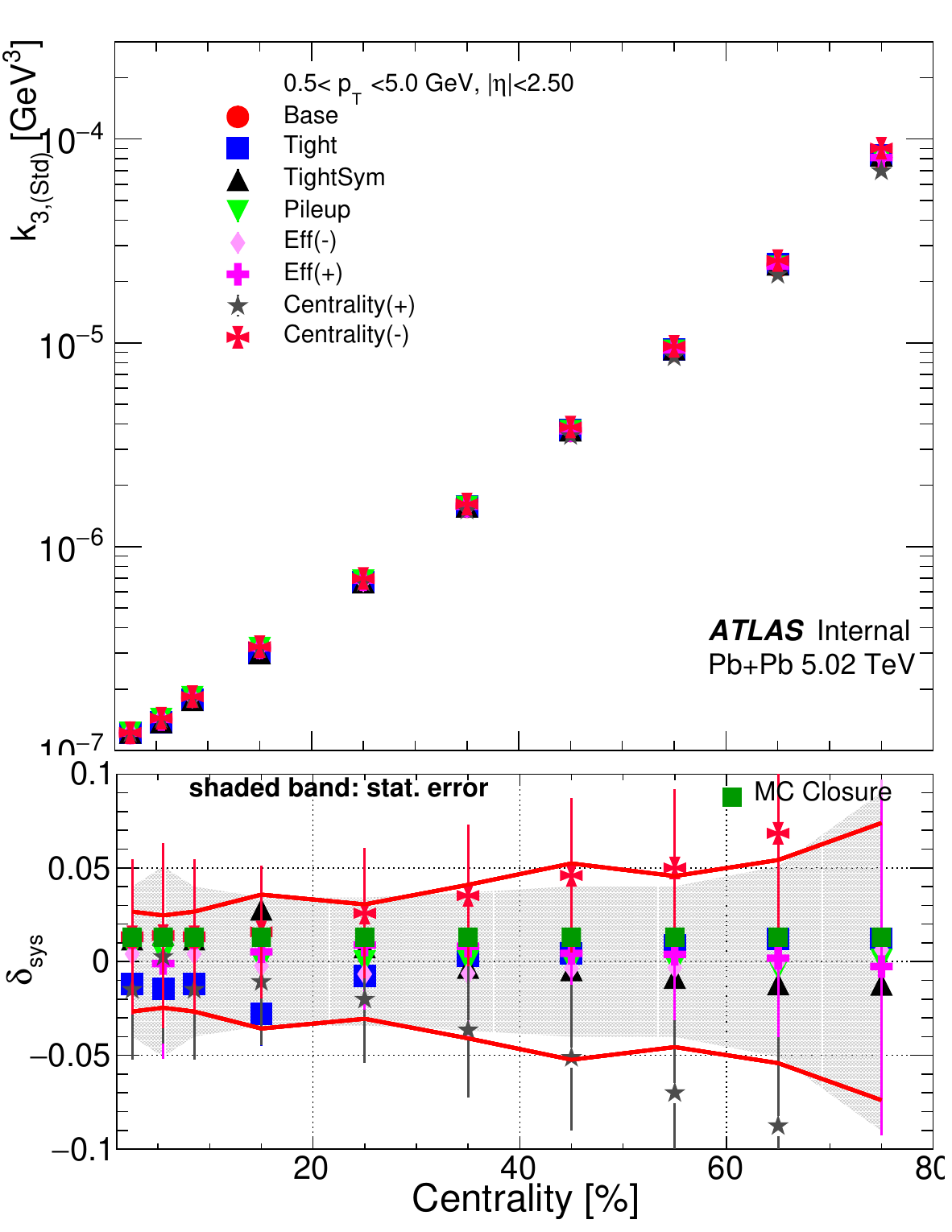}
\caption{Comparison between different systematic variations and default measurement for $\MpT$ (Left panel), $k_{2}$ (Central panel) and $k_{3}$ (Right panel) for 0.5-5 GeV particles in Xe+Xe collisions. The bottom panels show relative systematic uncertainties arising from systematic variations w.r.t default for respective observables on top panels. The error bars represent statistical uncertainties. The shaded area represents the ratio of the statistical uncertainty. Solid red-colored lines on the bottom panels reflect combined systematic uncertainty for the observables.}
\label{fig:PbSys_Fluc}
\end{figure}

Fig.~\ref{fig:PbSys_Fluc} shows the comparison of different systematic checks with the default case for $\MpT$, $k_{2}$ and $k_{3}$ in Pb+Pb collisions measured with the standard method. The bottom panels show relative difference w.r.t default along with the lower and upper bounds shown in lines and the ratio of statistical uncertainty of the default to its mean value shown as a shaded region. The systematic variation is within $10\%$ and is dominated by track selection and centrality definition. The systematic uncertainties are larger than statistical uncertainties for $\MpT$, $k_{2}$ and $k_{3}$.

\begin{figure}[htbp]
\centering

\includegraphics[width=0.325\linewidth]{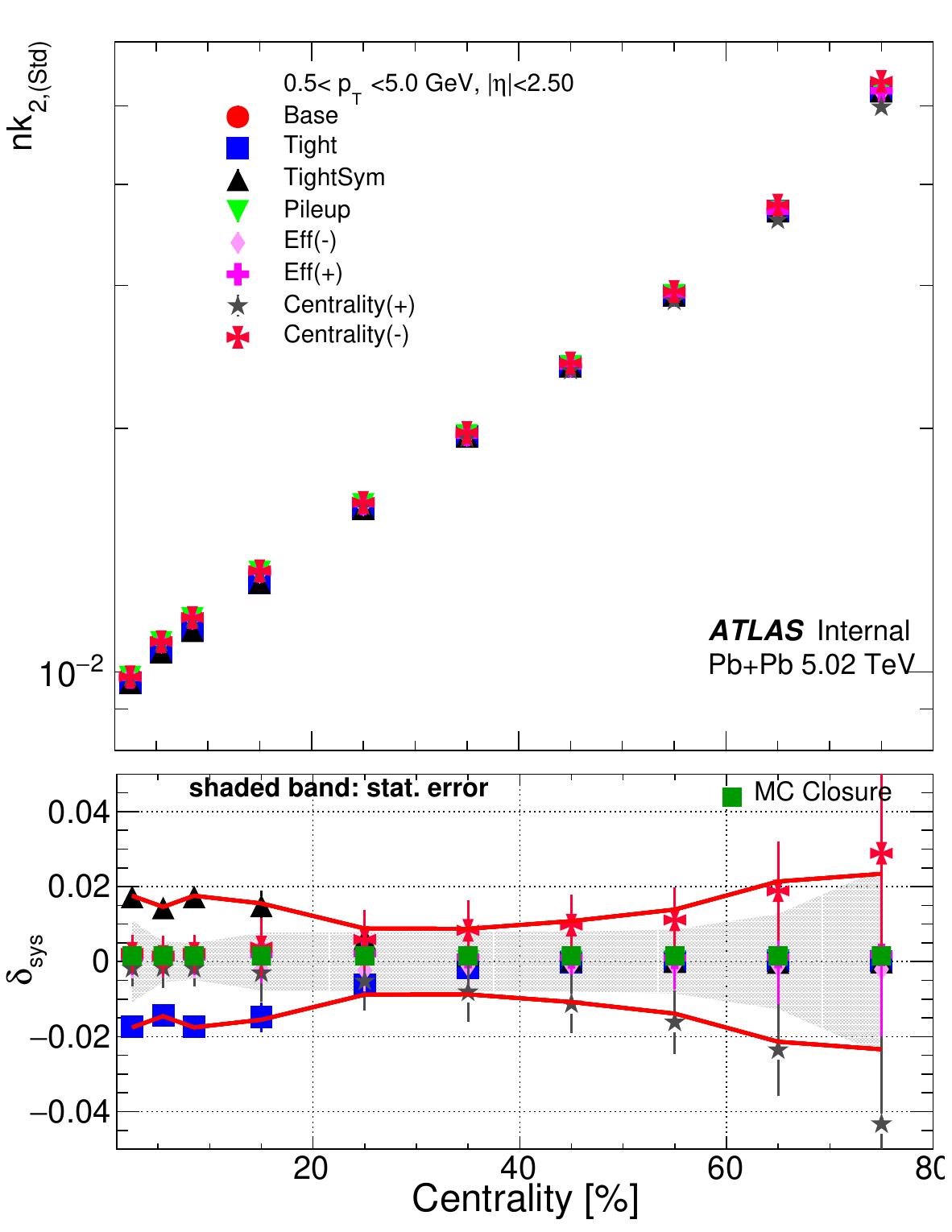}
\includegraphics[width=0.325\linewidth]{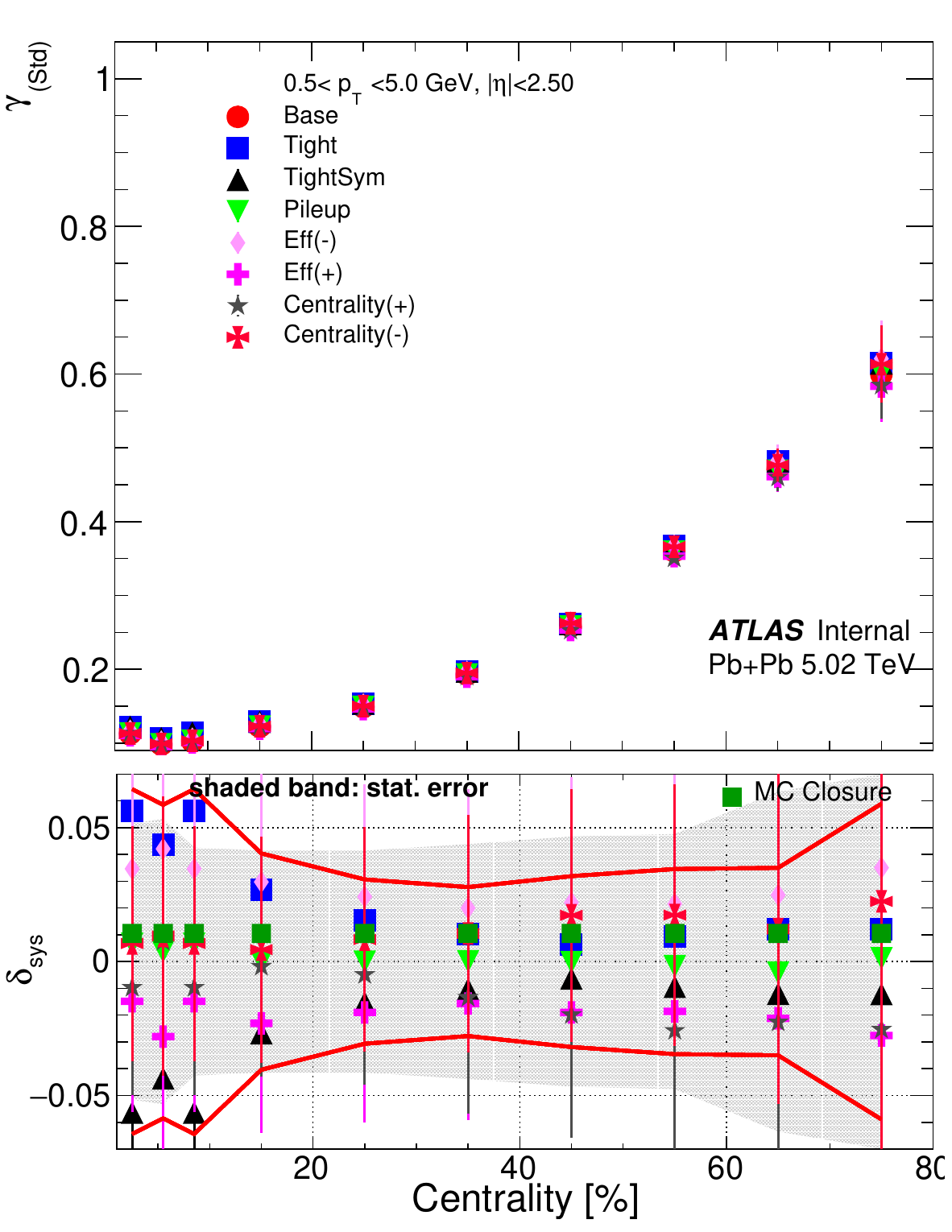}
\includegraphics[width=0.325\linewidth]{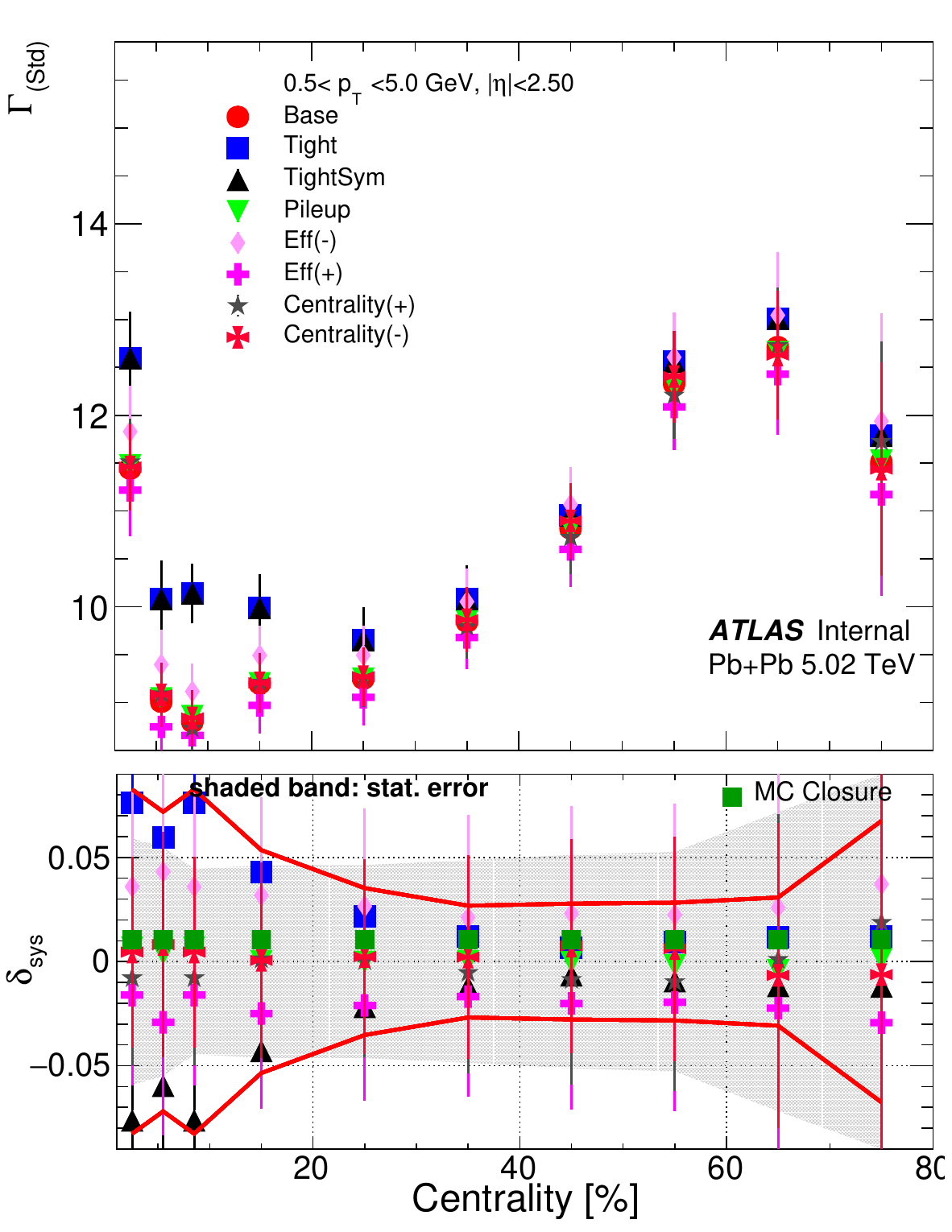}

\caption{Comparison between different systematic variations and default measurement for $nk_{2}$ (Left panel), $\gamma$ (Central panel) and $\Gamma$ (Right panel) for 0.5-5 GeV particles in Pb+Pb collisions. The bottom panels show relative systematic uncertainties arising from systematic variations w.r.t default for respective observables on top panels. The error bars represent statistical uncertainties. The shaded area represents the ratio of the statistical uncertainty. Solid red-colored lines on the bottom panels reflect combined systematic uncertainty for the observables.}
\label{fig:PbSys_NormFluc}
\end{figure}

Fig.~\ref{fig:PbSys_NormFluc} shows the comparison of different systematic checks with the default case for normalized $\lr{\pT}$-fluctuations in Pb+Pb (From left to right: $nk_{2}$, $\gamma$ and $\Gamma$). The bottom panels show relative difference w.r.t default along with the lower and upper bounds shown in lines and the ratio of statistical uncertainty of the default to its mean value shown as a shaded region. The systematic uncertainty partially cancels between the numerator and denominator. Therefore the final systematic uncertainty has weaker centrality dependence. Normalization also cancels systematics for the observables. The systematic uncertainty for normalized observables in Pb+Pb collisions is within 4\%.

To calculate the total systematics, the points that are above 0 in these figures are added in quadrature to find the upper bound. Similarly, all the relative error points that are below 0 are added in quadrature to find the lower bound.  

 For $n=2$, the dominant source of systematics arises from tracking quality variation and efficiency correction for all observables. For $n=3$, the systematic uncertainty arising from efficiency variations becomes comparable to those arising from track quality variation. Uncertainty on the centrality definition has the largest impact in the most peripheral region. Only in the peripheral region, for some observables, the systematics becomes significant, but this is mainly due to large statistical uncertainties in the data. For all observables, in general, the total systematic error is smaller than or comparable to the statistical error.

We carry out a closure check using the HIJING model as a useful validation of the analysis method. 

In the HIJING model, the $\lr{\pT}$ fluctuations arise purely from an independent superposition scenario. Therefore, the model predictions for $\lr{\pT}$ fluctuations are not expected to provide good estimates of these quantities. However, a study using HIJING events provides an estimate of the effect of short-range correlations, like jets or resonance decays, on the measured quantities~\cite{Aad:2019fgl}. 

\begin{figure}[htbp]
\centering
\includegraphics[width=0.325\linewidth]{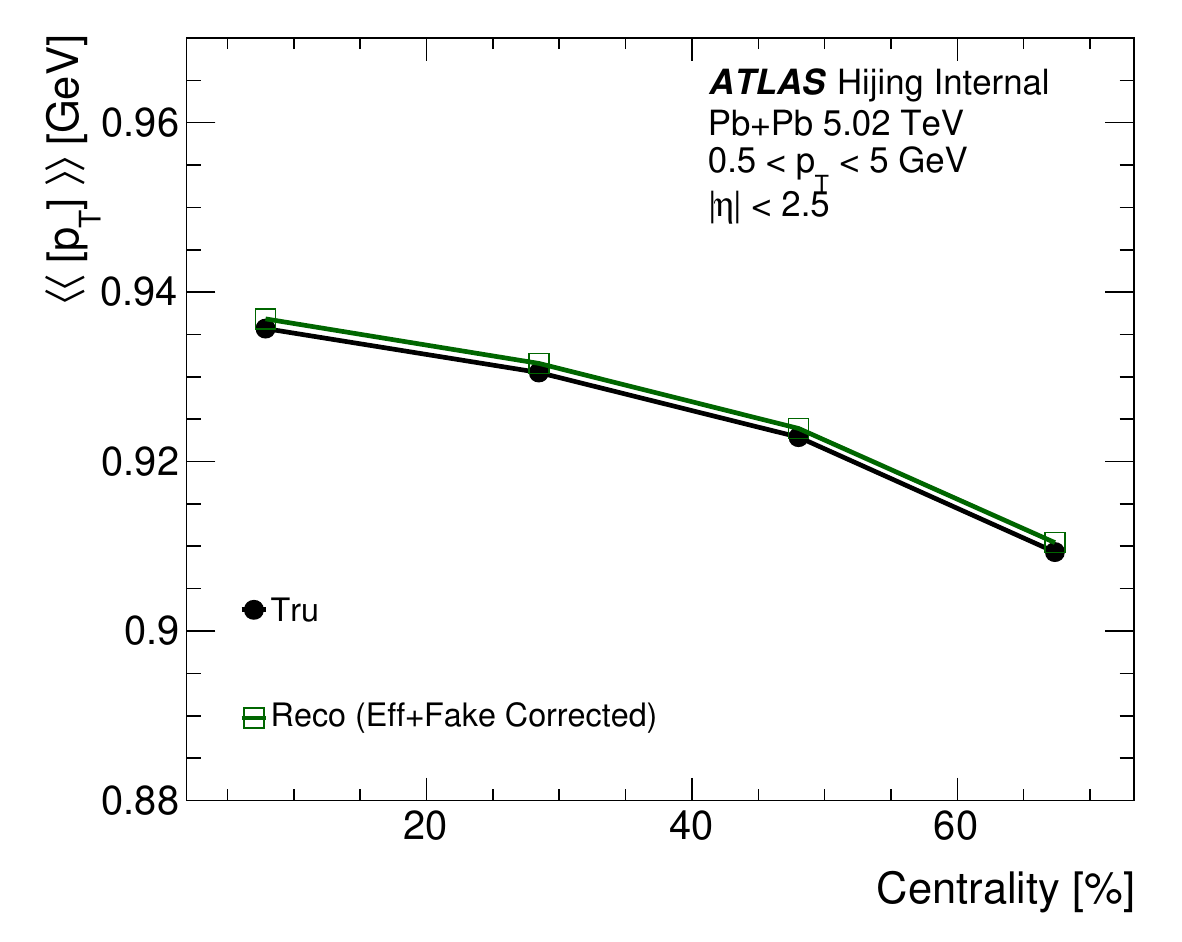}
\includegraphics[width=0.325\linewidth]{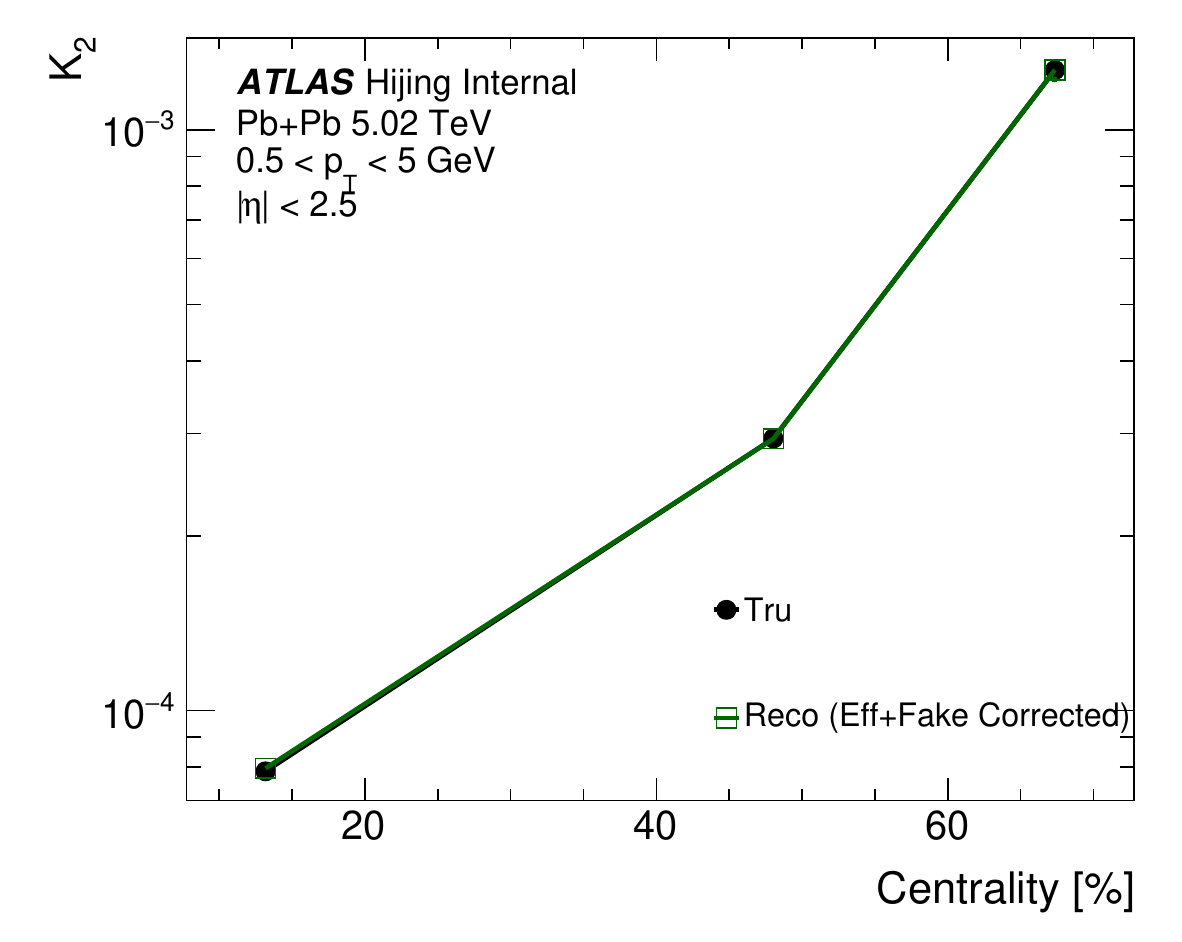}
\includegraphics[width=0.325\linewidth]{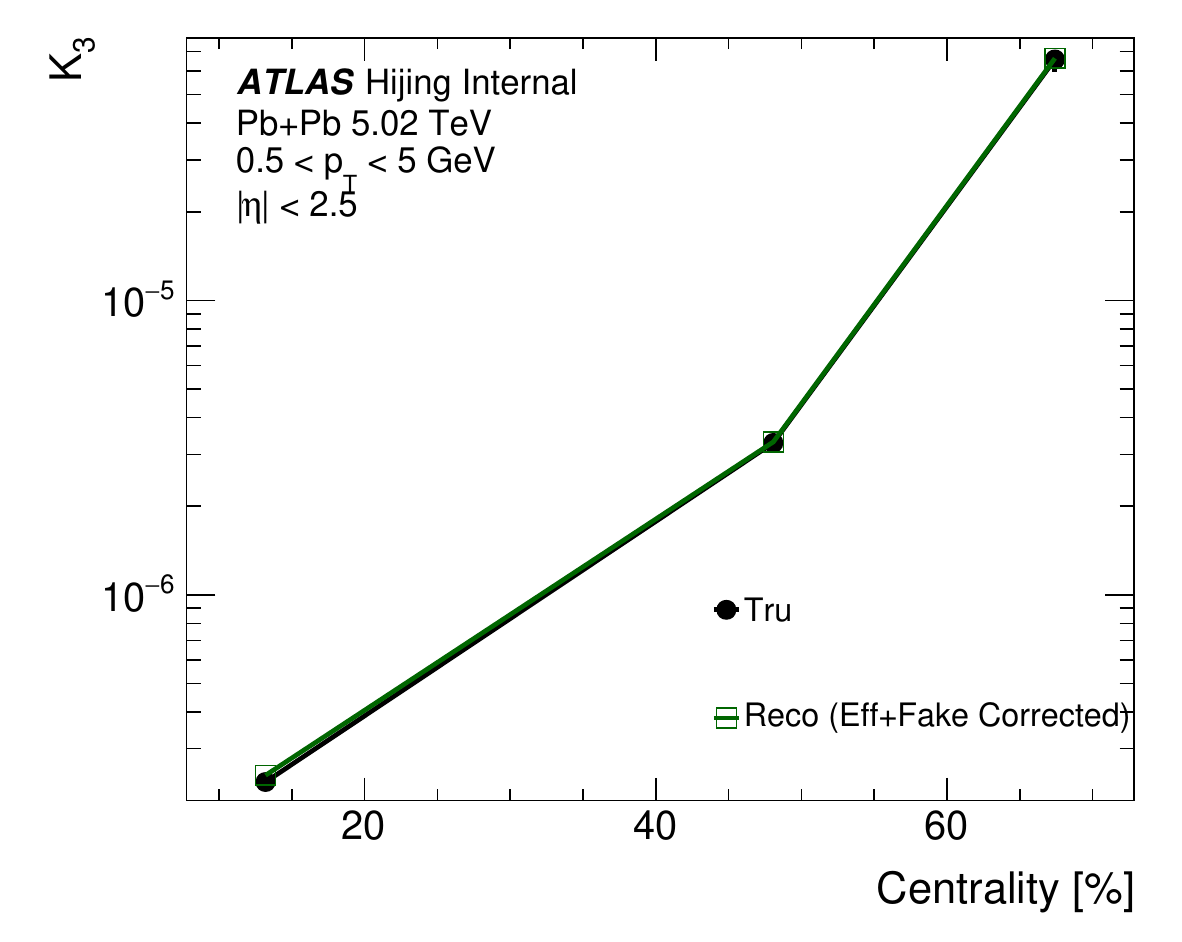}

\includegraphics[width=0.325\linewidth]{./FinPlotsV2/Closure/cC2_Cent.pdf}
\includegraphics[width=0.325\linewidth]{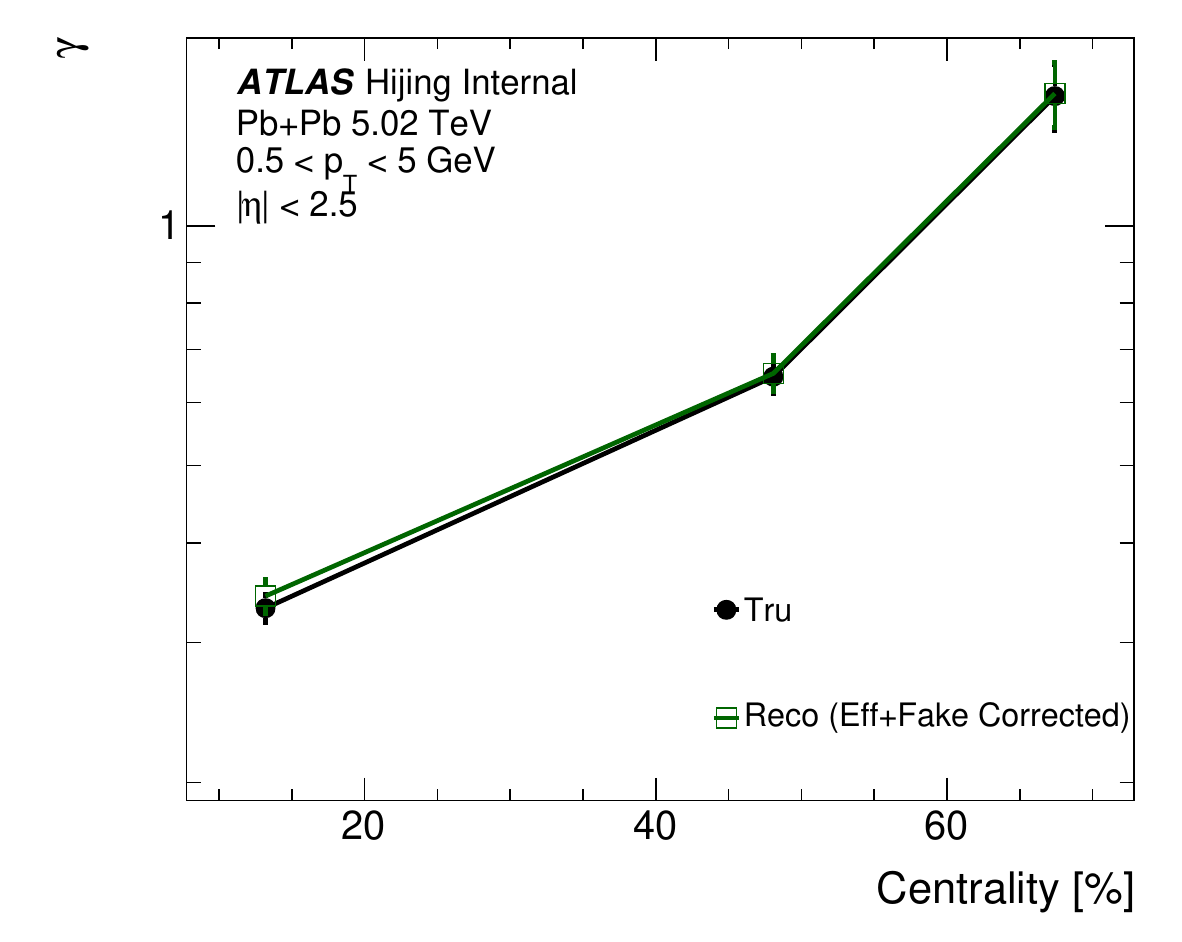}
\includegraphics[width=0.325\linewidth]{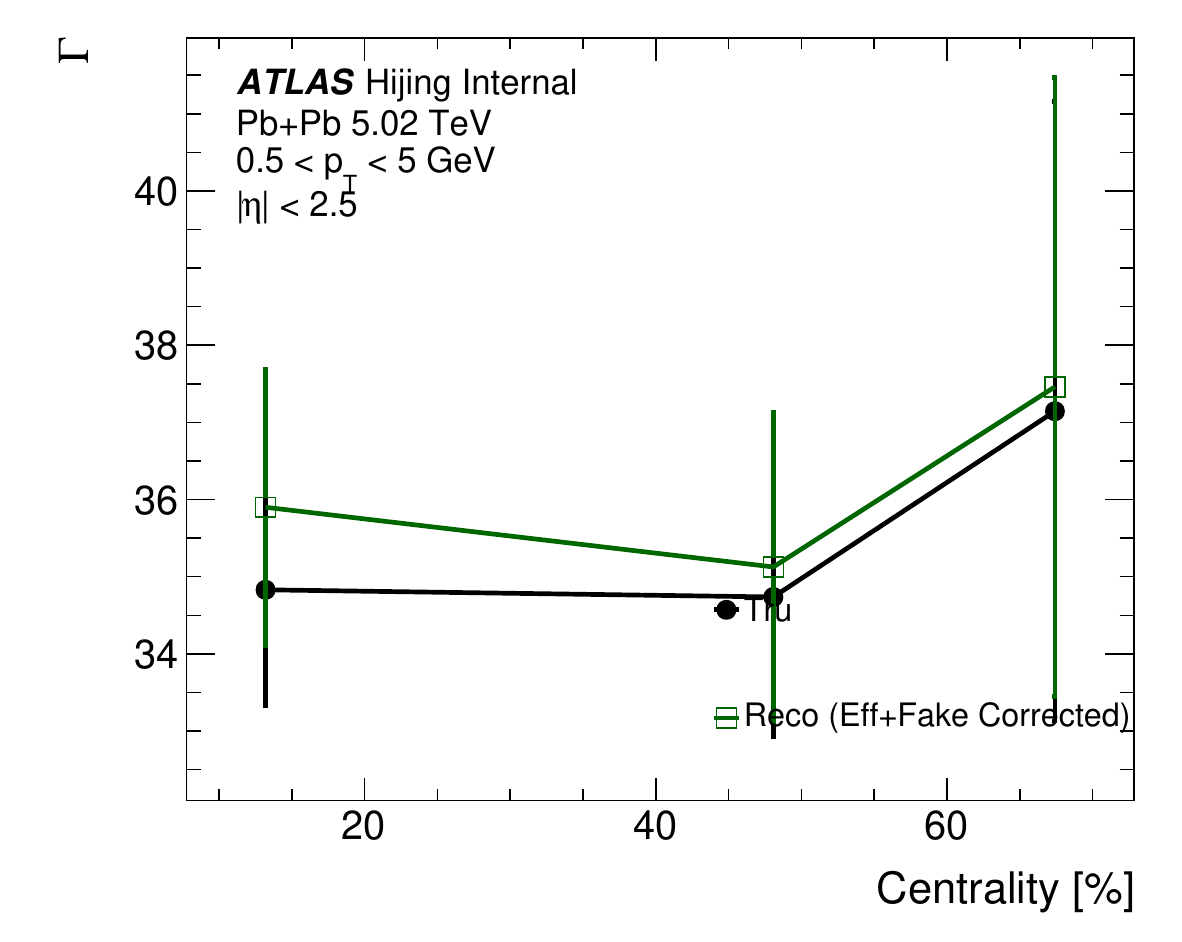}

\caption{Comparison between observables measured with reconstructed tracks and primary particles using HIJING MC sample for Pb+Pb collisions at 5 TeV. Top row shows $\MpT$, $k_{2}$, and $k_{3}$, Bottom row shows $nk_{2}$, $\gamma$, and $\Gamma$. The observables are plotted against $\NchR$ based centrality for particles with 0.5 $< p_{T} <$ 5.0 GeV and $|\eta| <$ 2.5. Error bars denote statistical uncertainties.}
\label{fig:Hijing_sys_Pb}
\end{figure}

 We compare the efficiency-corrected observables measured from reconstructed tracks with those measured with only truth level primary particles. A perfect closure would imply that measurements from both the aforementioned cases would agree exactly with each other. Fig.~\ref{fig:Hijing_sys_Pb} shows the comparisons for measurements carried out using reconstructed tracks (corrected for efficiency and fake tracks) and primary particles at the truth level from the Pb+Pb MC sample as a function of centrality. Owing to small event statistics ($\sim$1 Million) in the HIJING MC sample, the observables are binned in wide bins of centrality. The comparisons are carried out for particles falling within the acceptance  of 0.5$<p_{T}<$5.0 GeV and $|\eta|$ < 2.5. The measurements with generated particles are observed to be in good agreement with the observables measured with reconstructed tracks after using track weights.

\begin{figure}[htbp]
\centering
\includegraphics[width=0.325\linewidth]{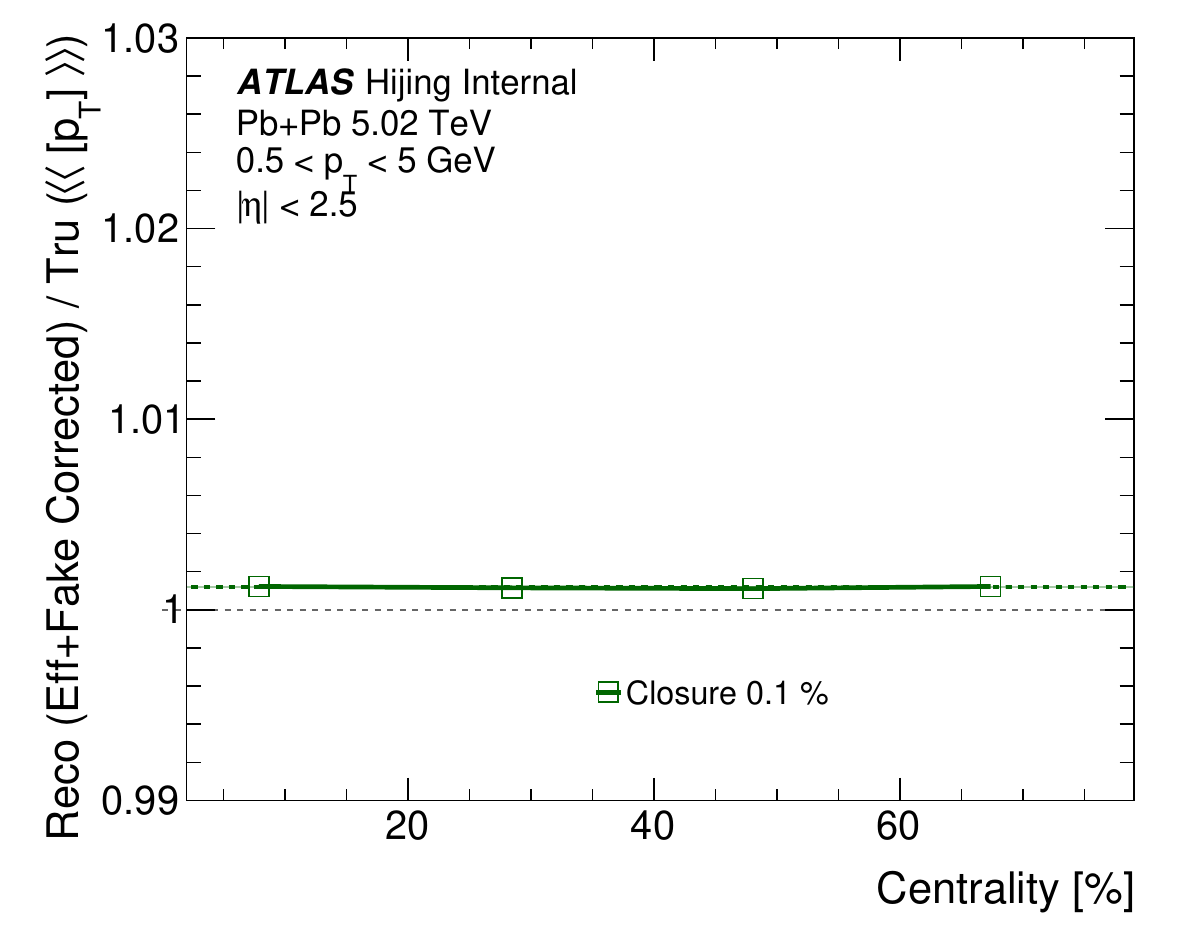}
\includegraphics[width=0.325\linewidth]{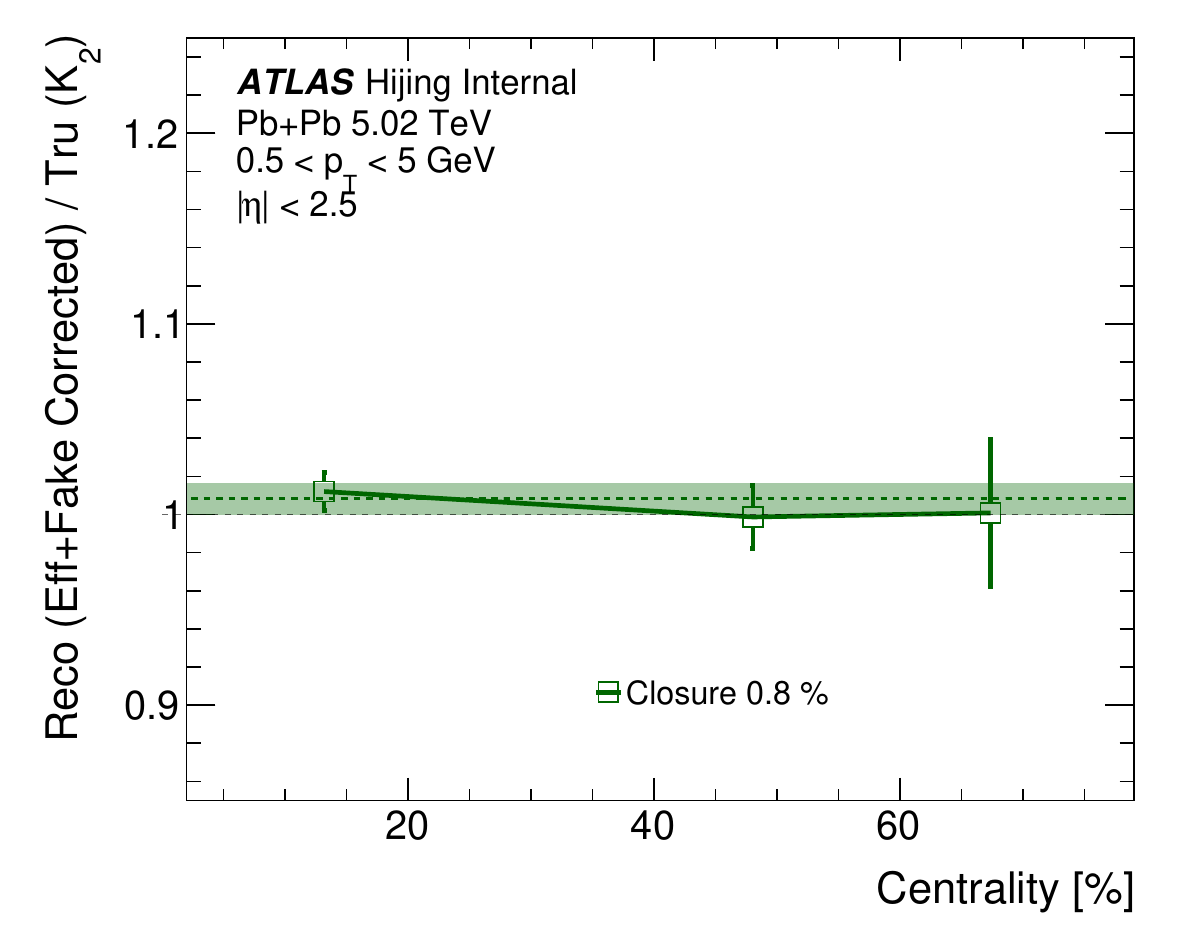}
\includegraphics[width=0.325\linewidth]{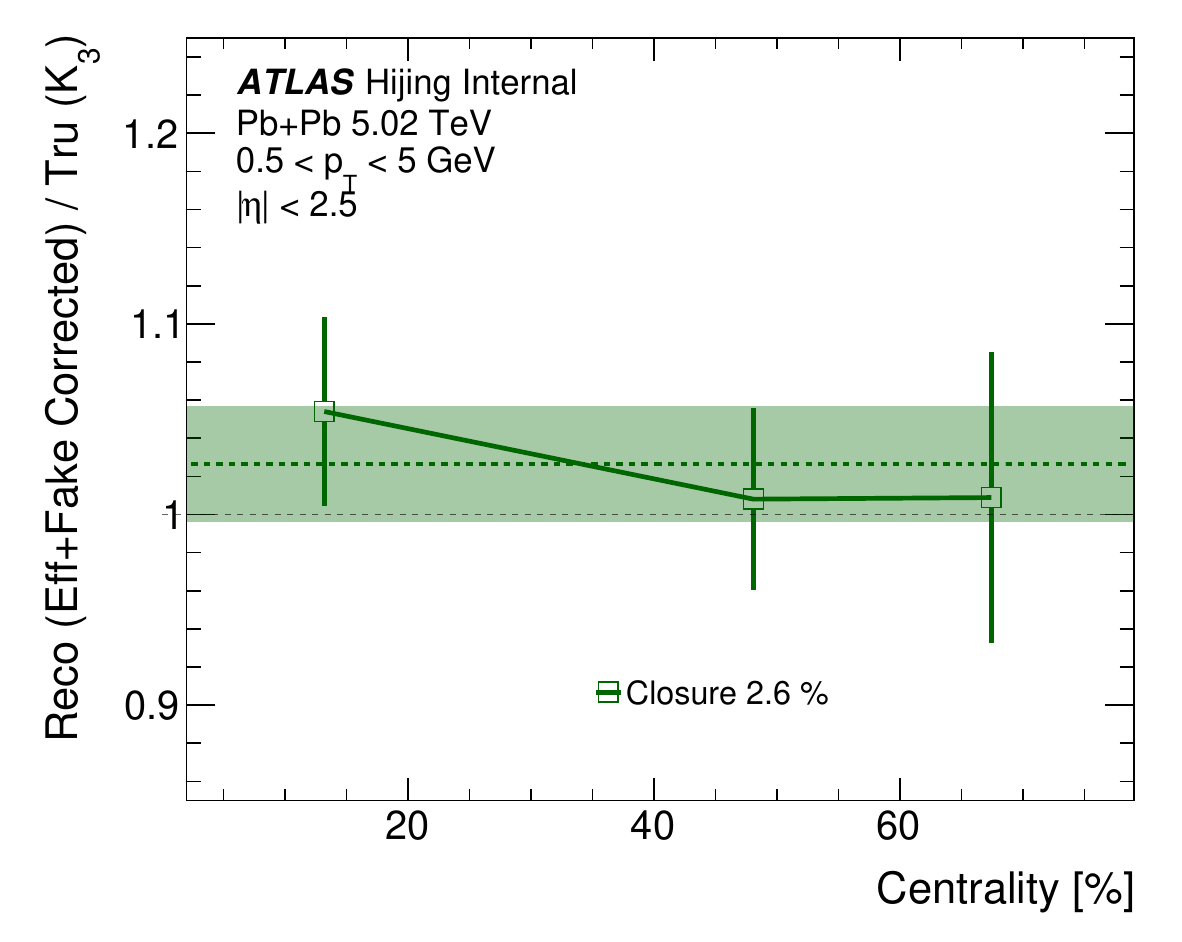}

\includegraphics[width=0.325\linewidth]{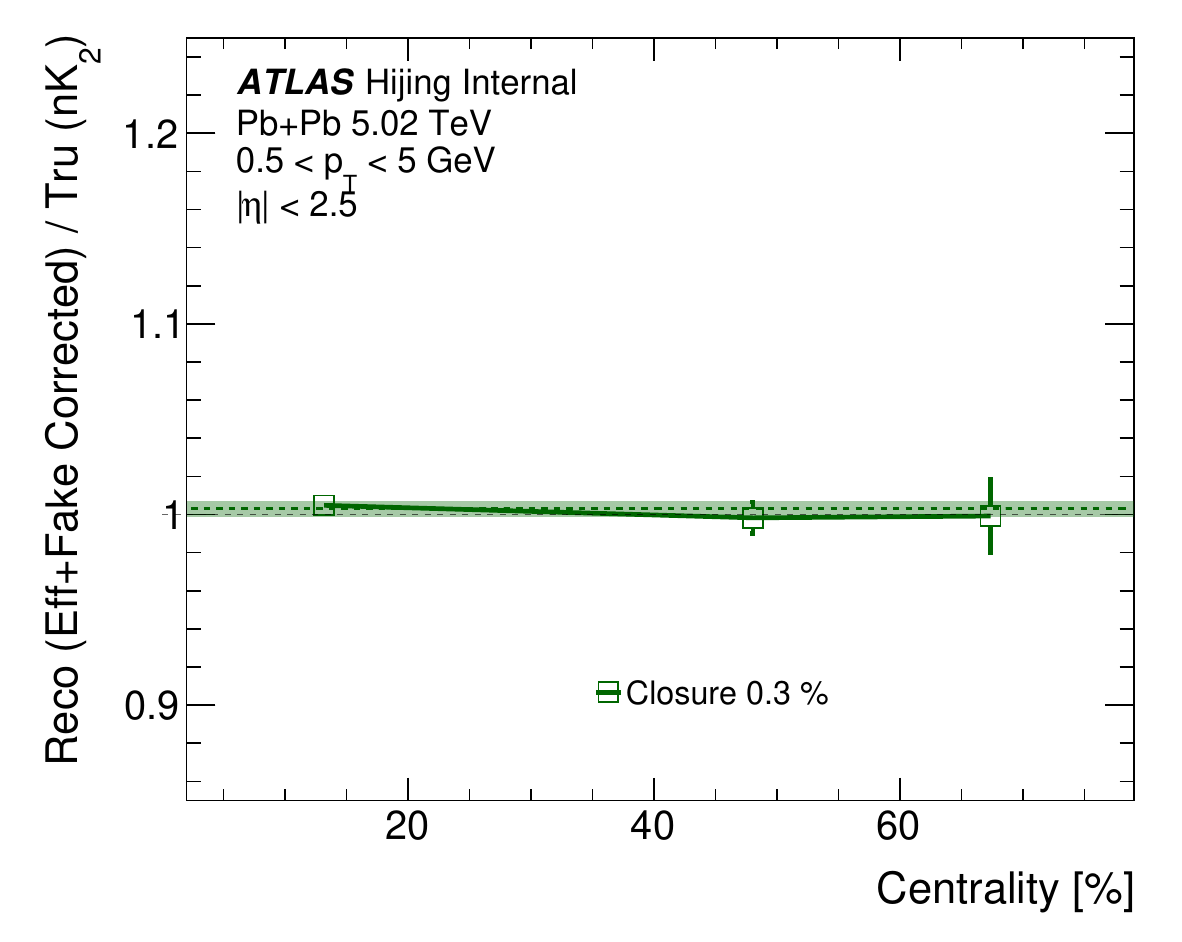}
\includegraphics[width=0.325\linewidth]{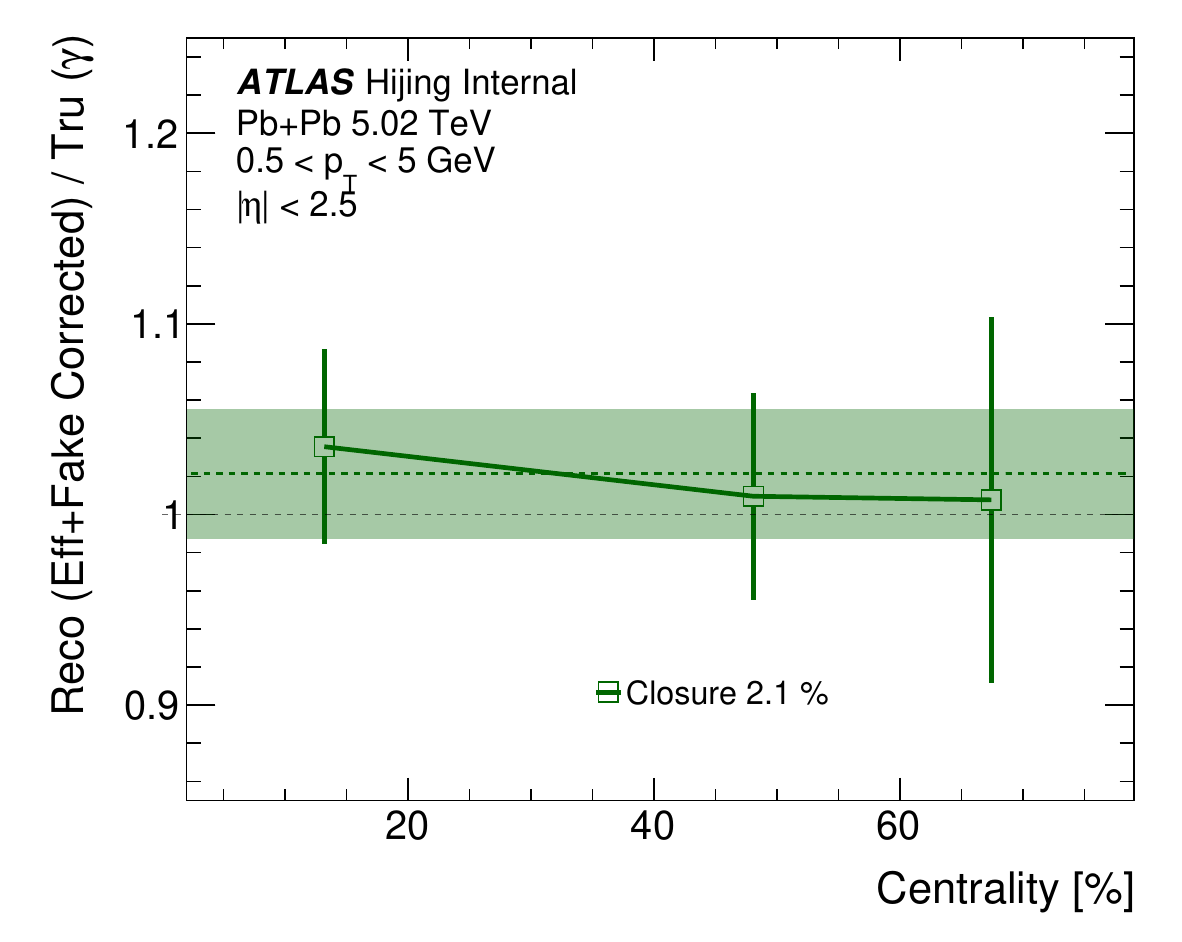}
\includegraphics[width=0.325\linewidth]{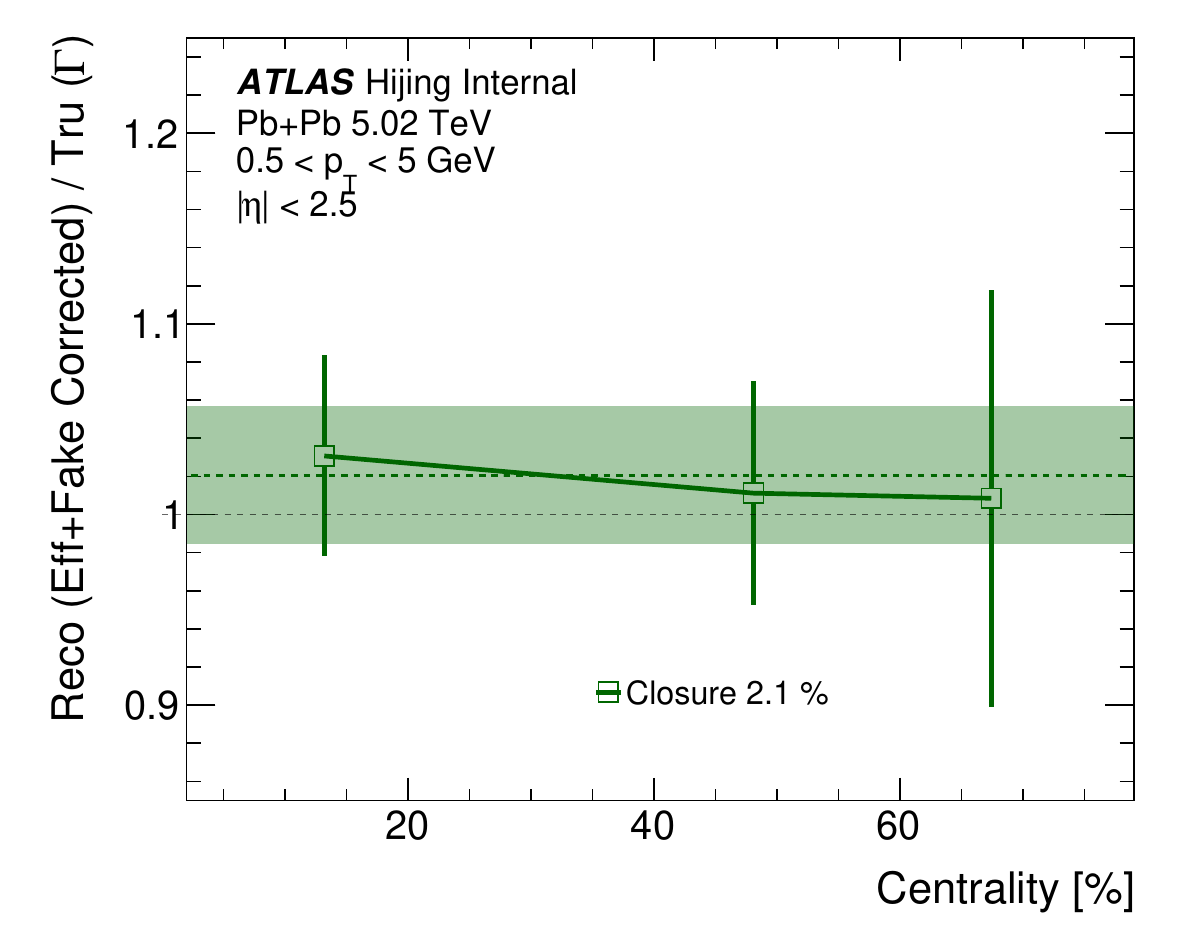}

\caption{Ratios between observables measured with reconstructed tracks and primary particles using HIJING MC sample for Pb+Pb collisions at 5 TeV. Top row shows ratios for $\MpT$, $k_{2}$, and $k_{3}$, Bottom row shows ratios for $nk_{2}$, $\gamma$, and $\Gamma$. The ratios are plotted against $\NchR$ based centrality for particles with 0.5 $< p_{T} <$ 5.0 GeV and $|\eta| <$ 2.5. Error bars denote statistical uncertainties. The dotted horizontal line denotes the average of all the points, whereas the faded band denotes the error on the calculated average of the points.}
\label{fig:Hijing_sys_Pb2a}
\end{figure}

 The level of agreement shown by the observables measured with reconstructed tracks and primary particles is not perfect. Therefore, to quantify the level of closure for the observables, Fig.~\ref{fig:Hijing_sys_Pb2a} shows the ratio of observables measured with reconstructed tracks (corrected for efficiency and fake tracks) and primary particles as a function of centrality. The ratio for $\MpT$, $k_{2}$, and $nk_{2}$ show closure within 0.1\%, 0.8\%, and 0.3\% on average, respectively, over the entire range of centrality. The statistical fluctuations in the ratio for $k_{3}$, $\gamma$, and $\Gamma$ are observed to be relatively large owing to limited event statistics. Nevertheless, on average, $k_{3}$, $\gamma$, and $\Gamma$ show closure within an averaged value of 2.6\%, 2.1\% and 2.1\% respectively over the entire centrality range. However, for the normalized observables, the error on the extracted average of the ratios for closure (shown in green faded bands) is also large and overlaps with 1.

To account for the level of non-closure in total systematics, we add a symmetric contribution to the total systematics in quadrature. Therefore, the quoted value of systematic contribution arising from non-closure in this study for $\MpT$, $k_{2}$, $k_{3}$ stand at $\pm$$0.05$\%, $\pm$$0.4$\%, $\pm$$1.3$\% respectively. For the normalized observables, $nk_{2}$, $\gamma$ and $\Gamma$, the systematic contribution from non-closure are $\pm$$0.15$\%, $\pm$$1.05$\%, $\pm$$1.05$\% respectively.  

The closure check in this study is carried out for the Pb+Pb collisions production leveraging the relatively larger event statistics in the HIJING MC sample compared to Xe+Xe MC sample. The systematic uncertainties extracted for the observables using the Pb+Pb HIJING sample are also used for observables measured in Xe+Xe collisions.

\subsubsection{In Xe+Xe Collisions}\label{sec:sys:Xe}
The systematics in Xe+Xe collisions are carried out in a very similar manner as is done for Pb+Pb collisions in Section~\ref{sec:sys:Pb}.

\begin{figure}[htbp]
\centering
\includegraphics[width=0.325\linewidth]{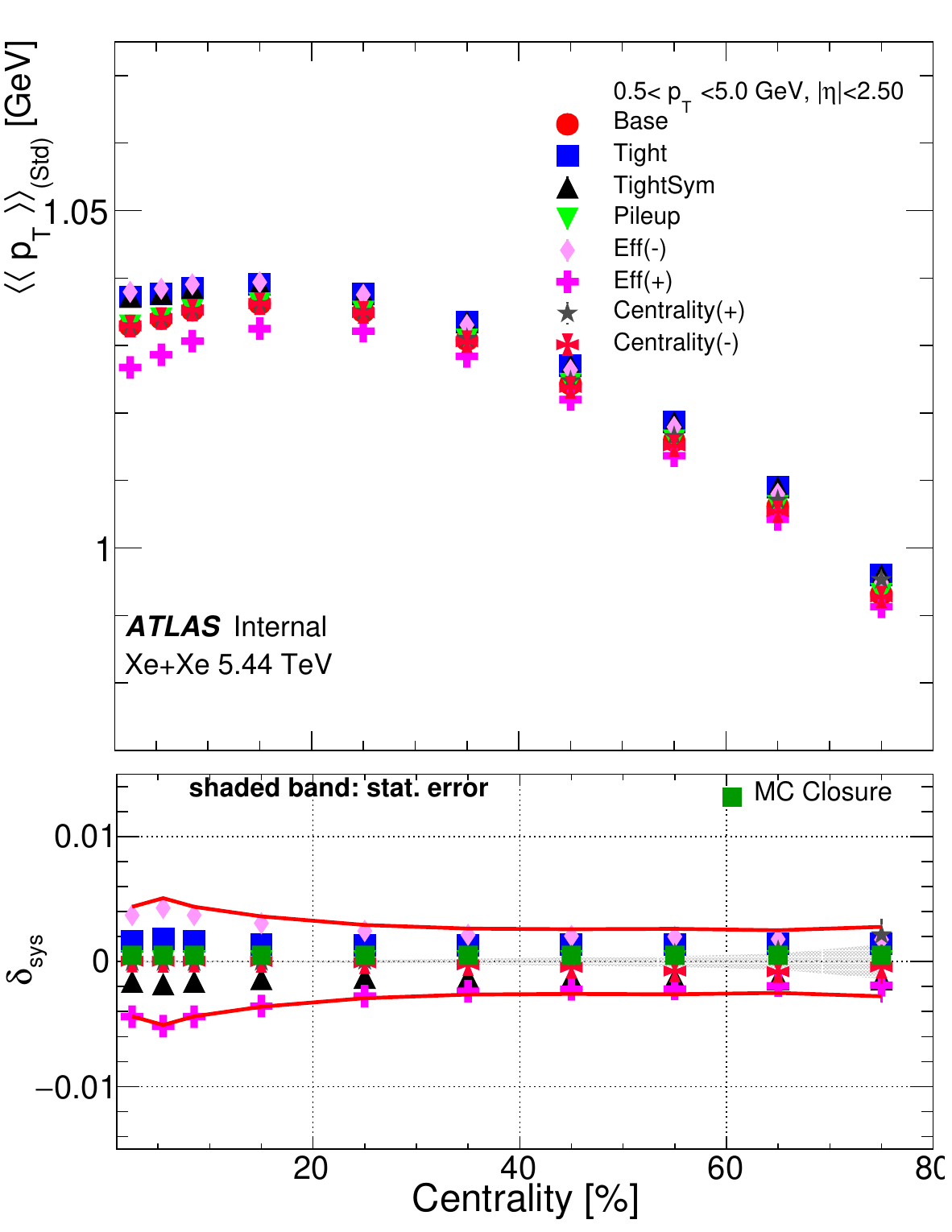}
\includegraphics[width=0.325\linewidth]{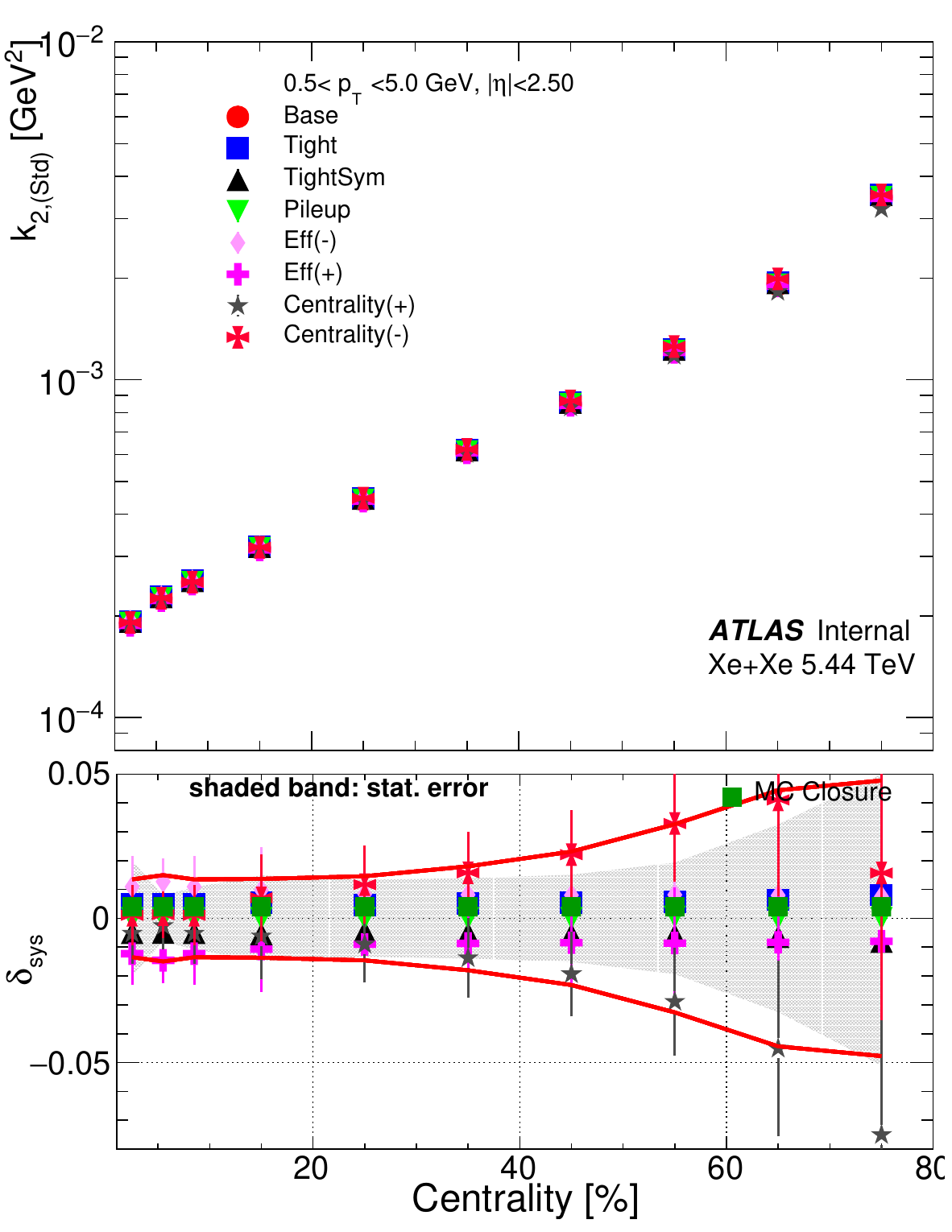}
\includegraphics[width=0.325\linewidth]{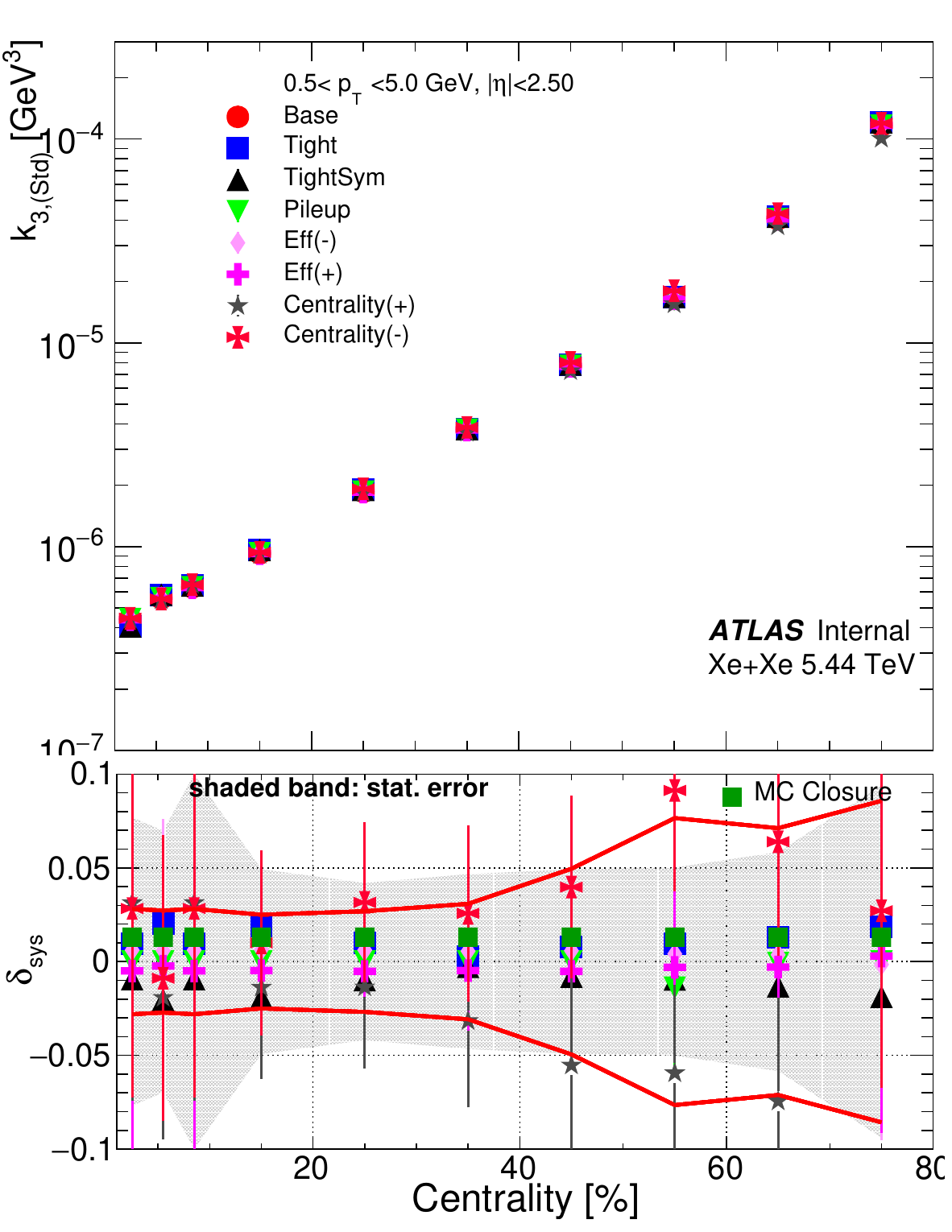}
\caption{Comparison between different systematic variations and default measurement for $\MpT$ (Left panel), $k_{2}$ (Central panel) and $k_{3}$ (Right panel) for 0.5-5 GeV particles in Xe+Xe collisions. The bottom panels show relative systematic uncertainties arising from systematic variations w.r.t default for respective observables on top panels. The error bars represent statistical uncertainties. The shaded area represents the ratio of the statistical uncertainty. Solid red-colored lines on the bottom panels reflect combined systematic uncertainty for the observables.}
\label{fig:XeSys_Fluc}
\end{figure}

Fig.~\ref{fig:XeSys_Fluc} shows the comparison of different systematic checks with the default case of $\MpT$ correlations in Xe+Xe. The bottom panels show relative difference w.r.t. default along with the lower and upper bounds shown in lines and the ratio of statistical uncertainty of the default to its mean value shown as a shaded region. The systematics are dominated by uncertainty in the track efficiency and track selection in the central and mid-central region. The uncertainty due to centrality is important in the peripheral region. The systematic uncertainty is generally larger than the statistical uncertainties for mean, variance, and skewness. 
Fig.~\ref{fig:XeSys_Fluc} also shows (bottom row) the relative difference w.r.t nominal cuts along with the lower and upper bounds shown in lines and the ratio of statistical uncertainty of the default to its mean value shown as a shaded region. The quadrature sum of the relative difference between the lower and upper limits from systematic variation in a given bin is chosen to be the systematic error. The total systematic error is comparable to or smaller than the statistical uncertainties for all centralities.

\begin{figure}[htbp]
\centering
\includegraphics[width=0.325\linewidth]{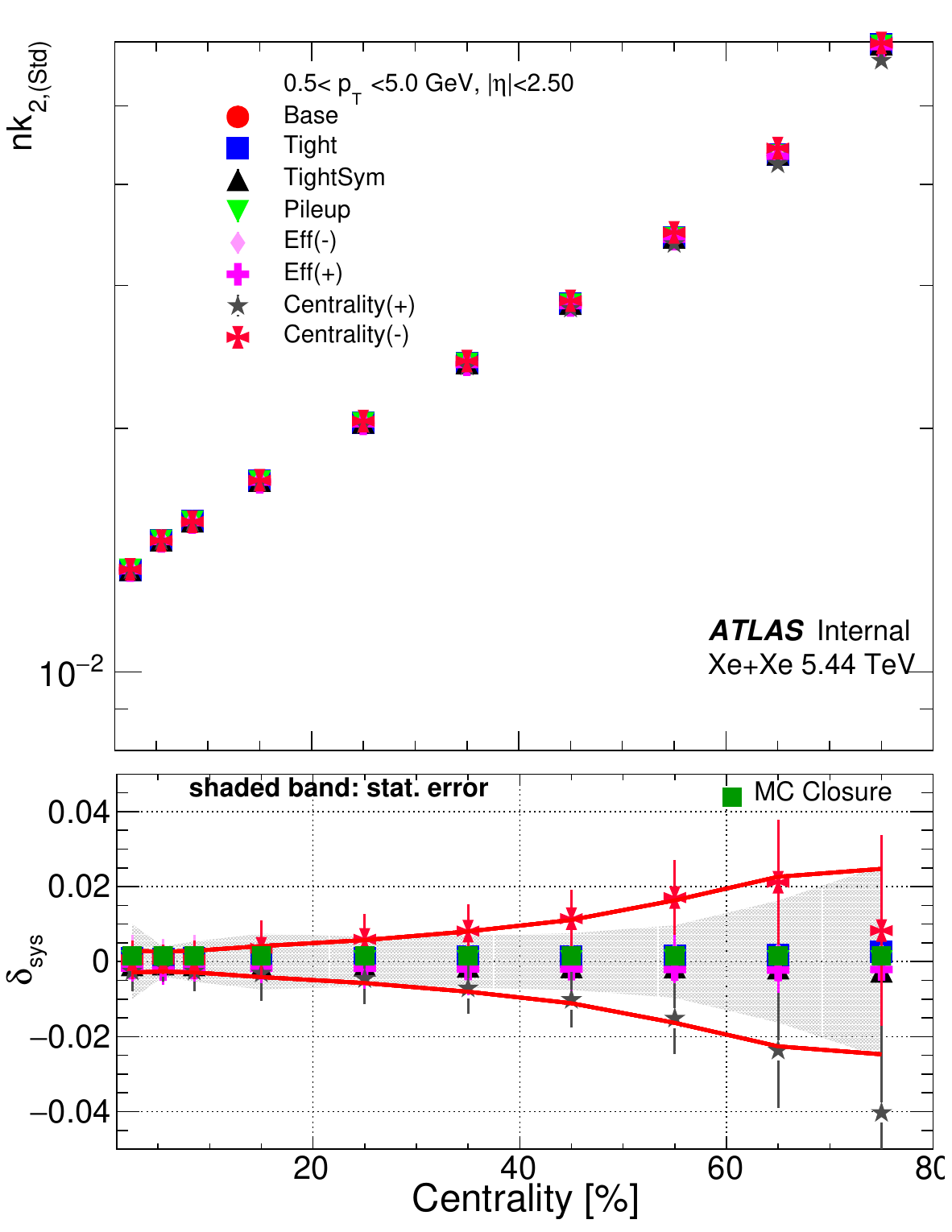}
\includegraphics[width=0.325\linewidth]{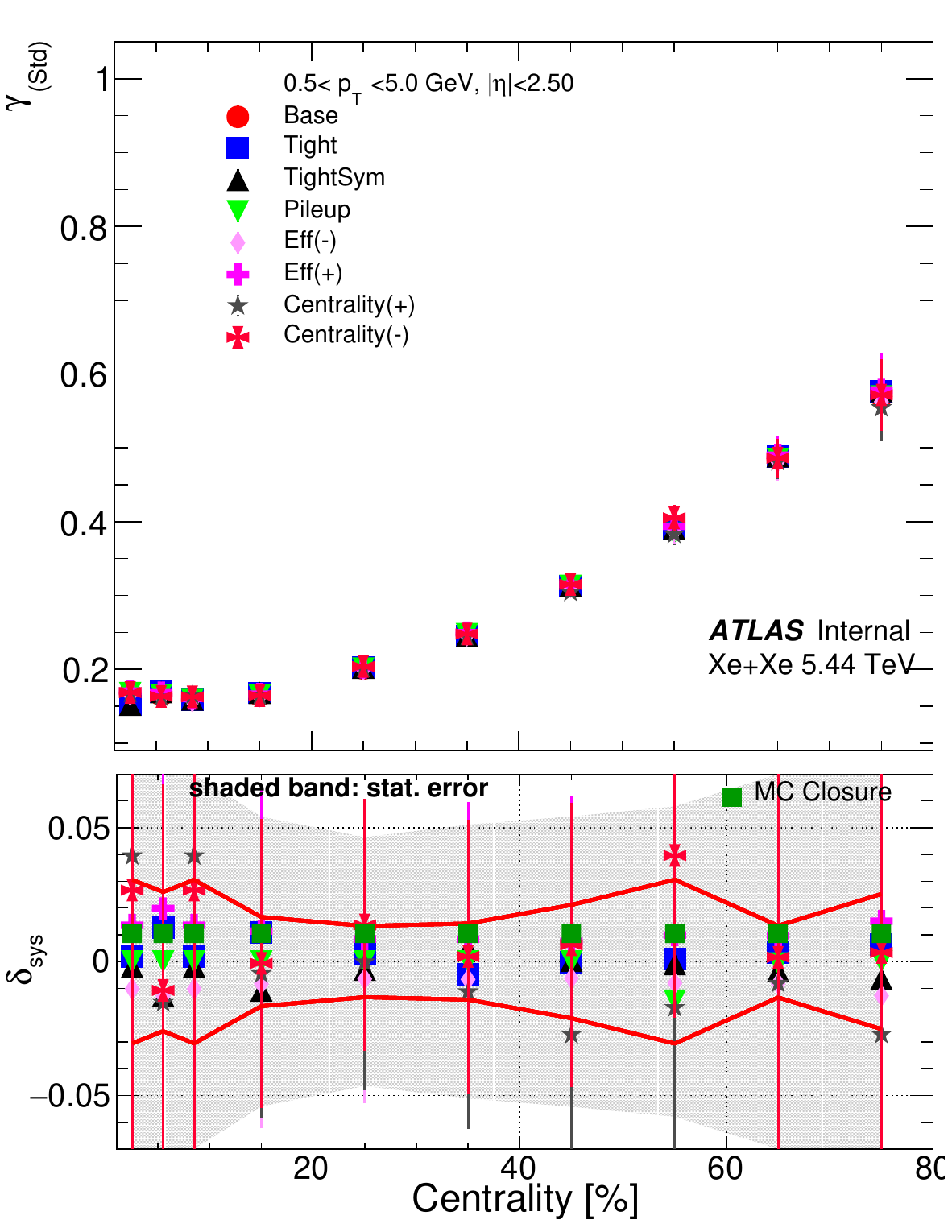}
\includegraphics[width=0.325\linewidth]{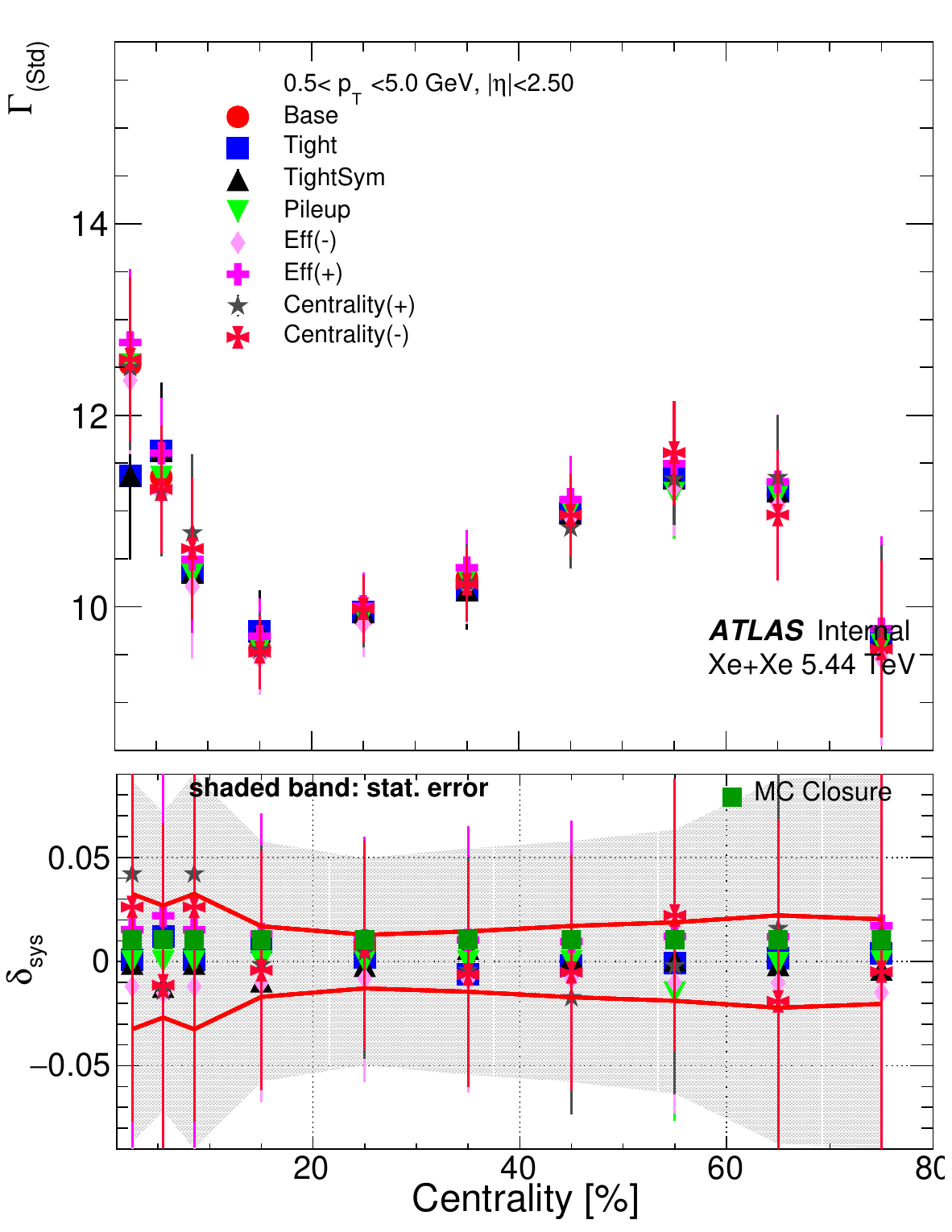}
\caption{Comparison between different systematic variations and default measurement for $nk_{2}$ (Left panel), $\gamma$ (Central panel) and $\Gamma$ (Right panel) for 0.5-5 GeV particles in Xe+Xe collisions. The bottom panels show relative systematic uncertainties arising from systematic variations w.r.t default for respective observables on top panels. The error bars represent statistical uncertainties. The shaded area represents the ratio of the statistical uncertainty. Solid red-colored lines on the bottom panels reflect combined systematic uncertainty for the observables.}
\label{fig:XeSys_NormFluc}
\end{figure}

 Fig.~\ref{fig:XeSys_NormFluc} shows the systematic checks for Normalized cumulants. The systematic uncertainty partially cancels between the numerator and denominator. Therefore, the final systematic uncertainty has weaker centrality dependence, especially for $n=2$ and 3. The uncertainty is less than 10\% for $n=3$.

For Xe+Xe, the centrality variation has the minimum contribution to systematics for most centrality intervals for all observables. For $n=2$, the dominant source of systematics arises from efficiency variation for all observables. For $n=3$ the systematic effect coming from track quality variations becomes larger in some cases.

\subsection{For $v_0(\pT)$}\label{sec:sysv0pt}

For the measurements of $\lr{\delta N(p_{T}) \delta p_{T} }$, $v_{0} (p_{\mathrm{T}})$, and $v_{0} (p_{\mathrm{T}})/v_{0}$ using tracks within $|\eta|<2.5$ and $0.5<p_T<5.0$ GeV in Pb+Pb collisions, the dominant systematic uncertainties arise from the track quality selection and tracking efficiency. The centrality definition uncertainty also plays a role, particularly when results are presented as a function of centrality. The HIJING closure check provides an additional estimate of potential systematic biases related to short-range correlations and detector effects. The level of non-closure observed in HIJING is accounted for by adding a symmetric contribution to the total systematic uncertainty in quadrature. For instance, for $\MpT$, $k_{2}$, and $k_{3}$, the systematic contributions from non-closure are $\pm 0.05\%$, $\pm 0.4\%$, and $\pm 1.3\%$, respectively. The systematic effect from pileup contamination is estimated to be less than $0.5\%$ for these observables across all centralities.

\begin{figure}[htbp!]
\centering
\includegraphics[width=0.41\linewidth]{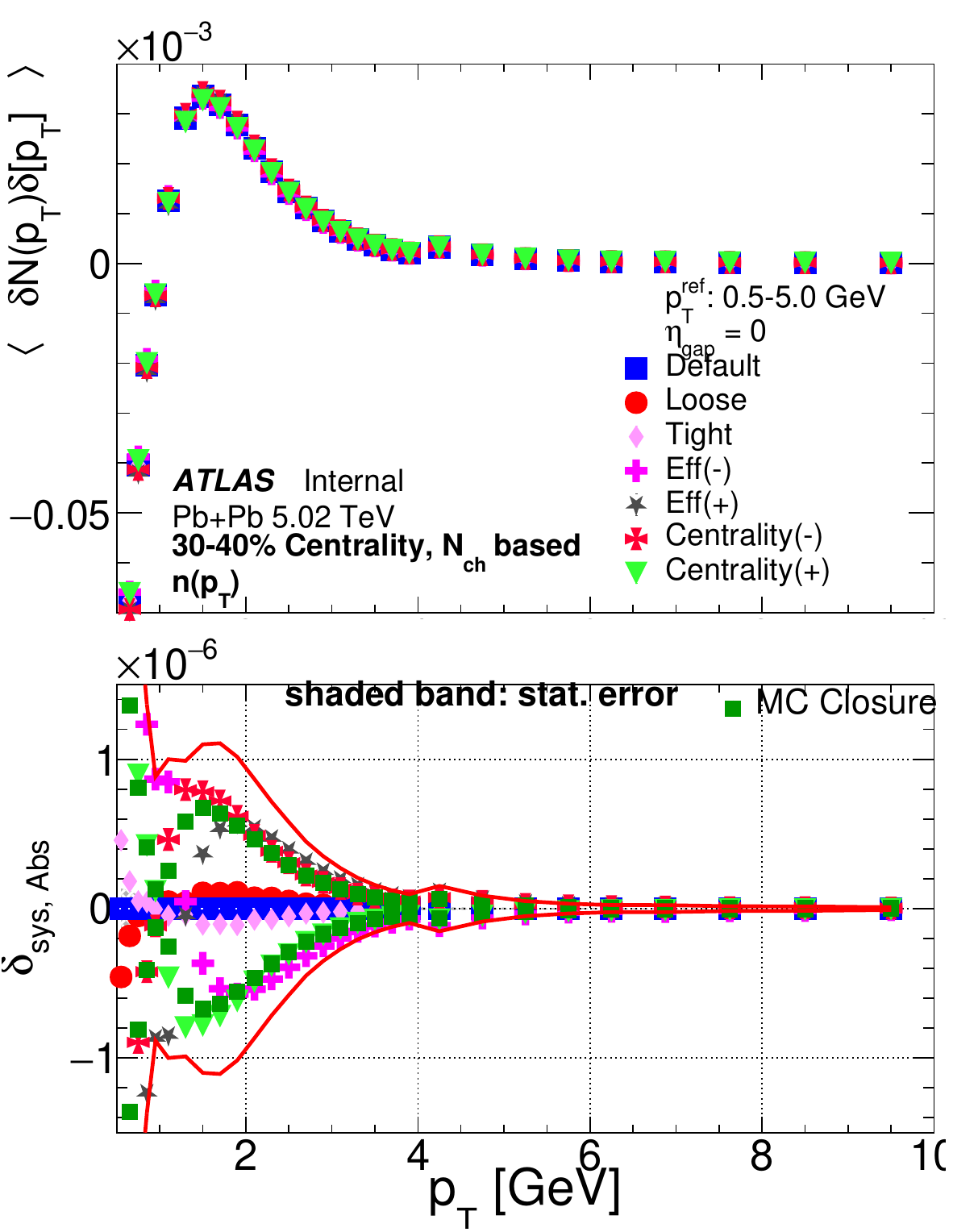}
\includegraphics[width=0.41\linewidth]{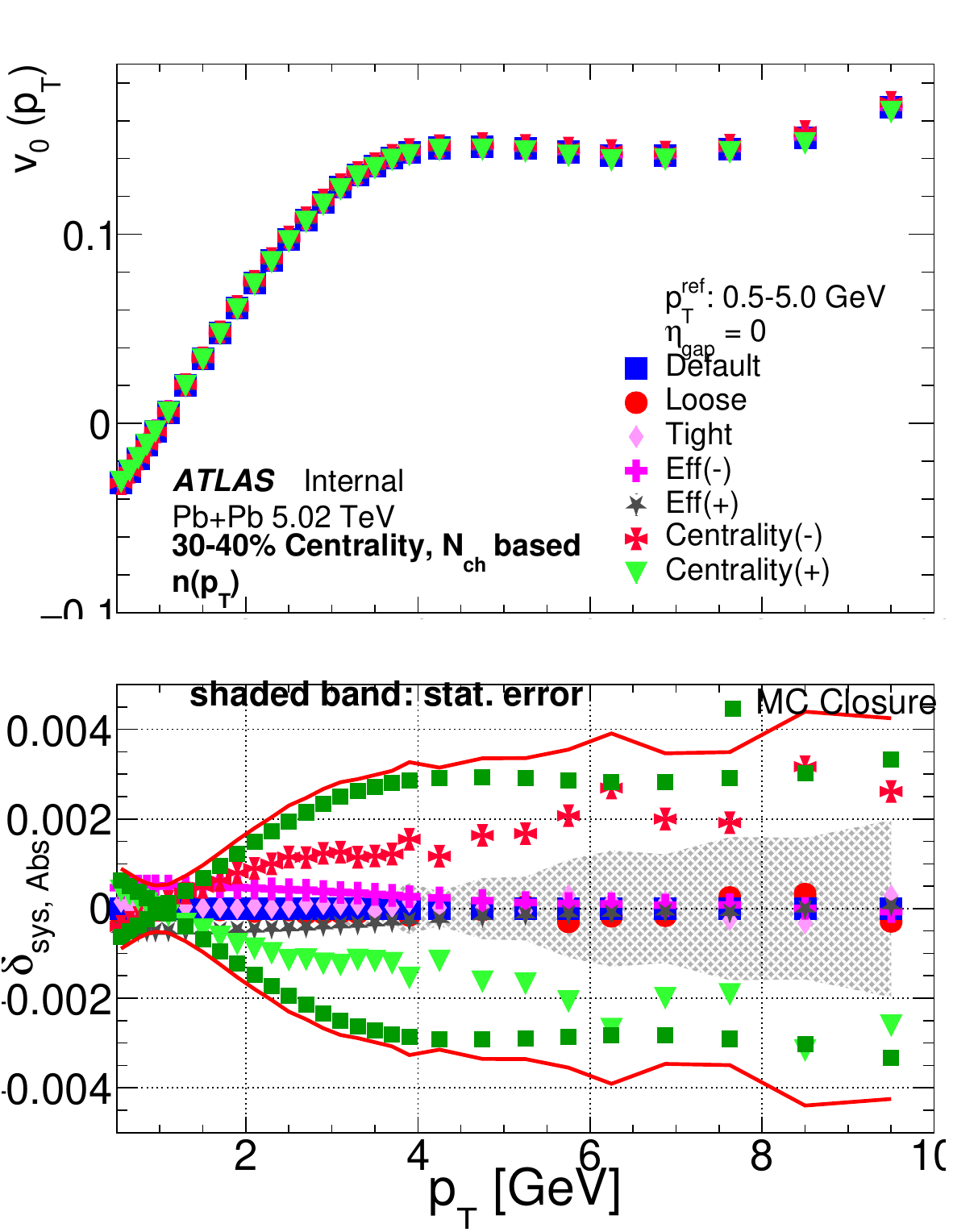}

\includegraphics[width=0.41\linewidth]{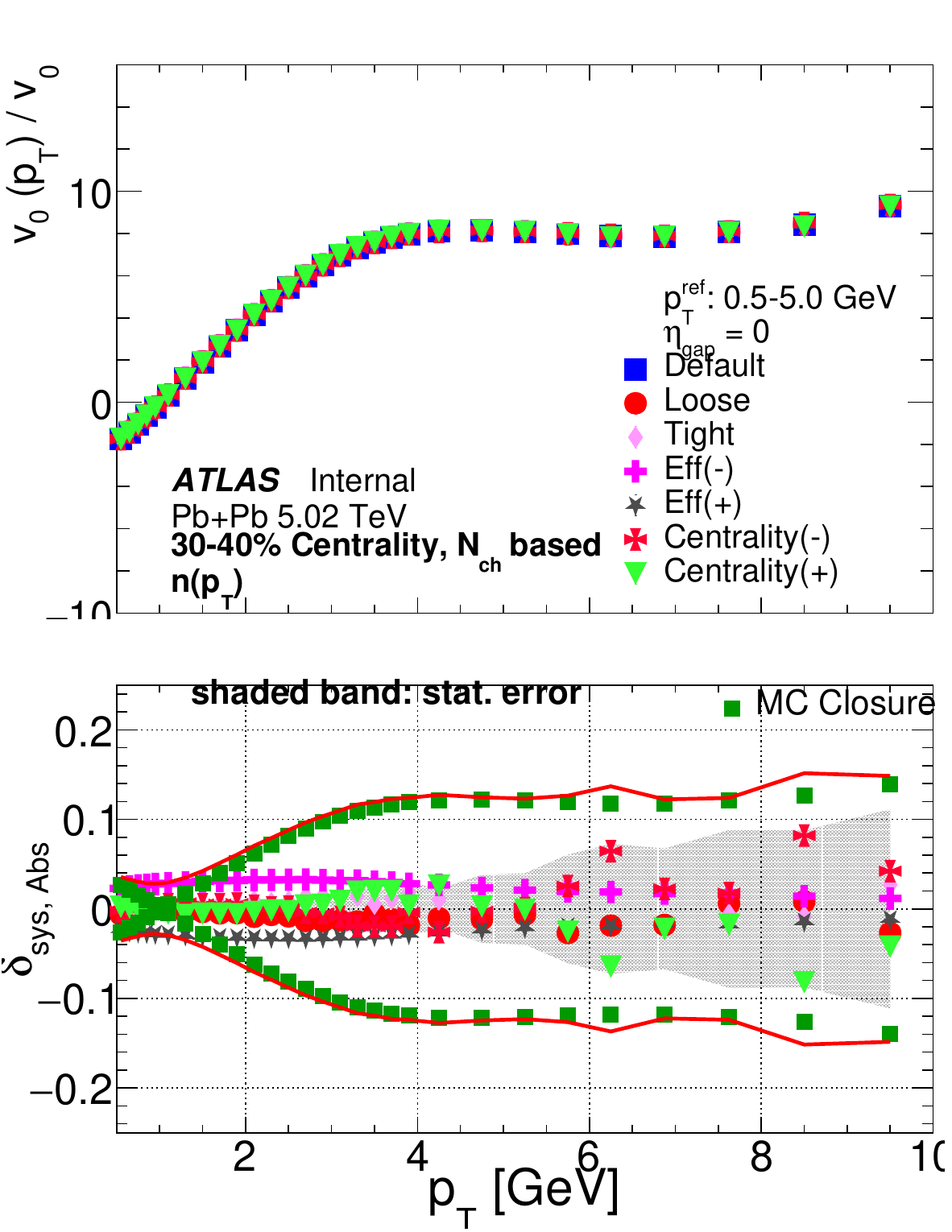}
\includegraphics[width=0.41\linewidth]{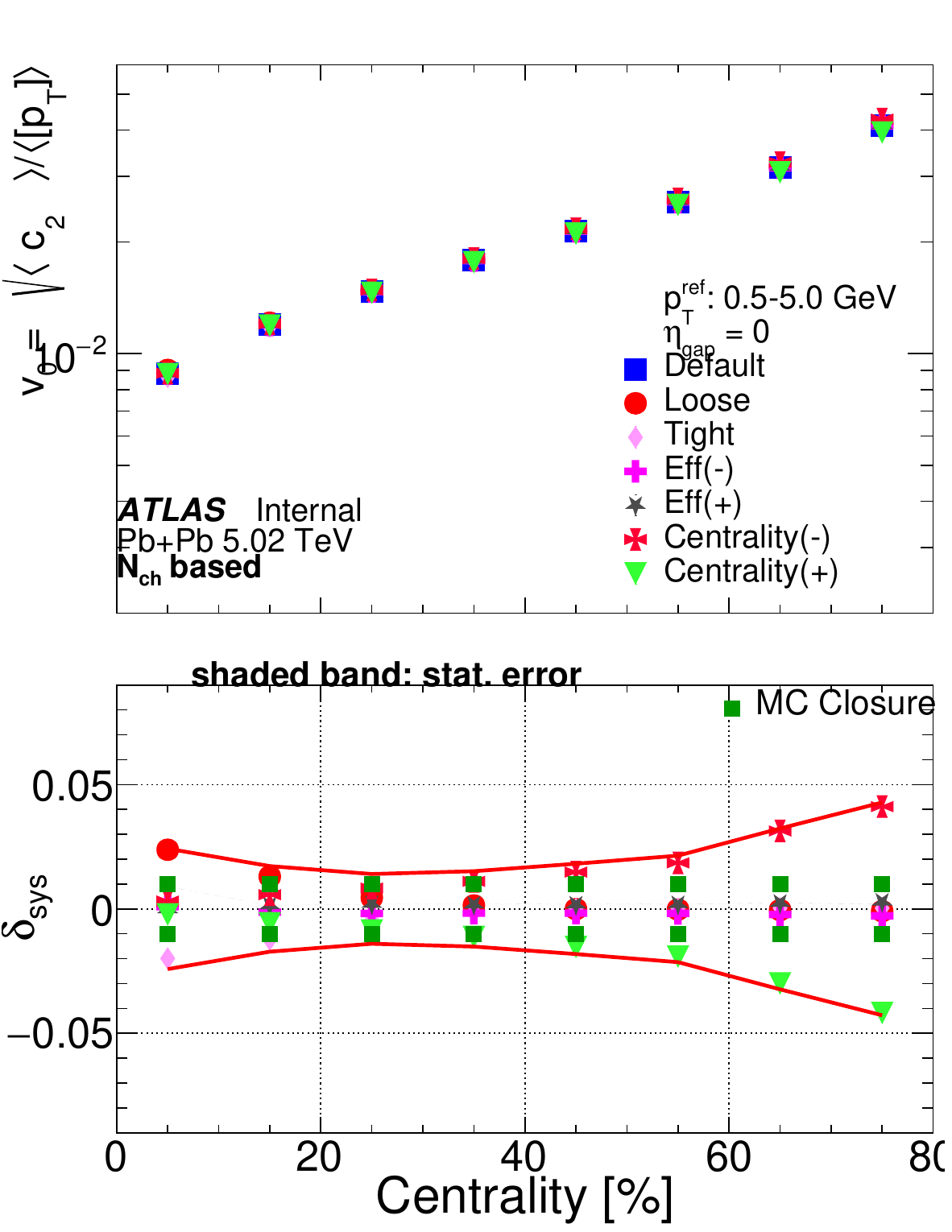}

    \caption{Comparison of different absolute systematic variations with the default measurement for $\Cov$, $v_0(\pT)$, and $v_0(\pT)/v_0$ as functions of $\pT$ in Pb+Pb collisions for 30–40\% centrality.  Additionally, a comparison of relative systematic variations with the default measurement for $v_0$ is shown as a function of centrality. The bottom panels depict the absolute uncertainties arising from these systematic variations with respect to the default measurement for the corresponding observables shown in the top panels. The solid red lines in the bottom panels denote the combined systematic uncertainty for the observables.}
\label{fig:PbSys_V0Cent}
\end{figure}

 Figure~\ref{fig:PbSys_V0Cent} compares various systematic checks with the default measurement for $\nCov$, $v_0(\pT)$ and $v_0(\pT)/v_0$ as functions of $\pT$ in Pb+Pb collisions for mid-central 30-40\% centrality. The bottom panels illustrate the absolute differences with respect to the default case, with the lower and upper bounds represented as lines, and the the statistical uncertainty of the default measurement as a shaded region. Additionally, it shows the relative systematic variations compared to the default measurement for $v_0$ as a function of centrality. 

To calculate the total systematics, the points that are above 0 in these figures are added in quadrature to find the upper bound. 
For $v_0$, the primary source of systematic uncertainty arises from tracking quality variations, which are most significant in central collisions, and from centrality selection variations, which dominate in peripheral collisions. For $\nCov$, $v_0(\pT)$, and $v_0(\pT)/v_0$, the observables change sign around $\pT \approx \MpT$. Since $\MpT$ is slightly larger for tight tracks compared to loose tracks, the systematic uncertainty related to tracking quality also changes sign near $\pT = \MpT$. In $\nCov$, tracking quality variation is the dominant source of systematic uncertainty, followed by efficiency corrections and centrality selection variations. For $v_0(\pT)$, the presence of a denominator suppresses systematic uncertainties from track quality criteria, making efficiency corrections dominant in the low $\pT$ region and centrality corrections dominant in the high $\pT$ region. Since $v_0(\pT)/v_0$ is centrality-independent, centrality variations contribute minimally to its systematic uncertainties. However, the inclusion of $v_0$ in the denominator causes tracking quality criteria to dominate, with efficiency corrections being a close secondary source.

 We also do one additional check to estimate the systematic effect from pileup contamination, we nominally vary the significance cut applied on ZDC-$\sumET$ correlation plot from 6$\sigma$ to 7$\sigma$ and estimate the systematic effect on the observables as a function of centrality. The effect is observed to be negligible for all the observables in this analysis.

We carry out a closure check using the HIJING model as a useful validation of the analysis method. In the HIJING model, the $[\pT]$ fluctuations arise purely from an independent superposition scenario. Therefore, the model predictions for $[\pT]$ fluctuations are not expected to provide good estimates of these quantities. 

\begin{figure}[htbp]
\centering
\includegraphics[width=0.49\linewidth]{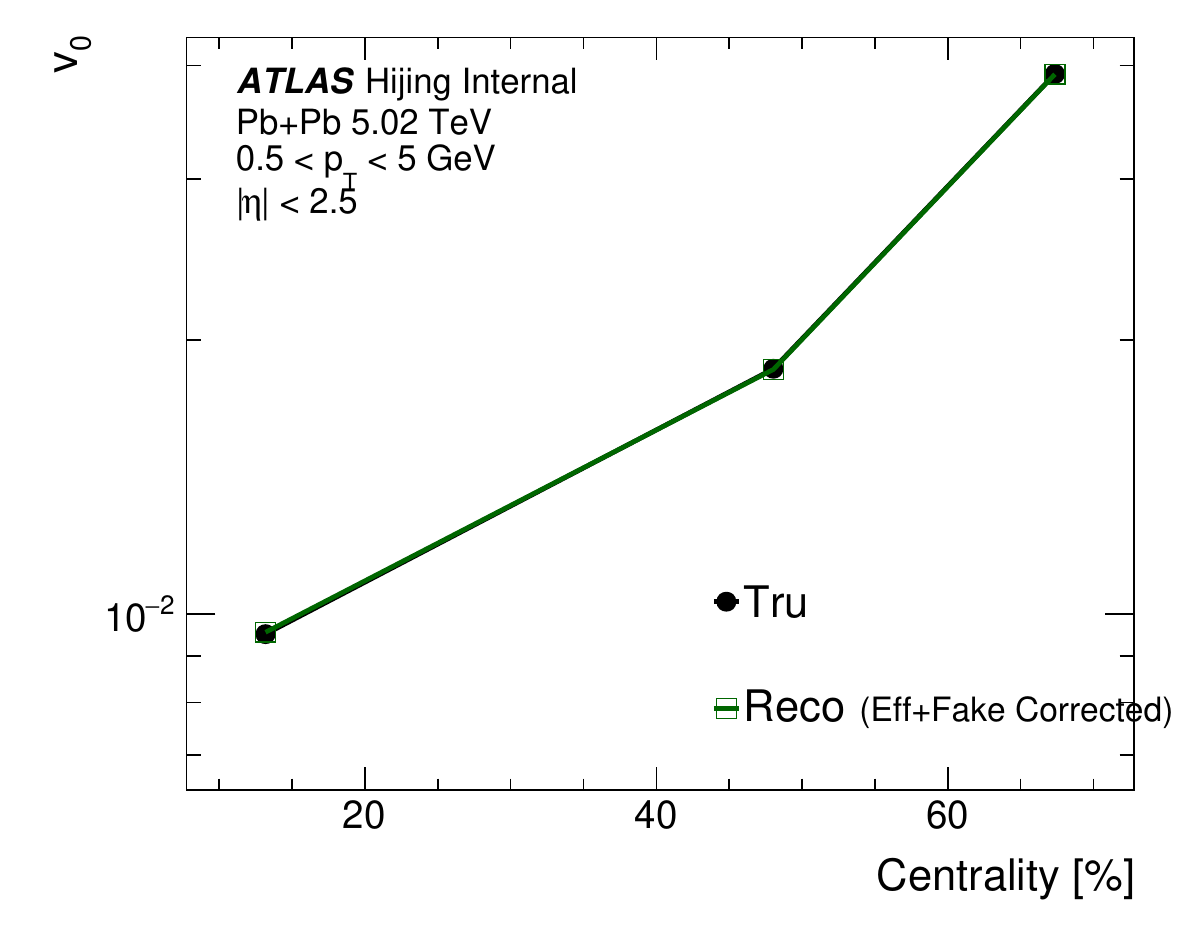}
\includegraphics[width=0.49\linewidth]{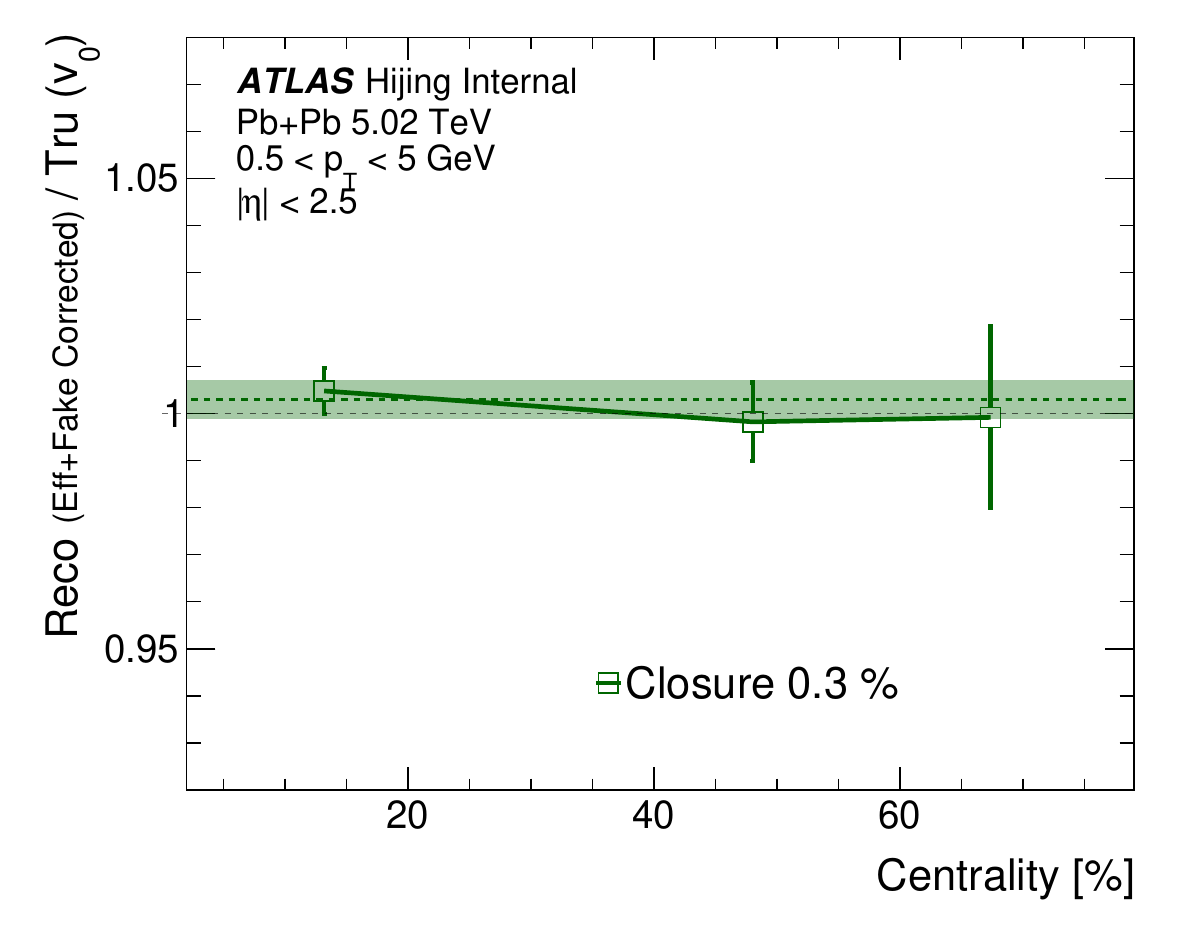}
\caption{Comparison between $v_{0}$ measured with reconstructed tracks and primary particles using HIJING MC sample for Pb+Pb collisions at 5 TeV.
The observables are plotted against $\Nch$ based centrality for particles Error bars denote statistical uncertainties. Ratios between $v_{0}$ measured with reconstructed tracks and primary particles are also shown on the right. The dotted horizontal line denotes the average of all the points, whereas the faded band denotes the error on the calculated average of the points.}
\label{fig:Hijing_sys_Pb0}
\end{figure}

\begin{figure}[htbp]
\centering
\includegraphics[width=0.49\linewidth]{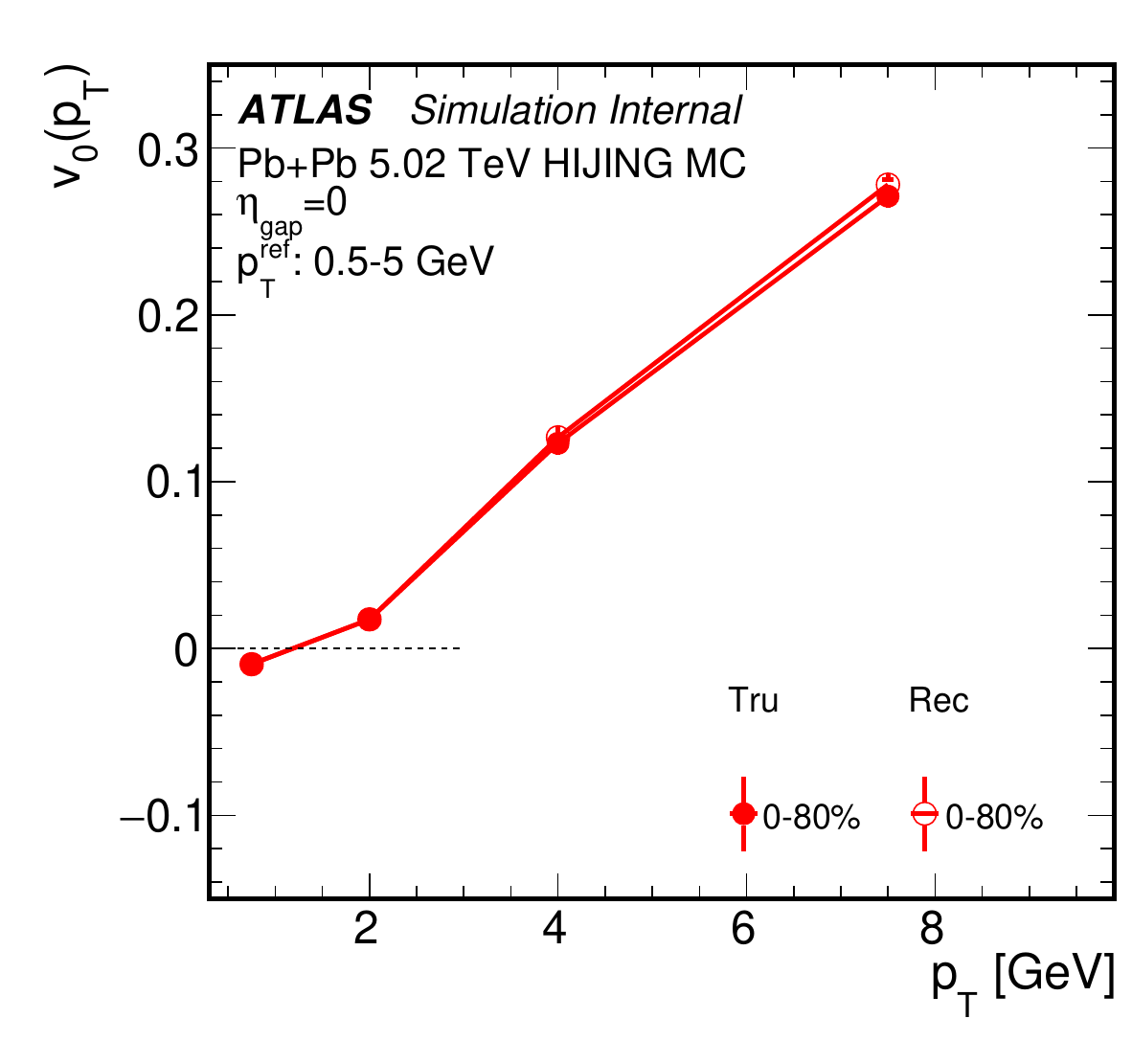}
\includegraphics[width=0.49\linewidth]{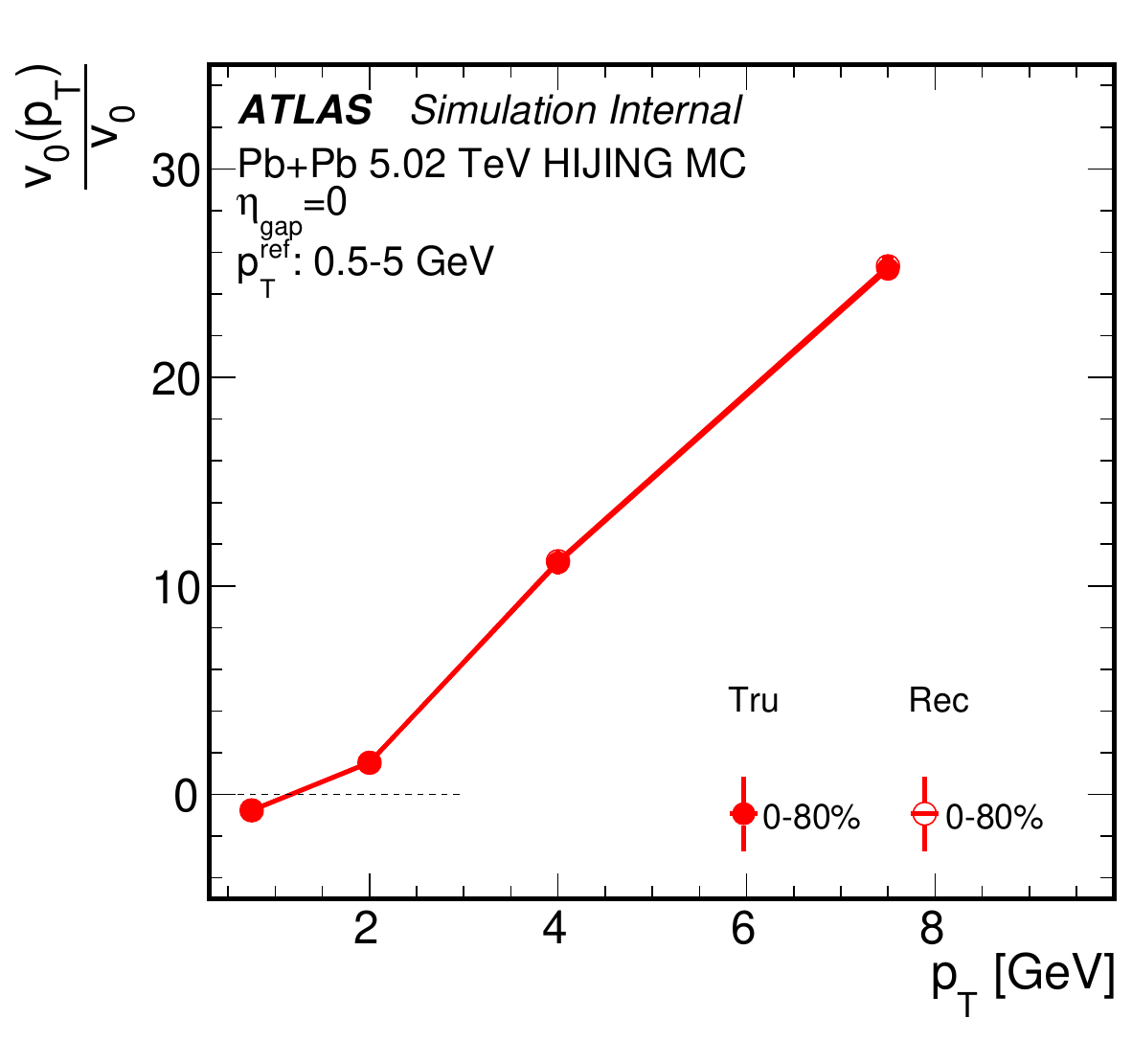}

\includegraphics[width=0.49\linewidth]{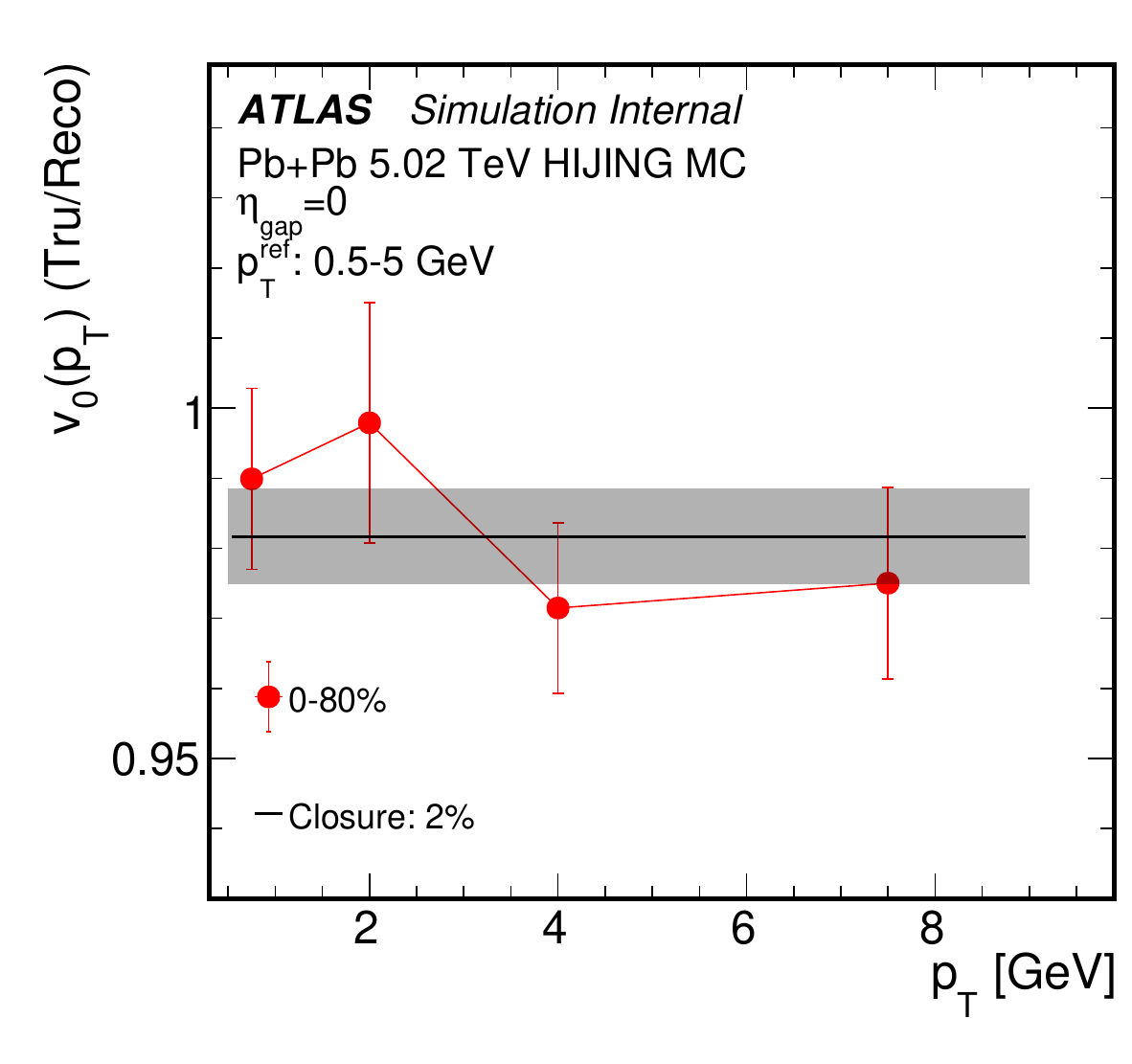}
\includegraphics[width=0.49\linewidth]{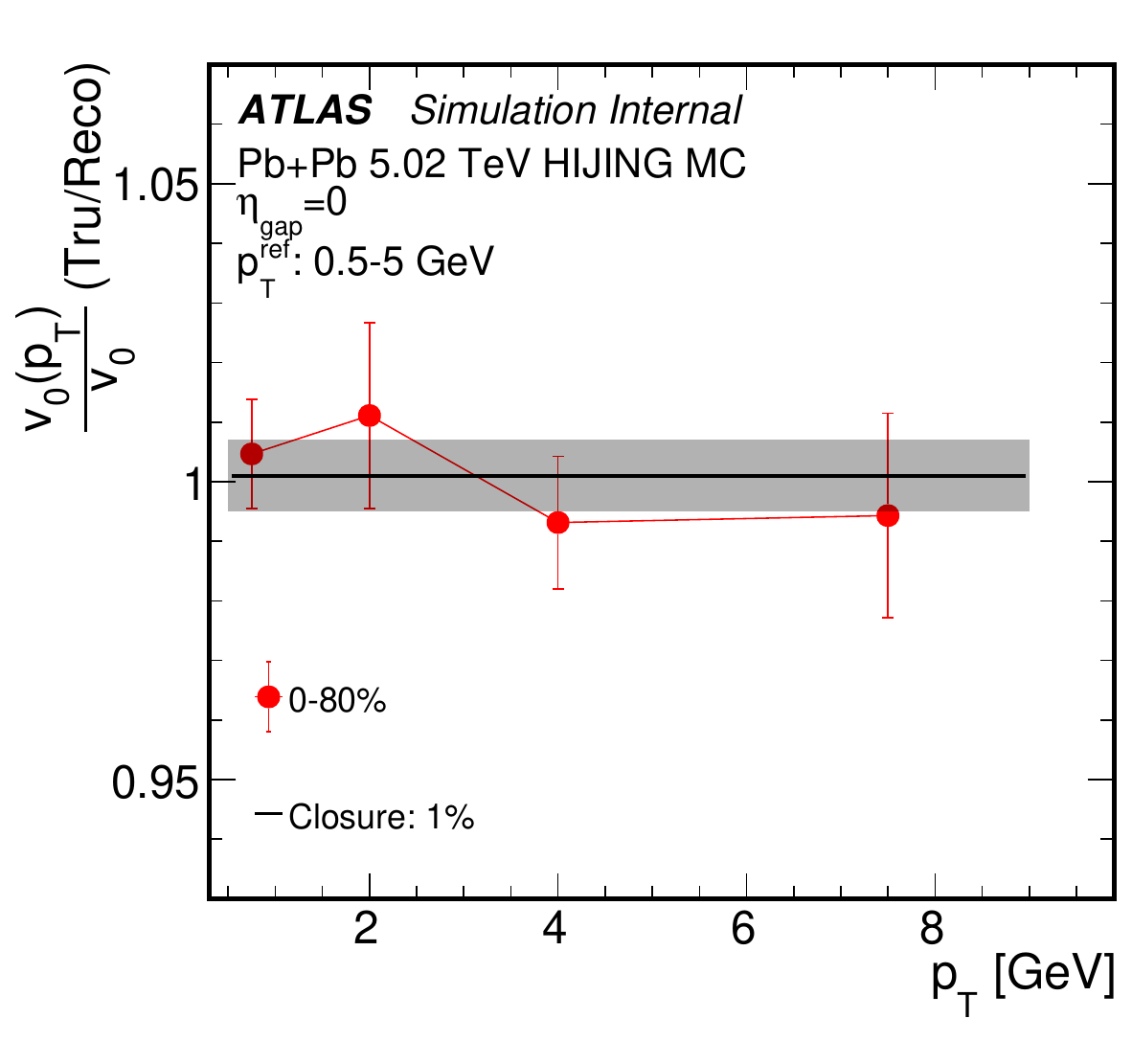}
\caption{Comparison between $v_{0}(\pT)$ measured with reconstructed tracks and primary particles using HIJING MC sample for Pb+Pb collisions at 5 TeV. The observables are plotted against $\pT$ for $p_{T}$ reference of 0.5-5.0 GeV and $|\eta| <$ 2.5 and for 0-80\% centrality. Error bars denote statistical uncertainties calculated with bootstrap method (Sec.~\ref{sec:app_bootstrap}).}
\label{fig:Hijing_sys_Pb2}
\end{figure}

 We compare the efficiency-corrected observables measured from reconstructed tracks with those measured with only truth level primary particles. A perfect closure would imply that measurements from both the aforementioned cases would agree exactly with each other. Figures.~\ref{fig:Hijing_sys_Pb0} and~\ref{fig:Hijing_sys_Pb2} show the comparisons of measurements carried out using reconstructed tracks (corrected for efficiency and Fake tracks) and primary particles at the truth level from the Pb+Pb MC sample as a function of centrality. Owing to small event statistics ($\approx$ 3 Million) in the HIJING MC sample, the observables are binned in wide bins of centrality for $v_{0}$ in Fig.~\ref{fig:Hijing_sys_Pb0} and in wide bins of $\pT$ for $v_{0}(\pT)$ and $v_{0}(\pT)/v_{0}$in Fig.~\ref{fig:Hijing_sys_Pb2}.

 The measurements with generated particles are observed to be in good agreement with the observables measured with reconstructed tracks after using track weights.

However, the level of agreement shown by the observables measured with reconstructed tracks and primary particles is not perfect. To account for the level of non-closure in total systematics, we add a symmetric contribution to the total systematics in quadrature. Therefore, the quoted value of systematic contribution arising from non-closure for $v_{0}$ stands at $\pm$ 1\%. 

Figure~\ref{fig:Hijing_sys_Pb2} presents the closure tests for $v_0(\pT)$ and $v_0(\pT)/v_0$ in Pb+Pb collisions at $\sqrt{s_{\mathrm{NN}}} = 5.02$ TeV. The results are consistent between measurements done with truth particles and reconstructed tracks across the entire $\pT$ range. However, minor discrepancies are observed, reflected by the ratios of measurements between truth particles and reconstructed  tracks, as shown in the bottom panels. Notably, the statistical uncertainties shown in Fig.~\ref{fig:Hijing_sys_Pb2} are derived using the Poisson bootstrap method, indicating that the observed variations in $v_{0}(\pT)$ between the two measurement approaches are well within expected statistical fluctuations. On an average, for $v_0(\pT)$, the non-closure is about $\pm 2\%$ and about $\pm 1.5\%$ for $v_0(\pT)/v_0$. These conservative estimates of non-closure are added to the total systematic uncertainty in quadrature.

Similarly, all the points that are below 0 are added in quadrature to find the lower bound.
\clearpage

\subsection{For $v_{n}$-$[\pT]$ Correlation}\label{sec:sysvnpt}

The systematic uncertainties for the covariance $\mathrm{cov}(v_n\{2\}^2,[\pT])$, dynamic variance $\mathrm{Var}(v_n\{2\}^2)_{\mathrm{dyn}}$, correlation coefficient $\rho(v_n\{2\}^2,[\pT])$ for $n=2, 3, 4$, and the two-particle $p_T$ correlator $c_k=\lr{\delta\pT\delta\pT}$ are also considered. The primary sources of systematic uncertainty include track quality selection, tracking efficiency, and centrality definition. Additionally, the $\phi$-flattening procedure introduces a systematic uncertainty, which is assessed by comparing results with and without this correction. The HIJING model is used as a cross-check to evaluate potential biases from short-range correlations and detector effects. The HIJING MC studies indicate negligible covariances between $v_2\{2\}^2$ and $[\pT]$, and the dynamic variance of $v_n\{2\}^2$ is also found to be small in the model. The $p_T$ cumulants $c_k$ show qualitative agreement between data and HIJING predictions. The systematic effect from pileup contamination is estimated to be less than $0.5\%$ for these observables across all centralities.

\subsubsection{In Pb+Pb Collisions}
\begin{figure}[htbp]
\centering
\includegraphics[width=0.32\linewidth]{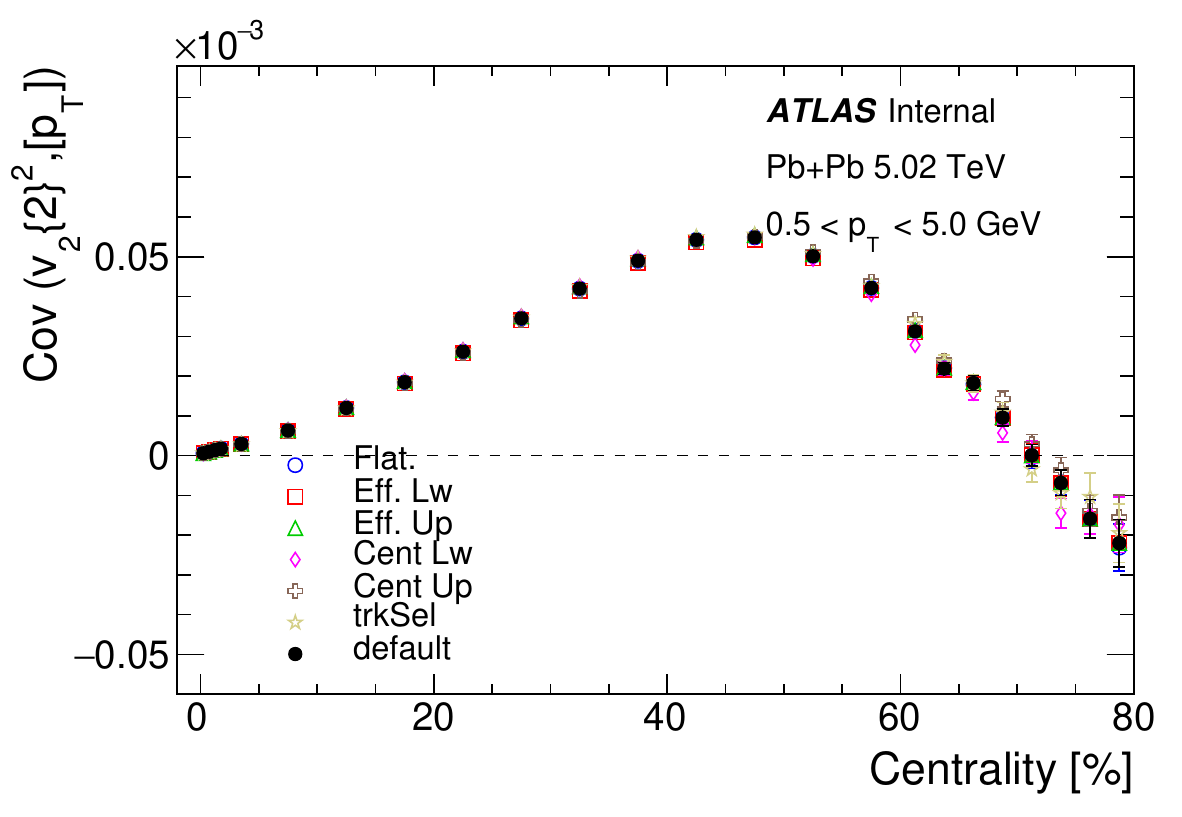}
\includegraphics[width=0.32\linewidth]{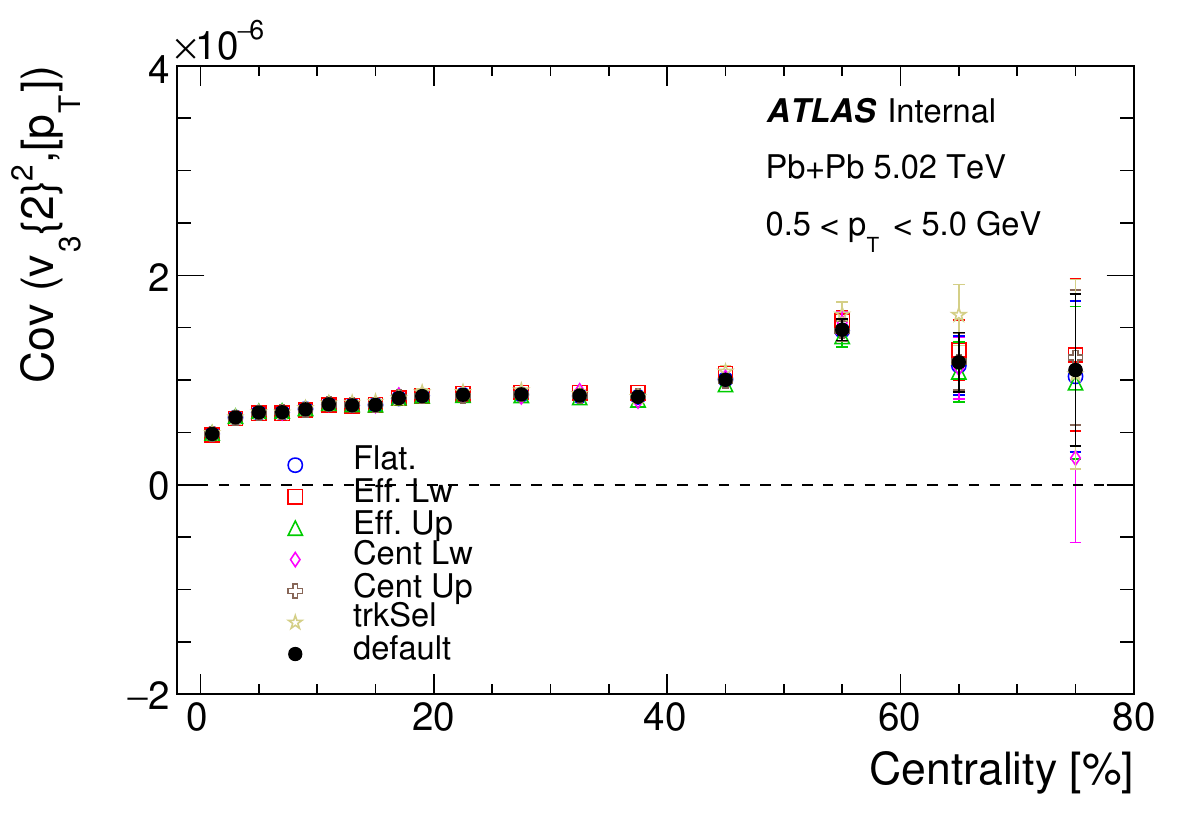}
\includegraphics[width=0.32\linewidth]{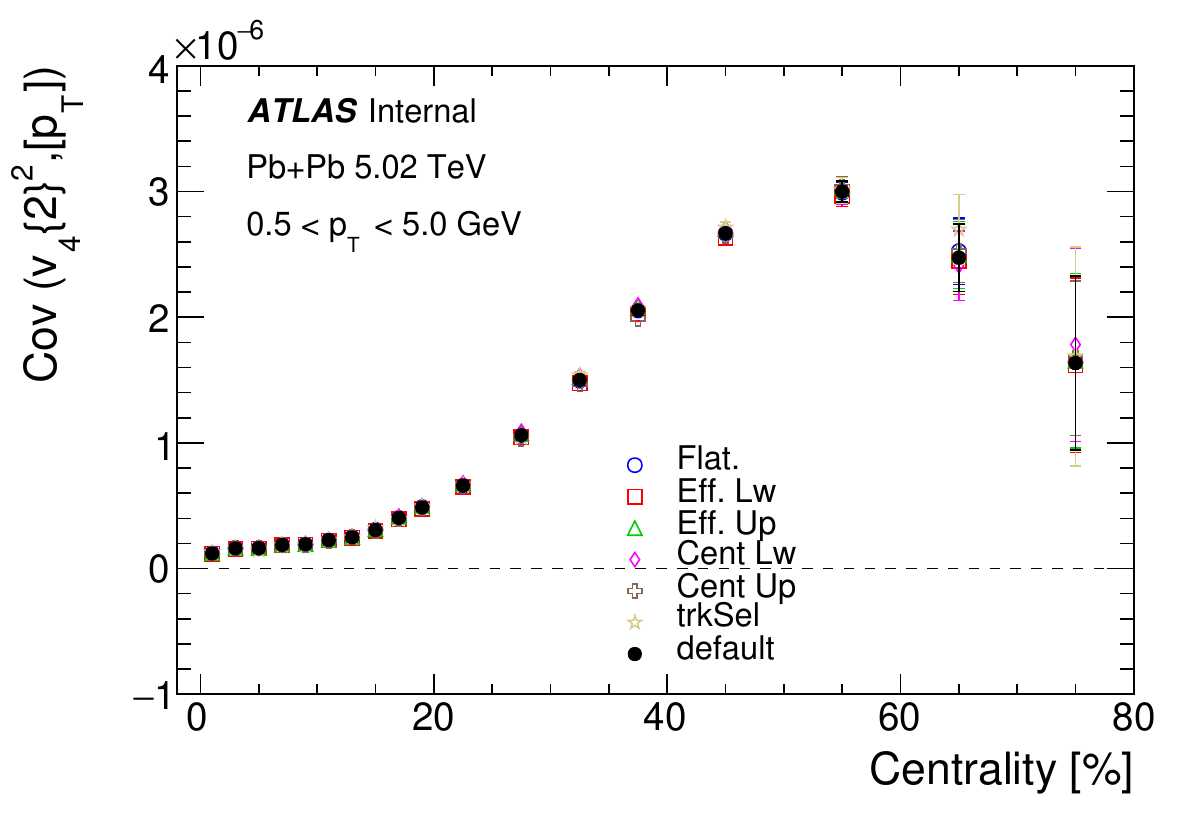}
\includegraphics[width=0.32\linewidth]{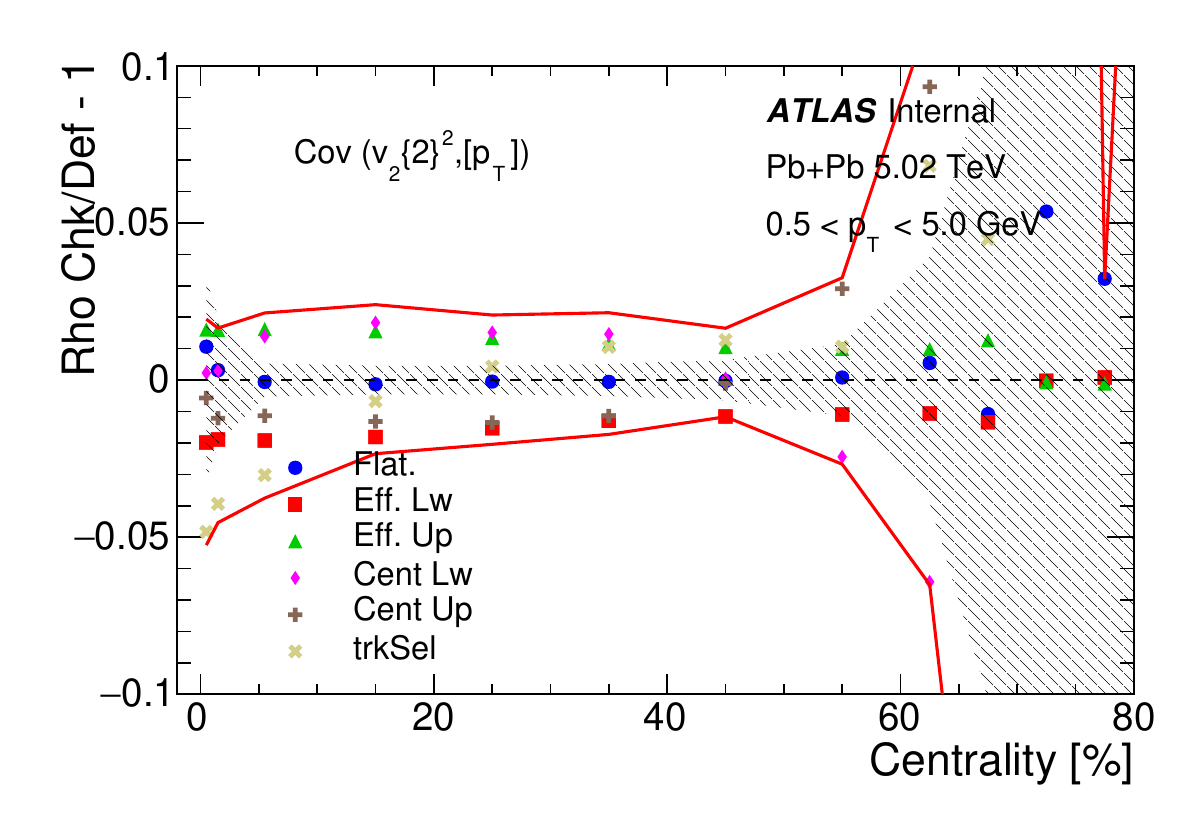}
\includegraphics[width=0.32\linewidth]{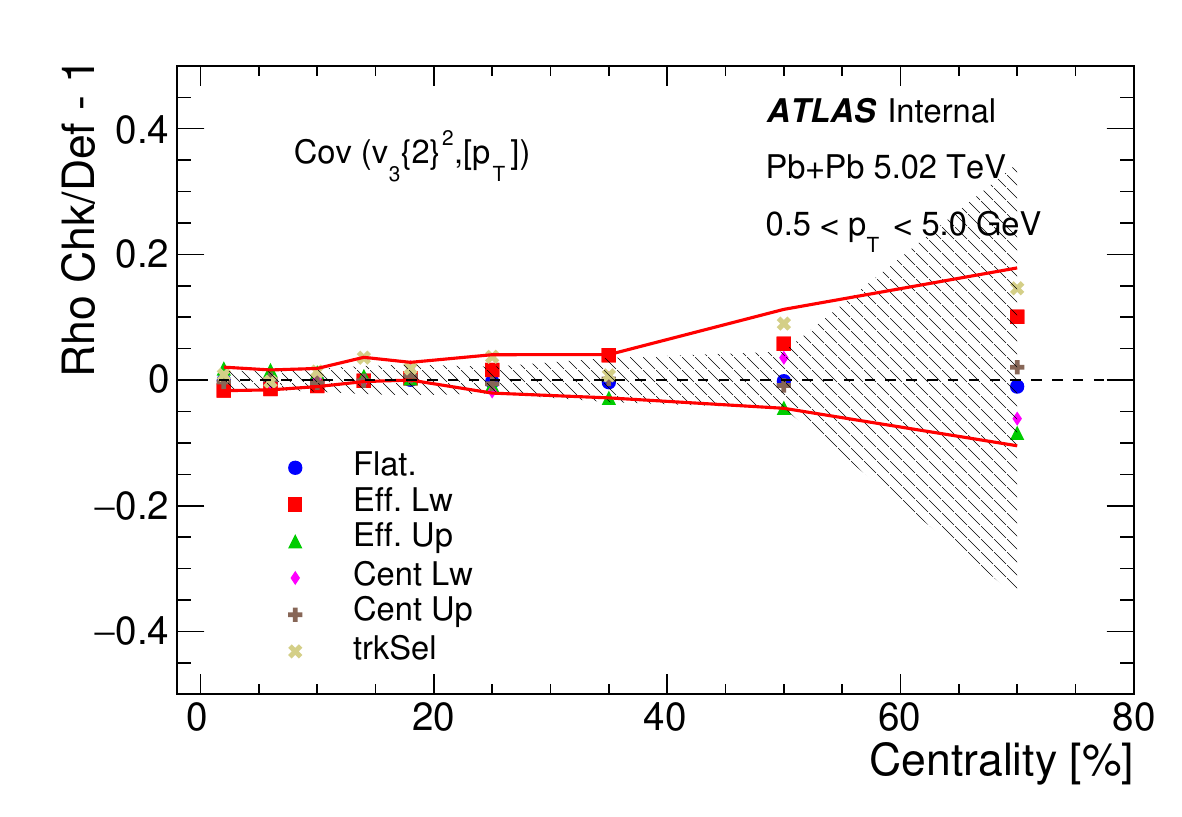}
\includegraphics[width=0.32\linewidth]{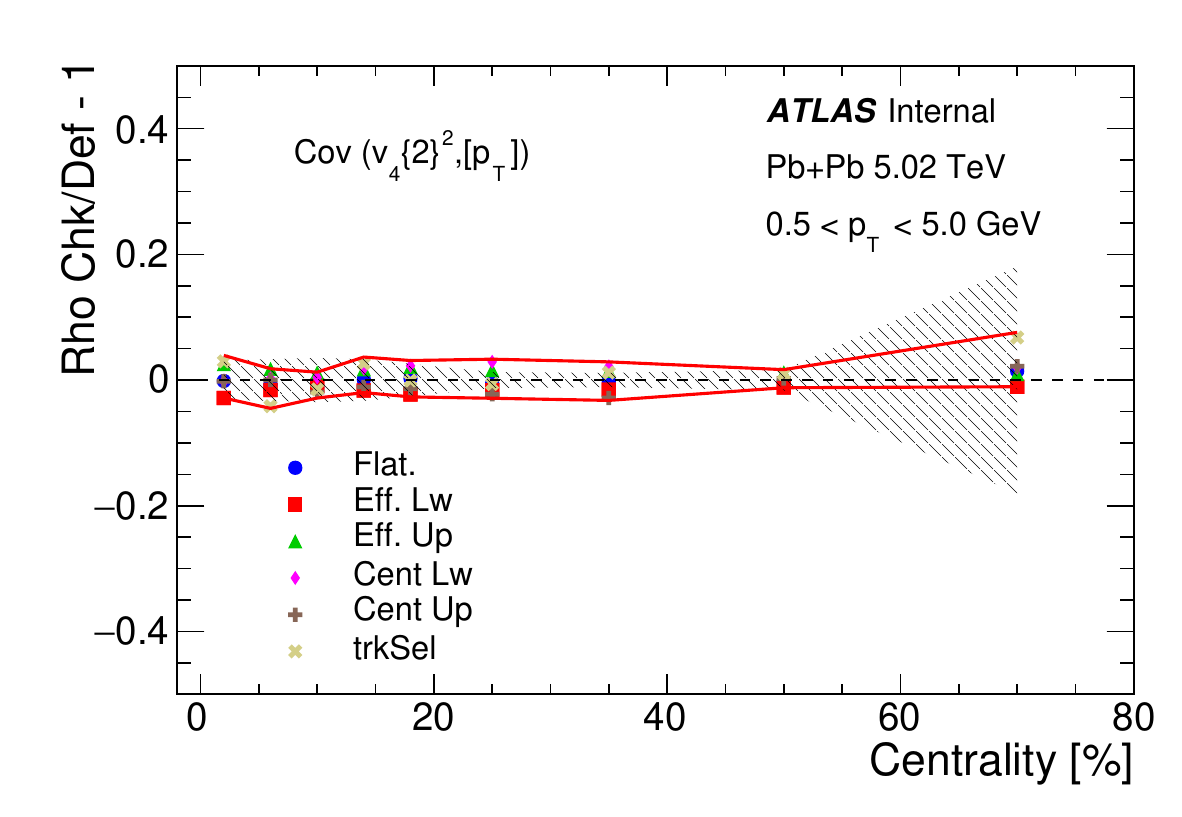}
\caption{(Top row) Comparison of $\mathrm{cov}(v_n\{2\}^2,[\pT])$ between different systematic checks  and default for $n=2$, 3 and 4. The error bars represent statistical uncertainties. (Bottom row) Comparison of the relative difference between systematic checks and default for $n=2$, 3 and 4. The shaded area represents the ratio of statistical uncertainty of the default to its mean value. The lines are the combined systematic uncertainty.}
\label{fig:Cov_sys_Pb}
\end{figure}

\begin{figure}[htbp]
\centering
\includegraphics[width=0.32\linewidth]{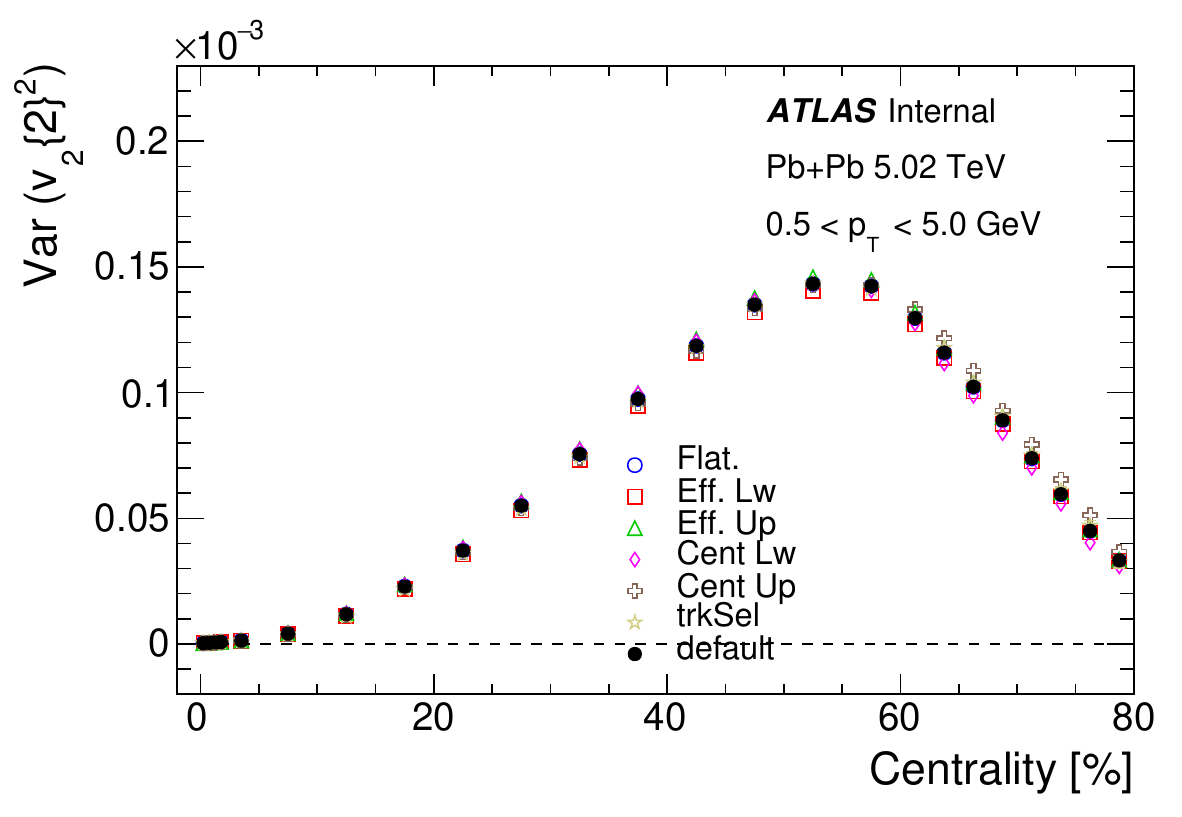}
\includegraphics[width=0.32\linewidth]{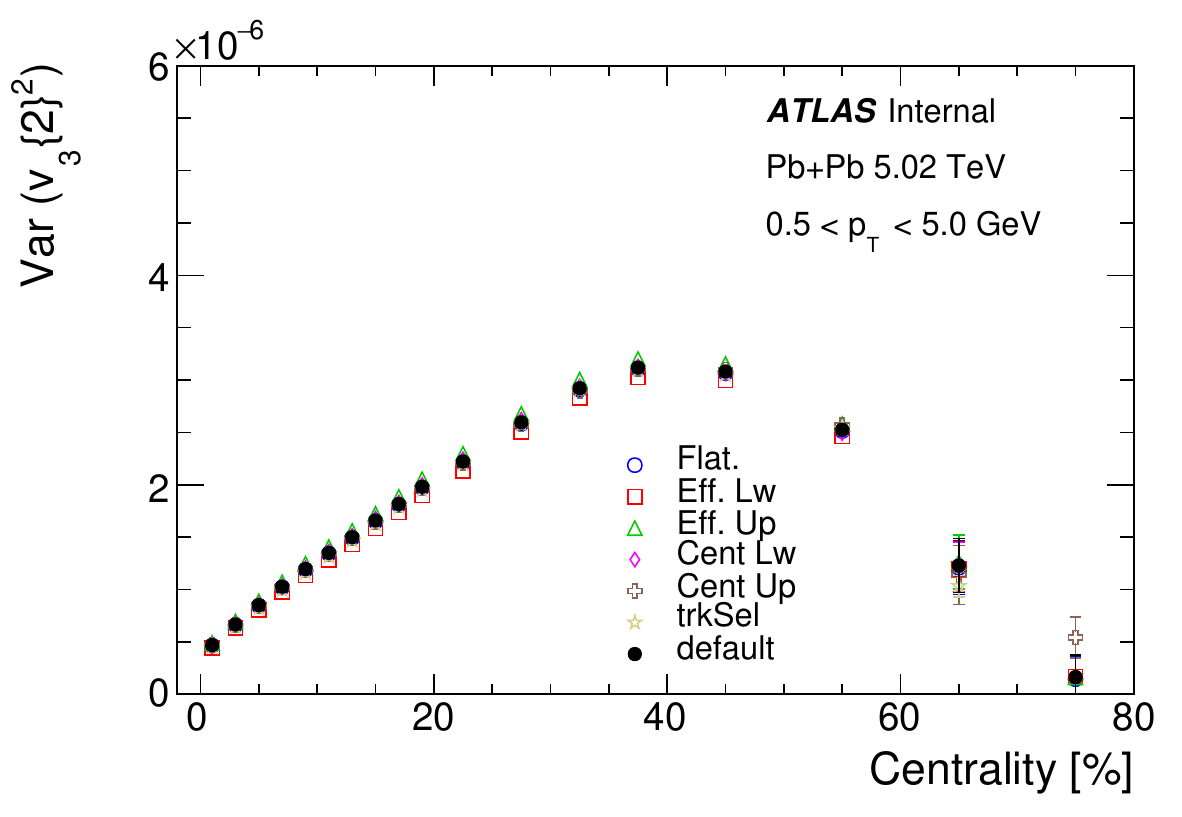}
\includegraphics[width=0.32\linewidth]{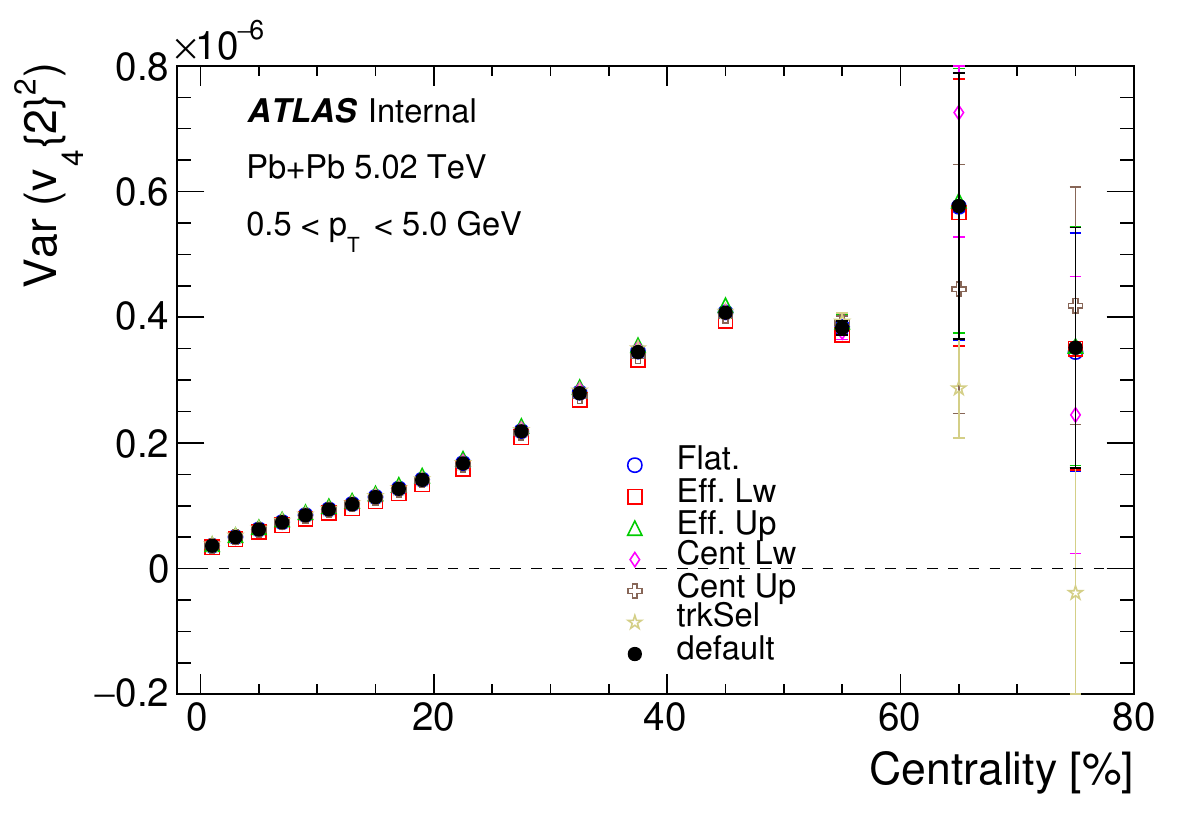}
\includegraphics[width=0.32\linewidth]{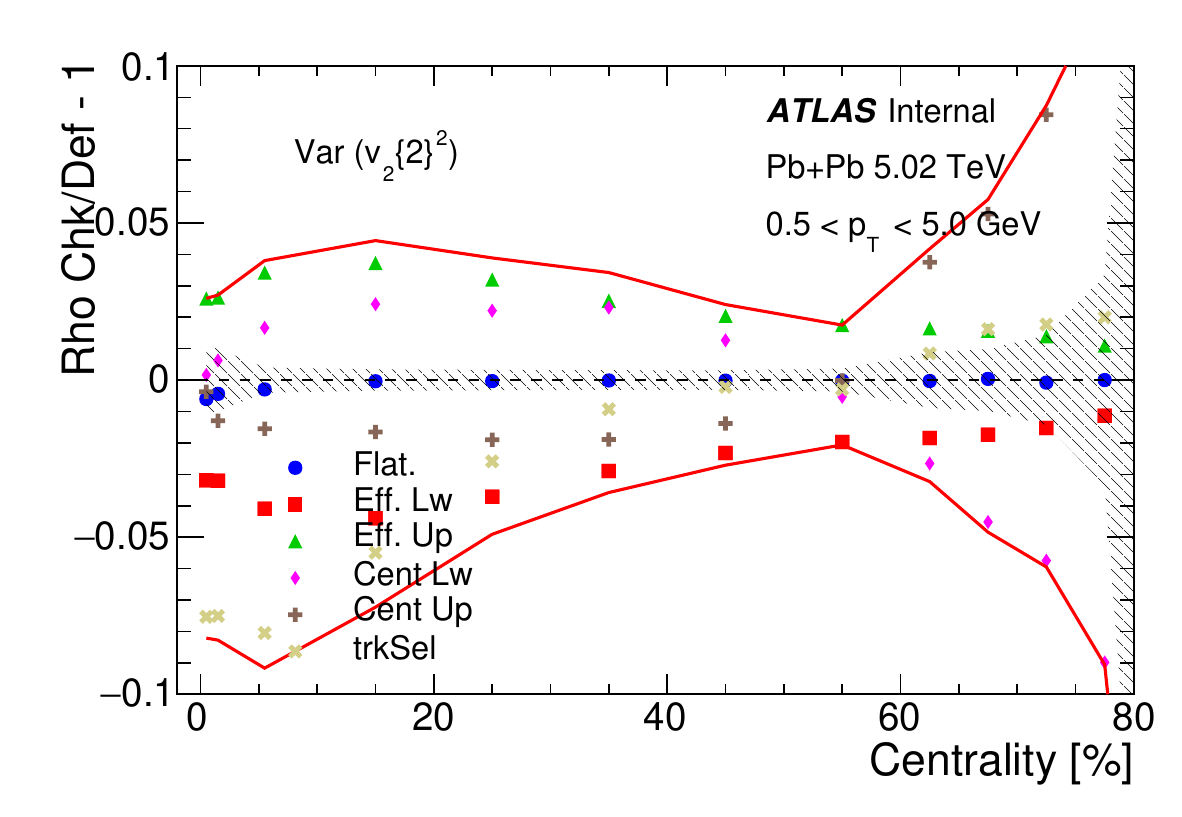}
\includegraphics[width=0.32\linewidth]{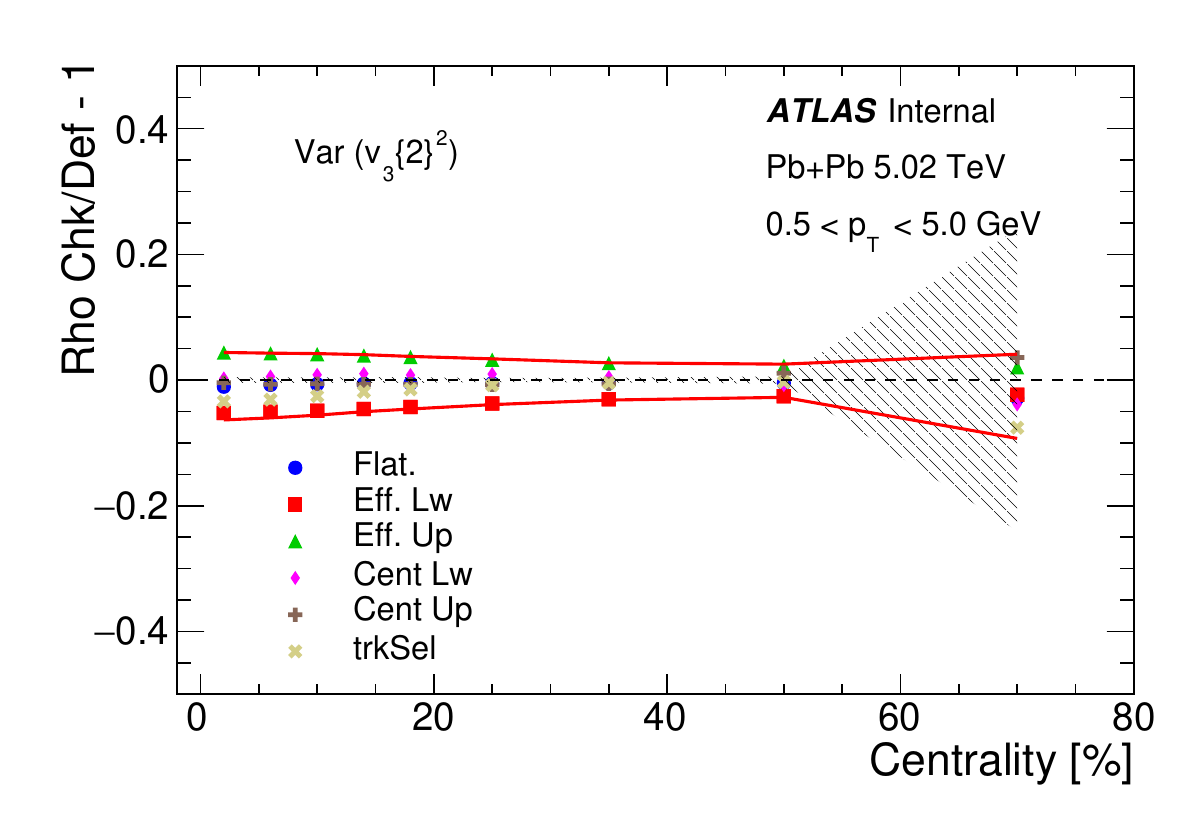}
\includegraphics[width=0.32\linewidth]{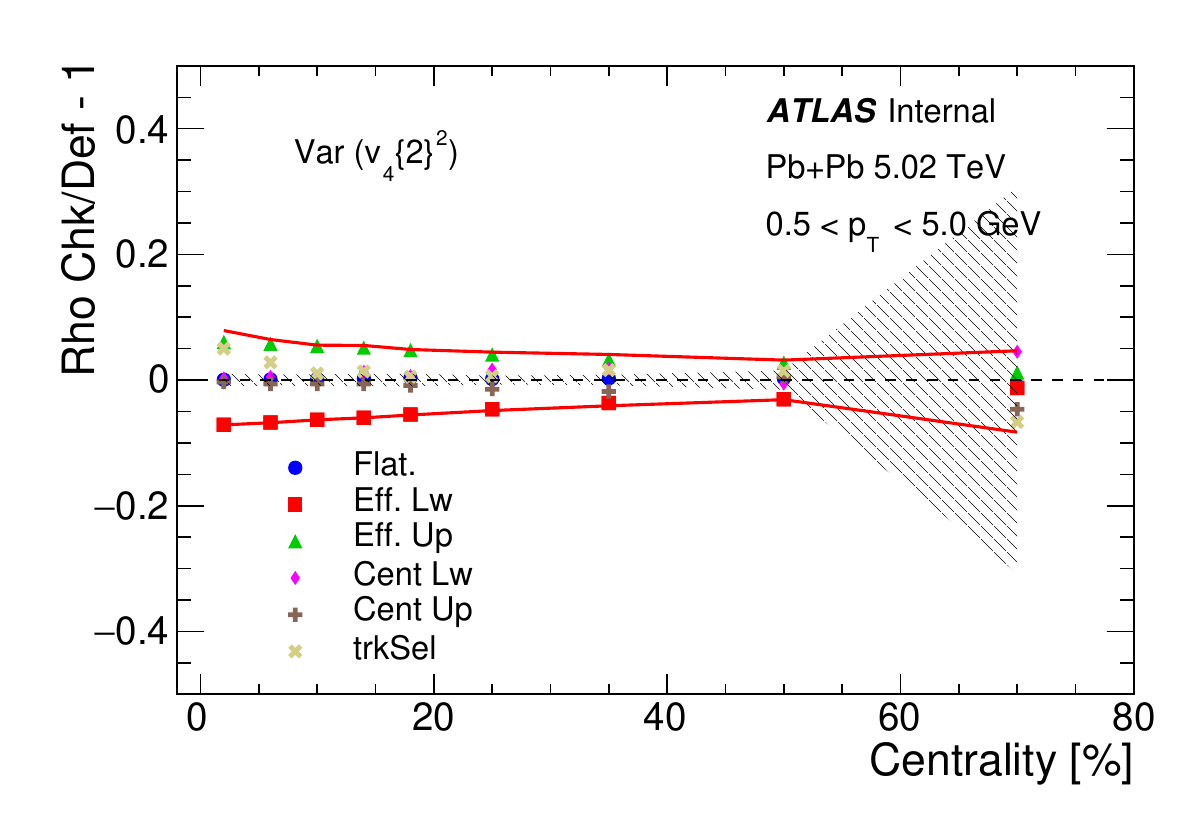}
\caption{(Top row) Comparison of $\mathrm{Var}(v_n\{2\}^2)_{\mathrm{dyn}}$ between different systematic checks and default for $n=2$, 3 and 4. The error bars represent statistical uncertainties. (Bottom row) Comparison of the relative difference between systematic checks and default for $n=2$, 3 and 4. The shaded area represents the ratio of statistical uncertainty of the default to its mean value. The lines are the combined systematic uncertainty.}
\label{fig:Var_sys_Pb}
\end{figure}

Figure~\ref{fig:Cov_sys_Pb} shows the comparison of different systematic checks with the default case of $\mathrm{cov}(v_n\{2\}^2,[\pT])$ for $n=2$, 3 and 4 in Pb+Pb. The bottom panels show relative difference w.r.t default along with the lower and upper bounds shown in lines and the ratio of statistical uncertainty of the default to its mean value shown as shaded region. The systematic variation is within 2--$5\%$ up to central and mid-central collisions. In the peripheral region, due to the sign-change for $n=2$, the relative uncertainty become very large. The systematic uncertainty are larger than statistical uncertainties for $n=2$, but are comparable or smaller for $n=3$ and 4.
Figure~\ref{fig:Var_sys_Pb} shows the comparison of different systematic checks with the default case of $\mathrm{Var}(v_n\{2\}^2)_{\mathrm{dyn}}$ for $n=2$, 3 and 4 in Pb+Pb. The bottom panels show relative difference w.r.t default along with the lower and upper bounds shown in lines and the ratio of statistical uncertainty of the default to its mean value shown as shaded region. The systematic variation is within 3--$18\%$ for $n=2$ and 3 over the entire centrality range. For n=4, the rise in systematics in peripheral region mostly arises from track selection. The systematics are dominated by uncertainty in the tracking efficiency and track selection in central and mid-central region. The uncertainty due to centrality is important in the peripheral region. The systematic uncertainty are generally larger than the statistical uncertainties for all harmonics. 

\begin{figure}[htbp]
\centering
\includegraphics[width=0.4\linewidth]{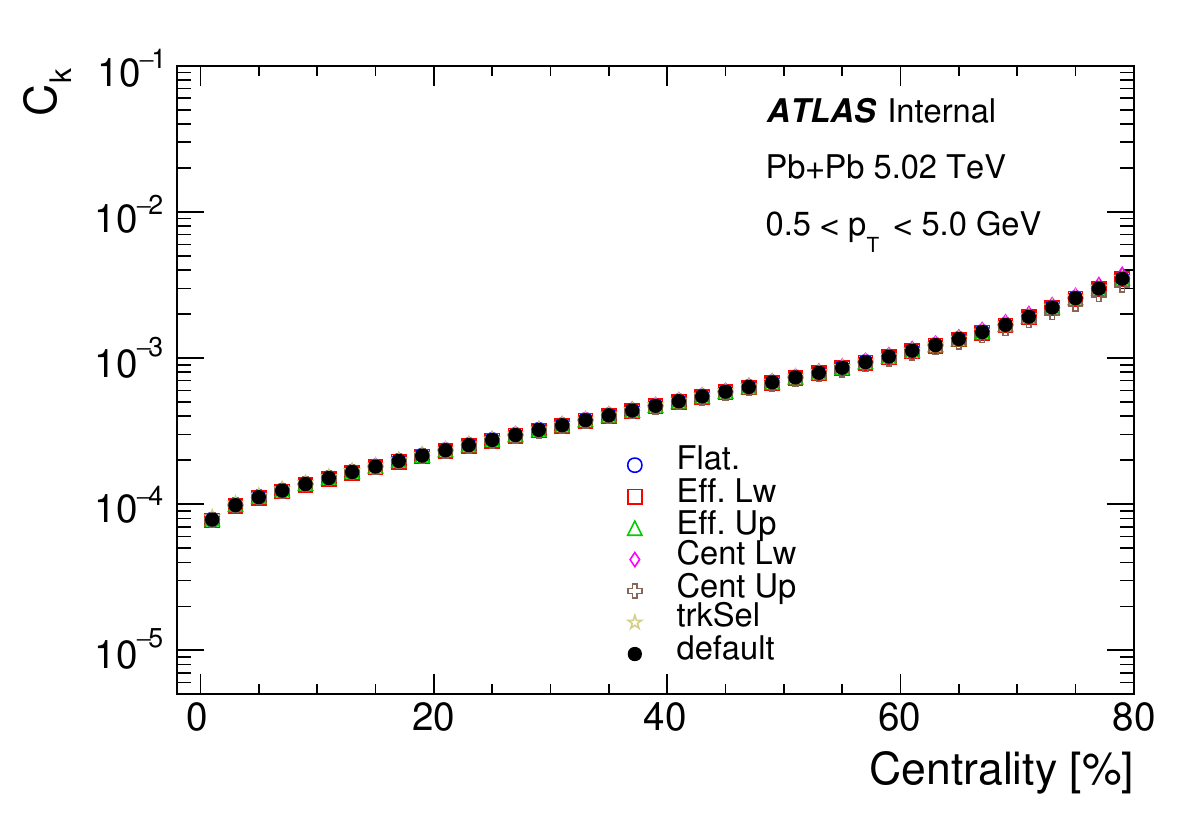}\includegraphics[width=0.4\linewidth]{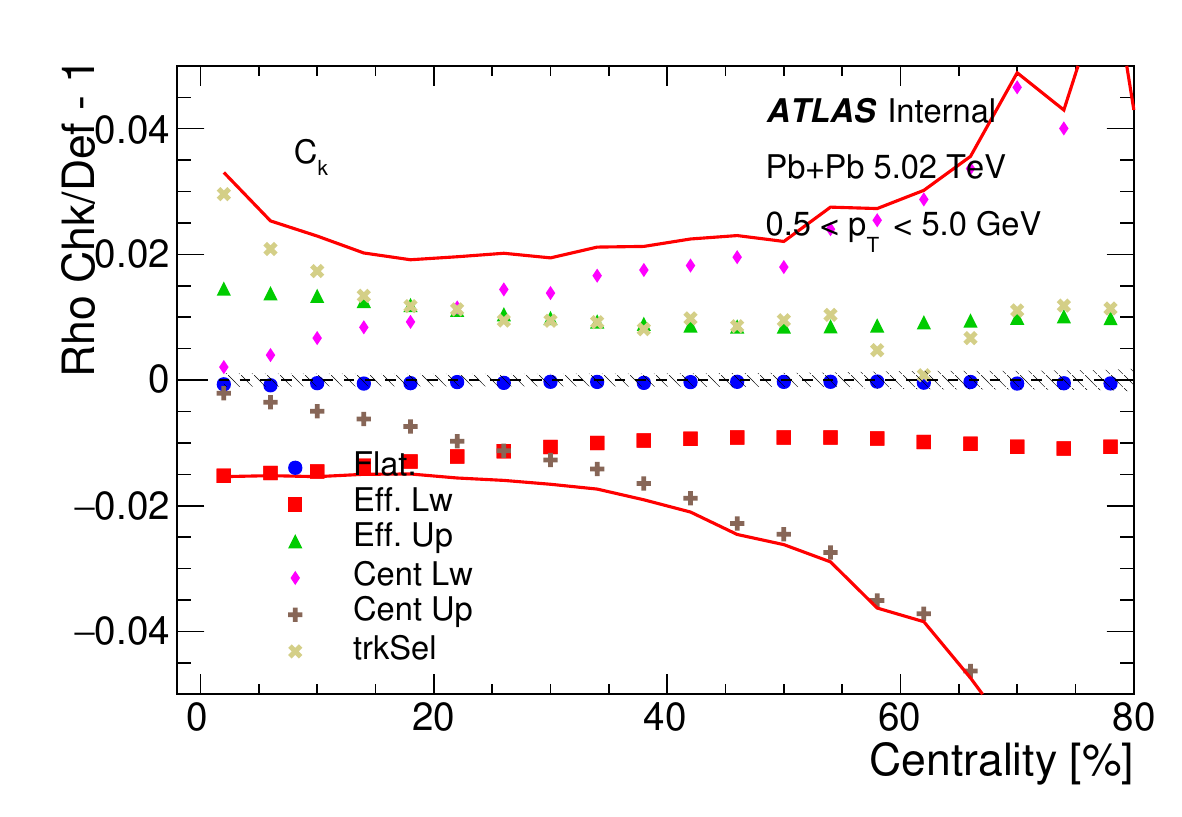}
\caption{(Left panel) Comparison of $[\pT]$ variance $c_k=\lr{\delta\pT\delta\pT}$ between different systematic checks and default. The error bars represent statistical uncertainties. (Right panel) Comparison of the relative difference between systematic checks and default for $n=2$, 3 and 4. The shaded area represents the ratio of statistical uncertainty of the default to its mean value. The lines are the combined systematic uncertainty.}
\label{fig:Ck_sys_Pb}
\end{figure}

Figure~\ref{fig:Ck_sys_Pb} shows the comparison of different systematic checks with the default for variance of $[\pT]$ fluctuations, $c_k$. The systematic uncertainty is 2--5\% from central to peripheral collisions, and is dominated by track selection and centrality definition.

\begin{figure}[htbp]
\centering
\includegraphics[width=0.32\linewidth]{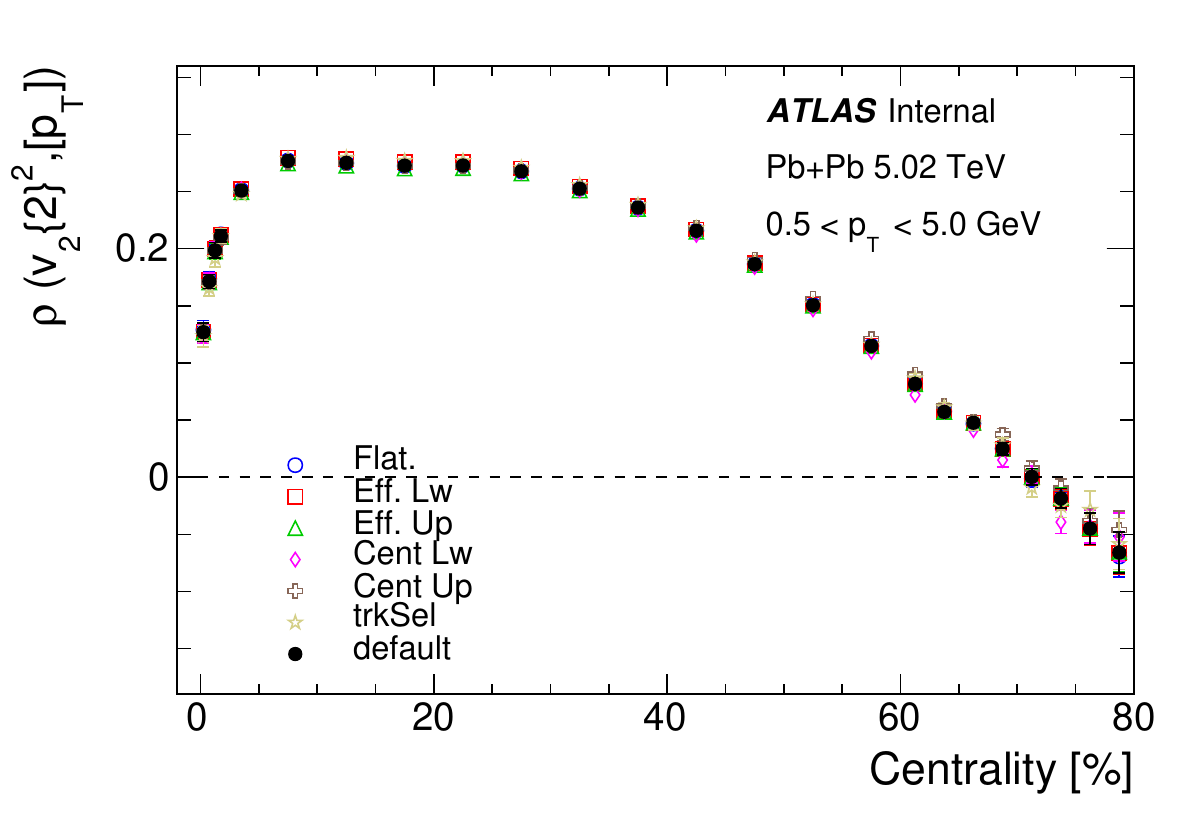}
\includegraphics[width=0.32\linewidth]{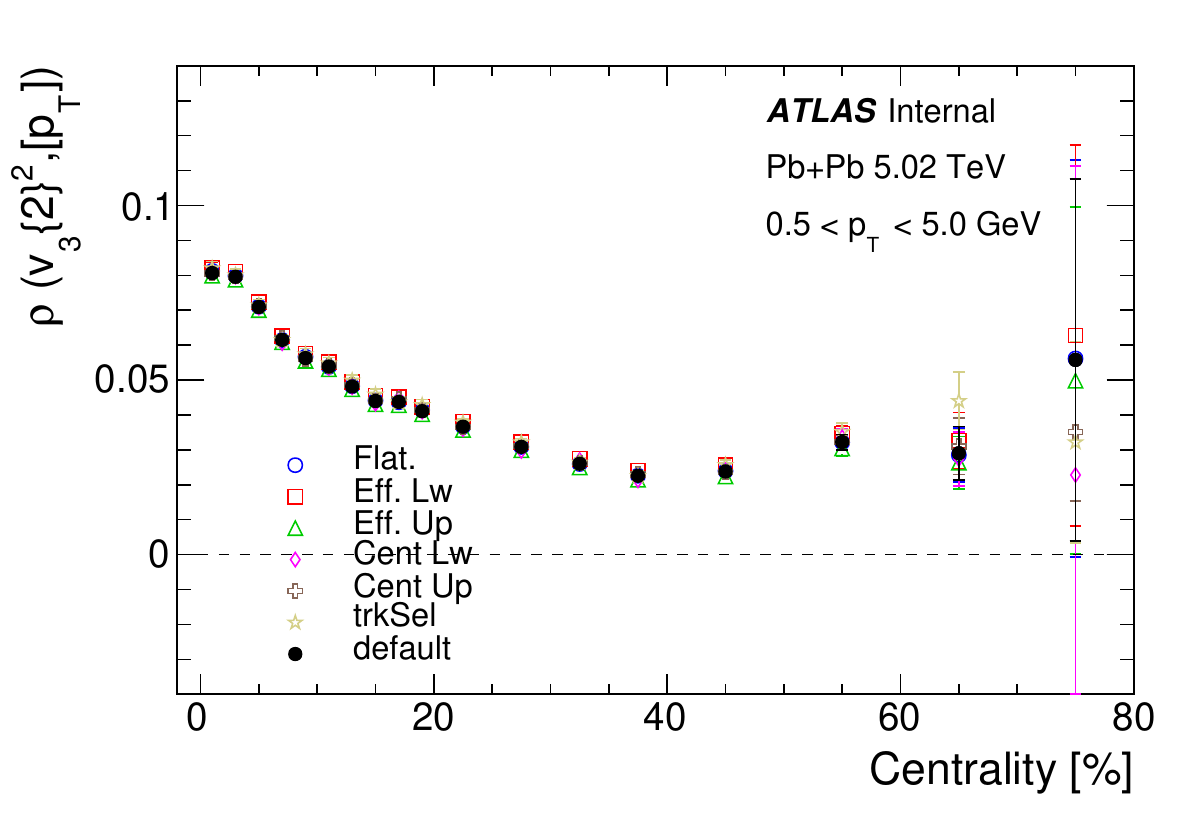}
\includegraphics[width=0.32\linewidth]{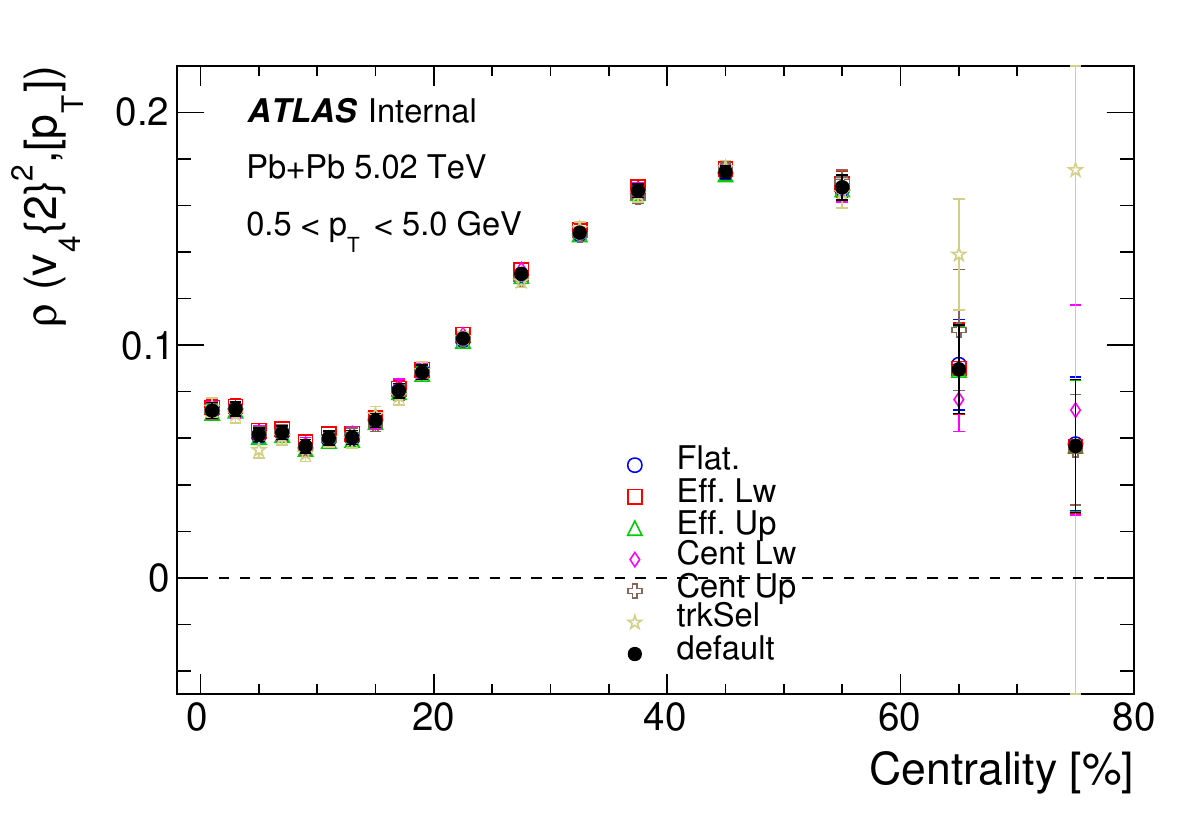}
\includegraphics[width=0.32\linewidth]{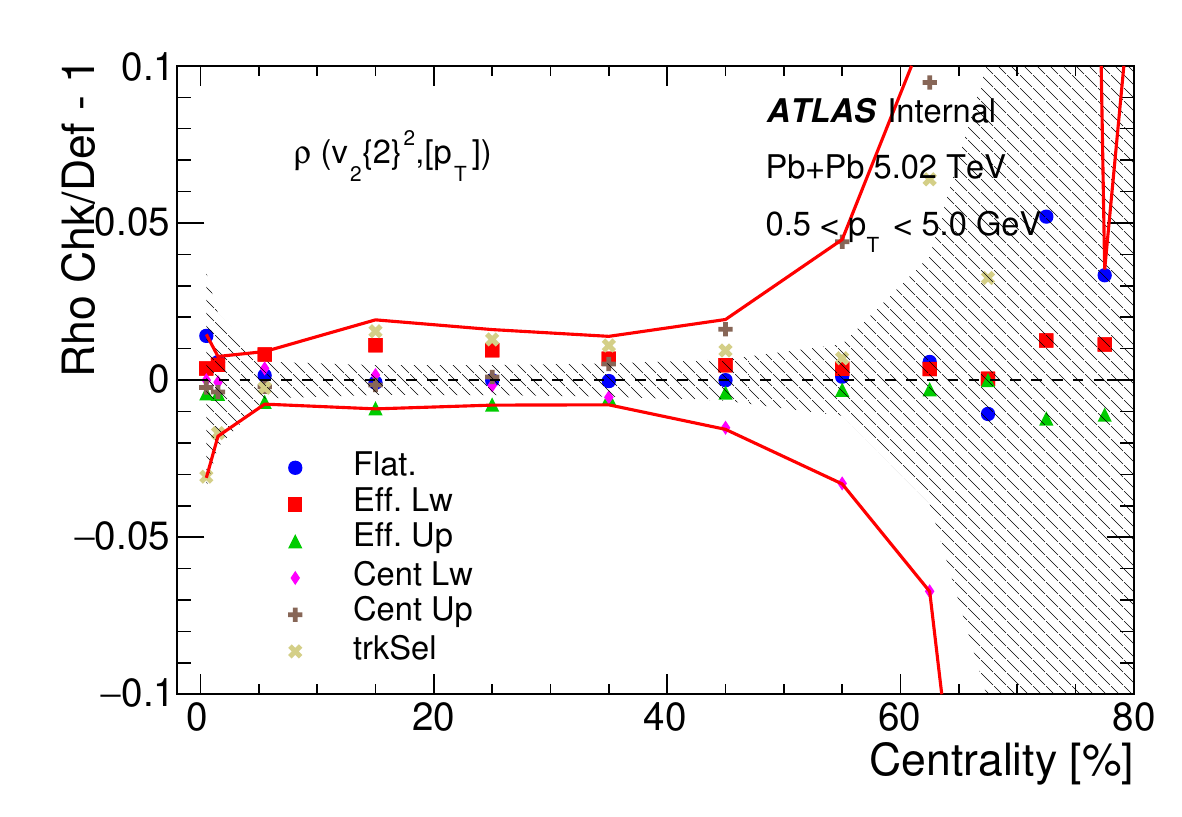}
\includegraphics[width=0.32\linewidth]{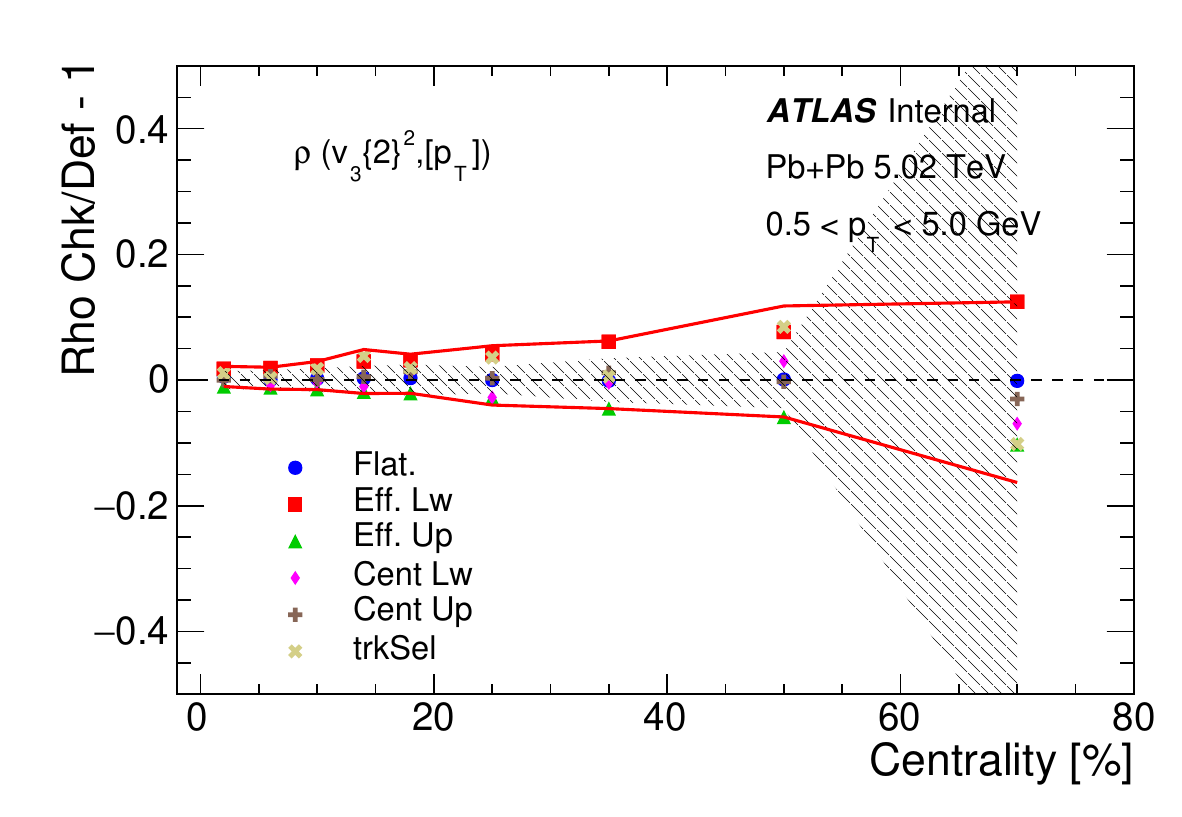}
\includegraphics[width=0.32\linewidth]{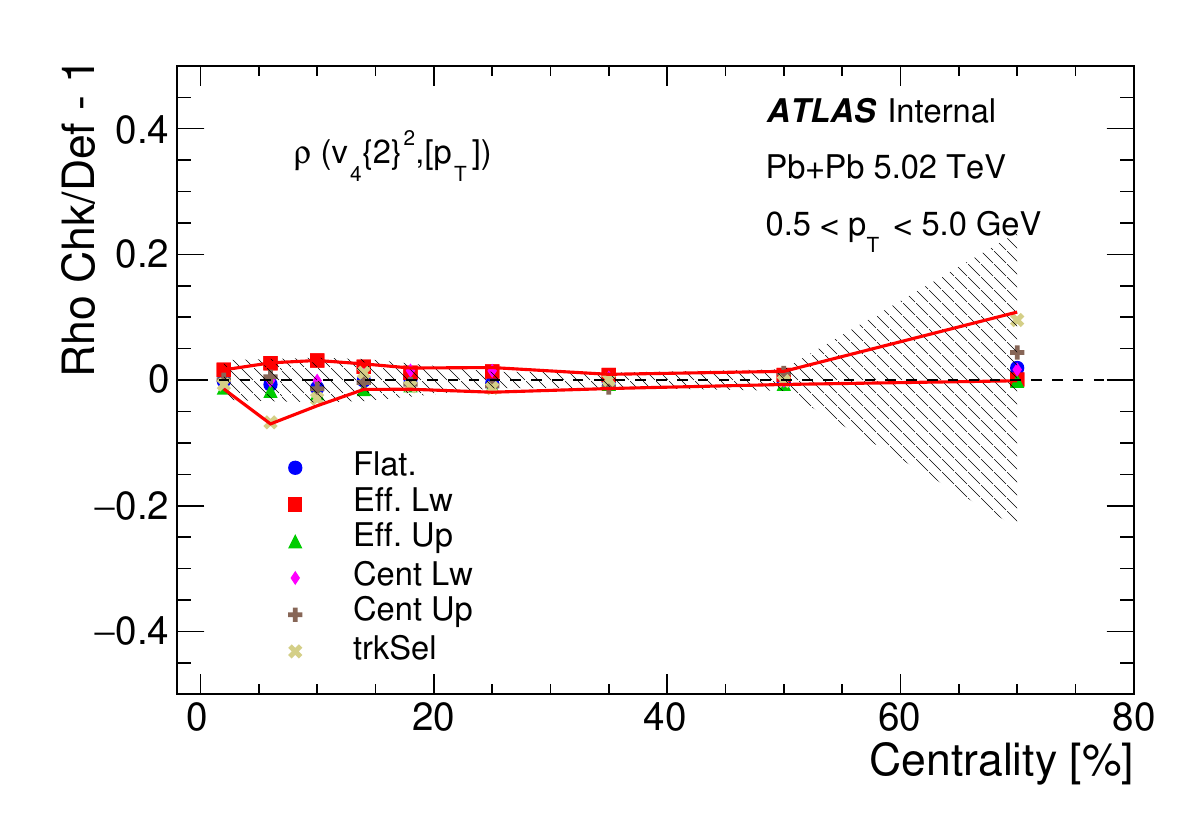}
\caption{(Top row)Comparison of $\rho(v_n\{2\}^2,[\pT])$ between different systematic checks and default for $n=2$, 3 and 4. The error bars represent statistical uncertainties. (Bottom row) Comparison of the relative difference between systematic checks and default for $n=2$, 3 and 4. The shaded area represents the ratio of statistical uncertainty of the default to its mean value. The lines are the combined systematic uncertainty.}
\label{fig:Rho_sys_Pb}
\end{figure}

Figure~\ref{fig:Rho_sys_Pb} shows the comparison of different systematic checks with the default case of $\rho(v_n\{2\}^2,[\pT])$ for $n=2$, 3 and 4 in Pb+Pb. The bottom panels show relative difference w.r.t default along with the lower and upper bounds shown in lines and the ratio of statistical uncertainty of the default to its mean value shown as shaded region. The systematic uncertainty partially cancel between the numerator and denominator. Therefore the final systematic uncertainty has weaker centrality dependence, especially for $n=2$. The uncertainty is less than 10\% for $n=2$ in central and mid-central collisions.
\clearpage

For all centrality ranges and for all observables, the flattening has the minimum contribution to systematics. For $n=2$, the dominant source of systematics arises from track quality variation and efficiency correlation for all observables. For $n=3$ and 4 the systematic uncertainty coming from efficiency variations becomes just as large if not greater in some cases. Uncertainty on the centrality percentile has largest impact in the most peripheral region. 

The major source of systematic in measurement of $\rho(v_n\{2\}^2,[\pT])$ arises from the contribution from the $\mathrm{cov}(v_n\{2\}^2,[\pT])$. The uncertainty in the tracking efficiency and track selection cuts also partially cancels. For $\mathrm{cov}(v_2\{2\}^2,[\pT])$, the systematics are within 5\% for all centrality. For $\mathrm{cov}(v_3\{2\}^2,[\pT])$ and $\mathrm{cov}(v_4\{2\}^2,[\pT])$, the systematics are within 5\% for central and mid-central events. 

In the more peripheral collisions, the values of the $v_n-[\pT]$ correlator either has very large statistical error or they changes sign. Nevertheless, the full systematic uncertainties are shown on all data points. 

\clearpage

\subsubsection{In Xe+Xe Collisions}
The systematics in Xe+Xe collisions are carried out in very similar manner. However due to the limited event statistics, the systematic uncertainty has significant statistical fluctuation, which require some smoothing as a function centrality. The relative uncertainty is calculated between the standard measurement and the varied checks defined by $(varied/nominal)-1$ which is the figure of merit for the systematics checks. 

\begin{figure}[htbp]
\centering
\includegraphics[width=0.32\linewidth]{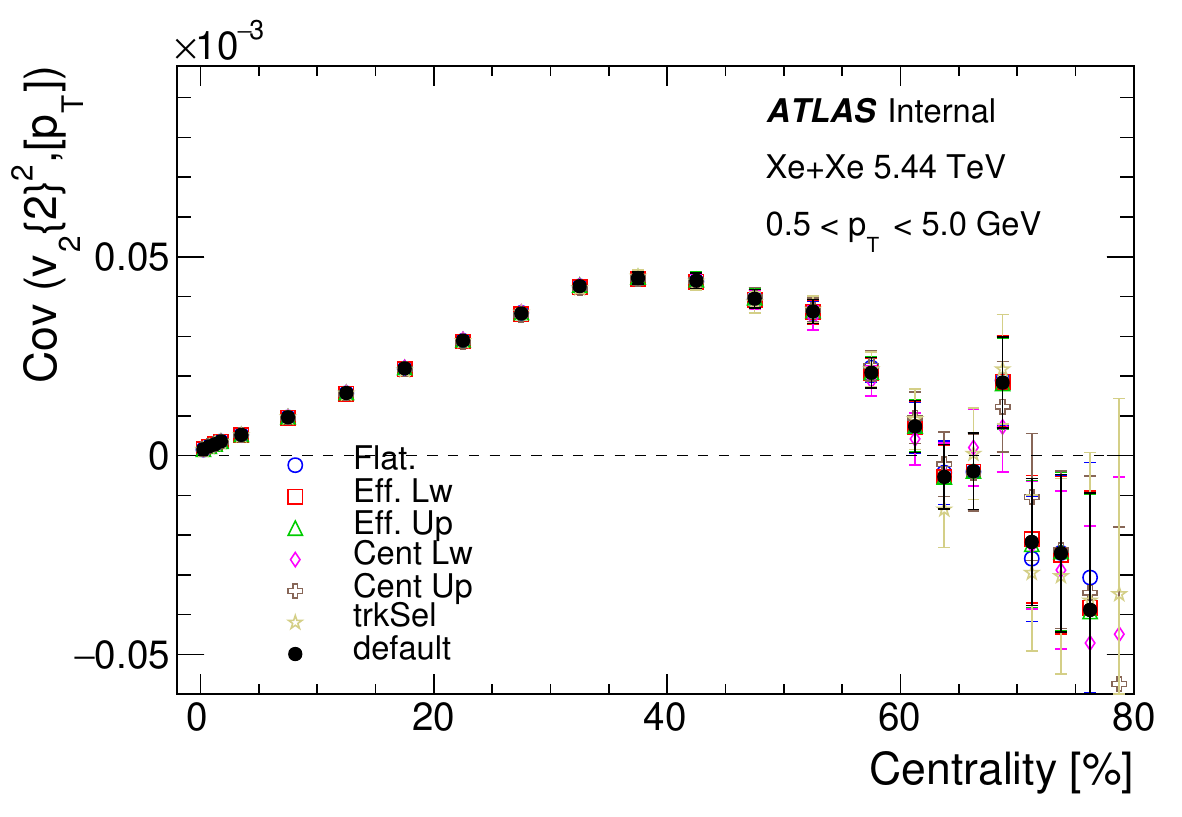}
\includegraphics[width=0.32\linewidth]{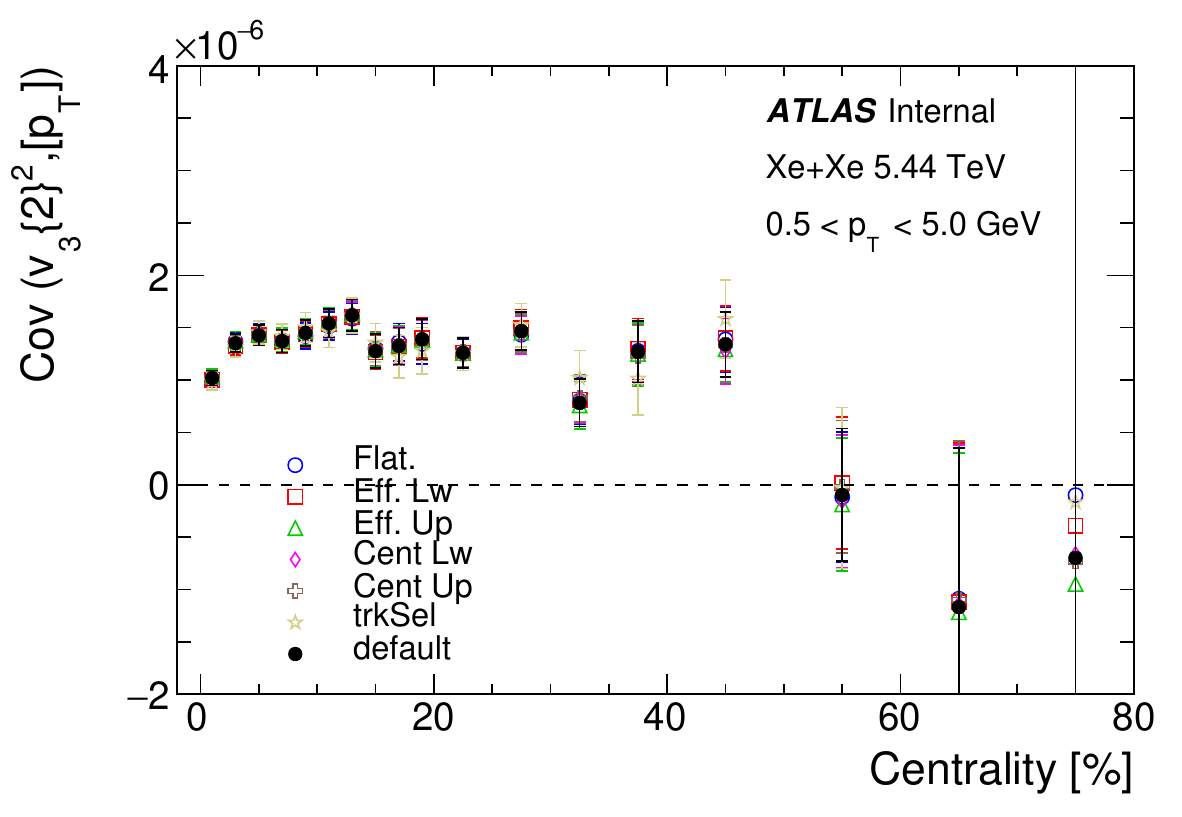}
\includegraphics[width=0.32\linewidth]{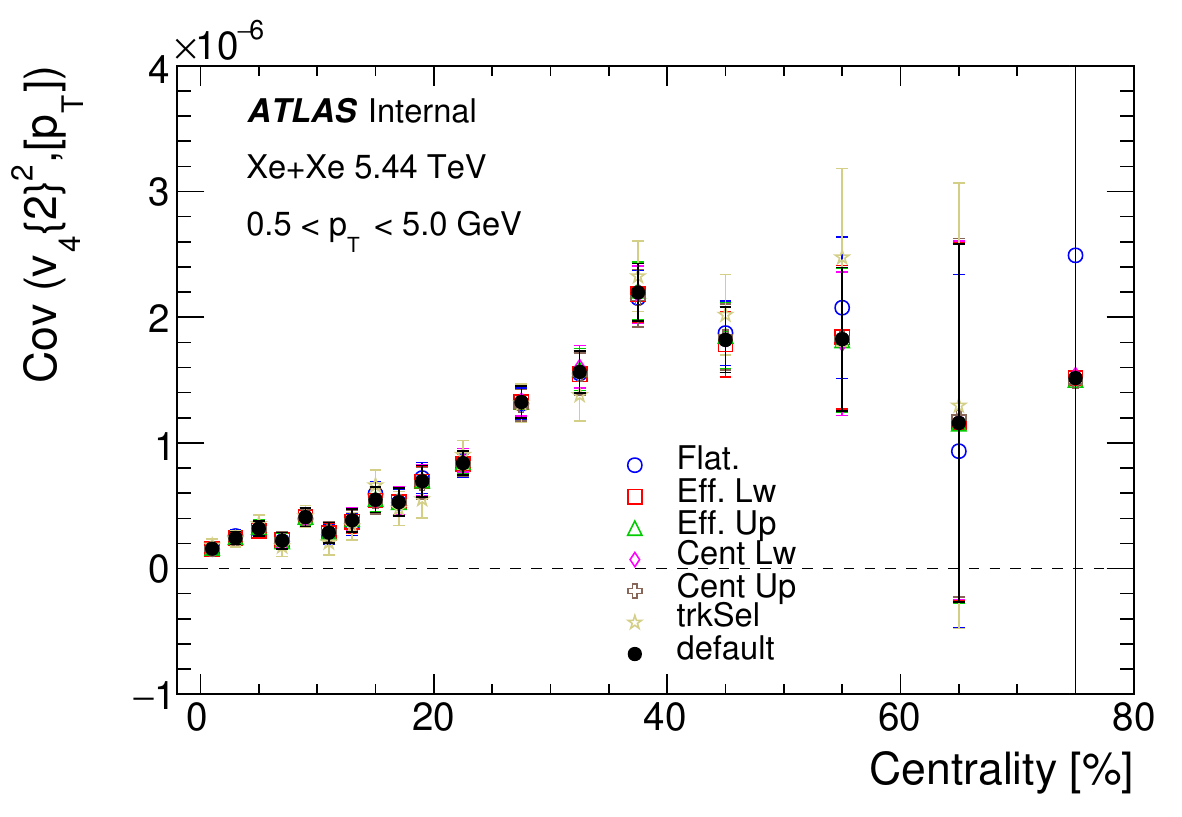}
\includegraphics[width=0.32\linewidth]{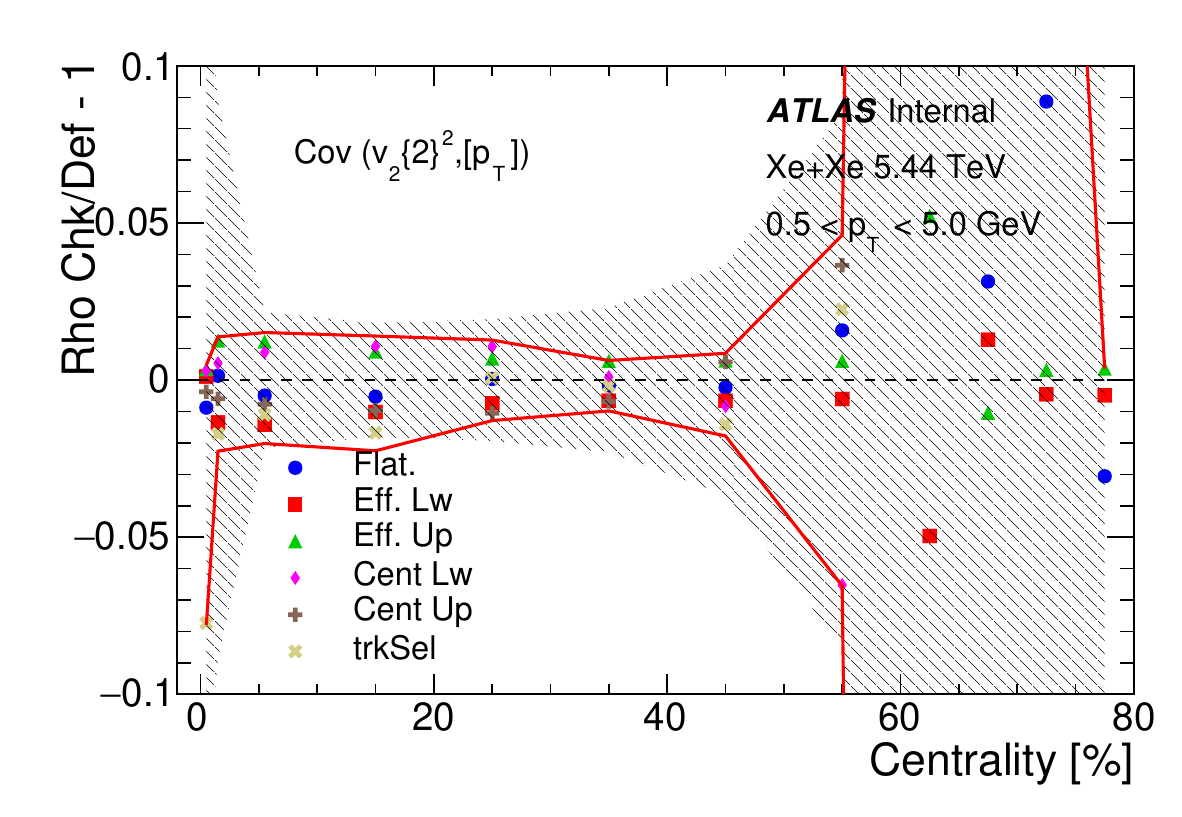}
\includegraphics[width=0.32\linewidth]{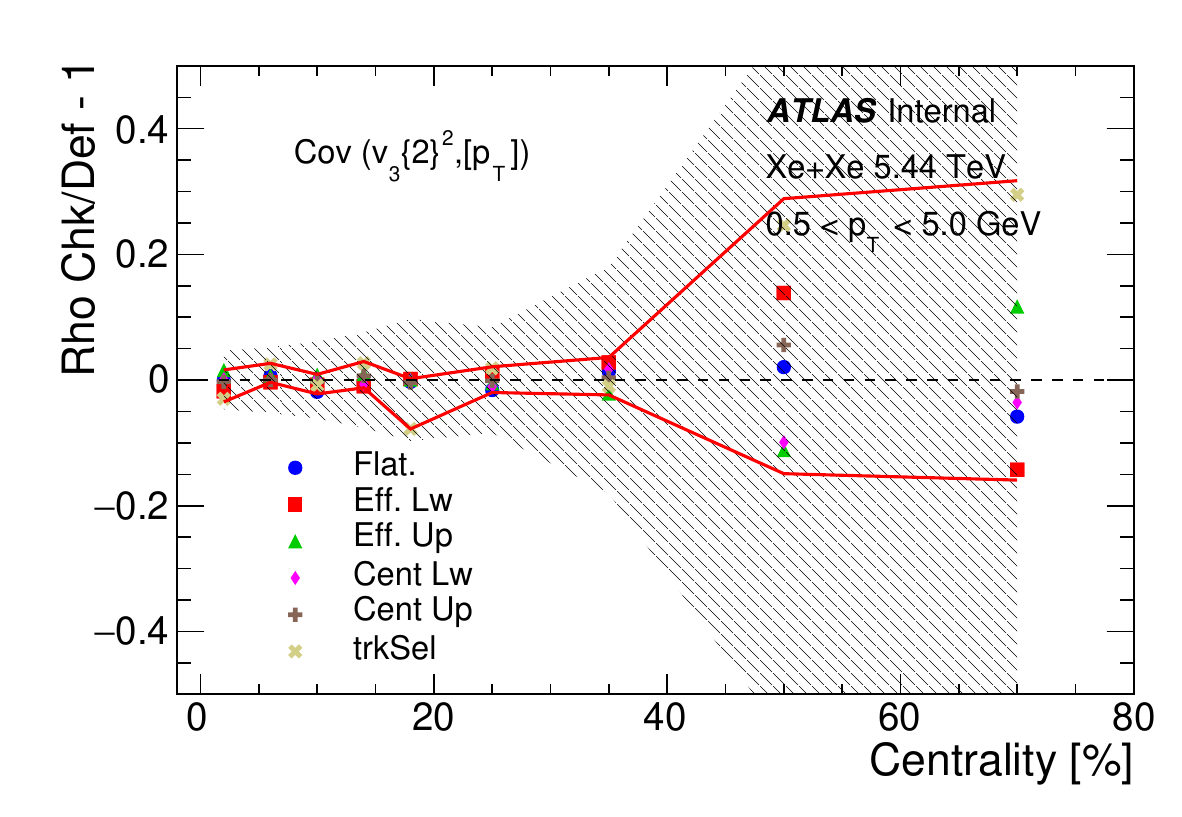}
\includegraphics[width=0.32\linewidth]{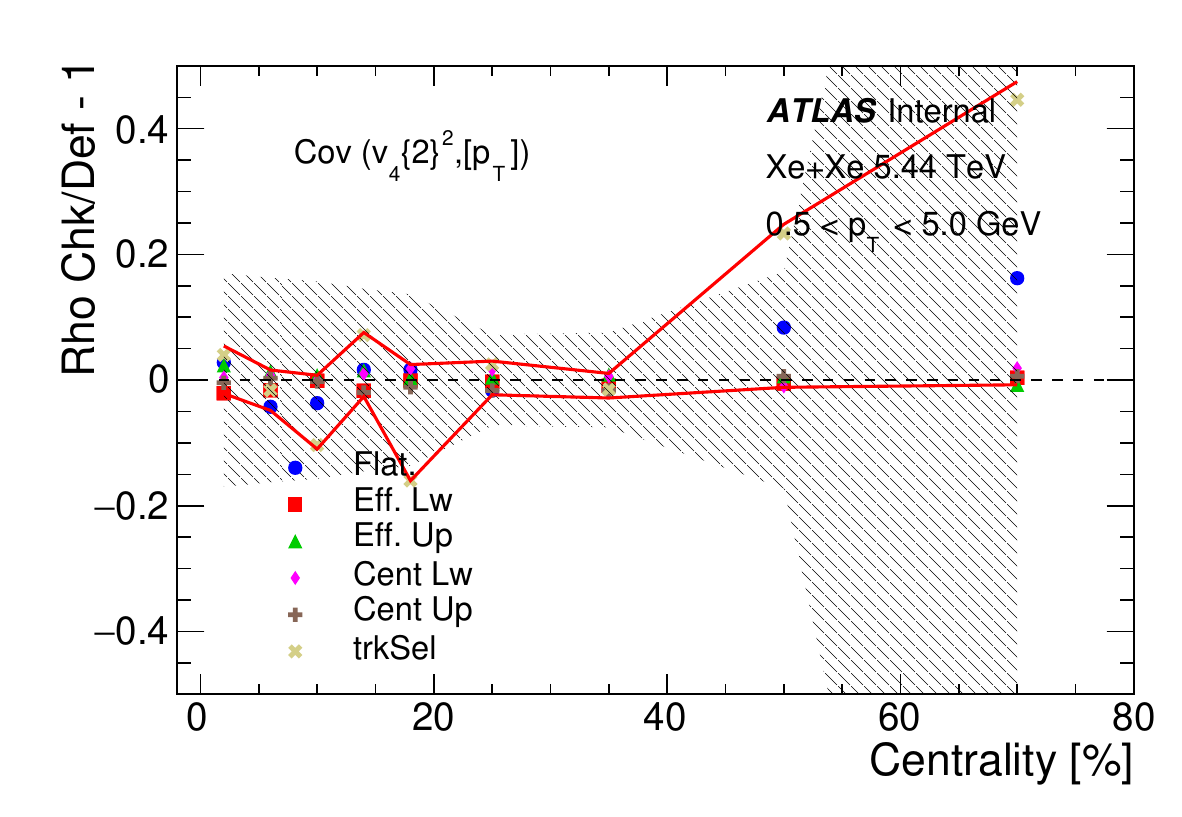}
\caption{(Top row) Comparison of $\mathrm{cov}(v_n\{2\}^2,[\pT])$ between different systematic checks and default for $n=2$, 3 and 4. The error bars represent statistical uncertainties. (Bottom row) Comparison of the relative difference between systematic checks and default for $n=2$, 3 and 4. The shaded area represents the ratio of statistical uncertainty of the default to its mean value. The lines are the combined systematic uncertainty.}
\label{fig:Cov_sys_Xe}
\end{figure}

Figure~\ref{fig:Cov_sys_Xe} shows the comparison of different systematic checks with the default case of $\mathrm{cov}(v_n\{2\}^2,[\pT])$  for $n=2$, 3 and 4 in Xe+Xe. The bottom panels show relative difference w.r.t default along with the lower and upper bounds shown in lines and the ratio of statistical uncertainty of the default to its mean value shown as shaded region. The systematic variation is within 0.5--$11\%$ over the entire centrality range except the peripheral centrality for $n=4$ where contribution from track selection dominates. The systematics are dominated by uncertainty in the track efficiency and track selection in central and mid-central region. The uncertainty due to centrality is important in the peripheral region. The systematic uncertainty are generally larger than the statistical uncertainties for all harmonics. 

Figure~\ref{fig:Cov_sys_Xe} also shows (bottom row) the relative difference w.r.t nominal cuts along with the lower and upper bounds shown in lines and the ratio of statistical uncertainty of the default to its mean value shown as shaded region. The quadrature sum of the relative difference between the lower and upper limits  from systematic variation in a given bin is chosen to be the systematic error. The total systematic error are comparable or smaller than the statistical uncertainties for all centralities.

\begin{figure}[htbp]
\centering
\includegraphics[width=0.32\linewidth]{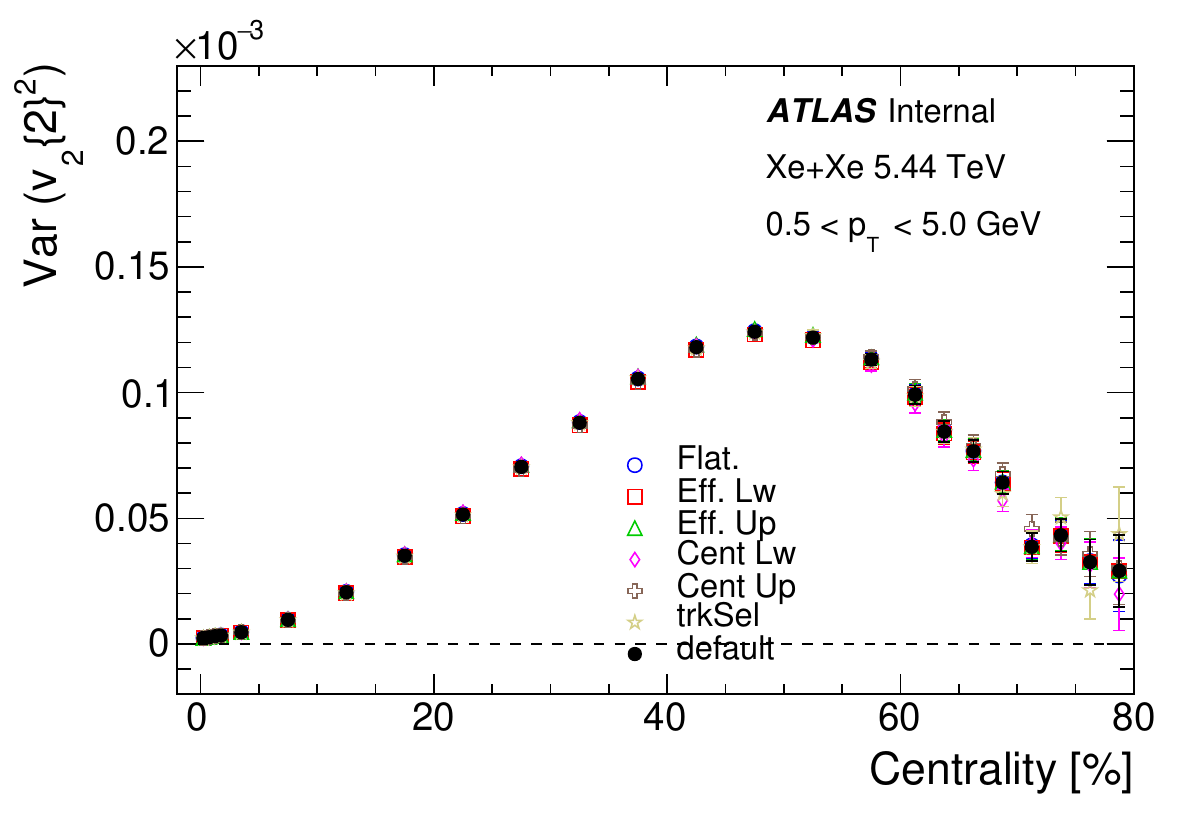}
\includegraphics[width=0.32\linewidth]{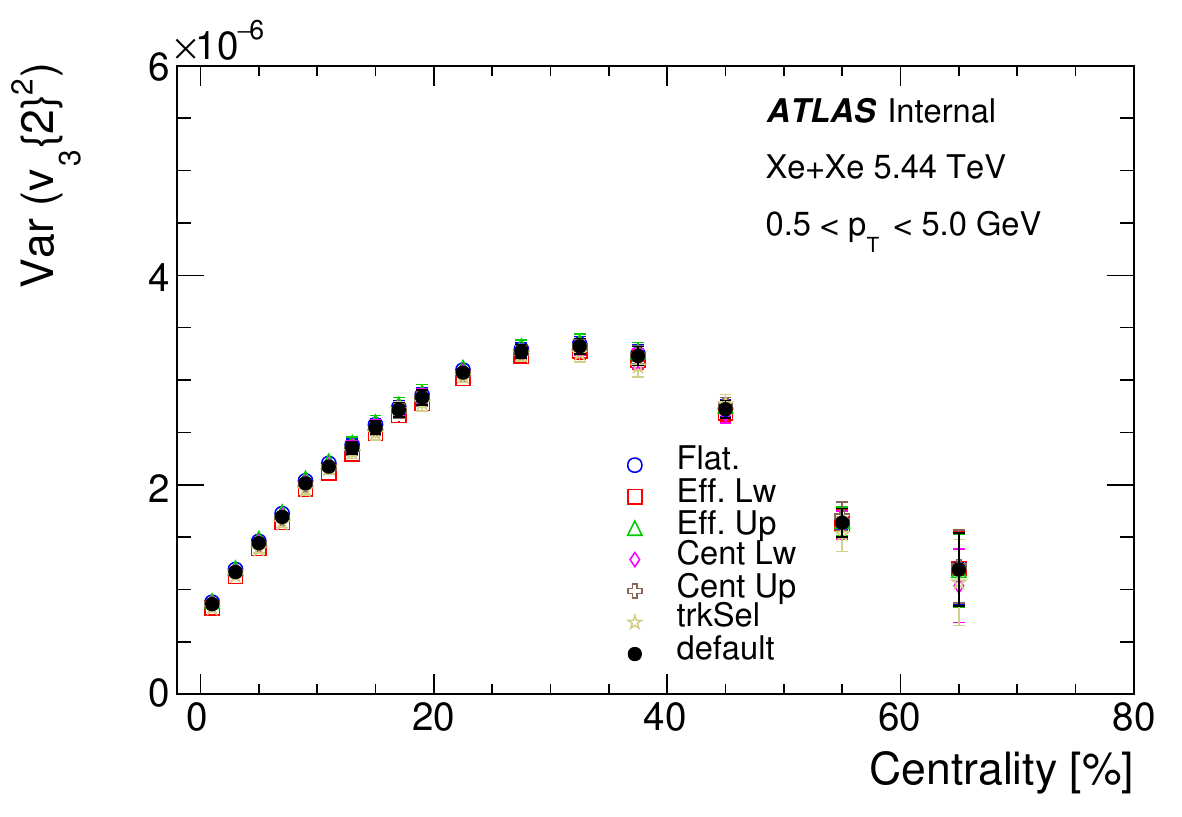}
\includegraphics[width=0.32\linewidth]{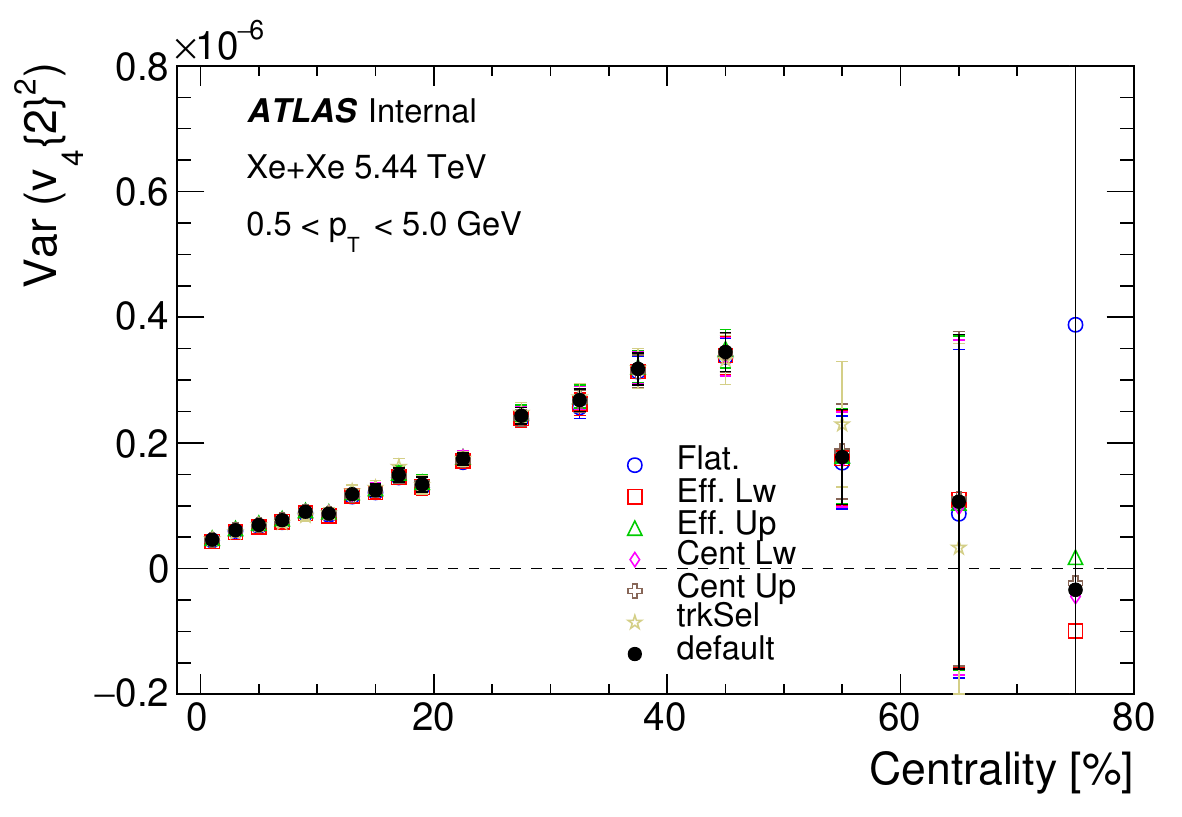}
\includegraphics[width=0.32\linewidth]{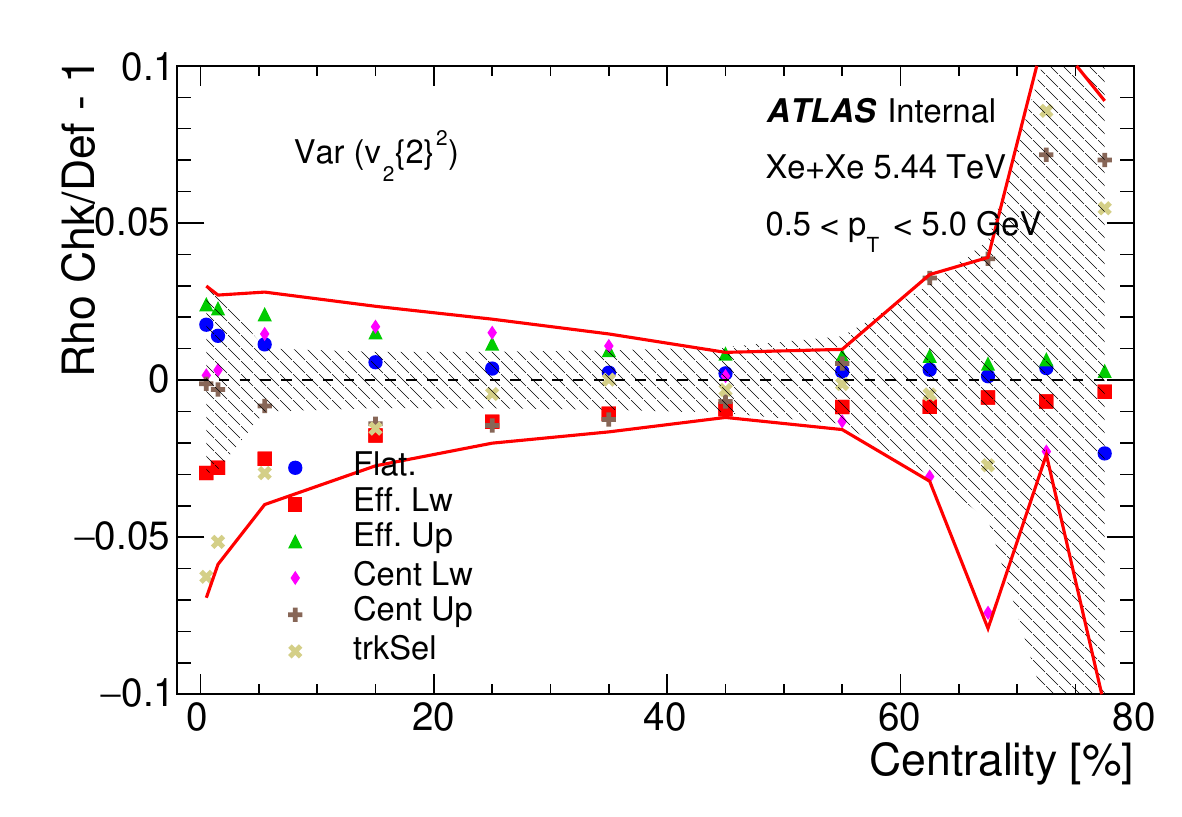}
\includegraphics[width=0.32\linewidth]{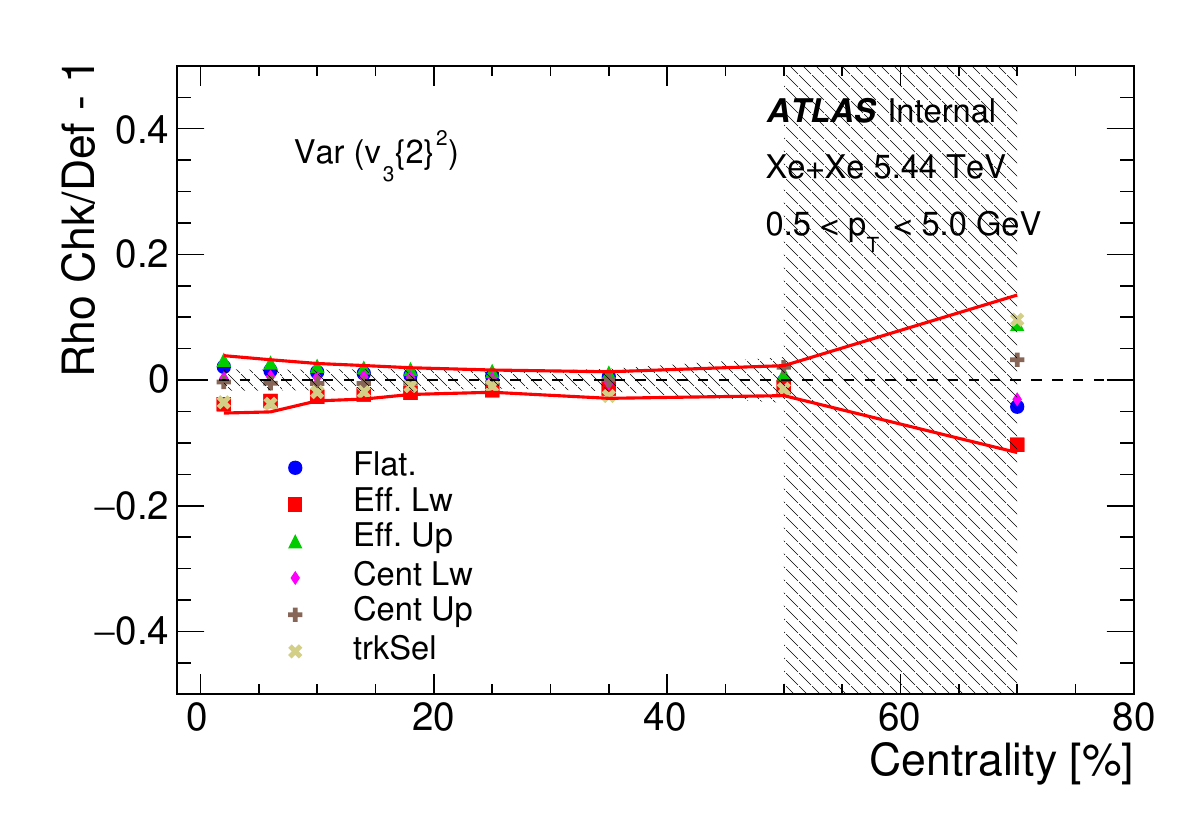}
\includegraphics[width=0.32\linewidth]{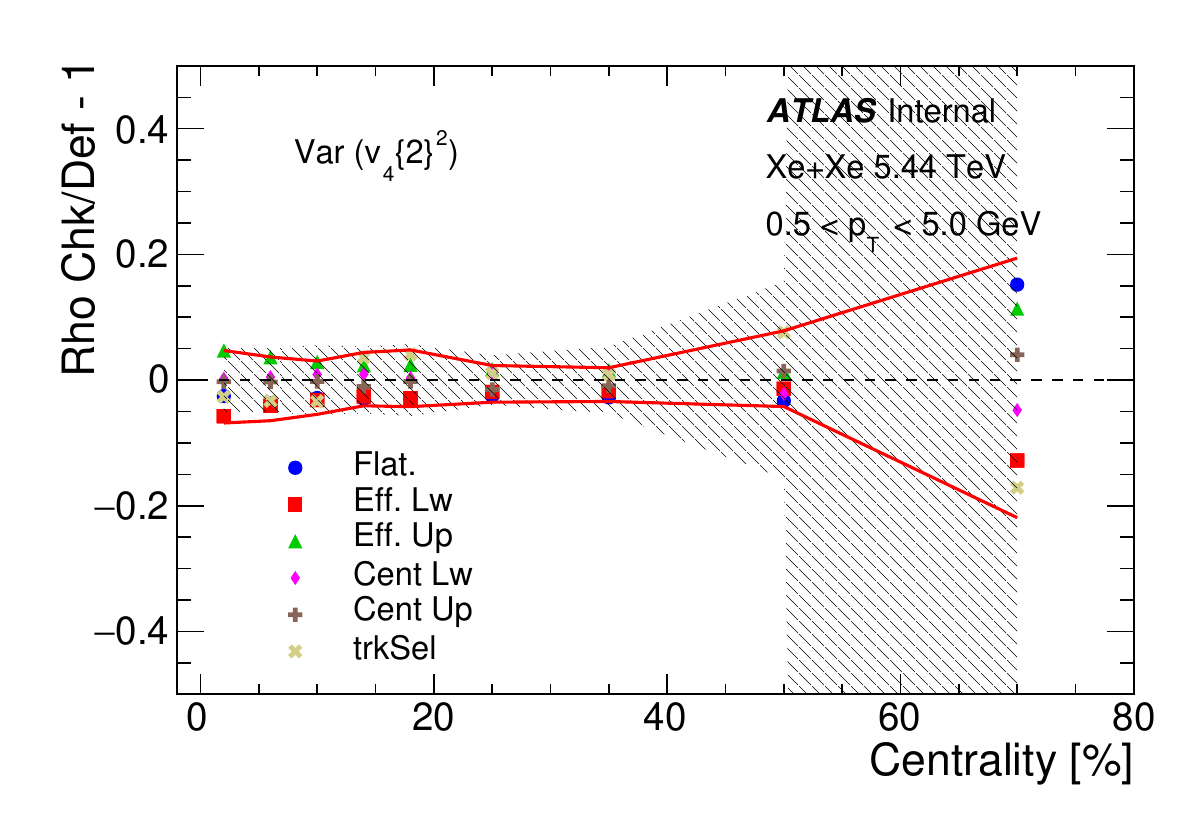}
\caption{(Top row)Comparison of $\mathrm{Var}(v_n\{2\}^2)$ between different systematic checks and default for $n=2$, 3 and 4. The error bars represent statistical uncertainties. (Bottom row) Comparison of the relative difference between systematic checks and default for $n=2$, 3 and 4. The shaded area represents the ratio of statistical uncertainty of the default to its mean value. The lines are the combined systematic uncertainty.}
\label{fig:Var_sys_Xe}
\end{figure}

 Figure~\ref{fig:Var_sys_Xe} show  the systematic checks for $\mathrm{Var}(v_n\{2\}^2)$. The total systematic lies within 8\% for most of the centrality intervals for all three harmonics and in most cases, are either comparable or smaller than the statistical uncertainty..

\begin{figure}[htbp]
\centering
\includegraphics[width=0.4\linewidth]{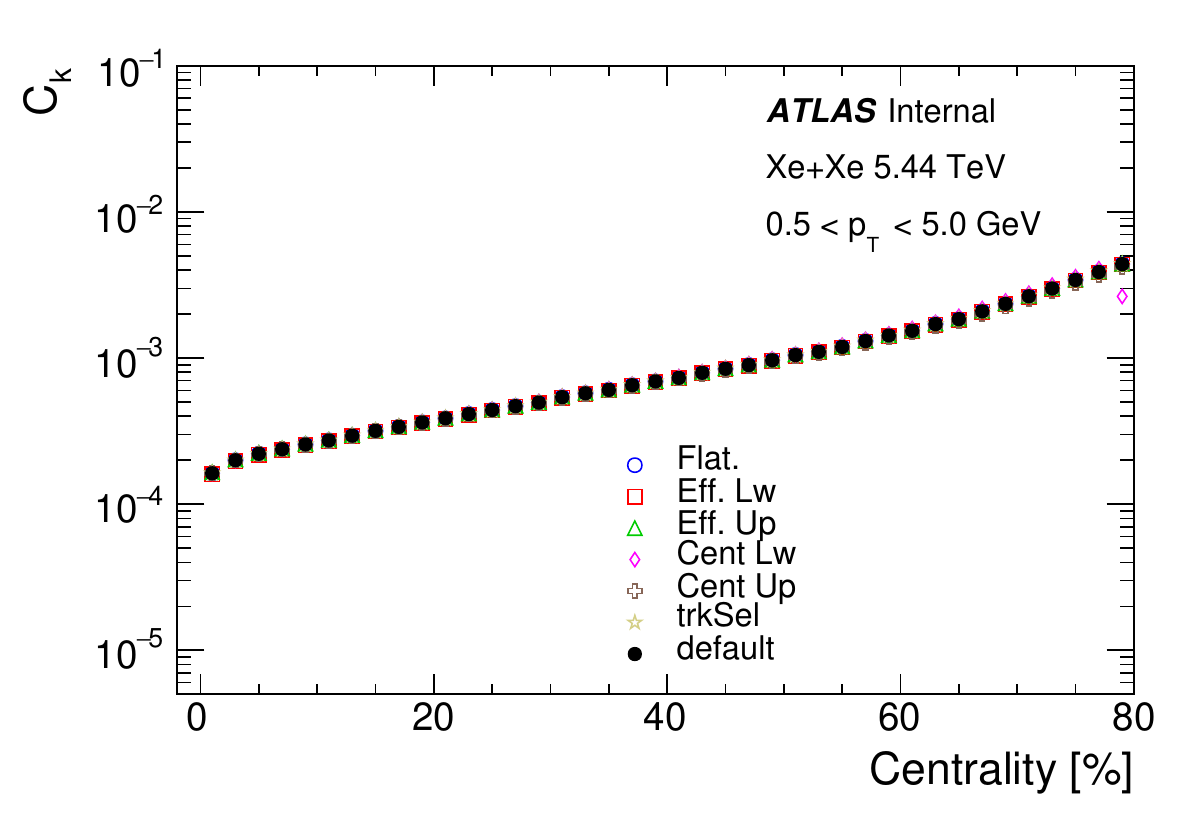}\includegraphics[width=0.4\linewidth]{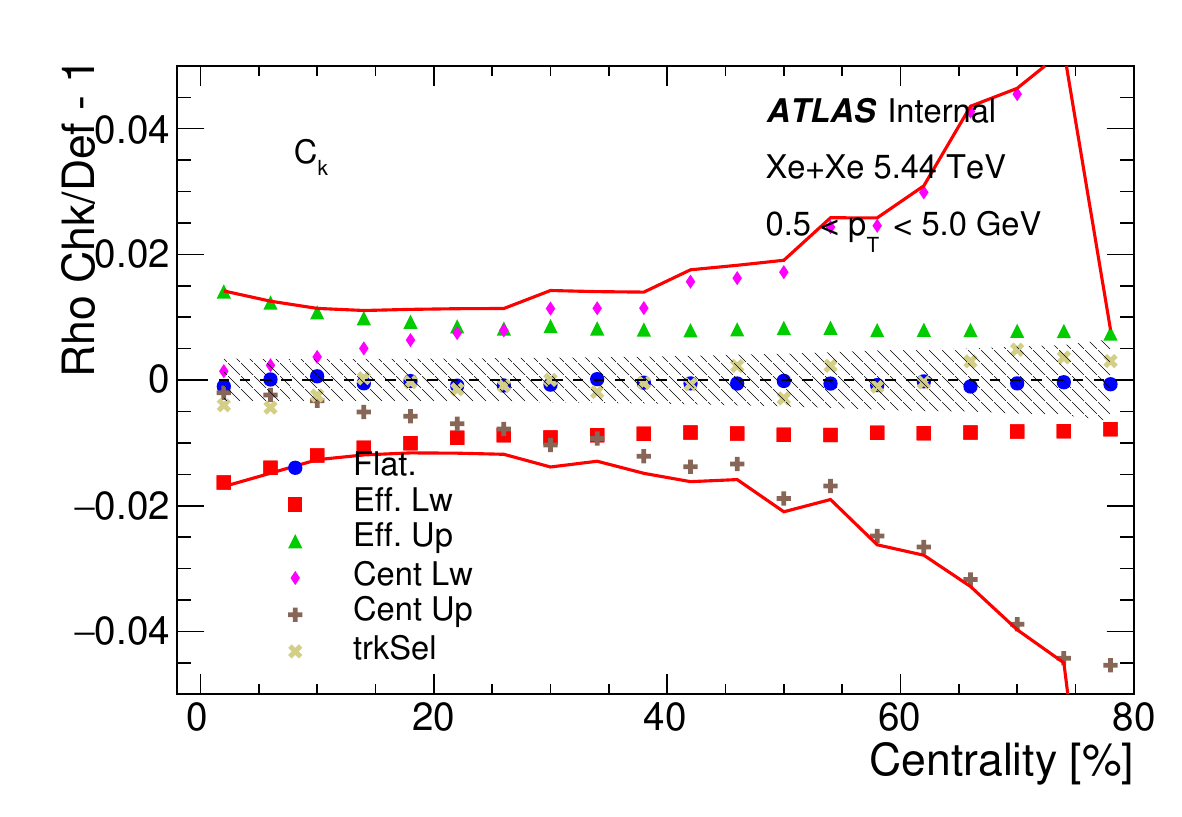}
\caption{(Left panel) Comparison of $[\pT]$ variance $c_k=\lr{\delta\pT\delta\pT}$ between different systematic checks and default. The error bars represent statistical uncertainties. (Right panel) Comparison of the relative difference between systematic checks and default for $n=2$, 3 and 4. The shaded area represents the ratio of statistical uncertainty of the default to its mean value. The lines are the combined systematic uncertainty.}
\label{fig:Ck_sys_Xe}
\end{figure}

Figure~\ref{fig:Ck_sys_Xe} shows the comparison of different systematic checks with the default for variance of $[\pT]$ fluctuations, $c_k$. The systematic uncertainty is $<3\%$ in the full centrality range, and is dominated by tracking efficiency.

\begin{figure}[htbp]
\centering
\includegraphics[width=0.32\linewidth]{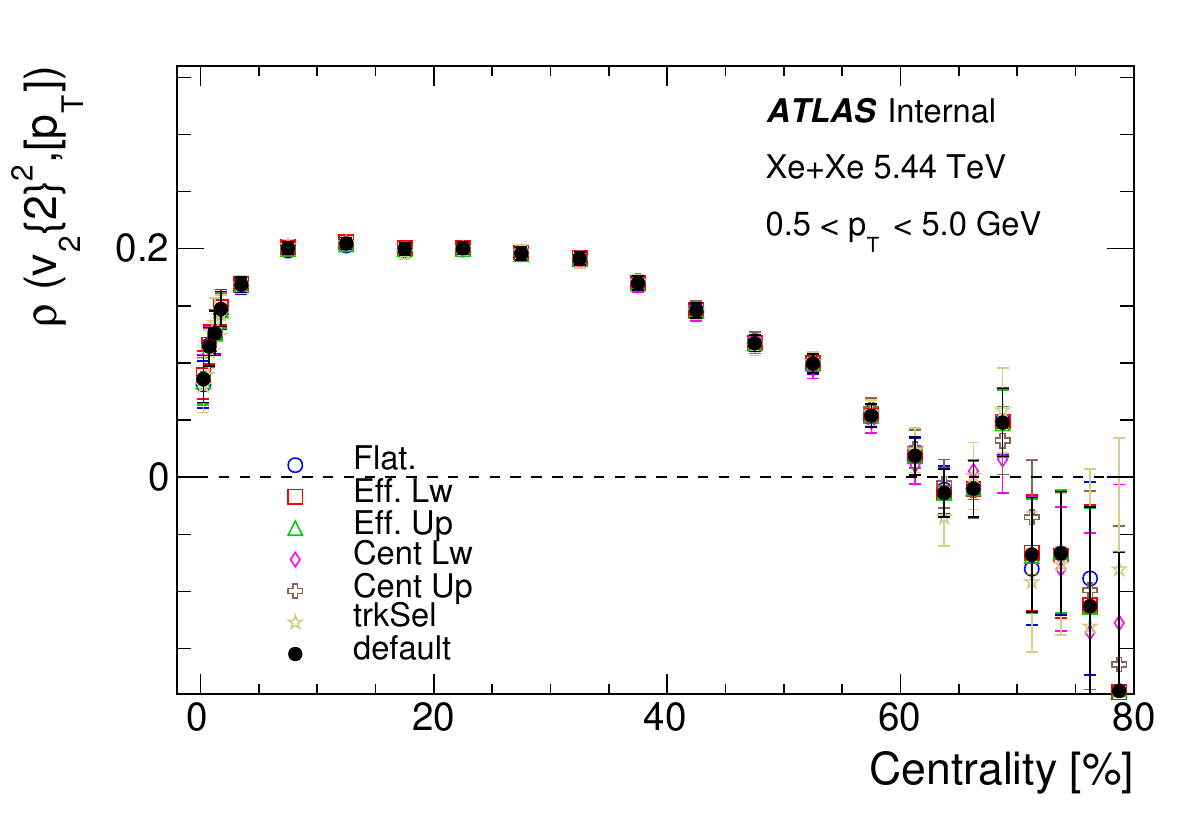}
\includegraphics[width=0.32\linewidth]{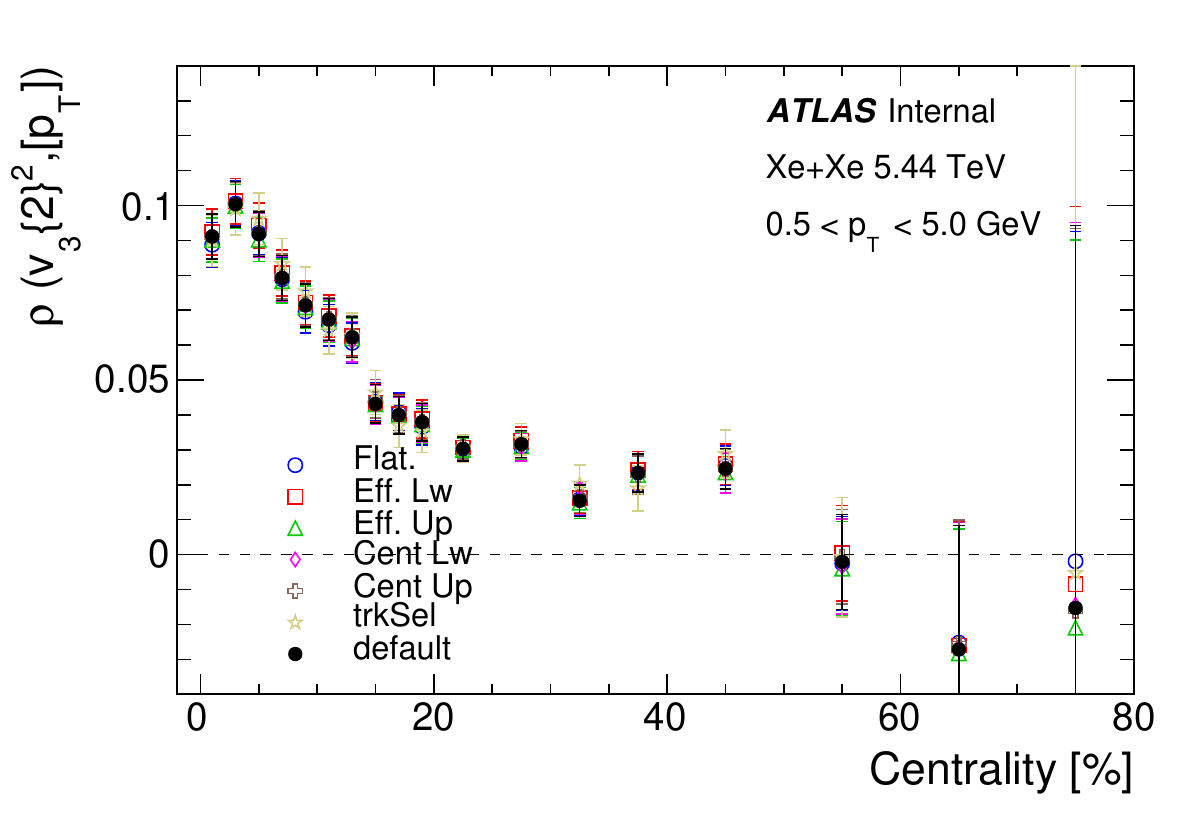}
\includegraphics[width=0.32\linewidth]{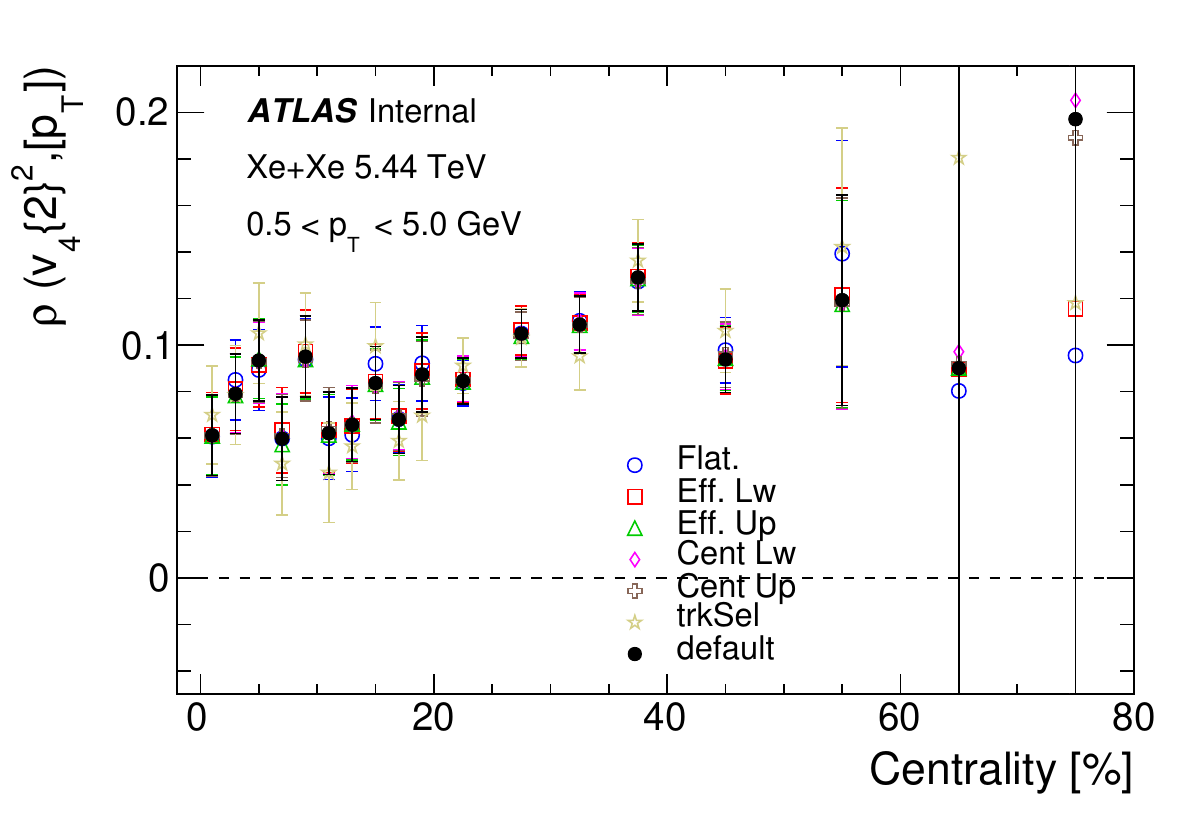}
\includegraphics[width=0.32\linewidth]{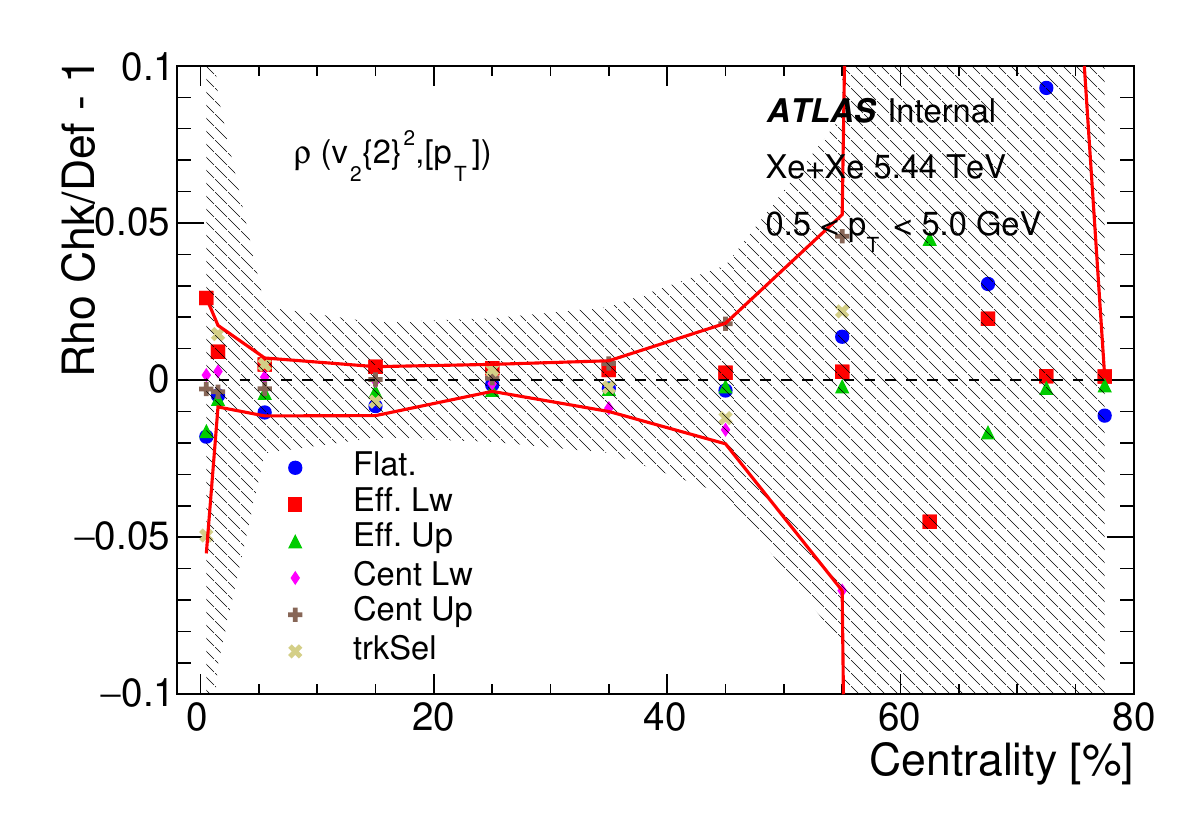}
\includegraphics[width=0.32\linewidth]{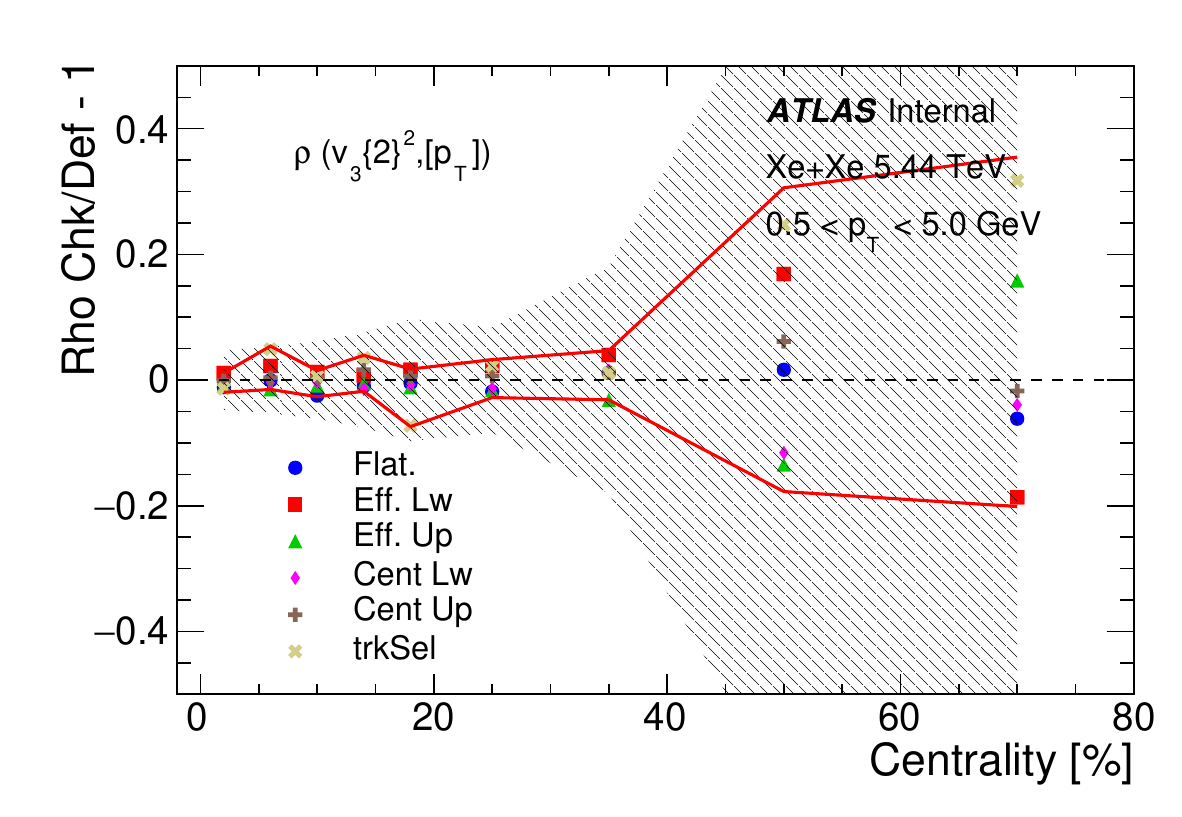}
\includegraphics[width=0.32\linewidth]{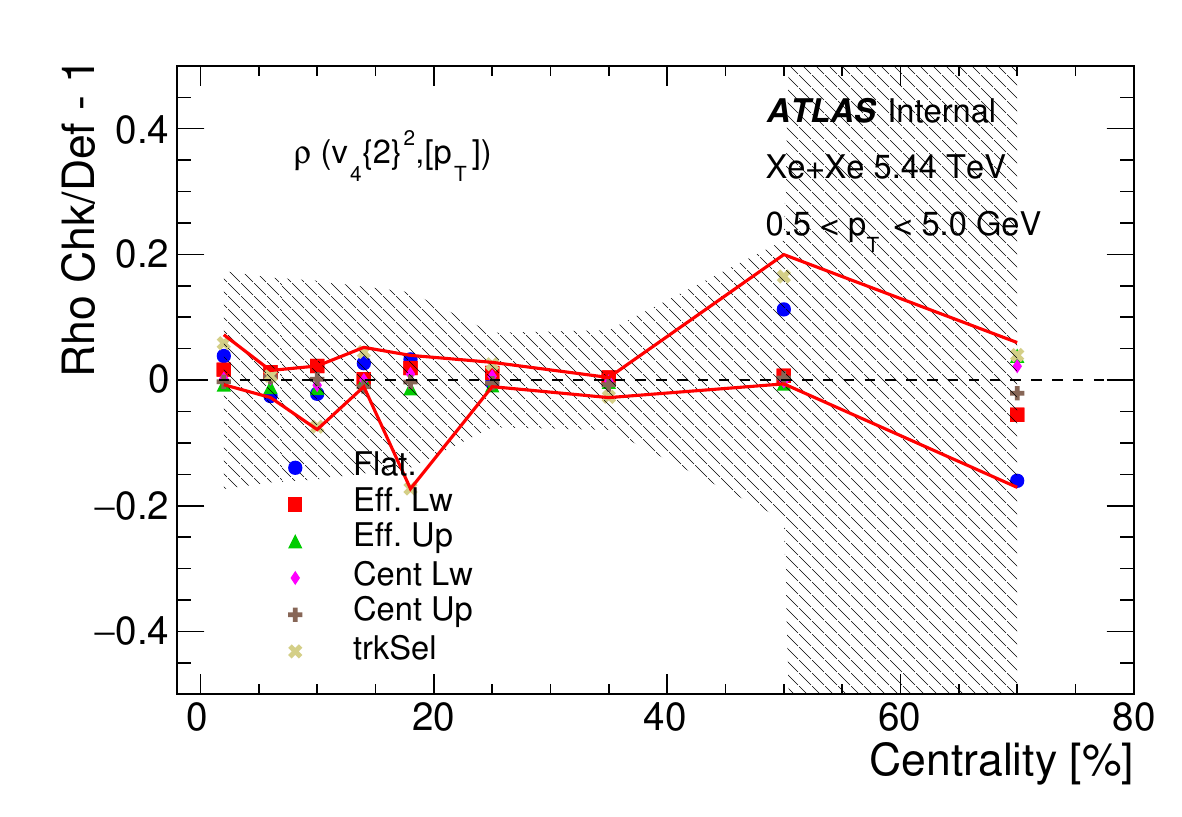}
\caption{(Top row)Comparison of $\rho(v_n\{2\}^2,[\pT])$ between different systematic checks and default for $n=2$, 3 and 4. The error bars represent statistical uncertainties. (Bottom row) Comparison of the relative difference between systematic checks and default for $n=2$, 3 and 4. The shaded area represents the ratio of statistical uncertainty of the default to its mean value. The lines are the combined systematic uncertainty.}
\label{fig:Rho_sys_Xe}
\end{figure}

Figure~\ref{fig:Rho_sys_Xe} shows the comparison of different systematic checks with the default case of $\rho(v_n\{2\}^2,[\pT])$ for $n=2$, 3 and 4 in Xe+Xe. The bottom panels show relative difference w.r.t default along with the lower and upper bounds shown in lines and the ratio of statistical uncertainty of the default to its mean value shown as shaded region. The systematic variation is within $13\%$ in the central and mid-central collisions.

For Xe+Xe, the centrality variation has the minimum contribution to systematics for most centrality intervals for all observables. For $n=2$, the dominant source of systematics arises from efficiency variation for all observables. For $n=3$, 4 the systematic effect coming from track quality variations becomes larger in some cases.  The major source of systematic in measurement of $\rho(v_n\{2\}^2,[\pT])$ arises from the contribution from the $\mathrm{cov}(v_n\{2\}^2,[\pT])$. For $\mathrm{cov}(v_2\{2\}^2,[\pT])$, the systematics are close to 10\% for all centralities. For $\mathrm{cov}(v_3\{2\}^2,[\pT])$ and $\mathrm{cov}(v_4\{2\}^2,[\pT])$, the systematics are within 10\% for central and mid-central events. In the peripheral region, the systematics for $\mathrm{cov}(v_4\{2\}^2,[\pT])$ rises to about 17\% for n=4.

The $\mathrm{Var}(v_n\{2\}^2)_{\mathrm{dyn}}$ systematics is well behaved for all centralities and for all harmonics except for peripheral centralities for n=3,4 where a sharp increase coming from track quality variation can be observed. But this sharp rise is well accommodated within the statistical error for this centrality which is quite large.

\subsubsection{Cross-checks}
There are several cross-checks that were carried out, as discussed here.
\begin{itemize}
\item {Contamination from pileup events}
Figure~\ref{fig:Pileup_sys_Xe} shows the relative contribution from Pileup in each centrality interval for the measurement of $\rho(v_n\{2\}^2,[\pT])$ for $n=2$, 3 and 4 in Xe+Xe collisions. It can be observed that the difference arising from pileup contribution is very close to zero for $n=2$ and 3. For $n=4$, this difference is close to zero for all centrality intervals except in most peripheral where the statistical uncertainty are very large. 
\begin{figure}[htbp]
\centering
\includegraphics[width=0.7\linewidth]{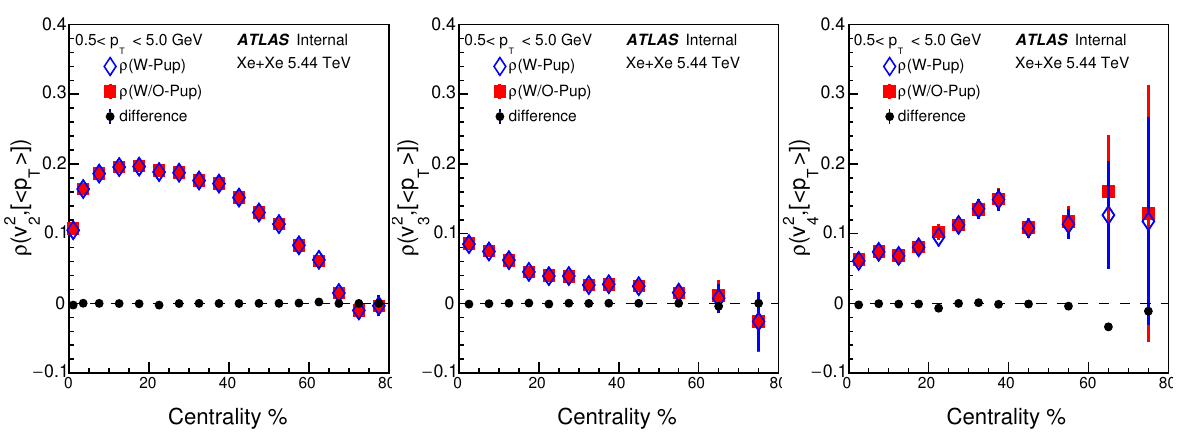}
\caption{Comparison of $\rho(v_n\{2\}^2,[\pT])$ with and without pileup contributions removed for $n=2$,3 and 4 for $\pT$-range of 0.5-5 GeV. For $\rho(v_4\{2\}^2,[\pT])$, the effect of pileup is minimal in all centralities for all harmonics.}
\label{fig:Pileup_sys_Xe}
\end{figure}

\item {Residual Sine terms}

For perfect detector operations, the imaginary part of the $\mathrm{cov}(v_n\{2\}^2,[\pT])$ and as a consequence that of $\rho(v_n\{2\}^2,[\pT])$ should equal to zero. A 'Perfect' flattening of the $q-$vectors as shown in Section~\ref{q-vecRec} should remove any nonzero sine term. Therefore, a systematic check of the Im($\mathrm{cov}(v_n\{2\}^2,[\pT])$) gives an idea of the residual detector non-uniformities and robustness of the analysis method. Figure~\ref{fig:Im_sys_Xe} shows the Imaginary component of $\rho(v_n\{2\}^2,[\pT])$ in each centrality interval along with the measurement of $\rho(v_n\{2\}^2,[\pT])$ for $n=2$, 3 and 4.  Figure~\ref{fig:Im_sys_Xe} indeed shows that the Im($\rho(v_n\{2\}^2,[\pT])$) are consistent with the zero within statistical errors.

\begin{figure}[htbp]
\centering
\includegraphics[width=0.7\linewidth]{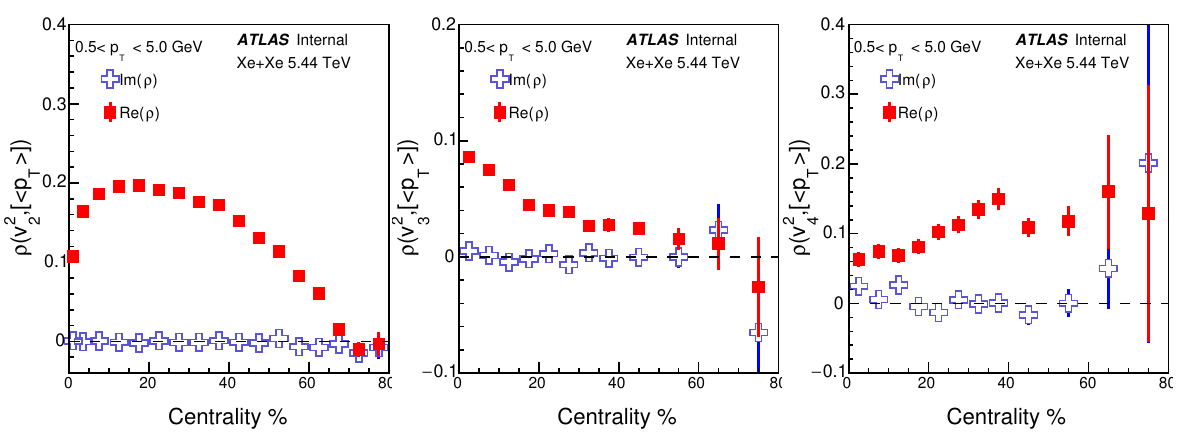}
\caption{Residual Sine term in measurement of $\rho(v_n\{2\}^2,[\pT])$ for $n=2$,3 and 4 for $\pT$-range of 0.5-5 GeV.The residual term is equal to zero within errors for $\rho(v_n\{2\}^2,[\pT])$ for all harmonics.}
\label{fig:Im_sys_Xe}
\end{figure}

\item{HIJING Closure check}

By construction, in the HIJING model the generated flow harmonics are not related either to the geometry of the initial condition of heavy ion collisions or to the transverse momentum fluctuations. Therefore, the model predictions for $\mathrm{Var}(v_n\{2\}^2)_{\mathrm{dyn}}$ and $\mathrm{cov}(v_n\{2\}^2,[\pT])$ are not expected to provide good estimates of these quantities even on qualitative basis. However, one can still study with the HIJING events the effect of non-flow correlations, like jets or resonance decays, on the measured quantities~\cite{Aad:2019fgl}. The single particle $v_n$ has been imposed on the HIJING, but no genuine higher-order flow correlations or $v_n-[\pT]$ correlations are expected.

It was found that much larger fluctuations of $v_{2}^2$ harmonic are present in data than in HIJING MC for all $\pT$ ranges of references particles. Similarly, comparing the covariances $\mathrm{cov}(v_2\{2\}^2,[\pT])$ in Figure~\ref{fig:Hijing_sys_Pb3} one can see that there are negligible covariances in the MC. Similarly, the magnitude of $\mathrm{Var}(v_n\{2\}^2)_{\mathrm{dyn}}$ observed in MC was also negligible. The $c_{k}$ for MC showed a qualitative agreement between data and HIJING predictions although on quantitative level small differences were observed. Although HIJING does not properly describe the investigated quantities we think its still useful to be considered as a validation of the analysis method. In fact it points to the fact that the correlation observed in data does not arise from a trivial upward $\pT$ fluctuation which would naively result in higher observed $v_n$ value and both of these effects are present in MC yet the correlation is effectively absent. We did not carry out a HIJING closure check for Xe+Xe system, although it is  expected the conclusion should be similar.
\begin{figure}[htbp]
\centering
\includegraphics[width=0.4\linewidth]{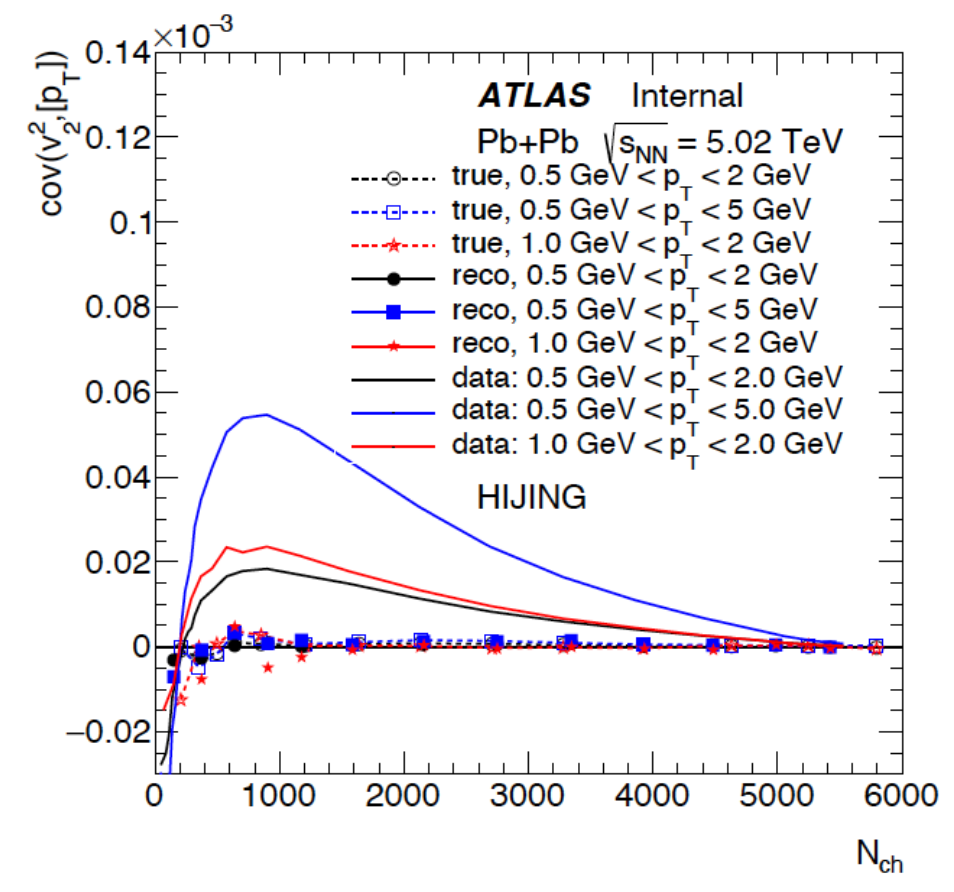}
\caption{  $\mathrm{cov}(v_n\{2\}^2,[\pT])$
 as a function of Nch in 5.02 TeV Pb+Pb collisions for different $\pT$ thresholds.}
\label{fig:Hijing_sys_Pb3}
\end{figure}
\end{itemize}
\setcounter{tocdepth}{2}%
\addtocontents{toc}{\protect\setcounter{tocdepth}{2}}%

\end{document}